%% file: Summer06.tex
\documentclass[12pt,twoside]{article}

\usepackage{amsmath}
\usepackage{graphicx}
\usepackage{epsfig}
\usepackage[figuresright]{rotating}
\usepackage{pstcol}
\usepackage{cite}
\usepackage[dvips]{hyperref}                              
\hypersetup{ pdfborder=0 0 0, urlbordercolor=1 0 0 }      
\usepackage{relsize}                                      
\input{b2charm-symbols}                                
\def\hfurl#1{http://hfag.phys.ntu.edu.tw/b2charm/#1} 
\input{symbols.tex}
\renewcommand{\mysection}[1]{\section[#1]{#1}} 

\begin{document}

\setcounter{page}{1}

\title{\begin{flushleft}
\end{flushleft}
\vskip 20pt
Averages of $b$-hadron Properties at the End of 2006
}
\author{Heavy Flavor Averaging Group (HFAG)\footnote{
The HFAG members involved in producing the averages for the end of 2006
update are:
   E.~Barberio, 
   R.~Bernhard, 
   S.~Blyth, 
   G.~Cavoto,     
   P.~Chang, 
   F.~Di~Lodovico, 
   S.~Eidelman, 
   T.~Gershon,    
   R.~Godang,     
   R.~Harr,       
   H.~Lacker, 
   A.~Limosani, 
   C.-J.~Lin, 
   O.~Long,       
   V.~Luth,    
   S.~Prell, 
   O.~Schneider, 
   J.~Smith, 
   S.~Tosi, 
   K.~Trabelsi,    
   R.~Van~Kooten, 
   C.~Voena, 
   and C.~Weiser. 
  }
 }
\date{April 26, 2007}
\maketitle
\thispagestyle{empty}
\begin{abstract}
This article reports world averages for measurements on $b$-hadron
properties obtained by the Heavy Flavor Averaging Group (HFAG) using the
available results at the end of 2006. In the
averaging, the input parameters used in the various analyses are
adjusted (rescaled) to common values, and all known correlations are
taken into account. The averages include lifetimes, neutral meson mixing
parameters, parameters of semileptonic decays, branching fractions of
$B$ decays to final states with open charm, charmonium and no charm, and
measurements related to \CP asymmetries. 
\end{abstract}

\newpage
\tableofcontents
\newpage

\input{hfag_intro.tex}

\input{methods.tex}

\clearpage
\input{life_mix.tex}

\clearpage
\input{cp_uta.tex}

\clearpage
\input{slbdecays.tex}

\clearpage 
\input{rare.tex}

\clearpage

\input{b2charm}                                        

\clearpage
\input{hfag_summary}

\input{hfag_ref}

\end{document}

%% file: b2charm-symbols.tex
\input{b2charm_summer2006_v009/b2charm_summer2006_v009_symbols.tex}

%% file: b2charm_summer2006_v009/b2charm_summer2006_v009_symbols.tex
\definecolor{hfviolet}{rgb}{0.5,0,0.5}
\definecolor{hflightcyan}{rgb}{0.1,1.0,1.0}

\def\hflmode{Mode}
\def\hflpdg#1{PDG #1}
\def\hflbabar{\mbox{\slshape B\kern-0.1em{\smaller A}\kern-0.1em B\kern-0.1em{\smaller A\kern-0.2em R}}}
\def\hflbelle{\mbox{Belle}}
\def\hflcdf{\mbox{CDF}}

\def\hfld0{\mbox{D0}}
\def\hflavg{Average}

\def\hfpubhotcolor{red}
\def\hfpubcolor{magenta}
\def\hfpuboldcolor{black}

\def\hfprehotcolor{blue}
\def\hfprecolor{cyan}
\def\hfpreoldcolor{hflightcyan}

\def\hfdefcolor{black}
\def\hferrcolor{yellow}
\def\hfsuperceededcolor{hfviolet}
\def\hfwaitingcolor{green}
\def\hfinactivecolor{hfviolet}
\def\hfnoquocolor{hfviolet}
\def\hfdeftext#1{\textcolor{\hfdefcolor}{#1}}
\def\hflabel#1{\textcolor{\hfdefcolor}{$#1$}}
\def\hfavg#1{\textcolor{\hfdefcolor}{$#1$}}
\def\hfnewavg#1{\textcolor{\hfdefcolor}{\boldmath$#1$}}

\def\hfpdg#1{\textcolor{\hfdefcolor}{$#1$}}
\def\hfwaiting#1{\textcolor{\hfwaitingcolor}{$#1$}}
\def\hfpubhot#1{\textcolor{\hfpubhotcolor}{$#1$}}
\def\hfprehot#1{\textcolor{\hfprehotcolor}{$#1$}}
\def\hfwaitingtext#1{\textcolor{\hfwaitingcolor}{#1}}
\def\hfpubhottext#1{\textcolor{\hfpubhotcolor}{#1}}
\def\hfprehottext#1{\textcolor{\hfprehotcolor}{#1}}
\def\hfpub#1{\textcolor{\hfpubcolor}{$#1$}}
\def\hfpre#1{\textcolor{\hfprecolor}{$#1$}}
\def\hfpubold#1{\textcolor{\hfpuboldcolor}{$#1$}}
\def\hfpreold#1{\textcolor{\hfpreoldcolor}{$#1$}}
\def\hfpubtext#1{\textcolor{\hfpubcolor}{#1}}
\def\hfpretext#1{\textcolor{\hfprecolor}{#1}}
\def\hfpuboldtext#1{\textcolor{\hfpuboldcolor}{#1}}
\def\hfpreoldtext#1{\textcolor{\hfpreoldcolor}{#1}}

\def\hfsuperceeded#1{\textcolor{\hfsuperceededcolor}{$#1$}}

\def\hferrortext#1{\textcolor{\hferrcolor}{#1}}
\def\hfsuperceededtext#1{\textcolor{\hfsuperceededcolor}{#1}}
\def\hfinactivetext#1{\textcolor{\hfinactivecolor}{#1}}
\def\hfnoquotext#1{\textcolor{\hfnoquocolor}{#1}}
\def\hfbb#1{#1M $B\bar{B}$ pairs}

\def\hffootnotemark#1{\tiny{$^{#1}$}}
\def\hffootnotetext#1{\tiny{#1}}

\def\hffootitemsep{-0.7}

\def\hfnewp{new particles }
\def\hfdstr{strange D mesons }
\def\hfbary{baryons }
\def\hfjpsi{$J/\psi(1S)$ }
\def\hfochm{charmonium other than $J/\psi(1S)$ }
\def\hfmuld{multiple $D$, $D^{*}$ or $D^{**}$ mesons }
\def\hfsgdx{a single $D^{*}$ or $D^{**}$ meson }
\def\hfsgld{a single D meson }

\def\hfsitebase{http://hfag.phys.ntu.edu.tw/b2charm/}
\def\hfurl#1{\href{\hfsitebase#1}{\hfsitebase#1}} 
\def\hfhref#1#2{\href{\hfsitebase#1}{#2}} 
\def\hftabletype{sidewaystable}
\def\hftableposn{!htbp}
\def\hfaftercapspace{\vspace*{2mm}}

\def\hfnewcaption#1#2#3#4#5#6#7{\caption{ #1 of #2 modes producing #3 #4, #5. The latest version is available at: \hfurl{#6} \label{#7}  }\hfaftercapspace}
\newcommand\hftabletlcell{\rule{0pt}{2.6ex}}  
\newcommand\hftableblcell{\rule[-1.2ex]{0pt}{0pt}}

\def\hfmetadata#1{}      

\def\hfcBR{{\cal{B}}}

%% file: symbols.tex
\newcommand{\mysection}[1]{\section{\boldmath #1}}
\newcommand{\mysubsection}[1]{\subsection[#1]{\boldmath #1}}
\newcommand{\mysubsubsection}[1]{\subsubsection[#1]{\boldmath #1}}
\newcommand{\mysubsubsubsection}[1]{\subsubsubsection{\boldmath #1}}

\newcommand{\lesssim}{\ensuremath{\raise-.5ex\hbox{$\buildrel<\over\sim$}\,}} 

\def\dof{{\rm dof}}

\newcommand\VCKM{{V}}
\newcommand\etacpf{{\eta_f}}
\newcommand\etacp{{\eta}}

\renewcommand\Im{{\rm Im}} 
\renewcommand\Re{{\rm Re}}

\newcommand\Abar{\kern 0.18em\overline{\kern -0.18em A}{}}
\newcommand\Af{A_f}
\newcommand\Abarf{\Abar_f}
\newcommand\Afbar{A_{\bar f}}
\newcommand\Abarfbar{\Abar_{\bar f}}
\newcommand\Acp{{\cal A}}
\newcommand\Adirnoncp{\ensuremath{\langle{\cal A}_{f\bar f}\rangle}\xspace}
\newcommand\mc{\multicolumn}
\newcommand\ph{\phantom}
\newcommand {\cbf}{\ensuremath{{\cal B}}}

\newcommand {\vcb}{\ensuremath{|V_{cb}|}}
\newcommand {\vub}{\ensuremath{|V_{ub}|}}

\newcommand {\Bxclnu}{\ensuremath{\Bb\to X_c\ell\nub}}

\newcommand {\breco}{\ensuremath{B_{reco}}}

\def\Bp      {\ensuremath{B^{+}}}
\def\Bm      {\ensuremath{B^{-}}}
\def\Bz      {\ensuremath{B^{0}}}
\def\Bs      {\ensuremath{B_{s}}}

\newcommand{\BzbDplnu}    {\ensuremath{\bar{B}^{0}\to D^{+}\ell^{-}\nub}}
\newcommand{\BzbDstarlnu} {\ensuremath{\bar{B}^{0}\to D^{*+}\ell^{-}\nub}}

\newcommand {\Bxulnu}{\ensuremath{B \to X_u \ell \bar{\nu}}\hbox{ }}
\newcommand {\Bxsg}{\ensuremath{B \to X_s \gamma}\hbox{ }}

\usepackage{relsize}

\def\beq{\begin{equation}}
\def\eeq#1{\label{#1}\end{equation}}
\def\eeqn{\end{equation}}

\def\beqa{\begin{eqnarray}}
\def\eeqa#1{\label{#1}\end{eqnarray}}
\def\eeqan{\end{eqnarray}}

\let\bar=\overbar

\def\etal{{\it et al.}}
\def\ie{{\it i.e.}}
\def\eg{{\it e.g.}}
\def\etc{{\it etc.}}
\def\cf{{\it cf.}}

\def\Dslash{\ensuremath{\not{\hbox{\kern-4pt $D$}}}\xspace}
\def\dslash{\not{\hbox{\kern-2pt $\del$}}}

\def\BR{\mbox{\rm BR}}
\def\ee{e^+e^-}

\def\msb{{\bar{\ssstyle M \kern -1pt S}}}

\RequirePackage{xspace}

\usepackage{relsize}
\def\babar{\mbox{\slshape B\kern-0.1em{\smaller A}\kern-0.1em
    B\kern-0.1em{\smaller A\kern-0.2em R}}\xspace}
\def\belle{\mbox{\normalfont Belle}\xspace}
\newcommand{\dzero}{D\O\xspace}

\def\ee         {\ensuremath{e^-e^-}\xspace}

\def\ellp       {\ensuremath{\ell^+}\xspace}

\def\nub        {\ensuremath{\overline{\nu}}\xspace}

\def\nub        {\ensuremath{\overline{\nu}}\xspace}

\def\nul        {\ensuremath{\nu_\ell}\xspace}

\def\Z      {\ensuremath{Z^0}\xspace}

\def\ubar  {\ensuremath{\overline u}\xspace}

\def\sbar  {\ensuremath{\overline s}\xspace}

\def\b  {\ensuremath{b}\xspace}

\def\t  {\ensuremath{t}\xspace}

\def\piz   {\ensuremath{\pi^0}\xspace}

\def\pip   {\ensuremath{\pi^+}\xspace}
\def\pim   {\ensuremath{\pi^-}\xspace}

\def\etapr {\ensuremath{\eta^{\prime}}\xspace}

\def\Kbar  {\kern 0.2em\overline{\kern -0.2em K}{}\xspace}

\def\Kpm   {\ensuremath{K^\pm}\xspace}
\def\Kmp   {\ensuremath{K^\mp}\xspace}
\def\Kp    {\ensuremath{K^+}\xspace}
\def\Km    {\ensuremath{K^-}\xspace}
\def\KS    {\ensuremath{K^0_{\scriptscriptstyle S}}\xspace} 
\def\KL    {\ensuremath{K^0_{\scriptscriptstyle L}}\xspace}

\def\Kstar   {\ensuremath{K^*}\xspace}

\def\Kstarpm   {\ensuremath{K^{*\pm}}\xspace}
\def\Kstarmp   {\ensuremath{K^{*\mp}}\xspace}
\def\Kz   {\ensuremath{K^0}\xspace}
\def\Kzb   {\ensuremath{\Kbar^0}\xspace}
\def\KzKzb {\ensuremath{K^0 \kern -0.16em \Kzb}\xspace}

\def\Dz    {\ensuremath{D^0}\xspace}
\def\Dbar  {\kern 0.2em\overline{\kern -0.2em D}{}\xspace}

\def\Dzb   {\ensuremath{\Dbar^0}\xspace}
\def\DzDzb {\ensuremath{D^0 {\kern -0.16em \Dzb}}\xspace}
\def\Dp    {\ensuremath{D^+}\xspace}

\def\Dstar   {\ensuremath{D^*}\xspace}

\def\Dstarz  {\ensuremath{D^{*0}}\xspace}

\def\Dstarp  {\ensuremath{D^{*+}}}
\def\Dstarm  {\ensuremath{D^{*-}}}
\def\DorDstar   {\ensuremath{D^{(*)}}\xspace}
\def\DorDstarz  {\ensuremath{D^{(*)0}}\xspace}
\def\DorDstarzb {\ensuremath{\Dbar^{(*)0}}\xspace}

\def\Ds    {\ensuremath{D^+_s}\xspace}

\def\Bz    {\ensuremath{B^0}\xspace}
\def\B     {\ensuremath{B}\xspace}
\def\Bbar  {\kern 0.18em\overline{\kern -0.18em B}{}\xspace}
\def\Bb    {\ensuremath{\Bbar}\xspace}
\def\Bzb   {\ensuremath{\Bbar^0}\xspace}
\def\Bu    {\ensuremath{B^+}\xspace}

\def\Bpm   {\ensuremath{B^\pm}\xspace}
\def\Bmp   {\ensuremath{B^\mp}\xspace}
\def\Bs    {\ensuremath{B_s}\xspace}
\def\Bsb   {\ensuremath{\Bbar_s}\xspace}
\def\BB    {\ensuremath{B\Bbar}\xspace} 
\def\BzBzb {\ensuremath{B^0 {\kern -0.16em \Bzb}}\xspace}

\def\jpsi  {\ensuremath{{J\mskip -3mu/\mskip -2mu\psi\mskip 2mu}}\xspace}

\mathchardef\Upsilon="7107
\def\Y#1S{\ensuremath{\Upsilon{(#1S)}}\xspace}

\def\FourS {\Y4S}

\mathchardef\Deltares="7101
\mathchardef\Xi="7104
\mathchardef\Lambda="7103
\mathchardef\Sigma="7106
\mathchardef\Omega="710A
\def\Deltabar   {\kern 0.25em\overline{\kern -0.25em \Deltares}{}\xspace}
\def\Lbar {\kern 0.2em\overline{\kern -0.2em\Lambda\kern 0.05em}\kern-0.05em{}\xspace}
\def\Sigbar{\kern 0.2em\overline{\kern -0.2em \Sigma}{}\xspace}
\def\Xibar{\kern 0.2em\overline{\kern -0.2em \Xi}{}\xspace}
\def\Obar{\kern 0.2em\overline{\kern -0.2em \Omega}{}\xspace}
\def\Nbar{\kern 0.2em\overline{\kern -0.2em N}{}\xspace}
\def\Xb{\kern 0.2em\overline{\kern -0.2em X}{}}

\def\BR{{\ensuremath{\cal B}}}

\newcommand{\tev}{\ensuremath{\mathrm{Te\kern -0.1em V}}\xspace}
\newcommand{\gev}{\ensuremath{\mathrm{Ge\kern -0.1em V}}\xspace}
\newcommand{\mev}{\ensuremath{\mathrm{Me\kern -0.1em V}}\xspace}
\newcommand{\kev}{\ensuremath{\mathrm{ke\kern -0.1em V}}\xspace}
\newcommand{\ev}{\ensuremath{\mathrm{e\kern -0.1em V}}\xspace}
\newcommand{\gevc}{\ensuremath{{\mathrm{Ge\kern -0.1em V\!/}c}}\xspace}
\newcommand{\mevc}{\ensuremath{{\mathrm{Me\kern -0.1em V\!/}c}}\xspace}
\newcommand{\gevcc}{\ensuremath{{\mathrm{Ge\kern -0.1em V\!/}c^2}}\xspace}
\newcommand{\mevcc}{\ensuremath{{\mathrm{Me\kern -0.1em V\!/}c^2}}\xspace}

\def\invfb   {\ensuremath{\mbox{\,fb}^{-1}}\xspace}
\def\mus  {\ensuremath{\rm \,\mus}\xspace}

\def\ps   {\ensuremath{\rm \,ps}\xspace}

\def\mus        {\ensuremath{\,\mu{\rm s}}\xspace}    
\def\ps         {\ensuremath{{\rm \,ps}}\xspace}  
\def\gsim{{~\raise.15em\hbox{$>$}\kern-.85em
          \lower.35em\hbox{$\sim$}~}\xspace}
\def\lsim{{~\raise.15em\hbox{$<$}\kern-.85em
          \lower.35em\hbox{$\sim$}~}\xspace}

\def\CP                 {\ensuremath{C\!P}\xspace}
\def\CPT                {\ensuremath{C\!PT}\xspace}

\def\pep2{PEP-II}

\def\rhobar {\ensuremath{\overline{\rho}}\xspace}
\def\etabar {\ensuremath{\overline{\eta}}\xspace}

\def\Vub  {\ensuremath{|V_{ub}|}\xspace}

\def\stwob{\ensuremath{\sin\! 2 \beta   }\xspace}

\def\deltamd{\ensuremath{{\rm \Delta}m_d}\xspace}

\xspace

\newcommand{\epjc}      [1]  {{Eur.\ Phys.\ Jour.\ {\bf C}{\bf #1}}}

\newcommand{\npb}       [1]  {{Nucl.\ Phys.\ {\bf B{\bf #1}}}}

\newcommand{\plb}       [1]  {{Phys.\ Lett.\ B~{\bf #1}}}   

\def\jetset74   {\mbox{\tt Jetset \hspace{-0.5em}7.\hspace{-0.2em}4}}

\newcommand{\aerr}[4]   {\mbox{${{#1}^{+ #2}_{- #3}\pm #4}$}}
\newcommand{\berr}[4]   {\mbox{${{#1}\pm #2^{+ #3}_{- #4}}$}}
\newcommand{\cerr}[3]   {\mbox{${{#1}^{+ #2}_{- #3}}$}}
\newcommand{\aerrsy}[5] {\mbox{${{#1}^{+ #2 + #4}_{- #3 - #5}}$}}

\newcommand{\berrsyt}[6] {\mbox{${{#1}\pm #2^{+ #3 + #5}_{- #4 - #6}}$}}
\newcommand{\err}[3]   {\mbox{${{#1}\pm{#2}\pm{#3}}$}}
\newcommand{\errede}[5] {\mbox{${#1}\pm{#2}^{+ #3}_{- #4}\pm{#5}$}}
\newcommand{\nodata}{$$}
\newcommand{\vs}{\mbox{$vs.$}}
\def\etapr{{\eta^{\prime}}}

\def\sgline{\noalign{\vskip 0.10truecm\hrule\vskip 0.10truecm}}
\def\sglinespt{\noalign{\vskip 0.05truecm\hrule}}
\def\sglinespb{\noalign{\hrule\vskip 0.05truecm}}
\newcommand{\prl}      [1]  {{Phys.\ Rev.\ Lett.\ {\bf #1}}}
\newcommand{\prd}      [1]  {{Phys.\ Rev.\ D~{\bf #1}}}

\newcommand{\kz}    {\mbox{$K^0$}}

\newcommand{\RPP}{}

%% file: hfag_intro.tex

\mysection{Introduction}
\label{sec:intro}

Flavor dynamics is an important element in understanding the nature of
particle physics.  The accurate knowledge of properties of heavy flavor
hadrons, especially $b$ hadrons, plays an essential role for
determination of the Cabibbo-Kobayashi-Maskawa (CKM)
matrix~\cite{ref_ckm}. Since asymmetric-energy $e^+e^-$ $B$ factories
started their operation, the size of available $B$ meson samples has
dramatically increased and the accuracies of measurements have been
improved. Tevatron experiments also started to provide rich results on
$B$ hadron decays with increased Run~II data samples. 
 
The Heavy Flavor Averaging Group (HFAG) has been formed in 2002,
continuing the activities of LEP Heavy Flavor Steering
group~\cite{LEPHFS}, to provide averages for measurements
of $b$-flavor related quantities. 

The HFAG is currently organized into five subgroups:
\begin{itemize}
\item the ``Lifetime and Mixing'' group provides averages for $b$-hadron
  lifetimes, $b$-hadron fractions in $\Upsilon(4S)$ decay and high
  energy collisions, and various parameters in $B^0$ and $B_s^0$
  oscillation (mixing);

\item the ``Semileptonic $B$ Decays'' group provides averages
for inclusive and exclusive $B$-decay branching fractions, and
estimates of the CKM matrix elements $|V_{cb}|$ and $|V_{ub}|$;

\item the ``$\CP(t)$ and Unitarity Triangle Angles'' group provides
  averages for time-dependent $\CP$ asymmetry parameters and angles of
  the $B$ unitarity triangle;

\item the ``Rare Decays'' group provides averages of branching fractions
  and their asymmetries between $B$ and $\bar B$ for charmless mesonic, 
  radiative, leptonic, and baryonic $B$ decays;

\item the ``$B$ to Charm Decays'' group provides averages of branching
  fractions for $B$ decays to final states involving open charm mesons
  or charmonium.
\end{itemize}

The first two subgroups continue the activities from LEP working groups
with some reorganization (merging four groups into two groups). The
latter three groups have been newly formed to provide averages for
results which are available from $B$ factory experiments. The five HFAG
subgroups consist of representatives and contact persons from the
experimental groups: \babar, \belle, CDF, CLEO, \dzero, and LEP. As of
the writing of this document a sixth HFAG subgroup which deals with the
physics of charm hadrons is being initiated. First averages from the
Charm subgroup are available for the Winter 2007 conferences on the HFAG
web page and will be included in the next update of this document.

This article is an update of the {\it End of 2005} HFAG
document~\cite{hfag_hepex_endof2005}, and we report the world averages
using the available results at the end of 2006. All results that are 
publicly available, including recent preliminary results, are used in
the averages.  We do not use preliminary results which remain
unpublished for a long time or for which no publication is planned. 
Close contacts have been established between representatives from
the experiments and members of different subgroups in charge of the
averages, to ensure that the data are prepared in a form suitable
for combinations.  

We do not scale the error of an average (as is presently done by the
Particle Data Group~\cite{Yao2006PDG}) in case $\chi^2/\dof > 1$,
where $\dof$ is the number of degrees of freedom in the average
calculation. In such a case, we examine the systematics of each
measurement and try to understand them. Unless we find possible
systematic discrepancies between the measurements, we do not make any
special treatment for the calculated error. We provide the confidence
level of the fit as an indicator for the consistency of the measurements
included in the average. We attach a warning message in case that some
special treatment was necessary to calculate the average or the
approximation used in the average calculation may not be good enough 
(\eg, Gaussian error is used in averaging although the likelihood 
indicates non-Gaussian behavior).

Section~\ref{sec:method} describes the methodology for calculating
averages for various quantities used by the HFAG. In the averaging, the
input parameters used in the various analyses are adjusted (rescaled) to
common values, and, where possible, known correlations are taken into
account. The general philosophy and tools for calculations of averages
are presented. 
Sections~\ref{sec:life_mix}--\ref{sec:BtoCharm} describe the averaging of 
the quantities from each of the subgroups mentioned above.
A brief summary of the averages described in this article is given in
Sec.~\ref{sec:summary}.   

 The complete listing of averages and plots described in this article
are also available on the HFAG web page:
 
 {\tt http://www.slac.stanford.edu/xorg/hfag } and 

 {\tt http://belle.kek.jp/mirror/hfag } (KEK mirror site).

%% file: methods.tex
\section{Methodology } \label{sec:method} 

The general averaging problem that HFAG faces is to combine the
information provided by different measurements of the same parameter,
to obtain our best estimate of the parameter's value and
uncertainty. The methodology described here focuses on the problems of
combining measurements performed with different systematic assumptions
and with potentially-correlated systematic uncertainties. Our methodology
relies on the close involvement of the people performing the
measurements in the averaging process.

Consider two hypothetical measurements of a parameter $x$, which might
be summarized as
\begin{align*}
x &= x_1 \pm \delta x_1 \pm \Delta x_{1,1} \pm \Delta x_{2,1} \ldots \\
x &= x_2 \pm \delta x_2 \pm \Delta x_{1,2} \pm \Delta x_{2,2} \ldots
\; ,
\end{align*}
where the $\delta x_k$ are statistical uncertainties, and
the $\Delta x_{i,k}$ are contributions to the systematic
uncertainty. One popular approach is to combine statistical and
systematic uncertainties in quadrature
\begin{align*}
x &= x_1 \pm \left(\delta x_1 \oplus \Delta x_{1,1} \oplus \Delta
x_{2,1} \oplus \ldots\right) \\
x &= x_2 \pm \left(\delta x_2 \oplus \Delta x_{1,2} \oplus \Delta
x_{2,2} \oplus \ldots\right)
\end{align*}
and then perform a weighted average of $x_1$ and $x_2$, using their
combined uncertainties, as if they were independent. This approach
suffers from two potential problems that we attempt to address. First,
the values of the $x_k$ may have been obtained using different
systematic assumptions. For example, different values of the \Bz
lifetime may have been assumed in separate measurements of the
oscillation frequency $\deltamd$. The second potential problem is that
some contributions of the systematic uncertainty may be correlated
between experiments. For example, separate measurements of $\deltamd$
may both depend on an assumed Monte-Carlo branching fraction used to
model a common background.

The problems mentioned above are related since, ideally, any quantity $y_i$
that $x_k$ depends on has a corresponding contribution $\Delta x_{i,k}$ to the
systematic error which reflects the uncertainty $\Delta y_i$ on $y_i$
itself. We assume that this is the case, and use the values of $y_i$ and
$\Delta y_i$ assumed by each measurement explicitly in our
averaging (we refer to these values as $y_{i,k}$ and $\Delta y_{i,k}$
below). Furthermore, since we do not lump all the systematics
together,
we require that each measurement used in an average have a consistent
definition of the various contributions to the systematic uncertainty.
Different analyses often use different decompositions of their systematic
uncertainties, so achieving consistent definitions for any potentially
correlated contributions requires close coordination between HFAG and
the experiments. In some cases, a group of
systematic uncertainties must be lumped to obtain a coarser
description that is consistent between measurements. Systematic uncertainties
that are uncorrelated with any other sources of uncertainty appearing
in an average are lumped with the statistical error, so that the only
systematic uncertainties treated explicitly are those that are
correlated with at least one other measurement via a consistently-defined
external parameter $y_i$. When asymmetric statistical or systematic
uncertainties are quoted, we symmetrize them since our combination
method implicitly assumes parabolic likelihoods for each measurement.

\begin{figure}
\begin{center}
\includegraphics[width=6.0in]{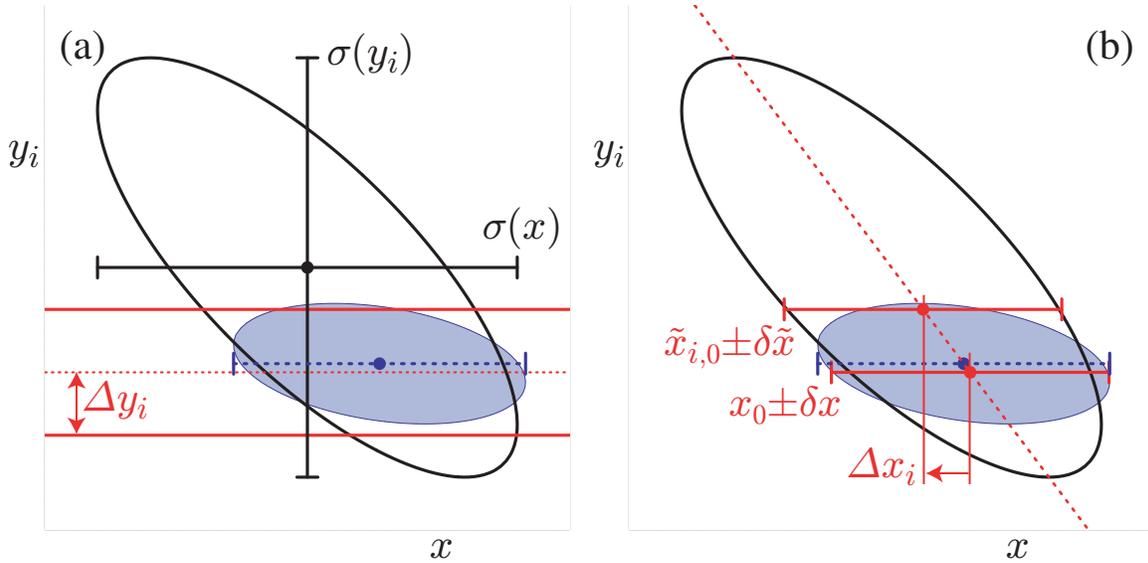}
\end{center}
\caption{The left-hand plot, (a), compares the 68\% confidence-level
  contours of a
  hypothetical measurement's unconstrained (large ellipse) and
  constrained (filled ellipse) likelihoods, using the Gaussian
  constraint on $y_i$ represented by the horizontal band. The solid
  error bars represent the statistical uncertainties, $\sigma(x)$ and
  $\sigma(y_i)$, of the unconstrained likelihood. The dashed
  error bar shows the statistical error on $x$ from a
  constrained simultaneous fit to $x$ and $y_i$. The right-hand plot,
  (b), illustrates the method described in the text of performing fits
  to $x$ only with $y_i$ fixed at different values. The dashed
  diagonal line between these fit results has the slope
  $\rho(x,y_i)\sigma(y_i)/\sigma(x)$ in the limit of a parabolic
  unconstrained likelihood. The result of the constrained simultaneous
  fit from (a) is shown as a dashed error bar on $x$.}
\label{fig:singlefit}
\end{figure}

The fact that a measurement of $x$ is sensitive to the value of $y_i$
indicates that, in principle, the data used to measure $x$ could
equally-well be used for a simultaneous measurement of $x$ and $y_i$, as
illustrated by the large contour in Fig.~\ref{fig:singlefit}(a) for a hypothetical
measurement. However, we often have an external constraint $\Delta
y_i$ on the value of $y_i$ (represented by the horizontal band in
Fig.~\ref{fig:singlefit}(a)) that is more precise than the constraint
$\sigma(y_i)$ from
our data alone. Ideally, in such cases we would perform a simultaneous
fit to $x$ and $y_i$, including the external constraint, obtaining the
filled $(x,y)$ contour and corresponding dashed one-dimensional estimate of
$x$ shown in Fig.~\ref{fig:singlefit}(a). Throughout, we assume that
the external constraint $\Delta y_i$ on $y_i$ is Gaussian.

In practice, the added technical complexity of a constrained fit with
extra free parameters is not justified by the small increase in
sensitivity, as long as the external constraints $\Delta y_i$ are
sufficiently precise when compared with the sensitivities $\sigma(y_i)$
to each $y_i$ of the data alone. Instead, the usual procedure adopted
by the experiments is to perform a baseline fit with all $y_i$ fixed
to nominal values $y_{i,0}$, obtaining $x = x_0 \pm \delta
x$. This baseline fit neglects the uncertainty due to $\Delta y_i$, but
this error can be mostly recovered by repeating the fit separately for
each external parameter $y_i$ with its value fixed at $y_i = y_{i,0} +
\Delta y_i$ to obtain $x = \tilde{x}_{i,0} \pm \delta\tilde{x}$, as
illustrated in Fig.~\ref{fig:singlefit}(b). The absolute shift,
$|\tilde{x}_{i,0} - x_0|$, in the central value of $x$ is what the
experiments usually quote as their systematic uncertainty $\Delta x_i$
on $x$ due to the unknown value of $y_i$. Our procedure requires that
we know not only the magnitude of this shift but also its sign. In the
limit that the unconstrained data is represented by a parabolic
likelihood, the signed shift is given by
\begin{equation}
\Delta x_i = \rho(x,y_i)\frac{\sigma(x)}{\sigma(y_i)}\,\Delta y_i \;,
\end{equation}
where $\sigma(x)$ and $\rho(x,y_i)$ are the statistical uncertainty on
$x$ and the correlation between $x$ and
$y_i$ in the unconstrained data.
While our procedure is not
equivalent to the constrained fit with extra parameters, it yields (in
the limit of a parabolic unconstrained likelihood) a central value
$x_0$ that agrees 
to ${\cal O}(\Delta y_i/\sigma(y_i))^2$ and an uncertainty $\delta x
\oplus \Delta x_i$ that agrees to ${\cal O}(\Delta y_i/\sigma(y_i))^4$.

In order to combine two or more measurements that share systematics
due to the same external parameters $y_i$, we would ideally perform a
constrained simultaneous fit of all data samples to obtain values of
$x$ and each $y_i$, being careful to only apply the constraint on each
$y_i$ once. This is not practical since we generally do not have
sufficient information to reconstruct the unconstrained likelihoods
corresponding to each measurement. Instead, we perform the two-step
approximate procedure described below.

\begin{figure}
\begin{center}
\includegraphics[width=6.0in]{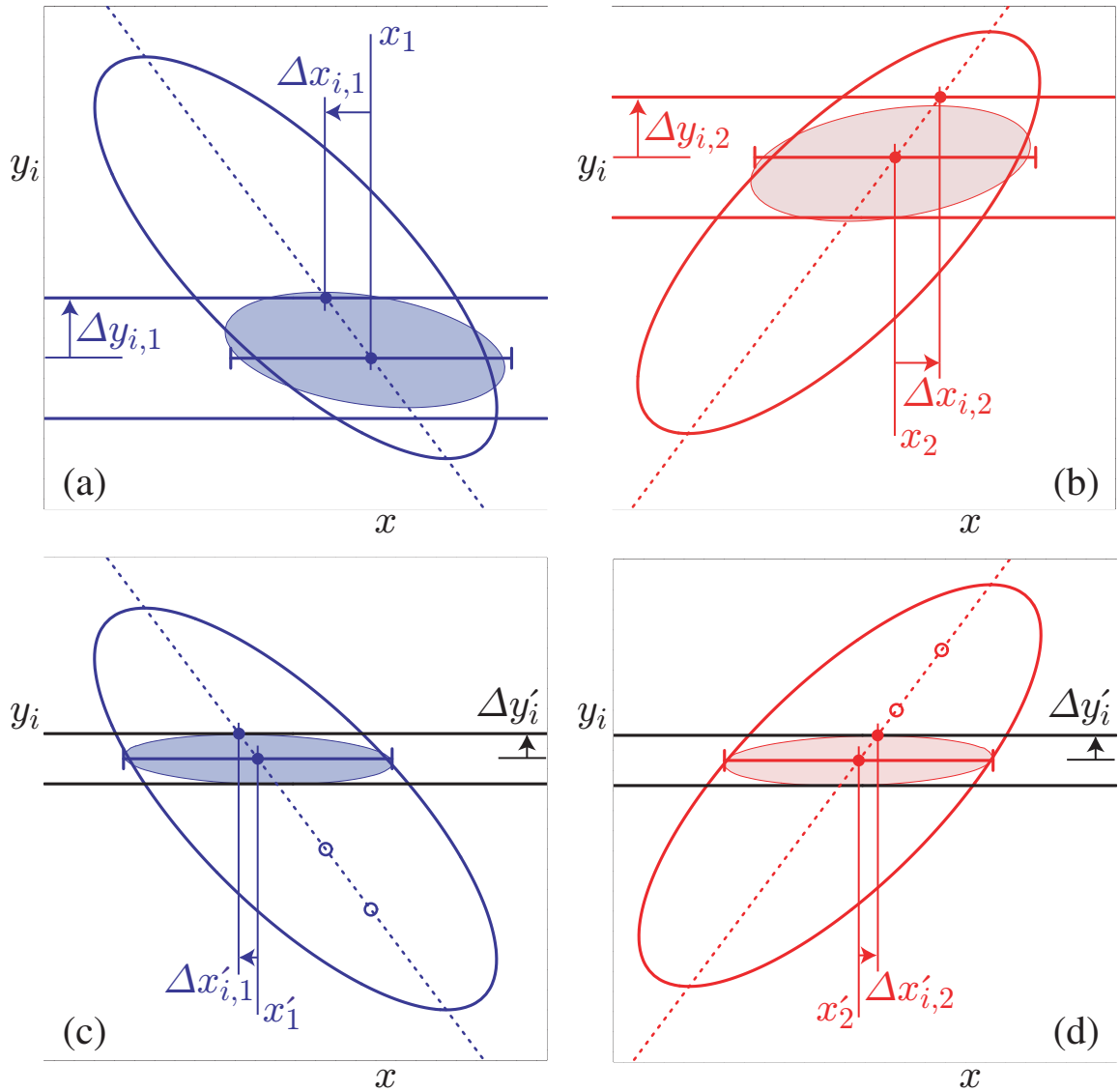}
\end{center}
\caption{The upper plots, (a) and (b), show examples of two individual
  measurements to be combined. The large ellipses represent their
  unconstrained likelihoods, and the filled ellipses represent their
  constrained likelihoods. Horizontal bands indicate the different
  assumptions about the value and uncertainty of $y_i$ used by each
  measurement. The error bars show the results of the approximate
  method described in the text for obtaining $x$ by performing fits
  with $y_i$ fixed to different values. The lower plots, (c) and (d),
  illustrate the adjustments to accommodate updated and consistent
  knowledge of $y_i$ described in the text. Hollow circles mark the
  central values of the unadjusted fits to $x$ with $y$ fixed, which
  determine the dashed line used to obtain the adjusted values. }
\label{fig:multifit}
\end{figure}

Figs.~\ref{fig:multifit}(a,b) illustrate two
statistically-independent measurements, $x_1 \pm (\delta x_1 \oplus
\Delta x_{i,1})$ and $x_2\pm(\delta x_i\oplus \Delta x_{i,2})$, of the same
hypothetical quantity $x$ (for simplicity, we only show the
contribution of a single correlated systematic due to an external
parameter $y_i$). As our knowledge of the external parameters $y_i$
evolves, it is natural that the different measurements of $x$ will
assume different nominal values and ranges for each $y_i$. The first
step of our procedure is to adjust the values of each measurement to
reflect the current best knowledge of the values $y_i'$ and ranges
$\Delta y_i'$ of the external parameters $y_i$, as illustrated in
Figs.~\ref{fig:multifit}(c,b). We adjust the
central values $x_k$ and correlated systematic uncertainties $\Delta
x_{i,k}$ linearly for each measurement (indexed by $k$) and each
external parameter (indexed by $i$):
\begin{align}
x_k' &= x_k + \sum_i\,\frac{\Delta x_{i,k}}{\Delta y_{i,k}}
\left(y_i'-y_{i,k}\right)\\
\Delta x_{i,k}'&= \Delta x_{i,k}\cdot \frac{\Delta y_i'}{\Delta
  y_{i,k}} \; .
\end{align}
This procedure is exact in the limit that the unconstrained
likelihoods of each measurement is parabolic.

\begin{figure}
\begin{center}
\includegraphics[width=3.5in]{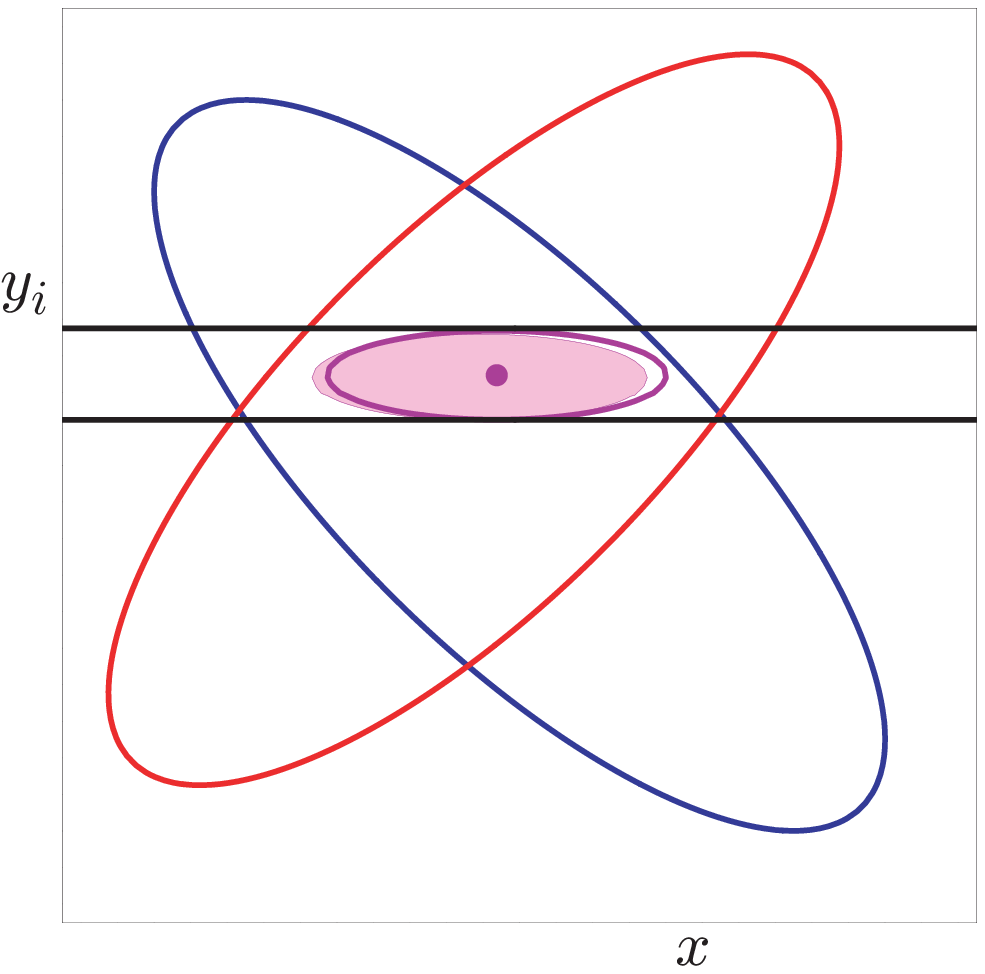}
\end{center}
\caption{An illustration of the combination of two hypothetical
  measurements of $x$ using the method described in the text. The
  ellipses represent the unconstrained likelihoods of each measurement
  and the horizontal band represents the latest knowledge about $y_i$ that
  is used to adjust the individual measurements. The filled small
  ellipse shows the result of the exact method using ${\cal
  L}_{\text{comb}}$ and the hollow small ellipse and dot show the
  result of the approximate method using $\chi^2_{\text{comb}}$.}
\label{fig:fit12}
\end{figure}

The second step of our procedure is to combine the adjusted
measurements, $x_k'\pm (\delta x_k\oplus \Delta x_{k,1}'\oplus \Delta
x_{k,2}'\oplus\ldots)$ using the chi-square 
\begin{equation}
\chi^2_{\text{comb}}(x,y_1,y_2,\ldots) \equiv \sum_k\,
\frac{1}{\delta x_k^2}\left[
x_k' - \left(x + \sum_i\,(y_i-y_i')\frac{\Delta x_{i,k}'}{\Delta y_i'}\right)
\right]^2 + \sum_i\,
\left(\frac{y_i - y_i'}{\Delta y_i'}\right)^2 \; ,
\end{equation}
and then minimize this $\chi^2$ to obtain the best values of $x$ and
$y_i$ and their uncertainties, as illustrated in
Fig.~\ref{fig:fit12}. Although this method determines new values for
the $y_i$, we do not report them since the $\Delta x_{i,k}$ reported
by each experiment are generally not intended for this purpose (for
example, they may represent a conservative upper limit rather than a
true reflection of a 68\% confidence level).

For comparison, the exact method we would
perform if we had the unconstrained likelihoods ${\cal L}_k(x,y_1,y_2,\ldots)$
available for each
measurement is to minimize the simultaneous constrained likelihood
\begin{equation}
{\cal L}_{\text{comb}}(x,y_1,y_2,\ldots) \equiv \prod_k\,{\cal
  L}_k(x,y_1,y_2,\ldots)\,\prod_{i}\,{\cal 
  L}_i(y_i) \; ,
\end{equation}
with an independent Gaussian external constraint on each $y_i$
\begin{equation}
{\cal L}_i(y_i) \equiv \exp\left[-\frac{1}{2}\,\left(\frac{y_i-y_i'}{\Delta
 y_i'}\right)^2\right] \; .
\end{equation}
The results of this exact method are illustrated by the filled ellipses
in Figs.~\ref{fig:fit12}(a,b), and agree with our method in the limit that
each ${\cal L}_k$ is parabolic and that each $\Delta
y_i' \ll \sigma(y_i)$. In the case of a non-parabolic unconstrained
likelihood, experiments would have to provide a description of ${\cal
  L}_k$ itself to allow an improved combination. In the case of some
$\sigma(y_i)\simeq \Delta y_i'$, experiments are advised to perform a
simultaneous measurement of both $x$ and $y$ so that their data will
improve the world knowledge about $y$. 

 The algorithm described above is used as a default in the averages
reported in the following sections.  For some cases, somewhat simplified
or more complex algorithms are used and noted in the corresponding 
sections. Some examples for extensions of the standard method for extracting
averages are given here. These include the case where measurement errors
depend on the measured value, i.e. are relative errors, unknown
correlation coefficients and the breakdown of error sources.

For measurements with Gaussian errors, the usual estimator for the
average of a set of measurements is obtained by minimizing the following
$\chi^2$:
\begin{equation}
\chi^2(t) = \sum_i^N \frac{\left(y_i-t\right)^2}{\sigma^2_i} ,
\label{eq:chi2t}
\end{equation}
where $y_i$ is the measured value for input $i$ and $\sigma_i^2$ is the
variance of the distribution from which $y_i$ was drawn.  The value $\hat{t}$
of $t$ at minimum $\chi^2$ is our estimator for the average.  (This
discussion is given for independent measurements for the sake of
simplicity; the generalization to correlated measurements is
straightforward, and has been used when averaging results.) 
The true $\sigma_i$ are unknown but typically the error as assigned by the
experiment $\sigma_i^{\mathrm{raw}}$ is used as an estimator for it.
Caution is advised,
however, in the case where $\sigma_i^{\mathrm{raw}}$
depends on the value measured for $y_i$. Examples of this include
an uncertainty in any multiplicative factor (like
an acceptance) that enters the determination of $y_i$, i.e. the $\sqrt{N}$
dependence of Poisson statistics, where $y_i \propto N$
and $\sigma_i \propto \sqrt{N}$.
Failing to account for this type of
dependence when averaging leads to a biased average.
Biases in the average can be avoided (or at least reduced)
by minimizing the following
$\chi^2$:
\begin{equation}
\chi^2(t) = \sum_i^N \frac{\left(y_i-t\right)^2}{\sigma^2_i(\hat{t})} .
\label{eq:chi2that}
\end{equation}
In the above $\sigma_i(\hat{t})$ is the uncertainty
assigned to input $i$ that includes the assumed dependence of the
stated error on the value measured.  As an example, consider 
a pure acceptance error, for which
$\sigma_i(\hat{t}) = (\hat{t} / y_i)\times\sigma_i^{\mathrm{raw}}$ .
It is easily verified that solving Eq.~\ref{eq:chi2that} 
leads to the correct behavior, namely
$$ 
\hat{t} = \frac{\sum_i^N y_i^3/(\sigma_i^{\mathrm{raw}})^2}{\sum_i^N y_i^2/(\sigma_i^{\mathrm{raw}})^2},
$$
i.e. weighting by the inverse square of the 
fractional uncertainty, $\sigma_i^{\mathrm{raw}}/y_i$.

It is sometimes difficult to assess the dependence of $\sigma_i^{\mathrm{raw}}$ on
$\hat{t}$ from the errors quoted by experiments.  

Another issue that needs careful treatment is the question of correlation
among different measurements, e.g. due to using the same theory for
calculating acceptances.  A common practice is to set the correlation
coefficient to unity to indicate full correlation.  However, this is
not a ``conservative'' thing to do, and can in fact lead to a significantly
underestimated uncertainty on the average.  In the absence of
better information, the most conservative choice of correlation coefficient
between two measurements $i$ and $j$
is the one that maximizes the uncertainty on $\hat{t}$
due to that pair of measurements:
\begin{equation}
\sigma_{\hat{t}(i,j)}^2 = \frac{\sigma_i^2\,\sigma_j^2\,(1-\rho_{ij}^2)}
   {\sigma_i^2 + \sigma_j^2 - 2\,\rho_{ij}\,\sigma_i\,\sigma_j} ,
\label{eq:correlij}
\end{equation}
namely
\begin{equation}
\rho_{ij} = \mathrm{min}\left(\frac{\sigma_i}{\sigma_j},\frac{\sigma_j}{\sigma_i}\right) ,
\label{eq:correlrho}
\end{equation}
which corresponds to setting $\sigma_{\hat{t}(i,j)}^2=\mathrm{min}(\sigma_i^2,\sigma_j^2)$.
Setting $\rho_{ij}=1$ when $\sigma_i\ne\sigma_j$ can lead to a significant
underestimate of the uncertainty on $\hat{t}$, as can be seen
from Eq.~\ref{eq:correlij}.

Finally, a note on the breakdown of the error sources
contributing to the overall uncertainty on the average.
The overall covariance matrix is constructed from a number of
individual sources, e.g.
$\mathbf{V} = \mathbf{V_{stat}+V_{sys}+V_{th}}$.
The variance on the average $\hat{t}$ can be written
\begin{eqnarray}
\sigma^2_{\hat{t}} 
 &=& 
\frac{ \sum_{i,j}\left(\mathbf{V^{-1}}\, 
\mathbf{[V_{stat}+V_{sys}+V_{th}]}\, \mathbf{V^{-1}}\right)_{ij}}
{\left(\sum_{i,j} V^{-1}_{ij}\right)^2}
= \sigma^2_{stat} + \sigma^2_{sys} + \sigma^2_{th} .
\end{eqnarray}
Written in this form, one can readily determine the 
contribution of each source of uncertainty to the overall uncertainty
on the average.  This breakdown of the uncertainties is used 
in the following sections.

Following the prescription described above, the central values and
errors are rescaled to a common set of input parameters in the averaging
procedures, according to the dependency on any of these input parameters.
We try to use the most up-to-date values for these common inputs and 
the same values among the HFAG subgroups. For the parameters whose
averages are produced by the HFAG, we use the updated values in the
current update cycle.  For other external parameters, we use the most
recent PDG values. 

  The parameters and values used in this update cycle are listed in
each subgroup section.

%% file: life_mix.tex
%
%
%
%

%

\input{life_mix_definitions}

\input{life_mix_averages}


\mysection{\b-hadron production fractions, lifetimes and mixing parameters}
\labs{life_mix}


Quantities such as \b-hadron production fractions, \b-hadron lifetimes, 
and neutral \B-meson oscillation frequencies have been studied
for many years at high-energy colliders, namely at LEP and SLC 
(\ee colliders at $\sqrt{s}=m_{\particle{Z}}$) as well as at the 
first version of the Tevatron
(\particle{p\bar{p}} collider at $\sqrt{s}=1.8\TeV$). In the last few years, 
precise measurements of the \Bd and \Bu lifetimes, as well as of the 
\Bd oscillation frequency, have also been performed at the 
asymmetric \B factories, KEKB and PEPII
(\ee colliders at $\sqrt{s}=m_{\Ups}$).
In most cases, these basic quantities, although interesting by themselves,
can now be seen as necessary ingredients for the more complicated and 
refined analyses being currently performed at the asymmetric \B factories
and at the upgraded Tevatron ($\sqrt{s}=1.96\TeV$),
in particular the time-dependent \CP asymmetry measurements.
It is therefore important that the best experimental
values of these quantities continue to be kept up-to-date and improved. 
Recently new measurements of \Bs mixing parameters have been 
performed at the Tevatron, with similar or sometimes better 
precision than the ``old'' measurements of \Bd mixing parameters. 

In several cases, the averages presented in this chapter are 
needed and used as input for the results given in the subsequent chapters. 
Within this chapter, some averages need the knowledge of other 
averages in a circular way. This coupling, which appears through the 
\b-hadron fractions whenever inclusive or semi-exclusive measurements 
have to be considered, has been reduced significantly in the last years 
with increasingly precise exclusive measurements becoming available. 
To cope with this circularity,
a rather involved averaging procedure had been developed, in the framework 
of the former LEP Heavy Flavour Steering Group. This is still in use now
(details can be found in~\cite{LEPHFS}), 
although simplifications can be envisaged in the future when even more 
precise exclusive measurements become available. 

\mysubsection{\b-hadron production fractions}
\labs{fractions}
 
We consider here the relative fractions of the different \b-hadron 
species found in an unbiased sample of weakly-decaying \b hadrons 
produced under some specific conditions. The knowledge of these fractions
is useful to characterize the signal composition in inclusive \b-hadron 
analyses, or to predict the background composition in exclusive analyses.
Many analyses in \B physics need these fractions as input. We distinguish 
here the following three conditions: \Ups decays, \Upsfive decays, and 
high-energy collisions (including \Z decays). 

\mysubsubsection{\b-hadron production fractions in \Ups decays}
\labs{fraction_Ups4S}

Only pairs of the two lightest (charged and neutral) \B mesons 
can be produced in \Ups decays, 
and it is enough to determine the following branching 
fractions:
\begin{eqnarray}
f^{+-} & = & \Gamma(\Ups \to \particle{B^+B^-})/
             \Gamma_{\rm tot}(\Ups)  \,, \\
f^{00} & = & \Gamma(\Ups \to \particle{B^0\bar{B}^0})/
             \Gamma_{\rm tot}(\Ups) \,.
\end{eqnarray}
In practice, most analyses measure their ratio
\begin{equation}
R^{+-/00} = f^{+-}/f^{00} = \Gamma(\Ups \to \particle{B^+B^-})/
             \Gamma(\Ups \to \particle{B^0\bar{B}^0}) \,,
\end{equation}
which is easier to access experimentally.
Since an inclusive (but separate) reconstruction of 
\Bu and \Bd is difficult, specific exclusive decay modes, 
${\Bu} \to x^+$ and ${\Bd} \to x^0$, are usually considered to perform 
a measurement of $R^{+-/00}$, whenever they can be related by 
isospin symmetry (for example \particle{\Bu \to J/\psi K^+} and 
\particle{\Bd \to J/\psi K^0}).
Under the assumption that $\Gamma(\Bu \to x^+) = \Gamma(\Bd \to x^0)$, 
\ie\ that isospin invariance holds in these \B decays,
the ratio of the number of reconstructed
$\Bu \to x^+$ and $\Bd \to x^0$ mesons is proportional to
\begin{equation}
\frac{f^{+-}\, \BR{\Bu\to x^+}}{f^{00}\, \BR{\Bd\to x^0}}
= \frac{f^{+-}\, \Gamma({\Bu}\to x^+)\, \tau(\Bu)}%
{f^{00}\, \Gamma({\Bd}\to x^0)\,\tau(\Bd)}
= \frac{f^{+-}}{f^{00}} \, \frac{\tau(\Bu)}{\tau(\Bd)}  \,, 
\end{equation} 
where $\tau(\Bu)$ and $\tau(\Bd)$ are the \Bu and \Bd 
lifetimes respectively.
Hence the primary quantity measured in these analyses 
is $R^{+-/00} \, \tau(\Bu)/\tau(\Bd)$, 
and the extraction of $R^{+-/00}$ with this method therefore 
requires the knowledge of the $\tau(\Bu)/\tau(\Bd)$ lifetime ratio. 

\begin{table}
\caption{Published measurements of the $\Bu/\Bd$ production ratio
in \Ups decays, together with their average (see text).
Systematic uncertainties due to the imperfect knowledge of 
$\tau(\Bu)/\tau(\Bd)$ are included.}
\labt{R_data}
\begin{center}
\begin{tabular}{lccll}
\hline
Experiment & Ref. & Decay modes & Published value of & Assumed value \\
and year & & or method & $R^{+-/00}=f^{+-}/f^{00}$ & of $\tau(\Bu)/\tau(\Bd)$ \\
\hline
CLEO,   2001 & \cite{CLEO_R2001}  & \particle{J/\psi K^{(*)}} 
             & $1.04 \pm0.07 \pm0.04$ & $1.066 \pm0.024$ \\
\babar, 2002 & \cite{BABAR_R2002} & \particle{(c\bar{c})K^{(*)}}
             & $1.10 \pm0.06 \pm0.05$ & $1.062 \pm0.029$\\ 
CLEO,   2002 & \cite{CLEO_R2002}  & \particle{D^*\ell\nu}
             & $1.058 \pm0.084 \pm0.136$ & $1.074 \pm0.028$\\
\belle, 2003 & \cite{BELLE_dmd_dilepton} & dilepton events 
             & $1.01 \pm0.03 \pm0.09$ & $1.083 \pm0.017$\\
\babar, 2004 & \cite{BABAR_R2004} & \particle{J/\psi K}
             & $1.006 \pm0.036 \pm0.031$ & $1.083 \pm0.017$ \\
\hline
Average      & & & \hfagFF~(tot) & \hfagRTAUBU \\
\hline
\end{tabular}
\end{center}
\end{table}

The published measurements of $R^{+-/00}$ are listed 
in \Table{R_data} together with the corresponding assumed values of 
$\tau(\Bu)/\tau(\Bd)$.
All measurements are based on the above-mentioned method, 
except the one from \belle, which is a by-product of the 
\Bd mixing frequency analysis using dilepton events
(but note that it also assumes isospin invariance, 
namely $\Gamma(\Bu \to \ell^+{\rm X}) = \Gamma(\Bd \to \ell^+{\rm X})$).
The latter is therefore treated in a slightly different 
manner in the following procedure used to combine 
these measurements:
\begin{itemize} 
\item each published value of $R^{+-/00}$ from CLEO and \babar
      is first converted back to the original measurement of 
      $R^{+-/00} \, \tau(\Bu)/\tau(\Bd)$, using the value of the 
      lifetime ratio assumed in the corresponding analysis;
\item a simple weighted average of these original
      measurements of $R^{+-/00} \, \tau(\Bu)/\tau(\Bd)$ from 
      CLEO and \babar (which do not depend on the assumed value 
      of the lifetime ratio) is then computed, assuming no 
      statistical or systematic correlations between them;


\item the weighted average of $R^{+-/00} \, \tau(\Bu)/\tau(\Bd)$ 
      is converted into a value of $R^{+-/00}$, using the latest 
      average of the lifetime ratios, $\tau(\Bu)/\tau(\Bd)=\hfagRTAUBU$ 
      (see \Sec{lifetime_ratio});
\item the \belle measurement of $R^{+-/00}$ is adjusted to the 
      current values of $\tau(\Bd)=\hfagTAUBD$ and 
      $\tau(\Bu)/\tau(\Bd)=\hfagRTAUBU$ (see \Sec{lifetime_ratio}),
      using the quoted systematic uncertainties due to these parameters;
\item the combined value of $R^{+-/00}$ from CLEO and \babar is averaged 
      with the adjusted value of $R^{+-/00}$ from \belle, assuming a 100\% 
      correlation of the systematic uncertainty due to the limited 
      knowledge on $\tau(\Bu)/\tau(\Bd)$; no other correlation is considered. 
\end{itemize} 
The resulting global average, 
\begin{equation}
R^{+-/00} = \frac{f^{+-}}{f^{00}} =  \hfagFF \,,
\labe{Rplusminus}
\end{equation}
is consistent with an equal production of charged and neutral \B mesons.

On the other hand, the \babar collaboration has 
performed a direct measurement of the $f^{00}$ fraction 
using a novel method, which does not rely on isospin symmetry nor requires 
the knowledge of $\tau(\Bu)/\tau(\Bd)$. Its analysis, 
based on a comparison between the number of events where a single 
$B^0 \to D^{*-} \ell^+ \nu$ decay could be reconstructed and the number 
of events where two such decays could be reconstructed, yields~\cite{BABAR_f00}
\begin{equation}
f^{00}= 0.487 \pm 0.010\,\mbox{(stat)} \pm 0.008\,\mbox{(syst)} \,.
\labe{fzerozero}
\end{equation}

The two results of \Eqss{Rplusminus}{fzerozero} are of very different natures 
and completely independent of each other. 
Their product is equal to $f^{+-} = \hfagFPROD$, 
while another combination of them gives $f^{+-} + f^{00}= \hfagFSUM$, 
compatible with unity.
Assuming\footnote{Two non-$\B\bar{B}$
decay modes of the $\Upsilon(4S)$, 
$\Upsilon(4S) \to \Upsilon(1S)\pi^+\pi^-$ and 
$\Upsilon(4S) \to \Upsilon(2S)\pi^+\pi^-$, have now been observed
with branching fractions
of the order of $10^{-4}$~\cite{BABAR_BELLE_Ups},
corresponding to a partial
width several times larger than that in the \ee channel.
However, this can still be
neglected and the assumption $f^{+-}+f^{00}=1$ remains valid
in the present context of the determination of $f^{+-}$ and $f^{00}$.}
 $f^{+-}+f^{00}= 1$, also consistent with 
CLEO's observation that the fraction of \Ups decays 
to \BB pairs is larger than 0.96 at \CL{95}~\cite{CLEO_frac_limit},
the results of \Eqss{Rplusminus}{fzerozero}
can be averaged (first converting \Eq{Rplusminus} 
into a value of $f^{00}=1/(R^{+-/00}+1)$) 
to yield the following more precise estimates:
\begin{equation}
f^{00} = \hfagFNW  \,,~~~ f^{+-} = 1 -f^{00} =  \hfagFCW \,,~~~
\frac{f^{+-}}{f^{00}} =  \hfagFFW \,.
\end{equation}

\mysubsubsection{\b-hadron production fractions in \Upsfive decays}
\labs{fraction_Ups5S}

Hadronic events produced in $e^+e^-$ collisions at the \Upsfive energy
can be classified into three categories: 
light-quark continuum events, $b\bar{b}$ continuum events,
and \Upsfive events. The latter two cannot be distinguished and 
are expected to always produce one of the following final states 
with a pair of $b$-flavored mesons:
$B\bar{B}$,
$B\bar{B}^*$, $B^*\bar{B}$, $B^*\bar{B}^*$, 
$B\bar{B}\pi$, $B\bar{B}^*\pi$, $B^*\bar{B}\pi$, 
$B^*\bar{B}^*\pi$, $B\bar{B}\pi\pi$,
$B_s^0\bar{B}_s^0$, $B_s^0\bar{B}_s^{0*}$, $B_s^{0*}\bar{B}_s^0$ or 
$B_s^{0*}\bar{B}_s^{0*}$, where 
$B$ denotes a $B^0$ or $B^+$ meson and 
$\bar{B}$ denotes a $\bar{B}^0$ or $B^-$ meson.
The excited states decay via $B^* \to B \gamma$ and
$B_s^{0*} \to B_s^0 \gamma$.
We define here $f_s(\Upsfive)$ as the fraction of 
$B_s^{0(*)}\bar{B}_s^{0(*)}$ events over all 
events with a pair of $b$-flavored mesons 
at the \Upsfive energy:
\begin{equation}
f_s(\Upsfive) =
\frac{\sigma(e^+e^- \to B_s^{0(*)}\bar{B}_s^{0(*)})}%
{\sigma(e^+e^- \to \Upsfive ~ \mbox{or} ~ b\bar{b}X)}
~~ \mbox{at} ~ \sqrt{s}=m(\Upsfive) \,.
\end{equation}

\begin{table}
\caption{Published values of $f_s(\Upsfive)$ and their average.}
\labt{fsFiveS}
\begin{center}
\begin{tabular}{lccl}
\hline
Experiment,   & Ref.                      & Decay modes              & Published value of          \\
year and dataset &                           & or method                & $f_s(\Upsfive)$         \\ \hline
CLEO, 2006, 0.42\invfb   & \cite{CLEO_huang2007}     & $\Upsfive\to D_{s}X$     & $0.168 \pm 0.026^{+0.067}_{-0.034}$  \\
             & \cite{CLEO_huang2007}     & $\Upsfive \to \phi X$    & $0.246 \pm 0.029^{+0.110}_{-0.053}$ \\
             & \cite{CLEO_huang2007}     & $\Upsfive \to B\bar{B}X$ & $0.411 \pm 0.100 \pm 0.092$ \\  
             & \cite{CLEO_huang2007}     & CLEO average of above 3  & $0.21^{+0.06}_{-0.03}$      \\  \hline
Belle, 2006, 1.86\invfb  & \cite{Belle_drutskoy2007} & $\Upsfive \to D_s X$     & $0.179 \pm 0.014 \pm 0.041$ \\
             & \cite{Belle_drutskoy2007} & $\Upsfive \to D^0 X$     & $0.181 \pm 0.036 \pm 0.075$ \\  
             & \cite{Belle_drutskoy2007} & Belle average of above 2 & $0.180 \pm 0.013 \pm 0.032$ \\  \hline 
\multicolumn{3}{l}{Average of all above
after adjustments to inputs of \Table{fsFiveS_external}} & 
\hfagFSFIVEfull             \\  \hline 
\end{tabular}
\end{center}
\end{table}

\begin{table}
\caption{External inputs on which the $f_s(\Upsfive)$ average
is based.}
\labt{fsFiveS_external}
\begin{center}
\begin{tabular}{lcc}
\hline
Branching fraction   & Value     & Explanation and reference \\
\hline
${\cal B}(B\to D_s X)\times {\cal B}(D_s \to \phi\pi)$ & 
$0.00381\pm 0.00015$ & world average~\cite{Yao2006PDG,CLEO_artuso2005} \\
${\cal B}(B^0_s \to D_s X)$ & 
$0.92\pm0.11$ & model-dependent estimate~\cite{CLEO_artuso2005} \\
${\cal B}(D_s \to \phi\pi)$ & 
$0.044\pm0.006$ & \cite{Yao2006PDG} \\
${\cal B}(B\to D^0 X)\times {\cal B}(D^0 \to K\pi)$ & 
$0.0243\pm0.0010$ & \cite{Yao2006PDG} \\
${\cal B}(B^0_s \to D^0 X)$ & 
$0.08\pm0.07$ & model-dependent estimate~\cite{CLEO_artuso2005} \\
${\cal B}(D^0 \to K\pi)$ & 
$0.0380\pm0.0007$ & \cite{Yao2006PDG} \\
${\cal B}(B \to \phi X)$ & 
$0.0344\pm0.0012$ & world average~\cite{Yao2006PDG,CLEO_huang2007} \\
${\cal B}(B^0_s \to \phi X)$ &
$0.161\pm0.024$ & model-dependent estimate~\cite{CLEO_huang2007} \\
\hline
\end{tabular}
\end{center}
\end{table}

The CLEO and Belle collaborations have recently published 
measurements of several inclusive \Upsfive branching fractions, 
${\cal B}(\Upsfive\to D_s X)$, 
${\cal B}(\Upsfive\to \phi X)$, 
${\cal B}(\Upsfive\to D^0 X)$, and 
${\cal B}(\Upsfive\to B\bar{B} X)$, 
from which they extract the
model-dependent estimates of $f_s(\Upsfive)$
reported in \Table{fsFiveS}.
This extraction requires the knowledge of several other 
branching fractions, which are listed in \Table{fsFiveS_external}
together with their most recent values. 
Before being averaged, the CLEO and Belle results are 
adjusted to these new external inputs. The world average of $f_s(\Upsfive)$
taking into account all systematic correlations introduced by the 
use of common external inputs, as well as the experiment-specific 
correlations due to the estimated number of $b\bar{b}$ events, is
\begin{equation}
f_s(\Upsfive) = \hfagFSFIVE \,.
\end{equation}
This production of $B^0_s$ mesons at the \Upsfive
is observed to be dominated by the $B_s^{0*}\bar{B}_s^{0*}$
channel~\cite{CLEO_bonvicini2006,Belle_drutskoy2007e}, with 
$\sigma(e^+e^- \to B_s^{0*}\bar{B}_s^{0*})/%
\sigma(e^+e^- \to B_s^{0(*)}\bar{B}_s^{0(*)})
= (93^{+7}_{-9}\pm 1)\%$~\cite{Belle_drutskoy2007e}.

\mysubsubsection{\b-hadron production fractions at high energy}
\labs{fractions_high_energy}
\labs{chibar}

At high energy, all species of weakly-decaying \b hadrons 
can be produced, either directly or in strong and electromagnetic 
decays of excited \b hadrons.
We assume here that the fractions of these different species 
are the same in unbiased samples of high-$p_{\rm T}$ \b jets 
originating from \particle{Z^0} decays or from \particle{p\bar{p}} 
collisions at the Tevatron.
This hypothesis is plausible considering that, in both cases, 
the last step of the jet hadronization is a non-perturbative
QCD process occurring at the scale of $\Lambda_{\rm QCD}$.
On the other hand, there is no strong argument to claim that these 
fractions should be strictly equal, so this assumption 
should be checked experimentally.
Although the available data is not sufficient at 
this time to perform a significant check, 
it is expected that more data from 
Tevatron Run~II may improve this situation and 
allow one to confirm or disprove this assumption with reasonable 
confidence. Meanwhile, the attitude adopted here is that these 
fractions are assumed to be equal at all high-energy colliders
until demonstrated otherwise by experiment.\footnote{It is not unlikely
that the \b-hadron fractions in low-$p_{\rm T}$ jets 
at a hadronic machine be different; in particular, beam-remnant effects may
enhance the \b-baryon production.}
However, as explained below, the measurements performed at LEP and at 
the Tevatron show slight discrepancies. Therefore we present two sets 
of averages: one set including only measurements performed at LEP, 
and a second set including measurements performed at both LEP and Tevatron. 

Contrary to what happens in the charm sector where the fractions of \particle{D^+} 
and \particle{D^0} are different, the relative amount of \Bu and \Bd is not affected by the 
electromagnetic decays of excited ${\Bu}^*$ and ${\Bd}^*$ states and strong decays of excited
${\Bu}^{**}$ and ${\Bd}^{**}$ states. Decays of the type \particle{{\Bs}^{**} \to B^{(*)}K}
also contribute to the \Bu and \Bd rates, but with the same magnitude if mass effects
can be neglected.
We therefore assume equal production of \Bu and \Bd. We also  
neglect the production of weakly-decaying states
made of several heavy quarks (like \Bc and other heavy baryons) 
which is known to be very small. Hence, for the purpose of determining 
the \b-hadron fractions, we use the constraints
\begin{equation}
\fBu = \fBd ~~~~\mbox{and}~~~ \fBu + \fBd + \fBs + \fbb = 1 \,,
\labe{constraints}
\end{equation}
where \fBu, \fBd, \fBs and \fbb
are the unbiased fractions of \Bu, \Bd, \Bs and \b baryons, respectively.

The LEP experiments have measured
$\fBs \times \BR{\Bs\to\particle{D_s^-} \ell^+ \nu_\ell \mbox{$X$}}$~\cite{LEP_fs}, 
$\BR{\b\to\Lb} \times \BR{\Lb\to\Lc\ell^-\bar{\nu}_\ell \mbox{$X$}}$~\cite{DELPHI_fla,ALEPH_fla}
and $\BR{\b\to\Xib^-} \times \BR{\Xi_b^- \to \Xi^-\ell^-\overline\nu_\ell 
\mbox{$X$}}$~\cite{ALEPH_fxi,DELPHI_fxi2}\footnote{The DELPHI result 
of \Ref{DELPHI_fxi2} is considered to supersede an older one~\cite{DELPHI_fxi}.}
from partially reconstructed final states 
including a lepton, \fbb
from protons identified in \b events~\cite{ALEPH-fbar}, and the 
production rate of charged \b hadrons~\cite{DELPHI-fch}. 
The various \b-hadron fractions 
have also been measured at CDF using electron-charm final states~\cite{CDF_f_ec}
and double semileptonic decays with \particle{\phi\ell} and 
\particle{K^*\ell} final states~\cite{CDF_f_phil_Kstl}.
All these published results\footnote{Preliminary 
measurements~\cite{CDF2_fractions} performed by CDF with Run~II data 
have not been included here.}
have been combined
following the procedure and assumptions described in~\cite{LEPHFS},
to yield $\fBu=\fBd=\hfagWFBDNOMIX$, 
$\fBs=\hfagWFBSNOMIX$ and $\fbb=\hfagWFBBNOMIX$
under the constraints of \Eq{constraints}. 
Following the PDG prescription, we have scaled the
combined uncertainties on these fractions 
by \hfagWFSFACTOR to account for slight 
discrepancies in the input data.
Repeating the combination using LEP data only, 
we obtain $\fBu=\fBd=\hfagLFBDNOMIX$,
$\fBs=\hfagLFBSNOMIX$ and $\fbb=\hfagLFBBNOMIX$
and find that no scaling factor is necessary. 
For these combinations other external inputs are used, \eg\ the branching 
ratios of \B mesons to final states with a \particle{D}, \particle{D^*} or 
\particle{D^{**}} in semileptonic decays, which are needed to evaluate the 
fraction of semileptonic \Bs decays with a \particle{D_s^-} in the final state.


Time-integrated mixing analyses performed with lepton pairs 
from \particle{b\bar{b}} 
events produced at high-energy colliders measure the quantity 
\begin{equation}
\chibar = f'_{\particle{d}} \,\chid + f'_{\particle{s}} \,\chis \,,
\labe{chibar}
\end{equation}
where $f'_{\particle{d}}$ and $f'_{\particle{s}}$ are 
the fractions of \Bd and \Bs hadrons 
in a sample of semileptonic \b-hadron decays, and where \chid and \chis 
are the \Bd and \Bs time-integrated mixing probabilities.
Assuming that all \b hadrons have the same semileptonic decay width implies 
$f'_i = f_i R_i$, where $R_i = \tau_i/\tau_{\particle{b}}$ is the ratio of the lifetime 
$\tau_i$ of species $i$ to the average \b-hadron lifetime 
$\tau_{\particle{b}} = \sum_i f_i \tau_i$.
Hence measurements of the mixing probabilities
\chibar, \chid and \chis can be used to improve our 
knowledge of \fBu, \fBd, \fBs and \fbb.
In practice, the above relations yield another determination of 
\fBs obtained from \fbb and mixing information, 
\begin{equation}
\fBs = \frac{1}{R_{\particle{s}}}
\frac{(1+r)\overline{\chi}-(1-\fbb R_{\rm baryon}) \chid}{(1+r)\chis - \chid} \,,
\labe{fBs-mixing}
\end{equation}
where $r=R_{\particle{u}}/R_{\particle{d}} = \tau(\Bu)/\tau(\Bd)$.

The published measurements of \chibar performed by the LEP
experiments have been combined by the LEP Electroweak Working Group to yield 
$\chibar = \hfagCHIBARLEP$~\cite{LEPEWWG}.
This can be compared with the Tevatron average, 
$\chibar = \hfagCHIBARTEV$,
obtained from a CDF measurement with Run~I data~\cite{CDF_chibar}
and from a recent \dzero measurement with Run~II data~\cite{D0_qp_chibar}.
The two averages deviate
from each other by $\hfagCHIBARSFACTOR\,\sigma$; 
this could be an indication that the production fractions of \b hadrons 
at the \particle{Z} peak or at the Tevatron are not the same. 
Although this discrepancy 
is not very significant it should be carefully monitored in the future. 
We choose to combine these two results in a simple weighted average,
assuming no correlations, and, following the PDG prescription, we 
multiply the combined uncertainty by \hfagCHIBARSFACTOR to account 
for the discrepancy. Our world average is then
$\chibar = \hfagCHIBAR$.

\begin{table}
\caption{Time-integrated mixing probability \chibar (defined in \Eq{chibar}), 
and fractions of the different \b-hadron species in an unbiased sample of 
weakly-decaying \b hadrons, obtained from both direct
and mixing measurements.
Measurements performed in $Z$ decays are included 
in both sets of averages.}
\labt{fractions}
\begin{center}
\begin{tabular}{llcc}
\hline
Quantity            &                      & in $Z$ decays   & at high energy   \\
\hline
Mixing probability  & $\overline{\chi}$    & \hfagCHIBARLEP  & \hfagCHIBAR \\
\Bu or \Bd fraction & $f_u = f_d$          & \hfagLFBD       & \hfagWFBD   \\
\Bs fraction        & $f_s$                & \hfagLFBS       & \hfagWFBS   \\
\b baryon fraction  & $f_{\rm baryon}$     & \hfagLFBB       & \hfagWFBB   \\
\multicolumn{2}{l}{Correlation between $f_s$ and $f_u = f_d$}            & \hfagLRHOFBDFBS & \hfagWRHOFBDFBS \\
\multicolumn{2}{l}{Correlation between $f_{\rm baryon}$ and $f_u = f_d$} & \hfagLRHOFBDFBB & \hfagWRHOFBDFBB \\
\multicolumn{2}{l}{Correlation between $f_{\rm baryon}$ and $f_s$}       & \hfagLRHOFBBFBS & \hfagWRHOFBBFBS \\
\hline
\end{tabular}
\end{center}
\end{table}

Introducing the \chibar average in \Eq{fBs-mixing}, together with our world average 
$\chid = \hfagCHIDWU$ (see \Eq{chid} of \Sec{dmd}), the assumption $\chis= 1/2$ 
(justified by \Eq{chis} in \Sec{dms}), the 
best knowledge of the lifetimes (see \Sec{lifetimes}) and the estimate of \fbb given above, 
yields $\fBs = \hfagWFBSMIX$ 
(or $\fBs = \hfagLFBSMIX$ using only LEP data),
an estimate dominated by the mixing information. 
Taking into account all known correlations (including the one introduced by \fbb), 
this result is then combined with the set of fractions obtained from direct measurements 
(given above), to yield the 
improved estimates of \Table{fractions}, 
still under the constraints of \Eq{constraints}. 
As can be seen, our knowledge on the mixing parameters 
substantially reduces the uncertainty on \fBs, and this 
even in the case of the world averages where a rather strong 
deweighting was introduced in the computation of \chibar.
It should be noted that the results 
are correlated, as indicated in \Table{fractions}.


\input{life_mix_tau1}  
\input{life_mix_tau2}  

\mysubsection{Neutral \B-meson mixing}
\labs{mixing}

The $\Bd-\Bdbar$ and $\Bs-\Bsbar$ systems
both exhibit the phenomenon of particle-antiparticle mixing. For each of them, 
there are two mass eigenstates which are linear combinations of the two flavour states,
\B and $\bar{\B}$. 
The heaviest (lightest) of the these mass states is denoted
$\B_{\rm H}$ ($\B_{\rm L}$),
with mass $m_{\rm H}$ ($m_{\rm L}$)
and total decay width $\Gamma_{\rm H}$ ($\Gamma_{\rm L}$). We define
\begin{eqnarray}
\Delta m &=& m_{\rm H} - m_{\rm L} \,, \labe{dm} \\
\Delta \Gamma &=& \Gamma_{\rm L} - \Gamma_{\rm H} \,, \labe{dg}
\end{eqnarray}
where $\Delta m$ is positive by definition, and 
$\Delta \Gamma$ is expected to be positive within
the Standard Model.\footnote{For reason of symmetry in 
\Eqss{dm}{dg}, $\Delta \Gamma$ is sometimes defined with 
the opposite sign. The definition adopted here, \ie\
\Eq{dg}, is the one used by most experimentalists and many
phenomenologists in \B physics.}

There are four different time-dependent probabilities describing the 
case of a neutral \B meson produced 
as a flavour state and decaying to a flavour-specific final state.
If \CPT is conserved (which  
will be assumed throughout), they can be written as 
\begin{equation}
\left\{
\begin{array}{rcl}
{\cal P}(\B\to\B) & = &  \frac{e^{-\Gamma t}}{2} 
\left[ \cosh\!\left(\frac{\Delta\Gamma}{2}t\right) + \cos\!\left(\Delta m t\right)\right]  \\
{\cal P}(\B\to\bar{\B}) & = &  \frac{e^{-\Gamma t}}{2} 
\left[ \cosh\!\left(\frac{\Delta\Gamma}{2}t\right) - \cos\!\left(\Delta m t\right)\right] 
\left|\frac{q}{p}\right|^2 \\
{\cal P}(\bar{\B}\to\B) & = &  \frac{e^{-\Gamma t}}{2} 
\left[ \cosh\!\left(\frac{\Delta\Gamma}{2}t\right) - \cos\!\left(\Delta m t\right)\right] 
\left|\frac{p}{q}\right|^2 \\
{\cal P}(\bar{\B}\to\bar{\B}) & = &  \frac{e^{-\Gamma t}}{2} 
\left[ \cosh\!\left(\frac{\Delta\Gamma}{2}t\right) + \cos\!\left(\Delta m t\right)\right] 
\end{array} \right. \,,
\labe{oscillations}
\end{equation}
where $t$ is the proper time of the system (\ie\ the time interval between the production 
and the decay in the rest frame of the \B meson) and $\Gamma = 
(\Gamma_{\rm H} + \Gamma_{\rm L})/2 =1/\bar{\tau}(\B)$ 
is the average decay width.
At the \B factories, only the proper-time difference $\Delta t$ between the decays
of the two neutral \B mesons from the \Ups can be determined, but, 
because the two \B mesons evolve coherently (keeping opposite flavours as long as
none of them has decayed), the 
above formulae remain valid 
if $t$ is replaced with $\Delta t$ and the production flavour is replaced by the flavour 
at the time of the decay of the accompanying \B meson in a flavour-specific state.
As can be seen in the above expressions,
the mixing probabilities 
depend on three mixing observables:
$\Delta m$, $\Delta\Gamma$,
and $|q/p|^2$ which signals \CP violation in the mixing if $|q/p|^2 \ne 1$.

In the next sections we review in turn the experimental knowledge
on these three parameters, separately 
for the \Bd meson (\dmd, \DGd, $|q/p|_{\particle{d}}$) 
and the \Bs meson (\dms, \DGs, $|q/p|_{\particle{s}}$). 

\mysubsubsection{\Bd mixing parameters}
\labs{qpd}
\labs{dmd}
\labs{DGd}

\subsubsubsection{\boldmath \CP violation parameter $|q/p|_{\particle{d}}$}

Evidence for \CP violation in \Bd mixing
has been searched for,
both with flavor-specific and inclusive \Bd decays, 
in samples where the initial 
flavor state is tagged. In the case of semileptonic 
(or other flavor-specific) decays, 
where the final state tag is 
also available, the following asymmetry
\begin{equation} 
\ASLd = \frac{
N(\hbox{\Bdbar}(t) \to \ell^+      \nu_{\ell} X) -
N(\hbox{\Bd}(t)    \to \ell^- \bar{\nu}_{\ell} X) }{
N(\hbox{\Bdbar}(t) \to \ell^+      \nu_{\ell} X) +
N(\hbox{\Bd}(t)    \to \ell^- \bar{\nu}_{\ell} X) } 
= \frac{|p/q|_{\particle{d}}^2 - |q/p|_{\particle{d}}^2}%
{|p/q|_{\particle{d}}^2 + |q/p|_{\particle{d}}^2}
\labe{ASL}
\end{equation} 
has been measured, either in time-integrated analyses at 
CLEO~\cite{CLEO_chid_CP,CLEO_chid_CP_y,CLEO_CP_semi},
CDF~\cite{CDF_CP_semi} and \dzero~\cite{D0_qp_chibar},
or in time-dependent analyses at 
OPAL~\cite{OPAL_CP_semi}, ALEPH~\cite{ALEPH_CP}, 
\babar~\cite{BABAR_DGd_qp,BABAR_CP_semi,BABAR_qp_dileptons,BABAR_qp_dstarlnu}
and \belle~\cite{BELLE_CP_semi}.
In the inclusive case, also investigated and published
at ALEPH~\cite{ALEPH_CP} and OPAL~\cite{OPAL_CP_incl},
no final state tag is used, and the asymmetry~\cite{incl_asym}
\begin{equation} 
\frac{
N(\hbox{\Bd}(t) \to {\rm all}) -
N(\hbox{\Bdbar}(t) \to {\rm all}) }{
N(\hbox{\Bd}(t) \to {\rm all}) +
N(\hbox{\Bdbar}(t) \to {\rm all}) } 
\simeq
\ASLd \left[ \frac{\dmd}{2\Gd} \sin(\dmd \,t) - 
\sin^2\left(\frac{\dmd \,t}{2}\right)\right] 
\labe{ASLincl}
\end{equation} 
must be measured as a function of the proper time to extract information 
on \CP violation.
In all cases asymmetries compatible with zero have been found,  
with a precision limited by the available statistics. 

A simple average of all measurements performed at 
\B factories~\cite{CLEO_chid_CP_y,CLEO_CP_semi,BABAR_DGd_qp,BABAR_qp_dileptons,BABAR_qp_dstarlnu,BELLE_CP_semi}
yields 
\begin{equation}
\ASLd = \hfagASLDB 
\labe{ASLDB}
\end{equation}
or, equivalently through \Eq{ASL},
\begin{equation}
|q/p|_{\particle{d}} = \hfagQPDB \,.
\end{equation}
Analyses performed at higher energy, either at LEP or at the Tevatron,
can't separate the contributions from the 
\Bd and \Bs mesons. Under the assumption of no \CP violation in \Bs mixing, a number of 
these analyses~\cite{D0_qp_chibar,OPAL_CP_semi,ALEPH_CP,OPAL_CP_incl}
quote a measurement of $\ASLd$ or $|q/p|_{\particle{d}}$ for the \Bd meson. Combining 
these results with the above \B factory averages lead to 
\begin{equation}
\left.
\begin{array}{l}
\ASLd = \hfagASLDA \\
|q/p|_{\particle{d}} = \hfagQPDA 
\end{array} \right\} ~~ \mbox{if $\ASLs =0$, $|q/p|_{\particle{s}}=1$.}
\end{equation}
These results\footnote{Early analyses and (perhaps hence) the PDG use the complex
parameter $\epsilon_{\B} = (p-q)/(p+q)$; if \CP violation in the mixing in small, 
$\ASLd \cong 4 {\rm Re}(\epsilon_{\B})/(1+|\epsilon_{\B}|^2)$ and our 
current averages  
are ${\rm Re}(\epsilon_{\B})/(1+|\epsilon_{\B}|^2)=\hfagREBDB$ (\B factory
measurements only) and $\hfagREBDA$ (all measurements).}, 
summarized in \Table{qoverp},
are compatible 
with no \CP violation in the \Bd mixing, an assumption we make for the rest 
of this section.

\begin{table}
\caption{Measurements of \CP violation in \Bd mixing and their average
in terms of both \ASLd and $|q/p|_{\particle{d}}$.
The individual results are listed as quoted in the original publications, 
or converted\addtocounter{footnote}{-1}\protect\footnotemark\
to an \ASLd value.
When two errors are quoted, the first one is statistical and the 
second one systematic. The second group of measurements, performed 
at high-energy colliders, assume no \CP violation in \Bs mixing, 
\ie\ $|q/p|_{\particle{s}}=1$.}
\labt{qoverp}
\begin{center}
\begin{tabular}{@{}rcl@{$\,\pm$}l@{$\pm$}ll@{$\,\pm$}l@{$\pm$}l@{}}
\hline
Exp.\ \& Ref. & Method & \multicolumn{3}{c}{Measured \ASLd} 
                       & \multicolumn{3}{c}{Measured $|q/p|_{\particle{d}}$} \\
\hline
CLEO   \cite{CLEO_chid_CP_y} & partial hadronic rec. 
                             & $+0.017$ & 0.070 & 0.014 
                             & \multicolumn{3}{c}{} \\
CLEO   \cite{CLEO_CP_semi}   & dileptons 
                             & $+0.013$ & 0.050 & 0.005 
                             & \multicolumn{3}{c}{} \\
CLEO   \cite{CLEO_CP_semi}   & average of above two 
                             & $+0.014$ & 0.041 & 0.006 
                             & \multicolumn{3}{c}{} \\
\babar \cite{BABAR_DGd_qp}   & full hadronic rec. 
                             & \multicolumn{3}{c}{}  
                             & 1.029 & 0.013 & 0.011  \\
\babar \cite{BABAR_qp_dileptons}  & dileptons
                             & \multicolumn{3}{c}{}
                             & 0.9992 & 0.0027 & 0.0019 \\ 
\babar \cite{BABAR_qp_dstarlnu}$^p$  & part.\ rec.\ $D^{*}\ell\nu$ 
                             & $-0.0130$ & 0.0068 & 0.0040 
                             & 1.0065 & 0.0034 & 0.0020 \\ 
\belle \cite{BELLE_CP_semi}  & dileptons 
                             & $-0.0011$ & 0.0079 & 0.0085 
                             & 1.0005 & 0.0040 & 0.0043 \\
\hline
& Average of  6 above        & \multicolumn{3}{l}{\hfagASLDB\ (tot)} 
                             & \multicolumn{3}{l}{\hfagQPDB\  (tot)} \\ 
\hline
OPAL   \cite{OPAL_CP_semi}   & leptons     
                             & $+0.008$ & 0.028 & 0.012 
                             & \multicolumn{3}{c}{} \\
OPAL   \cite{OPAL_CP_incl}   & inclusive (\Eq{ASLincl}) 
                             & $+0.005$ & 0.055 & 0.013 
                             & \multicolumn{3}{c}{} \\
ALEPH  \cite{ALEPH_CP}       & leptons 
                             & $-0.037$ & 0.032 & 0.007 
                             & \multicolumn{3}{c}{} \\
ALEPH  \cite{ALEPH_CP}       & inclusive (\Eq{ASLincl}) 
                             & $+0.016$ & 0.034 & 0.009 
                             & \multicolumn{3}{c}{} \\
ALEPH  \cite{ALEPH_CP}       & average of above two 
                             & $-0.013$ & \multicolumn{2}{l}{0.026 (tot)} 
                             & \multicolumn{3}{c}{} \\
\dzero  \cite{D0_qp_chibar}  & dimuons  
                             & $-0.0092$ & 0.0044 & 0.0032
                             & \multicolumn{3}{c}{} \\
\hline
& Average of 11 above        & \multicolumn{3}{l}{\hfagASLDA\ (tot)} 
                             & \multicolumn{3}{l}{\hfagQPDA\  (tot)} \\ 
\hline
\multicolumn{8}{l}{$^p$ {\footnotesize Preliminary.}}
\end{tabular}
\end{center}
\end{table}

\subsubsubsection{\boldmath Mass and decay width differences \dmd and \DGd}

\begin{table}
\caption{Time-dependent measurements included in the \dmd average.
The results obtained from multi-dimensional fits involving also 
the \Bd (and \Bu) lifetimes
as free parameter(s)~\cite{BABAR3,BABAR5,BELLE2} 
have been converted into one-dimensional measurements of \dmd.
All the measurements have then been adjusted to a common set of physics
parameters before being combined. 
The CDF results from Run~II are preliminary.}
\labt{dmd}
\begin{center}
\begin{tabular}{@{}rc@{}cc@{}c@{}cc@{}c@{}c@{}}
\hline
Experiment & \multicolumn{2}{c}{Method} & \multicolumn{3}{l}{\dmd in\invps}   
                                        & \multicolumn{3}{l}{\dmd in\invps}     \\
and Ref.   &  rec. & tag                & \multicolumn{3}{l}{before adjustment} 
                                        & \multicolumn{3}{l}{after adjustment} \\
\hline
\input{dmd_W_table.tex}
\\[-2.0ex]
\multicolumn{6}{l}{~~~ -- ALEPH, DELPHI, L3, OPAL and CDF1 only:}
     & \hfagDMDHval & \hfagDMDHsta & \hfagDMDHsys \\
\multicolumn{6}{l}{~~~ -- Above measurements of \babar and \belle only:}
     & \hfagDMDBval & \hfagDMDBsta & \hfagDMDBsys \\
\hline
\end{tabular}
\end{center}
\end{table}

Many time-dependent \Bd--\Bdbar oscillation analyses have been performed by the 
ALEPH, \babar, \belle, CDF, \dzero, DELPHI, L3 and OPAL collaborations. 
The corresponding measurements of \dmd are summarized in 
\Table{dmd},
where only the most recent results
are listed (\ie\ measurements superseded by more recent ones have been omitted). 
Although a variety of different techniques have been used, the 
individual \dmd
results obtained at high-energy colliders have remarkably similar precision.
Their average is compatible with the recent and more precise measurements 
from the asymmetric \B factories.
The systematic uncertainties are not negligible; 
they are often dominated by sample composition, mistag probability,
or \b-hadron lifetime contributions.
Before being combined, the measurements are adjusted on the basis of a 
common set of input values, including the averages of the 
\b-hadron fractions and lifetimes given in this report 
(see \Secss{fractions}{lifetimes}).
Some measurements are statistically correlated. 
Systematic correlations arise both from common physics sources 
(fractions, lifetimes, branching ratios of \b hadrons), and from purely 
experimental or algorithmic effects (efficiency, resolution, flavour tagging, 
background description). Combining all published measurements
listed in \Table{dmd}
and accounting for all identified correlations
as described in~\cite{LEPHFS} yields $\dmd = \hfagDMDWfull$.

On the other hand, ARGUS and CLEO have published 
measurements of the time-integrated mixing probability 
\chid~\cite{ARGUS_chid,CLEO_chid_CP,CLEO_chid_CP_y}, 
which average to $\chid =\hfagCHIDU$.
Following \Ref{CLEO_chid_CP_y}, 
the width difference \DGd could 
in principle be extracted from the
measured value of $\Gd=1/\tau(\Bd)$ and the above averages for 
\dmd and \chid 
(provided that \DGd has a negligible impact on 
the \dmd $\tau(\Bd)$ analyses that have assumed $\DGd=0$), 
using the relation
\begin{equation}
\chid = \frac{\xd^2+\yd^2}{2(\xd^2+1)} ~~~ \mbox{with} ~~ \xd=\frac{\dmd}{\Gd} 
~~~ \mbox{and} ~~ \yd=\frac{\DGd}{2\Gd} \,.
\labe{chid_definition}
\end{equation}
However, direct time-dependent studies provide much stronger constraints: 
$|\DGd|/\Gd < 18\%$ at \CL{95} from DELPHI~\cite{DELPHI_dmd_dms_vtx},
and $-6.8\% < {\rm sign}({\rm Re} \lambda_{\CP}) \DGGd < 8.4\%$
at \CL{90} from \babar~\cite{BABAR_DGd_qp},
where $\lambda_{\CP} = (q/p)_{\particle{d}} (\bar{A}_{\CP}/A_{\CP})$
is defined for a \CP-even final state 
(the sensitivity to the overall sign of 
${\rm sign}({\rm Re} \lambda_{\CP}) \DGGd$ comes
from the use of \Bd decays to \CP final states).
Combining these two results after adjustment to 
$1/\Gd=\tau(\Bd)=\hfagTAUBD$ yields
\begin{equation}
{\rm sign}({\rm Re} \lambda_{\CP}) \DGGd  = \hfagSDGDGD \,.
\end{equation}
The sign of ${\rm Re} \lambda_{\CP}$ is not measured,
but expected to be positive from the global fits
of the Unitarity Triangle within the Standard Model.

Assuming $\DGd=0$ 
and using $1/\Gd=\tau(\Bd)=\hfagTAUBD$,
the \dmd and \chid results are combined through \Eq{chid_definition} 
to yield the 
world average
\begin{equation} 
\dmd = \hfagDMDWU \,,
\labe{dmd}
\end{equation} 
or, equivalently,
\begin{equation} 
\xd= \hfagXDWU ~~~ \mbox{and} ~~~ \chid=\hfagCHIDWU \,.  
\labe{chid}
\end{equation}
\Figure{dmd} compares the \dmd values obtained by the different experiments.

\begin{figure}
\begin{center}
\epsfig{figure=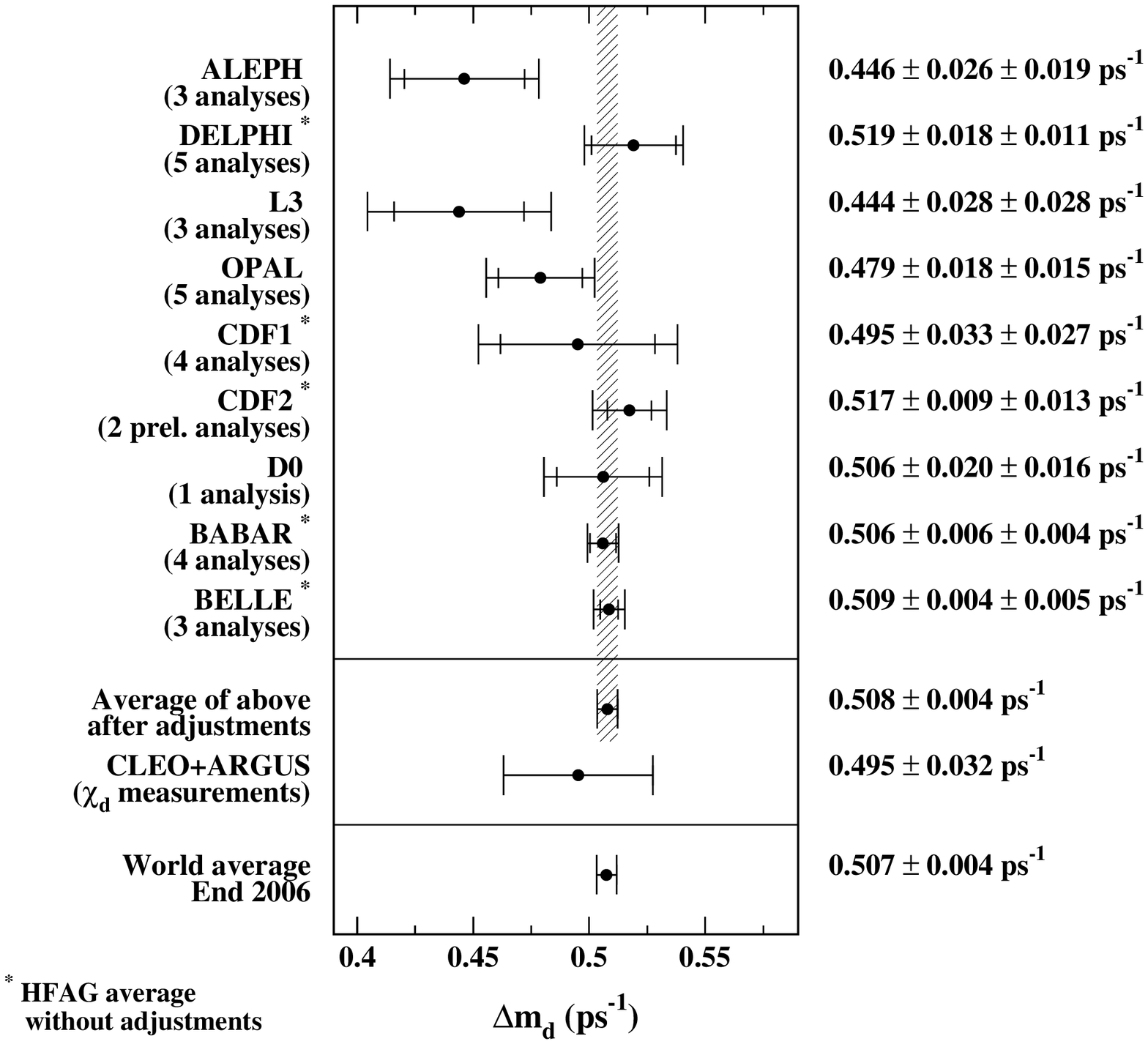,width=\textwidth}
\caption{The \Bd--\Bdbar oscillation frequency \dmd as measured by the different experiments. 
The averages quoted for ALEPH, L3 and OPAL are taken from the original publications, while the 
ones for DELPHI, CDF, \babar, and \belle have been computed from the individual results 
listed in \Table{dmd} without performing any adjustments. The time-integrated measurements 
of \chid from the symmetric \B factory experiments ARGUS and CLEO have been converted 
to a \dmd value using $\tau(\Bd)=\hfagTAUBD$. The two global averages have been obtained 
after adjustments of all the individual \dmd results of \Table{dmd} (see text).}
\labf{dmd}
\end{center}
\end{figure}

The \Bd mixing averages given in \Eqss{dmd}{chid}
and the \b-hadron fractions of \Table{fractions} have been obtained in a fully 
consistent way, taking into account the fact that the fractions are computed using 
the \chid value of \Eq{chid} and that many individual measurements of \dmd
at high energy depend on the assumed values for the \b-hadron fractions.
Furthermore, this set of averages is consistent with the lifetime averages 
of \Sec{lifetimes}.

\begin{table}
\caption{Simultaneous measurements of \dmd and $\tau(\Bd)$, and their average.
The \belle analysis also 
measures $\tau(\Bu)$ at the same time, but it is converted here into a two-dimensional measurement 
of \dmd and $\tau(\Bd)$, for an assumed value of $\tau(\Bu)$. 
The first quoted error on the measurements is statistical
and the second one systematic; in the case of adjusted measurements, the 
latter includes a contribution obtained from the variation of $\tau(\Bu)$ or 
$\tau(\Bu)/\tau(\Bd)$ in the indicated range. Units are\invps\ for \dmd
and\unit{ps} for lifetimes. 
The three different values of $\rho(\dmd,\tau(\Bd))$ correspond 
to the statistical, systematic and total correlation coefficients
between the adjusted measurements of \dmd and $\tau(\Bd)$.}
\labt{dmd2D}
\begin{center}
\begin{tabular}{@{}r@{~}c@{}c@{}c@{~}c@{}c@{}c@{~}c@{}c@{}c@{\hspace{0ex}}c@{}}
\hline
Exp.\ \& Ref.
& \multicolumn{3}{c}{Measured \dmd}   
& \multicolumn{3}{c}{Measured $\tau(\Bd)$}   
& \multicolumn{3}{c}{Measured $\tau(\Bu)$}   
&  Assumed $\tau(\Bu)$ \\
\hline
\babar \cite{BABAR3}  
      & $0.492$ & $\pm 0.018$ & $\pm 0.013$ 
      & $1.523$ & $\pm 0.024$ & $\pm 0.022$ 
      & \multicolumn{3}{c}{---}
      & $(1.083\pm 0.017)\tau(\Bd)$ \\  
\babar \cite{BABAR5}  
      & $0.511$ & $\pm 0.007$ & $^{+0.007}_{-0.006}$ 
      & $1.504$ & $\pm 0.013$ & $^{+0.018}_{-0.013}$
      & \multicolumn{3}{c}{---}
      & $1.671\pm 0.018$ \\  
\belle \cite{BELLE2}  
      & $0.511$ & $\pm 0.005$ & $\pm 0.006$
      & $1.534$ & $\pm 0.008$ & $\pm 0.010$
      & $1.635$ & $\pm 0.011$ & $\pm 0.011$
      & --- \\  
\cline{2-10}
& \multicolumn{3}{c}{Adjusted \dmd}   
& \multicolumn{3}{c}{Adjusted $\tau(\Bd)$}   
& \multicolumn{3}{c}{$\rho(\dmd,\Bd)$} 
&  Assumed $\tau(\Bu)$ \\
\cline{2-10}
\babar \cite{BABAR3}  
      & $0.492$ & $\pm 0.018$ & $\pm 0.013$  
      & $1.524$ & $\pm 0.025$ & $\pm 0.022$  
      & $-0.22$ & $+0.74$ & $+0.16$ 
      & $(\hfagRTAUBUval$$\hfagRTAUBUerr)\tau(\Bd)$ \\  
\babar \cite{BABAR5} 
      & $0.512$ & $\pm 0.007$ & $\pm 0.007$  
      & $1.506$ & $\pm 0.013$ & $\pm 0.018$ 
      & $+0.01$ & $-0.85$ & $-0.48$ 
      & $\hfagTAUBUval$$\hfagTAUBUerr$ \\  
\belle \cite{BELLE2}  
      & $0.510$ & $\pm 0.007$ & $\pm 0.005$ 
      & $1.535$ & $\pm 0.009$ & $\pm 0.009$ 
      & $-0.27$ & $-0.08$ & $-0.19$ 
      & $\hfagTAUBUval$$\hfagTAUBUerr$ \\  
\hline
\multicolumn{1}{l}{Average} 
      & \hfagDMDTWODval   & \hfagDMDTWODsta   & \hfagDMDTWODsys
      & \hfagTAUBDTWODval & \hfagTAUBDTWODsta & \hfagTAUBDTWODsys
      & \hfagRHOstaDMDTAUBD & \hfagRHOsysDMDTAUBD & \hfagRHODMDTAUBD 
      & $\hfagTAUBUval$$\hfagTAUBUerr$ \\  
\hline 
\end{tabular}
\end{center}
\end{table}

\begin{figure}
\begin{center}
\epsfig{figure=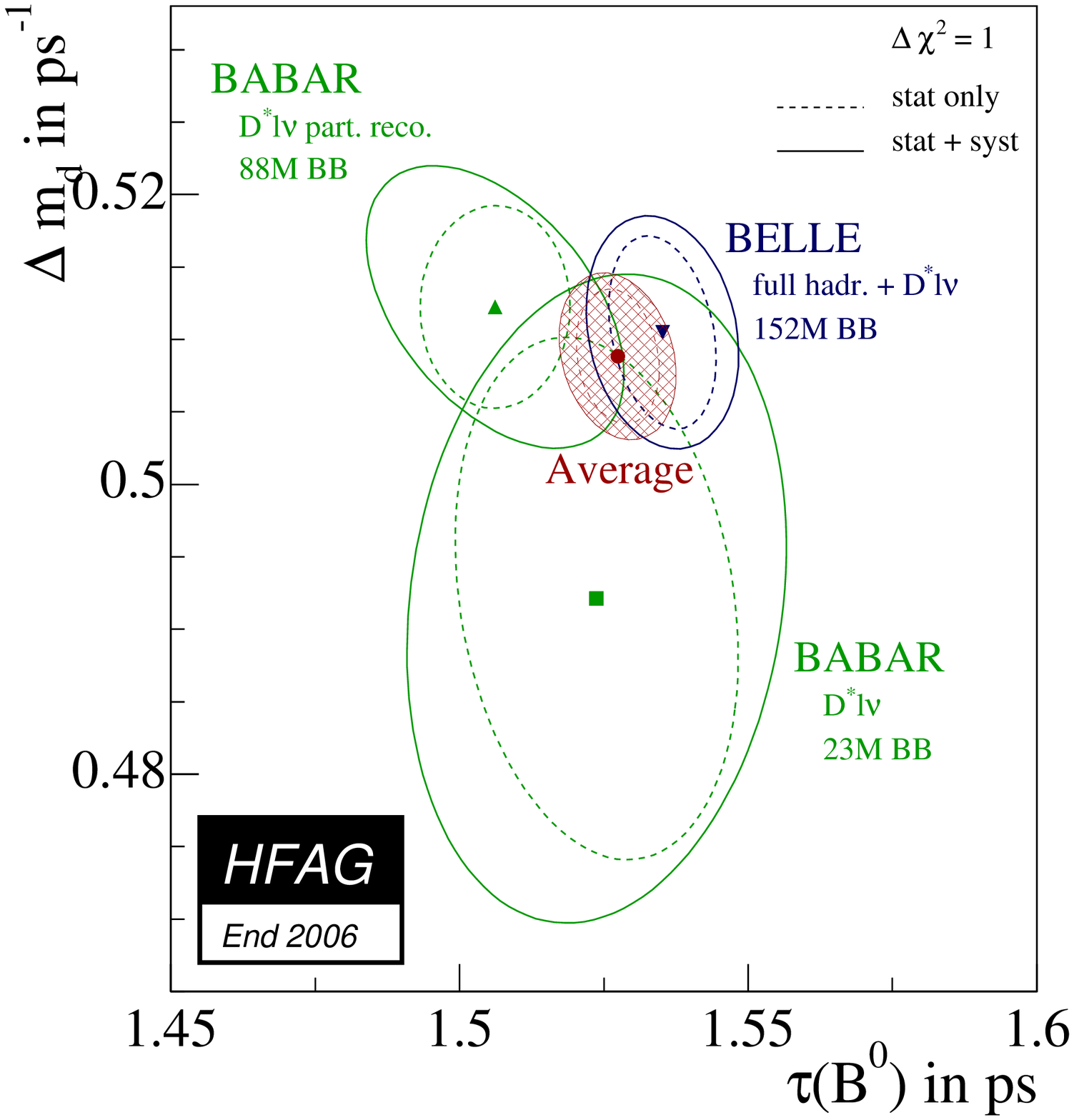,width=0.6\textwidth}
\caption{Simultaneous measurements of
\dmd and $\tau(\Bd)$~\cite{BABAR3,BABAR5,BELLE2}, 
after adjustment to a common set of parameters (see text). 
Statistical and total uncertainties are represented as dashed and
solid contours respectively.
The average of the three measurements
is indicated by a hatched ellipse.}
\labf{dmd2D}
\end{center}
\end{figure}

It should be noted that the most recent (and precise) analyses at the 
asymmetric \B factories measure \dmd
as a result of a multi-dimensional fit. 
Two \babar analyses~\cite{BABAR3,BABAR5},  
based on fully and partially reconstructed $\Bd \to D^*\ell\nu$ decays
respectively, 
extract simultaneously \dmd and $\tau(\Bd)$
while the latest \belle analysis~\cite{BELLE2},  
based on fully reconstructed hadronic \Bd decays and $\Bd \to D^*\ell\nu$ decays, 
extracts simultaneously \dmd, $\tau(\Bd)$ and $\tau(\Bu)$.
The measurements of \dmd and $\tau(\Bd)$ of these three analyses 
are displayed in \Table{dmd2D} and in \Fig{dmd2D}. Their two-dimensional average, 
taking into account all statistical and systematic correlations, and expressed
at $\tau(\Bu)=\hfagTAUBU$, is
\begin{equation}
\left.
\begin{array}{r@{}l}
\dmd = \hfagDMDTWODnounit & \invps \\
\tau(\Bd) = \hfagTAUBDTWODnounit & \ps
\end{array}
\right\}
~\mbox{with a total correlation of \hfagRHODMDTAUBD.}
\end{equation}

\mysubsubsection{\Bs mixing parameters}
\labs{qps}
\labs{DGs}
\labs{dms}

\subsubsubsection{\boldmath \CP violation parameter $|q/p|_{\particle{s}}$}

Constraints on a combination of $|q/p|_{\particle{d}}$ and 
$|q/p|_{\particle{s}}$ have been explicitly quoted by the Tevatron 
experiments, using inclusive semileptonic decays of \b hadrons:
\begin{eqnarray}
f'_{\particle{d}} \,\chid(1-|q/p|^2_{\particle{d}})+
 f'_{\particle{s}} \,\chis(1-|q/p|^2_{\particle{s}}) =
+0.006\pm 0.017 &~~~~& \mbox{CDF~\cite{CDF_CP_semi}} \,, 
\labe{CDF_ASLDS} \\
\frac{1}{4}\left(\ASLd +
\ASLs \frac{f'_\particle{s}\chis}{f'_\particle{d}\chid} \right) =
-0.0023 \pm 0.0011 \mbox{(stat)} \pm 0.0008 \mbox{(syst)}
&~~~~& \mbox{\dzero~\cite{D0_qp_chibar}} \,.
\labe{CDF_Dzero_ASLDS}
\end{eqnarray}
A first direct measurement of \ASLs and hence $|q/p|_{\particle{s}}$ has
been made by \dzero by measuring the charge asymmetry of
$\Bs \rightarrow D_s \mu \nu$ decays~\cite{D0_semi_asym}:
\begin{equation}
\ASLs =
+0.0245 \pm 0.0193 \mbox{(stat)} \pm 0.0035 \mbox{(syst)}\,.
\end{equation}

Given the average $\ASLd = \hfagASLDB$ of \Eq{ASLDB},
obtained from results at \B factories,
as well as other averages presented in this chapter
for the quantities appearing in 
\Eqss{CDF_ASLDS}{CDF_Dzero_ASLDS}, 
these three results are combined to yield
\begin{equation}
\ASLs = \hfagASLS 
\end{equation}
or, equivalently through \Eq{ASL},
\begin{equation}
|q/p|_{\particle{s}} = \hfagQPS \,.
\end{equation}

This result is compatible with no \CP violation in \Bs
mixing, an assumption made in almost all of the results described below. 


\input{life_mix_dgs}

\subsubsubsection{\boldmath Mass difference \dms}

Without doubt, the most striking \Bs physics result in 2006
is the first observation 
of \Bs oscillations by the CDF collaboration~\cite{CDF2_dms_observation},
based on large samples of flavour-tagged hadronic and semileptonic \Bs decays
(in flavour-specific final states), partially or fully reconstructed in 
$\rm 1~fb^{-1}$ of data collected during Tevatron's Run~II. 
From the proper-time dependence of these \Bs candidates, CDF
observe \Bs oscillations with a significance of at least $5\sigma$ 
and measure~\cite{CDF2_dms_observation}
\begin{equation}
\dms = \hfagDMSfull \,.  \labe{dms}
\end{equation}

Multiplying this result with the 
mean \Bs lifetime of \Eq{oneoverGs}, $1/\Gs=\hfagTAUBSMEANCON$,
yields
\begin{equation}
\xs = \frac{\dms}{\Gs} = \hfagXS \,. \labe{xs}
\end{equation}
With $2\ys = \DGGs=\hfagDGSGSCON$ (see \Eqss{DGGs_ave}{corrDGGs})
and under the assumption of no \CP violation in \Bs mixing,
this corresponds to
\begin{equation}
\chis = \frac{\xs^2+\ys^2}{2(\xs^2+1)} = \hfagCHIS \,. \labe{chis}
\end{equation}
The ratio of the \Bd and \Bs oscillation frequencies, 
obtained from \Eqss{dmd}{dms}, 
\begin{equation}
\frac{\dmd}{\dms} = \hfagRATIODMDDMS \,, \labe{dmd_over_dms}
\end{equation}
can be used to extract the following ratio of CKM matrix elements, 
\begin{equation}
\left|\frac{V_{td}}{V_{ts}}\right| =
\xi \sqrt{\frac{\dmd}{\dms}\frac{m(\Bs)}{m(\Bd)}} = 
\hfagVTDVTSfull \,, \labe{Vtd_over_Vts}
\end{equation}
where the first quoted error is from experimental uncertainties 
(with the masses $m(\Bs)$ and $m(\Bd)$ taken from~\cite{Yao2006PDG}),
and where the second quoted error is from theoretical uncertainties 
in the estimation of the SU(3) flavor-symmetry breaking factor
$\xi 
= \hfagXI$
obtained from lattice QCD calculations~\cite{xi_lattice_QCD}.

\Bs mesons were known to mix since many years. Indeed 
the time-integrated measurements of \chibar (see \Sec{chibar}),
when compared to our knowledge
of \chid and the \b-hadron fractions, indicated that \Bs mixing was large,
with a value of \chis close to its maximal possible value of $1/2$.
However, the time dependence of this mixing 
could not be observed until recently, 
mainly because of lack of proper-time resolution to resolve 
the small period of the \Bs\ oscillations.

The statistical significance ${\cal S}$ of a \Bs oscillation signal can be
approximated as~\cite{amplitude}
\begin{equation}
{\cal S} \approx \sqrt{\frac{N}{2}} \,f_{\rm sig}\, (1-2w)\,
\exp{\left(-\left(\dms\sigma_t\right)^2/2\right)}\,,
\labe{significance}
\end{equation}
where $N$ is 
the number of selected and tagged \Bs candidates, 
$f_{\rm sig}$ is the fraction of \Bs signal
in the selected and tagged sample, $w$ is the total mistag probability, 
and $\sigma_t$ is the resolution on proper time.
As can be seen, the quantity ${\cal S}$ decreases very quickly as 
\dms increases: this dependence is controlled by $\sigma_t$, 
which is therefore the most critical parameter for \dms analyses. 
The method widely used for \Bs oscillation searches
consists of measuring a \Bs oscillation amplitude ${\cal A}$
at several different test values of \dms, 
using a maximum likelihood fit based on the functions 
of \Eq{oscillations} where the cosine terms have been multiplied 
by ${\cal A}$.
One expects ${\cal A}=1$ at the true 
value of \dms and ${\cal A}=0$ at a test value of \dms 
(far) below the true value.
To a good approximation, the statistical uncertainty on ${\cal A}$
is Gaussian and equal to $1/{\cal S}$~\cite{amplitude}.
In any analysis, a particular value of \dms
can be excluded at \CL{95} if ${\cal A}+ 1.645\,\sigma_{\cal A} < 1$, 
where $\sigma_{\cal A}$ is the total uncertainty on ${\cal A}$.
Because of the proper time resolution, the quantity $\sigma_{\cal A}(\dms)$
is an increasing function of \dms (see \Eq{significance} which merely models  
$1/\sigma_{\cal A}(\dms)$ in an analysis limited 
by the available statistics). Therefore, 
if the true value of \dms were infinitely large, one 
expects to be able to exclude all values of \dms up to $\dms^{\rm sens}$, 
where $\dms^{\rm sens}$, called here the
sensitivity of the analysis, is defined by
$1.645\,\sigma_{\cal A}(\dms^{\rm sens}) = 1$. 

\begin{figure}
\begin{center}
\epsfig{figure=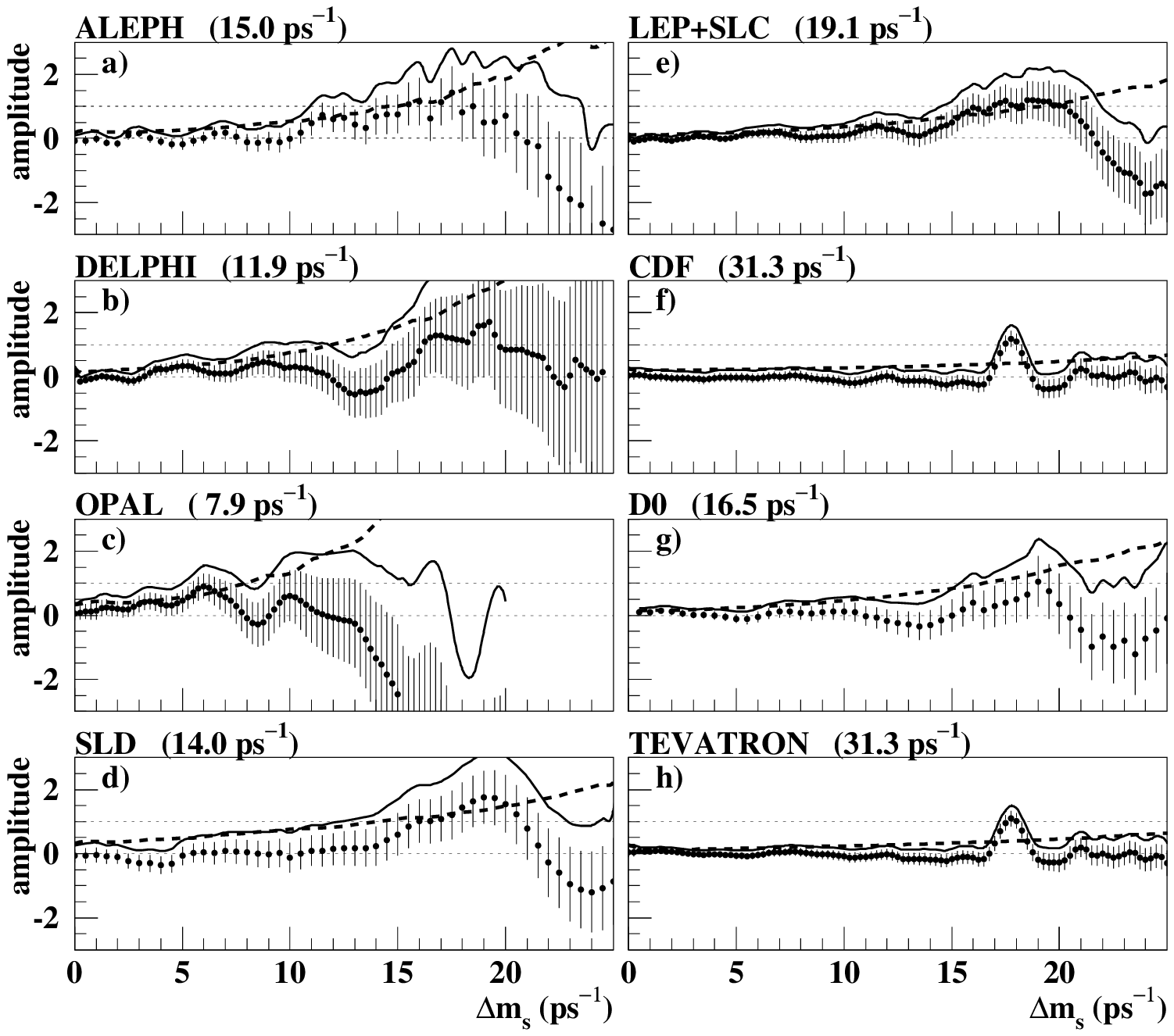,width=\textwidth,%
bbllx=55,bblly=55,bburx=490,bbury=490}
\caption{Combined \Bs-oscillation amplitude spectra, displayed separately
for each experiment. 
The points and error bars represent the measured
amplitude ${\cal A}$ and its total
uncertainty $\sigma_{\cal A}$, adjusted to a set of physics parameters
common to all analyses (including $\fBs=\hfagLFBS$ for LEP and SLC analyses).
Values of \dms for which the solid curve
(${\cal A}+1.645\,\sigma_{\cal A}$) is below 1 are excluded at \CL{95}.
The dashed curve shows $1.645\,\sigma_{\cal A}$; the number in parenthesis
indicates where this curve is equal to 1, and is a measure of the sensitivity 
for \CL{95} exclusion.
\newline
a) ALEPH~\cite{ALEPH_dms}, 
b) DELPHI~\cite{DELBS0_dms_dgs,DELBS1_dms_excl,DELPHI_dmd_dms_vtx,DELPHI_dms_last},
c) OPAL~\cite{OPAL_dms_l,OPAL_dms_dsl}, 
d) SLD~\cite{SLD_dms_dipole,SLD_dms_ds,SLD_dms_leptDvtx_unpublished},
\newline
e) ALEPH, DELPHI, OPAL and SLD combined,
f) CDF~\cite{CDF2_dms_observation,CDF1_dms}, 
g) \dzero~\cite{D0_dms_pub2006,D0_dms_prel2006}, 
h) CDF and \dzero combined.
}
\labf{individual_amplitudes_3}
\end{center}
\end{figure}

\begin{figure}
\begin{center}
\epsfig{figure=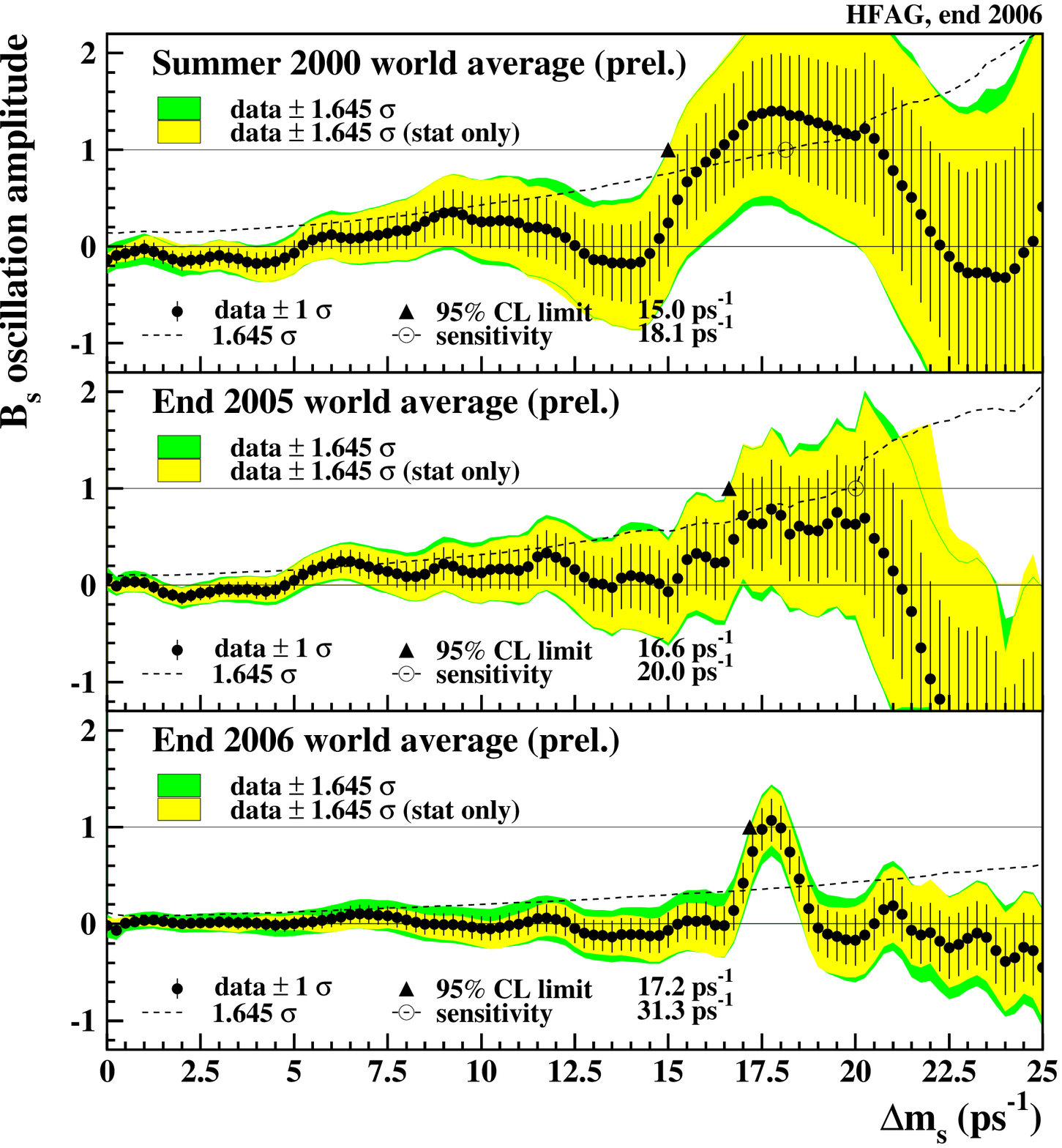,width=\textwidth}
\caption{World averages of all measurements
of the \Bs oscillation amplitude as a function of \dms.
\underline{Top:} situation in Summer 2000~\cite{LEPHFS}.
\underline{Middle:} situation at the end of 2005~\cite{hfag_hepex_endof2005}.
\underline{Bottom:} situation at the end of 2006, combining 
all published results~\cite{%
ALEPH_dms,DELBS0_dms_dgs,DELBS1_dms_excl,DELPHI_dmd_dms_vtx,DELPHI_dms_last,%
OPAL_dms_l,OPAL_dms_dsl,SLD_dms_dipole,SLD_dms_ds,%
CDF2_dms_observation,CDF1_dms,D0_dms_pub2006} 
as well as all recent preliminary results 
from Tevatron Run~II~\cite{D0_dms_prel2006}. 
The new results from CDF~\cite{CDF2_dms_observation} and
\dzero~\cite{D0_dms_pub2006,D0_dms_prel2006}, 
which became available in 2006,  
supersede all previous analyses
of Run~II data~\cite{CDF2_dms_fall2005_prel,D0_dms_dsmux_prel}. 
Statistical uncertainties dominate. 
Neighboring points are statistically correlated.}
\labf{amplitude}
\end{center}
\end{figure}

A large number of \Bs oscillation searches,
all based on the amplitude method, 
have been performed over the years by ALEPH~\cite{ALEPH_dms},
CDF~\cite{CDF2_dms_observation,CDF1_dms},
\dzero~\cite{D0_dms_pub2006,D0_dms_prel2006}, 
DELPHI~\cite{DELBS0_dms_dgs,DELBS1_dms_excl,DELPHI_dmd_dms_vtx,DELPHI_dms_last},
OPAL~\cite{OPAL_dms_l,OPAL_dms_dsl} and 
SLD~\cite{SLD_dms_dipole,SLD_dms_ds,SLD_dms_leptDvtx_unpublished}
(we omit references to searches that have been superseded
by more recent ones). They have been combined 
by averaging the measured amplitudes ${\cal A}$ at each test value 
of \dms.
The individual results have been adjusted to common physics inputs, 
and all known correlations have been accounted for; 
in the case of the inclusive (lepton) analyses, performed at LEP and SLC, 
the sensitivities (\ie\ 
the statistical uncertainties on ${\cal A}$), which depend directly 
through \Eq{significance} on the assumed fraction $f_{\rm sig}\sim\fBs$
of \Bs mesons in an unbiased sample of weakly-decaying \b hadrons, 
have also been rescaled to the LEP average $\fBs = \hfagLFBS$.

The combined amplitude spectra for the individual experiments are 
displayed in \Fig{individual_amplitudes_3}, and the world average spectrum is 
displayed in \Fig{amplitude}.
The appearance of the \Bs oscillation signal in 2006, which can clearly 
be seen by comparing the last two plots on \Fig{amplitude}, has been made 
possible by the latest analysis of the CDF data~\cite{CDF2_dms_observation},
which is, by far, the most sensitive one. 
It is interesting to note that a hint of a signal 
in the region 15--20\invps has been around since many years
at $e^+e^-\to\particle{Z}$ experiments,
and more recently at the \dzero experiment as well.

%% file: life_mix_definitions.tex
%
%
%
%

%

\renewcommand{\floatpagefraction}{0.8}
\renewcommand{\topfraction}{0.9}

\newcommand{\auth}[1]{#1,}
\newcommand{\coll}[1]{#1 Collaboration,}
\newcommand{\authcoll}[2]{#1 \etal\ (#2 Collaboration),}
\newcommand{\authgrp}[2]{#1 \etal\ (#2),}
\newcommand{\titl}[1]{``#1'',} 
\newcommand{\J}[4]{{#1} {\bf #2}, #3 (#4)}
\newcommand{\subJ}[1]{submitted to #1}
\newcommand{\PRL}[3]{\J{Phys.\ Rev.\ Lett.}{#1}{#2}{#3}}
\newcommand{\subPRL}{\subJ{Phys.\ Rev.\ Lett.}}
\newcommand{\PRD}[3]{\J{Phys.\ Rev.\ D}{#1}{#2}{#3}}
\newcommand{\subPRD}{\subJ{Phys.\ Rev.\ D}}
\newcommand{\PREP}[3]{\J{Phys.\ Reports}{#1}{#2}{#3}}
\newcommand{\ZPC}[3]{\J{Z.\ Phys.\ C}{#1}{#2}{#3}}
\newcommand{\PLB}[3]{\J{Phys.\ Lett.\ B}{#1}{#2}{#3}}
\newcommand{\subPLB}{\subJ{Phys.\ Lett.\ B}}
\newcommand{\EPJC}[3]{\J{Eur.\ Phys.\ J.\ C}{#1}{#2}{#3}}
\newcommand{\NPB}[3]{\J{Nucl.\ Phys.\ B}{#1}{#2}{#3}}
\newcommand{\subNPB}{\subJ{Nucl.\ Phys.\ B}}
\newcommand{\NIMA}[3]{\J{Nucl.\ Instrum.\ Methods A}{#1}{#2}{#3}}
\newcommand{\subNIMA}{\subJ{Nucl.\ Instrum.\ Methods A}}
\newcommand{\JHEP}[3]{\J{J.\ of High Energy Physics }{#1}{#2}{#3}}
\newcommand{\JPG}[3]{\J{J.\ of Physics G}{#1}{#2}{#3}}
\newcommand{\ARNS}[3]{\J{Ann.\ Rev.\ Nucl.\ Sci.}{#1}{#2}{#3}}
\newcommand{\newref}{\\}

\newcommand{\particle}[1]{\ensuremath{#1}\xspace}
\renewcommand{\ee}{\particle{e^+e^-}}
\newcommand{\Ups}{\particle{\Upsilon(4S)}}
\newcommand{\Upsfive}{\particle{\Upsilon(5S)}}
\renewcommand{\b}{\particle{b}}
\renewcommand{\B}{\particle{B}}
\newcommand{\Bd}{\particle{B^0}}
\renewcommand{\Bs}{\particle{B^0_s}}
\renewcommand{\Bu}{\particle{B^+}}
\newcommand{\Bc}{\particle{B^+_c}}
\newcommand{\Bdbar}{\particle{\bar{B}^0}}
\newcommand{\Bsbar}{\particle{\bar{B}^0_s}}
\newcommand{\Lb}{\particle{\Lambda_b^0}}
\newcommand{\Xib}{\particle{\Xi_b}}
\newcommand{\Lc}{\particle{\Lambda_c^+}}

\newcommand{\fBs}{\ensuremath{f_{\particle{s}}}\xspace}
\newcommand{\fBd}{\ensuremath{f_{\particle{d}}}\xspace}
\newcommand{\fBu}{\ensuremath{f_{\particle{u}}}\xspace}
\newcommand{\fbb}{\ensuremath{f_{\rm baryon}}\xspace}

\newcommand{\dmd}{\ensuremath{\Delta m_{\particle{d}}}\xspace}
\newcommand{\dms}{\ensuremath{\Delta m_{\particle{s}}}\xspace}
\newcommand{\xd}{\ensuremath{x_{\particle{d}}}\xspace}
\newcommand{\xs}{\ensuremath{x_{\particle{s}}}\xspace}
\newcommand{\yd}{\ensuremath{y_{\particle{d}}}\xspace}
\newcommand{\ys}{\ensuremath{y_{\particle{s}}}\xspace}
\newcommand{\chibar}{\ensuremath{\overline{\chi}}\xspace}
\newcommand{\chid}{\ensuremath{\chi_{\particle{d}}}\xspace}
\newcommand{\chis}{\ensuremath{\chi_{\particle{s}}}\xspace}
\newcommand{\Gd}{\ensuremath{\Gamma_{\particle{d}}}\xspace}
\newcommand{\DGd}{\ensuremath{\Delta\Gd}\xspace}
\newcommand{\DGGd}{\ensuremath{\DGd/\Gd}\xspace}
\newcommand{\Gs}{\ensuremath{\Gamma_{\particle{s}}}\xspace}
\newcommand{\DGs}{\ensuremath{\Delta\Gs}\xspace}
\newcommand{\DGGs}{\ensuremath{\Delta\Gs/\Gs}\xspace}
\newcommand{\ASLd}{\ensuremath{{\cal A}_{\rm SL}^\particle{d}}\xspace}
\newcommand{\ASLs}{\ensuremath{{\cal A}_{\rm SL}^\particle{s}}\xspace}

\renewcommand{\BR}[1]{\particle{{\cal B}(#1)}}
\newcommand{\CL}[1]{#1\%~\mbox{CL}}
\newcommand{\Qjet}{\ensuremath{Q_{\rm jet}}\xspace}

\newcommand{\labe}[1]{\label{equ:#1}}
\newcommand{\labs}[1]{\label{sec:#1}}
\newcommand{\labf}[1]{\label{fig:#1}}
\newcommand{\labt}[1]{\label{tab:#1}}
\newcommand{\refe}[1]{\ref{equ:#1}}
\newcommand{\refs}[1]{\ref{sec:#1}}
\newcommand{\reff}[1]{\ref{fig:#1}}
\newcommand{\reft}[1]{\ref{tab:#1}}
\newcommand{\Ref}[1]{Ref.~\cite{#1}}
\newcommand{\Refs}[1]{Refs.~\cite{#1}}
\newcommand{\Refss}[2]{Refs.~\cite{#1} and \cite{#2}}
\newcommand{\Refsss}[3]{Refs.~\cite{#1}, \cite{#2} and \cite{#3}}
\newcommand{\eq}[1]{(\refe{#1})}
\newcommand{\Eq}[1]{Eq.~(\refe{#1})}
\newcommand{\Eqs}[1]{Eqs.~(\refe{#1})}
\newcommand{\Eqss}[2]{Eqs.~(\refe{#1}) and (\refe{#2})}
\newcommand{\Eqssor}[2]{Eqs.~(\refe{#1}) or (\refe{#2})}
\newcommand{\Eqsss}[3]{Eqs.~(\refe{#1}), (\refe{#2}), and (\refe{#3})}
\newcommand{\Figure}[1]{Figure~\reff{#1}}
\newcommand{\Figuress}[2]{Figures~\reff{#1} and \reff{#2}}
\newcommand{\Fig}[1]{Fig.~\reff{#1}}
\newcommand{\Figs}[1]{Figs.~\reff{#1}}
\newcommand{\Figss}[2]{Figs.~\reff{#1} and \reff{#2}}
\newcommand{\Figsss}[3]{Figs.~\reff{#1}, \reff{#2}, and \reff{#3}}
\newcommand{\Section}[1]{Section~\refs{#1}}
\newcommand{\Sec}[1]{Sec.~\refs{#1}}
\newcommand{\Secs}[1]{Secs.~\refs{#1}}
\newcommand{\Secss}[2]{Secs.~\refs{#1} and \refs{#2}}
\newcommand{\Secsss}[3]{Secs.~\refs{#1}, \refs{#2}, and \refs{#3}}
\newcommand{\Table}[1]{Table~\reft{#1}}
\newcommand{\Tables}[1]{Tables~\reft{#1}}
\newcommand{\Tabless}[2]{Tables~\reft{#1} and \reft{#2}}
\newcommand{\Tablesss}[3]{Tables~\reft{#1}, \reft{#2}, and \reft{#3}}

\newcommand{\subsubsubsection}[1]{\vspace{2ex}\par\noindent {\bf\boldmath\em #1} \vspace{2ex}\par}

%% file: life_mix_averages.tex

\newcommand{\definemath}[2]{\newcommand{#1}{\ensuremath{#2}\xspace}}

\definemath{\hfagCHIBARLEPval}{0.1259}
\definemath{\hfagCHIBARLEPerr}{\pm0.0042}
\definemath{\hfagTAUBDval}{1.527}
\definemath{\hfagTAUBDerr}{\pm0.008}
\definemath{\hfagTAUBUval}{1.643}
\definemath{\hfagTAUBUerr}{\pm0.010}
\definemath{\hfagRTAUBUval}{1.076}
\definemath{\hfagRTAUBUerr}{\pm0.008}
\definemath{\hfagTAUBSval}{1.444}
\definemath{\hfagTAUBSerr}{\pm0.036}
\definemath{\hfagRTAUBSval}{0.945}
\definemath{\hfagRTAUBSerr}{\pm0.024}
\definemath{\hfagTAULBval}{1.393}
\definemath{\hfagTAULBerr}{\pm0.049}
\definemath{\hfagTAUXBval}{1.42}
\definemath{\hfagTAUXBerp}{^{+0.28}}
\definemath{\hfagTAUXBern}{_{-0.24}}
\definemath{\hfagTAUBBval}{1.325}
\definemath{\hfagTAUBBerr}{\pm0.039}
\definemath{\hfagRTAUBBval}{0.867}
\definemath{\hfagRTAUBBerr}{\pm0.026}
\definemath{\hfagTAUBval}{1.568}
\definemath{\hfagTAUBerr}{\pm0.009}
\definemath{\hfagTAUBCval}{0.460}
\definemath{\hfagTAUBCerr}{\pm0.066}
\definemath{\hfagTAUBSSLval}{1.440}
\definemath{\hfagTAUBSSLerr}{\pm0.036}
\definemath{\hfagTAUBSSLXval}{1.441}
\definemath{\hfagTAUBSSLXerr}{\pm0.037}
\definemath{\hfagTAUBSMEANCONval}{1.451}
\definemath{\hfagTAUBSMEANCONerp}{^{+0.029}}
\definemath{\hfagTAUBSMEANCONern}{_{-0.028}}
\definemath{\hfagTAUBSJFval}{1.404}
\definemath{\hfagTAUBSJFerr}{\pm0.066}
\definemath{\hfagRTAUBSSLval}{0.943}
\definemath{\hfagRTAUBSSLerr}{\pm0.024}
\definemath{\hfagRTAUBSMEANCONval}{0.950}
\definemath{\hfagRTAUBSMEANCONerr}{\pm0.019}
\definemath{\hfagRTAUBSMEANCONsig}{2.6}
\definemath{\hfagONEMINUSRTAUBSMEANCONpercent}{(5.0\pm1.9)\%}
\definemath{\hfagRTAULBval}{0.912}
\definemath{\hfagRTAULBerr}{\pm0.032}
\definemath{\hfagTAUBVTXval}{1.572}
\definemath{\hfagTAUBVTXerr}{\pm0.009}
\definemath{\hfagTAUBLEPval}{1.537}
\definemath{\hfagTAUBLEPerr}{\pm0.020}
\definemath{\hfagTAUBJPval}{1.533}
\definemath{\hfagTAUBJPerp}{^{+0.038}}
\definemath{\hfagTAUBJPern}{_{-0.034}}
\definemath{\hfagSDGDGDval}{0.009}
\definemath{\hfagSDGDGDerr}{\pm0.037}
\definemath{\hfagTAUBSMEANval}{1.494}
\definemath{\hfagTAUBSMEANerp}{^{+0.055}}
\definemath{\hfagTAUBSMEANern}{_{-0.054}}
\definemath{\hfagDGSGSval}{+0.206}
\definemath{\hfagDGSGSerp}{^{+0.106}}
\definemath{\hfagDGSGSern}{_{-0.111}}
\definemath{\hfagDGSGSlow}{-0.01}
\definemath{\hfagDGSGSupp}{+0.51}
\definemath{\hfagRHODGSGSTAUBSMEAN}{+0.39}
\definemath{\hfagDGSval}{+0.138}
\definemath{\hfagDGSerp}{^{+0.068}}
\definemath{\hfagDGSern}{_{-0.074}}
\definemath{\hfagDGSlow}{0.00}
\definemath{\hfagDGSupp}{+0.27}
\definemath{\hfagRHODGSTAUBSMEAN}{+0.33}
\definemath{\hfagTAUBSLval}{1.354}
\definemath{\hfagTAUBSLerp}{^{+0.065}}
\definemath{\hfagTAUBSLern}{_{-0.062}}
\definemath{\hfagTAUBSHval}{1.665}
\definemath{\hfagTAUBSHerp}{^{+0.143}}
\definemath{\hfagTAUBSHern}{_{+0.137}}
\definemath{\hfagDGSGSCONBDval}{N/A}
\definemath{\hfagDGSGSCONBDerp}{^{N/A}}
\definemath{\hfagDGSGSCONBDern}{_{N/A}}
\definemath{\hfagTAUBSMEANCONXval}{1.451}
\definemath{\hfagTAUBSMEANCONXerp}{^{+0.029}}
\definemath{\hfagTAUBSMEANCONXern}{_{-0.028}}
\definemath{\hfagDGSGSCONval}{+0.104}
\definemath{\hfagDGSGSCONerp}{^{+0.076}}
\definemath{\hfagDGSGSCONern}{_{-0.084}}
\definemath{\hfagDGSGSCONlow}{-0.07}
\definemath{\hfagDGSGSCONupp}{+0.25}
\definemath{\hfagRHODGSGSTAUBSMEANCON}{+0.08}
\definemath{\hfagDGSCONval}{0.071}
\definemath{\hfagDGSCONerp}{^{+0.053}}
\definemath{\hfagDGSCONern}{_{-0.057}}
\definemath{\hfagDGSCONlow}{-0.04}
\definemath{\hfagDGSCONupp}{+0.17}
\definemath{\hfagRHODGSTAUBSMEANCON}{+0.08}
\definemath{\hfagTAUBSLCONval}{1.380}
\definemath{\hfagTAUBSLCONerp}{^{+0.060}}
\definemath{\hfagTAUBSLCONern}{_{-0.057}}
\definemath{\hfagTAUBSHCONval}{1.531}
\definemath{\hfagTAUBSHCONerp}{^{+0.075}}
\definemath{\hfagTAUBSHCONern}{_{+0.069}}
\definemath{\hfagDGSGSCONBDCONval}{N/A}
\definemath{\hfagDGSGSCONBDCONerp}{^{N/A}}
\definemath{\hfagDGSGSCONBDCONern}{_{N/A}}
\definemath{\hfagBRDSDSval}{0.046}
\definemath{\hfagBRDSDSerr}{\pm0.022}
\definemath{\hfagDGSGSBRDSDSval}{+0.096}
\definemath{\hfagDGSGSBRDSDSerr}{\pm0.048}
\definemath{\hfagFCWval}{0.507}
\definemath{\hfagFCWerr}{\pm0.007}
\definemath{\hfagFNWval}{0.493}
\definemath{\hfagFNWerr}{\pm0.007}
\definemath{\hfagFFWval}{1.030}
\definemath{\hfagFFWerr}{\pm0.029}
\definemath{\hfagFCNval}{0.513}
\definemath{\hfagFCNerr}{\pm0.013}
\definemath{\hfagFNNval}{0.487}
\definemath{\hfagFNNerr}{\pm0.013}
\definemath{\hfagFFNval}{1.053}
\definemath{\hfagFFNerr}{\pm0.054}
\definemath{\hfagFCval}{0.505}
\definemath{\hfagFCerr}{\pm0.008}
\definemath{\hfagFNval}{0.495}
\definemath{\hfagFNerr}{\pm0.008}
\definemath{\hfagFFval}{1.021}
\definemath{\hfagFFerr}{\pm0.034}
\definemath{\hfagFPRODval}{0.497}
\definemath{\hfagFPRODerr}{\pm0.021}
\definemath{\hfagFSUMval}{0.984}
\definemath{\hfagFSUMerr}{\pm0.031}
\definemath{\hfagFSFIVEval}{0.199}
\definemath{\hfagFSFIVEsta}{\pm0.011}
\definemath{\hfagFSFIVEsys}{\pm0.030}
\definemath{\hfagFSFIVEerr}{\pm0.032}
\definemath{\hfagLFSFACTOR}{}
\definemath{\hfagLFBSNOMIXval}{0.088}
\definemath{\hfagLFBSNOMIXerr}{\pm0.016}
\definemath{\hfagLFBBNOMIXval}{0.100}
\definemath{\hfagLFBBNOMIXerr}{\pm0.017}
\definemath{\hfagLFBDNOMIXval}{0.406}
\definemath{\hfagLFBDNOMIXerr}{\pm0.010}
\definemath{\hfagWFSFACTOR}{1.1}
\definemath{\hfagWFBSNOMIXval}{0.093}
\definemath{\hfagWFBSNOMIXerr}{\pm0.018}
\definemath{\hfagWFBBNOMIXval}{0.099}
\definemath{\hfagWFBBNOMIXerr}{\pm0.019}
\definemath{\hfagWFBDNOMIXval}{0.404}
\definemath{\hfagWFBDNOMIXerr}{\pm0.011}
\definemath{\hfagCHIBARTEVval}{0.147}
\definemath{\hfagCHIBARTEVerr}{\pm0.011}
\definemath{\hfagCHIBARSFACTOR}{1.8}
\definemath{\hfagCHIBARval}{0.1284}
\definemath{\hfagCHIBARerr}{\pm0.0069}
\definemath{\hfagWFBSMIXval}{0.121}
\definemath{\hfagWFBSMIXerr}{\pm0.019}
\definemath{\hfagLFBSMIXval}{0.114}
\definemath{\hfagLFBSMIXerr}{\pm0.012}
\definemath{\hfagCHIDUval}{0.182}
\definemath{\hfagCHIDUerr}{\pm0.015}
\definemath{\hfagCHIDWUval}{0.1877}
\definemath{\hfagCHIDWUerr}{\pm0.0024}
\definemath{\hfagXDWval}{0.776}
\definemath{\hfagXDWerr}{\pm0.008}
\definemath{\hfagXDWUval}{0.775}
\definemath{\hfagXDWUerr}{\pm0.008}
\definemath{\hfagDMDWval}{0.508}
\definemath{\hfagDMDWsta}{\pm0.003}
\definemath{\hfagDMDWsys}{\pm0.003}
\definemath{\hfagDMDWerr}{\pm0.004}
\definemath{\hfagDMDWUval}{0.507}
\definemath{\hfagDMDWUerr}{\pm0.004}
\definemath{\hfagLFBSval}{0.104}
\definemath{\hfagLFBSerr}{\pm0.009}
\definemath{\hfagLFBBval}{0.091}
\definemath{\hfagLFBBerr}{\pm0.015}
\definemath{\hfagLFBDval}{0.402}
\definemath{\hfagLFBDerr}{\pm0.009}
\definemath{\hfagLRHOFBBFBS}{+0.029}
\definemath{\hfagLRHOFBDFBS}{-0.534}
\definemath{\hfagLRHOFBDFBB}{-0.860}
\definemath{\hfagWFBSval}{0.106}
\definemath{\hfagWFBSerr}{\pm0.013}
\definemath{\hfagWFBBval}{0.092}
\definemath{\hfagWFBBerr}{\pm0.018}
\definemath{\hfagWFBDval}{0.401}
\definemath{\hfagWFBDerr}{\pm0.010}
\definemath{\hfagWRHOFBBFBS}{-0.104}
\definemath{\hfagWRHOFBDFBS}{-0.517}
\definemath{\hfagWRHOFBDFBB}{-0.798}
\definemath{\hfagDMDHval}{0.496}
\definemath{\hfagDMDHsta}{\pm0.010}
\definemath{\hfagDMDHsys}{\pm0.009}
\definemath{\hfagDMDHerr}{\pm0.014}
\definemath{\hfagDMDBval}{0.508}
\definemath{\hfagDMDBsta}{\pm0.003}
\definemath{\hfagDMDBsys}{\pm0.003}
\definemath{\hfagDMDBerr}{\pm0.005}
\definemath{\hfagDMDTWODval}{0.509}
\definemath{\hfagDMDTWODsta}{\pm0.005}
\definemath{\hfagDMDTWODsys}{\pm0.003}
\definemath{\hfagDMDTWODerr}{\pm0.006}
\definemath{\hfagTAUBDTWODval}{1.527}
\definemath{\hfagTAUBDTWODsta}{\pm0.007}
\definemath{\hfagTAUBDTWODsys}{\pm0.007}
\definemath{\hfagTAUBDTWODerr}{\pm0.010}
\definemath{\hfagRHOstaDMDTAUBD}{-0.19}
\definemath{\hfagRHOsysDMDTAUBD}{-0.29}
\definemath{\hfagRHODMDTAUBD}{-0.23}
\definemath{\hfagQPDBval}{1.0024}
\definemath{\hfagQPDBerr}{\pm0.0023}
\definemath{\hfagQPDAval}{1.0033}
\definemath{\hfagQPDAerr}{\pm0.0017}
\definemath{\hfagASLDBval}{-0.0047}
\definemath{\hfagASLDBerr}{\pm0.0046}
\definemath{\hfagASLDAval}{-0.0064}
\definemath{\hfagASLDAerr}{\pm0.0034}
\definemath{\hfagREBDBval}{-0.0012}
\definemath{\hfagREBDBerr}{\pm0.0011}
\definemath{\hfagREBDAval}{-0.0016}
\definemath{\hfagREBDAerr}{\pm0.0009}
\definemath{\hfagASLSval}{+0.0003}
\definemath{\hfagASLSerr}{\pm0.0093}
\definemath{\hfagQPSval}{0.9998}
\definemath{\hfagQPSerr}{\pm0.0046}
\definemath{\hfagDMSWLIMval}{17.2}
\definemath{\hfagDMSWSENSval}{31.3}
\definemath{\hfagDMSXLIMval}{0.0}
\definemath{\hfagDMSXSENSval}{0.0}
\definemath{\hfagDMSDLIMval}{17.2}
\definemath{\hfagDMSDSENSval}{31.3}
\definemath{\hfagDMSWUPPval}{18.4}
\definemath{\hfagXSWLIMval}{24.5}
\definemath{\hfagCHISWLIMval}{0.49917}
\definemath{\hfagDMSval}{17.77}
\definemath{\hfagDMSsta}{\pm0.10}
\definemath{\hfagDMSsys}{\pm0.07}
\definemath{\hfagDMSerr}{\pm0.12}
\definemath{\hfagXSval}{25.8}
\definemath{\hfagXSerr}{\pm0.5}
\definemath{\hfagCHISval}{0.49925}
\definemath{\hfagCHISerr}{\pm0.00003}
\definemath{\hfagRATIODMDDMSval}{0.0286}
\definemath{\hfagRATIODMDDMSerr}{\pm0.0003}
\definemath{\hfagVTDVTSval}{0.2062}
\definemath{\hfagVTDVTSexx}{\pm0.0011}
\definemath{\hfagVTDVTSthp}{^{+0.0080}}
\definemath{\hfagVTDVTSthn}{_{-0.0060}}
\definemath{\hfagVTDVTSerp}{^{+0.0081}}
\definemath{\hfagVTDVTSern}{_{-0.0061}}
\definemath{\hfagXIval}{1.210}
\definemath{\hfagXIerp}{^{+0.047}}
\definemath{\hfagXIern}{_{-0.035}}

\newcommand{\unit}[1]{~\ensuremath{\rm #1}\xspace}
\renewcommand{\ps}{\unit{ps}}
\newcommand{\invps}{\unit{ps^{-1}}}
\newcommand{\TeV}{\unit{TeV}}
\newcommand{\MeVcc}{\unit{MeV/\mbox{$c$}^2}}
\newcommand{\MeV}{\unit{MeV}}

\definemath{\hfagCHIBARLEP}{\hfagCHIBARLEPval\hfagCHIBARLEPerr}
\definemath{\hfagTAUBD}{\hfagTAUBDval\hfagTAUBDerr\ps}
\definemath{\hfagTAUBDnounit}{\hfagTAUBDval\hfagTAUBDerr}
\definemath{\hfagTAUBU}{\hfagTAUBUval\hfagTAUBUerr\ps}
\definemath{\hfagTAUBUnounit}{\hfagTAUBUval\hfagTAUBUerr}
\definemath{\hfagRTAUBU}{\hfagRTAUBUval\hfagRTAUBUerr}
\definemath{\hfagTAUBS}{\hfagTAUBSval\hfagTAUBSerr\ps}
\definemath{\hfagTAUBSnounit}{\hfagTAUBSval\hfagTAUBSerr}
\definemath{\hfagRTAUBS}{\hfagRTAUBSval\hfagRTAUBSerr}
\definemath{\hfagTAULB}{\hfagTAULBval\hfagTAULBerr\ps}
\definemath{\hfagTAULBnounit}{\hfagTAULBval\hfagTAULBerr}
\definemath{\hfagTAUXBerr}{\hfagTAUXBerp\hfagTAUXBern}
\definemath{\hfagTAUXB}{\hfagTAUXBval\hfagTAUXBerr\ps}
\definemath{\hfagTAUXBnounit}{\hfagTAUXBval\hfagTAUXBerr}
\definemath{\hfagTAUBB}{\hfagTAUBBval\hfagTAUBBerr\ps}
\definemath{\hfagTAUBBnounit}{\hfagTAUBBval\hfagTAUBBerr}
\definemath{\hfagRTAUBB}{\hfagRTAUBBval\hfagRTAUBBerr}
\definemath{\hfagTAUB}{\hfagTAUBval\hfagTAUBerr\ps}
\definemath{\hfagTAUBnounit}{\hfagTAUBval\hfagTAUBerr}
\definemath{\hfagTAUBC}{\hfagTAUBCval\hfagTAUBCerr\ps}
\definemath{\hfagTAUBCnounit}{\hfagTAUBCval\hfagTAUBCerr}
\definemath{\hfagTAUBSSL}{\hfagTAUBSSLval\hfagTAUBSSLerr\ps}
\definemath{\hfagTAUBSSLnounit}{\hfagTAUBSSLval\hfagTAUBSSLerr}
\definemath{\hfagTAUBSSLX}{\hfagTAUBSSLXval\hfagTAUBSSLXerr\ps}
\definemath{\hfagTAUBSSLXnounit}{\hfagTAUBSSLXval\hfagTAUBSSLXerr}
\definemath{\hfagTAUBSMEANCONerr}{\hfagTAUBSMEANCONerp\hfagTAUBSMEANCONern}
\definemath{\hfagTAUBSMEANCON}{\hfagTAUBSMEANCONval\hfagTAUBSMEANCONerr\ps}
\definemath{\hfagTAUBSMEANCONnounit}{\hfagTAUBSMEANCONval\hfagTAUBSMEANCONerr}
\definemath{\hfagTAUBSJF}{\hfagTAUBSJFval\hfagTAUBSJFerr\ps}
\definemath{\hfagTAUBSJFnounit}{\hfagTAUBSJFval\hfagTAUBSJFerr}
\definemath{\hfagRTAUBSSL}{\hfagRTAUBSSLval\hfagRTAUBSSLerr}
\definemath{\hfagRTAUBSMEANCON}{\hfagRTAUBSMEANCONval\hfagRTAUBSMEANCONerr}
\definemath{\hfagRTAULB}{\hfagRTAULBval\hfagRTAULBerr}
\definemath{\hfagTAUBVTX}{\hfagTAUBVTXval\hfagTAUBVTXerr\ps}
\definemath{\hfagTAUBVTXnounit}{\hfagTAUBVTXval\hfagTAUBVTXerr}
\definemath{\hfagTAUBLEP}{\hfagTAUBLEPval\hfagTAUBLEPerr\ps}
\definemath{\hfagTAUBLEPnounit}{\hfagTAUBLEPval\hfagTAUBLEPerr}
\definemath{\hfagTAUBJPerr}{\hfagTAUBJPerp\hfagTAUBJPern}
\definemath{\hfagTAUBJP}{\hfagTAUBJPval\hfagTAUBJPerr\ps}
\definemath{\hfagTAUBJPnounit}{\hfagTAUBJPval\hfagTAUBJPerr}
\definemath{\hfagSDGDGD}{\hfagSDGDGDval\hfagSDGDGDerr}
\definemath{\hfagTAUBSMEANerr}{\hfagTAUBSMEANerp\hfagTAUBSMEANern}
\definemath{\hfagTAUBSMEAN}{\hfagTAUBSMEANval\hfagTAUBSMEANerr\ps}
\definemath{\hfagTAUBSMEANnounit}{\hfagTAUBSMEANval\hfagTAUBSMEANerr}
\definemath{\hfagDGSGSerr}{\hfagDGSGSerp\hfagDGSGSern}
\definemath{\hfagDGSGS}{\hfagDGSGSval\hfagDGSGSerr}
\definemath{\hfagDGSerr}{\hfagDGSerp\hfagDGSern}
\definemath{\hfagDGS}{\hfagDGSval\hfagDGSerr\invps}
\definemath{\hfagDGSnounit}{\hfagDGSval\hfagDGSerr}
\definemath{\hfagTAUBSLerr}{\hfagTAUBSLerp\hfagTAUBSLern}
\definemath{\hfagTAUBSL}{\hfagTAUBSLval\hfagTAUBSLerr\ps}
\definemath{\hfagTAUBSLnounit}{\hfagTAUBSLval\hfagTAUBSLerr}
\definemath{\hfagTAUBSHerr}{\hfagTAUBSHerp\hfagTAUBSHern}
\definemath{\hfagTAUBSH}{\hfagTAUBSHval\hfagTAUBSHerr\ps}
\definemath{\hfagTAUBSHnounit}{\hfagTAUBSHval\hfagTAUBSHerr}
\definemath{\hfagDGSGSCONBDerr}{\hfagDGSGSCONBDerp\hfagDGSGSCONBDern}
\definemath{\hfagDGSGSCONBD}{\hfagDGSGSCONBDval\hfagDGSGSCONBDerr}
\definemath{\hfagTAUBSMEANCONXerr}{\hfagTAUBSMEANCONXerp\hfagTAUBSMEANCONXern}
\definemath{\hfagTAUBSMEANCONX}{\hfagTAUBSMEANCONXval\hfagTAUBSMEANCONXerr\ps}
\definemath{\hfagTAUBSMEANCONXnounit}{\hfagTAUBSMEANCONXval\hfagTAUBSMEANCONXerr}
\definemath{\hfagDGSGSCONerr}{\hfagDGSGSCONerp\hfagDGSGSCONern}
\definemath{\hfagDGSGSCON}{\hfagDGSGSCONval\hfagDGSGSCONerr}
\definemath{\hfagDGSCONerr}{\hfagDGSCONerp\hfagDGSCONern}
\definemath{\hfagDGSCON}{\hfagDGSCONval\hfagDGSCONerr\invps}
\definemath{\hfagDGSCONnounit}{\hfagDGSCONval\hfagDGSCONerr}
\definemath{\hfagTAUBSLCONerr}{\hfagTAUBSLCONerp\hfagTAUBSLCONern}
\definemath{\hfagTAUBSLCON}{\hfagTAUBSLCONval\hfagTAUBSLCONerr\ps}
\definemath{\hfagTAUBSLCONnounit}{\hfagTAUBSLCONval\hfagTAUBSLCONerr}
\definemath{\hfagTAUBSHCONerr}{\hfagTAUBSHCONerp\hfagTAUBSHCONern}
\definemath{\hfagTAUBSHCON}{\hfagTAUBSHCONval\hfagTAUBSHCONerr\ps}
\definemath{\hfagTAUBSHCONnounit}{\hfagTAUBSHCONval\hfagTAUBSHCONerr}
\definemath{\hfagDGSGSCONBDCONerr}{\hfagDGSGSCONBDCONerp\hfagDGSGSCONBDCONern}
\definemath{\hfagDGSGSCONBDCON}{\hfagDGSGSCONBDCONval\hfagDGSGSCONBDCONerr}
\definemath{\hfagBRDSDS}{\hfagBRDSDSval\hfagBRDSDSerr}
\definemath{\hfagDGSGSBRDSDS}{\hfagDGSGSBRDSDSval\hfagDGSGSBRDSDSerr}
\definemath{\hfagFCW}{\hfagFCWval\hfagFCWerr}
\definemath{\hfagFNW}{\hfagFNWval\hfagFNWerr}
\definemath{\hfagFFW}{\hfagFFWval\hfagFFWerr}
\definemath{\hfagFCN}{\hfagFCNval\hfagFCNerr}
\definemath{\hfagFNN}{\hfagFNNval\hfagFNNerr}
\definemath{\hfagFFN}{\hfagFFNval\hfagFFNerr}
\definemath{\hfagFC}{\hfagFCval\hfagFCerr}
\definemath{\hfagFN}{\hfagFNval\hfagFNerr}
\definemath{\hfagFF}{\hfagFFval\hfagFFerr}
\definemath{\hfagFPROD}{\hfagFPRODval\hfagFPRODerr}
\definemath{\hfagFSUM}{\hfagFSUMval\hfagFSUMerr}
\definemath{\hfagFSFIVE}{\hfagFSFIVEval\hfagFSFIVEerr}
\definemath{\hfagFSFIVEfull}{\hfagFSFIVEval\hfagFSFIVEsta\hfagFSFIVEsys}
\definemath{\hfagLFBSNOMIX}{\hfagLFBSNOMIXval\hfagLFBSNOMIXerr}
\definemath{\hfagLFBBNOMIX}{\hfagLFBBNOMIXval\hfagLFBBNOMIXerr}
\definemath{\hfagLFBDNOMIX}{\hfagLFBDNOMIXval\hfagLFBDNOMIXerr}
\definemath{\hfagWFBSNOMIX}{\hfagWFBSNOMIXval\hfagWFBSNOMIXerr}
\definemath{\hfagWFBBNOMIX}{\hfagWFBBNOMIXval\hfagWFBBNOMIXerr}
\definemath{\hfagWFBDNOMIX}{\hfagWFBDNOMIXval\hfagWFBDNOMIXerr}
\definemath{\hfagCHIBARTEV}{\hfagCHIBARTEVval\hfagCHIBARTEVerr}
\definemath{\hfagCHIBAR}{\hfagCHIBARval\hfagCHIBARerr}
\definemath{\hfagWFBSMIX}{\hfagWFBSMIXval\hfagWFBSMIXerr}
\definemath{\hfagLFBSMIX}{\hfagLFBSMIXval\hfagLFBSMIXerr}
\definemath{\hfagCHIDU}{\hfagCHIDUval\hfagCHIDUerr}
\definemath{\hfagCHIDWU}{\hfagCHIDWUval\hfagCHIDWUerr}
\definemath{\hfagXDW}{\hfagXDWval\hfagXDWerr}
\definemath{\hfagXDWU}{\hfagXDWUval\hfagXDWUerr}
\definemath{\hfagDMDW}{\hfagDMDWval\hfagDMDWerr\invps}
\definemath{\hfagDMDWnounit}{\hfagDMDWval\hfagDMDWerr}
\definemath{\hfagDMDWfull}{\hfagDMDWval\hfagDMDWsta\hfagDMDWsys\invps}
\definemath{\hfagDMDWnounitfull}{\hfagDMDWval\hfagDMDWsta\hfagDMDWsys}
\definemath{\hfagDMDWU}{\hfagDMDWUval\hfagDMDWUerr\invps}
\definemath{\hfagDMDWUnounit}{\hfagDMDWUval\hfagDMDWUerr}
\definemath{\hfagLFBS}{\hfagLFBSval\hfagLFBSerr}
\definemath{\hfagLFBB}{\hfagLFBBval\hfagLFBBerr}
\definemath{\hfagLFBD}{\hfagLFBDval\hfagLFBDerr}
\definemath{\hfagWFBS}{\hfagWFBSval\hfagWFBSerr}
\definemath{\hfagWFBB}{\hfagWFBBval\hfagWFBBerr}
\definemath{\hfagWFBD}{\hfagWFBDval\hfagWFBDerr}
\definemath{\hfagDMDH}{\hfagDMDHval\hfagDMDHerr\invps}
\definemath{\hfagDMDHnounit}{\hfagDMDHval\hfagDMDHerr}
\definemath{\hfagDMDHfull}{\hfagDMDHval\hfagDMDHsta\hfagDMDHsys\invps}
\definemath{\hfagDMDHnounitfull}{\hfagDMDHval\hfagDMDHsta\hfagDMDHsys}
\definemath{\hfagDMDB}{\hfagDMDBval\hfagDMDBerr\invps}
\definemath{\hfagDMDBnounit}{\hfagDMDBval\hfagDMDBerr}
\definemath{\hfagDMDBfull}{\hfagDMDBval\hfagDMDBsta\hfagDMDBsys\invps}
\definemath{\hfagDMDBnounitfull}{\hfagDMDBval\hfagDMDBsta\hfagDMDBsys}
\definemath{\hfagDMDTWOD}{\hfagDMDTWODval\hfagDMDTWODerr\invps}
\definemath{\hfagDMDTWODnounit}{\hfagDMDTWODval\hfagDMDTWODerr}
\definemath{\hfagDMDTWODfull}{\hfagDMDTWODval\hfagDMDTWODsta\hfagDMDTWODsys\invps}
\definemath{\hfagDMDTWODnounitfull}{\hfagDMDTWODval\hfagDMDTWODsta\hfagDMDTWODsys}
\definemath{\hfagTAUBDTWOD}{\hfagTAUBDTWODval\hfagTAUBDTWODerr\ps}
\definemath{\hfagTAUBDTWODnounit}{\hfagTAUBDTWODval\hfagTAUBDTWODerr}
\definemath{\hfagTAUBDTWODfull}{\hfagTAUBDTWODval\hfagTAUBDTWODsta\hfagTAUBDTWODsys\ps}
\definemath{\hfagTAUBDTWODnounitfull}{\hfagTAUBDTWODval\hfagTAUBDTWODsta\hfagTAUBDTWODsys}
\definemath{\hfagQPDB}{\hfagQPDBval\hfagQPDBerr}
\definemath{\hfagQPDA}{\hfagQPDAval\hfagQPDAerr}
\definemath{\hfagASLDB}{\hfagASLDBval\hfagASLDBerr}
\definemath{\hfagASLDA}{\hfagASLDAval\hfagASLDAerr}
\definemath{\hfagREBDB}{\hfagREBDBval\hfagREBDBerr}
\definemath{\hfagREBDA}{\hfagREBDAval\hfagREBDAerr}
\definemath{\hfagASLS}{\hfagASLSval\hfagASLSerr}
\definemath{\hfagQPS}{\hfagQPSval\hfagQPSerr}
\definemath{\hfagDMSWLIM}{\hfagDMSWLIMval\invps}
\definemath{\hfagDMSWSENS}{\hfagDMSWSENSval\invps}
\definemath{\hfagDMSXLIM}{\hfagDMSXLIMval\invps}
\definemath{\hfagDMSXSENS}{\hfagDMSXSENSval\invps}
\definemath{\hfagDMSDLIM}{\hfagDMSDLIMval\invps}
\definemath{\hfagDMSDSENS}{\hfagDMSDSENSval\invps}
\definemath{\hfagDMSWUPP}{\hfagDMSWUPPval\invps}
\definemath{\hfagXSWLIM}{\hfagXSWLIMval}
\definemath{\hfagCHISWLIM}{\hfagCHISWLIMval}
\definemath{\hfagDMS}{\hfagDMSval\hfagDMSerr\invps}
\definemath{\hfagDMSnounit}{\hfagDMSval\hfagDMSerr}
\definemath{\hfagDMSfull}{\hfagDMSval\hfagDMSsta\hfagDMSsys\invps}
\definemath{\hfagDMSnounitfull}{\hfagDMSval\hfagDMSsta\hfagDMSsys}
\definemath{\hfagXS}{\hfagXSval\hfagXSerr}
\definemath{\hfagCHIS}{\hfagCHISval\hfagCHISerr}
\definemath{\hfagRATIODMDDMS}{\hfagRATIODMDDMSval\hfagRATIODMDDMSerr}
\definemath{\hfagVTDVTSerr}{\hfagVTDVTSerp\hfagVTDVTSern}
\definemath{\hfagVTDVTSthe}{\hfagVTDVTSthp\hfagVTDVTSthn}
\definemath{\hfagVTDVTS}{\hfagVTDVTSval\hfagVTDVTSerr}
\definemath{\hfagVTDVTSfull}{\hfagVTDVTSval\hfagVTDVTSexx\hfagVTDVTSthe}
\definemath{\hfagXIerr}{\hfagXIerp\hfagXIern}
\definemath{\hfagXI}{\hfagXIval\hfagXIerr}

%% file: life_mix_tau1.tex
%
%

\mysubsection{\b-hadron lifetimes}
\labs{lifetimes}

In the spectator model the decay of \b-flavored hadrons $H_b$ is
governed entirely by the flavor changing \particle{b\to Wq} transition
($\particle{q}=\particle{c,u}$).  For this very reason, lifetimes of all
\b-flavored hadrons are the same in the spectator approximation
regardless of the (spectator) quark content of the $H_b$.  In the early
1990's experiments became sophisticated enough to start seeing the
differences of the lifetimes among various $H_b$ species.  The first
theoretical calculations of the spectator quark effects on $H_b$
lifetime emerged only few years earlier.

Currently, most of such calculations are performed in the framework of
the Heavy Quark Expansion, HQE.  In the HQE, under certain assumptions
(most important of which is that of quark-hadron duality), the decay
rate of an $H_b$ to an inclusive final state $f$ is expressed as the sum
of a series of expectation values of operators of increasing dimension,
multiplied by the correspondingly higher powers of $\Lambda_{\rm
QCD}/m_b$:
\begin{equation}
\Gamma_{H_b\to f} = |CKM|^2\sum_n c_n^{(f)}
\Bigl(\frac{\Lambda_{\rm QCD}}{m_b}\Bigr)^n\langle H_b|O_n|H_b\rangle,
\labe{hqe}
\end{equation}
where $|CKM|^2$ is the relevant combination of the CKM matrix elements.
Coefficients $c_n^{(f)}$ of this expansion, known as Operator Product
Expansion~\cite{OPE}, can be calculated perturbatively.  Hence, the HQE
predicts $\Gamma_{H_b\to f}$ in the form of an expansion in both
$\Lambda_{\rm QCD}/m_{\b}$ and $\alpha_s(m_{\b})$.  The precision of
current experiments makes it mandatory to go to the next-to-leading
order in QCD, {\em i.e.}\ to include correction of the order of
$\alpha_s(m_{\b})$ to the $c_n^{(f)}$'s.  All non-perturbative physics
is shifted into the expectation values $\langle H_b|O_n|H_b\rangle$ of
operators $O_n$.  These can be calculated using lattice QCD or QCD sum
rules, or can be related to other observables via the
HQE~\cite{Bigi_1995}.  One may reasonably expect that powers of
$\Lambda_{\rm QCD}/m_{\b}$ provide enough suppression that only the
first few terms of the sum in \Eq{hqe} matter.

Theoretical predictions are usually made for the ratios of the lifetimes
(with $\tau(\Bd)$ chosen as the common denominator) rather than for the
individual lifetimes, for this allows several uncertainties to cancel.
The precision of the current HQE calculations (see
\Refs{nlo_lifetimes,tarantino,Gabbiani_et_al} for the latest updates)
is in some instances already surpassed by the measurements,
\eg\ in the case of $\tau(\Bu)/\tau(\Bd)$.  Also, HQE calculations are
not assumption-free.  More accurate predictions are a matter of progress
in the evaluation of the non-perturbative hadronic matrix elements and
verifying the assumptions that the calculations are based upon.
However, the HQE, even in its present shape, draws a number of important
conclusions, which are in agreement with experimental observations:
\begin{itemize}
\item The heavier the mass of the heavy quark the smaller is the
  variation in the lifetimes among different hadrons containing this
  quark, which is to say that as $m_{\b}\to\infty$ we retrieve the
  spectator picture in which the lifetimes of all $H_b$'s are the same.
   This is well illustrated by the fact that lifetimes are rather
   similar in the \b sector, while they differ by large factors
   in the \particle{c} sector ($m_{\particle{c}}<m_{\b}$).
\item The non-perturbative corrections arise only at the order of
  $\Lambda_{\rm QCD}^2/m_{\b}^2$, which translates into 
  differences among $H_b$ lifetimes of only a few percent.
\item It is only the difference between meson and baryon lifetimes that
  appears at the $\Lambda_{\rm QCD}^2/m_{\b}^2$ level.  The splitting of the
  meson lifetimes occurs at the $\Lambda_{\rm QCD}^3/m_{\b}^3$ level, yet it is
  enhanced by a phase space factor $16\pi^2$ with respect to the leading
  free \b decay.
\end{itemize}

To ensure that certain sources of systematic uncertainty cancel, 
lifetime analyses are sometimes designed to measure a 
ratio of lifetimes.  However, because of the differences in decay
topologies, abundance (or lack thereof) of decays of a certain kind,
{\em etc.}, measurements of the individual lifetimes are more 
common.  In the following section we review the most common
types of the lifetime measurements.  This discussion is followed by the
presentation of the averaging of the various lifetime measurements, each
with a brief description of its particularities.



\mysubsubsection{Lifetime measurements, uncertainties and correlations}

In most cases lifetime of an $H_b$ is estimated from a flight distance
and a $\beta\gamma$ factor which is used to convert the geometrical
distance into the proper decay time.  Methods of accessing lifetime
information can roughly be divided in the following five categories:
\begin{enumerate}
\item {\bf\em Inclusive (flavor blind) measurements}.  These
  measurements are aimed at extracting the lifetime from a mixture of
  \b-hadron decays, without distinguishing the decaying species.  Often
  the knowledge of the mixture composition is limited, which makes these
  measurements experiment-specific.  Also, these
  measurements have to rely on Monte Carlo for estimating the
  $\beta\gamma$ factor, because the decaying hadrons are not fully
  reconstructed.  On the bright side, these usually are the largest
  statistics \b-hadron lifetime measurements that are accessible to a
  given experiment, and can, therefore, serve as an important
  performance benchmark.
\item {\it\bf Measurements in semileptonic decays of a specific
  {\boldmath $H_b$\unboldmath}}.  \particle{W}from \particle{\b\to Wc}
  produces $\ell\nu_l$ pair (\particle{\ell=e,\mu}) in about 21\% of the
  cases.  Electron or muon from such decays is usually a well-detected
  signature, which provides for clean and efficient trigger.
  \particle{c} quark from \particle{b\to Wc} transition and the other
  quark(s) making up the decaying $H_b$ combine into a charm hadron,
  which is reconstructed in one or more exclusive decay channels.
  Knowing what this charmed hadron is allows one to separate, at least
  statistically, different $H_b$ species.  The advantage of these
  measurements is in statistics, which usually is superior to that of the
  exclusively reconstructed $H_b$ decays.  Some of the main
  disadvantages are related to the difficulty of estimating lepton+charm
  sample composition and Monte Carlo reliance for the $\beta\gamma$
  factor estimate.
\item {\bf\em Measurements in exclusively reconstructed hadronic decays}.
  These
  have the advantage of complete reconstruction of decaying $H_b$, which
  allows one to infer the decaying species as well as to perform precise
  measurement of the $\beta\gamma$ factor.  Both lead to generally
  smaller systematic uncertainties than in the above two categories.
  The downsides are smaller branching ratios, larger combinatoric
  backgrounds, especially in $H_b\rightarrow H_c\pi(\pi\pi)$ and
  multi-body $H_c$ decays, or in a hadron collider environment with
  non-trivial underlying event.  $H_b\to J/\psi H_s$ are relatively
  clean and easy to trigger on $J/\psi\to \ell^+\ell^-$, but their
  branching fraction is only about 1\%.
\item {\bf\em Measurements at asymmetric B factories}. In 
  the $\Ups\rightarrow B \bar{B}$ decay, the \B mesons (\Bu or \Bd) are
  essentially at rest in the \Ups rest frame.  This makes lifetime
  measurements impossible with experiments such as CLEO, in which \Ups
  produced at rest.  At asymmetric \B factories the \Ups meson is boosted
  resulting in \B and \particle{\bar{B}} moving nearly parallel to each
  other.  The lifetime is inferred from the distance $\Delta z$
  separating \B and \particle{\bar{B}} decay vertices and \Ups boost
  known from colliding beam energies.  
  In order to determine the charge of the \B mesons in each event, 
  one of the them is
  fully reconstructed in semileptonic or fully hadronic decay modes.
  The other \B is typically not fully reconstructed, only the position
  of its decay vertex is determined from the remaining tracks in the event.
  These measurements benefit from very large statistics, but suffer from
  poor $\Delta z$ resolution.
\item {\bf\em Direct measurement of lifetime ratios}.  This method has
  so far been only applied in the measurement of $\tau(\Bu)/\tau(\Bd)$.
  The ratio of the lifetimes is extracted from the dependence of the
  observed relative number of \Bu and \Bd candidates (both reconstructed
  in semileptonic decays) on the proper decay time.
\end{enumerate}

In some of the latest analyses, measurements of two (\eg\ $\tau(\Bu)$ and
$\tau(\Bu)/\tau(\Bd)$) or three (\eg\ $\tau(\Bu)$,
$\tau(\Bu)/\tau(\Bd)$, and \dmd) quantities are combined.  This
introduces correlations among measurements.  Another source of
correlations among the measurements are the systematic effects, which
could be common to an experiment or to an analysis technique across the
experiments.  When calculating the averages, such correlations are taken
into account per general procedure, described in
\Ref{lifetime_details}.

\mysubsubsection{Inclusive \b-hadron lifetimes}

The inclusive \b hadron lifetime is defined as $\tau_{\b} = \sum_i f_i
\tau_i$ where $\tau_i$ are the individual species lifetimes and $f_i$ are
the fractions of the various species present in an unbiased sample of
weakly-decaying \b hadrons produced at a high-energy
collider.\footnote{In principle such a quantity could be slightly
different in \particle{Z} decays and a the Tevatron, in case the
fractions of \b-hadron species are not exactly the same; see the
discussion in \Sec{fractions_high_energy}.}  This quantity is certainly
less fundamental than the lifetimes of the individual species, the
latter being much more useful in comparisons of the measurements with
the theoretical predictions.  Nonetheless, we perform the averaging of
the inclusive lifetime measurements for completeness as well as for the
reason that they might be of interest as ``technical numbers.''

\begin{table}[tp]
\caption{Measurements of average \b-hadron lifetimes.}
\labt{lifeincl}
\begin{center}
\begin{tabular}{lcccl} \hline
Experiment &Method           &Data set & $\tau_{\b}$ (ps)       &Ref.\\
\hline
ALEPH  &Dipole               &91     &$1.511\pm 0.022\pm 0.078$ &\cite{ALEIN2}\\
DELPHI &All track i.p.\ (2D) &91--92 &$1.542\pm 0.021\pm 0.045$ &\cite{DELIN0}$^a$\\
DELPHI &Sec.\ vtx            &91--93 &$1.582\pm 0.011\pm 0.027$ &\cite{DELIN}$^a$\\
DELPHI &Sec.\ vtx            &94--95 &$1.570\pm 0.005\pm 0.008$ &\cite{DELB04}\\
L3     &Sec.\ vtx + i.p.     &91--94 &$1.556\pm 0.010\pm 0.017$ &\cite{L3IN1}$^b$\\
OPAL   &Sec.\ vtx            &91--94 &$1.611\pm 0.010\pm 0.027$ &\cite{OPAIN2}\\
SLD    &Sec.\ vtx            &93     &$1.564\pm 0.030\pm 0.036$ &\cite{SLDIN}\\ 
\hline
\multicolumn{2}{l}{Average set 1 (\b vertex)} && \hfagTAUBVTXnounit &\\
\hline\hline
ALEPH  &Lepton i.p.\ (3D)    &91--93 &$1.533\pm 0.013\pm 0.022$ &\cite{ALEIN1}\\
L3     &Lepton i.p.\ (2D)    &91--94 &$1.544\pm 0.016\pm 0.021$ &\cite{L3IN1}$^b$\\
OPAL   &Lepton i.p.\ (2D)    &90--91 &$1.523\pm 0.034\pm 0.038$ &\cite{OPAIN1}\\ 
\hline
\multicolumn{2}{l}{Average set 2 ($\b\to\ell$)} && \hfagTAUBLEPnounit &\\
\hline\hline
CDF1   &\particle{J/\psi} vtx&92--95 &$1.533\pm 0.015^{+0.035}_{-0.031}$ &\cite{CDFIN_BS1} \\ 
\hline\hline
\multicolumn{2}{l}{Average of all above} && \hfagTAUBnounit & \\
\hline
\multicolumn{5}{l}{$^a$ \footnotesize The combined DELPHI result quoted in
\cite{DELIN} is 1.575 $\pm$ 0.010 $\pm$ 0.026 ps.} \\[-0.5ex]
\multicolumn{5}{l}{$^b$ \footnotesize The combined L3 result quoted in \cite{L3IN1} 
is 1.549 $\pm$ 0.009 $\pm$ 0.015 ps.}
\end{tabular}
\end{center}
\end{table}

In practice, an unbiased measurement of the inclusive lifetime is
difficult to achieve, because it would imply an efficiency which is
guaranteed to be the same across species.  So most of the measurements
are biased.  In an attempt to group analyses which are expected to
select the same mixture of \b hadrons, the available results (given in
\Table{lifeincl}) are divided into the following three sets:
\begin{enumerate}
\item measurements at LEP and SLD that accept any \b-hadron decay, based 
      on topological reconstruction (secondary vertex or track impact
      parameters);
\item measurements at LEP based on the identification
      of a lepton from a \b decay; and
\item measurements at the Tevatron based on inclusive 
      \particle{H_b\to J/\psi X} reconstruction, where the
      \particle{J/\psi} is fully reconstructed.
\end{enumerate}

The measurements of the first set are generally considered as estimates
of $\tau_{\b}$, although the efficiency to reconstruct a secondary
vertex most probably depends, in an analysis-specific way, on the number
of tracks coming from the vertex, thereby depending on the type of the
$H_b$.  Even though these efficiency variations can in principle be
accounted for using Monte Carlo simulations (which inevitably contain
assumptions on branching fractions), the $H_b$ mixture in that case can
remain somewhat ill-defined and could be slightly different among
analyses in this set.

On the contrary, the mixtures corresponding to the other two sets of
measurements are better defined in the limit where the reconstruction
and selection efficiency of a lepton or a \particle{J/\psi} from an
$H_b$ does not depend on the decaying hadron type.  These mixtures are
given by the production fractions and the inclusive branching fractions
for each $H_b$ species to give a lepton or a \particle{J/\psi}.  In
particular, under the assumption that all \b hadrons have the same
semileptonic decay width, the analyses of the second set should measure
$\tau(\b\to\ell) = (\sum_i f_i \tau_i^2) /(\sum_i f_i \tau_i)$ which is
necessarily larger than $\tau_{\b}$ if lifetime differences exist.
Given the present knowledge on $\tau_i$ and $f_i$,
$\tau(\b\to\ell)-\tau_{\b}$ is expected to be of the order of 0.01\ps.

Measurements by SLC and LEP experiments are subject to a number of
common systematic uncertainties, such as those due to (lack of knowledge
of) \b and \particle{c} fragmentation, \b and \particle{c} decay models,
\BR{B\to\ell}, \BR{B\to c\to\ell}, \BR{c\to\ell}, $\tau_{\particle{c}}$,
and $H_b$ decay multiplicity.  In the averaging, these systematic
uncertainties are assumed to be 100\% correlated.  The averages for the
sets defined above (also given in \Table{lifeincl}) are
\begin{eqnarray}
\tau(\b~\mbox{vertex}) &=& \hfagTAUBVTX \,,\\
\tau(\b\to\ell) &=& \hfagTAUBLEP  \,, \\
\tau(\b\to\particle{J/\psi}) &=& \hfagTAUBJP\,,
\end{eqnarray}
whereas an average of all measurements, ignoring mixture differences, 
yields \hfagTAUB.

\mysubsubsection{\Bd and \Bu lifetimes and their ratio}
\labs{taubd}
\labs{taubu}
\labs{lifetime_ratio}

\begin{table}[tp]
\caption{Measurements of the \Bd lifetime.}
\labt{lifebd}
\begin{center}
\begin{tabular}{lcccl} \hline
Experiment &Method                    &Data set &$\tau(\Bd)$ (ps)                  &Ref.\\
\hline
ALEPH  &\particle{D^{(*)} \ell}       &91--95 &$1.518\pm 0.053\pm 0.034$          &\cite{ALEB01}\\
ALEPH  &Exclusive                     &91--94 &$1.25^{+0.15}_{-0.13}\pm 0.05$     &\cite{ALEB0}\\
ALEPH  &Partial rec.\ $\pi^+\pi^-$    &91--94 &$1.49^{+0.17+0.08}_{-0.15-0.06}$   &\cite{ALEB0}\\
DELPHI &\particle{D^{(*)} \ell}       &91--93 &$1.61^{+0.14}_{-0.13}\pm 0.08$     &\cite{DELB01}\\
DELPHI &Charge sec.\ vtx              &91--93 &$1.63 \pm 0.14 \pm 0.13$           &\cite{DELB02}\\
DELPHI &Inclusive \particle{D^* \ell} &91--93 &$1.532\pm 0.041\pm 0.040$          &\cite{DELB03}\\
DELPHI &Charge sec.\ vtx              &94--95 &$1.531 \pm 0.021\pm0.031$          &\cite{DELB04}\\
L3     &Charge sec.\ vtx              &94--95 &$1.52 \pm 0.06 \pm 0.04$           &\cite{L3B01}\\
OPAL   &\particle{D^{(*)} \ell}       &91--93 &$1.53 \pm 0.12 \pm 0.08$           &\cite{OPAB0}\\
OPAL   &Charge sec.\ vtx              &93--95 &$1.523\pm 0.057\pm 0.053$          &\cite{OPAB1}\\
OPAL   &Inclusive \particle{D^* \ell} &91--00 &$1.541\pm 0.028\pm 0.023$          &\cite{OPAB2}\\
SLD    &Charge sec.\ vtx $\ell$       &93--95 &$1.56^{+0.14}_{-0.13} \pm 0.10$    &\cite{SLDB01}$^a$\\
SLD    &Charge sec.\ vtx              &93--95 &$1.66 \pm 0.08 \pm 0.08$           &\cite{SLDB01}$^a$\\
CDF1   &\particle{D^{(*)} \ell}       &92--95 &$1.474\pm 0.039^{+0.052}_{-0.051}$ &\cite{CDFB1}\\
CDF1  &Excl.\ \particle{J/\psi K^{*0}}&92--95 &$1.497\pm 0.073\pm 0.032$          &\cite{CDFB2}\\
CDF2  &Excl.\ \particle{J/\psi K^{*0}}&02--04 &$1.541\pm 0.050\pm0.020$           &\cite{CDF2_DGs}\\ 
CDF2   &Incl.\ \particle{D^{(*)} \ell}&02--04 &$1.473\pm 0.036\pm0.054$           &\cite{CDFB4}$^p$\\
CDF2   &Excl.\ \particle{D^-(3)\pi}   &02--04 &$1.511\pm 0.023\pm0.013$           &\cite{CDFB5}$^p$\\
CDF2   &Excl.\ \particle {J/\psi K_S} &02--06 &$1.524\pm 0.030\pm0.016$           &\cite{CDFLAM2} \\
\dzero &Excl.\ \particle{J/\psi K^{*0}}&02--05 &$1.530\pm0.043\pm0.023$ &\cite{D0B02,D0BS1}\\ 
\dzero &Excl.\ \particle {J/\psi K_S}  &02--06 &$1.492 \pm 0.075 \pm0.047$  &\cite{D0LAMB}$^p$ \\
\babar &Exclusive                     &99--00 &$1.546\pm 0.032\pm 0.022$          &\cite{BABAR1}\\
\babar &Inclusive \particle{D^* \ell} &99--01 &$1.529\pm 0.012\pm 0.029$          &\cite{BABAR2}\\
\babar &Exclusive \particle{D^* \ell} &99--02 &$1.523^{+0.024}_{-0.023}\pm 0.022$ &\cite{BABAR3}\\
\babar &Incl.\ \particle{D^*\pi}, \particle{D^*\rho} 
                                      &99--01 &$1.533\pm 0.034 \pm 0.038$         &\cite{BABAR4}\\
\babar &Inclusive \particle{D^* \ell}
&99--04 &$1.504\pm0.013^{+0.018}_{-0.013}$  &\cite{BABAR5} \\ 
\belle & Exclusive                     & 00--03 & $1.534\pm 0.008\pm0.010$        & \cite{BELLE2}\\
\hline
Average&                               &        & \hfagTAUBDnounit & \\
\hline\hline           
\multicolumn{5}{l}{$^a$ \footnotesize The combined SLD result 
quoted in \cite{SLDB01} is 1.64 $\pm$ 0.08 $\pm$ 0.08 ps.}\\[-0.5ex]
\multicolumn{5}{l}{$^p$ {\footnotesize Preliminary.}}
\end{tabular}
\end{center}
\end{table}

After a number of years of dominating these averages the LEP experiments
yielded the scene to the asymmetric \B~factories and
the Tevatron experiments.  The \B~factories have been very successful in
utilizing their potential -- in only a few years of running, \babar and,
to a greater extent, \belle, have struck a balance between the
statistical and the systematic uncertainties, with both being close to
(or even better than) the impressive 1\%.  In the meanwhile, CDF and
\dzero have emerged as significant contributors to the field as the
Tevatron Run~II data flowed in.  Both appear to enjoy relatively small
systematic effects, and while current statistical uncertainties of their
measurements are factors of 2 to 4 larger than those of their \B-factory
counterparts, both Tevatron experiments stand to increase their samples
by almost an order of magnitude.

\begin{table}[tbp]
\caption{Measurements of the \Bu lifetime.}
\labt{lifebu}
\begin{center}
\begin{tabular}{lcccl} \hline
Experiment &Method                 &Data set &$\tau(\Bu)$ (ps)                 &Ref.\\
\hline
ALEPH  &\particle{D^{(*)} \ell}    &91--95 &$1.648\pm 0.049\pm 0.035$          &\cite{ALEB01}\\
ALEPH  &Exclusive                  &91--94 &$1.58^{+0.21+0.04}_{-0.18-0.03}$   &\cite{ALEB0}\\
DELPHI &\particle{D^{(*)} \ell}    &91--93 &$1.61\pm 0.16\pm 0.12$             &\cite{DELB01}$^a$\\
DELPHI &Charge sec.\ vtx           &91--93 &$1.72\pm 0.08\pm 0.06$             &\cite{DELB02}$^a$\\
DELPHI &Charge sec.\ vtx           &94--95 &$1.624\pm 0.014\pm 0.018$          &\cite{DELB04}\\
L3     &Charge sec.\ vtx           &94--95 &$1.66\pm  0.06\pm 0.03$            &\cite{L3B01}\\
OPAL   &\particle{D^{(*)} \ell}    &91--93 &$1.52 \pm 0.14\pm 0.09$            &\cite{OPAB0}\\
OPAL   &Charge sec.\ vtx           &93--95 &$1.643\pm 0.037\pm 0.025$          &\cite{OPAB1}\\
SLD    &Charge sec.\ vtx $\ell$    &93--95 &$1.61^{+0.13}_{-0.12}\pm 0.07$     &\cite{SLDB01}$^b$\\
SLD    &Charge sec.\ vtx           &93--95 &$1.67\pm 0.07\pm 0.06$             &\cite{SLDB01}$^b$\\
CDF1   &\particle{D^{(*)} \ell}    &92--95 &$1.637\pm 0.058^{+0.045}_{-0.043}$ &\cite{CDFB1}\\
CDF1   &Excl.\ \particle{J/\psi K} &92--95 &$1.636\pm 0.058\pm 0.025$          &\cite{CDFB2}\\
CDF2   &Excl.\ \particle{J/\psi K} &02--04 &$1.662\pm 0.033\pm 0.008$          &\cite{CDFB3}$^p$\\
CDF2   &Incl.\ \particle{D^0 \ell} &02--04 &$1.653\pm 0.029^{+0.033}_{-0.031}$ &\cite{CDFB4}$^p$\\
CDF2   &Excl.\ \particle{D^0 \pi}  &02--04 &$1.661\pm 0.027\pm0.013$           &\cite{CDFB5}$^p$\\
\babar &Exclusive                  &99--00 &$1.673\pm 0.032\pm 0.023$          &\cite{BABAR1}\\
\belle &Exclusive                  &00--03 &$1.635\pm 0.011\pm 0.011$          &\cite{BELLE2}\\
\hline
Average&                           &       &\hfagTAUBUnounit &\\
\hline\hline
\multicolumn{5}{l}{$^a$ \footnotesize The combined DELPHI result quoted 
in~\cite{DELB02} is $1.70 \pm 0.09$ ps.} \\[-0.5ex]
\multicolumn{5}{l}{$^b$ \footnotesize The combined SLD result 
quoted in~\cite{SLDB01} is $1.66 \pm 0.06 \pm 0.05$ ps.}\\[-0.5ex]
\multicolumn{5}{l}{$^p$ {\footnotesize Preliminary.}}
\end{tabular}
\end{center}
\end{table}

At present time we are in an interesting position of having three sets
of measurements (from LEP/SLC, \B factories and the Tevatron) that
originate from different environments, obtained using substantially
different techniques and are precise enough for incisive comparison.


\begin{table}[tb]
\caption{Measurements of the ratio $\tau(\Bu)/\tau(\Bd)$.}
\labt{liferatioBuBd}
\begin{center}
\begin{tabular}{lcccl} 
\hline
Experiment &Method                 &Data set &Ratio $\tau(\Bu)/\tau(\Bd)$      &Ref.\\
\hline
ALEPH  &\particle{D^{(*)} \ell}    &91--95 &$1.085\pm 0.059\pm 0.018$          &\cite{ALEB01}\\
ALEPH  &Exclusive                  &91--94 &$1.27^{+0.23+0.03}_{-0.19-0.02}$   &\cite{ALEB0}\\
DELPHI &\particle{D^{(*)} \ell}    &91--93 &$1.00^{+0.17}_{-0.15}\pm 0.10$     &\cite{DELB01}\\
DELPHI &Charge sec.\ vtx           &91--93 &$1.06^{+0.13}_{-0.11}\pm 0.10$     &\cite{DELB02}\\
DELPHI &Charge sec.\ vtx           &94--95 &$1.060\pm 0.021 \pm 0.024$         &\cite{DELB04}\\
L3     &Charge sec.\ vtx           &94--95 &$1.09\pm 0.07  \pm 0.03$           &\cite{L3B01}\\
OPAL   &\particle{D^{(*)} \ell}    &91--93 &$0.99\pm 0.14^{+0.05}_{-0.04}$     &\cite{OPAB0}\\
OPAL   &Charge sec.\ vtx           &93--95 &$1.079\pm 0.064 \pm 0.041$         &\cite{OPAB1}\\
SLD    &Charge sec.\ vtx $\ell$    &93--95 &$1.03^{+0.16}_{-0.14} \pm 0.09$    &\cite{SLDB01}$^a$\\
SLD    &Charge sec.\ vtx           &93--95 &$1.01^{+0.09}_{-0.08} \pm0.05$     &\cite{SLDB01}$^a$\\
CDF1   &\particle{D^{(*)} \ell}    &92--95 &$1.110\pm 0.056^{+0.033}_{-0.030}$ &\cite{CDFB1}\\
CDF1   &Excl.\ \particle{J/\psi K} &92--95 &$1.093\pm 0.066 \pm 0.028$         &\cite{CDFB2}\\
CDF2   &Excl.\ \particle{J/\psi K} &02--04 &$1.080\pm 0.042$                   &\cite{CDFB3}$^p$\\
CDF2   &Incl.\ \particle{D \ell}   &02--04 &$1.123\pm0.040^{+0.041}_{-0.039}$  &\cite{CDFB4}$^p$\\
CDF2   &Excl.\ \particle{D \pi}    &02--04 &$1.10\pm 0.02\pm 0.01$             &\cite{CDFB5}$^p$\\
\dzero &\particle{D^{*+} \mu} \particle{D^0 \mu} ratio
	                           &02--04 &$1.080\pm 0.016\pm 0.014$          &\cite{D0B01}\\
\babar &Exclusive                  &99--00 &$1.082\pm 0.026\pm 0.012$          &\cite{BABAR1}\\
\belle &Exclusive                  &00--03 &$1.066\pm 0.008\pm 0.008$          &\cite{BELLE2}\\
\hline
Average&                           &       & \hfagRTAUBU & \\   
\hline\hline
\multicolumn{5}{l}{$^a$ \footnotesize The combined SLD result quoted
	   in~\cite{SLDB01} is $1.01 \pm 0.07 \pm 0.06$.} \\[-0.5ex] 
\multicolumn{5}{l}{$^p$ {\footnotesize Preliminary.}}
\end{tabular}
\end{center}
\end{table}

The averaging of $\tau(\Bu)$, $\tau(\Bd)$ and $\tau(\Bu)/\tau(\Bd)$
measurements is summarized in \Tablesss{lifebd}{lifebu}{liferatioBuBd}.
For $\tau(\Bu)/\tau(\Bd)$ we averaged only the measurements of this
quantity provided by experiments rather than using all available
knowledge, which would have included, for example, $\tau(\Bu)$ and
$\tau(\Bd)$ measurements which did not contribute to any of the ratio
measurements.

The following sources of correlated (within experiment/machine)
systematic uncertainties have been considered:
\begin{itemize}
\item for SLC/LEP measurements -- \particle{D^{**}} branching ratio uncertainties~\cite{LEPHFS},
momentum estimation of \b mesons from \particle{Z^0} decays
(\b-quark fragmentation parameter $\langle X_E \rangle = 0.702 \pm 0.008$~\cite{LEPHFS}),
\Bs and \b baryon lifetimes (see \Secss{taubs}{taulb}),
and \b-hadron fractions at high energy (see \Table{fractions}); 
\item for \babar measurements -- alignment, $z$ scale, PEP-II boost,
sample composition (where applicable);
\item for \dzero and CDF Run~II measurements -- alignment (separately
within each experiment).
\end{itemize}
The resultant averages are:
\begin{eqnarray}
\tau(\Bd) & = & \hfagTAUBD \,, \\
\tau(\Bu) & = & \hfagTAUBU \,, \\
\tau(\Bu)/\tau(\Bd) & = & \hfagRTAUBU \,.
\end{eqnarray}

%% file: life_mix_tau2.tex
%
%
%

\mysubsubsection{\Bs lifetime}
\labs{taubs}

Similar to the kaon system, neutral \B mesons contain
short- and long-lived components, since the
light (L) and heavy  (H)
eigenstates, $\B_{\rm L}$ and $\B_{\rm H}$, differ not only
in their masses, but also in their widths 
with $\Delta\Gamma = \Gamma_{\rm L} - \Gamma_{\rm H}$. 
In the case of the \Bs system, $\DGs$ can
be particularly large. The current theoretical
prediction in the Standard Model for
the fractional width difference is
$\DGs = 0.096 \pm 0.039$~\cite{delta_gams},
where $\Gs = (\Gamma_{\rm L} + \Gamma_{\rm H})/2$.
Specific measurements of \DGs and \Gs are explained
in \Sec{DGs}, but the result for
\Gs is quoted here.

Neglecting \CP violation in $\Bs-\Bsbar$ mixing, 
which is expected to be small~\cite{delta_gams}, the
\Bs mass eigenstates are also \CP eigenstates. In
the Standard Model assuming no \CP violation in
the \Bs system,
$\Gamma_{\rm L}$ is the width of
the \CP-even state and
$\Gamma_{\rm H}$ the width of
the \CP-odd state.
Final states can be decomposed into
\CP-even and \CP-odd components, each with a different
lifetime.

In view of a possibly substantial width difference,
and the fact that various
decay channels will have different proportions of 
the $\B_{\rm L}$ and $\B_{\rm H}$ eigenstates,
the straight average of all available 
\Bs lifetime measurements
is rather ill-defined.  Therefore,
the \Bs lifetime measurements are broken down into
four categories and averaged separately.

\begin{itemize}
\item 
{\bf\em Flavor-specific decays}, such as semileptonic
$\particle{B_s} \to \particle{D_s \ell \nu}$
or $\particle{B_s} \to \particle {D_s \pi}$, will
have equal 
fractions of $\B_{\rm L}$ and $\B_{\rm H}$ at time
zero, where
$\tau_{\rm L} = 1/\Gamma_{\rm L}$ 
is expected to be the shorter-lived component and
$\tau_{\rm H} = 1/\Gamma_{\rm H}$ 
expected
to be the longer-lived component.  A superposition
of two exponentials thus results with decay
widths $\Gs \pm \DGs /2$.
Fitting to a single exponential one obtains a
measure of the flavor-specific 
lifetime~\cite{Hartkorn_Moser}:
\begin{equation}
\tau(\Bs)_{\rm fs} = \frac{1}{\Gs}
\frac{{1+\left(\frac{\DGs}{2\Gs}\right)^2}}{{1-\left(\frac{\DGs}{2\Gs}\right)^2}
}.
\end{equation}
As given in \Table{lifebs}, the flavor-specific 
\Bs lifetime world average is:
\begin{equation}
\tau(\Bs)_{\rm fs} = \hfagTAUBSSL \,.
\end{equation}
This world average will be used later in \Sec{DGs} in combination
with other measurements to find
$\bar{\tau}(\Bs) = 1/\Gs$ and $\DGs$.

The following correlated systematic errors were considered:
average \B lifetime used in backgrounds,
\Bs decay multiplicity, and branching ratios used to determine 
backgrounds (\eg\ \BR{B\to D_s D}).
A knowledge of the multiplicity of \Bs decays is important for
measurements that partially reconstruct the final state such as 
\particle{\B\to D_s \mbox{$X$}} (where $X$ is not a lepton). 
The boost deduced from Monte Carlo simulation depends on the multiplicity used.
Since this is not well known, the multiplicity in the simulation is
varied and this range of values observed is taken to be a systematic.
Similarly not all the branching ratios for the potential background
processes are measured. Where they are available, the PDG values are
used for the error estimate. Where no measurements are available
estimates can usually be made by using measured branching ratios of
related processes and using some reasonable extrapolation.
\end{itemize}



\begin{table}[tb]
\caption{Measurements of the \Bs lifetime.}
\labt{lifebs}
\begin{center}
\begin{tabular}{lcccl} \hline
Experiment & Method           & Data set & $\tau(\Bs)$ (ps)               & Ref. \\
\hline
ALEPH  & \particle{D_s \ell}  & 91--95 & $1.54^{+0.14}_{-0.13}\pm 0.04$   & \cite{ALEBS1}          \\
CDF1   & \particle{D_s \ell}  & 92--96 & $1.36\pm 0.09 ^{+0.06}_{-0.05}$  & \cite{CDFBS}           \\
DELPHI & \particle{D_s \ell}  & 91--95 & $1.42^{+0.14}_{-0.13}\pm 0.03$   & \cite{DELBS0_dms_dgs}          \\
OPAL   & \particle{D_s \ell}  & 90--95 & $1.50^{+0.16}_{-0.15}\pm 0.04$   & \cite{OPABS1_OPALAM2}  \\
\dzero & \particle{D_s \mu}  & 02--04 & $1.398 \pm 0.044 ^{+0.028}_{-0.025}   $   & \cite{D0BS2}       \\ 
CDF2   & \particle{D_s \pi, D_s \pi \pi \pi} 
                              & 02--04 & $1.60 \pm 0.10 \pm 0.02      $   & \cite{CDFBS2}$^p$      \\
CDF2   & \particle{D_s \ell}  & 02--04 & $1.381 \pm 0.055 ^{+0.052}_{-0.046} $ &
\cite{CDFBS3}$^p$ \\ \hline
\multicolumn{3}{l}{Average of flavor-specific measurements} &  \hfagTAUBSSLnounit & \\  
\hline
ALEPH  & \particle{D_s h}     & 91--95 & $1.47\pm 0.14\pm 0.08$           & \cite{ALEBS2}          \\
DELPHI & \particle{D_s h}     & 91--95 & $1.53^{+0.16}_{-0.15}\pm 0.07$   & \cite{DELBS1_dms_excl} \\
OPAL   & \particle{D_s} incl. & 90--95 & $1.72^{+0.20+0.18}_{-0.19-0.17}$ & \cite{OPABS2}          \\ 
\hline
\multicolumn{3}{l}{Average of all above \particle{D_s} measurements} &  \hfagTAUBSnounit & \\ 
\hline\hline
CDF1     & \particle{J/\psi\phi} & 92--95  & $1.34^{+0.23}_{-0.19}    \pm 0.05$ & \cite{CDFIN_BS1} \\
CDF2     & \particle{J/\psi\phi} & 02--04  & $1.369 \pm 0.100 ^{+0.008}_{-0.010}$ & \cite{CDFB3}$^p$ \\
\dzero   & \particle{J/\psi\phi} & 02--04  & $1.444^{+0.098}_{-0.090} \pm 0.02$ & \cite{D0BS1}  \\ \hline 
\multicolumn{3}{l}{Average of \particle{J/\psi \phi} measurements} &  \hfagTAUBSJFnounit & \\ 
\hline
\multicolumn{5}{l}{$^p$ \footnotesize Preliminary.}
\end{tabular}
\end{center}
\end{table}

\begin{itemize}
\item
{\bf\em \boldmath $\Bs\to\Ds X$ decays}.
Included in \Table{lifebs} are measurements
of lifetimes using samples of \particle{\Bs} decays to
\particle{D_s} plus
hadrons, and hence into a less known mixture
of \CP-states.  A lifetime
weighted this way can still be a useful input
for analyses examining such an inclusive sample.
These are separated in \Table{lifebs} and combined
with the semileptonic lifetime to obtain:
\begin{equation}
\tau(\Bs)_{\particle{D_s {\rm X}}} = \hfagTAUBS \,.
\end{equation}

\item
{\bf\em Fully exclusive 
{\boldmath \Bs $\to J/\psi \phi$ \unboldmath}decays}
are expected to be
dominated by the \CP-even state and its lifetime.
First measurements of the \CP mix for this decay mode
are outlined in \Sec{DGs}.
CDF and \dzero measurements from this particular mode
\particle{\Bs\to J/\psi\phi} are combined into an
average
given in \Table{lifebs}.  There are no correlations
between the measurements for this fully exclusive
channel, and the world average for this 
specific decay is:
\begin{equation}
\tau(\Bs)_{\particle{J/\psi \phi}} = \hfagTAUBSJF \,.
\end{equation}
A caveat is that different experimental acceptances
will likely lead to different admixtures of the 
\CP-even and \CP-odd states, and fits to a single
exponential may result in inherently different 
measurements of these quantities.

\item
{\bf\em Fully exclusive 
{\boldmath \Bs $\to K^+ K^-$ \unboldmath}decays}
are expected to be
\CP even to within 5\%, and hence measures the lifetime
of the ``light" mass eigenstate
$\tau_L = 1/\Gamma_L$.  The measurement of this 
lifetime from CDF in Run~II~\cite{CDFBS4}
is:
\begin{equation}
\tau(\Bs)_{\particle{K^+K^-}} = 1.53 \pm 0.18 \pm 0.02 
\thinspace {\mathrm ps}, 
\end{equation}
and will be used as an input in \Sec{DGs} for the
average described below.
\end{itemize}

Finally, as will be shown in \Sec{DGs}, measurements
of $\DGs$, including separation into
\CP-even and \CP-odd components, give
\begin{equation}
\bar{\tau}(\Bs) = 1/\Gs = \hfagTAUBSMEAN \,,
\end{equation}
and when combined with the flavor-specific lifetime
measurements:
\begin{equation}
\bar{\tau}(\Bs) = 1/\Gs = \hfagTAUBSMEANCON \,.
\end{equation}

\mysubsubsection{\Bc lifetime}
\labs{taubc}

There are currently three measurements of the lifetime of the \Bc meson
from CDF~\cite{CDFBC1,CDFBC2} and \dzero~\cite{D0BC1} using the semileptonic decay
mode \particle{\Bc \to J/\psi \ell} and fitting
simultaneously to the mass and lifetime using the vertex formed
with the leptons from the decay of the \particle{J/\psi} and
the third lepton. Correction factors
to estimate the boost due to the missing neutrino are used.
Mass values of
$6.40 \pm 0.39 \pm 0.13$~GeV/$c^2$ for the CDF Run~I 
result~\cite{CDFBC1} and 
$5.95^{+0.14}_{-0.13} \pm 0.34$~GeV/$c^2$ for the \dzero Run~II 
result~\cite{D0BC1} are
found by fitting
to the tri-lepton invariant mass spectrum. 
In the CDF Run~II result~\cite{CDFBC2}, the mass is fixed
to 6.271~GeV/$c^2$, but then varied between 
6.2 and 6.4~GeV/$c^2$ to assess the systematic error on the
lifetime due to the \Bc mass value.
These mass measurements
are consistent within uncertainties, and also consistent with the
precision mass measurement from CDF of 
$6285.7 \pm 5.3 \pm 1.2$~MeV/$c^2$~\cite{BCmass}.
Correlated systematic errors include the impact
of the uncertainty of the \Bc $p_T$ spectrum on the correction
factors, the level of feed-down from $\psi(2S)$, 
MC modeling of the decay model varying from phase space
to the ISGW model, and mass variations.
Values of the \particle{\Bc} lifetime are given
in \Table{lifebc} and the world average is
determined to be:
\begin{equation}
\tau(\Bc) = \hfagTAUBC \,.
\end{equation}

\begin{table}[tb]
\caption{Measurements of the \Bc lifetime.}
\labt{lifebc}
\begin{center}
\begin{tabular}{lcccl} \hline
Experiment & Method                    & Data set  & $\tau(\Bc)$ (ps)
      & Ref.\\   \hline
CDF1       & \particle{J/\psi \ell} & 92--95  & $0.46^{+0.18}_{-0.16} \pm
 0.03$   & \cite{CDFBC1}  \\ 
CDF2       & \particle{J/\psi e} & 02--04  & $0.463^{+0.073}_{-0.065} \pm
 0.036$   & \cite{CDFBC2} \\
 \dzero & \particle{J/\psi \mu} & 02--04  & $0.448^{+0.123}_{-0.096} 
\pm  0.121$   & \cite{D0BC1}$^p$  \\ \hline
  \multicolumn{2}{l}{Average} &   &  \hfagTAUBCnounit
                 &    \\   \hline
\multicolumn{5}{l}{$^p$ \footnotesize Preliminary.}
\end{tabular}
\end{center}
\end{table}

\mysubsubsection{\Lb and \b-baryon lifetimes}
\labs{taulb}

The most precise measurements of the \b-baryon lifetime
originate from two classes of partially reconstructed decays.
In the first class, decays with an exclusively 
reconstructed \Lc baryon
and a lepton of opposite charge are used. These products are
more likely to occur in the decay of \Lb baryons.
In the second class, more inclusive final states with a baryon
(\particle{p}, \particle{\bar{p}}, $\Lambda$, or $\bar{\Lambda}$) 
and a lepton have been used, and these final states can generally
arise from any \b baryon.

The following sources of correlated systematic uncertainties have 
been considered:
experimental time resolution within a given experiment, \b-quark
fragmentation distribution into weakly decaying \b baryons,
\Lb polarization, decay model,
and evaluation of the \b-baryon purity in the selected event samples.
In computing the averages
the central values of the masses are scaled to 
$M(\Lb) = 5620 \pm 2\MeVcc$~\cite{PDGmass} and
$M(\mbox{\b-baryon}) = 5670 \pm 100\MeVcc$.

The meaning of decay model and the correlations are not always clear.
Uncertainties related to the decay model are dominated by
assumptions on the fraction of $n$-body decays.
To be conservative it is assumed
that it is correlated whenever given as an error.
DELPHI varies the fraction of 4-body decays from 0.0 to 0.3. 
In computing the average, the DELPHI
result is corrected for $0.2 \pm 0.2$.

Furthermore, in computing the average,
the semileptonic decay results from LEP are corrected for a polarization of 
$-0.45^{+0.19}_{-0.17}$~\cite{LEPHFS} and  a 
\Lb fragmentation parameter
$\langle X_E \rangle =0.70\pm 0.03$~\cite{LBFRAG}.




\begin{table}[t]
\caption{Measurements of the \b-baryon lifetimes.
}
\labt{lifelb}
\begin{center}
\begin{tabular}{lcccl} 
\hline
Experiment&Method                &Data set& Lifetime (ps) & Ref. \\\hline
ALEPH  &$\Lc\ell$             & 91--95 &$1.18^{+0.13}_{-0.12} \pm 0.03$ & \cite{ALEPH_fla}$^a$\\
ALEPH  &$\Lambda\ell^-\ell^+$ & 91--95 &$1.30^{+0.26}_{-0.21} \pm 0.04$ & \cite{ALEPH_fla}$^a$\\
CDF1   &$\Lc\ell$             & 91--95 &$1.32 \pm 0.15        \pm 0.07$ & \cite{CDFLAM}\\
CDF2   &$J/\psi \Lambda$      & 02--06 &$1.593^{+0.083}_{-0.078} \pm 0.033$ & \cite{CDFLAM2} \\
\dzero &$J/\psi \Lambda$      & 02--06 &$1.298 \pm 0.137 \pm 0.050$ & \cite{D0LAMB}$^p$ \\
\dzero &$\Lc\mu$              & 02--06 &$1.28^{+0.12}_{-0.11} \pm 0.09$ & 
\cite{D0LAMB2}$^p$ \\
DELPHI &$\Lc\ell$             & 91--94 &$1.11^{+0.19}_{-0.18} \pm 0.05$ & 
\cite{DELLAM0}$^b$\\
OPAL   &$\Lc\ell$, $\Lambda\ell^-\ell^+$ 
                                 & 90--95 & $1.29^{+0.24}_{-0.22} \pm 0.06$ & \cite{OPABS1_OPALAM2}\\ 
\hline
\multicolumn{3}{l}{Average of above 8 (\Lb lifetime)} & \hfagTAULBnounit & \\
\hline
ALEPH  &$\Lambda\ell$         & 91--95 &$1.20 \pm 0.08 \pm 0.06$ & \cite{ALEPH_fla}\\
DELPHI &$\Lambda\ell\pi$ vtx  & 91--94 &$1.16 \pm 0.20 \pm 0.08$        & \cite{DELLAM0}$^b$\\
DELPHI &$\Lambda\mu$ i.p.     & 91--94 &$1.10^{+0.19}_{-0.17} \pm 0.09$ & \cite{DELLAM1}$^b$ \\
DELPHI &\particle{p\ell}      & 91--94 &$1.19 \pm 0.14 \pm 0.07$        & \cite{DELLAM0}$^b$\\
OPAL   &$\Lambda\ell$ i.p.    & 90--94 &$1.21^{+0.15}_{-0.13} \pm 0.10$ & \cite{OPALAM1}$^c$  \\
OPAL   &$\Lambda\ell$ vtx     & 90--94 &$1.15 \pm 0.12 \pm 0.06$        & \cite{OPALAM1}$^c$ \\ 
\hline
\multicolumn{3}{l}{Average of above 14 (\b-baryon lifetime)} & \hfagTAUBBnounit & \\  
\hline\hline
ALEPH  &$\Xi\ell$             & 90--95 &$1.35^{+0.37+0.15}_{-0.28-0.17}$ & \cite{ALEPH_fxi}\\
DELPHI &$\Xi\ell$             & 91--93 &$1.5 ^{+0.7}_{-0.4} \pm 0.3$     & \cite{DELPHI_fxi}$^d$ \\
DELPHI &$\Xi\ell$             & 92--95 &$1.45 ^{+0.55}_{-0.43} \pm 0.13$     & \cite{DELPHI_fxi2}$^d$ \\
\hline
\multicolumn{3}{l}{Average of above 3 (\Xib lifetime)} & \hfagTAUXBnounit & \\
\hline
\multicolumn{5}{l}{$^a$ \footnotesize The combined ALEPH result quoted 
in \cite{ALEPH_fla} is $1.21 \pm 0.11$ ps.} \\[-0.5ex]
\multicolumn{5}{l}{$^b$ \footnotesize The combined DELPHI result quoted 
in \cite{DELLAM0} is $1.14 \pm 0.08 \pm 0.04$ ps.} \\[-0.5ex]
\multicolumn{5}{l}{$^c$ \footnotesize The combined OPAL result quoted 
in \cite{OPALAM1} is $1.16 \pm 0.11 \pm 0.06$ ps.} \\[-0.5ex]
\multicolumn{5}{l}{$^d$ \footnotesize The combined DELPHI result quoted 
in \cite{DELPHI_fxi2} is $1.48 ^{+0.40}_{-0.31} \pm 0.12$ ps.} \\[-0.5ex]
\multicolumn{5}{l}{$^p$ \footnotesize Preliminary.}
\end{tabular}
\end{center}
\end{table}

Inputs to the averages are given in \Table{lifelb}.
Note that before averaging, the CDF input of $\Lambda_b$ lifetime
measured exclusively in the decay mode $J/\psi \Lambda$~\cite{CDFLAM2}  is
$3.1\sigma$ larger than the world average. It is combined with the rest 
without adjustment of input errors.
The world average lifetime of \b baryons is then:
\begin{equation}
\langle\tau(\mbox{\b-baryon})\rangle = \hfagTAUBB \,.
\end{equation}
Keeping only \particle{\Lambda^{\pm}_c \ell^{\mp}}, 
$\Lambda \ell^- \ell^+$, and fully exclusive
final states, as representative of
the \Lb baryon, the following lifetime is obtained:
\begin{equation}
\tau(\Lb) = \hfagTAULB \,. 
\end{equation}

Averaging the measurements based on the $\Xi^{\mp} \ell^{\mp}$
final states~\cite{ALEPH_fxi,DELPHI_fxi,DELPHI_fxi2} gives
a lifetime value for a sample of events
containing $\Xib^0$ and $\Xib^-$ baryons:
\begin{equation}
\langle\tau(\Xib)\rangle = \hfagTAUXB \,.
\end{equation}

\mysubsubsection{Summary and comparison with theoretical predictions}
\labs{lifesummary}

Averages of lifetimes of specific \b-hadron species are collected
in \Table{sumlife}.
\begin{table}[t]
\caption{Summary of lifetimes of different \b-hadron species.}
\labt{sumlife}
\begin{center}
\begin{tabular}{lc} \hline
\b-hadron species & Measured lifetime \\ \hline
\Bu                         & \hfagTAUBU   \\
\Bd                         & \hfagTAUBD   \\
\Bs ($\to$ flavor specific) & \hfagTAUBSSL \\
\Bs ($\to J/\psi\phi$)      & \hfagTAUBSJF \\
\Bs ($1/\Gs$)               & \hfagTAUBSMEANCON \\
\Bc                         & \hfagTAUBC   \\ 
\Lb                         & \hfagTAULB   \\
\Xib mixture                & \hfagTAUXB   \\
\b-baryon mixture           & \hfagTAUBB   \\
\b-hadron mixture           & \hfagTAUB    \\
\hline
\end{tabular}
\end{center}
\caption{Measured ratios of \b-hadron lifetimes relative to
the \Bd lifetime and ranges predicted
by theory~\cite{tarantino,Gabbiani_et_al}.}
\labt{liferatio}
\begin{center}
\begin{tabular}{lcc} \hline
Lifetime ratio & Measured value & Predicted range \\ \hline
$\tau(\Bu)/\tau(\Bd)$ & \hfagRTAUBU & 1.04 -- 1.08 \\
$\bar{\tau}(\Bs)/\tau(\Bd)^a$ & \hfagRTAUBSMEANCON & 0.99 -- 1.01 \\
$\tau(\Lb)/\tau(\Bd)$ & \hfagRTAULB & 0.86 -- 0.95    \\
$\tau(\mbox{\b-baryon})/\tau(\Bd)$  & \hfagRTAUBB & 0.86 -- 0.95 \\
\hline
\multicolumn{3}{l}{$^a$ \footnotesize 
Using $\bar{\tau}(\Bs) = 1/\Gs = 2/(\Gamma_{\rm L} + \Gamma_{\rm H})$.
}
\end{tabular}
\end{center}
\end{table}
As described in \Sec{lifetimes},
Heavy Quark Effective Theory
can be employed to explain the hierarchy of
$\tau(\Bc) \ll \tau(\Lb) < \bar{\tau}(\Bs) \approx \tau(\Bd) < \tau(\Bu)$,
and used to predict the ratios between lifetimes.
Typical predictions are compared to the measured 
lifetime ratios in \Table{liferatio}.
A recent prediction of the ratio between the \Bu and \Bd lifetimes,
is $1.06 \pm 0.02$~\cite{tarantino}, in good agreement with experiment. 


The total widths of the \Bs and \Bd mesons
are expected to be very close and differ by at most 
1\%~\cite{equal_lifetimes,Gabbiani_et_al}.
However, the experimental ratio $\bar{\tau}(\Bs)/\tau(\Bd)$,
where $\bar{\tau}(\Bs)=1/\Gs$ is obtained from \DGs and 
flavour-specific lifetime measurements, appears to be 
smaller than 1 by 
\hfagONEMINUSRTAUBSMEANCONpercent, 
at deviation with respect to the prediction. 

The ratio $\tau(\Lb)/\tau(\Bd)$ has particularly
been the source of theoretical
scrutiny since earlier calculations~\cite{OPE,lblife_early}
predicted a value greater than 0.90, almost two sigma higher
than the world average at the time. 
Many predictions cluster around a most likely central value
of 0.94~\cite{lblife_mid}.
More recent calculations
of this ratio that include higher-order effects predict a
lower ratio between the
\Lb and \Bd lifetimes~\cite{tarantino,Gabbiani_et_al}
and reduce this difference.
References~\cite{tarantino,Gabbiani_et_al} present probability density functions
of their predictions with variation of theoretical inputs, and the
indicated ranges in \Table{liferatio}
are the RMS of the distributions from the most probable values.
Again, the CDF measurement of the $\Lambda_b$ lifetime
in the exclusive decay mode $J/\psi \Lambda$~\cite{CDFLAM2} is significantly 
higher than the world average before inclusion, with a ratio
to the $\tau(\Bd)$ world average of 
$\tau(\Lb)/\tau(\Bd) = 1.042 \pm 0.057$, resulting in continued interest
in lifetimes of $b$ baryons.


%% file: dmd_W_table.tex
 ALEPH~\cite{ALEPH_dmd}  & \particle{ \ell  } & \particle{ \Qjet  } & $  0.404 $ & $ \pm  0.045 $ & $ \pm  0.027 $ & & & \\
 ALEPH~\cite{ALEPH_dmd}  & \particle{ \ell  } & \particle{ \ell  } & $  0.452 $ & $ \pm  0.039 $ & $ \pm  0.044 $ & & & \\
 ALEPH~\cite{ALEPH_dmd}  & \multicolumn{2}{c}{above two combined} & $  0.422 $ & $ \pm  0.032 $ & $ \pm  0.026 $ & $  0.441 $ & $ \pm  0.032 $ & $ \pm  0.020 $ \\
 ALEPH~\cite{ALEPH_dmd}  & \particle{ D^*  } & \particle{ \ell,\Qjet  } & $  0.482 $ & $ \pm  0.044 $ & $ \pm  0.024 $ & $  0.482 $ & $ \pm  0.044 $ & $ \pm  0.024 $ \\
 DELPHI~\cite{DELPHI_dmd}  & \particle{ \ell  } & \particle{ \Qjet  } & $  0.493 $ & $ \pm  0.042 $ & $ \pm  0.027 $ & $  0.502 $ & $ \pm  0.042 $ & $ \pm  0.024 $ \\
 DELPHI~\cite{DELPHI_dmd}  & \particle{ \pi^*\ell  } & \particle{ \Qjet  } & $  0.499 $ & $ \pm  0.053 $ & $ \pm  0.015 $ & $  0.501 $ & $ \pm  0.053 $ & $ \pm  0.015 $ \\
 DELPHI~\cite{DELPHI_dmd}  & \particle{ \ell  } & \particle{ \ell  } & $  0.480 $ & $ \pm  0.040 $ & $ \pm  0.051 $ & $  0.498 $ & $ \pm  0.040 $ & $ ^{+  0.042 }_{-  0.041 } $ \\
 DELPHI~\cite{DELPHI_dmd}  & \particle{ D^*  } & \particle{ \Qjet  } & $  0.523 $ & $ \pm  0.072 $ & $ \pm  0.043 $ & $  0.518 $ & $ \pm  0.072 $ & $ \pm  0.043 $ \\
 DELPHI~\cite{DELPHI_dmd_dms_vtx}  & \particle{ \mbox{vtx}  } & \particle{ \mbox{comb}  } & $  0.531 $ & $ \pm  0.025 $ & $ \pm  0.007 $ & $  0.528 $ & $ \pm  0.025 $ & $ \pm  0.006 $ \\
 L3~\cite{L3_dmd}  & \particle{ \ell  } & \particle{ \ell  } & $  0.458 $ & $ \pm  0.046 $ & $ \pm  0.032 $ & $  0.465 $ & $ \pm  0.046 $ & $ \pm  0.028 $ \\
 L3~\cite{L3_dmd}  & \particle{ \ell  } & \particle{ \Qjet  } & $  0.427 $ & $ \pm  0.044 $ & $ \pm  0.044 $ & $  0.438 $ & $ \pm  0.044 $ & $ \pm  0.042 $ \\
 L3~\cite{L3_dmd}  & \particle{ \ell  } & \particle{ \ell\mbox{(IP)}  } & $  0.462 $ & $ \pm  0.063 $ & $ \pm  0.053 $ & $  0.472 $ & $ \pm  0.063 $ & $ ^{+  0.045 }_{-  0.044 } $ \\
 OPAL~\cite{OPAL_dmd_dilepton}  & \particle{ \ell  } & \particle{ \ell  } & $  0.430 $ & $ \pm  0.043 $ & $ ^{+  0.028 }_{-  0.030 } $ & $  0.463 $ & $ \pm  0.043 $ & $ ^{+  0.018 }_{-  0.016 } $ \\
 OPAL~\cite{OPAL_dmd_lepton}  & \particle{ \ell  } & \particle{ \Qjet  } & $  0.444 $ & $ \pm  0.029 $ & $ ^{+  0.020 }_{-  0.017 } $ & $  0.471 $ & $ \pm  0.029 $ & $ ^{+  0.014 }_{-  0.013 } $ \\
 OPAL~\cite{OPAL_dmd_dstar}  & \particle{ D^*\ell  } & \particle{ \Qjet  } & $  0.539 $ & $ \pm  0.060 $ & $ \pm  0.024 $ & $  0.544 $ & $ \pm  0.060 $ & $ \pm  0.023 $ \\
 OPAL~\cite{OPAL_dmd_dstar}  & \particle{ D^*  } & \particle{ \ell  } & $  0.567 $ & $ \pm  0.089 $ & $ ^{+  0.029 }_{-  0.023 } $ & $  0.571 $ & $ \pm  0.089 $ & $ ^{+  0.028 }_{-  0.022 } $ \\
 OPAL~\cite{OPAL_dmd_slowpion}  & \particle{ \pi^*\ell  } & \particle{ \Qjet  } & $  0.497 $ & $ \pm  0.024 $ & $ \pm  0.025 $ & $  0.495 $ & $ \pm  0.024 $ & $ \pm  0.025 $ \\
 CDF1~\cite{CDF1_dmd_dlepton_SST}  & \particle{ D\ell  } & \particle{ \mbox{SST}  } & $  0.471 $ & $ ^{+  0.078 }_{-  0.068 } $ & $ ^{+  0.033 }_{-  0.034 } $ & $  0.470 $ & $ ^{+  0.078 }_{-  0.068 } $ & $ ^{+  0.033 }_{-  0.034 } $ \\
 CDF1~\cite{CDF1_dmd_dimuon}  & \particle{ \mu  } & \particle{ \mu  } & $  0.503 $ & $ \pm  0.064 $ & $ \pm  0.071 $ & $  0.515 $ & $ \pm  0.064 $ & $ \pm  0.070 $ \\
 CDF1~\cite{CDF1_dmd_lepton}  & \particle{ \ell  } & \particle{ \ell,\Qjet  } & $  0.500 $ & $ \pm  0.052 $ & $ \pm  0.043 $ & $  0.541 $ & $ \pm  0.052 $ & $ \pm  0.036 $ \\
 CDF1~\cite{CDF1_dmd_dstarlepton}  & \particle{ D^*\ell  } & \particle{ \ell  } & $  0.516 $ & $ \pm  0.099 $ & $ ^{+  0.029 }_{-  0.035 } $ & $  0.523 $ & $ \pm  0.099 $ & $ ^{+  0.028 }_{-  0.035 } $ \\
 CDF2~\cite{CDF2_dmd_dlepton_preliminary}  & \particle{ D^{(*)}\ell  } & \particle{ \mbox{OST}  } & $  0.509 $ & $ \pm  0.010 $ & $ \pm  0.016 $ & $  0.509 $ & $ \pm  0.010 $ & $ \pm  0.016 $ \\
 CDF2~\cite{CDF2_dmd_exclusive_preliminary}  & \particle{ \Bd  } & \particle{ \mbox{comb}  } & $  0.536 $ & $ \pm  0.028 $ & $ \pm  0.006 $ & $  0.536 $ & $ \pm  0.028 $ & $ \pm  0.006 $ \\
 \dzero~\cite{D0_dmd}  & \particle{ D^{(*)}\mu  } & \particle{ \mbox{OST}  } & $  0.506 $ & $ \pm  0.020 $ & $ \pm  0.016 $ & $  0.506 $ & $ \pm  0.020 $ & $ \pm  0.016 $ \\
 \babar~\cite{BABAR_dmd_full}  & \particle{ \Bd  } & \particle{ \ell,K,\mbox{NN}  } & $  0.516 $ & $ \pm  0.016 $ & $ \pm  0.010 $ & $  0.520 $ & $ \pm  0.016 $ & $ \pm  0.008 $ \\
 \babar~\cite{BABAR_dmd_dilepton}  & \particle{ \ell  } & \particle{ \ell  } & $  0.493 $ & $ \pm  0.012 $ & $ \pm  0.009 $ & $  0.489 $ & $ \pm  0.012 $ & $ \pm  0.006 $ \\
 \babar~\cite{BABAR5}  & \particle{ D^*\ell\nu\mbox{(part)}  } & \particle{ \ell  } & $  0.511 $ & $ \pm  0.007 $ & $ \pm  0.007 $ & $  0.512 $ & $ \pm  0.007 $ & $ \pm  0.007 $ \\
 \babar~\cite{BABAR3}  & \particle{ D^*\ell\nu  } & \particle{ \ell,K,\mbox{NN}  } & $  0.492 $ & $ \pm  0.018 $ & $ \pm  0.014 $ & $  0.491 $ & $ \pm  0.018 $ & $ \pm  0.013 $ \\
 \belle~\cite{BELLE_dmd_dstarpi_partial}  & \particle{ D^*\pi\mbox{(part)}  } & \particle{ \ell  } & $  0.509 $ & $ \pm  0.017 $ & $ \pm  0.020 $ & $  0.512 $ & $ \pm  0.017 $ & $ \pm  0.019 $ \\
 \belle~\cite{BELLE_dmd_dilepton}  & \particle{ \ell  } & \particle{ \ell  } & $  0.503 $ & $ \pm  0.008 $ & $ \pm  0.010 $ & $  0.506 $ & $ \pm  0.008 $ & $ \pm  0.009 $ \\
 \belle~\cite{BELLE2}  & \particle{ \Bd,D^*\ell\nu  } & \particle{ \mbox{comb}  } & $  0.511 $ & $ \pm  0.005 $ & $ \pm  0.006 $ & $  0.512 $ & $ \pm  0.005 $ & $ \pm  0.006 $ \\
 \hline \\[-2.0ex]
 \multicolumn{6}{l}{World average (all above measurements included):} & $  0.508 $ & $ \pm  0.003 $ & $ \pm  0.003 $ \\

%% file: life_mix_dgs.tex
%
%
%

\subsubsubsection{Decay width difference \DGs}



Definitions and an introduction to \DGs can also 
be found in \Sec{taubs}.
Neglecting \CP violation, the mass eigenstates are
also \CP eigenstates, with the short-lived state being
\CP-even and the long-lived one being \CP-odd.
Information on \DGs can be obtained by studying the proper time 
distribution of untagged data samples enriched in 
\Bs mesons~\cite{Hartkorn_Moser}.
In the case of an inclusive \Bs selection~\cite{L3B01} or a semileptonic 
\Bs decay selection~\cite{DELBS0_dms_dgs,CDFBS,D0BS2}, 
both the short- and long-lived
components are present, and the proper time distribution is a superposition 
of two exponentials with decay constants 
$\Gs\pm\DGs/2$.
In principle, this provides sensitivity to both \Gs and 
$(\DGGs)^2$. Ignoring \DGs and fitting for 
a single exponential leads to an estimate of \Gs with a 
relative bias proportional to $(\DGGs)^2$. 
An alternative approach, which is directly sensitive to first order in \DGGs, 
is to determine the lifetime of \Bs candidates decaying to \CP
eigenstates; measurements exist for 
\particle{\Bs\to J/\psi\phi}~\cite{CDFIN_BS1,CDFB3,D0BS1} and 
\particle{\Bs\to D_s^{(*)+} D_s^{(*)-}}, discussed later, which are 
mostly \CP-even states~\cite{Aleksan}. 
However, more recent
time-dependent angular analyses of \particle{\Bs\to J/\psi\phi} 
allow the simultaneous extraction of \DGGs and the \CP-even and \CP-odd 
amplitudes~\cite{CDF2_DGs,D01_DGs}. 

\begin{table}
\caption{Experimental constraints on \DGGs from lifetime analyses,
assuming no \CP violation.
The upper limits,
which have been obtained by the working group, are quoted at the \CL{95}.}
\labt{dgammat}
\begin{center}
\begin{tabular}{l|c|c|c}
\hline
Experiment & Method            & $\Delta \Gs/\Gs$ & Ref.  \\
\hline
L3         & lifetime of inclusive \b-sample              
           & $<0.67$   & \cite{L3B01}      \\
DELPHI     & $\Bsb\to D_s^+\ell^- \overline{\nu_{\ell}} X$, lifetime
	   & $<0.46$   & \cite{DELBS0_dms_dgs} \\
DELPHI     & $\Bsb \to D_s^+$ hadron, lifetime
           & $<0.69$ & \cite{DELBS1_dms_excl}   \\
CDF1       & $\Bs \to J/\psi\phi$, lifetime
	   & $0.33^{+0.45}_{-0.42}$ & \cite{CDFIN_BS1} \\ \hline
CDF2       & $\Bs \to J/\psi\phi$, time-dependent angular analysis
           & $0.65^{+0.25}_{-0.33} \pm 0.01$ & \cite{CDF2_DGs} \\ \hline
           &                   & $ \DGs$    \\
\hline
\dzero     & $\Bs \to J/\psi\phi$, time-dependent angular analysis
           & $0.12{^{+0.08}_{-0.10}}{\pm 0.02}$ & \cite{D01_DGs} \\
	 \hline
	 \end{tabular}
	 \end{center}
	 \end{table}

Measurements quoting \DGs results from lifetime analyses 
are listed in \Table{dgammat} under the hypothesis of no
\CP violation.
There is significant correlation
between \DGs and $1/\Gamma_s$. In order to combine these measurements,
the two-dimensional log-likelihood for each measurement
in the $(1/\Gs,\,\DGs)$ plane is summed and the total
normalized with respect to its minimum.  The one-sigma contour (corresponding
to 0.5 units of log-likelihood greater than the minimum) and
95\% contour are found. 
Inputs as indicated in \Table{dgammat} were used in the combination, 
with the exception of the L3~\cite{L3B01} result since the likelihood
in this case was not available.

Results of the combination are shown as the one-sigma contour
labeled ``Direct" in both plots of \Fig{DGs}.  Transformation
of variables from $(1/\Gs,\,\DGs)$ space to other pairs
of variables such as $(1/\Gs,\,\DGGs)$ and 
$(\tau_{\rm L} = 1/\Gamma_{\rm L},\,\tau_{\rm H} = 1/\Gamma_{\rm H})$
are also made.
The resulting one-sigma contour for the latter is shown in
\Fig{DGs}(b). 

\begin{figure}
\begin{center}
\epsfig{figure=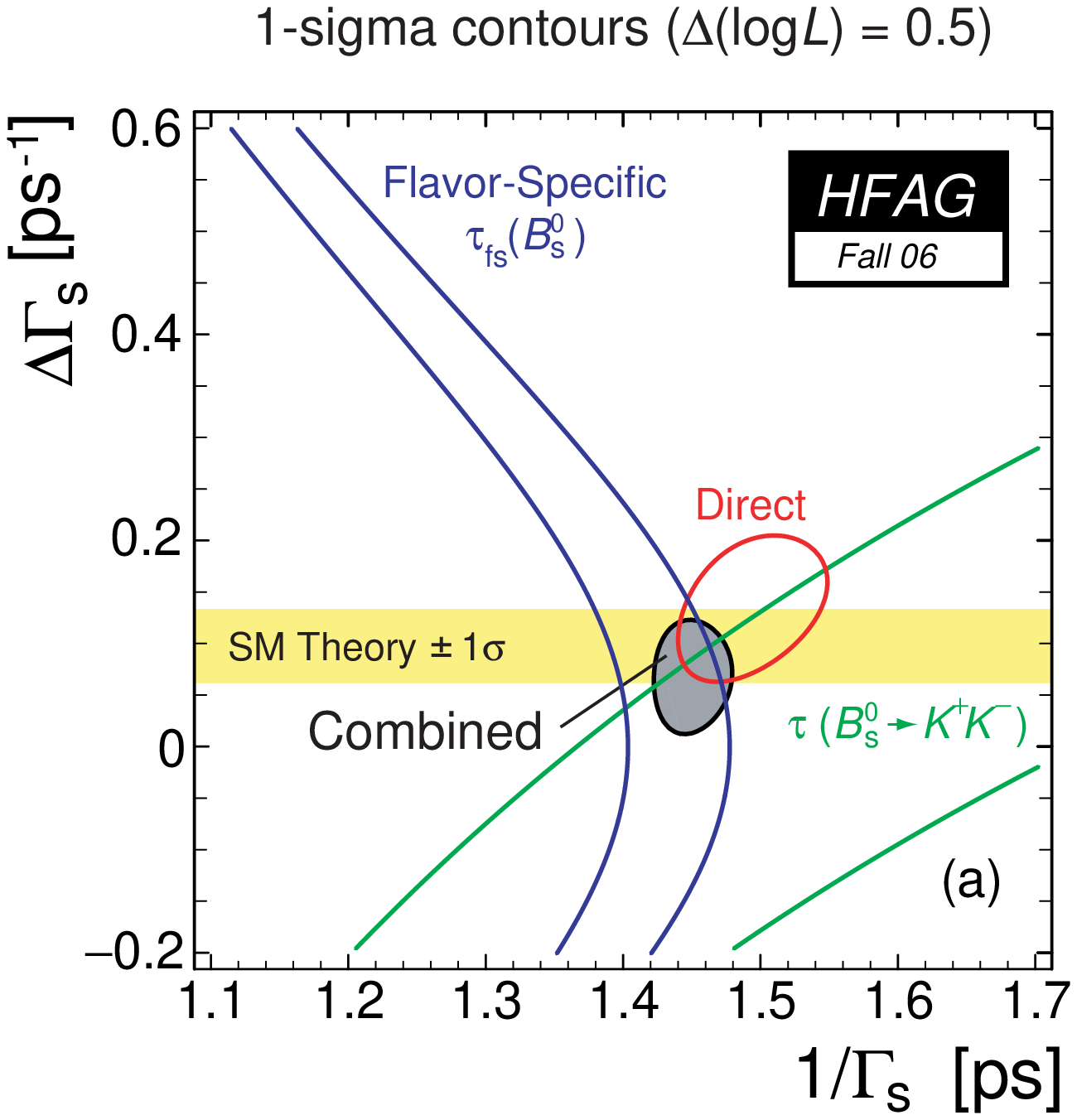,width=0.45\textwidth}
\hfill
\epsfig{figure=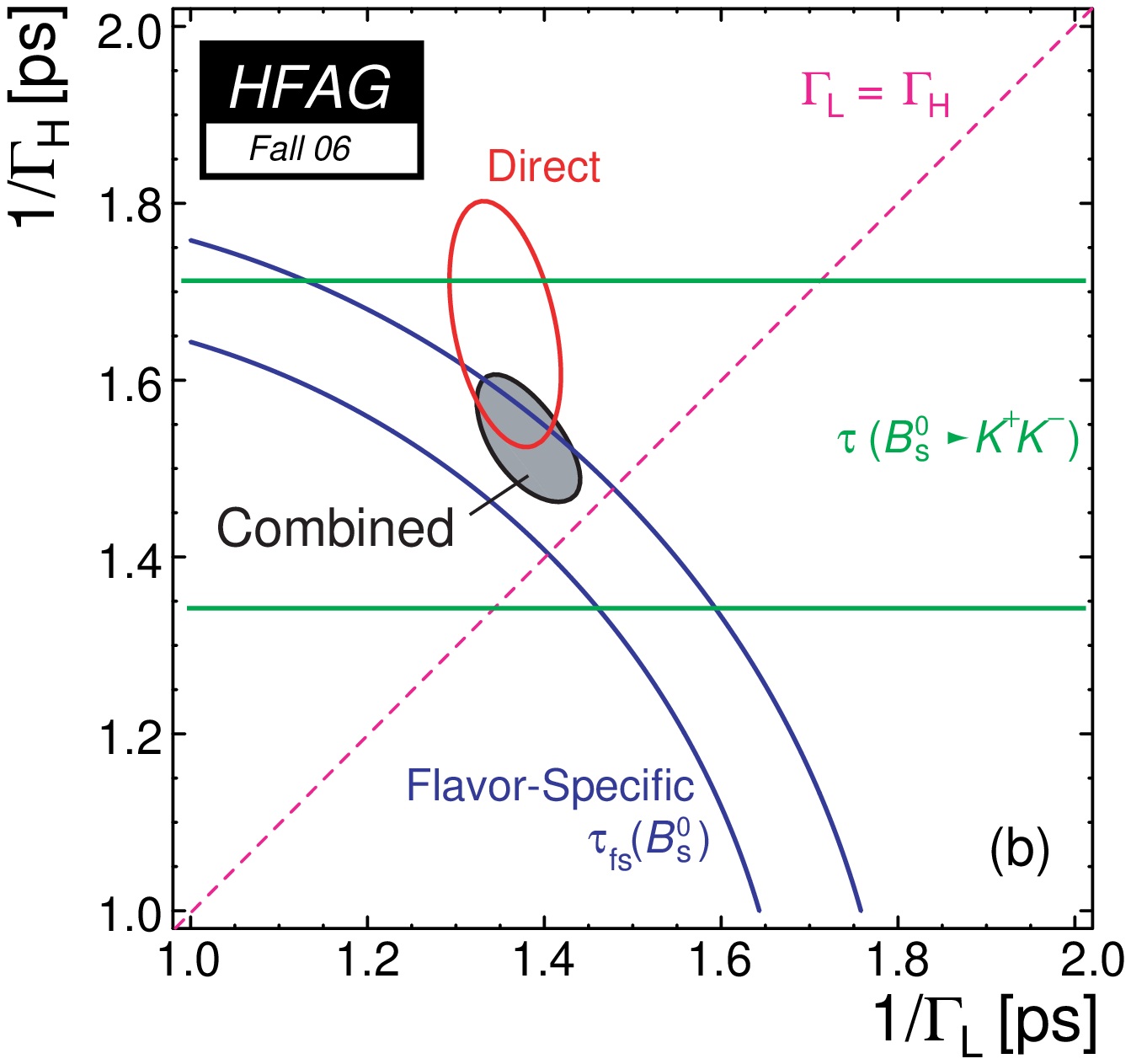,width=0.45\textwidth}
\caption{\DGs combination results with one-sigma contours
($\Delta\log\mathcal{L} = 0.5$) shown for (a) \DGs versus
$\bar{\tau}(\Bs) = 1/\Gs$  and (b)
$\tau_{\rm H} = 1/\Gamma_{\rm H}$ versus $\tau_{\rm L} = 1/\Gamma_{\rm L}$.
The red contours labeled ``Direct" are the result of the combination of
most measurements of \Table{dgammat}, the blue bands are the one-sigma
contours due to the world average of flavor-specific measurements,
the green bands are the one-sigma contour of the $\Bs \to K^+K-$ 
lifetime measurement, 
and the shaded region the combination of all three.
In (b), the diagonal dashed line indicates 
$\Gamma_{\rm L} = \Gamma_{\rm H}$, \ie, where $\DGs = 0$.}
\labf{DGs}
\end{center}
\end{figure}




\newcommand{\comment}[1]{}\comment{

{\em This text below taken by Donatella from previous \DGs notes \ldots}

\newcommand{\Bh}{B^{\rm heavy}_{d,s}}
\newcommand{\Bl}{B^{\rm light}_{d,s}}
\newcommand{\Mh}{m^{\rm heavy}_{d,s}}
\newcommand{\Ml}{m^{\rm light}_{d,s}}
\newcommand{\Gh}{\Gamma^{\rm heavy}_{d,s}}
\newcommand{\Gl}{\Gamma^{\rm light}_{d,s}}
\newcommand{\Gsho}{\Gamma^{\rm short}_{d,s}}
\newcommand{\Glon}{\Gamma^{\rm long}_{d,s}}
\newcommand{\G}{\Gamma_{d,s}}
\newcommand{\tb}{\tau(B^0_{d,s})}
\newcommand{\dg}{\Delta\Gamma_{d,s}}
\newcommand{\tbssemi}{\tau(B_s)_{\rm semi}}
\newcommand{\Bssh}{B^{\rm short}_s}
\newcommand{\tbsshort}{\tau(\Bssh)}

The Standard Model predicts that the \Bs and \Bd can mix before decay. 
The phenomenology of this interaction can be 
described in terms of a $2\times 2$ effective Hamiltonian matrix, 
$M - i\Gamma /2$.
This results in new states called heavy and light, 
$\Bh$  and $\Bl$, for \Bs and \Bd with masses $\Mh$ ,
$\Ml$. Also the  widths  $\Gh$ and $\Gl$ could be different.

Neglecting \CP violation, the mass eigenstates are also \CP eigenstates, the ``long''  state being 
\CP even and the short one being \CP odd.  For convenience of notation, in the following
we therefore substitute 
$\Gl \equiv \Gsho$ and $\Gh \equiv \Glon$, and 
define $\G=1/\tb=(\Glon+\Gsho)/2$ and 
$\Delta \G = \Gsho-\Glon$ which is positive.

$\dg$ is related to the off-diagonal matrix elements, which have been recently 
calculated at the NLO including NLO QCD correction~\cite{Ciuchini2}. 
The theoretical values are:
\begin{equation}
\DGGs = (7.4\pm 2.4 )\times 10^{-2} \,, \hspace{1truecm} \DGGd = (2.42 \pm 0.59 ) \times 10^{-3} \,.
\end{equation}

In the same work the ratio $\DGd/\DGs$ is evaluated since the uncertainties 
coming from higher orders of QCD and $\Lambda_{\rm QCD}/m_{\b}$ corrections cancel out:
\begin{equation}
\DGd/\DGs = (3.2\pm 0.8)\times10^{-2}
\end{equation}

Experimentally \DGs can be measured fitting the lifetime of the 
light and heavy component of the \Bs.
An alternative method is based on the measurement of the  
branching fraction \particle{\Bs\to D_s^{(*)+}D_s^{(*)-}}.
Methods based on lifetime measurements have two different approaches.
Double exponential lifetime fits to samples containing a mixture of \CP eigenstates like
inclusive or semileptonic \Bs decays or $\Bs\to D_s$-hadron have a quadratic sensitivity 
to \DGs.
Whereas the isolation of a single \CP eigenstate as $\Bs\to\phi\phi$ or 
\particle{\Bs\to J/\psi\phi} to extract the lifetime of the \CP-even or odd state have 
a linear dependence on \DGs and it is more sensitive to \DGs but tend 
to suffer from reduced statistics.
The branching fraction method, exploited by ALEPH~\cite{ALEPH-phiphi}, 
is based on several theoretical assumptions~\cite{theoBR}, and allows to have 
information on \DGs only through the branching fraction measurement:
\begin{equation}
\BR{B_s\to D_s^{(*)+}
D_s^{(*)-}} = \frac{\DGs}{\Gs\left(1+\frac{\DGs}{2\Gs}\right)} \,.
\label{eq:dg_ratio}
\end{equation}

The available results are summarized in \Table{dgammat}. 
The values of the limit on \DGGs quoted in the last column of this 
table have been obtained by the working group.

Details on how these measurements are included in the average can be found  
in the previous summaries~\cite{Pcomb}.

\begin{table}
\caption{Experimental constraints on \DGGs. The upper limits,
which have been obtained by the working group, are quoted at the \CL{95}.}
\labt{dgammat}
\begin{center}
\begin{tabular}{|l|c|c|c|} 
\hline
Experiment & Selection        & Measurement            & $\Delta \Gs/\Gs$ \\ 
\hline
L3~\cite{L3B01}         & inclusive \b-sample              &                               & $<0.67$         \\
DELPHI~\cite{DELBS0_dms_dgs}     & $\Bsb\to D_s^+\ell^- \overline{\nu_{\ell}} X$ & $\tbssemi=(1.42^{+0.14}_{-0.13}\pm0.03)$~ps  & $<0.46$ \\
others~\cite{ref:others}& $\Bsb\to D_s^+\ell^-  \overline{\nu_{\ell}} X$  & $\tbssemi=(1.46\pm{0.07})$~ps & $<0.30$ \\
ALEPH~\cite{ALEPH-phiphi}      & $\Bs\to\phi\phi X$      & 
$\BR{\Bssh \to D_s^{(*)+} D_s^{(*)-}} =(23\pm10^{+19}_{-~9})\%$       & $0.26^{+0.30}_{-0.15}$ \\
ALEPH~\cite{ALEPH-phiphi}      & $\Bs\to\phi\phi X$      & $\tbsshort=(1.27\pm0.33\pm0.07)$~ps           & 
$0.45^{+0.80}_{-0.49}$ \\ 
DELPHI~\cite{DELBS0_dms_dgs}$^a$    & $\Bsb \to D_s^+$ hadron    
&  $\tau_{\rm B^{D_s-had.}_s}=(1.53^{+0.16}_{-0.15}\pm0.07)$~ps                          & $<0.69$         \\
CDF~\cite{CDFB01} & $\Bs \to {\rm J}/\psi\phi$        
& $\tau_{\rm B^{{\rm J}/\psi \phi}_s}=(1.34^{+0.23}_{-0.19}\pm0.05)$~ps & $0.33^{+0.45}_{-0.42}$ \\ 
\hline
\multicolumn{4}{l}{$^a$ \footnotesize 
The value quoted for the measured lifetime differs
slightly from the one quoted in \Table{bs} because it} \\[-1ex]
\multicolumn{4}{l}{~~ \footnotesize 
corresponds to the present status of the analysis in which the information
on \DGs has been obtained.}
\end{tabular}
\end{center}
\end{table}

Here only a short description will be given.

L3 and DELPHI use inclusively reconstructed
\Bs and $\Bs\to \particle{D_s} \ell\nu X$ events respectively.
If those sample are fitted assuming a single exponential lifetime then,
assuming \DGGs is small, the measured lifetime is given by:
\begin{equation}
\tau(\Bs)_{\rm incl.} = \frac{1}{\Gs} \frac{1}{1-\left(\frac{\DGs}{2\Gs}\right)^2}
\quad \quad ; \quad \quad
\tau(\Bs)_{\rm semi.} = \frac{1}{\Gs} 
\frac{{1+\left(\frac{\DGs}{2\Gs}\right)^2}}{{1-\left(\frac{\DGs}{2\Gs}\right)^2}}.     
\end{equation}

The single lifetime fit is thus more sensitive to the effects of 
\DGs in the semileptonic case than in the fully inclusive case.

The same method is used for the \Bs world average lifetime
(recomputed without the DELPHI measurement) obtained
by using only the semileptonic decays and referenced in \Table{dgammat} as {\it others}.

The technique of reconstructing only decays at defined \CP
has been exploited by ALEPH, DELPHI and CDF.

ALEPH reconstructs the decay
$\Bs\to \particle{D_s^{(*)+}D_s^{(*)-}} \to \phi\phi X$
which is predominantly \CP even.
The proper time dependence of the \Bs component is a simple
exponential and the lifetime is related to \DGs via
\begin{equation}
\frac{\Delta \Gs}{\Gs}=2(\frac{1}{\Gs~\tbsshort}-1).  
\end{equation} 
 The same data have been used by ALEPH to exploit the branching fraction method.

DELPHI uses a sample of $\Bs\to D_s$-hadron,
which is expected to have an increased \CP-even component as the contribution
due to \particle{D_s^{(*)+}_s^{(*)-}} events is enhanced by
selection criteria.

CDF reconstructs \particle{\Bs\to J/\psi\phi} with
\particle{J/\psi\to\mu^+\mu^-} and \particle{\phi\to K^+K^-}
where the \CP-even component is equal to $0.84\pm 0.16$ obtained by
combining CLEO~\cite{cleo} measurement of \CP-even fraction in
\particle{\Bd\to J/\psi K^{*0}} and possible SU(3) symmetry
correction.

In order to combine all the measurements~\footnote{L3 is not 
included since the likelihood for the results
was not available} the two-dimensional log-likelihood in the ($1/\Gs$, \DGGs) 
plane is summed and normalized with respect to its minimum.
The 68\%, 95\% and \CL{99} contours of the combined negative 
log-likelihood are shown in \Fig{dgplot} (left)
The corresponding limit on \DGGs is:
\begin{eqnarray}
\DGGs & = & 0.16^{+0.15}_{-0.16}  \,, \\
\DGGs & < & 0.54~\mbox{at \CL{95}} \,. 
\end{eqnarray}

\begin{figure}
\begin{center}
\epsfig{figure=figures/osc/dg_w_notaubd_bw_2d.eps,width=\textwidth}
\epsfig{figure=figures/osc/dg_w_taubd_bw_2d.eps,width=\textwidth}
\end{center}
\caption{Top: 68\%, 95\% and \CL{99} contours of the negative log-likelihood 
distribution in the plane ($1/\Gs$, \DGGs).
Bottom: Same, but with the constraint $1/\Gs \equiv\tau_{\Bd}$} 
\labf{dgplot}
\end{figure}

\begin{figure}
\begin{center}
\epsfig{figure=figures/osc/dg_w_taubd_col_1d,width=\textwidth}
\end{center}
\caption{Probability density distribution for \DGGs after applying the constraint; 
the three shaded regions show the limits at the 68\%, 95\% and \CL{99} respectively.} 
\labf{dgprobplot}
\end{figure}

An improved limit on \DGGs can be obtained by applying the $\tau_{\Bd}=\HFAGtauBd$ constraint.
The world average \Bs lifetime is not used, as its meaning 
is not clear if $\Delta \Gs$ is non-zero.
This is well motivated theoretically, as 
the total widths of the \Bs and \Bd mesons
are expected to be 
equal within less than one percent~\cite{bigilife}, \cite{Beneke}
and \DGd is expected to be small. 
 
The two-dimensional log-likelihood obtained, after including the constraint is shown in 
\Fig{dgplot} (right). The resulting probability density distribution for \DGGs is 
shown in \Fig{dgprobplot}. The corresponding limit on \DGGs is:
\begin{eqnarray}
\DGGs & = & 0.07^{+0.09}_{-0.07} \,, \\
\DGGs & < & 0.29~~\mbox{at \CL{95}} \,.
\end{eqnarray}

} 




Numerical results of the combination of the described inputs
of \Table{dgammat} are:
\begin{eqnarray}
\DGGs &\in& [\hfagDGSGSlow,\hfagDGSGSupp] ~ \mbox{at \CL{95}} \,, \\
\DGGs &=& \hfagDGSGS \,, \\
\DGs &=& \hfagDGS \,, \\
\bar{\tau}(\Bs) = 1/\Gs &=& \hfagTAUBSMEAN \,, \\
\rho(\DGGs, 1/\DGs) &=& \hfagRHODGSGSTAUBSMEAN \,, \\
1/\Gamma_{\rm L} = \tau_{\rm short} &=& \hfagTAUBSL \,, \\
1/\Gamma_{\rm H} = \tau_{\rm long}  &=& \hfagTAUBSH \,. 
\end{eqnarray}

Flavor-specific lifetime measurements are of an equal mix
of \CP-even and \CP-odd states at time zero, and  
if a single exponential function is used in the likelihood
lifetime fit of such a sample~\cite{Hartkorn_Moser}, 
\begin{equation}
\tau(\Bs)_{\rm fs} = \frac{1}{\Gs}
\frac{{1+\left(\frac{\DGs}{2\Gs}\right)^2}}{{1-\left(\frac{\DGs}{2\Gs}\right)^2}
} \,.
\end{equation}
Using the world average flavor-specific 
lifetime\footnote{The world average of all \Bs lifetime 
measurements using flavour-specific final states is \hfagTAUBSSL; however,
for the purpose of the \DGs extraction, we remove from this average one
DELPHI analysis~\cite{DELBS0_dms_dgs} 
that is already included in the set of ``direct 
measurements'' and obtain \hfagTAUBSSLX, 
shown as the blue bands on the two plots of \Fig{DGs}.} of \Sec{taubs}
the one-sigma blue bands shown in \Fig{DGs} are obtained. 
Higher-order corrections were checked to be negligible in the
combination.

As described earlier, \Bs $\to K^+ K^-$ decays
can be used to
measure the lifetime
of the ``light" mass eigenstate
$\tau_L = 1/\Gamma_L = 
\tau(\Bs)_{\particle{K^+K^-}} = 1.53 \pm 0.18 \pm 0.02 
\thinspace {\mathrm ps}$~\cite{CDFBS4}, and this 
additional constraint is shown by the green bands in
\Fig{DGs}.

When the flavor-specific lifetime measurements and $\tau_L$ 
measurements
are combined with the measurements of \Table{dgammat}, the shaded
regions of \Fig{DGs} are obtained, with numerical results:
\begin{eqnarray}
\DGGs &\in& [\hfagDGSGSCONlow,\hfagDGSGSCONupp] ~ \mbox{at \CL{95}} \,, \\
\DGGs &=& \hfagDGSGSCON \,, \labe{DGGs_ave} \\
\DGs &=& \hfagDGSCON \,, \\
\bar{\tau}(\Bs) = 1/\Gs &=& \hfagTAUBSMEANCON \,, \labe{oneoverGs} \\
\rho(\DGGs, 1/\DGs) &=& \hfagRHODGSGSTAUBSMEANCON \,, \labe{corrDGGs} \\
1/\Gamma_{\rm L} = \tau_{\rm short} &=& \hfagTAUBSLCON \,, \\
1/\Gamma_{\rm H} = \tau_{\rm long}  &=& \hfagTAUBSHCON \,. 
\end{eqnarray}
These results can
be compared with the theoretical prediction of 
$\DGs/\Gs = 0.096 \pm 0.039$
(or $\DGGs = 0.088 \pm 0.017$ if there is no new physics in
\dms)~\cite{delta_gams}.

Measurements of $\BR{B^0_s \rightarrow D_s^{(*)+} D_s^{(*)-}}$ can 
also be sensitive to \DGs.
The decay $\Bs \rightarrow D_s^{+} D_s^{-}$ is into
a final state that is purely \CP even. 
Under various theoretical assumptions~\cite{Aleksan}, the
inclusive decay into this plus the excited states
$\Bs \rightarrow D_s^{(*)+} D_s^{(*)-}$ is also \CP even
to within 5\%, and 
$\Bs \rightarrow D_s^{(*)+} D_s^{(*)-}$ saturates
$\Gamma_s^{\CP \thinspace {\mathrm{even}}}$.
Under these assumptions, for no \CP violation, we have: 
\begin{equation}
\DGGs \approx
\frac{2 \BR{\Bs \rightarrow D_s^{(*)+} D_s^{(*)-}}}
{1 - \BR{\Bs \rightarrow D_s^{(*)+} D_s^{(*)-}}} \,.
\labe{dGsBr}
\end{equation}
However, there are concerns~\cite{nierste} that the assumptions needed
for the above are overly restrictive and that the inclusive branching
ratio may be \CP even to only 30\%.
Due to this uncertainty, extracted values of \DGGs 
from this branching ratio are not included in the overall combination 
but are only extracted here to compare with the world average result.

\begin{table}
\caption{Measurements of $\BR{\Bs \rightarrow D_s^{(*)+} D_s^{(*)-}}$.}
\labt{dGsBr}
\begin{center}
\begin{tabular}{l|c|c|c}
\hline
Experiment & Method & Value & Ref.  \\
\hline
\belle         & \Bs-pair production at $\Upsilon(5S)$
              & $< 0.257$ at 90\% CL & \cite{Belle_drutskoy2007e}$^a$  \\
ALEPH         & $\phi$-$\phi$ correlations              
           & $0.077 \pm 0.034^{+0.038}_{-0.026}$  & \cite{ALEPH_DGs}$^b$     \\
\dzero        & $D_s \rightarrow \phi \pi$, $D_s \rightarrow \phi \mu \nu$            
           & $0.039^{+0.019 + 0.016}_{-0.017 - 0.015}$  & \cite{D0_DsDs}   \\
	 \hline
Average       &      &   \hfagBRDSDS  &   \\
      \hline
\multicolumn{4}{l}{
$^a$ \footnotesize This limit is for $\Bs \rightarrow D_s^{*+} D_s^{*-}$.} \\[-0.5ex]
\multicolumn{4}{l}{
$^b$ \footnotesize Recalculated using the PDG 2006 
value of $\BR{D_s \rightarrow \phi \pi}$.} 
\end{tabular}
\end{center}
\end{table}

Measurements for the branching fraction for this
decay channel are shown in \Table{dGsBr}.
Using their average value of \hfagBRDSDS with \Eq{dGsBr} yields
\begin{equation}
\DGGs = \hfagDGSGSBRDSDS \,,
\end{equation}
consistent with the value given in \Eq{DGGs_ave}, but with the above
caveat.
CDF has also measured the ratio of exclusive modes 
${\BR{B^0_s \rightarrow D^+_s D^-_s}}/
{\BR{B^0_s \rightarrow D^+_s D^-}}$~\cite{CDF_DsDs}, 
and they continue work to use this ratio to extract \DGs.

Again, note that the above combination and average was found assuming
no \CP violation in \Bs mixing, \ie, that the \CP-violating phase
$\phi_s$ is zero:
\begin{equation}
\phi_s = \arg \left( -\frac{M_{12}}{\Gamma_{12}} \right) = 0 \,.
\end{equation}
Under the assumption of non-zero $\phi_s$, in addition to the result listed
in \Table{dgammat}, the \dzero collaboration~\cite{D01_DGs}  has also made simultaneous
fits allowing $\phi_s$ to float finding: 
\begin{eqnarray}
\DGs &=& +0.17 \pm 0.09 \pm 0.03~{\mathrm{ps}}^{-1}\,,  \\ 
\bar{\tau}(\Bs) &= &1/\Gs = 1.49 \pm 0.08^{+0.01}_{-0.03}~{\mathrm{ps}}\,,  \\
\phi_s &=& -0.79 \pm 0.56 \pm 0.01 \,. 
\end{eqnarray}


The {\em average} \Bs and \Bd
lifetimes are predicted to be equal within
1\%~\cite{equal_lifetimes,Gabbiani_et_al}
and in the past, an additional constraint was applied 
by setting
$\Gs = \Gd$, \ie, $1/\Gs = \tau(\Bd)$,
where $\tau(\Bd) = \hfagTAUBD$ is the world average of experimental results,
including a relative 1\% theoretical uncertainty added
in quadrature with the indicated experimental error.
However, with the increased inconsistency of
the measured values of $1/\Gs= \bar{\tau}(\Bs)$ and $\tau(\Bd)$
at the level of $\hfagRTAUBSMEANCONsig\,\sigma$,
this constraint is no longer applied. 


%% file: cp_uta.tex


\mysection{Measurements related to Unitarity Triangle angles
}
\label{sec:cp_uta}

The charge of the ``$\CP(t)$ and Unitarity Triangle angles'' group
is to provide averages of measurements 
from time-dependent asymmetry analyses,
and other quantities that are related 
to the angles of the Unitarity Triangle (UT).
In cases where considerable theoretical input is required to 
extract the fundamental quantities, no attempt is made to do so at 
this stage. However, straightforward interpretations of the averages 
are given, where possible.

In Sec.~\ref{sec:cp_uta:introduction} 
a brief introduction to the relevant phenomenology is given.
In Sec.~\ref{sec:cp_uta:notations}
an attempt is made to clarify the various different notations in use.
In Sec.~\ref{sec:cp_uta:common_inputs}
the common inputs to which experimental results are rescaled in the
averaging procedure are listed. 
We also briefly introduce the treatment of experimental errors. 
In the remainder of this section,
the experimental results and their averages are given,
divided into subsections based on the underlying quark-level decays.

\mysubsection{Introduction
}
\label{sec:cp_uta:introduction}

The Standard Model Cabibbo-Kobayashi-Maskawa (CKM) quark mixing matrix $\VCKM$ 
must be unitary. A $3 \times 3$ unitary matrix has four free parameters,\footnote{
  In the general case there are nine free parameters,
  but five of these are absorbed into unobservable quark phases.}
and these are conventionally written by the product
of three (complex) rotation matrices~\cite{ref:cp_uta:chau}, where the rotations are 
characterized by the Euler angles $\theta_{12}$, $\theta_{13}$ 
and $\theta_{23}$, which are the mixing angles
between the generations, and one overall phase $\delta$,
\begin{equation}
\label{eq:ckmPdg}
\VCKM =
        \left(
          \begin{array}{ccc}
            V_{ud} & V_{us} & V_{ub} \\
            V_{cd} & V_{cs} & V_{cb} \\
            V_{td} & V_{ts} & V_{tb} \\
          \end{array}
        \right)
        =
        \left(
        \begin{array}{ccc}
        c_{12}c_{13}    
                &    s_{12}c_{13}   
                        &   s_{13}e^{-i\delta}  \\
        -s_{12}c_{23}-c_{12}s_{23}s_{13}e^{i\delta} 
                &  c_{12}c_{23}-s_{12}s_{23}s_{13}e^{i\delta} 
                        & s_{23}c_{13} \\
        s_{12}s_{23}-c_{12}c_{23}s_{13}e^{i\delta}  
                &  -c_{12}s_{23}-s_{12}c_{23}s_{13}e^{i\delta} 
                        & c_{23}c_{13} 
        \end{array}
        \right)
\end{equation}
where $c_{ij}=\cos\theta_{ij}$, $s_{ij}=\sin\theta_{ij}$ for 
$i<j=1,2,3$. 

Following the observation of a hierarchy between the different
matrix elements, the Wolfenstein parameterization~\cite{ref:cp_uta:wolfenstein}
is an expansion of $\VCKM$ in terms of the four real parameters $\lambda$
(the expansion parameter), $A$, $\rho$ and $\eta$. Defining to 
all orders in $\lambda$~\cite{ref:cp_uta:buras}
\begin{eqnarray}
  \label{eq:burasdef}
  s_{12}             &\equiv& \lambda,\nonumber \\ 
  s_{23}             &\equiv& A\lambda^2, \\
  s_{13}e^{-i\delta} &\equiv& A\lambda^3(\rho -i\eta),\nonumber
\end{eqnarray}
and inserting these into the representation of Eq.~(\ref{eq:ckmPdg}), 
unitarity of the CKM matrix is achieved to all orders.
A Taylor expansion of $\VCKM$ leads to the familiar approximation
\begin{equation}
  \label{eq:cp_uta:ckm}
  \VCKM
  = 
  \left(
    \begin{array}{ccc}
      1 - \lambda^2/2 & \lambda & A \lambda^3 ( \rho - i \eta ) \\
      - \lambda & 1 - \lambda^2/2 & A \lambda^2 \\
      A \lambda^3 ( 1 - \rho - i \eta ) & - A \lambda^2 & 1 \\
    \end{array}
  \right) + {\cal O}\left( \lambda^4 \right) \, .
\end{equation}
At order $\lambda^{5}$, the obtained CKM matrix in this extended
Wolfenstein parametrization is:
{\small
  \begin{equation}
    \label{eq:cp_uta:ckm_lambda5}
    \VCKM
    =
    \left(
      \begin{array}{ccc}
        1 - \frac{1}{2}\lambda^{2} - \frac{1}{8}\lambda^4 &
        \lambda &
        A \lambda^{3} (\rho - i \eta) \\
        - \lambda + \frac{1}{2} A^2 \lambda^5 \left[ 1 - 2 (\rho + i \eta) \right] &
        1 - \frac{1}{2}\lambda^{2} - \frac{1}{8}\lambda^4 (1+4A^2) &
        A \lambda^{2} \\
        A \lambda^{3} \left[ 1 - (1-\frac{1}{2}\lambda^2)(\rho + i \eta) \right] &
        -A \lambda^{2} + \frac{1}{2}A\lambda^4 \left[ 1 - 2(\rho + i \eta) \right] &
        1 - \frac{1}{2}A^2 \lambda^4
      \end{array} 
    \right) + {\cal O}\left( \lambda^{6} \right).
  \end{equation}
}
The non-zero imaginary part of the CKM matrix,
which is the origin of $\CP$ violation in the Standard Model,
is encapsulated in a non-zero value of $\eta$.



The unitarity relation $\VCKM^\dagger\VCKM = {\mathit 1}$
results in a total of nine expressions,
that can be written as
$\sum_{i=u,c,t} V^*_{ij}V_{ik} = \delta_{jk}$,
where $\delta_{jk}$ is the Kronecker symbol.
Of the off-diagonal expressions ($j \neq k$),
three can be trivially transformed into the other three 
(under $j \leftrightarrow k$),
leaving six relations, in which three complex numbers sum to zero,
which therefore can be expressed as triangles in the complex plane.

\begin{figure}[t]
  \begin{center}
    \resizebox{0.55\textwidth}{!}{\includegraphics{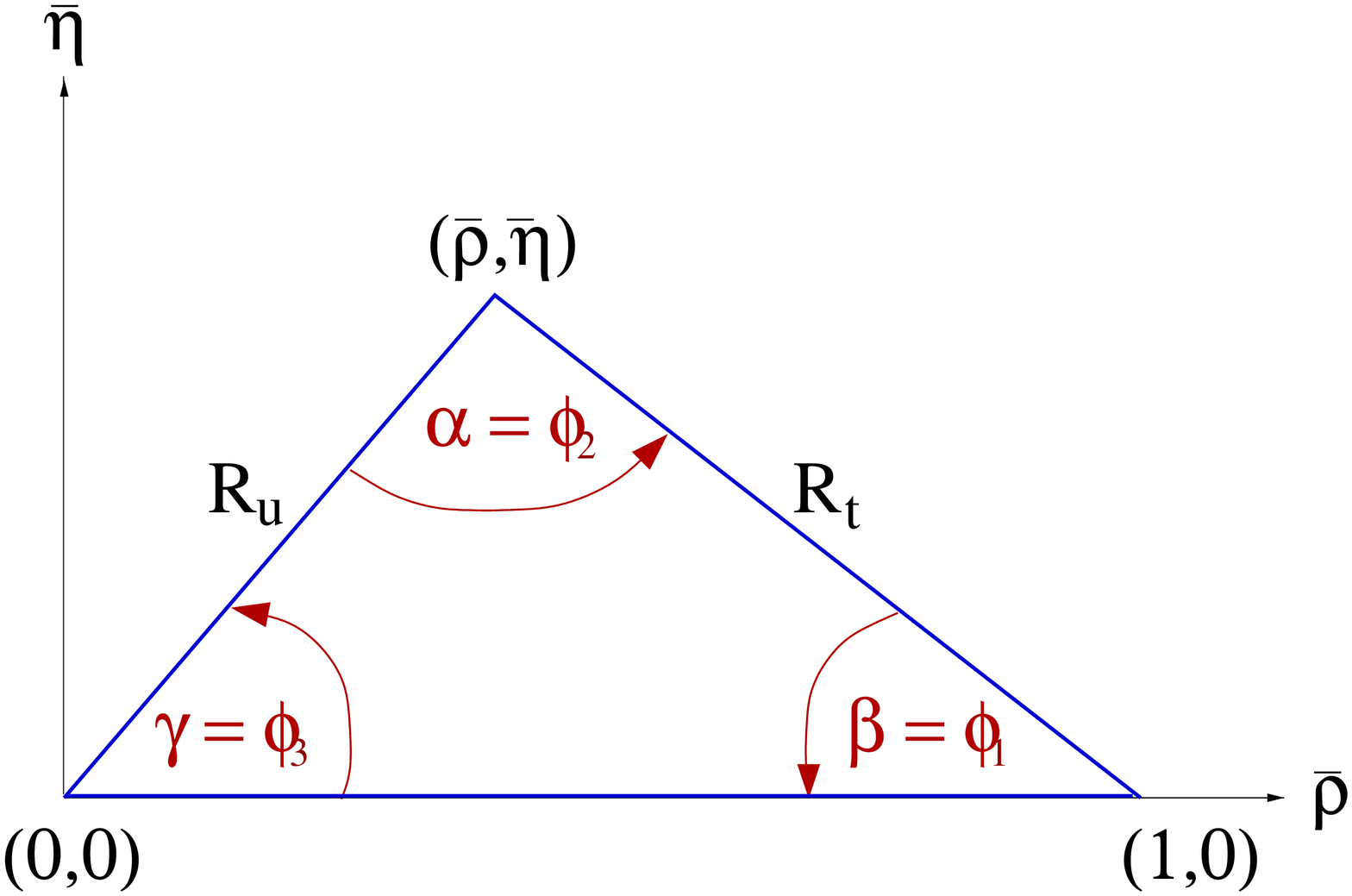}}
    \caption{The Unitarity Triangle.}
    \label{fig:cp_uta:ut}
  \end{center}
\end{figure}

One of these,
\begin{equation}
  \label{eq:cp_uta:ut}
  V_{ud}V^*_{ub} + V_{cd}V^*_{cb} + V_{td}V^*_{tb} = 0,
\end{equation}
is specifically related to $\B$ decays.
The three terms in Eq.~(\ref{eq:cp_uta:ut}) are of the same order 
(${\cal O}\left( \lambda^3 \right)$),
and this relation is commonly known as the Unitarity Triangle.
For presentational purposes,
it is convenient to rescale the triangle by $(V_{cd}V^*_{cb})^{-1}$,
as shown in Fig.~\ref{fig:cp_uta:ut}.

Two popular naming conventions for the UT angles exist in the literature:
\begin{equation}
  \label{eq:cp_uta:abc}
  \alpha  \equiv  \phi_2  = 
  \arg\left[ - \frac{V_{td}V_{tb}^*}{V_{ud}V_{ub}^*} \right],
  \hspace{0.5cm}
  \beta   \equiv   \phi_1 =  
  \arg\left[ - \frac{V_{cd}V_{cb}^*}{V_{td}V_{tb}^*} \right],
  \hspace{0.5cm}
  \gamma  \equiv   \phi_3  =  
  \arg\left[ - \frac{V_{ud}V_{ub}^*}{V_{cd}V_{cb}^*} \right].
\end{equation}
In this document the $\left( \alpha, \beta, \gamma \right)$ set is used.
The sides $R_u$ and $R_t$ of the Unitarity Triangle 
(the third side being normalized to unity) 
are given by
\begin{equation}
  \label{eq:ru_rt}
  R_u =
  \left|\frac{V_{ud}V_{ub}^*}{V_{cd}V_{cb}^*} \right|
  = \sqrt{\rhobar^2+\etabar^2},
  \hspace{0.5cm}
  R_t = 
  \left|\frac{V_{td}V_{tb}^*}{V_{cd}V_{cb}^*}\right| 
  = \sqrt{(1-\rhobar)^2+\etabar^2}.
\end{equation} 
where $\rhobar$ and $\etabar$ define 
the apex of the Unitarity Triangle~\cite{ref:cp_uta:buras} 
\begin{equation}
  \label{eq:rhoetabar}
  \rhobar + i\etabar
  \equiv -\frac{V_{ud}V_{ub}^*}{V_{cd}V_{cb}^*}
  \equiv 1 + \frac{V_{td}V_{tb}^*}{V_{cd}V_{cb}^*}
  = \frac{\sqrt{1-\lambda^2}\,(\rho + i \eta)}{\sqrt{1-A^2\lambda^4}+\sqrt{1-\lambda^2}A^2\lambda^4(\rho+i\eta)}
\end{equation}
The exact relation between $\left( \rho, \eta \right)$ and 
$\left( \rhobar, \etabar \right)$ is
\begin{equation}
  \label{eq:rhoetabarinv}
  \rho + i\eta \;=\; 
  \frac{ 
    \sqrt{ 1-A^2\lambda^4 }(\rhobar+i\etabar) 
  }{
    \sqrt{ 1-\lambda^2 } \left[ 1-A^2\lambda^4(\rhobar+i\etabar) \right]
  } \, .
\end{equation}

By expanding in powers of $\lambda$, several useful approximate expressions
can be obtained, including
\begin{equation}
  \label{eq:rhoeta_approx}
  \rhobar = \rho (1 - \frac{1}{2}\lambda^{2}) + {\cal O}(\lambda^4) \, ,
  \hspace{0.5cm}
  \etabar = \eta (1 - \frac{1}{2}\lambda^{2}) + {\cal O}(\lambda^4) \, ,
  \hspace{0.5cm}
  V_{td} = A \lambda^{3} (1-\rhobar -i\etabar) + {\cal O}(\lambda^6) \, .
\end{equation}

\mysubsection{Notations
}
\label{sec:cp_uta:notations}

Several different notations for $\CP$ violation parameters
are commonly used.
This section reviews those found in the experimental literature,
in the hope of reducing the potential for confusion, 
and to define the frame that is used for the averages.

In some cases, when $\B$ mesons decay into 
multibody final states via broad resonances ($\rho$, $\Kstar$, \etc),
the experimental analyses ignore the effects of interference 
between the overlapping structures.
This is referred to as the quasi-two-body (Q2B) approximation
in the following.

\mysubsubsection{$\CP$ asymmetries
}
\label{sec:cp_uta:notations:pra}

The $\CP$ asymmetry is defined as the difference between the rate 
involving a $b$ quark and that involving a $\bar b$ quark, divided 
by the sum. For example, the partial rate (or charge) asymmetry for 
a charged $\B$ decay would be given as 
\begin{equation}
  \label{eq:cp_uta:pra}
  \Acp_{f} \;\equiv\; 
  \frac{\Gamma(\Bm \to f)-\Gamma(\Bp \to \bar{f})}{\Gamma(\Bm \to f)+\Gamma(\Bp \to \bar{f})}.
\end{equation}

\mysubsubsection{Time-dependent \CP asymmetries in decays to $\CP$ eigenstates
}
\label{sec:cp_uta:notations:cp_eigenstate}

If the amplitudes for $\Bz$ and $\Bzb$ to decay to a final state $f$, 
which is a $\CP$ eigenstate with eigenvalue $\etacpf$,
are given by $\Af$ and $\Abarf$, respectively, 
then the decay distributions for neutral $\B$ mesons, 
with known flavour at time $\Delta t =0$,
are given by
\begin{eqnarray}
  \label{eq:cp_uta:td_cp_asp1}
  \Gamma_{\Bzb \to f} (\Delta t) & = &
  \frac{e^{-| \Delta t | / \tau(\Bz)}}{4\tau(\Bz)}
  \left[ 
    1 +
    \frac{2\, \Im(\lambda_f)}{1 + |\lambda_f|^2} \sin(\Delta m \Delta t) -
    \frac{1 - |\lambda_f|^2}{1 + |\lambda_f|^2} \cos(\Delta m \Delta t)
  \right], \\
  \label{eq:cp_uta:td_cp_asp2}
  \Gamma_{\Bz \to f} (\Delta t) & = &
  \frac{e^{-| \Delta t | / \tau(\Bz)}}{4\tau(\Bz)}
  \left[ 
    1 -
    \frac{2\, \Im(\lambda_f)}{1 + |\lambda_f|^2} \sin(\Delta m \Delta t) +
    \frac{1 - |\lambda_f|^2}{1 + |\lambda_f|^2} \cos(\Delta m \Delta t)
  \right].
\end{eqnarray}
Here $\lambda_f = \frac{q}{p} \frac{\Abarf}{\Af}$ 
contains terms related to $\Bz$\textendash$\Bzb$ mixing and to the decay amplitude
(the eigenstates of the effective Hamiltonian in the $\BzBzb$ system 
are $\left| B_\pm \right> = p \left| \Bz \right> \pm q \left| \Bzb \right>$).
This formulation assumes $\CPT$ invariance, 
and neglects possible lifetime differences 
(between the eigenstates of the effective Hamiltonian;
see Section~\ref{sec:mixing} where the mass difference $\Delta m$ is also defined)
in the neutral $\B$ meson system.
The case where non-zero lifetime differences are taken into account is 
discussed in Section~\ref{sec:cp_uta:notations:Bs}.
The time-dependent $\CP$ asymmetry,
again defined as the difference between the rate 
involving a $b$ quark and that involving a $\bar b$ quark,
is then given by
\begin{equation}
  \label{eq:cp_uta:td_cp_asp}
  \Acp_{f} \left(\Delta t\right) \; \equiv \;
  \frac{
    \Gamma_{\Bzb \to f} (\Delta t) - \Gamma_{\Bz \to f} (\Delta t)
  }{
    \Gamma_{\Bzb \to f} (\Delta t) + \Gamma_{\Bz \to f} (\Delta t)
  } \; = \;
  \frac{2\, \Im(\lambda_f)}{1 + |\lambda_f|^2} \sin(\Delta m \Delta t) -
  \frac{1 - |\lambda_f|^2}{1 + |\lambda_f|^2} \cos(\Delta m \Delta t).
\end{equation}

While the coefficient of the $\sin(\Delta m \Delta t)$ term in 
Eq.~(\ref{eq:cp_uta:td_cp_asp}) is everywhere\footnote
{
  Occasionally one also finds Eq.~(\ref{eq:cp_uta:td_cp_asp}) written as
  $\Acp_{f} \left(\Delta t\right) = 
  {\cal A}^{\rm mix}_f \sin(\Delta m \Delta t) + {\cal A}^{\rm dir}_f \cos(\Delta m \Delta t)$,
  or similar.
} denoted $S_f$:
\begin{equation}
  \label{eq:cp_uta:s_def}
  S_f \;\equiv\; \frac{2\, \Im(\lambda_f)}{1 + \left|\lambda_f\right|^2},
\end{equation}
different notations are in use for the
coefficient of the $\cos(\Delta m \Delta t)$ term:
\begin{equation}
  \label{eq:cp_uta:c_def}
  C_f \;\equiv\; - A_f \;\equiv\; \frac{1 - \left|\lambda_f\right|^2}{1 + \left|\lambda_f\right|^2}.
\end{equation}
The $C$ notation is used by the \babar\  collaboration 
(see \eg~\cite{ref:cp_uta:ccs:babar_obs}), 
and also in this document.
The $A$ notation is used by the \belle\ collaboration
(see \eg~\cite{ref:cp_uta:ccs:belle_obs}).

Neglecting effects due to $\CP$ violation in mixing 
(by taking $|q/p| = 1$),
if the decay amplitude contains terms with 
a single weak (\ie, $\CP$ violating) phase
then $\left|\lambda_f\right| = 1$ and one finds
$S_f = -\etacpf \sin(\phi_{\rm mix} + \phi_{\rm dec})$, $C_f = 0$,
where $\phi_{\rm mix}=\arg(q/p)$ and $\phi_{\rm dec}=\arg(\Abarf/\Af)$.
Note that the $\Bz$--$\Bzb$ mixing phase $\phi_{\rm mix}\approx2\beta$
in the Standard Model (in the usual phase convention). 

If amplitudes with different weak phases contribute to the decay, 
no clean interpretation of $S_f$ is possible. If the decay amplitudes
have in addition different $\CP$ conserving strong phases,
then $\left| \lambda_f \right| \neq 1$ and no clean interpretation is possible.
The coefficient of the cosine term becomes non-zero,
indicating direct $\CP$ violation.
The sign of $A_f$ as defined above is consistent with that of $\Acp_{f}$ in 
Eq.~(\ref{eq:cp_uta:pra}).

Frequently, we are interested in combining measurements 
governed by similar or identical short-distance physics,
but with different final states
(\eg, $\Bz \to \jpsi \KS$ and $\Bz \to \jpsi \KL$).
In this case, we remove the dependence on the $\CP$ eigenvalue 
of the final state by quoting $-\etacp S_f$.
In cases where the final state is not a $\CP$ eigenstate but has
an effective $\CP$ content (see below),
the reported $-\etacp S$ is corrected by the effective $\CP$.

\mysubsubsection{Time-dependent \CP asymmetries in decays to vector-vector final states
}
\label{sec:cp_uta:notations:vv}

Consider \B decays to states consisting of two spin-1 particles,
such as $\jpsi K^{*0}(\to\KS\piz)$, $D^{*+}D^{*-}$ and $\rho^+\rho^-$,
which are eigenstates of charge conjugation but not of parity.\footnote{
  \noindent
  This is not true of all vector-vector final states,
  \eg, $D^{*\pm}\rho^{\mp}$ is clearly not an eigenstate of 
  charge conjugation.
}
In fact, for such a system, there are three possible final states;
in the helicity basis these can be written $h_{-1}, h_0, h_{+1}$.
The $h_0$ state is an eigenstate of parity, and hence of $\CP$;
however, $\CP$ transforms $h_{+1} \leftrightarrow h_{-1}$ (up to 
an unobservable phase). In the transversity basis, these states 
are transformed into  $h_\parallel =  (h_{+1} + h_{-1})/2$ and 
$h_\perp = (h_{+1} - h_{-1})/2$.
In this basis all three states are $\CP$ eigenstates, 
and $h_\perp$ has the opposite $\CP$ to the others.

The amplitudes to these states are usually given by $A_{0,\perp,\parallel}$
(here we use a normalization such that 
$| A_0 |^2 + | A_\perp |^2 + | A_\parallel |^2 = 1$).
Then the effective $\CP$ of the vector-vector state is known if 
$| A_\perp |^2$ is measured.
An alternative strategy is to measure just the longitudinally polarized 
component,  $| A_0 |^2$
(sometimes denoted by $f_{\rm long}$), 
which allows a limit to be set on the effective $\CP$ since
$| A_\perp |^2 \leq | A_\perp |^2 + | A_\parallel |^2 = 1 - | A_0 |^2$.
The most complete treatment for 
neutral $\B$ decays to vector-vector final states
is time-dependent angular analysis 
(also known as time-dependent transversity analysis).
In such an analysis, 
the interference between the $\CP$ even and $\CP$ odd states 
provides additional sensitivity to the weak and strong phases involved.

\mysubsubsection{Time-dependent asymmetries in decays to self-conjugate multiparticle final states
}
\label{sec:cp_uta:notations:dalitz}

Amplitudes for neutral \B decays into 
self-conjugate multiparticle final states
such as $\pi^+\pi^-\pi^0$, $K^+K^-\KS$,
$\jpsi \pi^+\pi^-$ or $D\pi^0$ with $D \to \KS\pi^+\pi^-$
may be written in terms of \CP-even and \CP-odd amplitudes.
As above, the interference between these terms 
provides additional sensitivity to the weak and strong phases
involved in the decay,
and the time-dependence depends on both the sine and cosine
of the weak phase difference.
In order to perform unbinned maximum likelihood fits,
and thereby extract as much information as possible from the distributions,
it is necessary to select a model for the multiparticle decay,
and therefore the results acquire some model dependence
(binned, model independent methods are also possible,
though are not as statistically powerful).
The number of observables depends on the final state (and on the model used);
the key feature is that as long as there are regions where both
\CP-even and \CP-odd amplitudes contribute,
the interference terms will be sensitive to the cosine 
of the weak phase difference.
Therefore, these measurements allow distinction between multiple solutions
for, \eg, the four values of $\beta$ from the measurement of $\sin(2\beta)$.

We now consider the various notations which have been used 
in experimental studies of
time-dependent asymmetries in decays to self-conjugate multiparticle final states.

\mysubsubsubsection{$\Bz \to D^{(*)}h^0$ with $D \to \KS\pi^+\pi^-$
}
\label{sec:cp_uta:notations:dalitz:dh0}

The states $D\pi^0$, $D^*\pi^0$, $D\eta$, $D^*\eta$, $D\omega$
are collectively denoted $D^{(*)}h^0$.
When the $D$ decay model is fixed,
fits to the time-dependent decay distributions can be performed
to extract the weak phase difference.
However, it is experimentally advantageous to use the sine and cosine of 
this phase as fit parameters, since these behave as essentially 
independent parameters, with low correlations and (potentially)
rather different uncertainties.
A parameter representing direct $\CP$ violation in the $B$ decay 
can also be floated.  
For consistency with other analyses, this could be chosen to be $C_f$,
but could equally well be $\left| \lambda_f \right|$, or other possibilities.

\belle\ performed an analysis of these channels
with $\sin(2\phi_1)$ and $\cos(2\phi_1)$ as free parameters~\cite{ref:cp_uta:cud_beta:belle}.
\babar\ have performed an analysis floating also $\left| \lambda_f \right|$~\cite{ref:cp_uta:cud_beta:babar}
(and, of course, replacing $\phi_1 \Leftrightarrow \beta$).

\mysubsubsubsection{$\Bz \to D^{*+}D^{*-}\KS$
}
\label{sec:cp_uta:notations:dalitz:dstardstarks}

The hadronic structure of the $\Bz \to D^{*+}D^{*-}\KS$ decay
is not sufficiently well understood to perform a full 
time-dependent Dalitz plot analysis.
Instead, following Browder {\it et al.}~\cite{ref:cp_uta:ccs:ddks:browder},
\babar~\cite{ref:cp_uta:ccs:ddks:babar} divide the Dalitz plane in two:
$m(D^{*+}\KS)^2 > m(D^{*-}\KS)^2$ $(\eta_y = +1)$ and 
$m(D^{*+}\KS)^2 < m(D^{*-}\KS)^2$ $(\eta_y = -1)$;
and then fit to a decay time distribution with asymmetry given by
\begin{equation}
  \Acp_{f} \left(\Delta t\right) =
  \eta_y \frac{J_c}{J_0} \cos(\Delta m \Delta t) -  
  \left[ 
    \frac{2J_{s1}}{J_0} \sin(2\beta) + \eta_y \frac{2J_{s2}}{J_0} \cos(2\beta) 
  \right] \sin(\Delta m \Delta t) \, .
\end{equation}
The measured values are $\frac{J_c}{J_0}$, $\frac{2J_{s1}}{J_0} \sin(2\beta)$
and $\frac{2J_{s2}}{J_0} \cos(2\beta)$, 
where the parameters $J_0$, $J_c$, $J_{s1}$ and $J_{s2}$ are the integrals 
over the half Dalitz plane $m(D^{*+}\KS)^2 < m(D^{*-}\KS)^2$ 
of the functions $|a|^2 + |\bar{a}|^2$, $|a|^2 - |\bar{a}|^2$, 
$\Re(\bar{a}a^*)$ and $\Im(\bar{a}a^*)$ respectively, 
where $a$ and $\bar{a}$ are the decay amplitudes of 
$\Bz \to D^{*+}D^{*-}\KS$ and $\Bzb \to D^{*+}D^{*-}\KS$ respectively. 
The parameter $J_{s2}$ (and hence $J_{s2}/J_0$) is predicted to be positive;
with this assumption is it possible to determine the sign of $\cos(2\beta)$.

\mysubsubsubsection{$\Bz \to K^+K^-\Kz$
}
\label{sec:cp_uta:notations:dalitz:kkk0}

Studies of $\Bz \to K^+K^-\Kz$~\cite{ref:cp_uta:qqs:babar:kkk0_tddp} 
and of the related decay 
$\Bp \to K^+K^-K^+$~\cite{ref:cp_uta:qqs:belle:kkk,ref:cp_uta:qqs:babar:kkk},
show that the decay is dominated by components from the 
intermediate $K^+K^-$ resonances $\phi(1020)$, $f_0(980)$,
a poorly understood scalar structure that peaks near 
$m(K^+K^-) \sim 1550 \ {\rm MeV}/c^2$ and is denoted $X_0(1550)$,
as well as a large nonresonant contribution.
There is also a contribution from $\chi_{c0}$.

The full time-dependent Dalitz plot analysis allows 
the complex amplitudes of each contributing term to be determined from data,
including $\CP$ violation effects
(\ie\ allowing the complex amplitude for the $\Bz$ decay to be independent
from that for $\Bzb$ decay), although one amplitude must be fixed 
to give a reference point.
There are several choices for parametrization of the complex amplitudes 
(\eg\ real and imaginary part, or magnitude and phase).
Similarly, there are various approaches to include $\CP$ violation effects.
Note that positive definite parameters such as magnitudes are
disfavoured in certain circumstances 
(they inevitably lead to biases for small values).
In order to compare results between analyses,
it is useful for each experiment to present results in terms of the 
parameters that can be measured in a Q2B analysis
(such as $\Acp_{f}$, $S_f$, $C_f$, 
$\sin(2\beta^{\rm eff})$, $\cos(2\beta^{\rm eff})$, \etc)

In the \babar\ analysis of $\Bz \to K^+K^-\Kz$~\cite{ref:cp_uta:qqs:babar:kkk0_tddp},
the complex amplitude for each resonant contribution is written as
\begin{equation}
  A_f = c_f ( 1 + b_f ) e^{i ( \phi_f + \delta_f )} 
  \ , \ \ \ \ 
  \bar{A}_f = c_f ( 1 - b_f ) e^{i ( \phi_f - \delta_f )} \, ,
\end{equation}
where $b_f$ and $\delta_f$ introduce $\CP$ violation in the magnitude 
and phase respectively.
[The weak phase in $B^0$--$\bar{B}^0$ mixing ($2\beta$) also appears 
in the full formula for the time-dependent decay distribution.]
The Q2B direct $\CP$ violation parameter is directly related to $b_f$
\begin{equation}
  \Acp_{f} = \frac{-2b_f}{1+b_f^2} \, .
\end{equation}

\babar\ present results for $c_f$, $\phi_f$, $\Acp_{f}$ and $\beta^{\rm eff}$
for each resonant contribution,
as well as averaged values of $\Acp_{f}$ and $\beta^{\rm eff}$
for the entire $K^+K^-\Kz$ Dalitz plot.

\mysubsubsubsection{$\Bz \to \pi^+\pi^-\pi^0$
}
\label{sec:cp_uta:notations:dalitz:pipipi0}

The $\Bz \to \pi^+\pi^-\pi^0$ decay is dominated by 
intermediate $\rho$ resonances.
Though it is possible, as above, 
to determine directly the complex amplitudes for each component,
an alternative approach, 
suggested by Quinn and Silva~\cite{ref:cp_uta:uud:quinnsilva},
has been used by both \babar~\cite{ref:cp_uta:uud:babar:pipipi0}
and \belle~\cite{ref:cp_uta:uud:belle:pipipi0}.
The amplitudes for $\Bz$ and $\Bzb$ to $\pi^+\pi^-\pi^0$ are written
\begin{equation}
  A_{3\pi} = f_+ A_+ + f_- A_- + f_0 A_0
  \ , \ \ \ 
  \bar{A}_{3\pi} = f_+ \bar{A}_+ + f_- \bar{A}_- + f_0 \bar{A}_0
\end{equation}
respectively.
$A_+$, $A_-$ and $A_0$
represent the complex decay amplitudes for 
$\Bz \to \rho^+\pi^-$, $\Bz \to \rho^-\pi^+$ and $\Bz \to \rho^0\pi^0$
while 
$\bar{A}_+$, $\bar{A}_-$ and $\bar{A}_0$
represent those for 
$\Bzb \to \rho^+\pi^-$, $\Bzb \to \rho^-\pi^+$ and $\Bzb \to \rho^0\pi^0$
respectively.
$f_+$, $f_-$ and $f_0$ incorporate kinematical and dynamical factors
and depend on the Dalitz plot coordinates.
The full time-dependent decay distribution can then be written 
in terms of 27 free parameters,
one for each coefficient of the form factor bilinears,
as listed in Table~\ref{tab:cp_uta:pipipi0:uandi}.
These parameters are often referred to as ``the $U$s and $I$s'',
and can be expressed in terms of 
$A_+$, $A_-$, $A_0$, $\bar{A}_+$, $\bar{A}_-$ and $\bar{A}_0$.
If the full set of parameters is determined,
together with their correlations,
other parameters, such as weak and strong phases,
direct $\CP$ violation parameters, \etc, 
can be subsequently extracted.
Note that one of the parameters (typically $U_+^+$)
is often fixed to unity to provide a reference point;
this does not affect the analysis.


\begin{table}[htb]
  \begin{center}
    \setlength{\tabcolsep}{0.3pc}
    \begin{tabular}{l@{\extracolsep{5mm}}l}
      \hline
      Parameter   & Description \\
      \hline
      $U_+^+$          & Coefficient of $|f_+|^2$ \\
      $U_0^+$          & Coefficient of $|f_0|^2$ \\
      $U_-^+$          & Coefficient of $|f_-|^2$ \\
      [0.15cm]
      $U_0^-$          & Coefficient of $|f_0|^2\cos(\Delta m\Delta t)$ \\
      $U_-^-$          & Coefficient of $|f_-|^2\cos(\Delta m\Delta t)$ \\
      $U_+^-$          & Coefficient of $|f_+|^2\cos(\Delta m\Delta t)$ \\
      [0.15cm]
      $I_0$            & Coefficient of $|f_0|^2\sin(\Delta m\Delta t)$ \\
      $I_-$            & Coefficient of $|f_-|^2\sin(\Delta m\Delta t)$ \\
      $I_+$            & Coefficient of $|f_+|^2\sin(\Delta m\Delta t)$ \\
      [0.15cm]
      $U_{+-}^{+,\Im}$ & Coefficient of $\Im[f_+f_-^*]$ \\
      $U_{+-}^{+,\Re}$ & Coefficient of $\Re[f_+f_-^*]$ \\
      $U_{+-}^{-,\Im}$ & Coefficient of $\Im[f_+f_-^*]\cos(\Delta m\Delta t)$ \\
      $U_{+-}^{-,\Re}$ & Coefficient of $\Re[f_+f_-^*]\cos(\Delta m\Delta t)$ \\
      $I_{+-}^{\Im}$   & Coefficient of $\Im[f_+f_-^*]\sin(\Delta m\Delta t)$ \\
      $I_{+-}^{\Re}$   & Coefficient of $\Re[f_+f_-^*]\sin(\Delta m\Delta t)$ \\
      [0.15cm]
      $U_{+0}^{+,\Im}$ & Coefficient of $\Im[f_+f_0^*]$ \\
      $U_{+0}^{+,\Re}$ & Coefficient of $\Re[f_+f_0^*]$ \\
      $U_{+0}^{-,\Im}$ & Coefficient of $\Im[f_+f_0^*]\cos(\Delta m\Delta t)$ \\
      $U_{+0}^{-,\Re}$ & Coefficient of $\Re[f_+f_0^*]\cos(\Delta m\Delta t)$ \\
      $I_{+0}^{\Im}$   & Coefficient of $\Im[f_+f_0^*]\sin(\Delta m\Delta t)$ \\
      $I_{+0}^{\Re}$   & Coefficient of $\Re[f_+f_0^*]\sin(\Delta m\Delta t)$ \\
      [0.15cm]
      $U_{-0}^{+,\Im}$ & Coefficient of $\Im[f_-f_0^*]$ \\
      $U_{-0}^{+,\Re}$ & Coefficient of $\Re[f_-f_0^*]$ \\
      $U_{-0}^{-,\Im}$ & Coefficient of $\Im[f_-f_0^*]\cos(\Delta m\Delta t)$ \\
      $U_{-0}^{-,\Re}$ & Coefficient of $\Re[f_-f_0^*]\cos(\Delta m\Delta t)$ \\
      $I_{-0}^{\Im}$   & Coefficient of $\Im[f_-f_0^*]\sin(\Delta m\Delta t)$ \\
      $I_{-0}^{\Re}$   & Coefficient of $\Re[f_-f_0^*]\sin(\Delta m\Delta t)$ \\     
      \hline
    \end{tabular}
    \caption{
      Definitions of the $U$ and $I$ coefficients.
      Modified from~\cite{ref:cp_uta:uud:babar:pipipi0}.
    }
    \label{tab:cp_uta:pipipi0:uandi}
  \end{center}
\end{table}

\mysubsubsection{Time-dependent \CP asymmetries in decays to non-$\CP$ eigenstates
}
\label{sec:cp_uta:notations:non_cp}

Consider a non-$\CP$ eigenstate $f$, and its conjugate $\bar{f}$. 
For neutral $\B$ decays to these final states,
there are four amplitudes to consider:
those for $\Bz$ to decay to $f$ and $\bar{f}$
($\Af$ and $\Afbar$, respectively),
and the equivalents for $\Bzb$
($\Abarf$ and $\Abarfbar$).
If $\CP$ is conserved in the decay, then
$\Af = \Abarfbar$ and $\Afbar = \Abarf$.


The time-dependent decay distributions can be written in many different ways.
Here, we follow Sec.~\ref{sec:cp_uta:notations:cp_eigenstate}
and define $\lambda_f = \frac{q}{p}\frac{\Abarf}{\Af}$ and
$\lambda_{\bar f} = \frac{q}{p}\frac{\Abarfbar}{\Afbar}$.
The time-dependent \CP asymmetries then follow Eq.~(\ref{eq:cp_uta:td_cp_asp}):
\begin{eqnarray}
\label{eq:cp_uta:non-cp-obs}
  {\cal A}_f (\Delta t) \; \equiv \;
  \frac{
    \Gamma_{\Bzb \to f} (\Delta t) - \Gamma_{\Bz \to f} (\Delta t)
  }{
    \Gamma_{\Bzb \to f} (\Delta t) + \Gamma_{\Bz \to f} (\Delta t)
  } & = & S_f \sin(\Delta m \Delta t) - C_f \cos(\Delta m \Delta t), \\
  {\cal A}_{\bar{f}} (\Delta t) \; \equiv \;
  \frac{
    \Gamma_{\Bzb \to \bar{f}} (\Delta t) - \Gamma_{\Bz \to \bar{f}} (\Delta t)
  }{
    \Gamma_{\Bzb \to \bar{f}} (\Delta t) + \Gamma_{\Bz \to \bar{f}} (\Delta t)
  } & = & S_{\bar{f}} \sin(\Delta m \Delta t) - C_{\bar{f}} \cos(\Delta m \Delta t),
\end{eqnarray}
with the definitions of the parameters 
$C_f$, $S_f$, $C_{\bar{f}}$ and $S_{\bar{f}}$,
following Eqs.~(\ref{eq:cp_uta:s_def}) and~(\ref{eq:cp_uta:c_def}).

The time-dependent decay rates are given by
\begin{eqnarray}
  \Gamma_{\Bzb \to f} (\Delta t) & = &
  \frac{e^{-\left| \Delta t \right| / \tau(\Bz)}}{8\tau(\Bz)} 
  ( 1 + \Adirnoncp ) 
  \left\{ 
    1 + S_f \sin(\Delta m \Delta t) - C_f \cos(\Delta m \Delta t) 
  \right\},
  \\
  \Gamma_{\Bz \to f} (\Delta t) & = &
  \frac{e^{-\left| \Delta t \right| / \tau(\Bz)}}{8\tau(\Bz)} 
  ( 1 + \Adirnoncp ) 
  \left\{ 
    1 - S_f \sin(\Delta m \Delta t) + C_f \cos(\Delta m \Delta t) 
  \right\},
  \\
  \Gamma_{\Bzb \to \bar{f}} (\Delta t) & = &
  \frac{e^{-\left| \Delta t \right| / \tau(\Bz)}}{8\tau(\Bz)} 
  ( 1 - \Adirnoncp ) 
  \left\{ 
    1 + S_{\bar{f}} \sin(\Delta m \Delta t) - C_{\bar{f}} \cos(\Delta m \Delta t) 
  \right\},
  \\
  \Gamma_{\Bz \to \bar{f}} (\Delta t) & = &
    \frac{e^{-\left| \Delta t \right| / \tau(\Bz)}}{8\tau(\Bz)} 
  ( 1 - \Adirnoncp ) 
  \left\{ 
    1 - S_{\bar{f}} \sin(\Delta m \Delta t) + C_{\bar{f}} \cos(\Delta m \Delta t) 
  \right\},
\end{eqnarray}
where the time-independent parameter \Adirnoncp
represents an overall asymmetry in the production of the 
$f$ and $\bar{f}$ final states,\footnote{
  This parameter is often denoted ${\cal A}_f$ (or ${\cal A}_{\CP}$),
  but here we avoid this notation to prevent confusion with the
  time-dependent $\CP$ asymmetry.
}
\begin{equation}
  \Adirnoncp = 
  \frac{
    \left( 
      \left| \Af \right|^2 + \left| \Abarf \right|^2
    \right) - 
    \left( 
      \left| \Afbar \right|^2 + \left| \Abarfbar \right|^2
    \right)
  }{
    \left( 
      \left| \Af \right|^2 + \left| \Abarf \right|^2
    \right) +
    \left( 
      \left| \Afbar \right|^2 + \left| \Abarfbar \right|^2
    \right)
  }.
\end{equation}
Assuming $|q/p| = 1$,
the parameters $C_f$ and $C_{\bar{f}}$
can also be written in terms of the decay amplitudes as follows:
\begin{equation}
  C_f = 
  \frac{
    \left| \Af \right|^2 - \left| \Abarf \right|^2 
  }{
    \left| \Af \right|^2 + \left| \Abarf \right|^2
  }
  \hspace{5mm}
  {\rm and}
  \hspace{5mm}
  C_{\bar{f}} = 
  \frac{
    \left| \Afbar \right|^2 - \left| \Abarfbar \right|^2
  }{
    \left| \Afbar \right|^2 + \left| \Abarfbar \right|^2
  },
\end{equation}
giving asymmetries in the decay amplitudes of $\Bz$ and $\Bzb$
to the final states $f$ and $\bar{f}$ respectively.
In this notation, the direct $\CP$ invariance conditions are
$\Adirnoncp = 0$ and $C_f = - C_{\bar{f}}$.
Note that $C_f$ and $C_{\bar{f}}$ are typically non-zero;
\eg, for a flavour-specific final state, 
$\Abarf = \Afbar = 0$ ($\Af = \Abarfbar = 0$), they take the values
$C_f = - C_{\bar{f}} = 1$ ($C_f = - C_{\bar{f}} = -1$).

The coefficients of the sine terms
contain information about the weak phase. 
In the case that each decay amplitude contains only a single weak phase
(\ie, no direct $\CP$ violation),
these terms can be written
\begin{equation}
  S_f = 
  \frac{ 
    - 2 \left| \Af \right| \left| \Abarf \right| 
    \sin( \phi_{\rm mix} + \phi_{\rm dec} - \delta_f )
  }{
    \left| \Af \right|^2 + \left| \Abarf \right|^2
  } 
  \hspace{5mm}
  {\rm and}
  \hspace{5mm}
  S_{\bar{f}} = 
  \frac{
    - 2 \left| \Afbar \right| \left| \Abarfbar \right| 
    \sin( \phi_{\rm mix} + \phi_{\rm dec} + \delta_f )
  }{
    \left| \Afbar \right|^2 + \left| \Abarfbar \right|^2
  },
\end{equation}
where $\delta_f$ is the strong phase difference between the decay amplitudes.
If there is no $\CP$ violation, the condition $S_f = - S_{\bar{f}}$ holds.
If decay amplitudes with different weak and strong phases contribute,
no clean interpretation of $S_f$ and $S_{\bar{f}}$ is possible.

Since two of the $\CP$ invariance conditions are 
$C_f = - C_{\bar{f}}$ and $S_f = - S_{\bar{f}}$,
there is motivation for a rotation of the parameters:
\begin{equation}
\label{eq:cp_uta:non-cp-s_and_deltas}
  S_{f\bar{f}} = \frac{S_{f} + S_{\bar{f}}}{2},
  \hspace{4mm}
  \Delta S_{f\bar{f}} = \frac{S_{f} - S_{\bar{f}}}{2},
  \hspace{4mm}
  C_{f\bar{f}} = \frac{C_{f} + C_{\bar{f}}}{2},
  \hspace{4mm}
  \Delta C_{f\bar{f}} = \frac{C_{f} - C_{\bar{f}}}{2}.
\end{equation}
With these parameters, the $\CP$ invariance conditions become
$S_{f\bar{f}} = 0$ and $C_{f\bar{f}} = 0$. 
The parameter $\Delta C_{f\bar{f}}$ gives a measure of the ``flavour-specificity''
of the decay:
$\Delta C_{f\bar{f}}=\pm1$ corresponds to a completely flavour-specific decay,
in which no interference between decays with and without mixing can occur,
while $\Delta C_{f\bar{f}} = 0$ results in 
maximum sensitivity to mixing-induced $\CP$ violation.
The parameter $\Delta S_{f\bar{f}}$ is related to the strong phase difference 
between the decay amplitudes of $\Bz$ to $f$ and to $\bar f$. 
We note that the observables of Eq.~(\ref{eq:cp_uta:non-cp-s_and_deltas})
exhibit experimental correlations 
(typically of $\sim 20\%$, depending on the tagging purity, and other effects)
between $S_{f\bar{f}}$ and  $\Delta S_{f\bar{f}}$, 
and between $C_{f\bar{f}}$ and $\Delta C_{f\bar{f}}$. 
On the other hand, 
the final state specific observables of Eq.~(\ref{eq:cp_uta:non-cp-obs})
tend to have low correlations.

Alternatively, if we recall that the $\CP$ invariance
conditions at the decay amplitude level are
$\Af = \Abarfbar$ and $\Afbar = \Abarf$,
we are led to consider the parameters~\cite{ref:cp_uta:ckmfitter}
\begin{equation}
  \label{eq:cp_uta:non-cp-directcp}
  {\cal A}_{f\bar{f}} = 
  \frac{
    \left| \Abarfbar \right|^2 - \left| \Af \right|^2 
  }{
    \left| \Abarfbar \right|^2 + \left| \Af \right|^2
  }
  \hspace{5mm}
  {\rm and}
  \hspace{5mm}
  {\cal A}_{\bar{f}f} = 
  \frac{
    \left| \Abarf \right|^2 - \left| \Afbar \right|^2
  }{
    \left| \Abarf \right|^2 + \left| \Afbar \right|^2
  }.
\end{equation}
These are sometimes considered more physically intuitive parameters
since they characterize direct $\CP$ violation 
in decays with particular topologies.
For example, in the case of $\Bz \to \rho^\pm\pi^\mp$
(choosing $f =  \rho^+\pi^-$ and $\bar{f} = \rho^-\pi^+$),
${\cal A}_{f\bar{f}}$ (also denoted ${\cal A}^{+-}_{\rho\pi}$)
parameterizes direct $\CP$ violation
in decays in which the produced $\rho$ meson does not contain the 
spectator quark,
while ${\cal A}_{\bar{f}f}$ (also denoted ${\cal A}^{-+}_{\rho\pi}$)
parameterizes direct $\CP$ violation 
in decays in which it does.
Note that we have again followed the sign convention that the asymmetry 
is the difference between the rate involving a $b$ quark and that
involving a $\bar{b}$ quark, \cf\ Eq.~(\ref{eq:cp_uta:pra}). 
Of course, these parameters are not independent of the 
other sets of parameters given above, and can be written
\begin{equation}
  {\cal A}_{f\bar{f}} =
  - \frac{
    \Adirnoncp + C_{f\bar{f}} + \Adirnoncp \Delta C_{f\bar{f}} 
  }{
    1 + \Delta C_{f\bar{f}} + \Adirnoncp C_{f\bar{f}} 
  }
  \hspace{5mm}
  {\rm and}
  \hspace{5mm}
  {\cal A}_{\bar{f}f} =
  \frac{
    - \Adirnoncp + C_{f\bar{f}} + \Adirnoncp \Delta C_{f\bar{f}} 
  }{
    - 1 + \Delta C_{f\bar{f}} + \Adirnoncp C_{f\bar{f}}  
  }.
\end{equation}
They usually exhibit strong correlations.

We now consider the various notations which have been used 
in experimental studies of
time-dependent $\CP$ asymmetries in decays to non-$\CP$ eigenstates.

\mysubsubsubsection{$\Bz \to D^{*\pm}D^\mp$
}
\label{sec:cp_uta:notations:non_cp:dstard}

The above set of parameters 
($\Adirnoncp$, $C_f$, $S_f$, $C_{\bar{f}}$, $S_{\bar{f}}$),
has been used by both
\babar~\cite{ref:cp_uta:ccd:babar:dd_dstard} and
\belle~\cite{ref:cp_uta:ccd:belle:dstard} in the $D^{*\pm}D^{\mp}$ system
($f = D^{*+}D^-$, $\bar{f} = D^{*-}D^+$).
However, slightly different names for the parameters are used:
\babar\ uses 
(${\cal A}$, $C_{+-}$, $S_{+-}$, $C_{-+}$, $S_{-+}$);
\belle\ uses
(${\cal A}$, $C_{+}$,  $S_{+}$,  $C_{-}$,  $S_{-}$).
In this document, we follow the notation used by \babar.

\mysubsubsubsection{$\Bz \to \rho^{\pm}\pi^\mp$
}
\label{sec:cp_uta:notations:non_cp:rhopi}

In the $\rho^\pm\pi^\mp$ system, the 
($\Adirnoncp$, $C_{f\bar{f}}$, $S_{f\bar{f}}$, $\Delta C_{f\bar{f}}$, 
$\Delta S_{f\bar{f}}$)
set of parameters has been used 
originally by 
\babar~\cite{ref:cp_uta:uud:babar:rhopi_old} and
\belle~\cite{ref:cp_uta:uud:belle:rhopi_old}, in the Q2B approximation;
the exact names\footnote{
  \babar\ has used the notations
  $A_{\CP}^{\rho\pi}$~\cite{ref:cp_uta:uud:babar:rhopi_old} and 
  ${\cal A}_{\rho\pi}$~\cite{ref:cp_uta:uud:babar:pipipi0}
  in place of ${\cal A}_{\CP}^{\rho\pi}$.
}
used in this case are
$\left( 
  {\cal A}_{\CP}^{\rho\pi}, C_{\rho\pi}, S_{\rho\pi}, \Delta C_{\rho\pi}, \Delta S_{\rho\pi}
\right)$,
and these names are also used in this document.

Since $\rho^\pm\pi^\mp$ is reconstructed in the final state $\pi^+\pi^-\pi^0$,
the interference between the $\rho$ resonances
can provide additional information about the phases 
(see Sec.~\ref{sec:cp_uta:notations:dalitz}).
Both \babar~\cite{ref:cp_uta:uud:babar:pipipi0} 
and \belle~\cite{ref:cp_uta:uud:belle:pipipi0}
have performed time-dependent Dalitz plot analyses, 
from which the weak phase $\alpha$ is directly extracted.
In such an analysis, the measured Q2B parameters are 
also naturally corrected for interference effects.
See Sec.~\ref{sec:cp_uta:notations:dalitz:pipipi0}.

\mysubsubsubsection{$\Bz \to D^{\pm}\pi^{\mp}, D^{*\pm}\pi^{\mp}, D^{\pm}\rho^{\mp}$
}
\label{sec:cp_uta:notations:non_cp:dstarpi}

Time-dependent $\CP$ analyses have also been performed for the
final states $D^{\pm}\pi^{\mp}$, $D^{*\pm}\pi^{\mp}$ and $D^{\pm}\rho^{\mp}$.
In these theoretically clean cases, no penguin contributions are possible,
so there is no direct $\CP$ violation.
Furthermore, due to the smallness of the ratio of the magnitudes of the 
suppressed ($b \to u$) and favoured ($b \to c$) amplitudes (denoted $R_f$),
to a very good approximation, $C_f = - C_{\bar{f}} = 1$
(using $f = D^{(*)-}h^+$, $\bar{f} = D^{(*)+}h^-$ $h = \pi,\rho$),
and the coefficients of the sine terms are given by
\begin{equation}
  S_f = - 2 R_f \sin( \phi_{\rm mix} + \phi_{\rm dec} - \delta_f )
  \hspace{5mm}
  {\rm and}
  \hspace{5mm}
  S_{\bar{f}} = - 2 R_f \sin( \phi_{\rm mix} + \phi_{\rm dec} + \delta_f ).
\end{equation}
Thus weak phase information can be cleanly obtained from measurements
of $S_f$ and $S_{\bar{f}}$, 
although external information on at least one of $R_f$ or $\delta_f$ is necessary.
(Note that $\phi_{\rm mix} + \phi_{\rm dec} = 2\beta + \gamma$ for all the decay modes 
in question, while $R_f$ and $\delta_f$ depend on the decay mode.)

Again, different notations have been used in the literature.
\babar\xspace\cite{ref:cp_uta:cud:babar:full,ref:cp_uta:cud:babar:partial}
defines the time-dependent probability function by
\begin{equation}
  f^\pm (\eta, \Delta t) = \frac{e^{-|\Delta t|/\tau}}{4\tau} 
  \left[  
    1 \mp S_\zeta \sin (\Delta m \Delta t) \mp \eta C_\zeta \cos(\Delta m \Delta t) 
  \right],
\end{equation} 
where the upper (lower) sign corresponds to 
the tagging meson being a $\Bz$ ($\Bzb$). 
[Note here that a tagging $\Bz$ ($\Bzb$) corresponds to $-S_\xi$ ($+S_\xi$).]
The parameters $\eta$ and $\zeta$ take the values $+1$ and $+$ ($-1$ and $-$) 
when the final state is, \eg, $D^-\pi^+$ ($D^+\pi^-$). 
However, in the fit, the substitutions $C_\zeta = 1$ and 
$S_\zeta = a \mp \eta b_i - \eta c_i$ are made.\footnote{
  The subscript $i$ denotes tagging category.
}
[Note that, neglecting $b$ terms, $S_+ = a - c$ and $S_- = a + c$, 
so that $a = (S_+ + S_-)/2$, $c = (S_- - S_+)/2$, in analogy to 
the parameters of Eq.~(\ref{eq:cp_uta:non-cp-s_and_deltas}).] 
The subscript $i$ denotes the tagging category. 
These are motivated by the possibility of 
$\CP$ violation on the tag side~\cite{ref:cp_uta:cud:tagside}, 
which is absent for semileptonic $\B$ decays (mostly lepton tags). 
The parameter $a$ is not affected by tag side $\CP$ violation. 
The parameter $b$ only depends on tag side $\CP$ violation parameters 
and is not directly useful for determining UT angles.
A clean interpretation of the $c$ parameter is only possible for 
lepton-tagged events,
so the \babar\ measurements report $c$ measured with those events only.

The parameters used by \belle\ in the analysis using 
partially reconstructed $\B$ decays~\cite{ref:cp_uta:cud:belle:partial}, 
are similar to the $S_\zeta$ parameters defined above. 
However, in the \belle\ convention, 
a tagging $\Bz$ corresponds to a $+$ sign in front of the sine coefficient; 
furthermore the correspondence between the super/subscript 
and the final state is opposite, so that $S_\pm$ (\babar) = $- S^\mp$ (\belle). 
In this analysis, only lepton tags are used, 
so there is no effect from tag side $\CP$ violation. 
In the \belle\ analysis using 
fully reconstructed $\B$ decays~\cite{ref:cp_uta:cud:belle}, 
this effect is measured and taken into account using $\Dstar l \nu$ decays; 
in neither \belle\ analysis are the $a$, $b$ and $c$ parameters used. 
In the latter case, the measured parameters are 
$2 R_{D^{(*)}\pi} \sin( 2\phi_1 + \phi_3 \pm \delta_{D^{(*)}\pi} )$; 
the definition is such that 
$S^\pm$ (\belle) = $- 2 R_{\Dstar \pi} \sin( 2\phi_1 + \phi_3 \pm \delta_{\Dstar \pi} )$. 
However, the definition includes an 
angular momentum factor $(-1)^L$~\cite{ref:cp_uta:cud:fleischer}, 
and so for the results in the $D\pi$ system, 
there is an additional factor of $-1$ in the conversion.

Explicitly, the conversion then reads as given in 
Table~\ref{tab:cp_uta:notations:non_cp:dstarpi}, 
where we have neglected the $b_i$ terms used by \babar
(which are zero in the absence of tag side $\CP$ violation).
For the averages in this document,
we use the $a$ and $c$ parameters,
and give the explicit translations used in 
Table~\ref{tab:cp_uta:notations:non_cp:dstarpi2}.
It is to be fervently hoped that the experiments will
converge on a common notation in future.

\begin{table}
  \begin{center} 
    \caption{
      Conversion between the various notations used for 
      $\CP$ violation parameters in the 
      $D^{\pm}\pi^{\mp}$, $D^{*\pm}\pi^{\mp}$ and $D^{\pm}\rho^{\mp}$ systems.
      The $b_i$ terms used by \babar\ have been neglected.
      Recall that $\left( \alpha, \beta, \gamma \right) = \left( \phi_2, \phi_1, \phi_3 \right)$.
    }
    \vspace{0.2cm}
    \setlength{\tabcolsep}{0.0pc}
    \begin{tabular*}{\textwidth}{@{\extracolsep{\fill}}cccc} \hline 
      & \babar\ & \belle\ partial rec. & \belle\ full rec. \\
      \hline
      $S_{D^+\pi^-}$    & $- S_- = - (a + c_i)$ &  N/A  &
      $2 R_{D\pi} \sin( 2\phi_1 + \phi_3 + \delta_{D\pi} )$ \\
      $S_{D^-\pi^+}$    & $- S_+ = - (a - c_i)$ &  N/A  &
      $2 R_{D\pi} \sin( 2\phi_1 + \phi_3 - \delta_{D\pi} )$ \\
      $S_{D^{*+}\pi^-}$ & $- S_- = - (a + c_i)$ & $S^+$ &   
      $- 2 R_{\Dstar \pi} \sin( 2\phi_1 + \phi_3 + \delta_{\Dstar \pi} )$ \\
      $S_{D^{*-}\pi^+}$ & $- S_+ = - (a - c_i)$ & $S^-$ &
      $- 2 R_{\Dstar \pi} \sin( 2\phi_1 + \phi_3 - \delta_{\Dstar \pi} )$ \\
      $S_{D^+\rho^-}$    & $- S_- = - (a + c_i)$ &  N/A  &  N/A  \\
      $S_{D^-\rho^+}$    & $- S_+ = - (a - c_i)$ &  N/A  &  N/A  \\
      \hline 
    \end{tabular*}
    \label{tab:cp_uta:notations:non_cp:dstarpi}
  \end{center}
\end{table}
   
\begin{table}
  \begin{center} 
    \caption{
      Translations used to convert the parameters measured by \belle
      to the parameters used for averaging in this document.
      The angular momentum factor $L$ is $-1$ for $\Dstar\pi$ and $+1$ for $D\pi$.
      Recall that $\left( \alpha, \beta, \gamma \right) = \left( \phi_2, \phi_1, \phi_3 \right)$.
    }
    \vspace{0.2cm}
    \setlength{\tabcolsep}{0.0pc}
    \begin{tabular*}{\textwidth}{@{\extracolsep{\fill}}ccc} \hline 
        & $\Dstar\pi$ partial rec. & $D^{(*)}\pi$ full rec. \\
        \hline
        $a$ & $- (S^+ + S^-)$ &
        $\frac{1}{2} (-1)^{L+1}
        \left(
          2 R_{D^{(*)}\pi} \sin( 2\phi_1 + \phi_3 + \delta_{D^{(*)}\pi} ) + 
          2 R_{D^{(*)}\pi} \sin( 2\phi_1 + \phi_3 - \delta_{D^{(*)}\pi} )
        \right)$ \\
        $c$ & $- (S^+ - S^-)$ & 
        $\frac{1}{2} (-1)^{L+1}
        \left(
          2 R_{D^{(*)}\pi} \sin( 2\phi_1 + \phi_3 + \delta_{D^{(*)}\pi} ) -
          2 R_{D^{(*)}\pi} \sin( 2\phi_1 + \phi_3 - \delta_{D^{(*)}\pi} )
        \right)$ \\
        \hline 
      \end{tabular*}
    \label{tab:cp_uta:notations:non_cp:dstarpi2}
  \end{center}
\end{table}

\mysubsubsubsection{Time-dependent asymmetries in radiative $\B$ decays
}
\label{sec:cp_uta:notations:non_cp:radiative}

As a special case of decays to non-$\CP$ eigenstates,
let us consider radiative $\B$ decays.
Here, the emitted photon has a distinct helicity,
which is in principle observable, but in practice is not usually measured.
Thus the measured time-dependent decay rates 
are given by~\cite{ref:cp_uta:bsg:ags,ref:cp_uta:bsg:aghs}
\begin{eqnarray}
  \Gamma_{\Bzb \to X \gamma} (\Delta t) & = &
  \Gamma_{\Bzb \to X \gamma_L} (\Delta t) + \Gamma_{\Bzb \to X \gamma_R} (\Delta t) \\ \nonumber
  & = &
  \frac{e^{-\left| \Delta t \right| / \tau(\Bz)}}{4\tau(\Bz)} 
  \left\{ 
    1 + 
    \left( S_L + S_R \right) \sin(\Delta m \Delta t) - 
    \left( C_L + C_R \right) \cos(\Delta m \Delta t) 
  \right\},
  \\
  \Gamma_{\Bz \to X \gamma} (\Delta t) & = & 
  \Gamma_{\Bz \to X \gamma_L} (\Delta t) + \Gamma_{\Bz \to X \gamma_R} (\Delta t) \\ \nonumber 
  & = &
  \frac{e^{-\left| \Delta t \right| / \tau(\Bz)}}{4\tau(\Bz)} 
  \left\{ 
    1 - 
    \left( S_L + S_R \right) \sin(\Delta m \Delta t) + 
    \left( C_L + C_R \right) \cos(\Delta m \Delta t) 
  \right\},
\end{eqnarray}
where in place of the subscripts $f$ and $\bar{f}$ we have used $L$ and $R$
to indicate the photon helicity.
In order for interference between decays with and without $\Bz$-$\Bzb$ mixing
to occur, the $X$ system must not be flavour-specific,
\eg, in case of $\Bz \to K^{*0}\gamma$, the final state must be $\KS \pi^0 \gamma$.
The sign of the sine term depends on the $C$ eigenvalue of the $X$ system.
At leading order, the photons from 
$b \to q \gamma$ ($\bar{b} \to \bar{q} \gamma$) are predominantly
left (right) polarized, with corrections of order of $m_q/m_b$,
thus interference effects are suppressed.
Higher order effects can lead to corrections of order 
$\Lambda_{\rm QCD}/m_b$~\cite{ref:cp_uta:bsg:gglp}.
The predicted smallness of the $S$ terms in the Standard Model
results in sensitivity to new physics contributions.

\mysubsubsection{Time-dependent \CP asymmetries in the $B_s$ System}
\label{sec:cp_uta:notations:Bs}

A complete analysis of the time-dependent decay rates of 
neutral $B$ mesons must also take into account the lifetime difference
between the eigenstates of the effective Hamiltonian, 
denoted by $\Delta \Gamma$.
This is particularly important in the $B_s$ system,
since non-negligible values of  $\Delta \Gamma_s$ are expected 
(see Section~\ref{sec:mixing} for the latest experimental constraints).
Neglecting $\CP$ violation in mixing,
the relevant replacements for 
Eqs.~\ref{eq:cp_uta:td_cp_asp1}~\&~\ref{eq:cp_uta:td_cp_asp2} 
are~\cite{ref:cp_uta:dfn2000}
\begin{equation}
  \label{eq:cp_uta:td_cp_bs_asp1}
  \begin{array}{lcr}
    \mc{2}{l}{
      \Gamma_{\Bsb \to f} (\Delta t) = 
      {\cal N} 
      \frac{e^{-| \Delta t | / \tau(\Bs)}}{4\tau(\Bs)}
      \Big[ 
      \cosh(\frac{\Delta \Gamma \Delta t}{2}) +
    } & \hspace{40mm} \\
    \hspace{40mm} &
    \mc{2}{r}{
      \frac{2\, \Im(\lambda_f)}{1 + |\lambda_f|^2} \sin(\Delta m \Delta t) -
      \frac{1 - |\lambda_f|^2}{1 + |\lambda_f|^2} \cos(\Delta m \Delta t) -
      \frac{2\, \Re(\lambda_f)}{1 + |\lambda_f|^2} \sinh(\frac{\Delta \Gamma \Delta t}{2})
      \Big],
    } \\
  \end{array}
\end{equation}
and
\begin{equation}
  \label{eq:cp_uta:td_cp_bs_asp2}
  \begin{array}{lcr}
    \mc{2}{l}{
      \Gamma_{\Bs \to f} (\Delta t) =
      {\cal N} 
      \frac{e^{-| \Delta t | / \tau(\Bs)}}{4\tau(\Bs)}
      \Big[ 
      \cosh(\frac{\Delta \Gamma \Delta t}{2}) -
    } & \hspace{40mm} \\
    \hspace{40mm} & 
    \mc{2}{r}{
      \frac{2\, \Im(\lambda_f)}{1 + |\lambda_f|^2} \sin(\Delta m \Delta t) +
      \frac{1 - |\lambda_f|^2}{1 + |\lambda_f|^2} \cos(\Delta m \Delta t) -
      \frac{2\, \Re(\lambda_f)}{1 + |\lambda_f|^2} \sinh(\frac{\Delta \Gamma \Delta t}{2})
      \Big]. 
    } \\
  \end{array}
\end{equation}

The {\it untagged} time-dependent decay rate is given by
\begin{equation}
  \Gamma_{\Bsb \to f} (\Delta t) + \Gamma_{\Bs \to f} (\Delta t)
  = 
  {\cal N} 
  \frac{e^{-| \Delta t | / \tau(\Bs)}}{2\tau(\Bs)}
  \Big[ 
  \cosh(\frac{\Delta \Gamma \Delta t}{2}) -
  \frac{2\, \Re(\lambda_f)}{1 + |\lambda_f|^2} \sinh(\frac{\Delta \Gamma \Delta t}{2})
  \Big] \, .
\end{equation}
With the requirement
$\int_{-\infty}^{+\infty} \Gamma_{\Bsb \to f} (\Delta t) + \Gamma_{\Bs \to f} (\Delta t) d(\Delta t) = 1$,
the normalization factor ${\cal N}$ 
is fixed to $1 - (\frac{\Delta \Gamma}{2\Gamma})^2$.
Note that an untagged time-dependent analysis can probe
$\lambda_f$, through $\Re(\lambda_f)$, when $\Delta \Gamma \neq 0$.
The tagged analysis is, of course, more sensitive.

To be consistent with our earlier notation,\footnote{
  As ever, alternative and conflicting notations appear in the literature.
  One popular alternative notation for this parameter is 
  ${\cal A}_{\Delta \Gamma}$.
  Particular care must be taken over the signs.
}
we write here the coefficient of the $\sinh$ term as
\begin{equation}
  A^{\Delta \Gamma}_f = - \frac{2\, \Re(\lambda_f)}{1 + |\lambda_f|^2} \, .
\end{equation}
Note that 
\begin{equation}
  \left( S_f \right)^2 + \left( C_f \right)^2 + \left( A^{\Delta \Gamma}_f \right)^2 = 1 \, .
\end{equation}

Other expressions can be similarly modified to take into account 
non-zero lifetime differences.
Note that when the final state contains 
a mixture of $\CP$ even and $\CP$ odd states
(as, for example, for vector-vector or multibody self-conjugate states),
that $\Re(\lambda_f)$ contains terms proportional to 
both the sine and cosine of the weak phase difference, 
albeit with rather different sensitivities.

\mysubsubsection{Asymmetries in $\B \to \DorDstar K^{(*)}$ decays
}
\label{sec:cp_uta:notations:cus}

$\CP$ asymmetries in $\B \to \DorDstar K^{(*)}$ decays are sensitive to $\gamma$.
The neutral $D^{(*)}$ meson produced 
is an admixture of $\DorDstarz$ (produced by a $b \to c$ transition) and 
$\DorDstarzb$ (produced by a colour-suppressed $b \to u$ transition) states.
If the final state is chosen so that both $\DorDstarz$ and $\DorDstarzb$ 
can contribute, the two amplitudes interfere,
and the resulting observables are sensitive to $\gamma$, 
the relative weak phase between 
the two $\B$ decay amplitudes~\cite{ref:cp_uta:cus:bs}.
Various methods have been proposed to exploit this interference,
including those where the neutral $D$ meson is reconstructed 
as a $\CP$ eigenstate (GLW)~\cite{ref:cp_uta:cus:glw},
in a suppressed final state (ADS)~\cite{ref:cp_uta:cus:ads},
or in a self-conjugate three-body final state, 
such as $\KS \pi^+\pi^-$ (Dalitz)~\cite{ref:cp_uta:cus:dalitz}.
It should be emphasised that while each method 
differs in the choice of $D$ decay,
they are all sensitive to the same parameters of the $B$ decay,
and can be considered as variations of the same technique.

Consider the case of $\Bmp \to D \Kmp$,
with $D$ decaying to a final state $f$,
which is accessible to both $\Dz$ and $\Dzb$.
We can write the decay rates for $\Bm$ and $\Bp$ ($\Gamma_\mp$), 
the charge averaged rate ($\Gamma = (\Gamma_- + \Gamma_+)/2$)
and the charge asymmetry 
(${\cal A} = (\Gamma_- - \Gamma_+)/(\Gamma_- + \Gamma_+)$, see Eq.~(\ref{eq:cp_uta:pra})) as 
\begin{eqnarray}
  \label{eq:cp_uta:dk:rate_def}
  \Gamma_\mp  & \propto & 
  r_B^2 + r_D^2 + 2 r_B r_D \cos \left( \delta_B + \delta_D \mp \gamma \right), \\
  \label{eq:cp_uta:dk:av_rate_def}
  \Gamma & \propto &  
  r_B^2 + r_D^2 + 2 r_B r_D \cos \left( \delta_B + \delta_D \right) \cos \left( \gamma \right), \\
  \label{eq:cp_uta:dk:acp_def}
  {\cal A} & = & 
  \frac{
    2 r_B r_D \sin \left( \delta_B + \delta_D \right) \sin \left( \gamma \right)
  }{
    r_B^2 + r_D^2 + 2 r_B r_D \cos \left( \delta_B + \delta_D \right) \cos \left( \gamma \right),  
  }
\end{eqnarray}
where the ratio of $\B$ decay amplitudes\footnote{
  Note that here we use the notation $r_B$ to denote the ratio
  of $\B$ decay amplitudes, 
  whereas in Sec.~\ref{sec:cp_uta:notations:non_cp:dstarpi} 
  we used, \eg, $R_{D\pi}$, for a rather similar quantity.
  The reason is that here we need to be concerned also with 
  $D$ decay amplitudes,
  and so it is convenient to use the subscript to denote the decaying particle.
  Hopefully, using $r$ in place of $R$ will help reduce potential confusion.
} 
is usually defined to be less than one,
\begin{equation}
  \label{eq:cp_uta:dk:rb_def}
  r_B = 
  \frac{
    \left| A\left( \Bm \to \Dzb K^- \right) \right|
  }{
    \left| A\left( \Bm \to \Dz  K^- \right) \right|
  },
\end{equation}
and the ratio of $D$ decay amplitudes is correspondingly defined by
\begin{equation}
  \label{eq:cp_uta:dk:rd_def}
  r_D = 
  \frac{
    \left| A\left( \Dz  \to f \right) \right|
  }{
    \left| A\left( \Dzb \to f \right) \right|
  }.
\end{equation}
The strong phase differences between the $\B$ and $D$ decay amplitudes 
are given by $\delta_B$ and $\delta_D$, respectively.
The values of $r_D$ and $\delta_D$ depend on the final state $f$:
for the GLW analysis, $r_D = 1$ and $\delta_D$ is trivial (either zero or $\pi$),
in the Dalitz plot analysis $r_D$ and $\delta_D$ vary across the Dalitz plot,
and depend on the $D$ decay model used,
for the ADS analysis, the values of $r_D$ and $\delta_D$ are not trivial.

Note that, for given values of $r_B$ and $r_D$, 
the maximum size of ${\cal A}$ (at $\sin \left( \delta_B + \delta_D \right) = 1$)
is $2 r_B r_D \sin \left( \gamma \right) / \left( r_B^2 + r_D^2 \right)$.
Thus even for $D$ decay modes with small $r_D$, 
large asymmetries, and hence sensitivity to $\gamma$, 
may occur for $B$ decay modes with similar values of $r_B$.
For this reason, the ADS analysis of the decay $B^\mp \to D \pi^\mp$ 
is also of interest.

In the GLW analysis, the measured quantities are the 
partial rate asymmetry, and the charge averaged rate,
which are measured both for $\CP$ even and $\CP$ odd $D$ decays.
For the latter, it is experimentally convenient to measure a double ratio,
\begin{equation}
  \label{eq:cp_uta:dk:double_ratio}
  R_{\CP} = 
  \frac{
    \Gamma\left( \Bm \to D_{\CP} \Km  \right) \, / \, \Gamma\left( \Bm \to \Dz \Km \right)
  }{
    \Gamma\left( \Bm \to D_{\CP} \pim \right) \, / \, \Gamma\left( \Bm \to \Dz \pim \right)
  }
\end{equation}
that is normalized both to the rate for the favoured $\Dz \to \Km\pip$ decay, 
and to the equivalent quantities for $\Bm \to D\pim$ decays
(charge conjugate modes are implicitly 
included in Eq.~(\ref{eq:cp_uta:dk:double_ratio})).
In this way the constant of proportionality drops out of 
Eq.~(\ref{eq:cp_uta:dk:av_rate_def}).

For the ADS analysis, using a suppressed $D \to f$ decay,
the measured quantities are again the partial rate asymmetry, 
and the charge averaged rate.
In this case it is sufficient to measure the rate in a single ratio
(normalized to the favoured $D \to \bar{f}$ decay)
since detection systematics cancel naturally;
the observed quantity is then
\begin{equation}
  \label{eq:cp_uta:dk:r_ads}
  R_{\rm ADS} = 
  \frac{
    \Gamma \left( \Bm \to \left[ f \right]_D \Km \right)
  }{
    \Gamma \left( \Bm \to \left[ \bar{f} \right]_D \Km \right)
  }.
\end{equation}
In the ADS analysis, there are an additional two unknowns ($r_D$ and $\delta_D$)
compared to the GLW case.  
However, the value of $r_D$ can be measured using 
decays of $D$ mesons of known flavour.

In the Dalitz plot analysis,
once a model is assumed for the $D$ decay, 
which gives the values of $r_D$ and $\delta_D$ across the Dalitz plot,
it is possible to perform a simultaneous fit to the $B^+$ and $B^-$ samples 
and directly extract $\gamma$, $r_B$ and $\delta_B$.
However, the uncertainties on the phases depend inversely on $r_B$.
Furthermore, $r_B$ is positive definite (and small), 
and therefore tends to be overestimated,
which can lead to an underestimation of the uncertainty.
Some statistical treatment is necessary to correct for this bias.
An alternative approach is to extract from the data the ``Cartesian''
variables
\begin{equation}
  \left( x_\pm, y_\pm \right) = 
  \left( \Re(r_B e^{i(\delta_B\pm\gamma)}), \Im(r_B e^{i(\delta_B\pm\gamma)}) \right) = 
  \left( r_B \cos(\delta_B\pm\gamma), r_B \sin(\delta_B\pm\gamma) \right).
\end{equation}
These are (a) approximately statistically uncorrelated 
and (b) almost Gaussian.
Use of these variables makes the combination of results much simpler.

The relations between the measured quantities and the
underlying parameters are summarized in Table~\ref{tab:cp_uta:notations:dk}.
Note carefully that the hadronic factors $r_B$ and $\delta_B$ 
are different, in general, for each $\B$ decay mode.

\begin{table}
  \begin{center} 
    \caption{
      Summary of relations between measured and physical parameters 
      in GLW, ADS and Dalitz analyses of $\B \to \DorDstar K^{(*)}$.
    }
    \vspace{0.2cm}
    \setlength{\tabcolsep}{1.0pc}
    \begin{tabular}{cc} \hline 
      \mc{2}{c}{GLW analysis} \\
      $R_{\CP\pm}$ & $1 + r_B^2 \pm 2 r_B \cos \left( \delta_B \right) \cos \left( \gamma \right)$ \\
      $A_{\CP\pm}$ & $\pm 2 r_B \sin \left( \delta_B \right) \sin \left( \gamma \right) / R_{\CP\pm}$ \\
      \hline
      \mc{2}{c}{ADS analysis} \\
      $R_{\rm ADS}$ & $r_B^2 + r_D^2 + 2 r_B r_D \cos \left( \delta_B + \delta_D \right) \cos \left( \gamma \right)$ \\
      $A_{\rm ADS}$ & $2 r_B r_D \sin \left( \delta_B + \delta_D \right) \sin \left( \gamma \right) / R_{\rm ADS}$ \\
      \hline
      \mc{2}{c}{Dalitz analysis} \\
      $x_\pm$ & $r_B \cos(\delta_B\pm\gamma)$ \\
      $y_\pm$ & $r_B \sin(\delta_B\pm\gamma)$ \\
      \hline
    \end{tabular}
    \label{tab:cp_uta:notations:dk}
  \end{center}
\end{table}

\mysubsection{Common inputs and error treatment
}
\label{sec:cp_uta:common_inputs}

The common inputs used for rescaling are listed in 
Table~\ref{tab:cp_uta:common_inputs}.
The $\Bz$ lifetime ($\tau(\Bz)$) and mixing parameter ($\Delta m_d$)
averages are provided by the HFAG Lifetimes and Oscillations 
subgroup (Sec.~\ref{sec:life_mix}).
The fraction of the perpendicularly polarized component 
($\left| A_{\perp} \right|^2$) in $\B \to \jpsi \Kstar(892)$ decays,
which determines the $\CP$ composition, 
is averaged from results by 
\babar~\cite{ref:cp_uta:ccs:babar:psi_kstar} and
\belle~\cite{ref:cp_uta:ccs:belle:psi_kstar}.
See also HFAG $B$ to Charm Decay Parameters subgroup 
(Sec.~\ref{sec:BtoCharm}).

At present, we only rescale to a common set of input parameters
for modes with reasonably small statistical errors
($b \to c\bar{c}s$ transitions).
Correlated systematic errors are taken into account
in these modes as well.
For all other modes, the effect of such a procedure is 
currently negligible.

\begin{table}
  \begin{center}
    \caption{
      Common inputs used in calculating the averages.
    }
    \vspace{0.2cm}
    \setlength{\tabcolsep}{1.0pc}
    \begin{tabular}{cc} \hline 
      $\tau(\Bz)$ $({\rm ps})$  & $1.527 \pm 0.008$  \\
      $\Delta m_d$ $({\rm ps}^{-1})$ & $0.508 \pm 0.004$ \\
      $\left| A_{\perp} \right|^2 (\jpsi \Kstar)$ & $0.219 \pm 0.009$ \\
      \hline
    \end{tabular}
    \label{tab:cp_uta:common_inputs}
  \end{center}
\end{table}

As explained in Sec.~\ref{sec:intro},
we do not apply a rescaling factor on the error of an average
that has $\chi^2/\dof > 1$ 
(unlike the procedure currently used by the PDG~\cite{Yao2006PDG}).
We provide a confidence level of the fit so that
one can know the consistency of the measurements included in the average,
and attach comments in case some care needs to be taken in the interpretation.
Note that, in general, results obtained from data samples with low statistics
will exhibit some non-Gaussian behaviour.
We average measurements with asymmetric errors 
using the PDG~\cite{Yao2006PDG} prescription.
In cases where several measurements are correlated
(\eg\ $S_f$ and $C_f$ in measurements of time-dependent $\CP$ violation
in $B$ decays to a particular $\CP$ eigenstate)
we take these into account in the averaging procedure
if the uncertainties are sufficiently Gaussian.
For measurements where one error is given, 
it represents the total error, 
where statistical and systematic uncertainties have been added in quadrature.
If two errors are given, the first is statistical and the second systematic.
If more than two errors are given,
the origin of the additional uncertainty will be explained in the text.


\mysubsection{Time-dependent asymmetries in $b \to c\bar{c}s$ transitions
}
\label{sec:cp_uta:ccs}

\mysubsubsection{Time-dependent $\CP$ asymmetries in $b \to c\bar{c}s$ decays to $\CP$ eigenstates
}
\label{sec:cp_uta:ccs:cp_eigen}

\begin{figure}[htb]
  \begin{center}
    \resizebox{0.55\textwidth}{!}{
      \includegraphics{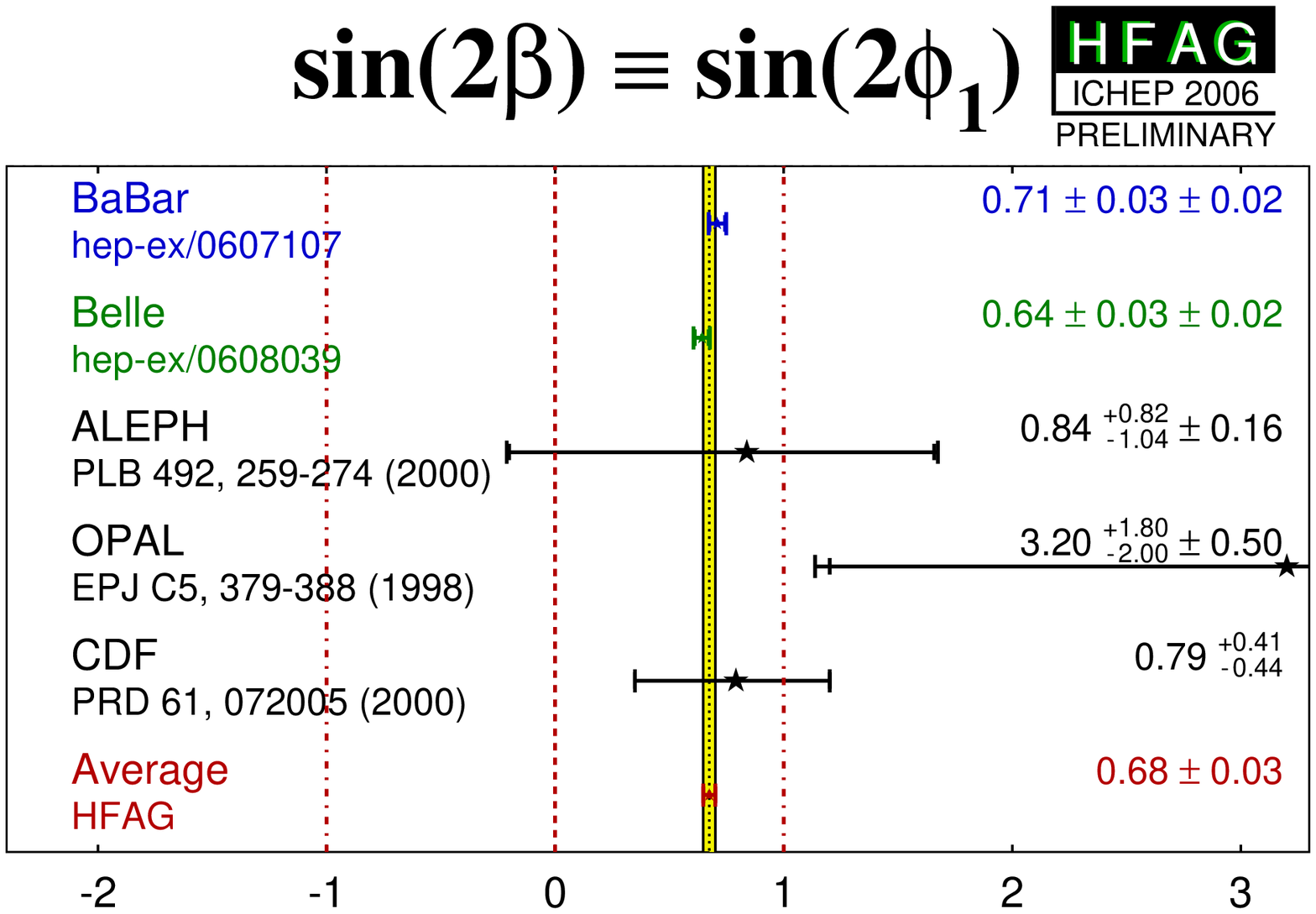}
    }
    \hfill
    \resizebox{0.44\textwidth}{!}{
      \includegraphics{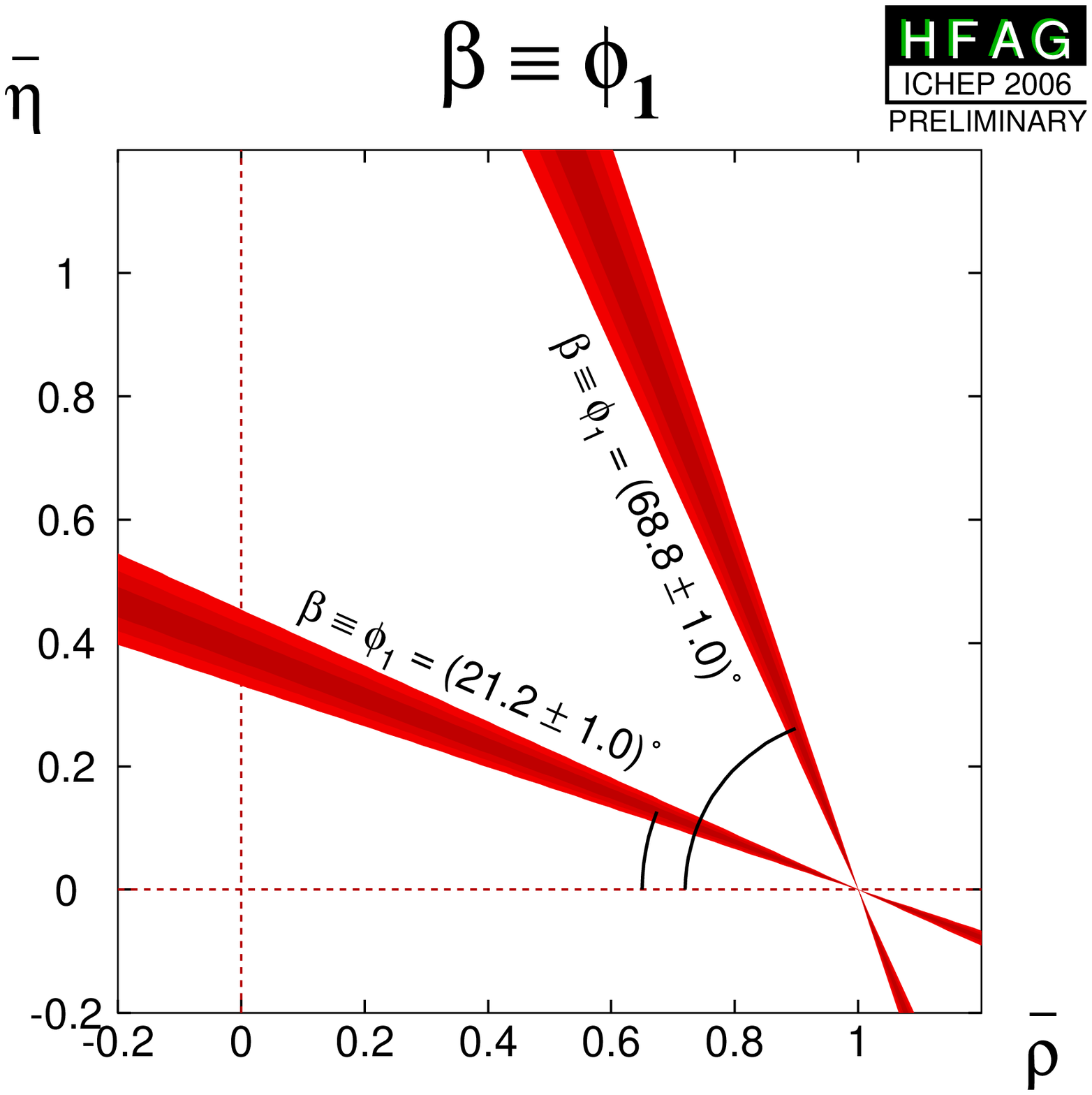}
    }
  \end{center}
  \vspace{-0.5cm}
  \caption{
    (Left) Average of measurements of $S_{b \to c\bar c s}$.
    (Right) Constraints on the $(\rhobar,\etabar)$ plane,
    obtained from the average of $-\etacp S_{b \to c\bar c s}$ 
    and Eq.~\ref{eq:cp_uta:sin2beta}.
  }
  \label{fig:cp_uta:ccs}
\end{figure}

In the Standard Model, the time-dependent parameters for
$b \to c\bar c s$ transitions are predicted to be: 
$S_{b \to c\bar c s} = - \etacp \sin(2\beta)$,
$C_{b \to c\bar c s} = 0$ to very good accuracy.
The averages for $-\etacp S_{b \to c\bar c s}$ and $C_{b \to c\bar c s}$
are provided in Table~\ref{tab:cp_uta:ccs}.
The averages for $-\etacp S_{b \to c\bar c s}$ 
are shown in Fig.~\ref{fig:cp_uta:ccs}.

Both \babar\  and \belle\ have used the $\etacp = -1$ modes
$\jpsi \KS$, $\psi(2S) \KS$, $\chi_{c1} \KS$ and $\eta_c \KS$, 
as well as $\jpsi \KL$, which has $\etacp = +1$
and $\jpsi K^{*0}(892)$, which is found to have $\etacp$ close to $+1$
based on the measurement of $\left| A_\perp \right|$ 
(see Sec.~\ref{sec:cp_uta:common_inputs}).
ALEPH, OPAL and CDF use only the $\jpsi \KS$ final state.
In the latest result from \belle, 
only $\jpsi \KS$ and $\jpsi \KL$ are used.
In future updates, it is hoped to perform separate averages for 
each charmonium-kaon final state.

\input{tables/cp_uta/btoccs.tex}

It should be noted that, while the uncertainty in the average for 
$-\etacp S_{b \to c\bar c s}$ is still limited by statistics,
that for $C_{b \to c\bar c s}$ is close to being dominated by systematics.
This occurs due to the possible effect of tag side interference on the
$C_{b \to c\bar c s}$ measurement, an effect which is correlated between
the different experiments.
Understanding of this effect may continue to improve in future,
allowing the uncertainty to reduce.

From the average for $-\etacp S_{b \to c\bar c s}$ above, 
we obtain the following solutions for $\beta$
(in $\left[ 0, \pi \right]$):
\begin{equation}
  \beta = \left( 21.2 \pm 1.0 \right)^\circ
  \hspace{5mm}
  {\rm or}
  \hspace{5mm}
  \beta = \left( 68.8 \pm 1.0 \right)^\circ
  \label{eq:cp_uta:sin2beta}
\end{equation}
In radians, these values are 
$\beta = \left( 0.37 \pm 0.02 \right)$, $\beta = \left( 1.20 \pm 0.02 \right)$.

This result gives a precise constraint on the $(\rhobar,\etabar)$ plane,
as shown in Fig.~\ref{fig:cp_uta:ccs}.
The measurement is in remarkable agreement with other constraints from 
$\CP$ conserving quantities, 
and with $\CP$ violation in the kaon system, in the form of the parameter $\epsilon_K$.
Such comparisons have been performed by various phenomenological groups,
such as CKMfitter~\cite{ref:cp_uta:ckmfitter} 
and UTFit~\cite{ref:cp_uta:utfit}.

\mysubsubsection{Time-dependent transversity analysis of $\Bz \to J/\psi K^{*0}$
}
\label{sec:cp_uta:ccs:vv}

$\B$ meson decays to the vector-vector final state $J/\psi K^{*0}$
are also mediated by the $b \to c \bar c s$ transition.
When a final state which is not flavour-specific ($K^{*0} \to \KS \pi^0$) is used,
a time-dependent transversity analysis can be performed 
allowing sensitivity to both 
$\sin(2\beta)$ and $\cos(2\beta)$~\cite{ref:cp_uta:ccs:dqstl}.
Such analyses have been performed by both $\B$ factory experiments.
In principle, the strong phases between the transversity amplitudes
are not uniquely determined by such an analysis, 
leading to a discrete ambiguity in the sign of $\cos(2\beta)$.
The \babar\ collaboration resolves 
this ambiguity using the known variation~\cite{ref:cp_uta:ccs:lass}
of the P-wave phase (fast) relative to the S-wave phase (slow) 
with the invariant mass of the $K\pi$ system 
in the vicinity of the $K^*(892)$ resonance. 
The result is in agreement with the prediction from 
$s$ quark helicity conservation,
and corresponds to Solution II defined by Suzuki~\cite{ref:cp_uta:ccs:suzuki}.
We use this phase convention for the averages given in 
Table~\ref{tab:cp_uta:ccs:psi_kstar}.

\input{tables/cp_uta/jpsiKstar.tex}

At present the results are dominated by 
large and non-Gaussian statistical errors,
and exhibit significant correlations.
We perform uncorrelated averages, 
the interpretation of which has to be done with the greatest care. 
Nonetheless, it is clear that $\cos(2\beta)>0$ is preferred 
by the experimental data in $J/\psi \Kstar$.
[\babar~\cite{ref:cp_uta:ccs:babar:psi_kstar2} 
find a confidence level for $\cos(2\beta)>0$ of $89\%$.] 

\mysubsubsection{Time-dependent $\CP$ asymmetries in $\Bz \to \Dstarp \Dstarm \KS$ decays
}
\label{sec:cp_uta:ccs:DstarDstarKs}

\babar~\cite{ref:cp_uta:ccs:ddks:babar} have performed a 
time-dependent analysis of the $\Bz \to \Dstarp \Dstarm \KS$ decay,
to obtain information on the sign of $\cos(2\beta)$.
More information can be found in 
Sec.~\ref{sec:cp_uta:notations:dalitz:dstardstarks}.
The results are shown in Table~\ref{tab:cp_uta:ccs:dstardstarks}.

\input{tables/cp_uta/dstardstarks.tex}

From the above result and the assumption that $J_{s2}>0$, 
\babar\ infer that $\cos(2\beta)>0$ at the $94\%$ confidence level.

\mysubsubsection{Time-dependent analysis of $\Bs \to J/\psi \phi$
}
\label{sec:cp_uta:ccs:jpsiphi}

As described in Sec.~\ref{sec:cp_uta:notations:Bs},
an untagged time-dependent analysis of $\Bs \to J/\psi \phi$
can probe the $\CP$ violating phase of $\Bs$--$\Bsb$ oscillations, $\phi_s$.
Within the Standard Model, this parameter is predicted to be small.

\dzero~\cite{D01_DGs} have performed such an analysis.
They simultaneously measure the average $\Bs$ lifetime $\tau(\Bs)$, 
$\Delta \Gamma$, $\phi_s$, 
the magnitude of the perpendicularly polarized component $A_\perp$, 
the difference in the fractions of the two $CP$-even components $|A_0|^2 - |A|_\parallel^2$, 
and the strong phases associated with the two $CP$-even components 
$\delta_0$ and $\delta_\parallel$.
The results are given in Table~\ref{tab:cp_uta:ccs:psiphi} below.
See also Section~\ref{sec:qps}.

\input{tables/cp_uta/psiphi.tex}

Note the implicit convention above is that 
$|A_\perp|^2 + |A_0|^2 + |A_\parallel|^2 = 1$, 
and the strong phases are measured relative to that of the $A_\perp$ component 
(which is set to zero). 
The polarization components are defined at time $t=0$, 
\ie\ at the production (primary) vertex of the $\Bs$. 
Note also that there is an ambiguity in the result for $\phi_s$.

\dzero~\cite{ref:cp_uta:ccs:psiphi:dzero2}
have combined the contour in the $(\phi_s,\Delta\Gamma)$ plane 
obtained above with a constraint obtained from the charge asymmetry in 
$\Bs$--$\Bsb$ oscillations
(see also HFAG Lifetimes and Oscillations, Sec.~\ref{sec:life_mix})
to obtain the result $\phi_s = -0.70 \, ^{+0.47}_{-0.39}$.

\mysubsection{Time-dependent $\CP$ asymmetries in colour-suppressed $b \to c\bar{u}d$ transitions
}
\label{sec:cp_uta:cud_beta}

Decays of $\B$ mesons to final states such as $D\pi^0$ are 
governed by $b \to c\bar{u}d$ transitions. 
If the final state is a $\CP$ eigenstate, \eg\ $D_{\CP}\pi^0$, 
the usual time-dependence formulae are recovered, 
with the sine coefficient sensitive to $\sin(2\beta)$. 
Since there is no penguin contribution to these decays, 
there is even less associated theoretical uncertainty 
than for $b \to c\bar{c}s$ decays like $\B \to \jpsi \KS$.
Such measurements therefore allow to test the Standard Model prediction
that the $\CP$ violation parameters in $b \to c\bar{u}d$ transitions
are the same as those in $b \to c\bar{c}s$~\cite{ref:cp_uta:cud_beta:gw}.

Note that there is an additional contribution from CKM suppressed
$b \to u \bar{c} d$ decays.
The effect of this contribution is small, and can be taken into 
account in the analysis~\cite{ref:cp_uta:cud_beta:fleischer}.

When multibody $D$ decays, such as $D \to \KS\pi^+\pi^-$ are used, 
a time-dependent analysis of the Dalitz plot of the neutral $D$ decay 
allows a direct determination of the weak phase: $2\beta$. 
(Equivalently, both $\sin(2\beta)$ and $\cos(2\beta)$ can be measured.)
This information allows to resolve the ambiguity in the 
measurement of $2\beta$ from $\sin(2\beta)$~\cite{ref:cp_uta:cud_beta:bgk}.

\begin{figure}[htb]
  \begin{center}
    \begin{tabular}{cc}
      \resizebox{0.46\textwidth}{!}{
        \includegraphics{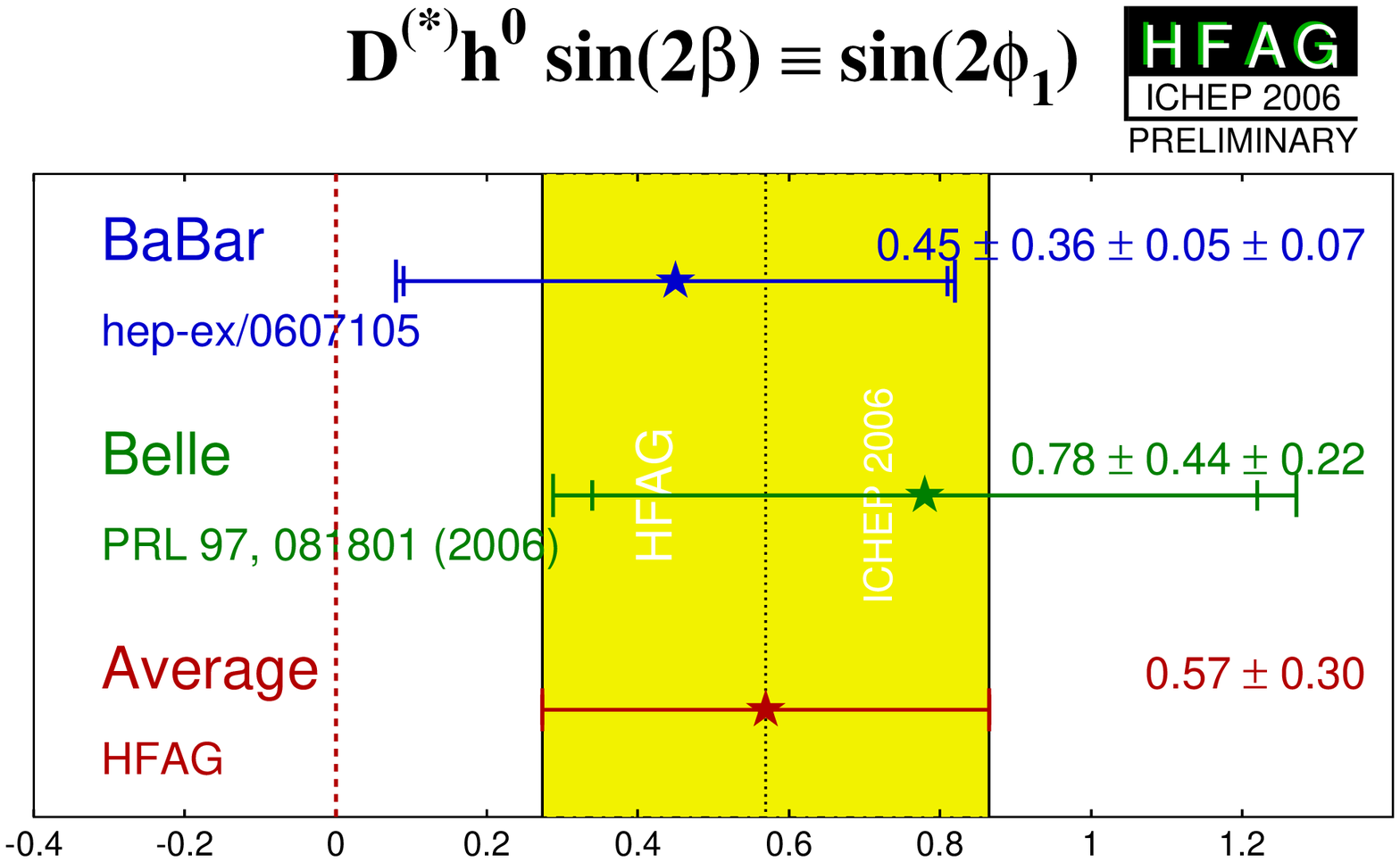}
      }
      &
      \resizebox{0.46\textwidth}{!}{
        \includegraphics{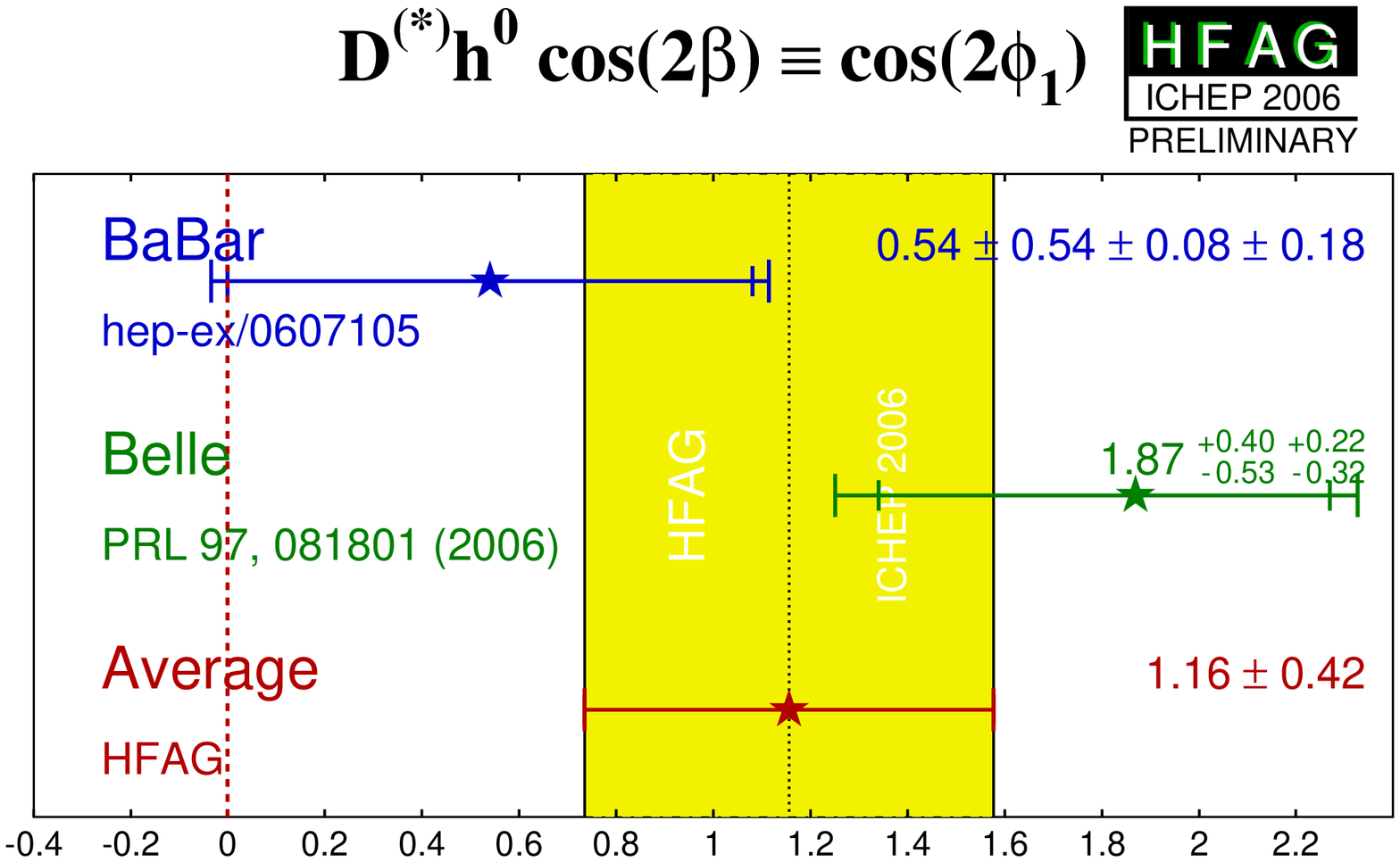}
      }
    \end{tabular}
  \end{center}
  \vspace{-0.8cm}
  \caption{
    Averages of 
    (left) $\sin(2\beta)$ and (right) $\cos(2\beta)$
    measured in colour-suppressed $b \to c\bar{u}d$ transitions.
  }
  \label{fig:cp_uta:cud_beta}
\end{figure}

Results of such analyses are available from both 
\belle~\cite{ref:cp_uta:cud_beta:belle} and
\babar~\cite{ref:cp_uta:cud_beta:babar}.
The decays $\B \to D\pi^0$, $\B \to D\eta$, $\B \to D\omega$, 
$\B \to D^*\pi^0$ and $\B \to D^*\eta$ are used. 
[This collection of states is denoted by $D^{(*)}h^0$.]
The daughter decays are $D^* \to D\pi^0$ and $D \to \KS\pi^+\pi^-$.
The results are shown in Table~\ref{tab:cp_uta:cud_beta},
and Fig.~\ref{fig:cp_uta:cud_beta}.
Note that \babar\ quote uncertainties due to the $D$ decay model 
separately from other systematic errors, while \belle\ do not.

\input{tables/cp_uta/dh0.tex}

Again, it is clear that the data prefer $\cos(2\beta)>0$.
Indeed, \belle~\cite{ref:cp_uta:cud_beta:belle} 
determine the sign of $\cos(2\phi_1)$ to be positive at $98.3\%$ confidence level,
while \babar~\cite{ref:cp_uta:cud_beta:babar} 
favour the solution of $\beta$ with $\cos(2\beta)>0$ at $87\%$ confidence level.
Note, however, that the Belle measurement has strongly non-Gaussian behaviour. 
Therefore, we perform uncorrelated averages, 
from which any interpretation has to be done with the greatest care. 

\mysubsection{Time-dependent $\CP$ asymmetries in charmless $b \to q\bar{q}s$ transitions
}
\label{sec:cp_uta:qqs}

The flavour changing neutral current $b \to s$ penguin
can be mediated by any up-type quark in the loop, 
and hence the amplitude can be written as
\begin{equation}
  \label{eq:cp_uta:b_to_s}
  \begin{array}{ccccc}
    A_{b \to s} & = & 
    \mc{3}{l}{F_u V_{ub}V^*_{us} + F_c V_{cb}V^*_{cs} + F_t V_{tb}V^*_{ts}} \\
    & = & (F_u - F_c) V_{ub}V^*_{us} & + & (F_t - F_c) V_{tb}V^*_{ts} \\
    & = & {\cal O}(\lambda^4) & + & {\cal O}(\lambda^2) \\
  \end{array}
\end{equation}
using the unitarity of the CKM matrix.
Therefore, in the Standard Model, 
this amplitude is dominated by $V_{tb}V^*_{ts}$, 
and to within a few degrees 
($\delta\beta \lesssim 2^\circ$ for $\beta \approx 20^\circ$) 
the time-dependent parameters can be written\footnote
{
  The presence of a small (${\cal O}(\lambda^2)$) weak phase in 
  the dominant amplitude of the $s$ penguin decays introduces 
  a phase shift given by
  $S_{b \to q\bar q s} = -\eta\sin(2\beta)\cdot(1 + \Delta)$. 
  Using the CKMfitter results for the Wolfenstein 
  parameters~\cite{ref:cp_uta:ckmfitter}, one finds: 
  $\Delta \simeq 0.033$, which corresponds to a shift of 
  $2\beta$ of $+2.1$ degrees. 
  Nonperturbative contributions can alter this result.
}
$S_{b \to q\bar q s} \approx - \etacp \sin(2\beta)$,
$C_{b \to q\bar q s} \approx 0$,
assuming $b \to s$ penguin contributions only ($q = u,d,s$).

Due to the large virtual mass scales occurring in the penguin loops,
additional diagrams from physics beyond the Standard Model,
with heavy particles in the loops, may contribute.
In general, these contributions will affect the values of 
$S_{b \to q\bar q s}$ and $C_{b \to q\bar q s}$.
A discrepancy between the values of 
$S_{b \to c\bar c s}$ and $S_{b \to q\bar q s}$
can therefore provide a clean indication of new physics.

However, there is an additional consideration to take into account.
The above argument assumes only the $b \to s$ penguin contributes
to the $b \to q\bar q s$ transition.
For $q = s$ this is a good assumption, 
which neglects only rescattering effects.
However, for $q = u$ there is a colour-suppressed $b \to u$ tree diagram
(of order ${\cal O}(\lambda^4)$), which has a different weak 
(and possibly strong) phase.
In the case $q = d$, any light neutral meson that is formed from
$d \bar{d}$ also has a $u \bar{u}$ component,
and so again there is ``tree pollution''. The \Bz decays to 
$\piz\KS$ and $\omega\KS$ belong to this category.
The mesons $f_0$ and $\etapr$ are expected to have predominant $s\bar s$ parts,
which reduces the possible tree pollution. If the inclusive 
decay $\Bz\to\Kp\Km\Kz$ (excluding $\phi\Kz$) is dominated by
a non-resonant three-body transition, an OZI-rule suppressed 
tree-level diagram can occur through insertion of an 
$s\sbar$ pair. The corresponding penguin-type transition 
proceeds via insertion of a $u\ubar$ pair, which is expected
to be favored over the $s\sbar$ insertion by fragmentation models.
Neglecting rescattering, the final state $\Kz\Kzb\Kz$ 
(reconstructed as $\KS\KS\KS$) has no tree pollution.
Various estimates, using different theoretical approaches,
of the values of $\Delta S = S_{b \to q\bar q s} - S_{b \to c\bar c s}$
exist in the literature~\cite{ref:cp_uta:qqs:theory}.
In general, there is agreement that the modes
$\phi\Kz$, $\etapr\Kz$ and $\Kz\Kzb\Kz$ are the cleanest,
with values of $\left| \Delta S \right|$ at or below the few percent level 
($\Delta S$ is usually positive).

\mysubsubsection{Time-dependent $\CP$ asymmetries in charmless $b \to q\bar{q}s$ decays to $\CP$ eigenstates
}
\label{sec:cp_uta:qqs:cp_eigen}

\begin{figure}[htb]
  \begin{center}
    \resizebox{0.45\textwidth}{!}{
      \includegraphics{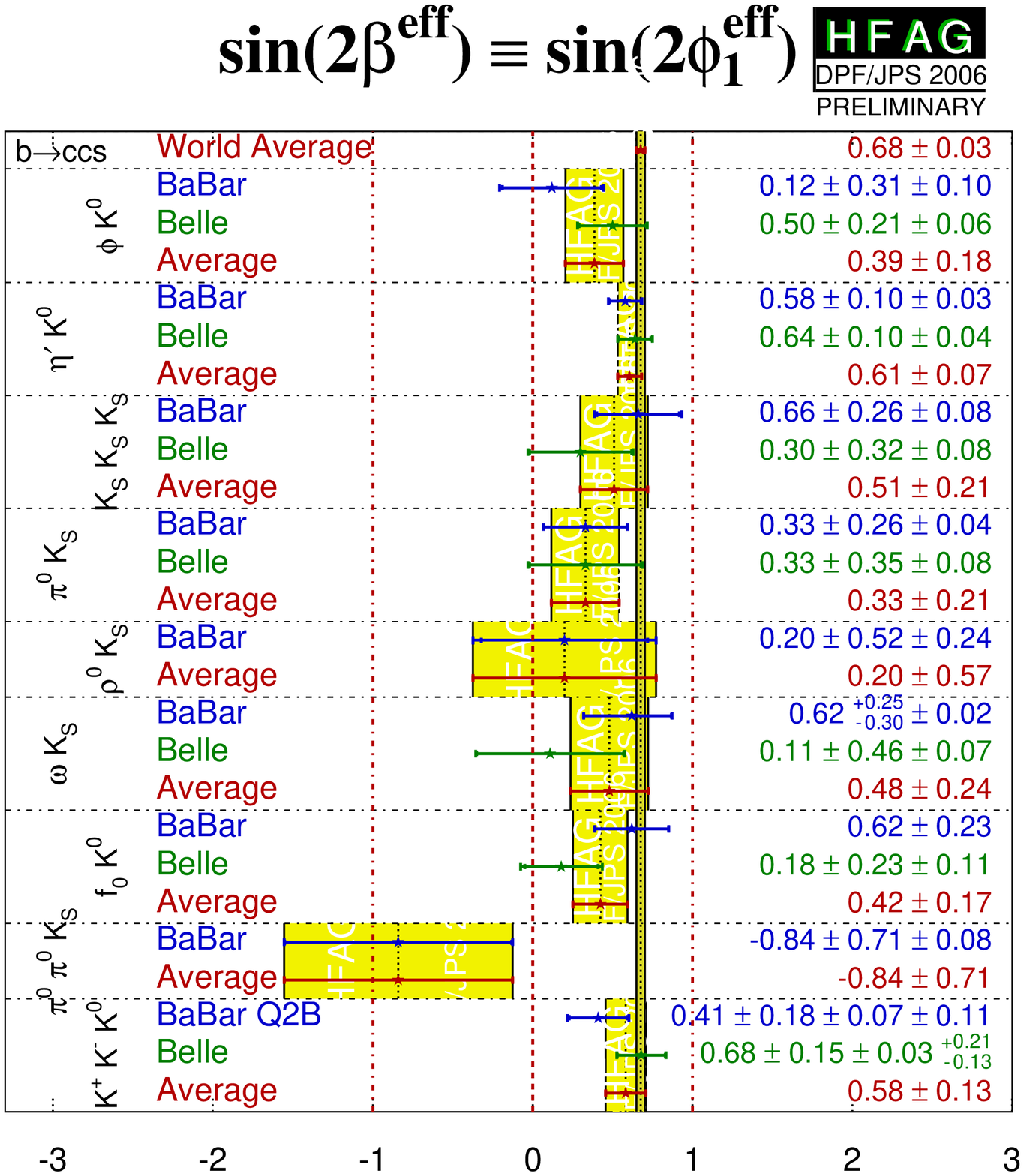}
    }
    \hfill
    \resizebox{0.45\textwidth}{!}{
      \includegraphics{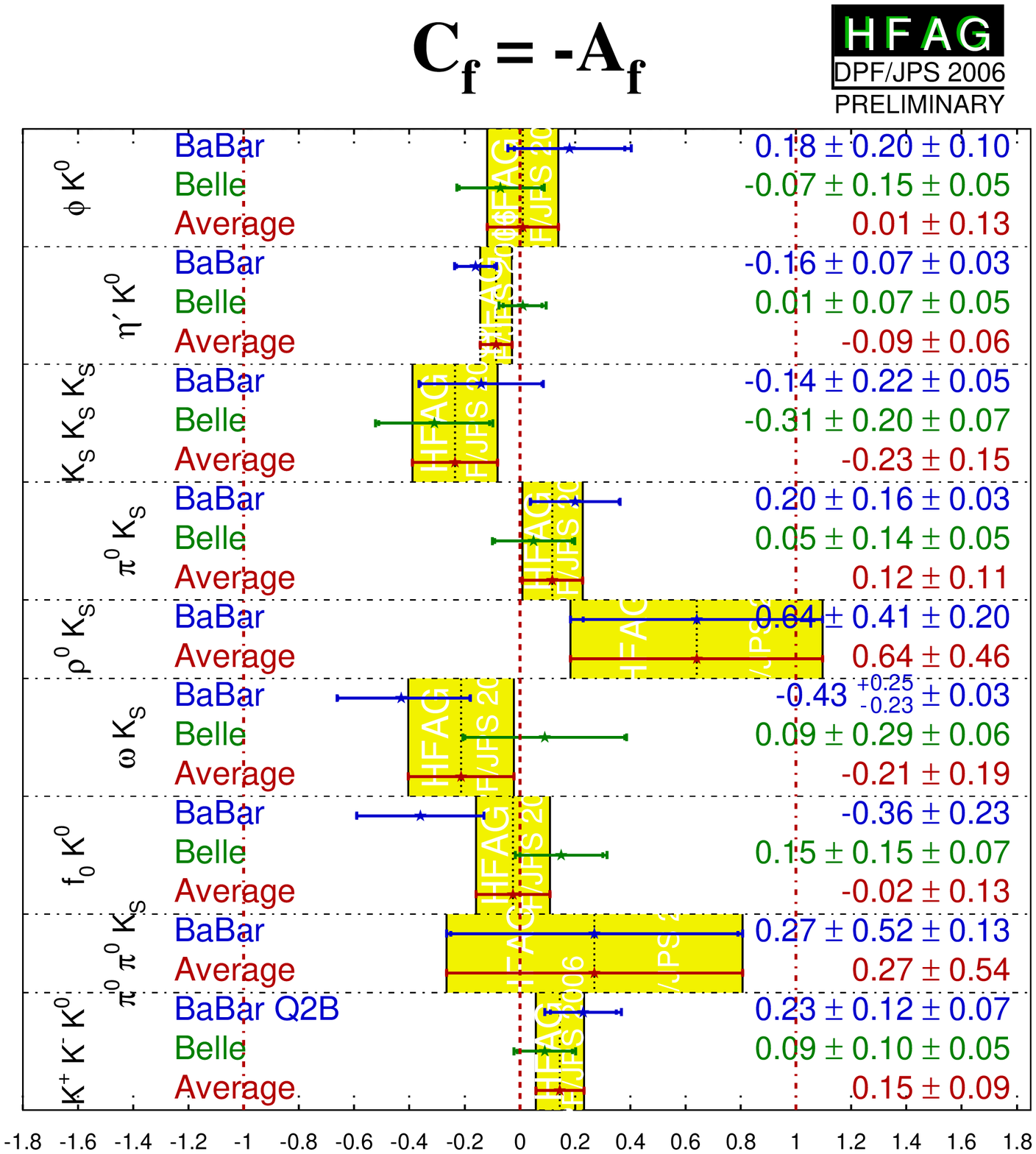}
    }
    \\
    \resizebox{0.45\textwidth}{!}{
      \includegraphics{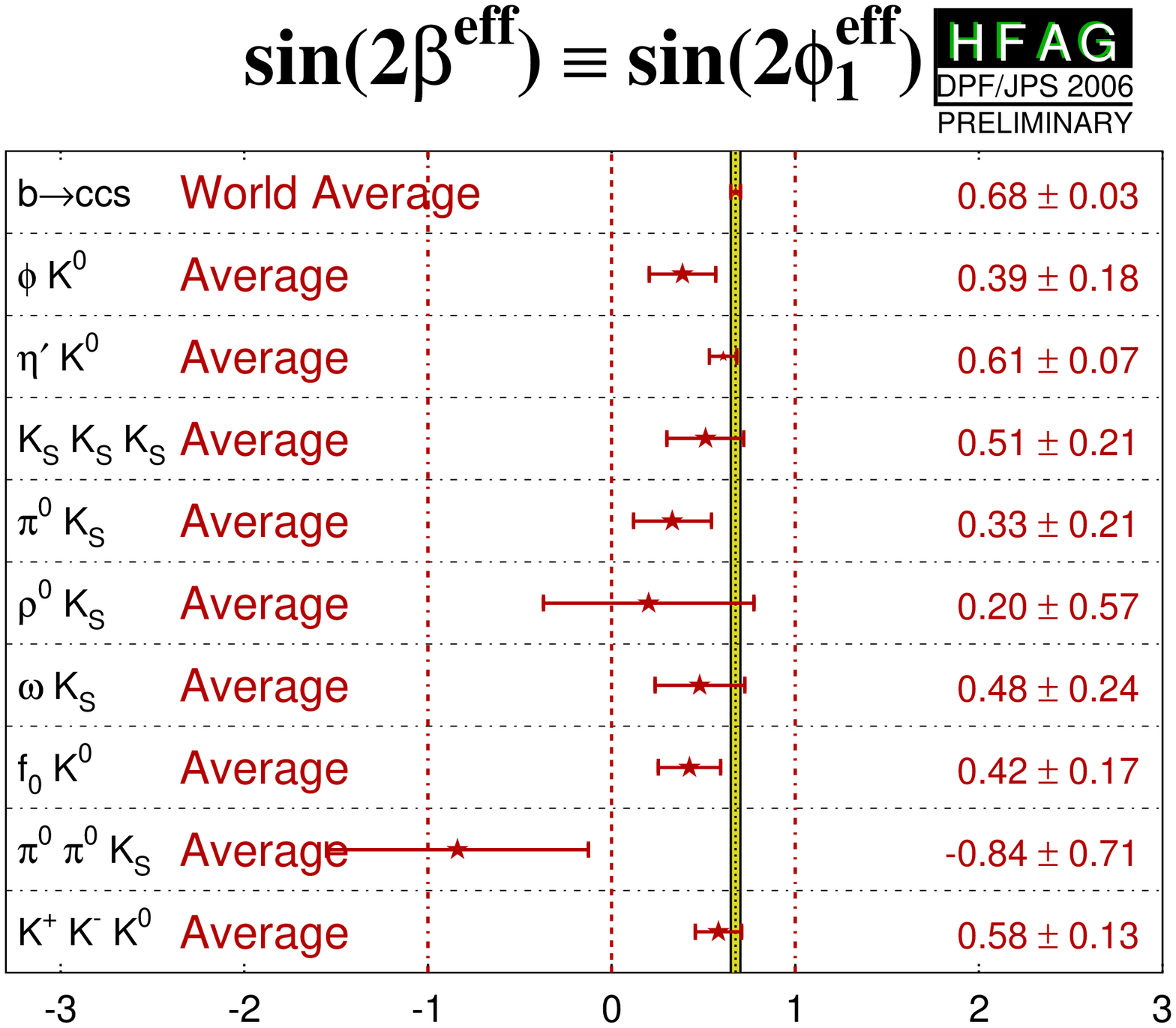}
    }
    \hfill
    \resizebox{0.45\textwidth}{!}{
      \includegraphics{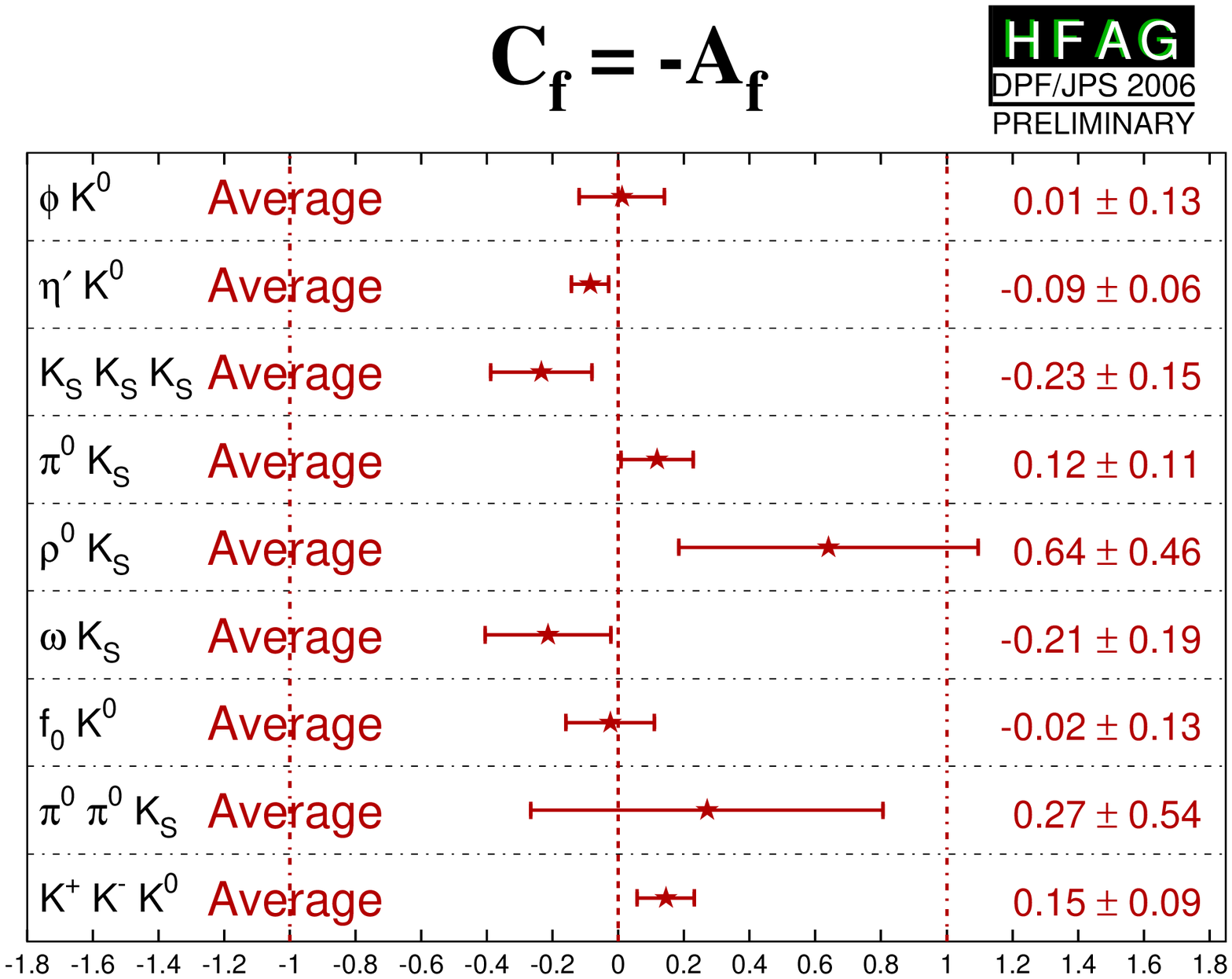}
    }
  \end{center}
  \vspace{-0.8cm}
  \caption{
    (Top)
    Averages of 
    (left) $-\etacp S_{b \to q\bar q s}$ and (right) $C_{b \to q\bar q s}$.
    The $-\etacp S_{b \to q\bar q s}$ figure compares the results to 
    the world average 
    for $-\etacp S_{b \to c\bar c s}$ (see Section~\ref{sec:cp_uta:ccs:cp_eigen}).
    (Bottom) Same, but only averages for each mode are shown.
    More figures are available from the HFAG web pages.
  }
  \label{fig:cp_uta:qqs}
\end{figure}

\begin{figure}[htb]
  \begin{center}
    \resizebox{0.33\textwidth}{!}{
      \includegraphics{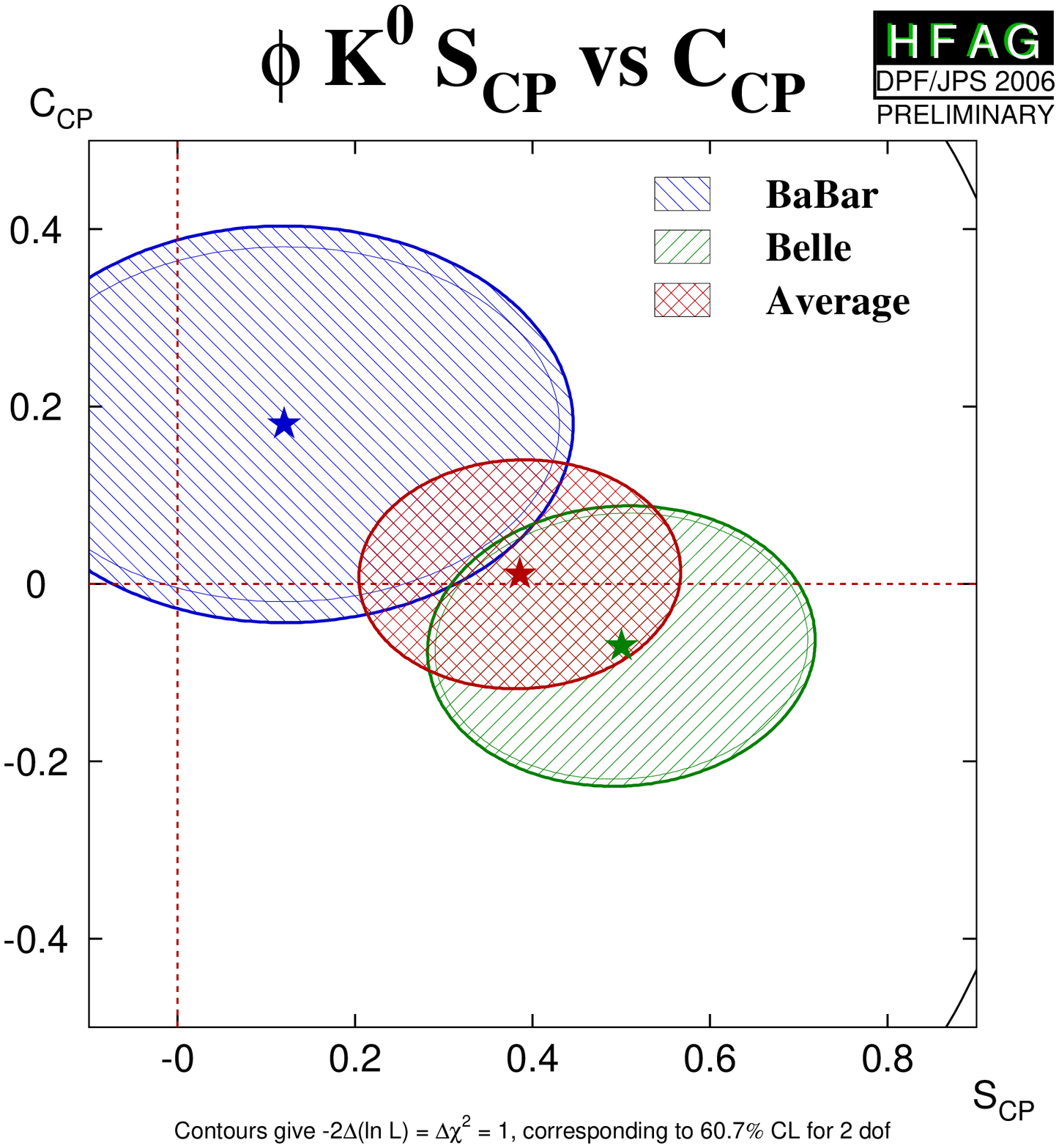}
    }
    \hspace{0.08\textwidth}
    \resizebox{0.33\textwidth}{!}{
      \includegraphics{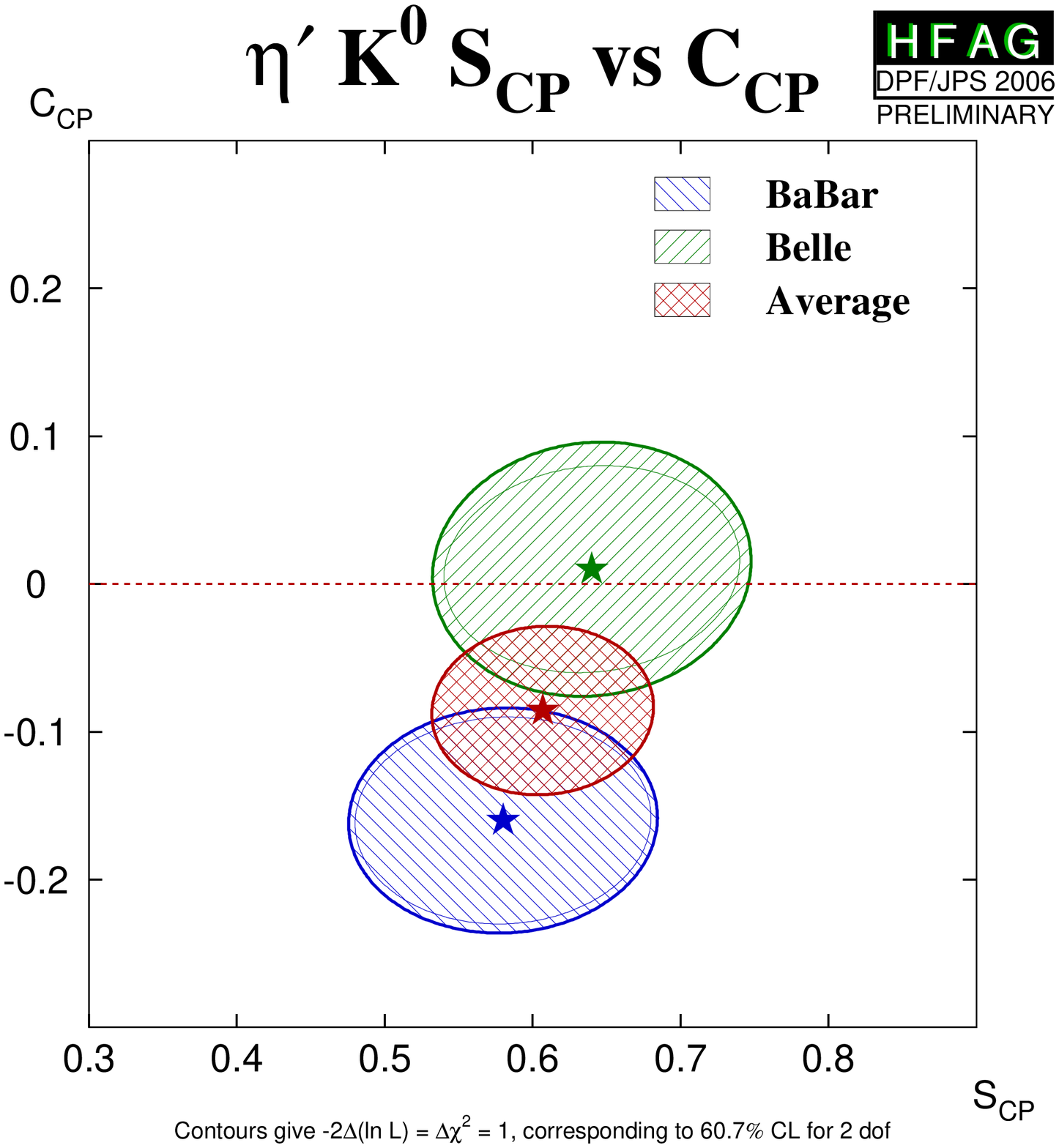}
    }
    \\
    \resizebox{0.33\textwidth}{!}{
      \includegraphics{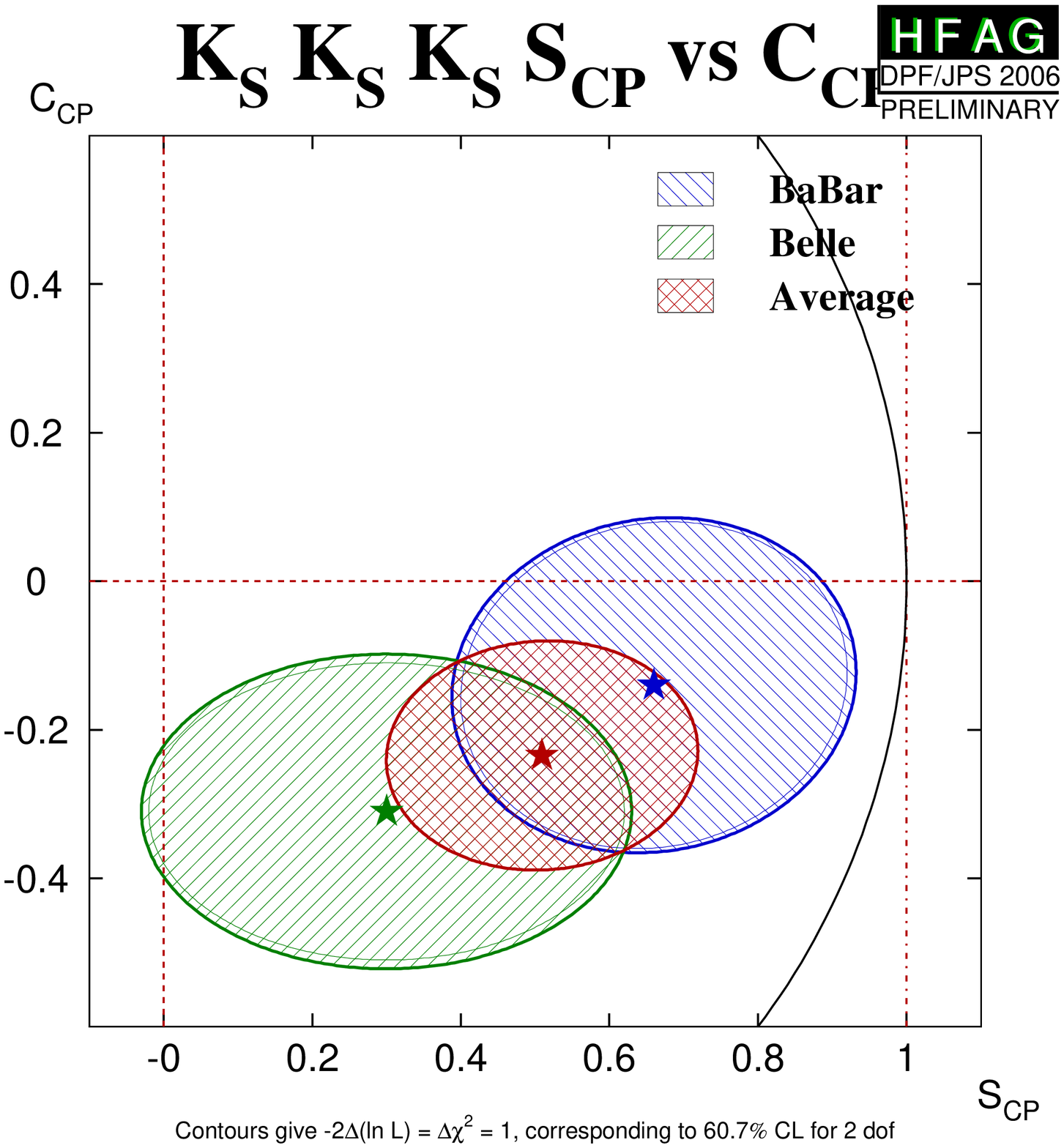}
    }
    \hspace{0.08\textwidth}
    \resizebox{0.33\textwidth}{!}{
      \includegraphics{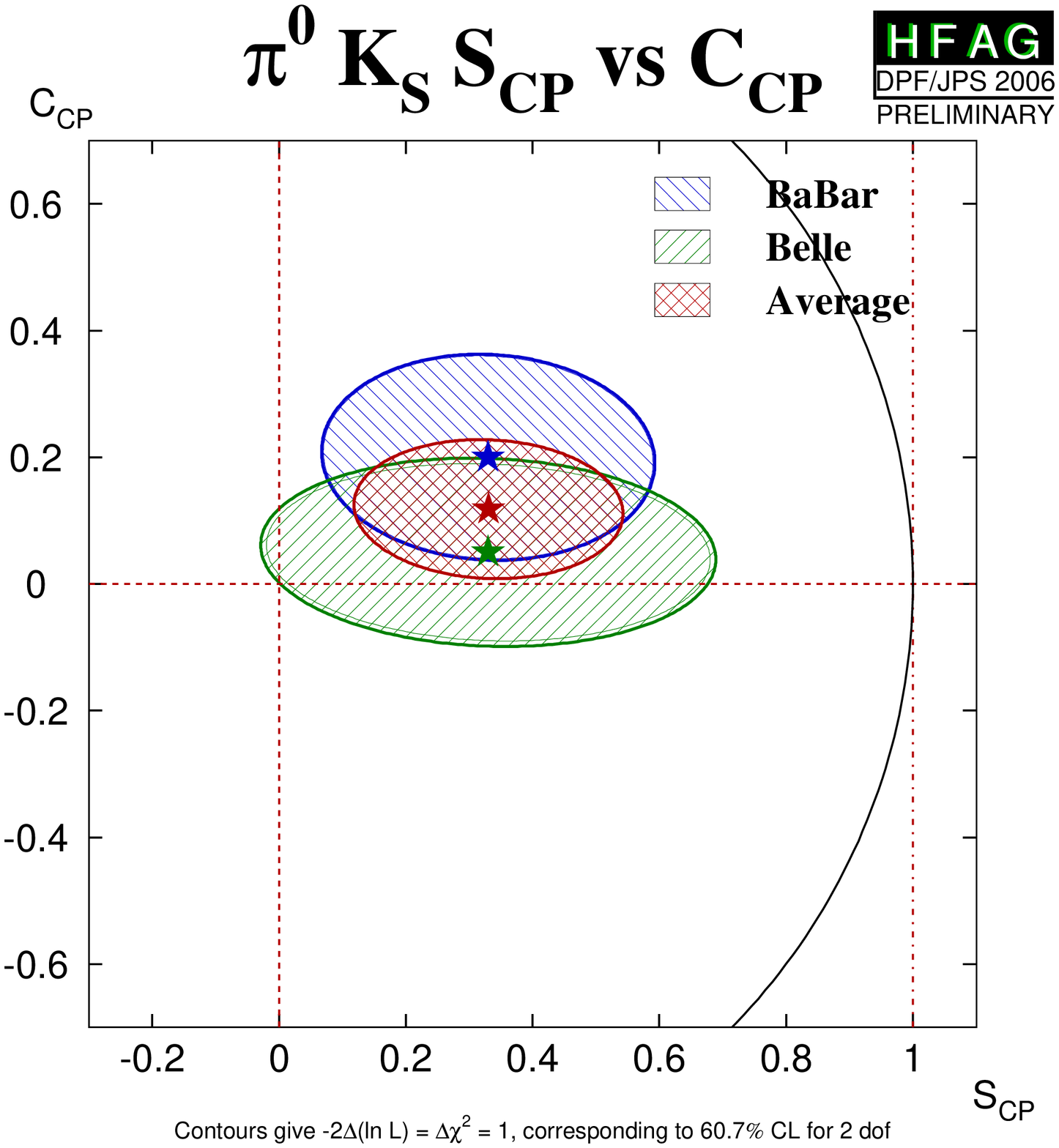}
    }
  \end{center}
  \vspace{-0.8cm}
  \caption{
    (Top)
    Averages of four $b \to q\bar q s$ dominated channels,
    for which correlated averages are performed,
    in the $S_{\CP}$ \vs\ $C_{\CP}$ plane,
    where $S_{\CP}$ has been corrected by the $\CP$ eigenvalue to give
    $\sin(2\beta^{\rm eff})$.
    (Top left) $\Bz \to \phi\Kz$,
    (top right) $\Bz \to \eta^\prime\Kz$,
    (bottom left) $\Bz \to \KS\KS\KS$,
    (bottom right) $\Bz \to \pi^0\KS$.
    More figures are available from the HFAG web pages.
  }
  \label{fig:cp_uta:qqs_SvsC}
\end{figure}

The averages for $-\etacp S_{b \to q\bar q s}$ and $C_{b \to q\bar q s}$
can be found in Table~\ref{tab:cp_uta:qqs},
and are shown in Fig.~\ref{fig:cp_uta:qqs} and Fig.~\ref{fig:cp_uta:qqs_SvsC}.
Results from both \babar\  and \belle\ are averaged for the modes
$\phi\Kz$, $\etapr\Kz$, $f_0\Kz$ and $K^+K^-\Kz$
($\Kz$ indicates that both $\KS$ and $\KL$ are used, 
although \belle\ use neither $f_0\KL$ nor $K^+K^-\KL$), 
$\pi^0 \KS$, $\omega\KS$ and $\KS\KS\KS$. 
\babar\ also has results using $\pi^0\pi^0\KS$ and $\rho^0\KS$.
Of these modes,
$\phi\KS$, $\etapr\KS$, $\pi^0 \KS$, $\omega\KS$ and $\rho^0\KS$
have $\CP$ eigenvalue $\etacp = -1$, 
while $\phi\KL$, $\etapr\KL$, $f_0 \KS$, $\pi^0\pi^0\KS$ and $\KS\KS\KS$ 
have $\etacp = +1$.

The final state $K^+K^-\Kz$ 
(contributions from $\phi\Kz$ are implicitly excluded) 
is not a $\CP$ eigenstate.
However, the $\CP$ composition can be determined using either an 
isospin argument (used by \belle\ to determine a $\CP$ even fraction of 
$0.93 \pm 0.09 \pm 0.05$~\cite{ref:cp_uta:qqs:belle:otherqqs})
or a moments analysis (used by \babar\ to find $\CP$ even fractions of 
$0.89 \pm 0.08 \pm 0.06$ in $K^+K^-\KS$~\cite{ref:cp_uta:qqs:babar:phik0_old}
and
$0.92 \pm 0.07 \pm 0.06$ in $K^+K^-\KL$~\cite{ref:cp_uta:qqs:babar:kkk0_q2b}).
The uncertainty in the $\CP$ even fraction leads to an 
asymmetric error on $S_{b \to q\bar q s}$, which is taken to be 
correlated among the experiments.
To combine, we rescale the results to the 
average $\CP$ even fraction of $0.91 \pm 0.07$.

\babar\ have performed a time-dependent Dalitz plot analysis of
$\Bz \to K^+K^-\Kz$ (see subsection~\ref{sec:cp_uta:qqs:dp}).
Their results for $\phi\Kz$ are determined from that analysis.
Their results for $f^0\Kz$ are a combination of results from the 
Dalitz plot analysis ($-\etacp S_{b \to q\bar q s} = 0.31 \pm 0.32 \pm 0.07$, 
$C_{b \to q\bar q s} = -0.45 \pm 0.28 \pm 0.10$~\cite{ref:cp_uta:qqs:babar:kkk0_tddp}), 
with those from the quasi-two-body analysis of 
$B^0 \to f^0\KS$, $f^0 \to \pi^+\pi^-$
($-\etacp S_{b \to q\bar q s} = 0.95\, ^{+0.23}_{-0.32} \pm 0.10$, 
$C_{b \to q\bar q s} = -0.24 \pm 0.31 \pm 0.15$~\cite{ref:cp_uta:qqs:babar:f0ks}). 
The \babar\ results for $K^+K^-\Kz$ 
are taken from their previous quasi-two-body analysis~\cite{ref:cp_uta:qqs:babar:kkk0_q2b}.
Note that for both \babar\ and \belle\ results for $K^+K^-\Kz$,
uncertainty in the $\CP$ composition of the final state leads to 
a third source of uncertainty on the results for $-\etacp S_{K^+K^-\Kz}$.
 
\input{tables/cp_uta/sPenguinAverages.tex}

As explained above,
each of the modes listed in Table~\ref{tab:cp_uta:qqs} has
different uncertainties within the Standard Model,
and so each may have a different value of $-\etacp S_{b \to q\bar q s}$.
Therefore, there is no strong motivation to make a combined average
over the different modes.
We refer to such an average as a ``na\"\i ve $s$-penguin average.''
It is na\"\i ve not only because of the neglect of the theoretical uncertainty,
but also since possible correlations of systematic effects 
between different modes are neglected.
In spite of these caveats, there remains substantial interest 
in the value of this quantity,
and therefore it is given here:
$\langle -\etacp S_{b \to q\bar q s} \rangle = 0.53 \pm 0.05$,
with confidence level $0.59~(0.5\sigma)$.
Again treating the uncertainties as Gaussian and neglecting correlations,
this value is found to be $2.6\sigma$ below the average $-\etacp S_{b \to c\bar c s}$
given in Sec.~\ref{sec:cp_uta:ccs:cp_eigen}.
(The average for $C_{b \to q\bar q s}$ is 
$\langle C_{b \to q\bar q s} \rangle = -0.01 \pm 0.04$
with confidence level $0.13~(1.5\sigma)$).
However, we do not advocate the use of these averages,
and we emphasise that the values should be treated with 
{\it extreme caution}, if at all.
What is unambiguous (although only qualitative) 
is that there is a trend that the values 
of $-\etacp S_{b \to q\bar q s}$ in different modes 
are below the average for $-\etacp S_{b \to c\bar c s}$.

From Table~\ref{tab:cp_uta:qqs} it may be noted 
that the average for $-\etacp S_{b \to q\bar q s}$ in $\etapr \Kz$ 
($0.61 \pm 0.07$),
is now more than $5\sigma$ away from zero, 
so that $\CP$ violation in this mode is well established.
Amongst other modes,
$\CP$ violation in both $f_0 \KS$ and $K^+K^- \Kz$ is near the $3\sigma$ level,
although due to possible non-Gaussian errors in these results
it may be prudent to defer any strong conclusion on these modes.
There is no evidence (above $2\sigma$) for direct $\CP$ violation 
in any $b \to q \bar q s$ mode.

\mysubsubsection{Time-dependent Dalitz plot analysis of $\Bz \to K^+K^-\Kz$ decays
}
\label{sec:cp_uta:qqs:dp}

As mentioned in Sec.~\ref{sec:cp_uta:notations:dalitz:kkk0},
\babar\ have performed a time-dependent Dalitz plot analysis of 
the $\Bz \to K^+K^-\Kz$ decay~\cite{ref:cp_uta:qqs:babar:kkk0_tddp}.
The results are summarized in Tab.~\ref{tab:cp_uta:kkk0_tddp}.
They are presented in terms of the effective weak phase (from mixing and decay)
difference $\beta^{\rm eff}$ and the direct $\CP$ violation parameter $\Acp$
($\Acp = -C$) for each of the resonant contributions 
$\phi\Kz$ and $f_0\Kz$, together with averaged values of those parameters 
(taking $\CP$ properties into account) over the entire Dalitz plot.

\input{tables/cp_uta/kkk0.tex}

From the above results \babar\ infer that the trigonometric reflection 
at $\pi/2 - \beta^{\rm eff}$,
which is inconsistent with the Standard Model expectation,
is disfavoured at $4.6\sigma$.

\mysubsection{Time-dependent $\CP$ asymmetries in $b \to c\bar{c}d$ transitions
}
\label{sec:cp_uta:ccd}

The transition $b \to c\bar c d$ can occur via either a $b \to c$ tree
or a $b \to d$ penguin amplitude.  
Similarly to Eq.~(\ref{eq:cp_uta:b_to_s}), the amplitude for 
the $b \to d$ penguin can be written
\begin{equation}
  \label{eq:cp_uta:b_to_d}
  \begin{array}{ccccc}
    A_{b \to d} & = & 
    \mc{3}{l}{F_u V_{ub}V^*_{ud} + F_c V_{cb}V^*_{cd} + F_t V_{tb}V^*_{td}} \\
    & = & (F_u - F_c) V_{ub}V^*_{ud} & + & (F_t - F_c) V_{tb}V^*_{td} \\
    & = & {\cal O}(\lambda^3) & + & {\cal O}(\lambda^3). \\
  \end{array}
\end{equation}
From this it can be seen that the $b \to d$ penguin amplitude 
contains terms with different weak phases at the same order of
CKM suppression.

In the above, we have followed Eq.~(\ref{eq:cp_uta:b_to_s}) 
by eliminating the $F_c$ term using unitarity.
However, we could equally well write
\begin{equation}
  \label{eq:cp_uta:b_to_d_alt}
  \begin{array}{ccccc}
    A_{b \to d} 
    & = & (F_u - F_t) V_{ub}V^*_{ud} & + & (F_c - F_t) V_{cb}V^*_{cd}, \\
    & = & (F_c - F_u) V_{cb}V^*_{cd} & + & (F_t - F_u) V_{tb}V^*_{td}. \\
  \end{array}
\end{equation}
Since the $b \to c\bar{c}d$ tree amplitude 
has the weak phase of $V_{cb}V^*_{cd}$,
either of the above expressions allow the penguin to be decomposed into 
parts with weak phases the same and different to the tree amplitude
(the relative weak phase can be chosen to be either $\beta$ or $\gamma$).
However, if the tree amplitude dominates,
there is little sensitivity to any phase 
other than that from $\Bz$\textendash$\Bzb$ mixing.

\begin{figure}[htb]
  \begin{center}
    \begin{tabular}{cc}
      \resizebox{0.46\textwidth}{!}{
        \includegraphics{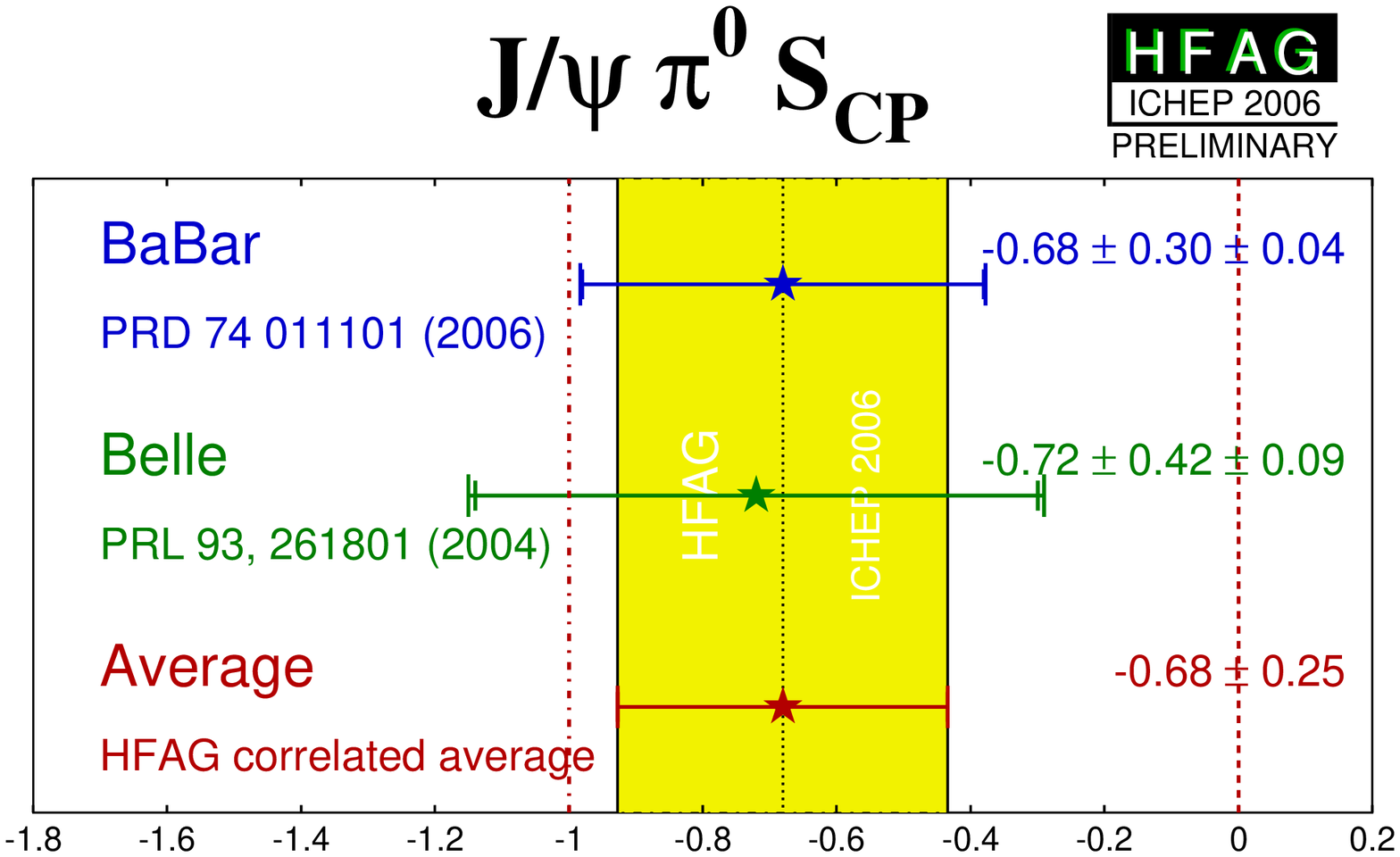}
      }
      &
      \resizebox{0.46\textwidth}{!}{
        \includegraphics{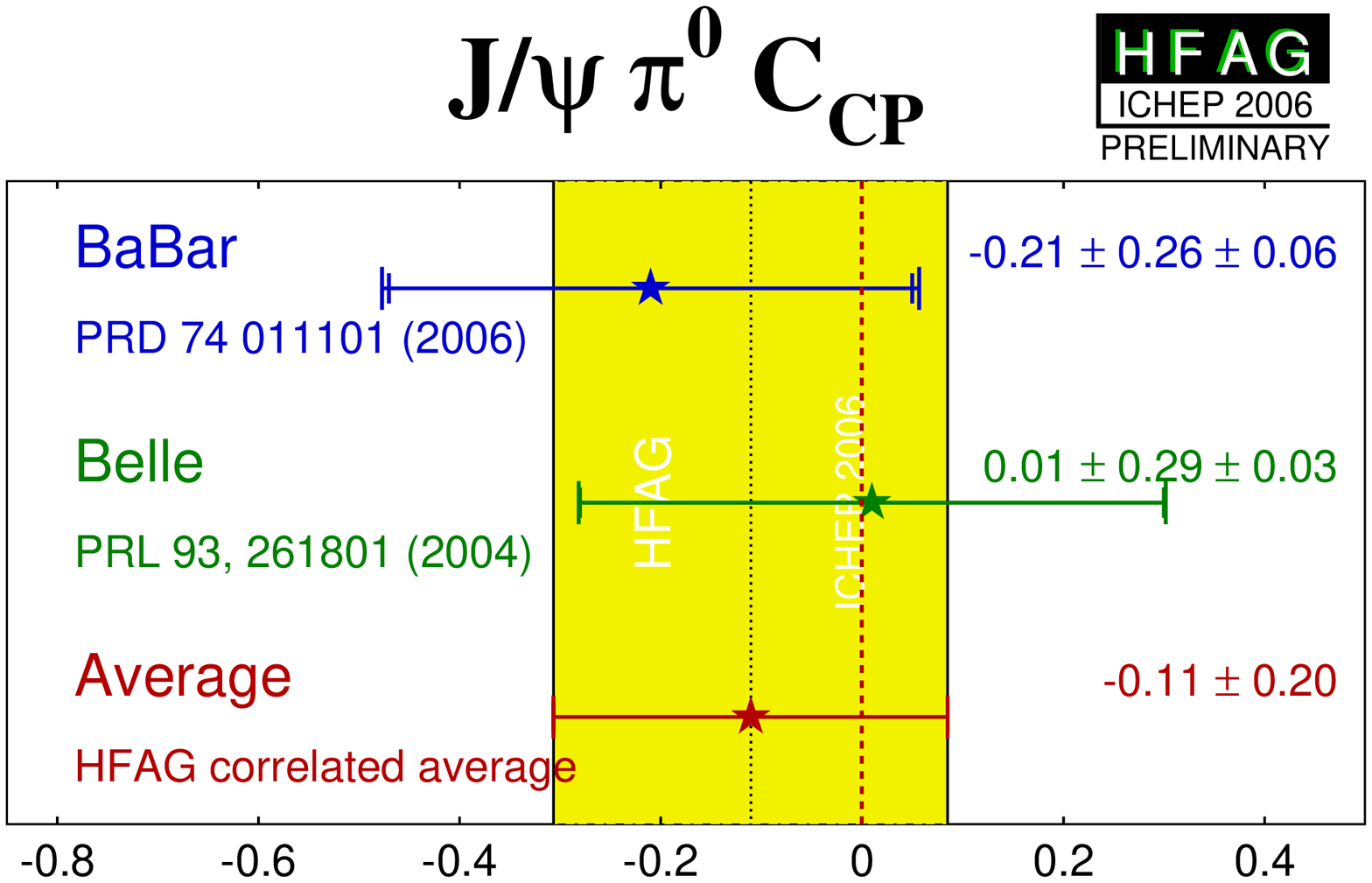}
      }
    \end{tabular}
  \end{center}
  \vspace{-0.8cm}
  \caption{
    Averages of 
    (left) $S_{b \to c\bar c d}$ and (right) $C_{b \to c\bar c d}$ 
    for the mode $\Bz \to J/ \psi \pi^0$.
  }
  \label{fig:cp_uta:ccd:psipi0}
\end{figure}

\begin{figure}[htb]
  \begin{center}
    \begin{tabular}{cc}
      \resizebox{0.46\textwidth}{!}{
        \includegraphics{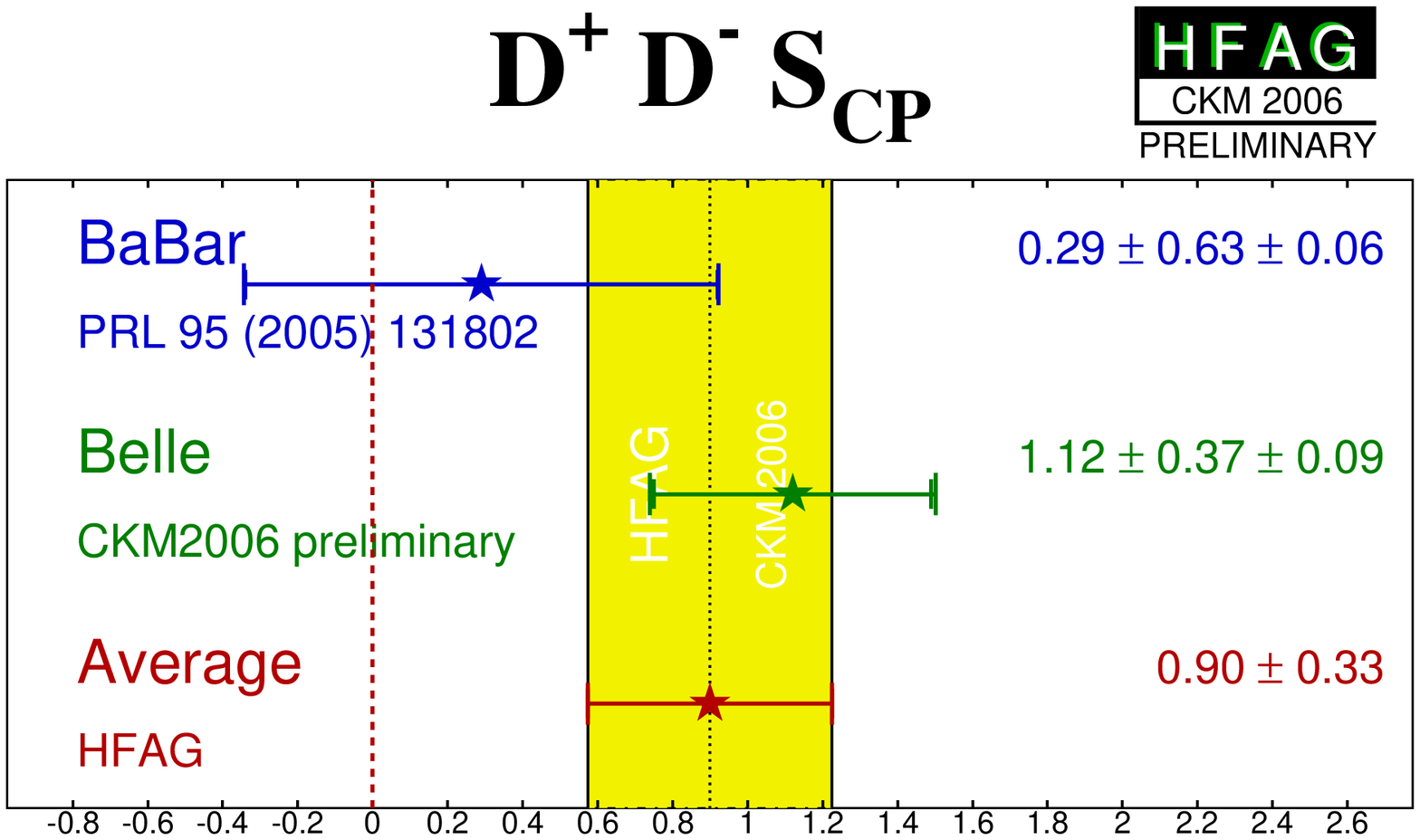}
      }
      &
      \resizebox{0.46\textwidth}{!}{
        \includegraphics{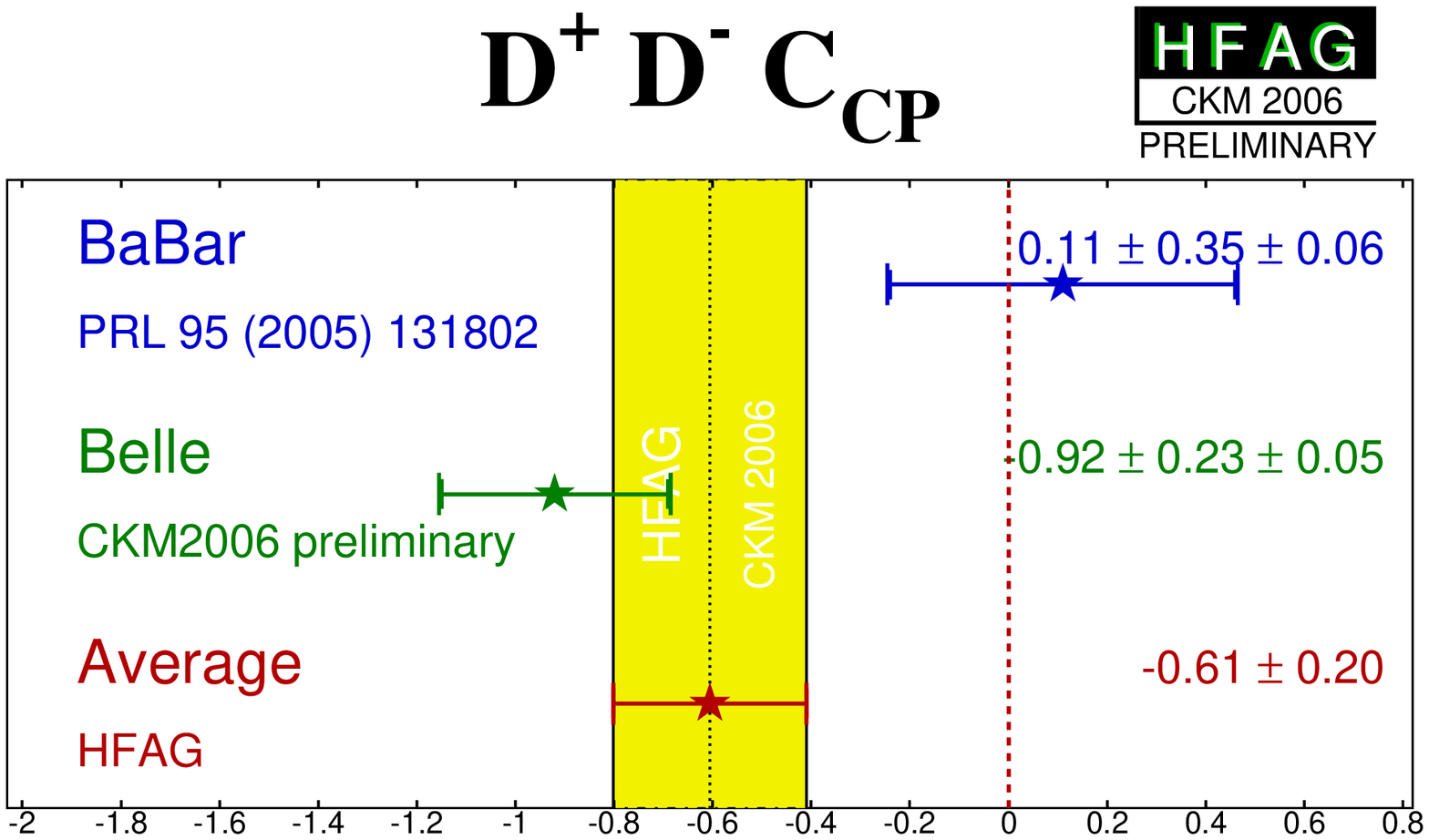}
      }
    \end{tabular}
  \end{center}
  \vspace{-0.8cm}
  \caption{
    Averages of 
    (left) $S_{b \to c\bar c d}$ and (right) $C_{b \to c\bar c d}$ 
    for the mode $\Bz \to D^+D^-$.
  }
  \label{fig:cp_uta:ccd:dd}
\end{figure}

The $b \to c\bar{c}d$ transitions can be investigated with studies 
of various different final states. 
Results are available from both \babar\  and \belle\ 
using the final states $\jpsi \pi^0$, $D^+D^-$, 
$D^{*+}D^{*-}$ and $D^{*\pm}D^{\mp}$,
the averages of these results are given in Table~\ref{tab:cp_uta:ccd}.
The results using the $\CP$ eigenstate ($\etacp = +1$) modes
$\jpsi \pi^0$ and $D^+D^-$
are shown in Fig.~\ref{fig:cp_uta:ccd:psipi0} and 
Fig.~\ref{fig:cp_uta:ccd:dd} respectively.

\begin{figure}[htb]
  \begin{center}
    \begin{tabular}{cc}
      \resizebox{0.46\textwidth}{!}{
        \includegraphics{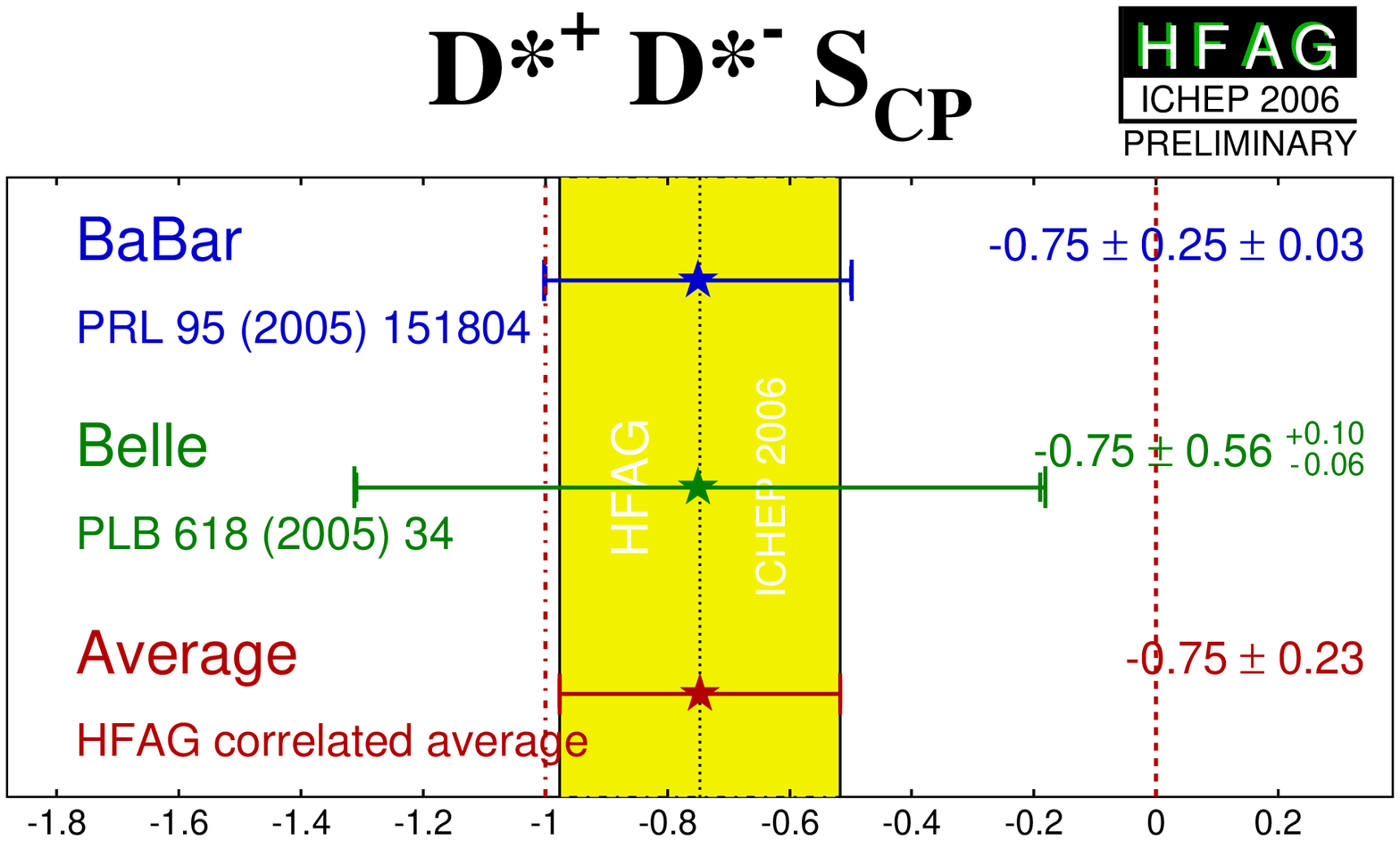}
      }
      &
      \resizebox{0.46\textwidth}{!}{
        \includegraphics{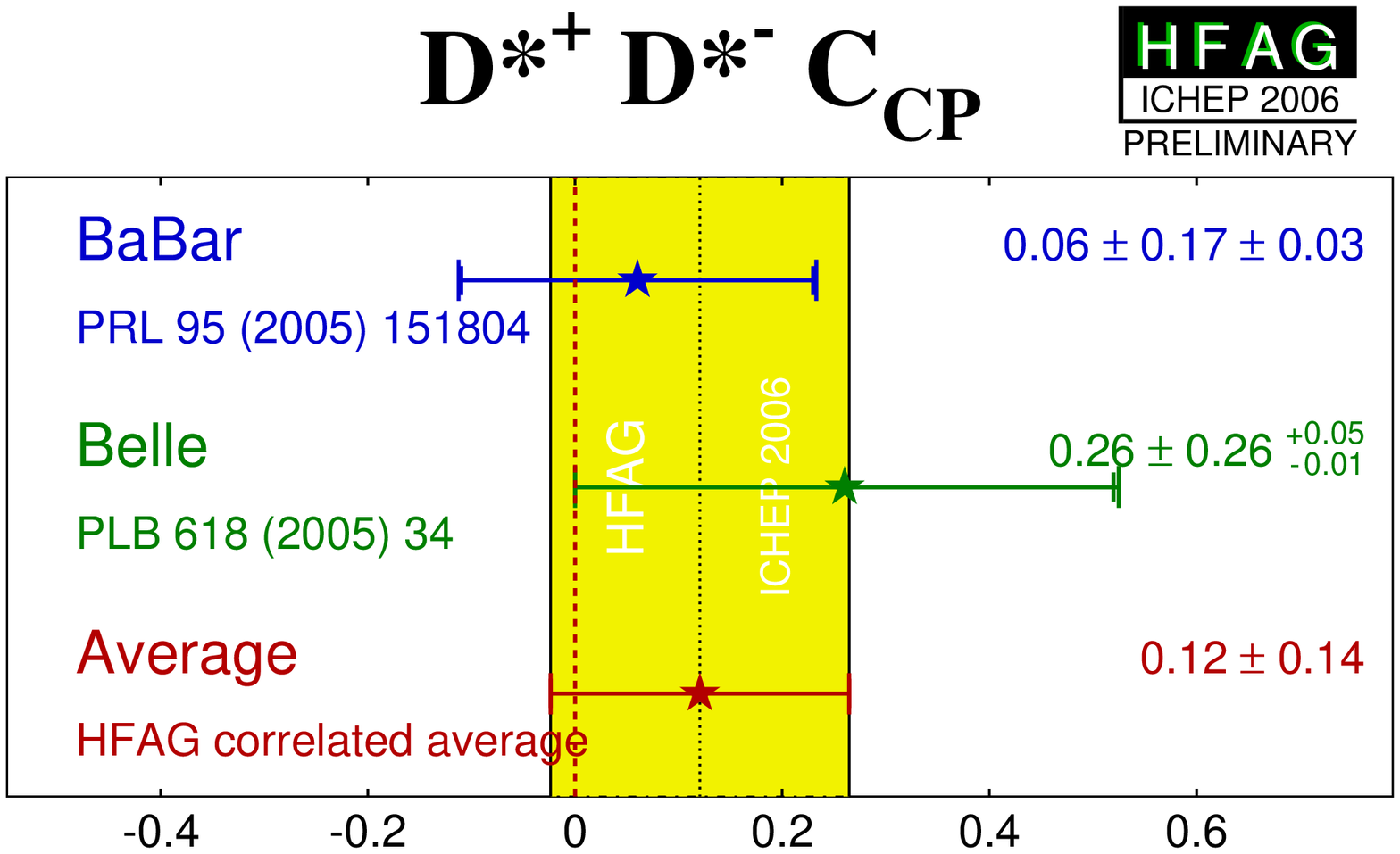}
      }
    \end{tabular}
  \end{center}
  \vspace{-0.8cm}
  \caption{
    Averages of 
    (left) $S_{b \to c\bar c d}$ and (right) $C_{b \to c\bar c d}$ 
    for the mode $\Bz \to D^{*+}D^{*-}$.
  }
  \label{fig:cp_uta:ccd:dstardstar}
\end{figure}

The vector-vector mode $D^{*+}D^{*-}$ 
is found to be dominated by the $\CP$ even longitudinally polarized component;
\babar\ measures a $\CP$ odd fraction of 
$0.125 \pm 0.044 \pm 0.007$~\cite{ref:cp_uta:ccd:babar:dstardstar} while
\belle\ measures a $\CP$ odd fraction of 
$0.19  \pm 0.08  \pm 0.01 $~\cite{ref:cp_uta:ccd:belle:dstardstar}
(here we do not average these fractions and rescale the inputs,
however the average is almost independent of the treatment).
We treat the uncertainty due to the error in the $\CP$-odd fractions
(quoted as a third uncertainty) as a correlated systematic error.
Results using $D^{*+}D^{*-}$ are shown in Fig.~\ref{fig:cp_uta:ccd:dstardstar}.

For the non-$\CP$ eigenstate mode $D^{*\pm}D^{\mp}$
\babar\ uses fully reconstructed events while 
\belle\ combines both fully and partially reconstructed samples.
The most recent results from \babar\ do not include 
a measurement of the overall asymmetry ${\cal A}$.
At present we perform uncorrelated averages of the parameters in the 
$D^{*\pm}D^{\mp}$ system, using only the information from \belle\ on ${\cal A}$.

\input{tables/cp_uta/btoccd.tex}

In the absence of the penguin contribution (tree dominance),
the time-dependent parameters would be given by
$S_{b \to c\bar c d} = - \etacp \sin(2\beta)$,
$C_{b \to c\bar c d} = 0$,
$S_{+-} = \sin(2\beta + \delta)$,
$S_{-+} = \sin(2\beta - \delta)$,
$C_{+-} = - C_{-+}$ and 
${\cal A} = 0$,
where $\delta$ is the strong phase difference between the 
$D^{*+}D^-$ and $D^{*-}D^+$ decay amplitudes.
In the presence of the penguin contribution,
there is no clean interpretation in terms of CKM parameters,
however
direct $\CP$ violation may be observed as any of
$C_{b \to c\bar c d} \neq 0$, $C_{+-} \neq - C_{-+}$ or $A_{+-} \neq 0$.

\begin{figure}[htb]
  \begin{center}
    \begin{tabular}{cc}
      \resizebox{0.46\textwidth}{!}{
        \includegraphics{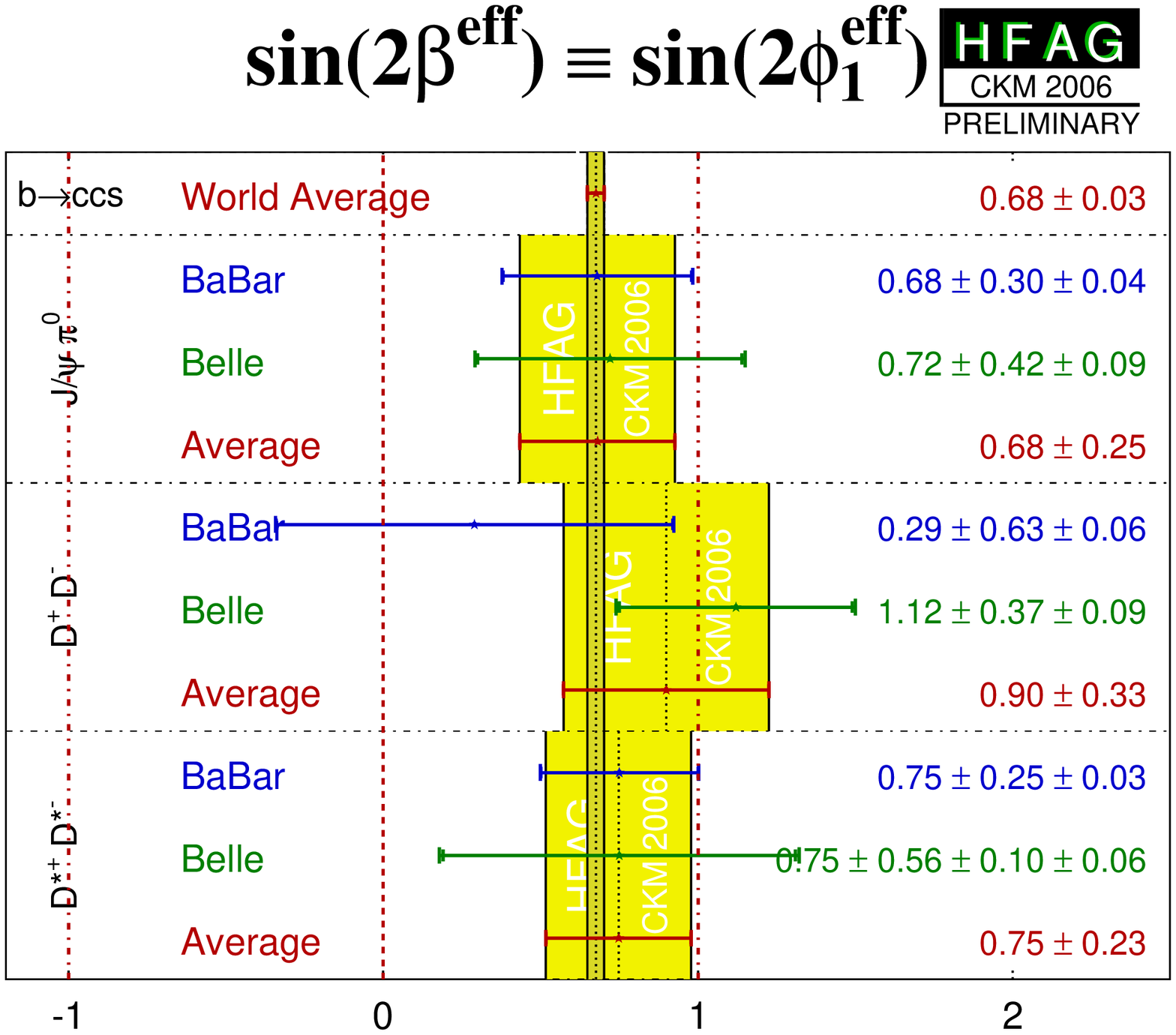}
      }
      &
      \resizebox{0.46\textwidth}{!}{
        \includegraphics{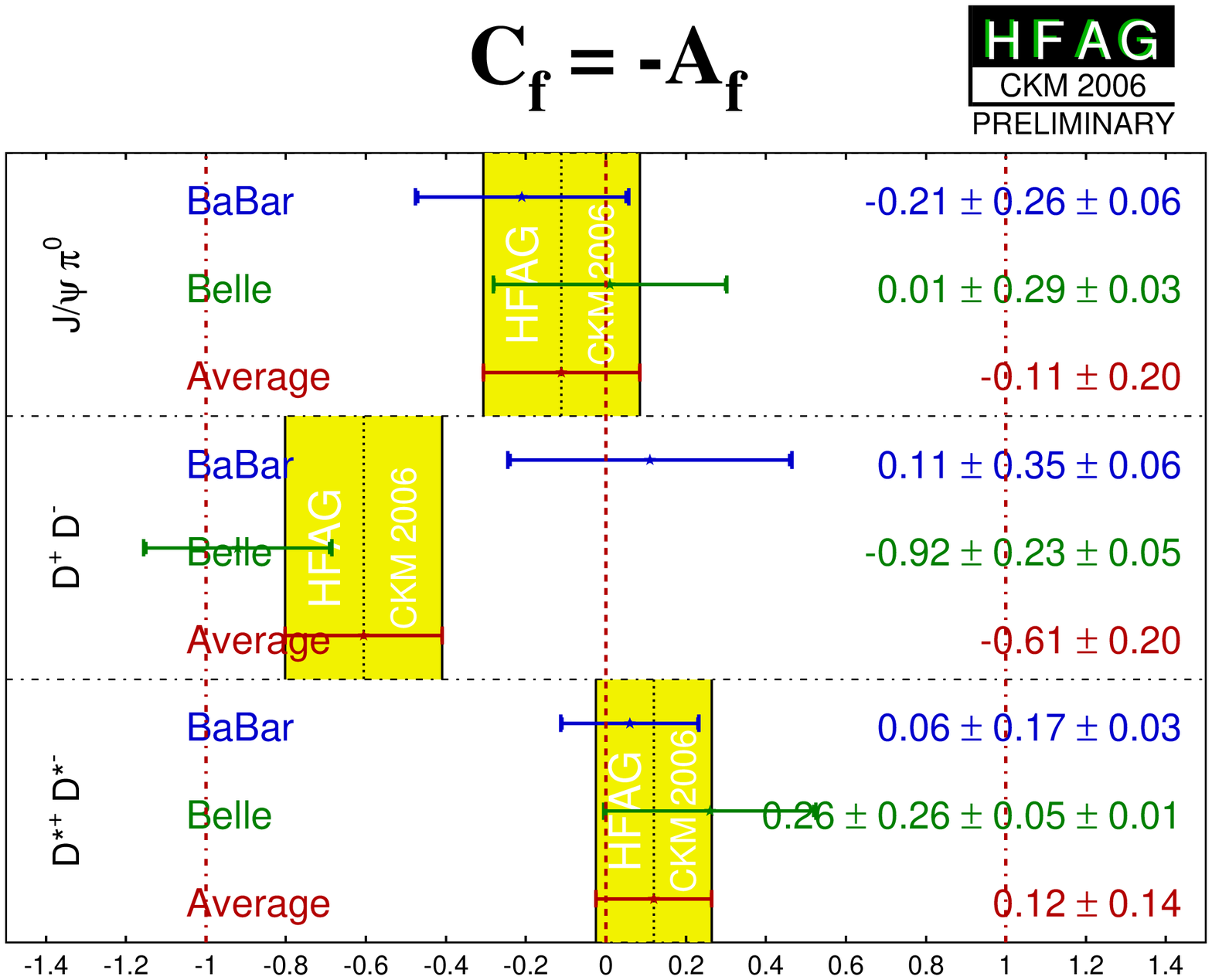}
      }
    \end{tabular}
  \end{center}
  \vspace{-0.8cm}
  \caption{
    Averages of 
    (left) $-\etacp S_{b \to c\bar c d}$ and (right) $C_{b \to c\bar c d}$.
    The $-\etacp S_{b \to q\bar q s}$ figure compares the results to 
    the world average 
    for $-\etacp S_{b \to c\bar c s}$ (see Section~\ref{sec:cp_uta:ccs:cp_eigen}).
  }
  \label{fig:cp_uta:ccd}
\end{figure}

The averages for the $b \to c\bar c d$ modes 
are shown in Fig.~\ref{fig:cp_uta:ccd}.
Results are consistent with tree dominance,
and with the Standard Model,
though the \belle\ results in $\Bz \to D^+D^-$~\cite{ref:cp_uta:ccd:belle:dd}
show an indication of direct $\CP$ violation,
and hence a non-zero penguin contribution.
The average of $S_{b \to c\bar c d}$ in the $D^{*+}D^{*-}$ final state
is about $3\sigma$ from zero;
however, due to the large uncertainty and possible non-Gaussian effects,
any strong conclusion should be deferred.


\mysubsection{Time-dependent $\CP$ asymmetries in $b \to q\bar{q}d$ transitions
}
\label{sec:cp_uta:qqd}

Decays such as $\Bz\to\KS\KS$ are pure $b \to q\bar{q}d$ penguin transitions.
As shown in Eq.~\ref{eq:cp_uta:b_to_d},
this diagram has different contributing weak phases,
and therefore the observables are sensitive to the difference 
(which can be chosen to be either $\beta$ or $\gamma$).
Note that if the contribution with the top quark in the loop dominates,
the weak phase from the decay amplitudes should cancel that from mixing,
so that no $\CP$ violation (neither mixing-induced nor direct) occurs.
Non-zero contributions from loops with intermediate up and charm quarks
can result in both types of effect 
(as usual, a strong phase difference is required for direct $\CP$ violation
to occur).

\babar~\cite{ref:cp_uta:qqd:babar:ksks} 
have performed a time-dependent analysis of $\Bz\to\KS\KS$.
The results are shown in Table~\ref{tab:cp_uta:qqd}.

\input{tables/cp_uta/qqdAverages.tex}

\mysubsection{Time-dependent asymmetries in $b \to s\gamma$ transitions
}
\label{sec:cp_uta:bsg}

The radiative decays $b \to s\gamma$ produce photons 
which are highly polarized in the Standard Model.
The decays $\Bz \to F \gamma$ and $\Bzb \to F \gamma$ 
produce photons with opposite helicities, 
and since the polarization is, in principle, observable,
these final states cannot interfere.
The finite mass of the $s$ quark introduces small corrections
to the limit of maximum polarization,
but any large mixing induced $\CP$ violation would be a signal for new physics.
Since a single weak phase dominates the $b \to s \gamma$ transition in the 
Standard Model, the cosine term is also expected to be small.

Atwood {\it et al.}~\cite{ref:cp_uta:bsg:aghs} have shown that 
an inclusive analysis with respect to $\KS\pi^0\gamma$ can be performed,
since the properties of the decay amplitudes 
are independent of the angular momentum of the $\KS\pi^0$ system. 
However, if non-dipole operators contribute significantly to the amplitudes, 
then the Standard Model mixing-induced $\CP$ violation could be larger 
than the na\"\i ve expectation 
$S \simeq -2 (m_s/m_b) \sin \left(2\beta\right)$~\cite{ref:cp_uta:bsg:gglp}.
In this case, 
the $\CP$ parameters may vary over the $\KS\pi^0\gamma$ Dalitz plot, 
for example as a function of the $\KS\pi^0$ invariant mass.

With the above in mind, 
we quote two averages: one for $K^*(892)$ candidates only, 
and the other one for the inclusive $\KS\pi^0\gamma$ decay (including the $K^*(892)$).
If the Standard Model dipole operator is dominant, 
both should give the same quantities 
(the latter naturally with smaller statistical error). 
If not, care needs to be taken in interpretation of the inclusive parameters, 
while the results on the $K^*(892)$ resonance remain relatively clean.
Results from \babar\ and \belle\ are used for both averages;
both experiments use the invariant mass range 
$0.60 \ {\rm GeV}/c^2 < M_{\KS\pi^0} < 1.80 \ {\rm GeV}/c^2$
in the inclusive analysis.

\input{tables/cp_uta/bsgAverages.tex}

\begin{figure}[htb]
  \begin{center}
    \begin{tabular}{cc}
      \resizebox{0.46\textwidth}{!}{
        \includegraphics{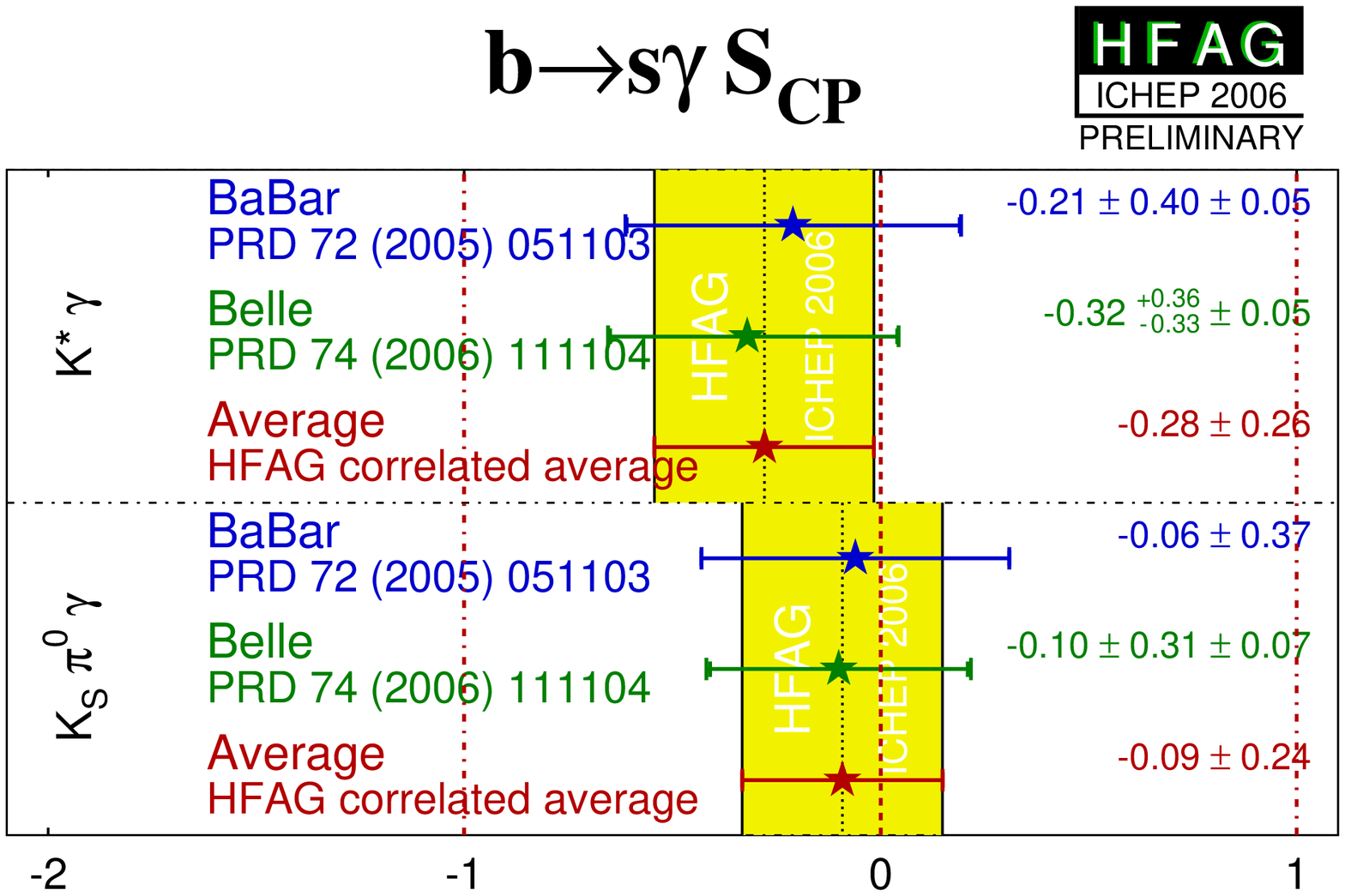}
      }
      &
      \resizebox{0.46\textwidth}{!}{
        \includegraphics{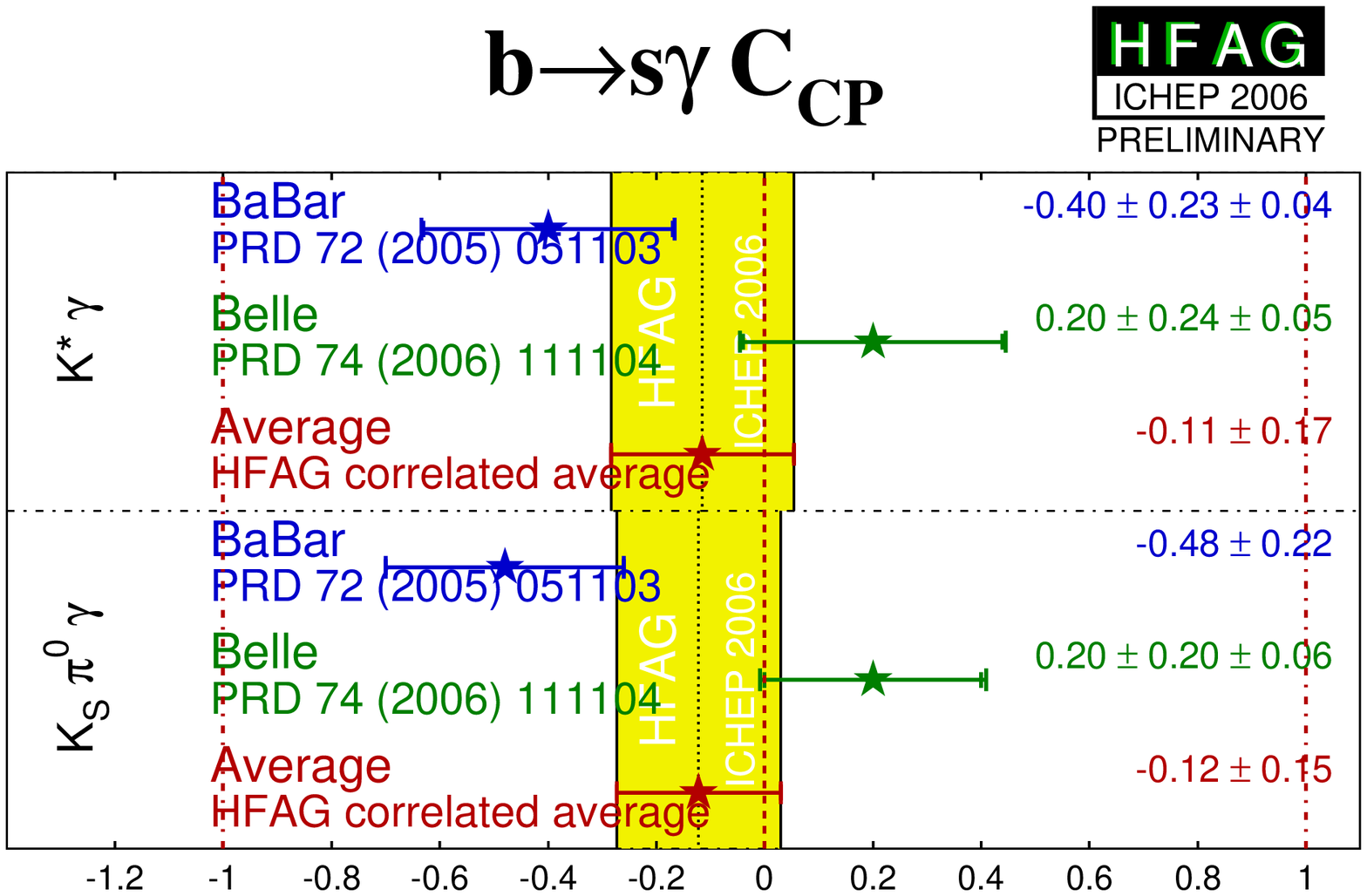}
      }
    \end{tabular}
  \end{center}
  \vspace{-0.8cm}
  \caption{
    Averages of (left) $S_{b \to s \gamma}$ and (right) $C_{b \to s \gamma}$.
    Recall that the data for $K^*\gamma$ is a subset of that for $\KS\pi^0\gamma$.
  }
  \label{fig:cp_uta:bsg}
\end{figure}

The results are shown in Table~\ref{tab:cp_uta:bsg},
and in Fig.~\ref{fig:cp_uta:bsg}.
No significant $\CP$ violation results are seen;
the results are consistent with the Standard Model
and with other measurements in the $b \to s\gamma$ system (see Sec.~\ref{sec:rare}).

\mysubsection{Time-dependent $\CP$ asymmetries in $b \to u\bar{u}d$ transitions
}
\label{sec:cp_uta:uud}

\begin{figure}[htb]
  \begin{center}
    \begin{tabular}{ccc}
      \resizebox{0.34\textwidth}{!}{\includegraphics{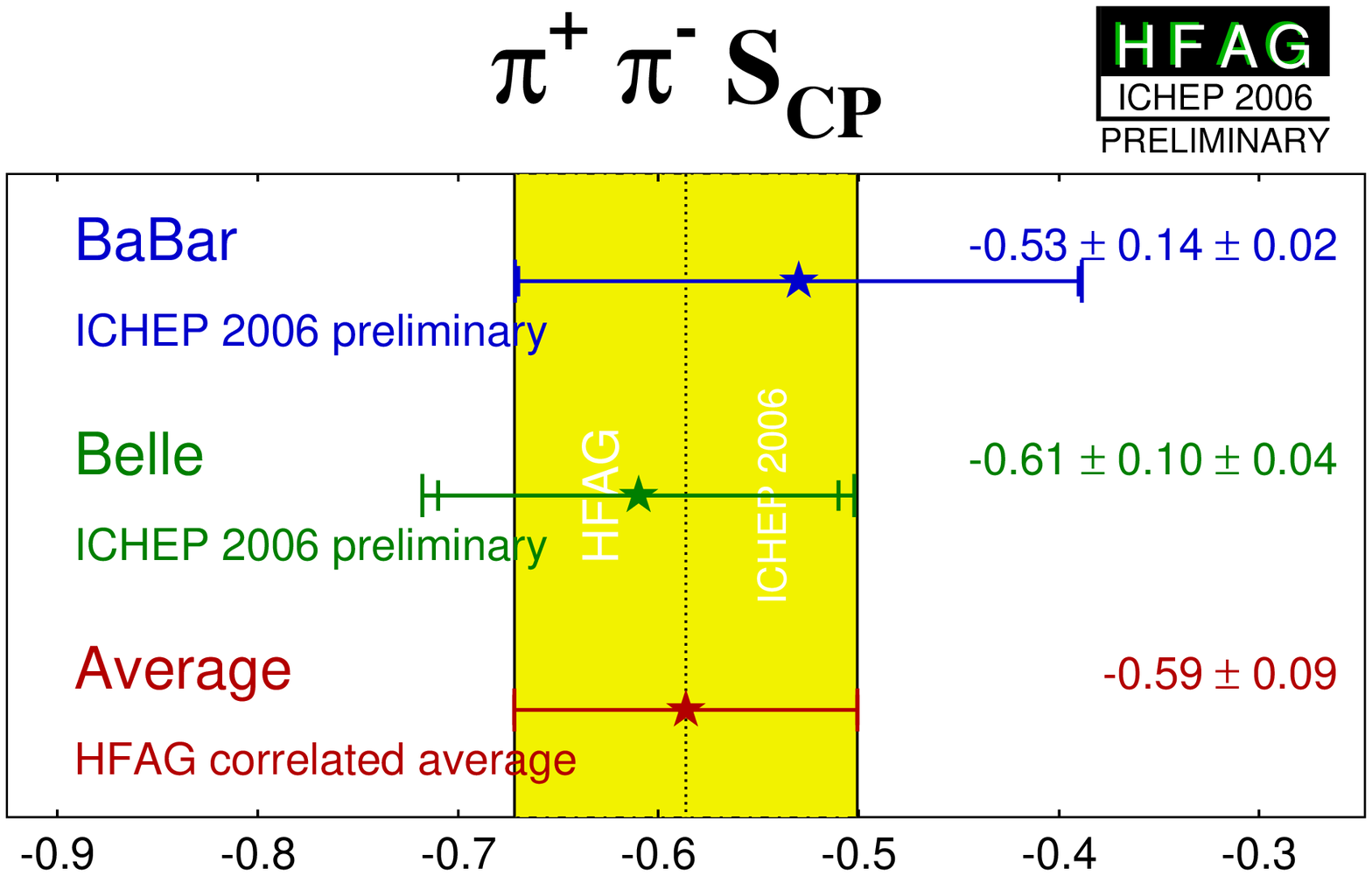}}
      &
      \resizebox{0.34\textwidth}{!}{\includegraphics{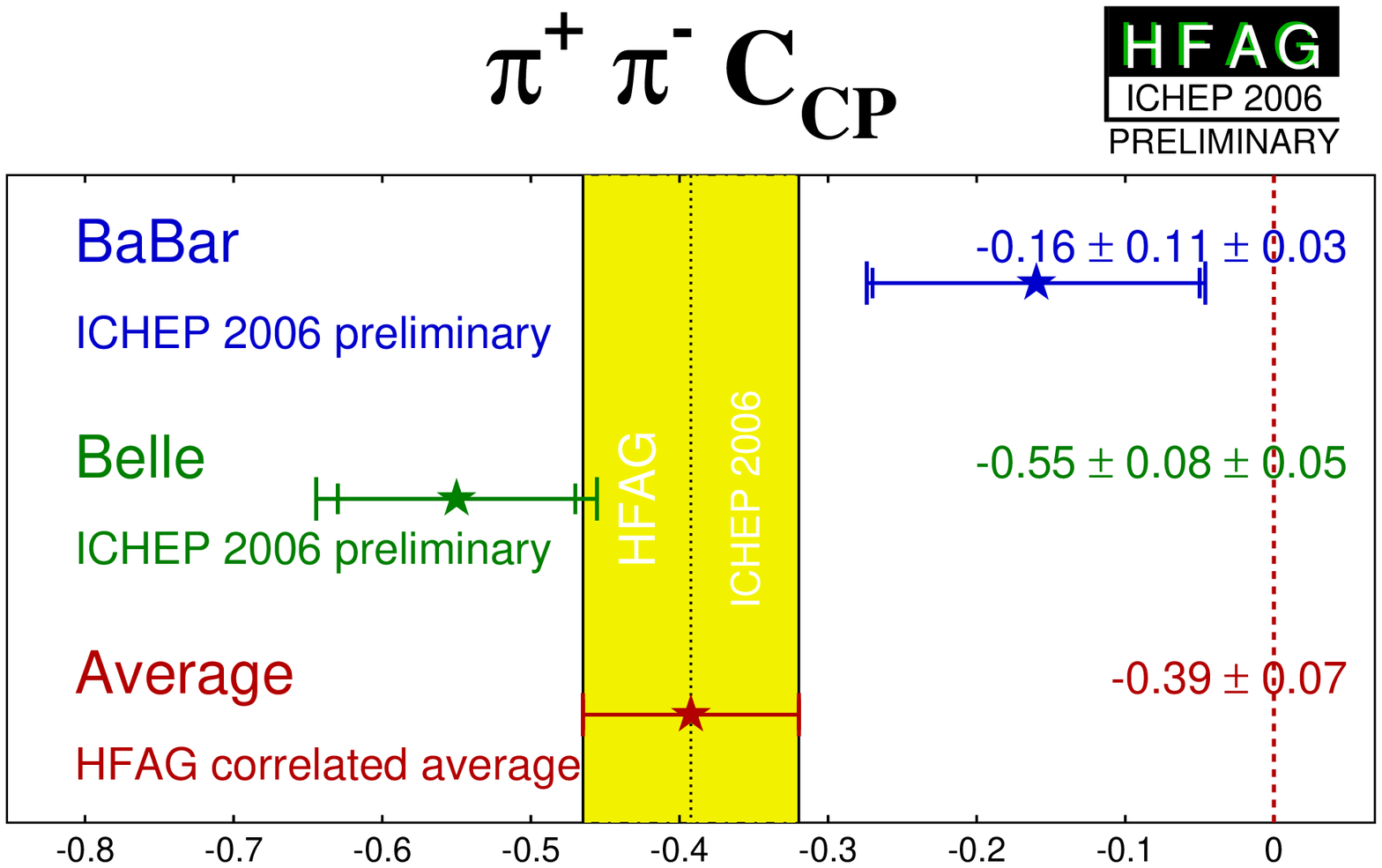}}
      &
      \resizebox{0.22\textwidth}{!}{\includegraphics{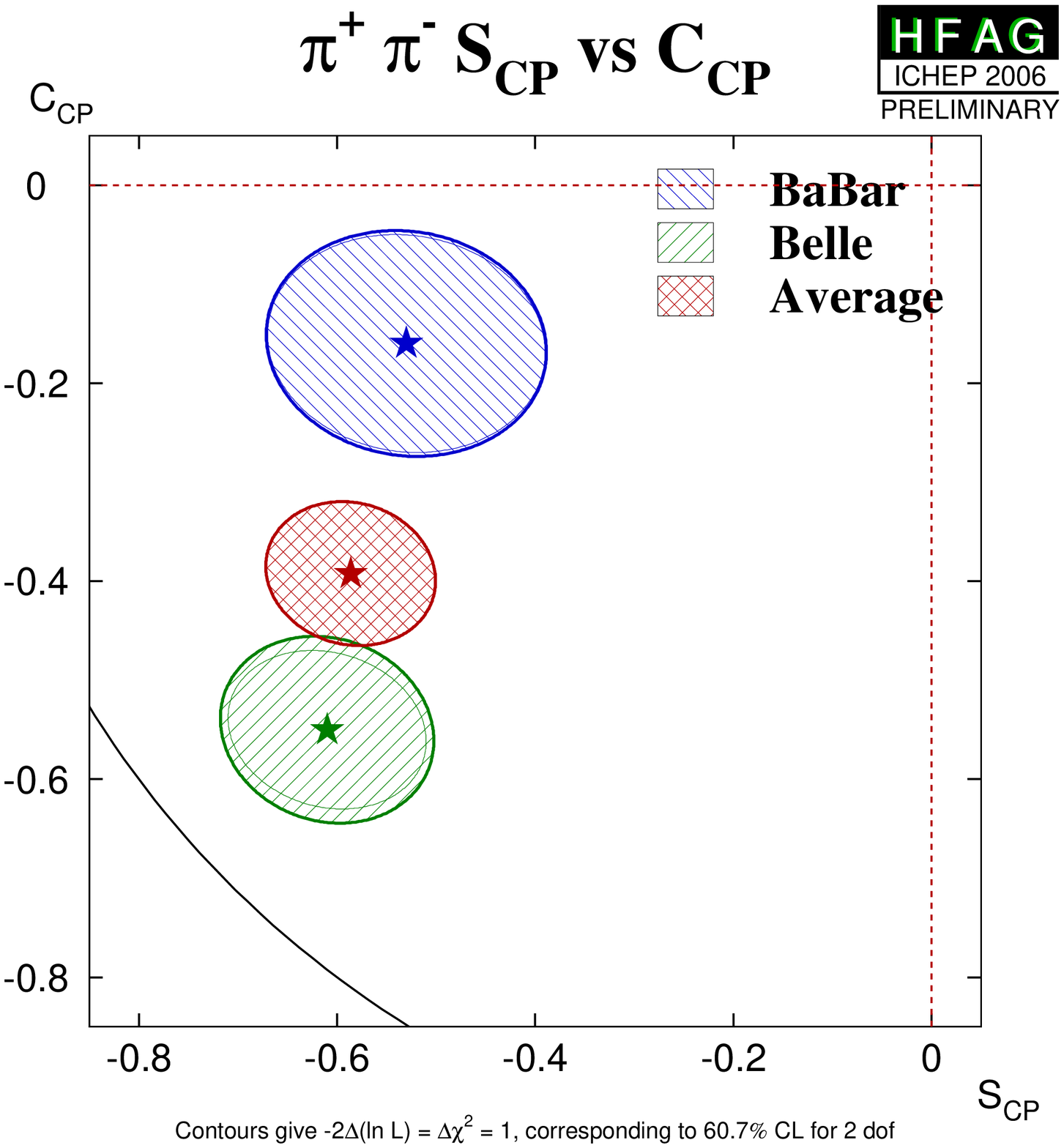}}      
    \end{tabular}
  \end{center}
  \vspace{-0.8cm}
  \caption{
    Averages of 
    (left) $S_{b \to u\bar u d}$,
    (middle) $C_{b \to u\bar u d}$ and
    (right) $S_{b \to u\bar u d}$ \vs\ $C_{b \to u\bar u d}$
    for the mode $\Bz \to \pi^+\pi^-$.
  }
  \label{fig:cp_uta:uud:pipi}
\end{figure}

\begin{figure}[htb]
  \begin{center}
    \begin{tabular}{ccc}
      \resizebox{0.34\textwidth}{!}{\includegraphics{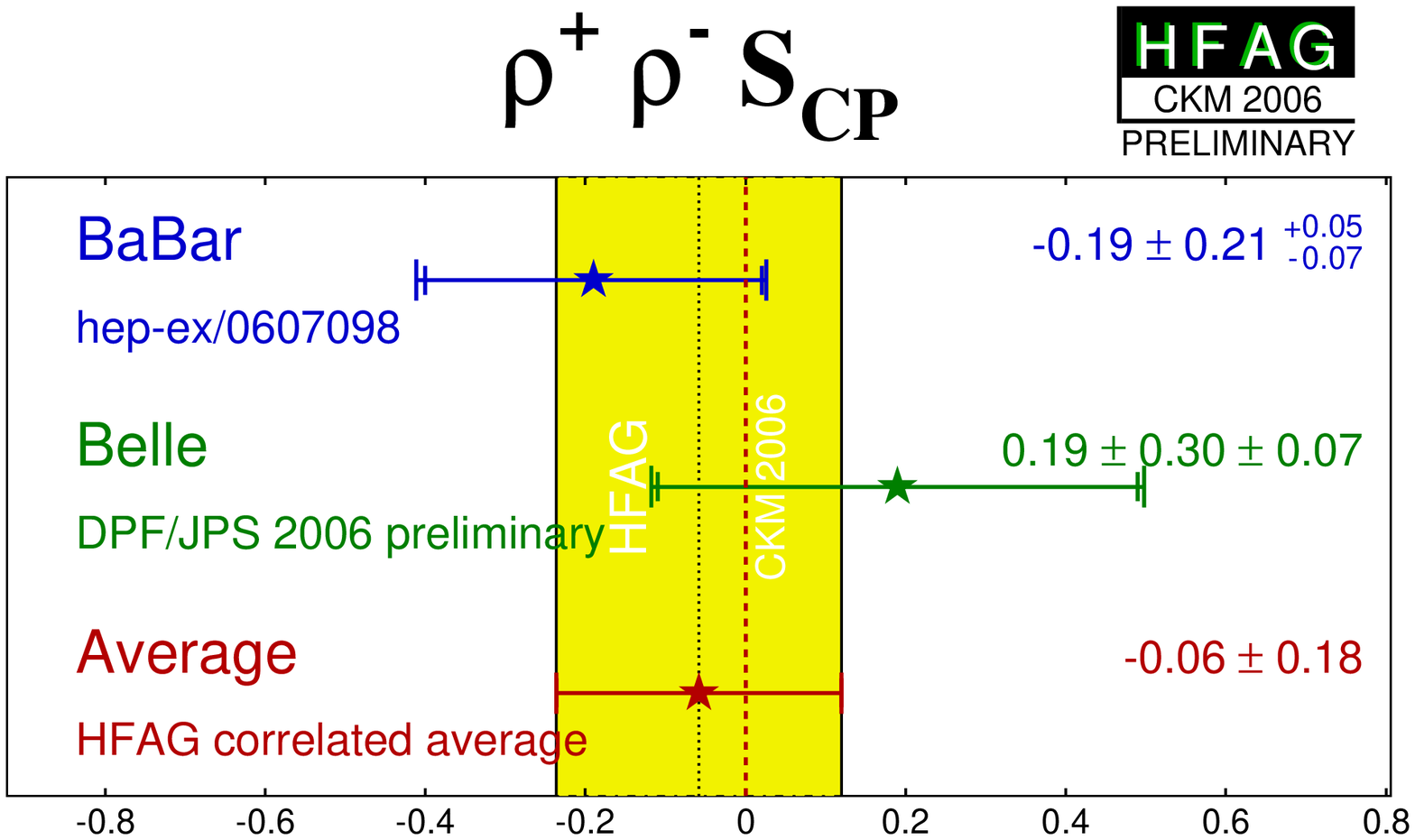}}
      &
      \resizebox{0.34\textwidth}{!}{\includegraphics{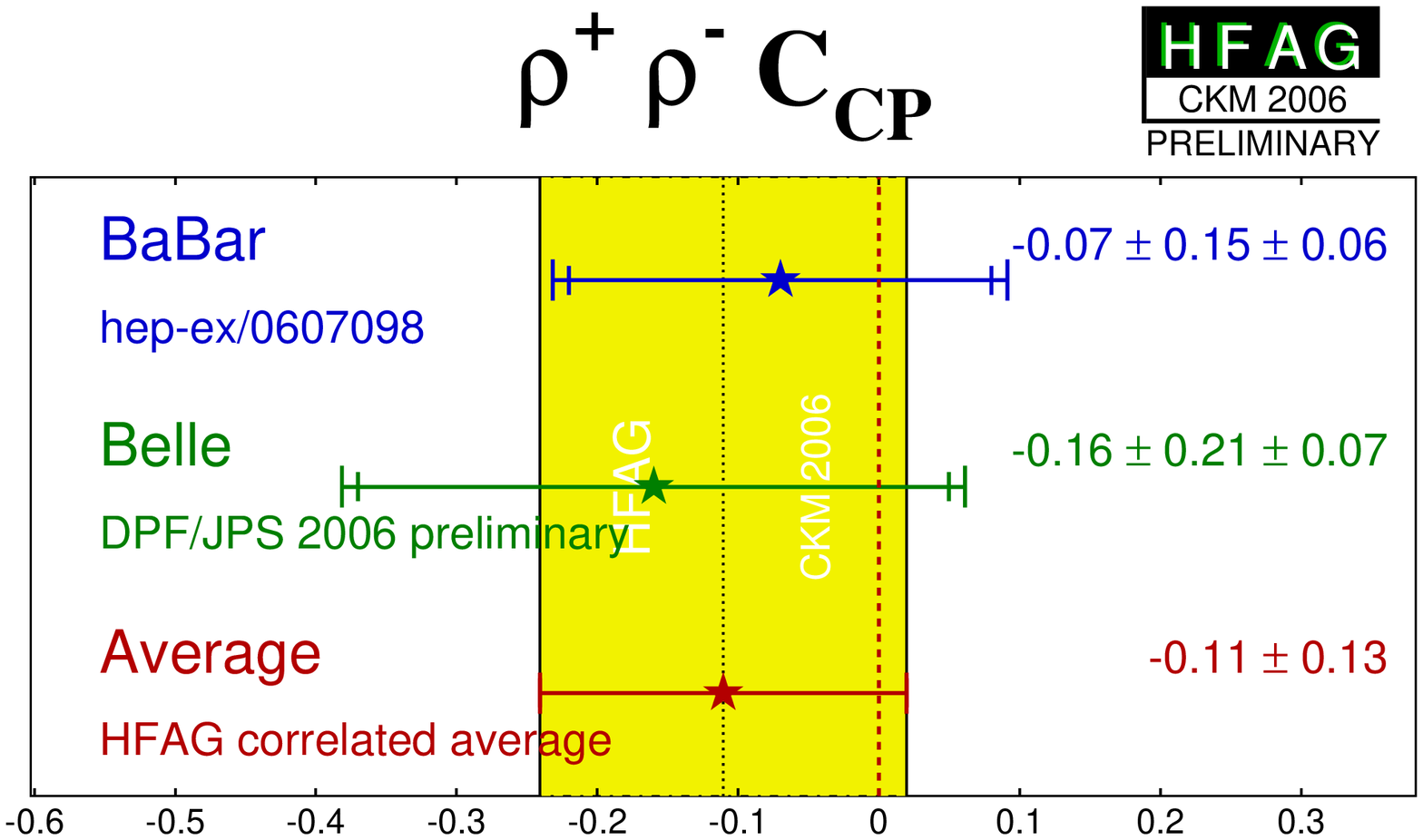}}
      &
      \resizebox{0.22\textwidth}{!}{\includegraphics{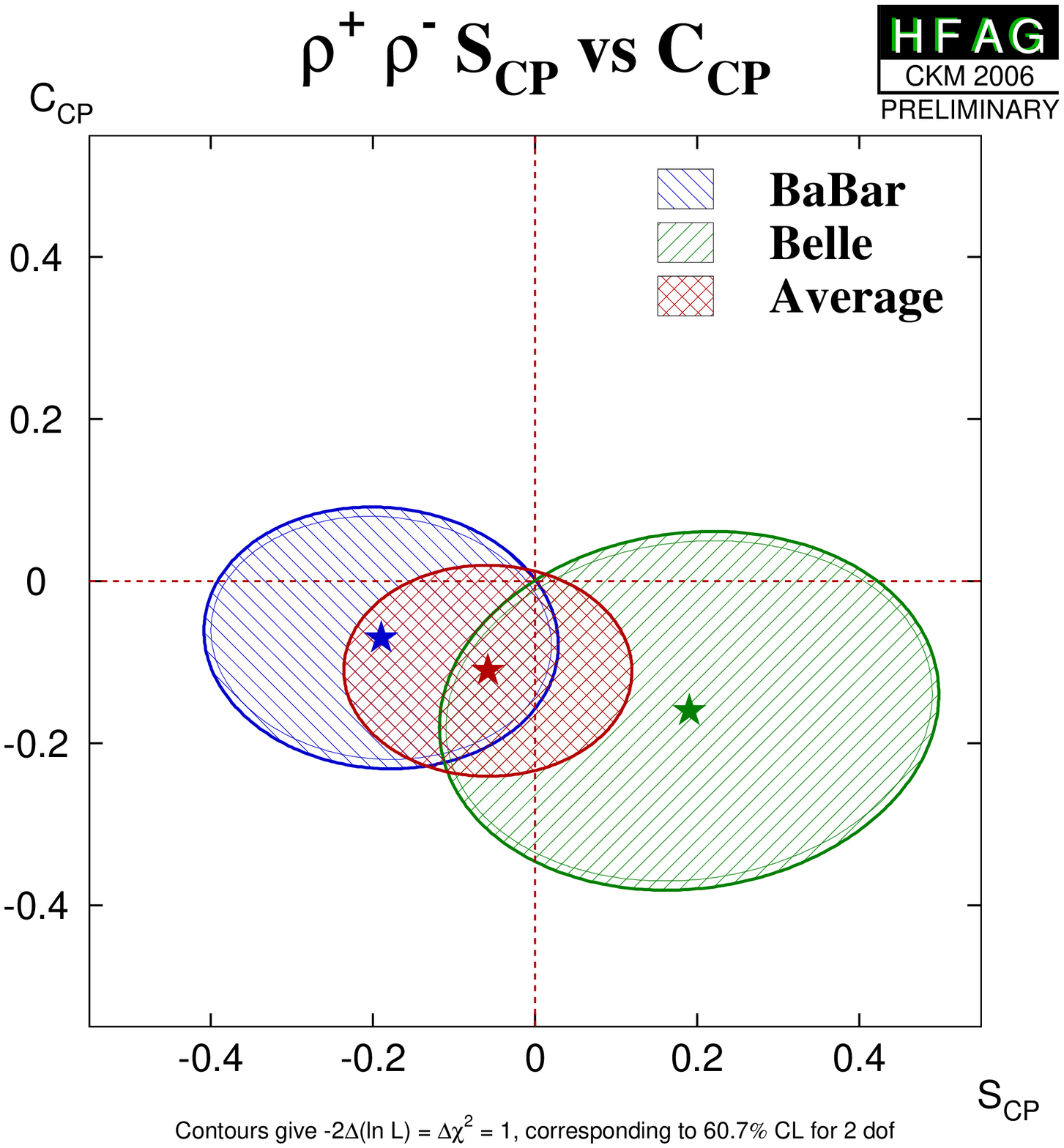}}
    \end{tabular}
  \end{center}
  \vspace{-0.8cm}
  \caption{
    Averages of 
    (left) $S_{b \to u\bar u d}$,
    (middle) $C_{b \to u\bar u d}$ and
    (right) $S_{b \to u\bar u d}$ \vs\ $C_{b \to u\bar u d}$
    for the mode $\Bz \to \rho^+\rho^-$.
  }
  \label{fig:cp_uta:uud:rhorho}
\end{figure}

The $b \to u \bar u d$ transition can be mediated by either 
a $b \to u$ tree amplitude or a $b \to d$ penguin amplitude.
These transitions can be investigated using 
the time dependence of $\Bz$ decays to final states containing light mesons.
Results are available from both \babar\ and \belle\ for the 
$\CP$ eigenstate ($\etacp = +1$) $\pi^+\pi^-$ final state
and for the vector-vector final state $\rho^+\rho^-$,
which is found to be dominated by the $\CP$ even
longitudinally polarized component
(\babar\ measure $f_{\rm long} = 
0.977 \pm 0.024 \, ^{+0.015}_{-0.013}$~\cite{ref:cp_uta:uud:babar:rhorho}
while \belle\ measure $f_{\rm long} = 
0.941 \, ^{+0.034}_{-0.040} \pm 0.030$~\cite{ref:cp_uta:uud:belle:rhorho}).
\babar\ have also performed a time-dependent analysis of the 
$\Bz \to a_1^\pm \pi^\mp$ decay~\cite{ref:cp_uta:uud:babar:a1pi}.
These results, and averages, are listed in Table~\ref{tab:cp_uta:uud}.
The averages for $\pi^+\pi^-$ are shown in Fig.~\ref{fig:cp_uta:uud:pipi},
and those for $\rho^+\rho^-$ are shown in Fig.~\ref{fig:cp_uta:uud:rhorho}.

\input{tables/cp_uta/uudAverages.tex}

If the penguin contribution is negligible, 
the time-dependent parameters for $\Bz \to \pi^+\pi^-$ 
and $\Bz \to \rho^+\rho^-$ are given by
$S_{b \to u\bar u d} = \etacp \sin(2\alpha)$ and
$C_{b \to u\bar u d} = 0$.
In the presence of the penguin contribution, 
direct $\CP$ violation may arise, 
and there is no straightforward interpretation 
of $S_{b \to u\bar u d}$ and $C_{b \to u\bar u d}$.
An isospin analysis~\cite{ref:cp_uta:uud:gronaulondon} 
can be used to disentangle the contributions and extract $\alpha$.

\begin{figure}[htb]
  \begin{center}
    \begin{tabular}{cc}
      \resizebox{0.46\textwidth}{!}{
        \includegraphics{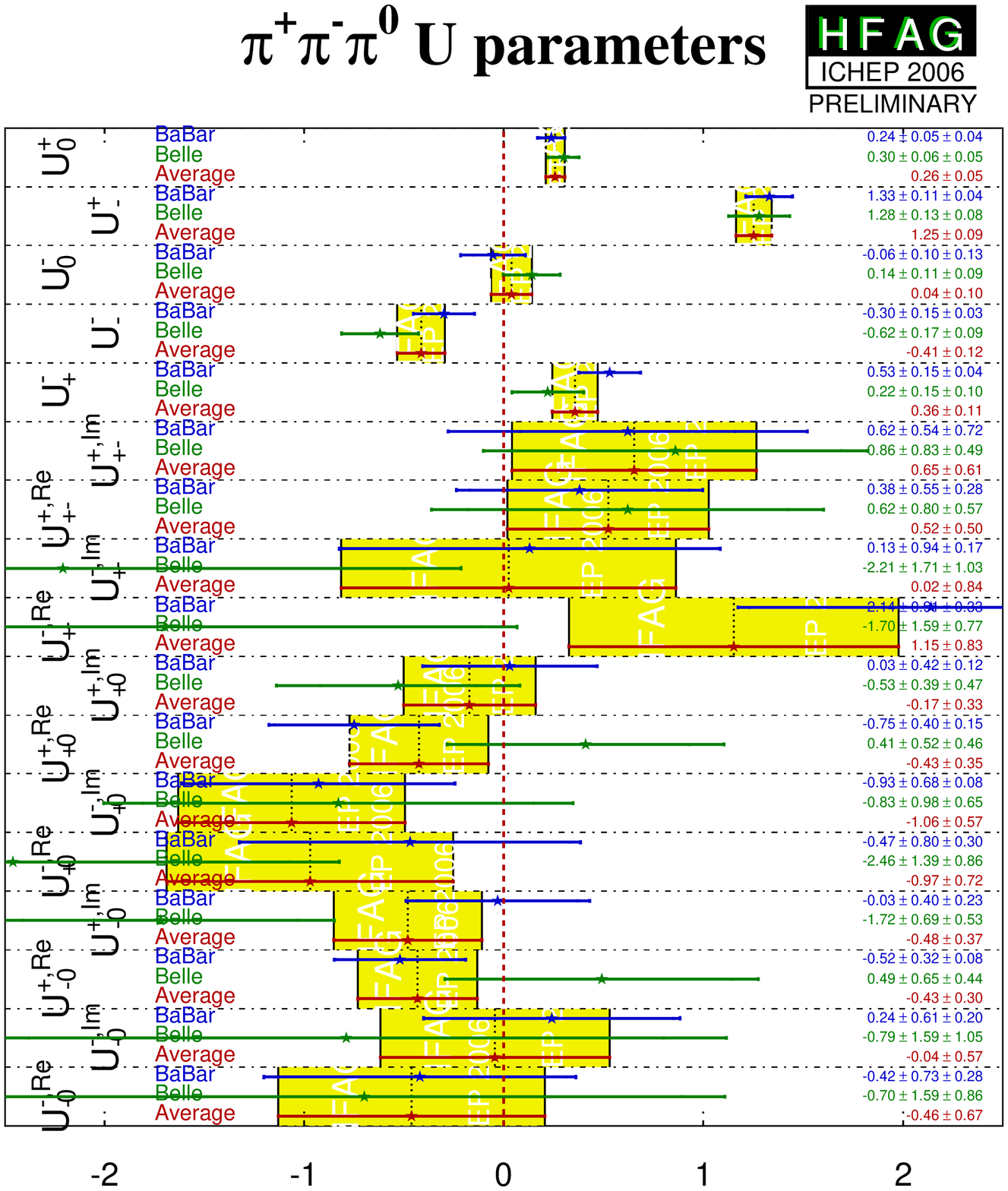}
      }
      &
      \resizebox{0.46\textwidth}{!}{
        \includegraphics{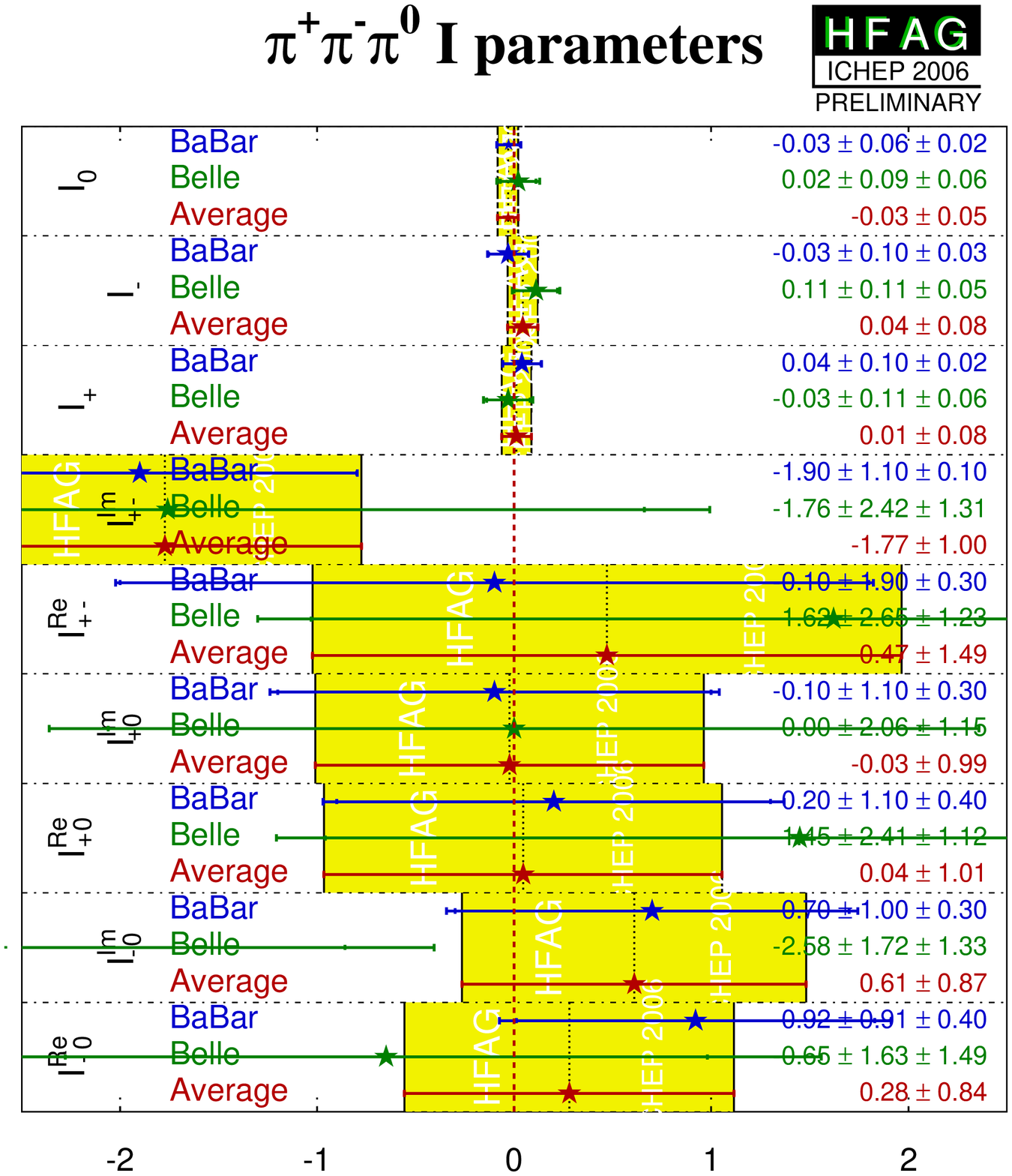}
      }
    \end{tabular}
  \end{center}
  \vspace{-0.8cm}
  \caption{
    Summary of the $U$ and $I$ parameters measured in the 
    time-dependent $\Bz \to \pi^+\pi^-\pi^0$ Dalitz plot analysis.
  }
  \label{fig:cp_uta:uud:uandi}
\end{figure}

For the non-$\CP$ eigenstate $\rho^{\pm}\pi^{\mp}$, 
both \babar~\cite{ref:cp_uta:uud:babar:pipipi0} 
and \belle~\cite{ref:cp_uta:uud:belle:pipipi0} have performed 
time-dependent Dalitz plot (DP) analyses
of the $\pi^+\pi^-\pi^0$ final state~\cite{ref:cp_uta:uud:snyderquinn};
such analyses allow direct measurements of the phases.
Both experiments have measured the $U$ and $I$ parameters discussed in 
Sec.~\ref{sec:cp_uta:notations:dalitz:pipipi0} and defined in 
Table~\ref{tab:cp_uta:pipipi0:uandi}.
We have performed a full correlated average of these parameters,
the results of which are summarized in Fig.~\ref{fig:cp_uta:uud:uandi}.

Both experiments have also extracted the Q2B parameters.
We have performed a full correlated average of these parameters,
which is equivalent to determining the values from the 
averaged $U$ and $I$ parameters.
The results are shown in Table.~\ref{tab:cp_uta:uud:rhopi_q2b}.
Note that only \belle\ has extracted the Q2B $\CP$ violation parameters
for $\Bz \to \rho^0\pi^0$.

\input{tables/cp_uta/pipipi0_q2b.tex}

\begin{figure}[htb]
  \begin{center}
    \resizebox{0.46\textwidth}{!}{
      \includegraphics{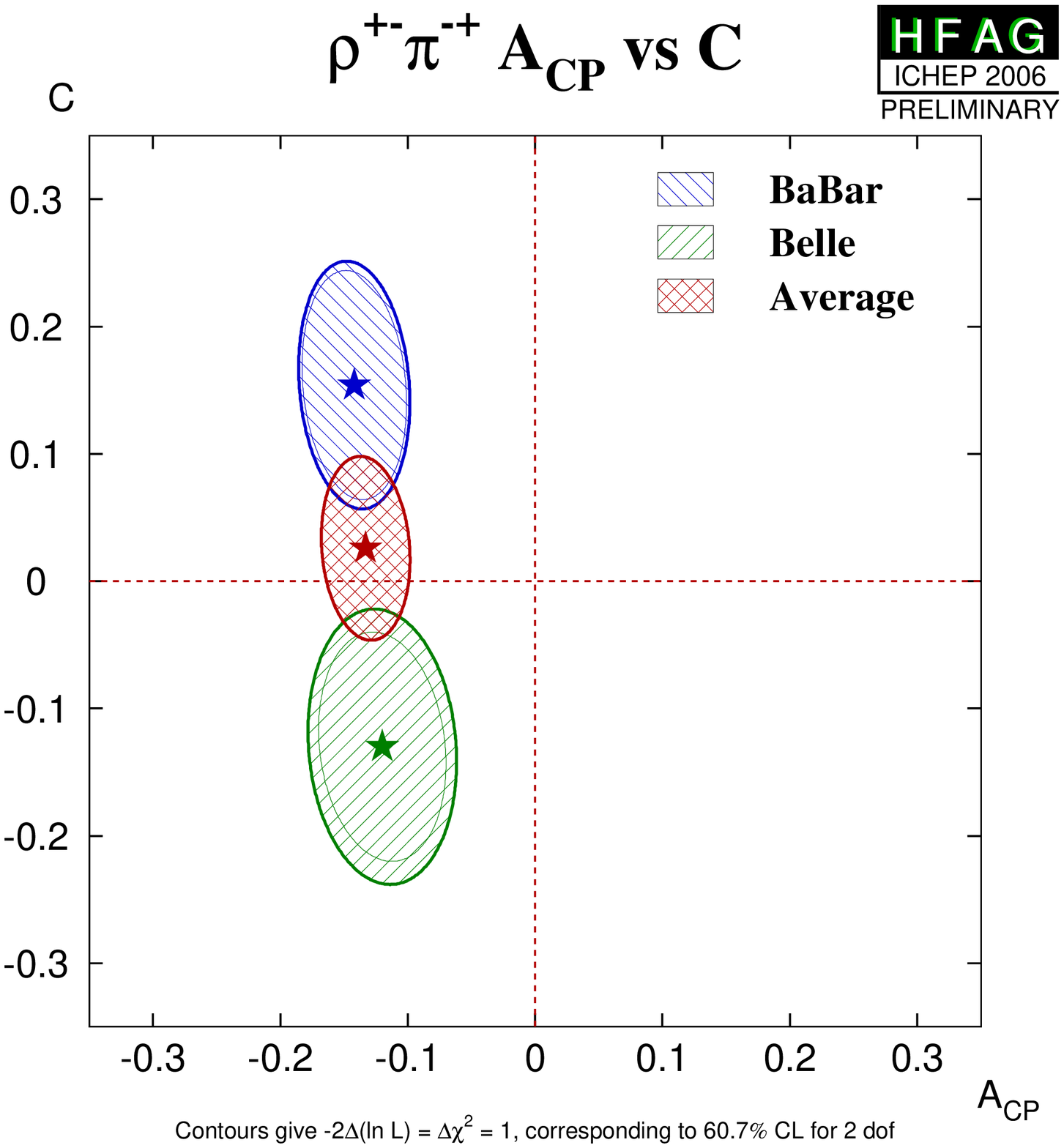}
    }
    \hfill
    \resizebox{0.46\textwidth}{!}{
      \includegraphics{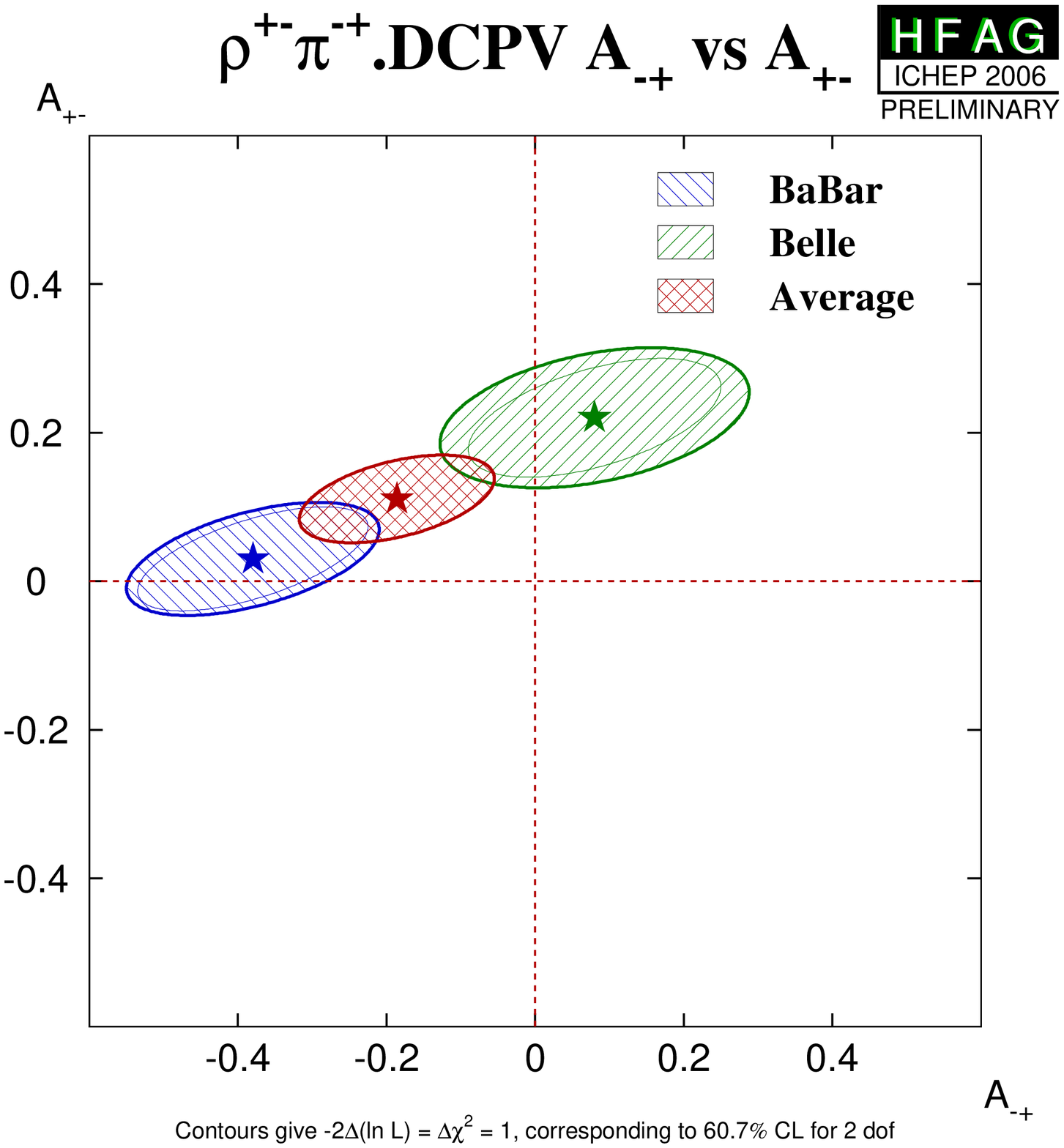}
    }    
  \end{center}
  \vspace{-0.8cm}
  \caption{
    Direct $\CP$ violation in $\Bz\to\rho^\pm\pi^\mp$.
    (Left) ${\cal A}^{\rho\pi}_{\CP}$ \vs\ $C_{\rho\pi}$ space,
    (right) ${\cal A}^{-+}_{\rho\pi}$ \vs\ ${\cal A}^{+-}_{\rho\pi}$ space.
  }
  \label{fig:cp_uta:uud:rhopi-dircp}
\end{figure}

With the notation described in Sec.~\ref{sec:cp_uta:notations}
(Eq.~(\ref{eq:cp_uta:non-cp-s_and_deltas})), 
the time-dependent parameters for the Q2B $\Bz \to \rho^\pm\pi^\mp$ analysis are,
neglecting penguin contributions, given by
\begin{equation}
  S_{\rho\pi} = 
  \sqrt{1 - \left(\frac{\Delta C}{2}\right)^2}\sin(2\alpha)\cos(\delta)
  \ , \ \ \ 
  \Delta S_{\rho\pi} = 
  \sqrt{1 - \left(\frac{\Delta C}{2}\right)^2}\cos(2\alpha)\sin(\delta)
\end{equation} 
and $C_{\rho\pi} = {\cal A}_{\CP}^{\rho\pi} = 0$,
where $\delta=\arg(A_{-+}A^*_{+-})$ is the strong phase difference 
between the $\rho^-\pi^+$ and $\rho^+\pi^-$ decay amplitudes.
In the presence of the penguin contribution, there is no straightforward 
interpretation of the Q2B observables in the $\Bz \to \rho^\pm\pi^\mp$ system
in terms of CKM parameters.
However direct $\CP$ violation may arise,
resulting in either or both of $C_{\rho\pi} \neq 0$ and ${\cal A}_{\CP}^{\rho\pi} \neq 0$.
Equivalently,
direct $\CP$ violation may be seen by either of
the decay-type-specific observables ${\cal A}^{+-}_{\rho\pi}$ 
and ${\cal A}^{-+}_{\rho\pi}$, defined in Eq.~(\ref{eq:cp_uta:non-cp-directcp}), 
deviating from zero.
Results and averages for these parameters
are also given in Table~\ref{tab:cp_uta:uud:rhopi_q2b}.
Averages of the direct $\CP$ violation effect in $\Bz \to \rho^\pm\pi^\mp$
are shown in Fig.~\ref{fig:cp_uta:uud:rhopi-dircp},
both in 
${\cal A}^{\rho\pi}_{\CP}$ \vs\ $C_{\rho\pi}$ space and in 
${\cal A}^{-+}_{\rho\pi}$ \vs\ ${\cal A}^{+-}_{\rho\pi}$ space.

Some difference is seen between the 
\babar\ and \belle\ measurements in the $\pi^+\pi^-$ system.
The confidence level of the average is $0.024$,
which corresponds to a $2.3\sigma$ discrepancy.  Since there is no
evidence of systematic problems in either analysis,
we do not rescale the errors of the averages.
The averages for $S_{b \to u\bar u d}$ and $C_{b \to u\bar u d}$ 
in $\Bz \to \pi^+\pi^-$ are both more than $5\sigma$ away from zero,
suggesting that both mixing-induced and direct $\CP$ violation 
are well-established in this channel.
Nonetheless, due to the possible discrepancy mentioned above,
only a cautious interpretation should be made 
on the significance of direct $\CP$ violation.

In $\Bz \to \rho^\pm\pi^\mp$, however,
both experiments see an indication of direct $\CP$ violation in the 
${\cal A}^{\rho\pi}_{\CP}$ parameter 
(as seen in Fig.~\ref{fig:cp_uta:uud:rhopi-dircp}).
The average is more than $3\sigma$ from zero,
providing evidence of direct $\CP$ violation in this channel.

\vspace{3ex}

\noindent
\underline{\large Constraints on $\alpha$}

The precision of the measured $\CP$ violation parameters in
$b \to u\bar{u}d$ transitions allows constraints to be set on the UT angle $\alpha$. 
In addition to the value of $\alpha$ from the \babar\ time-dependent DP analysis,
given in Table~\ref{tab:cp_uta:uud},
constraints have been obtained with various methods:
\begin{itemize}\setlength{\itemsep}{0.5ex}
\item 
  Both \babar~\cite{ref:cp_uta:uud:babar:pipi}
  and  \belle~\cite{ref:cp_uta:uud:belle:pipi} have performed 
  isospin analyses in the $\pi\pi$ system.
  \belle\ exclude $9^\circ < \phi_2 < 81^\circ$ at the $95.4\%$  C.L. while
  \babar\ give a confidence level interpretation for $\alpha$. 
  In both cases, only solutions in $0^\circ$--$180^\circ$ are considered.

\item
  Both experiments have also performed isospin analyses in the $\rho\rho$ system.
  \babar~\cite{ref:cp_uta:uud:babar:rhorho} find
  $\alpha \in \left[74, 117\right]^\circ$ at $68\%$ C.L.
  while \belle~\cite{ref:cp_uta:uud:belle:rhorho} obtain 
  $\phi_2 = (88 \pm 17)^\circ$ or $59^\circ < \phi_2 < 117^\circ$ 
  at $90\%$ confidence level. 
  The largest contribution to the uncertainty is due to the 
  possible penguin contribution, limited by the knowledge of the 
  $\Bz \to \rho^0\rho^0$ branching fraction~\cite{ref:cp_uta:uud:babar:rho0rho0},
  and is correlated between the measurements.  

\item
  The time-dependent Dalitz plot analysis of the $\Bz \to \pi^+\pi^-\pi^0$
  decay allows a determination of $\alpha$ without input from any other 
  channels.
  \babar~\cite{ref:cp_uta:uud:babar:pipipi0} obtain the constraint 
  $75^\circ < \alpha < 152^\circ$ at $68\%$ C.L.
  \belle~\cite{ref:cp_uta:uud:belle:pipipi0} have performed a similar analysis,
  and in addition have included information from the SU(2) partners of 
  $B \to \rho\pi$, which can be used to constrain $\alpha$
  via an isospin pentagon relation~\cite{ref:cp_uta:uud:lnqs}. 
  With this analysis,
  \belle\ obtain the tighter constraint $\phi_2 = (83 \, ^{+12}_{-23})^\circ$
  (where the errors correspond to $1\sigma$, \ie\ $68.3\%$ confidence level.

\item 
  Each experiment has obtained a value of $\alpha$ from combining its 
  results in the different $b \to u \bar{u} d$ modes 
  (with some input also from HFAG).
  These values have appeared in talks, but not in publications,
  and are not listed here.

\item 
  The CKMfitter~\cite{ref:cp_uta:ckmfitter} and 
  UTFit~\cite{ref:cp_uta:utfit} groups use the measurements 
  from \belle\ and \babar\ given above
  with other branching fractions and \CP asymmetries in 
  $\B\to\pi\pi$, $\rho\pi$ and $\rho\rho$ modes, 
  to perform isospin analyses for each system, 
  and to make combined constraints on $\alpha$.
\end{itemize}

Note that methods based on isospin symmetry make extensive use of 
measurements of branching fractions and direct $\CP$ asymmetries,
as averaged by the HFAG Rare Decays subgroup (Sec.~\ref{sec:rare}).
Note also that each method suffers from discrete ambiguities in the solutions.
The model assumption in the $\Bz \to \pi^+\pi^-\pi^0$ analysis 
allows to resolve some of the multiple solutions, 
and results in a single preferred value for $\alpha$ in $\left[ 0, \pi \right]$.
All the above measurements correspond to the choice
that is in agreement with the global CKM fit.

At present we make no attempt to provide an HFAG average for $\alpha$.
More details on procedures to calculate a best fit value for $\alpha$ 
can be found in Refs.~\cite{ref:cp_uta:ckmfitter,ref:cp_uta:utfit}.

\mysubsection{Time-dependent $\CP$ asymmetries in $b \to c\bar{u}d / u\bar{c}d$ transitions
}
\label{sec:cp_uta:cud}

Non-$\CP$ eigenstates such as $D^\pm\pi^\mp$, $D^{*\pm}\pi^\mp$ and $D^\pm\rho^\mp$ can be produced 
in decays of $\Bz$ mesons either via Cabibbo favoured ($b \to c$) or
doubly Cabibbo suppressed ($b \to u$) tree amplitudes. 
Since no penguin contribution is possible,
these modes are theoretically clean.
The ratio of the magnitudes of the suppressed and favoured amplitudes, $R$,
is sufficiently small (predicted to be about $0.02$),
that terms of ${\cal O}(R^2)$ can be neglected, 
and the sine terms give sensitivity to the combination of UT angles $2\beta+\gamma$.

\begin{figure}[htb]
  \begin{center}
    \begin{tabular}{cc}
      \resizebox{0.46\textwidth}{!}{
        \includegraphics{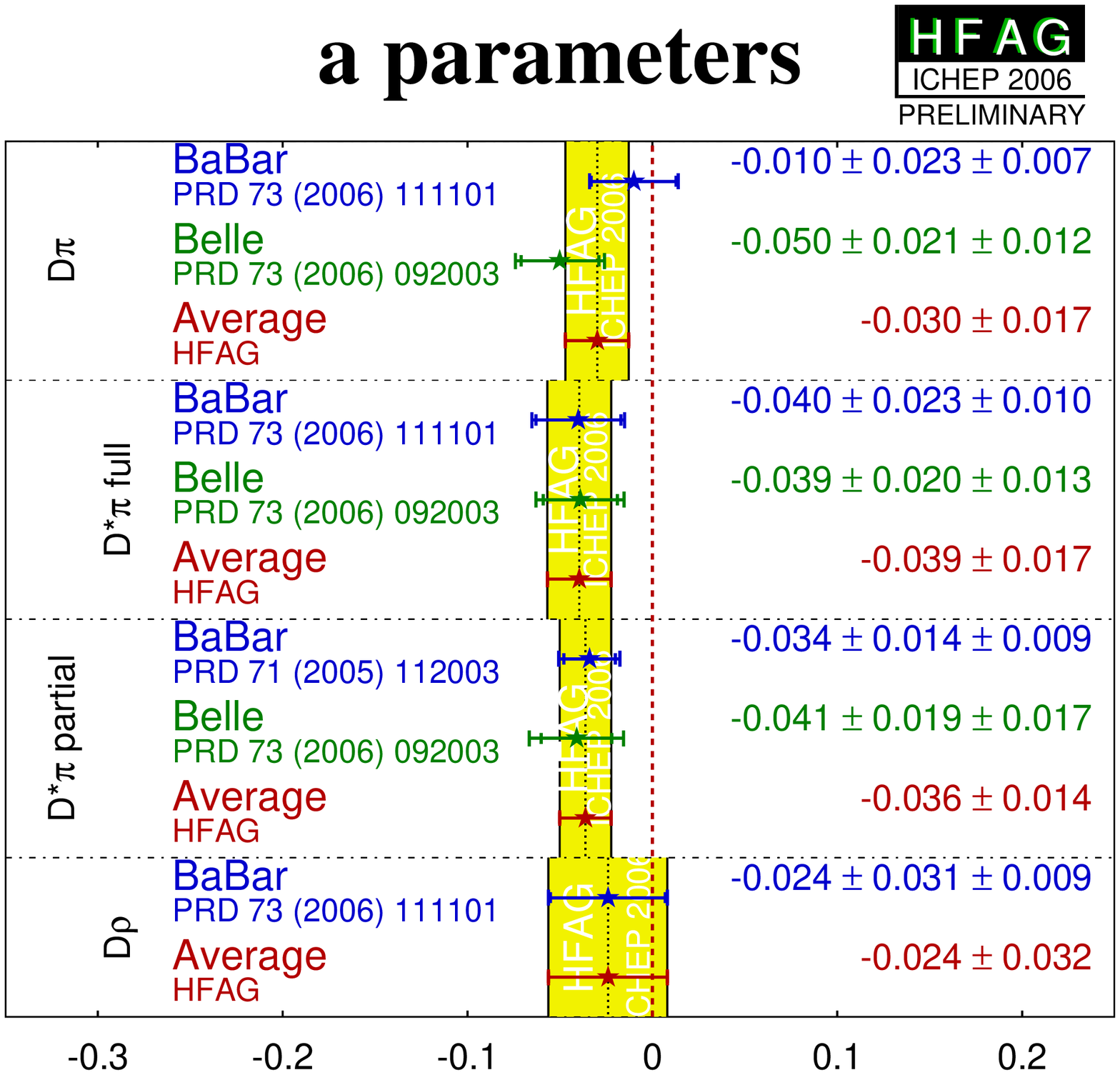}
      }
      &
      \resizebox{0.46\textwidth}{!}{
        \includegraphics{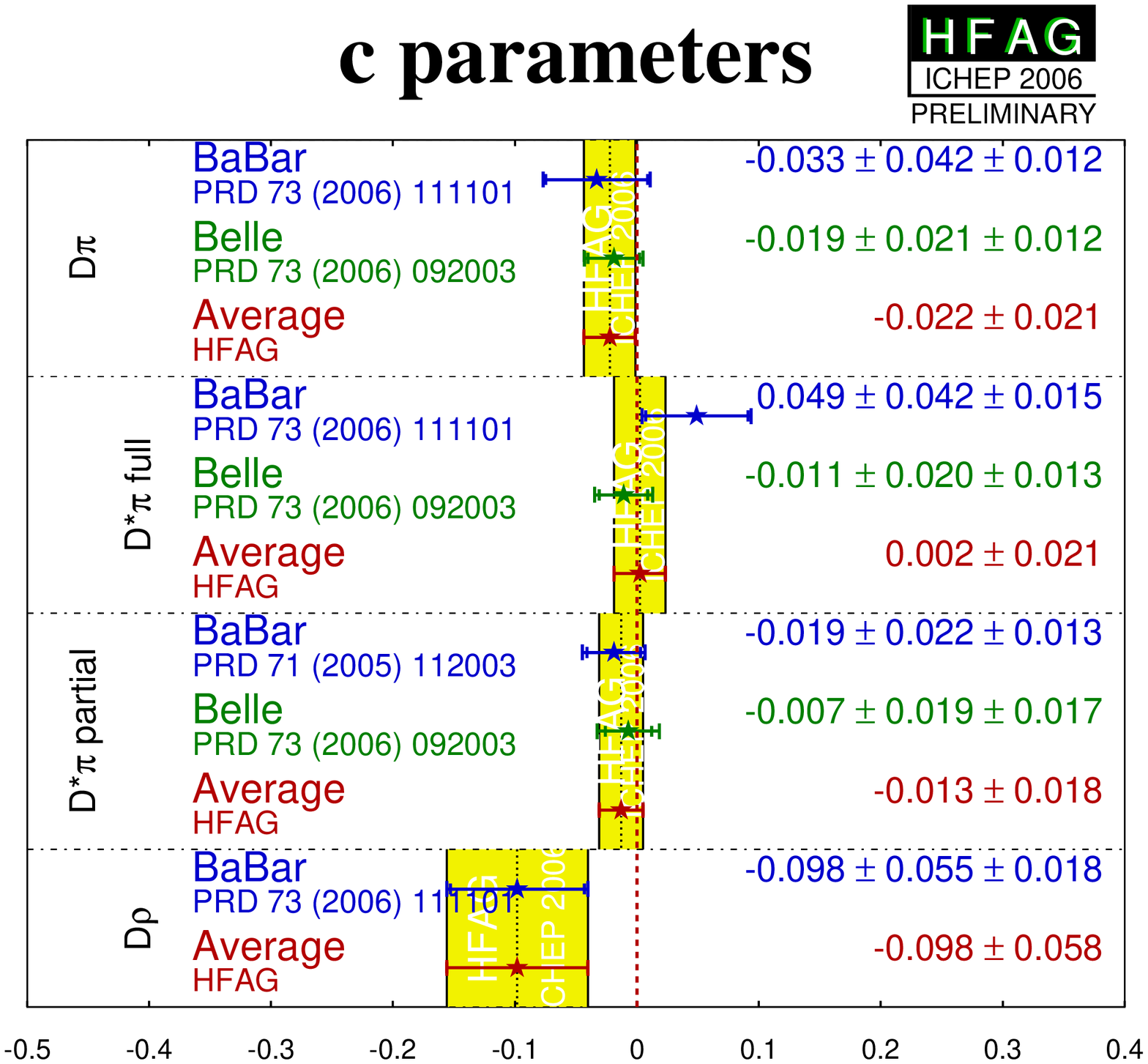}
      }
    \end{tabular}
  \end{center}
  \vspace{-0.8cm}
  \caption{
    Averages for $b \to c\bar{u}d / u\bar{c}d$ modes.
  }
  \label{fig:cp_uta:cud}
\end{figure}

\begin{figure}[htb]
  \begin{center}
    \begin{tabular}{cc}
      \resizebox{0.46\textwidth}{!}{
        \includegraphics{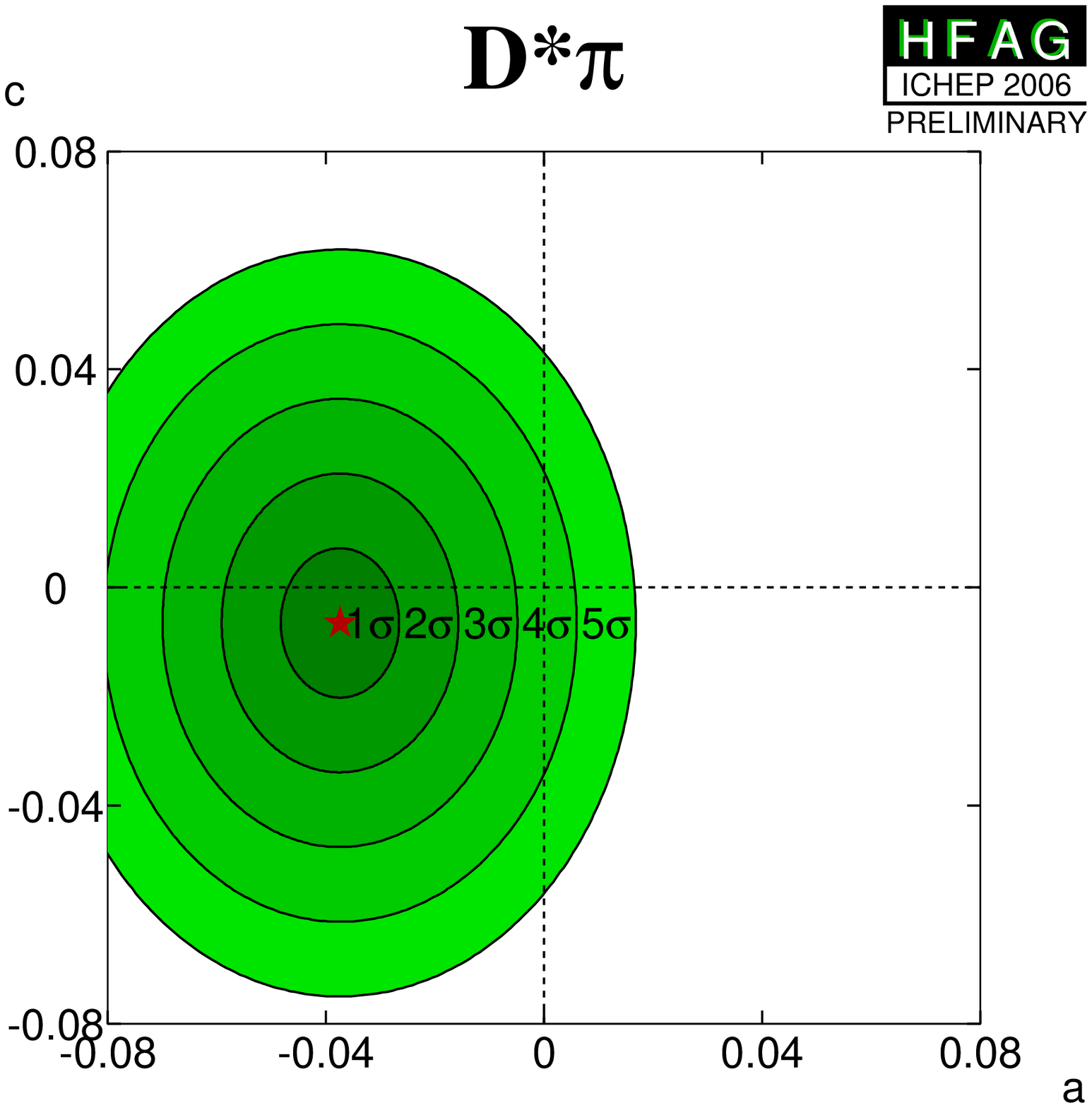}
      }
      &
      \resizebox{0.46\textwidth}{!}{
        \includegraphics{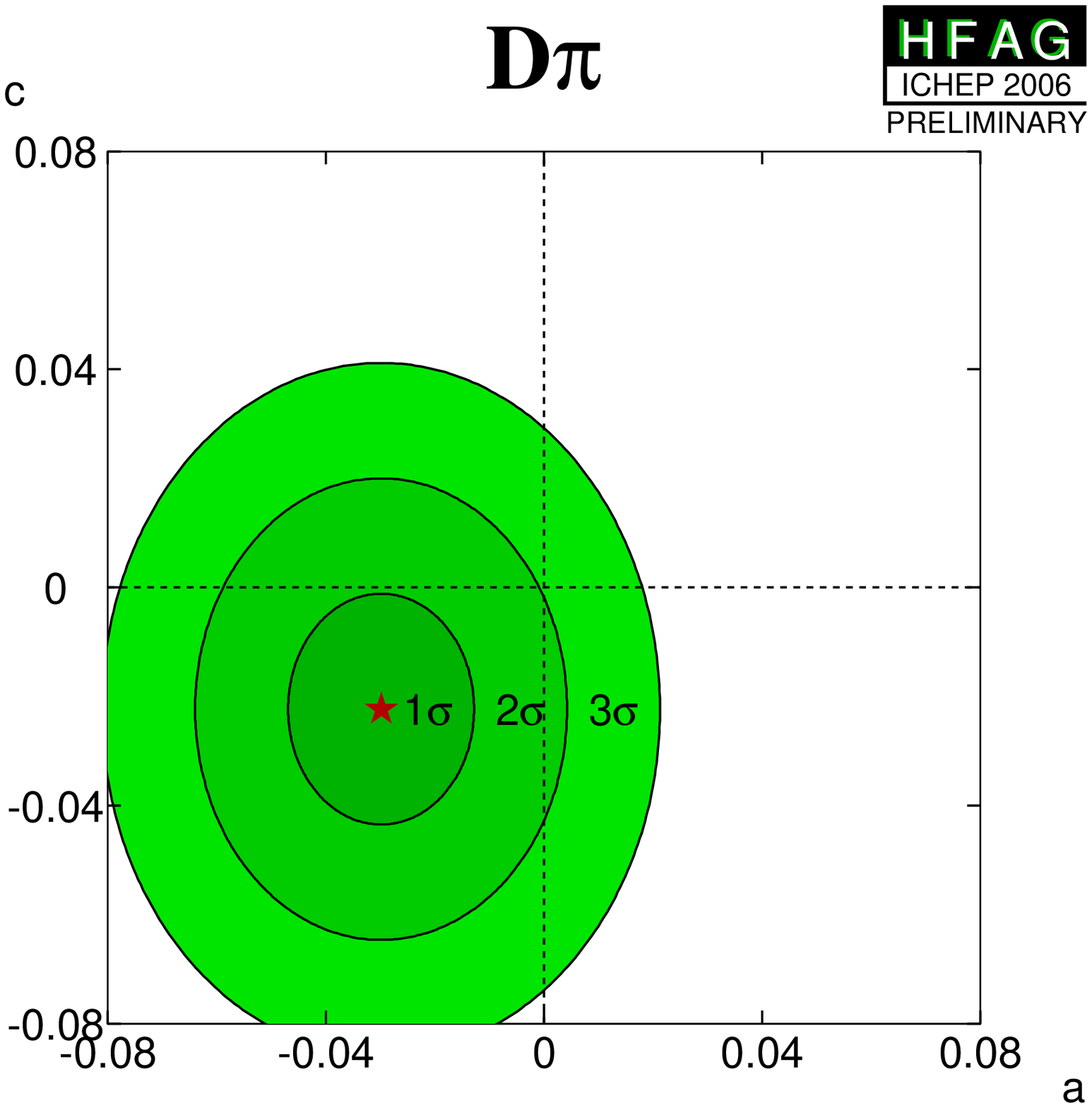}
      } \\
      \resizebox{0.46\textwidth}{!}{
        \includegraphics{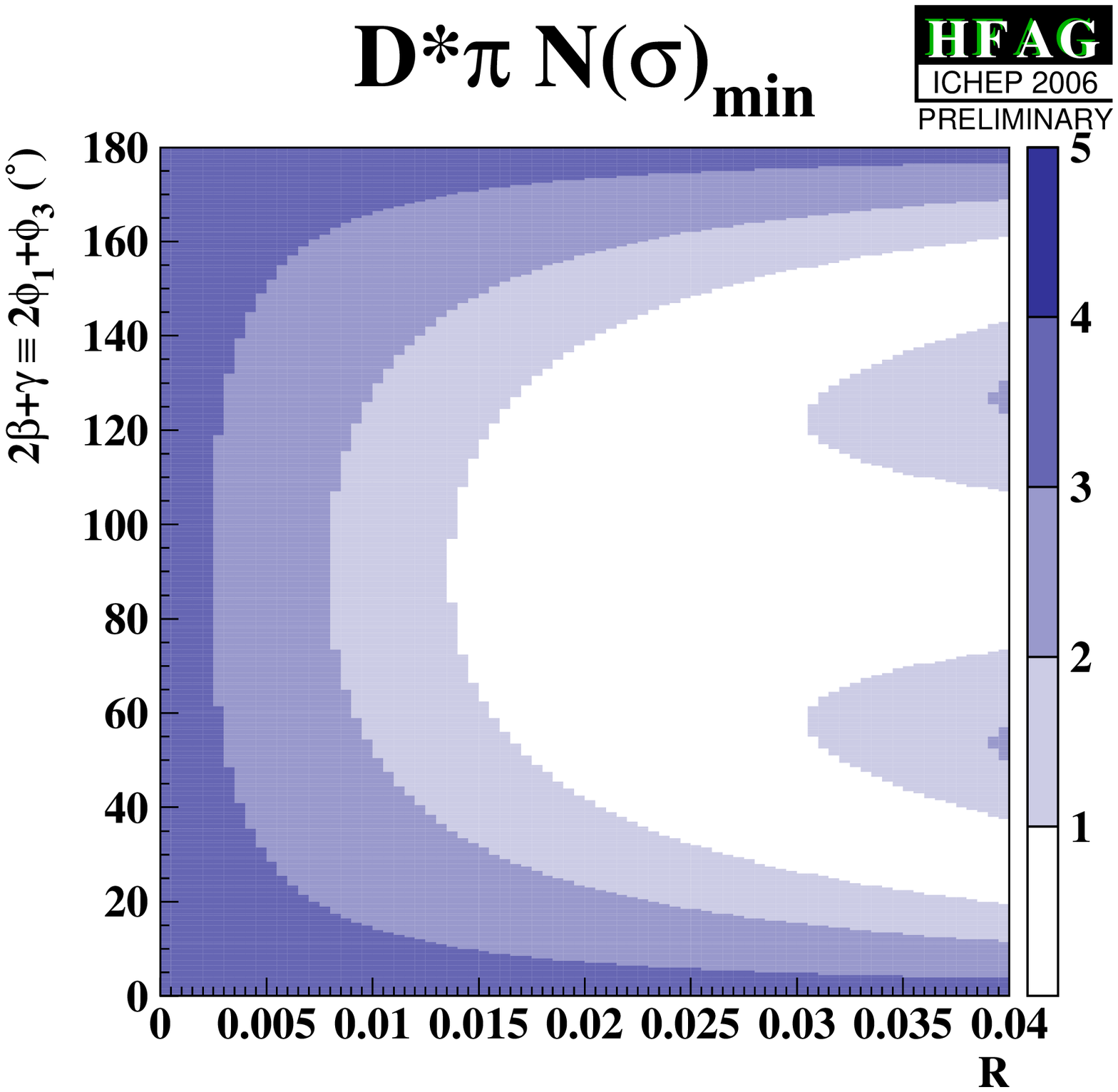}
      }
      &
      \resizebox{0.46\textwidth}{!}{
        \includegraphics{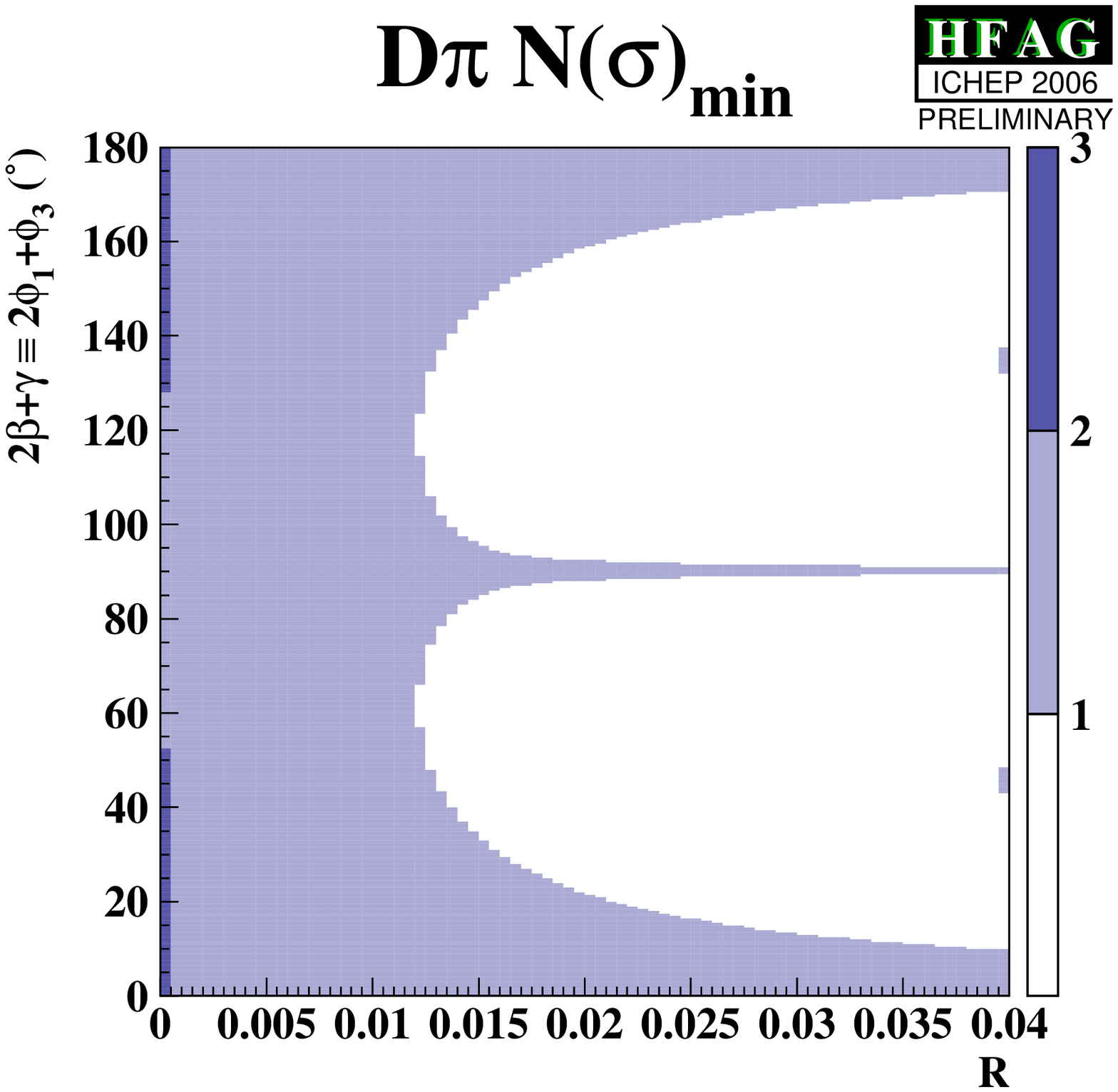}
      }          
    \end{tabular}
  \end{center}
  \vspace{-0.8cm}
  \caption{
    Results from $b \to c\bar{u}d / u\bar{c}d$ modes.
    (Top) Constraints in $c$ {\it vs.} $a$ space.
    (Bottom) Constraints in $2\beta+\gamma$ {\it vs.} $R$ space.
    (Left) $D^*\pi$ and (right) $D\pi$ modes.
  }
  \label{fig:cp_uta:cud_constraints}
\end{figure}

As described in Sec.~\ref{sec:cp_uta:notations:non_cp:dstarpi},
the averages are given in terms of parameters $a$ and $c$.
$\CP$ violation would appear as $a \neq 0$.
Results are available from both \babar\ and \belle\ in the modes
$D^\pm\pi^\mp$ and $D^{*\pm}\pi^\mp$; for the latter mode both experiments 
have used both full and partial reconstruction techniques.
(\babar\ have provided separate results with each technique,
while \belle\ have in addition provided a combined result.)
Results are also available from \babar\ using $D^\pm\rho^\mp$.
These results, and their averages, are listed in Table~\ref{tab:cp_uta:cud},
and are shown in Fig.~\ref{fig:cp_uta:cud}.
The constraints in $c$ \vs\ $a$ space for the $D\pi$ and $D^*\pi$ modes
are shown in Fig.~\ref{fig:cp_uta:cud_constraints}.
It is notable that the average value of $a$ from $D^*\pi$ is more than
$3\sigma$ from zero, providing evidence of $\CP$ violation in this channel.

\input{tables/cp_uta/cud.tex}

For each of $D\pi$, $D^*\pi$ and $D\rho$, 
there are two measurements ($a$ and $c$, or $S^+$ and $S^-$) 
which depend on three unknowns ($R$, $\delta$ and $2\beta+\gamma$), 
of which two are different for each decay mode. 
Therefore, there is not enough information to solve directly for $2\beta+\gamma$. 
However, for each choice of $R$ and $2\beta+\gamma$, 
one can find the value of $\delta$ that allows $a$ and $c$ to be closest 
to their measured values, 
and calculate the distance in terms of numbers of standard deviations.
(We currently neglect experimental correlations in this analysis.) 
These values of $N(\sigma)_{\rm min}$ can then be plotted 
as a function of $R$ and $2\beta+\gamma$
(and can trivially be converted to confidence levels). 
These plots are given for the $D\pi$ and $D^*\pi$ modes 
in Figure~\ref{fig:cp_uta:cud_constraints}; 
the uncertainties in the $D\rho$ mode are currently too large 
to give any meaningful constraint.

The constraints can be tightened if one is willing 
to use theoretical input on the values of $R$ and/or $\delta$. 
One popular choice is the use of SU(3) symmetry to obtain 
$R$ by relating the suppressed decay mode to $\B$ decays 
involving $D_s$ mesons. 
More details can be found 
in Refs.~\cite{ref:cp_uta:ckmfitter,ref:cp_uta:utfit}.

\mysubsection{Rates and asymmetries in $\Bmp \to \DorDstar K^{(*)\mp}$ decays
}
\label{sec:cp_uta:cus}

As explained in Sec.~\ref{sec:cp_uta:notations:cus},
rates and asymmetries in $\Bmp \to \DorDstar K^{(*)\mp}$ decays
are sensitive to $\gamma$.
Various methods using different $\DorDstar$ final states exist.

\mysubsubsection{$D$ decays to $\CP$ eigenstates
}
\label{sec:cp_uta:cus:glw}

\begin{figure}[htb]
  \begin{center}
    \begin{tabular}{cc}
      \resizebox{0.46\textwidth}{!}{
        \includegraphics{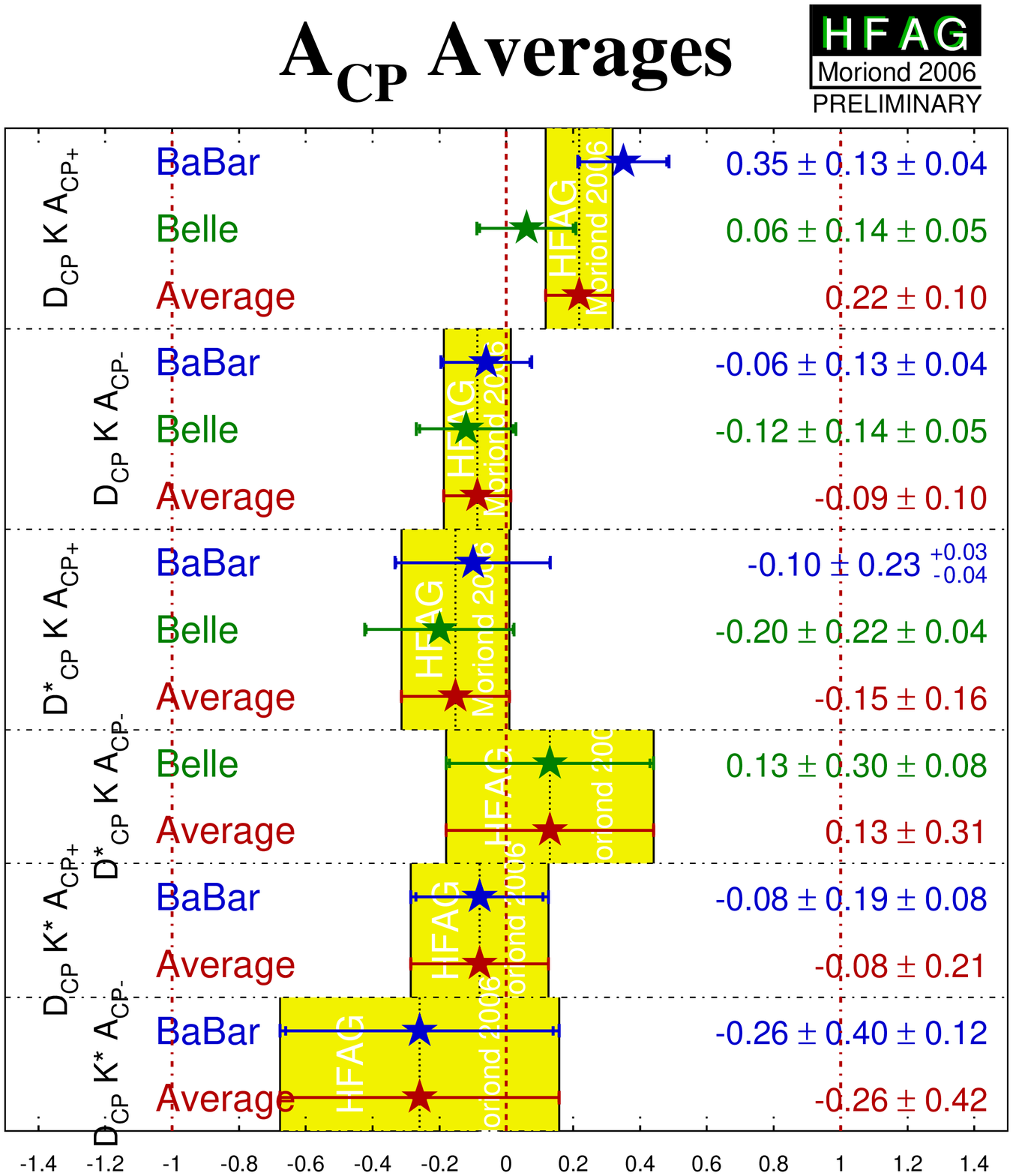}
      }
      &
      \resizebox{0.46\textwidth}{!}{
        \includegraphics{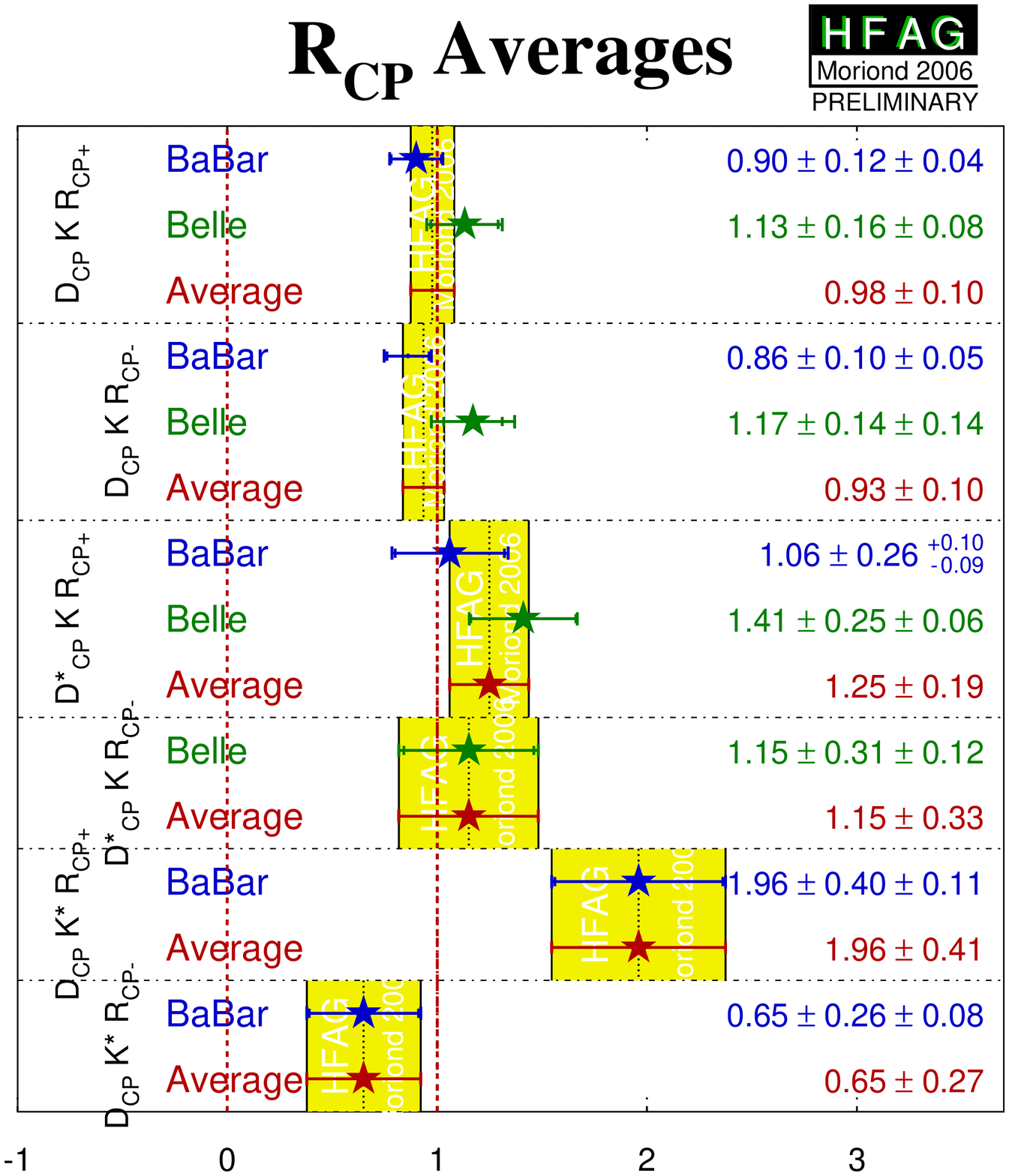}
      }
    \end{tabular}
  \end{center}
  \vspace{-0.8cm}
  \caption{
    Averages of $A_{\CP}$ and $R_{\CP}$ from GLW analyses.
  }
  \label{fig:cp_uta:cus:glw}
\end{figure}

Results are available from both \babar\ and \belle\ on GLW analyses
in the decay modes $\Bmp \to D\Kmp$, $\Bmp \to \Dstar\Kmp$ and $\Bmp \to D\Kstarmp$.
Both experiments use the 
$\CP$ even $D$ decay final states $K^+K^-$ and $\pi^+\pi^-$ in all three modes; 
both experiments also use only the $\Dstar \to D\pi^0$ decay, 
which gives $\CP(\Dstar) = \CP(D)$. 
For $\CP$ odd $D$ decay final states, 
\belle\ uses $\KS\pi^0$, $\KS\eta$ and $\KS\phi$ in all three analyses, 
and also use $\KS\omega$ in $D\Kmp$ and $\Dstar\Kmp$ analyses. 
\babar\ uses $\KS\pi^0$ only for $D\Kmp$ analysis; 
for $D\Kstarmp$ analysis they also use $\KS\phi$ and $\KS\omega$
(and assign an asymmetric systematic error due to $\CP$ even pollution 
in these $\CP$ odd channels~\cite{ref:cp_uta:cus:babar:dstarcpk}).
The results and averages are given in Table~\ref{tab:cp_uta:cus:glw}
and shown in Fig.~\ref{fig:cp_uta:cus:glw}.

\input{tables/cp_uta/cus_glw.tex}

\mysubsubsection{$D$ decays to suppressed final states}
\label{sec:cp_uta:cus:ads}

\begin{figure}[htb]
  \begin{center}
    \resizebox{0.46\textwidth}{!}{
      \includegraphics{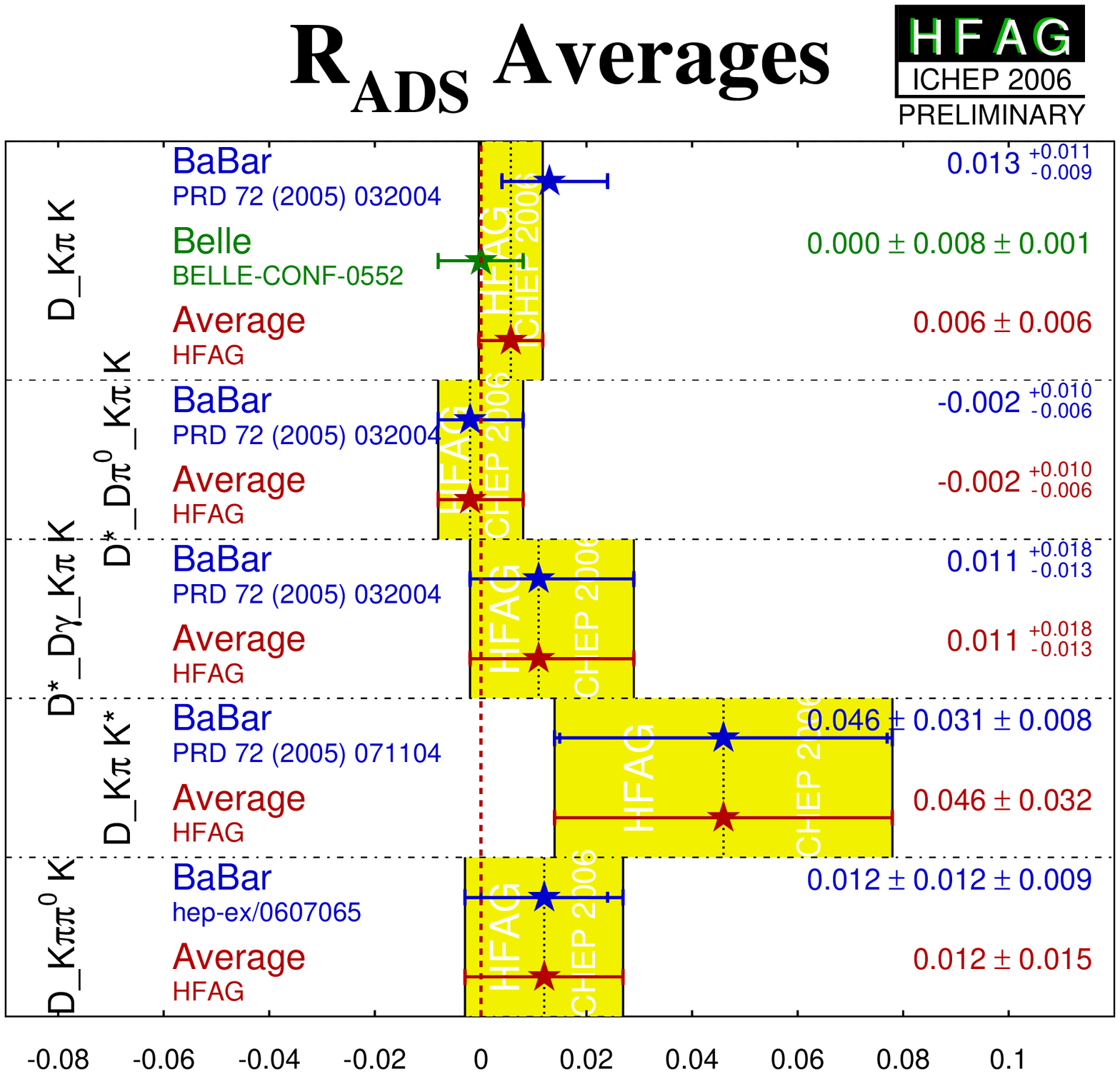}
    }
  \end{center}
  \vspace{-0.8cm}
  \caption{
    Averages of $R_{\rm ADS}$.
  }
  \label{fig:cp_uta:cus:ads}
\end{figure}

For ADS analysis, both \babar\ and \belle\ have studied
the mode $\Bmp \to D\Kmp$;
\belle\ has also studied $\Bmp \to D\pi^\mp$
and \babar\ has also analyzed the $\Bmp \to \Dstar\Kmp$ 
and $\Bmp \to D\Kstarmp$ modes
($\Dstar \to D\pi^0$ and $\Dstar \to D\gamma$ are studied separately;
$\Kstarmp$ is reconstructed as $\KS\pi^\mp$).
In all cases the suppressed decay $D \to K^+\pi^-$ has been used.
\babar\ also has results using $\Bmp \to D\Kmp$ with $D \to K^+\pi^-\pi^0$.
The results and averages are given in Table~\ref{tab:cp_uta:cus:ads}
and shown in Fig.~\ref{fig:cp_uta:cus:ads}.
Note that although no clear signals for these modes have yet been seen,
the central values are given.
In $\Bm \to \Dstar\Km$ decays there is an effective shift of $\pi$
in the strong phase difference between the cases that the $\Dstar$ is 
reconstructed as $D\pi^0$ and $D\gamma$~\cite{ref:cp_uta:cus:bg}.
As a consequence, the different $D^*$ decay modes are treated separately.

\input{tables/cp_uta/cus_ads.tex}

\mysubsubsection{$D$ decays to multiparticle self-conjugate final states}
\label{sec:cp_uta:cus:dalitz}

For the Dalitz plot analysis, both 
\babar~\cite{ref:cp_uta:cus:babar:dk_dalitz} and 
\belle~\cite{ref:cp_uta:cus:belle:dk_dalitz} have studied
the modes $\Bmp \to D\Kmp$, $\Bmp \to \Dstar\Kmp$ and $\Bmp \to D\Kstarmp$.
For $\Bmp \to \Dstar\Kmp$,
\belle\ has used only $\Dstar \to D\pi^0$,
while \babar\ has used both $\Dstar$ decay modes and 
taken the effective shift in the strong phase difference into account.
In all cases the decay $D \to \KS\pi^+\pi^-$ has been used.
Results and averages are given in Table~\ref{tab:cp_uta:cus:dalitz}.
The third error on each measurement is due to $D$ decay model uncertainty.

The parameters measured in the analyses are explained in
Sec.~\ref{sec:cp_uta:notations:cus}.
Both \babar\ and \belle\ have measured the ``Cartesian''
$(x_\pm,y_\pm)$ variables,
and perform frequentist statistical procedures,
to convert these into measurements of $\gamma$, $r_B$ and $\delta_B$.

Both experiments reconstruct $\Kstarmp$ as $\KS\pi^\mp$,
but the treatment of possible nonresonant $\KS\pi^\mp$ differs:
\belle\ assign an additional model uncertainty,
while \babar\ use a reparametrization 
suggested by Gronau~\cite{ref:cp_uta:cus:gronau}.
The parameters $r_B$ and $\delta_B$ are replaced with 
effective parameters $\kappa r_s$ and $\delta_s$;
no attempt is made to extract the true hadronic parameters 
of the $\Bmp \to D\Kstarmp$ decay.

We perform averages using the following procedure, 
which is based on a set of reasonable, though imperfect, assumptions.

\begin{itemize}\setlength{\itemsep}{0.5ex}
\item 
  It is assumed that both experiments use the same $D$ decay model. 
  Therefore, we do not rescale the results to a common model.
\item 
  It is further assumed that the model uncertainty is $100\%$ 
  correlated between experiments, 
  and therefore this source of error is not used in the averaging procedure.
\item 
  We include in the average the effect of correlations 
  within each experiments set of measurements.
\item 
  At present it is unclear how to assign an average model uncertainty. 
  We have not attempted to do so. 
  Our average includes only statistical and systematic error. 
  An unknown amount of model uncertainty should be added to the final error.
\item 
  We follow the suggestion of Gronau~\cite{ref:cp_uta:cus:gronau} 
  in making the $DK^*$ averages. 
  Explicitly, we assume that the selection of $K^{*\pm} \to \KS\pi^\pm$
  is the same in both experiments 
  (so that $\kappa$, $r_s$ and $\delta_s$ are the same), 
  and drop the additional source of model uncertainty 
  assigned by Belle due to possible nonresonant decays.
\item 
  We do not consider common systematic errors, 
  other than the $D$ decay model. 
\end{itemize}


\input{tables/cp_uta/cus_dalitz.tex}

\begin{figure}[htb]
  \begin{center}
    \resizebox{0.30\textwidth}{!}{
      \includegraphics{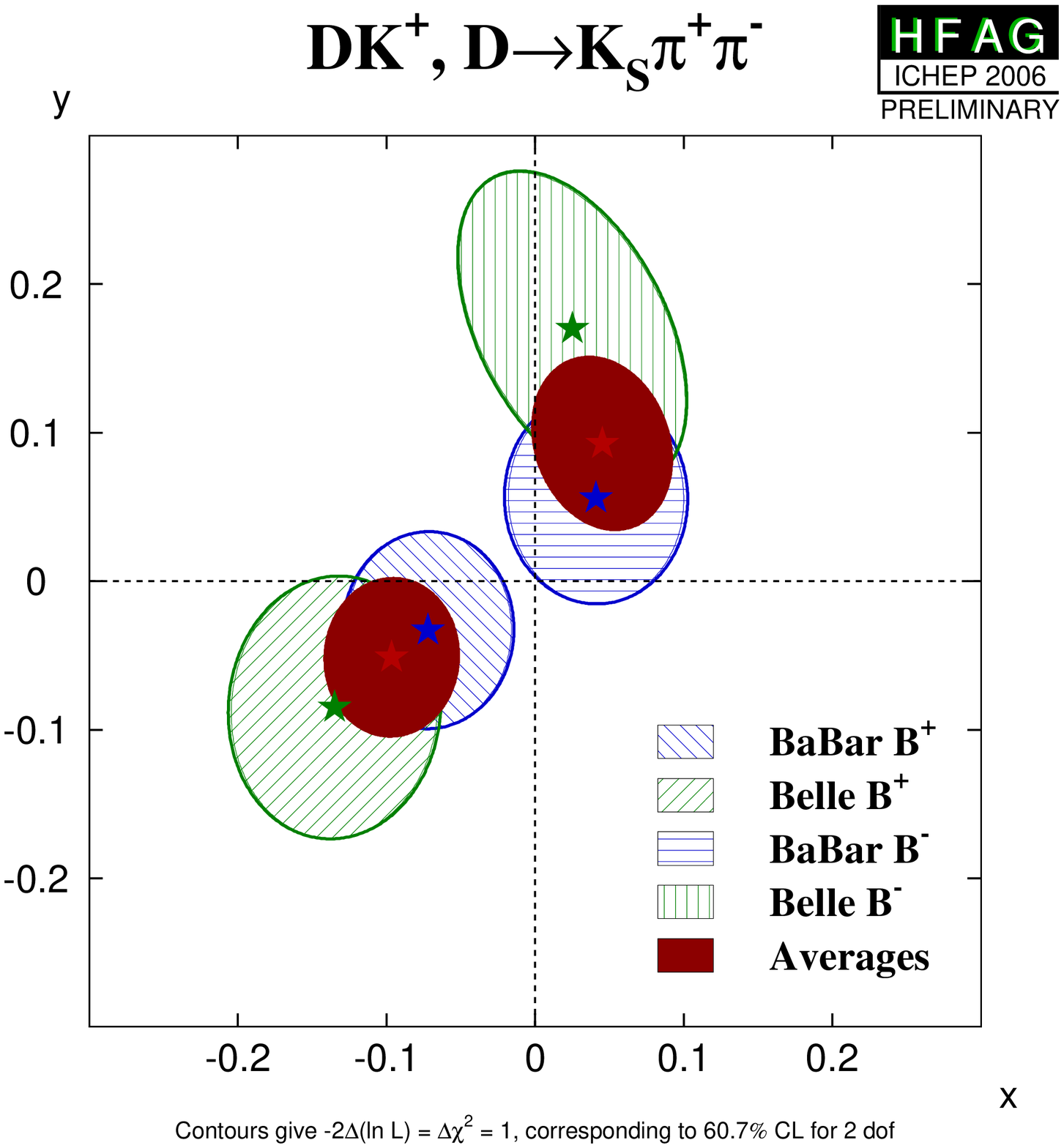}
    }
    \hfill
    \resizebox{0.30\textwidth}{!}{
      \includegraphics{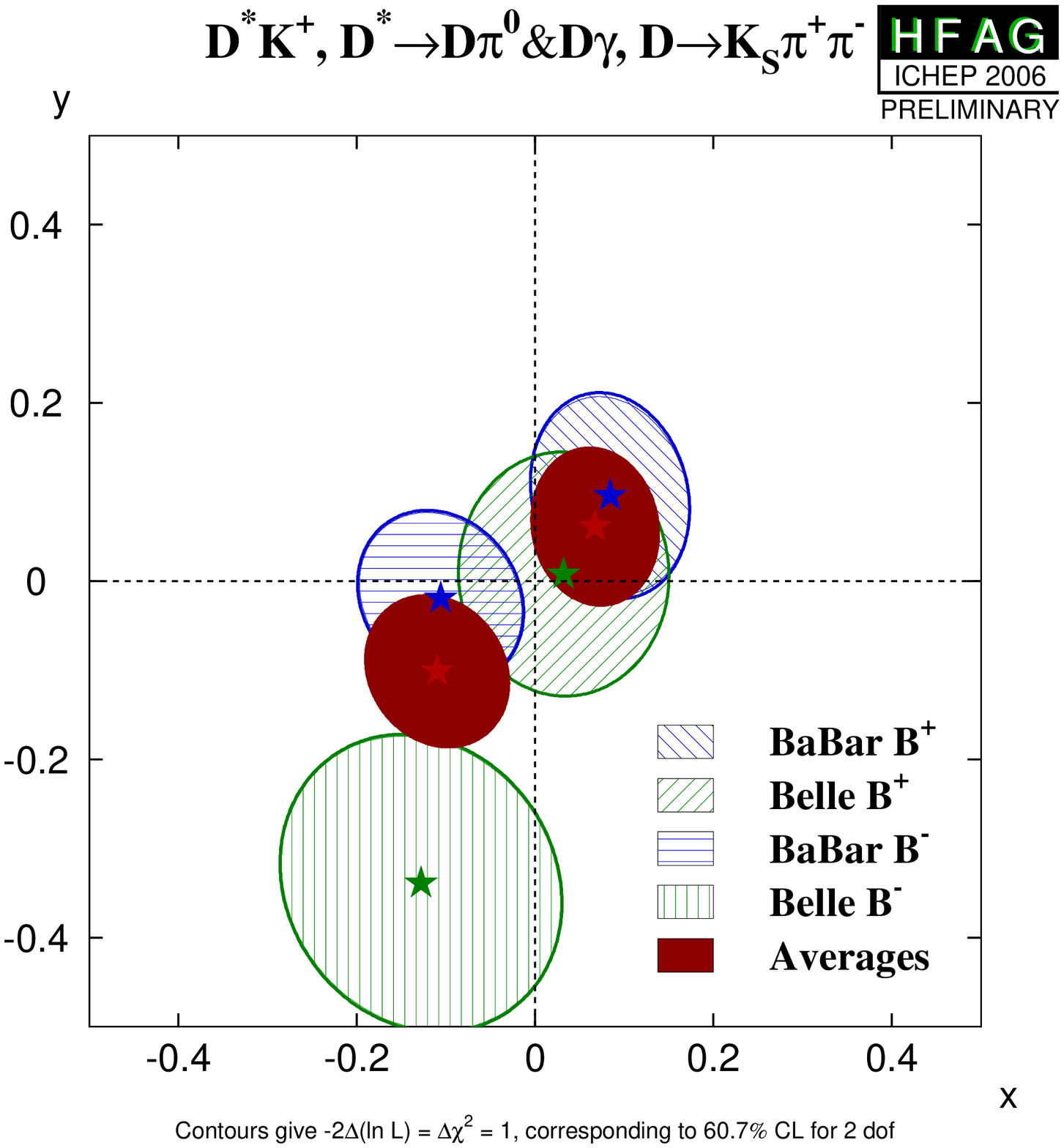}
    }
    \hfill
    \resizebox{0.30\textwidth}{!}{
      \includegraphics{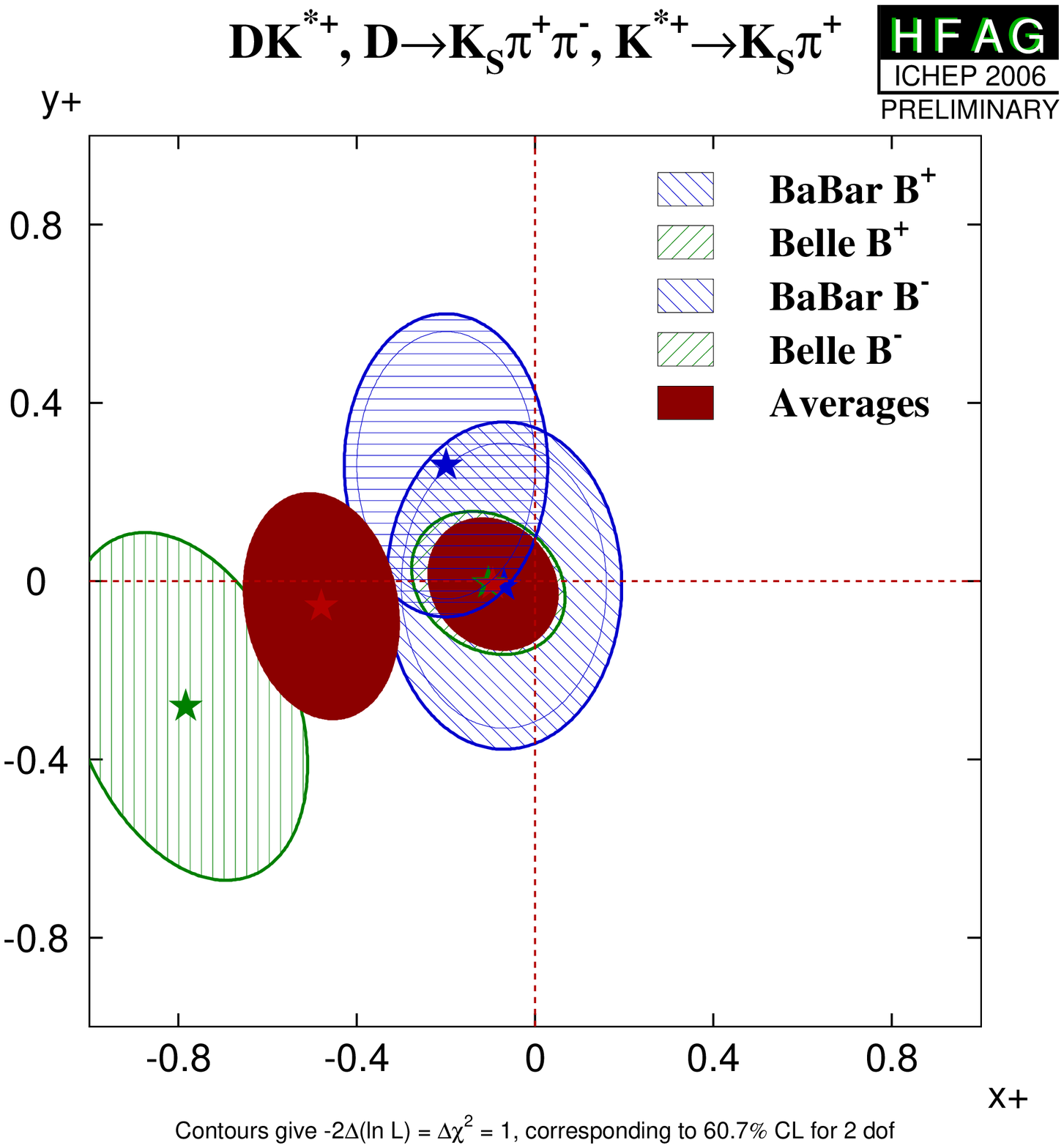}
    }
  \end{center}
  \vspace{-0.8cm}
  \caption{
    Contours in the $(x_\pm, y_\pm)$ from $\Bmp \to D^{(*)}K^{(*)\pm}$.
    (Left) $\Bmp \to D\Kmp$, 
    (middle) $\Bmp \to \Dstar\Kmp$,
    (right) $\Bmp \to D\Kstarmp$.
    Note that the uncertainities assigned to the averages given in these plots
    do not include model errors.        
  }
  \label{fig:cp_uta:cus:dalitz_2d}
\end{figure}

\begin{figure}[htb]
  \begin{center}
    \resizebox{0.40\textwidth}{!}{
      \includegraphics{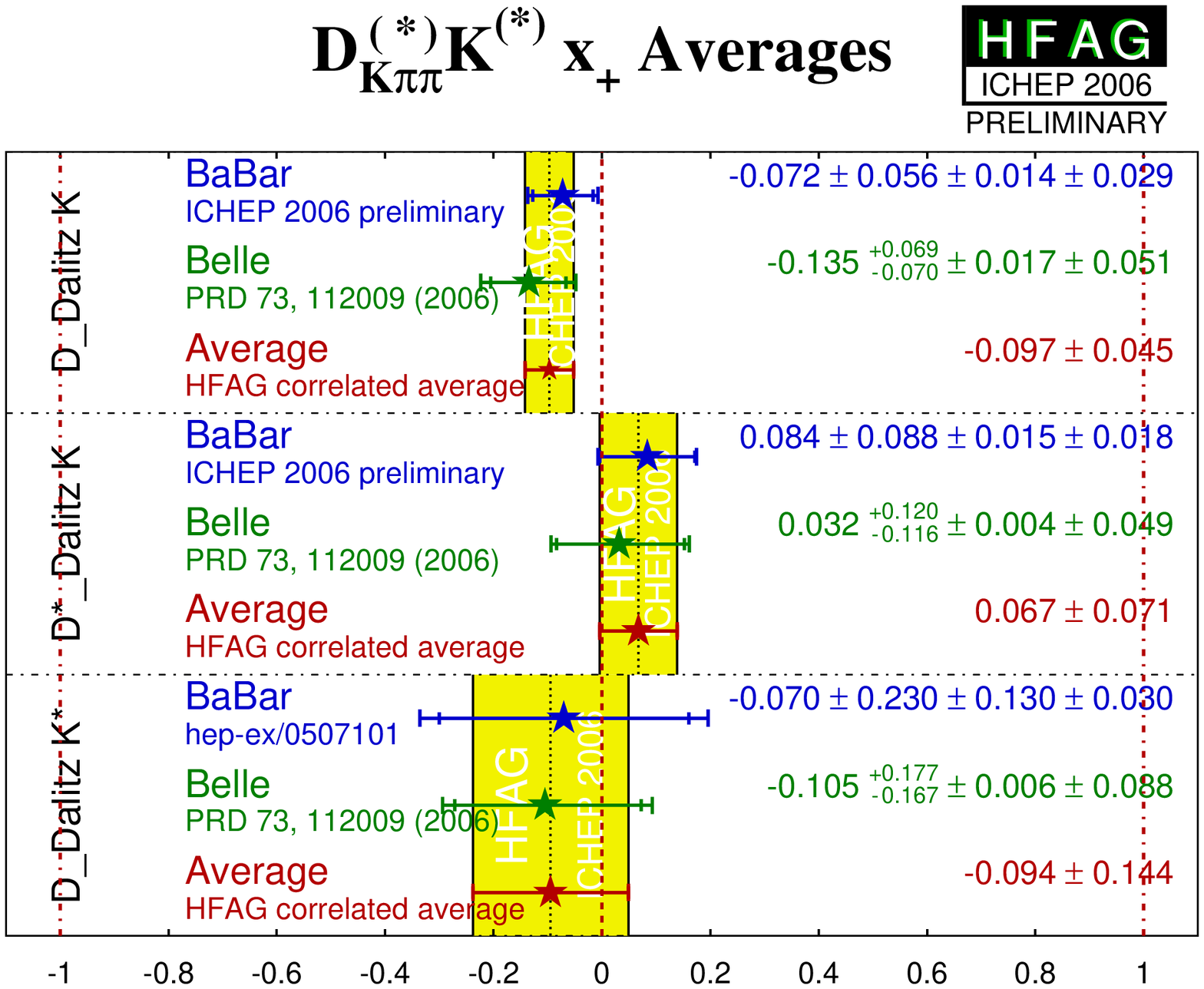}
    }
    \hspace{0.1\textwidth}
    \resizebox{0.40\textwidth}{!}{
      \includegraphics{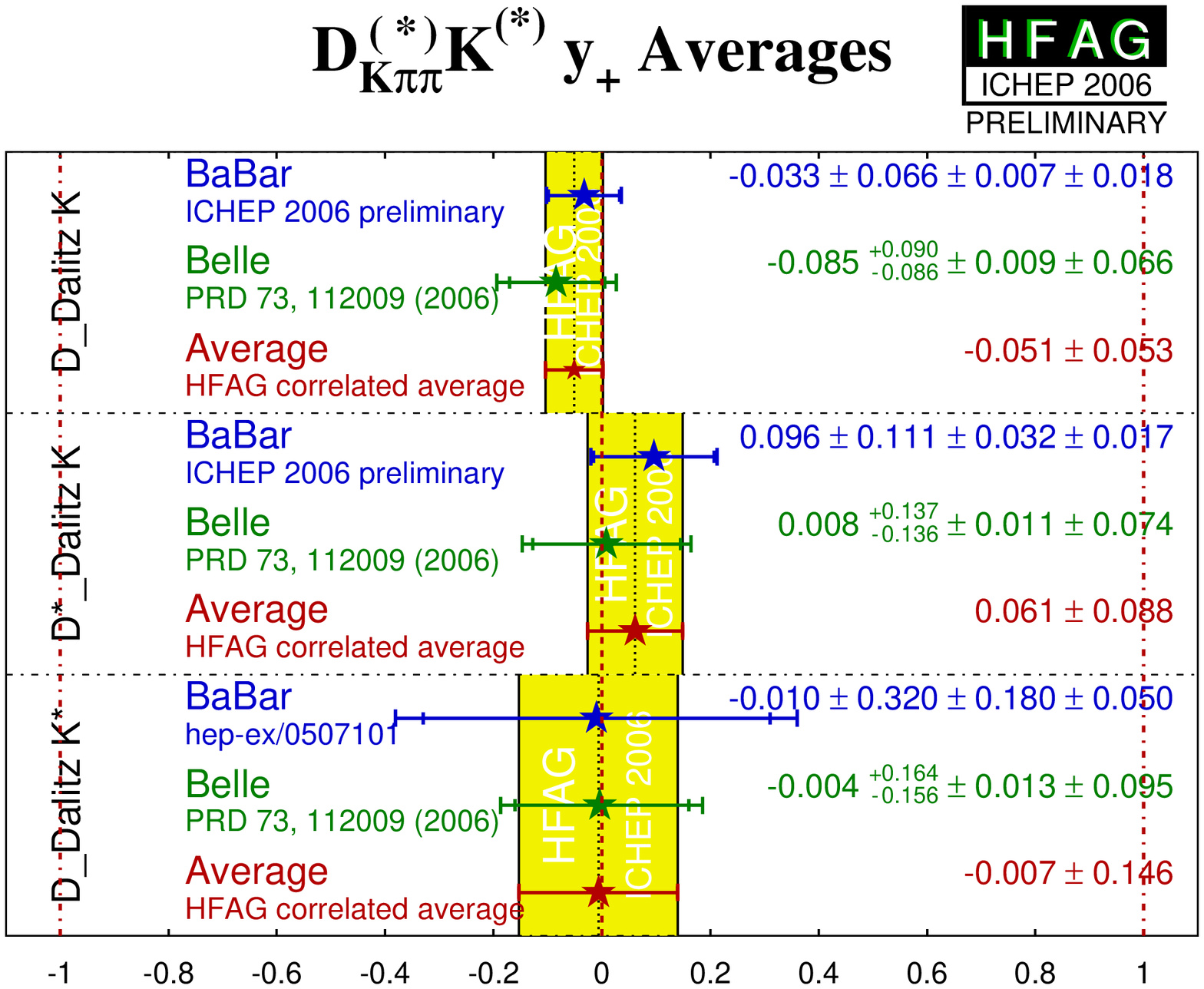}
    }
    \\
    \resizebox{0.40\textwidth}{!}{
      \includegraphics{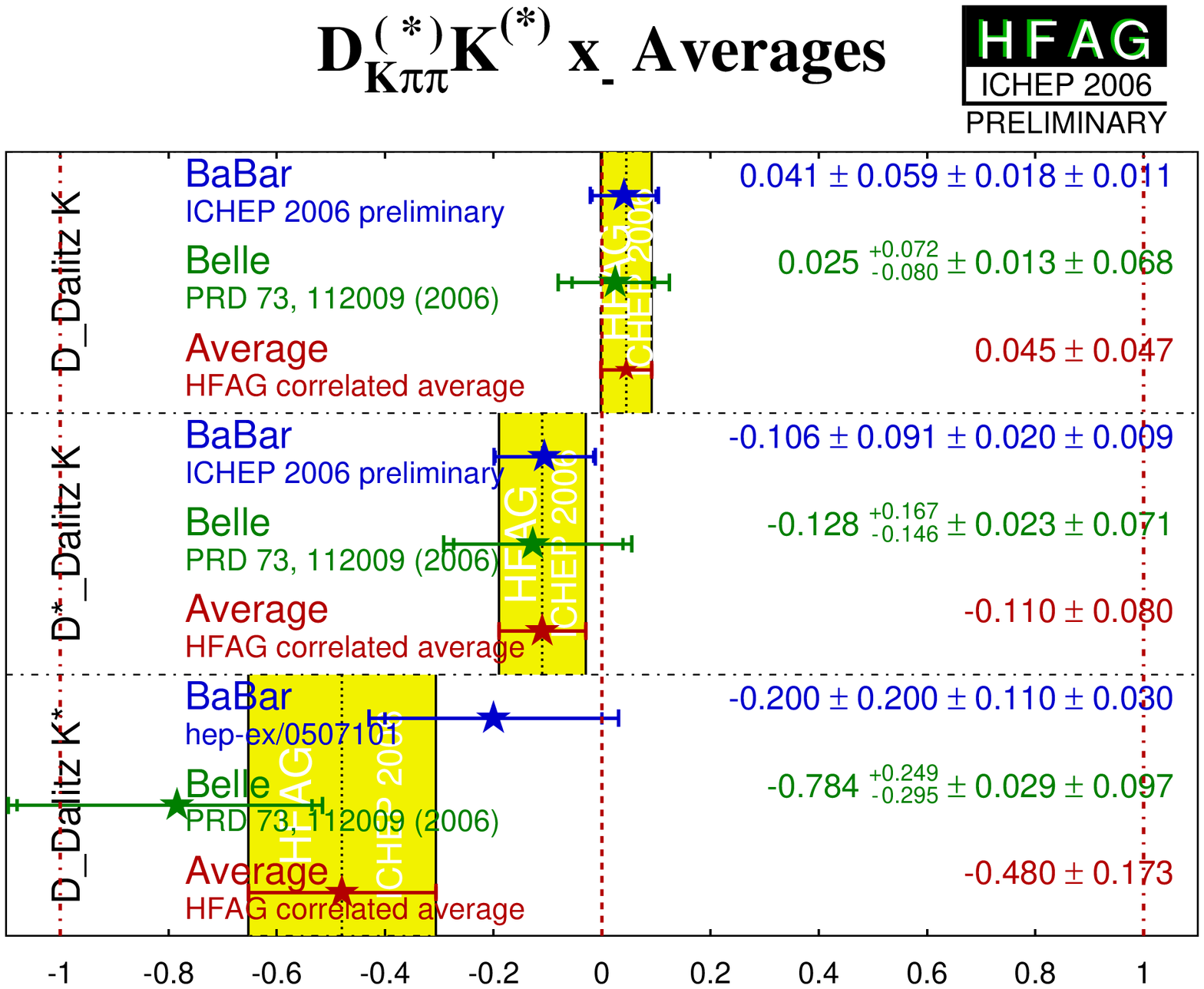}
    }
    \hspace{0.1\textwidth}
    \resizebox{0.40\textwidth}{!}{
      \includegraphics{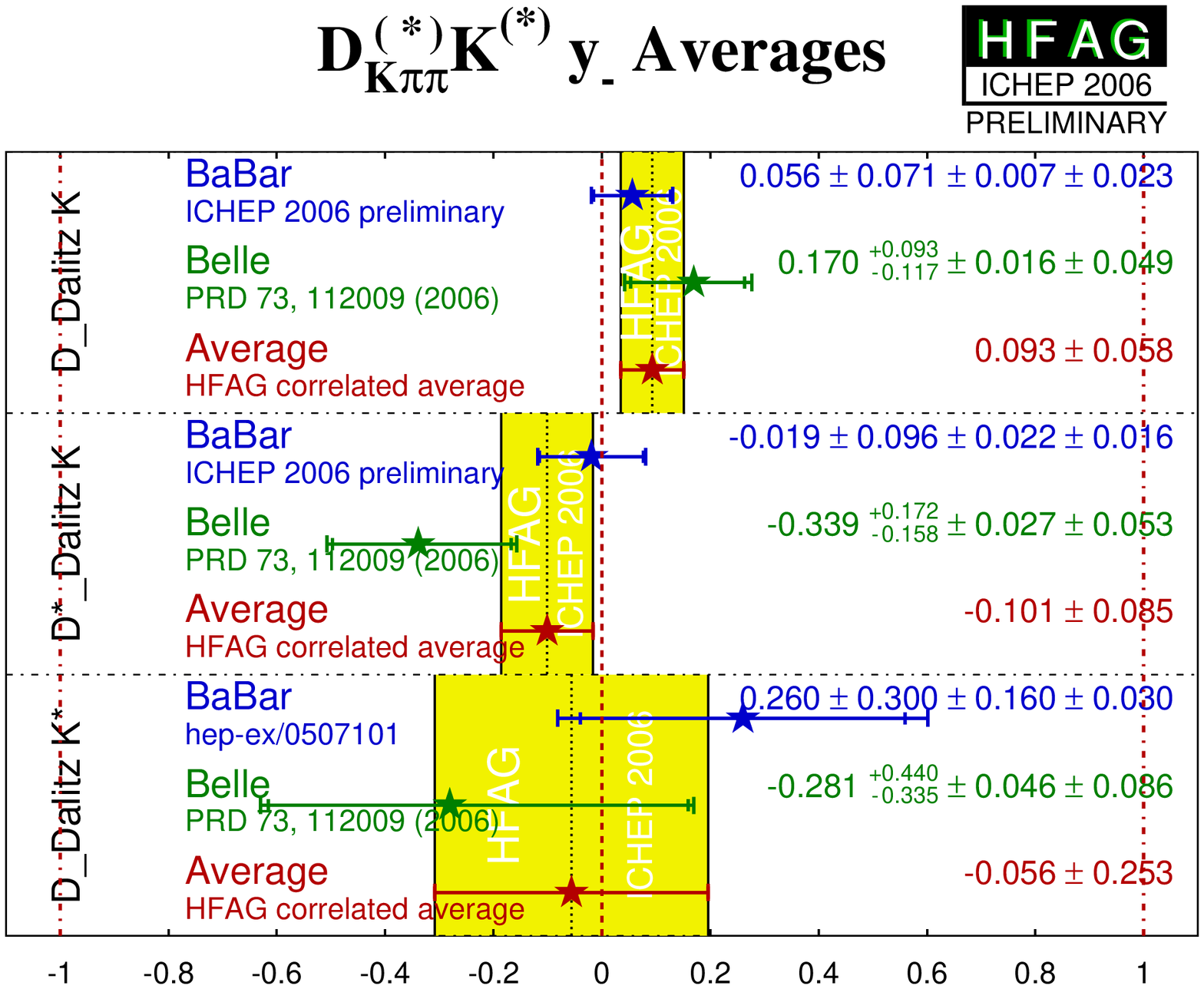}
    }
  \end{center}
  \vspace{-0.8cm}
  \caption{
    Averages of $(x_\pm, y_\pm)$ from $\Bmp \to D^{(*)}K^{(*)\pm}$.
    (Top left) $x_+$, (top right) $y_+$,
    (bottom left) $x_-$, (bottom right) $y_-$.
    Note that the uncertainities assigned to the averages given in these plots
    do not include model errors.        
  }
  \label{fig:cp_uta:cus:dalitz_1d}
\end{figure}

\vspace{3ex}

\noindent
\underline{\large Constraints on $\gamma$}

The measurements of $(x_\pm, y_\pm)$ can be used to obtain constraints on 
$\gamma$, as well as the hadronic parameters $r_B$ and $\delta_B$.
Both
\babar~\cite{ref:cp_uta:cus:babar:dk_dalitz} and 
\belle~\cite{ref:cp_uta:cus:belle:dk_dalitz} 
have done so using a frequentist procedure 
(there are some differences in the details of the techniques used).

\begin{itemize}\setlength{\itemsep}{0.5ex}

\item 
  \babar\ obtain $\gamma = (92 \pm 41 \pm 11 \pm 12)^\circ$
  from $D\Kpm$ and $\Dstar\Kpm$

\item
  \belle\ obtain $\phi_3 = (53 \, ^{+15}_{-18} \pm 3 \pm 9)^\circ$
  from $D\Kpm$, $\Dstar\Kpm$ and $D\Kstarpm$

\item
  The experiments also obtain values for the hadronic parameters. \\
  In $D\Kpm$ \\
  \babar\ obtain 
  $r_B (D\Kpm) < 0.140 (1\sigma)$ and
  $\delta_B (D\Kpm) = (118 \pm 63 \pm 19 \pm 36)^\circ$ \\
  \belle\ obtain 
  $r_B (D\Kpm) = 0.16 \pm 0.05 \pm 0.01 \pm 0.05$ and
  $\delta_B (D\Kpm) = (146 \, ^{+19}_{-20} \pm 3 \pm 23)^\circ$. \\
  In $\Dstar\Kpm$ \\
  \babar\ obtain   
  $0.017 < r_B (\Dstar\Kpm) < 0.203$ and
  $\delta_B (\Dstar\Kpm) = (298 \pm 59 \pm 18 \pm 10)^\circ$ \\
  \belle\ obtain 
  $r_B (\Dstar\Kpm) = 0.18 \, ^{+0.11}_{-0.10} \pm 0.01 \pm 0.05$ and
  $\delta_B (\Dstar\Kpm) = (302 \, ^{+34}_{-35} \pm 6 \pm 23)^\circ$. \\
  In $D\Kstarpm$ \\
  \belle\ obtain 
  $r_B (D\Kstarpm) = 0.56 \, ^{+0.22}_{-0.16} \pm 0.04 \pm 0.08$ and
  $\delta_B (D\Kstarpm) = (243 \, ^{+20}_{-23} \pm 3 \pm 50)^\circ$.
  \babar\ do not obtain a constraint on the hadronic parameters in 
  $D\Kstarpm$ due to the reparametrization described above.

\item 
  Improved constraints can be achieved combining the information from
  $\Bpm \to D\Kpm$ analysis with different $D$ decay modes.
  The experiments have not yet published such results,
  and none are listed here.

\item 
  The CKMfitter~\cite{ref:cp_uta:ckmfitter} and 
  UTFit~\cite{ref:cp_uta:utfit} groups use the measurements 
  from \belle\ and \babar\ given above
  to make combined constraints on $\gamma$.

\end{itemize}

At present we make no attempt to provide an HFAG average for $\gamma$.
More details on procedures to calculate a best fit value for $\gamma$ 
can be found in Refs.~\cite{ref:cp_uta:ckmfitter,ref:cp_uta:utfit}.


%% file: tables/cp_uta/btoccs.tex
\begin{table}
  \begin{center}
    \caption{
      $S_{b \to c\bar c s}$ and $C_{b \to c\bar c s}$.
    }
    \vspace{0.2cm}
    \setlength{\tabcolsep}{0.0pc}
    \begin{tabular*}{\textwidth}{@{\extracolsep{\fill}}lrcc} \hline 
      \mc{2}{l}{Experiment} & 
      $- \etacp S_{b \to c\bar c s}$ & $C_{b \to c\bar c s}$ \\
      \hline
      \babar & \cite{ref:cp_uta:ccs:babar} & 
      $0.710 \pm 0.034 \pm 0.019$ & $0.070 \pm 0.028 \pm 0.018$ \\
      \belle & \cite{ref:cp_uta:ccs:belle} & 
      $0.642 \pm 0.031 \pm 0.017$ & $-0.018 \pm 0.021 \pm 0.014$ \\
      \hline
      \mc{2}{l}{\bf \boldmath $\B$ factory average} & 
      $0.674 \pm 0.026 $ & $0.012 \pm 0.022 $ \\
      \mc{2}{l}{\small Confidence level} & 
      \small $0.18$ & \small $0.02$ \\
      \hline
      ALEPH & \cite{ref:cp_uta:ccs:aleph} & 
      $0.84 \, ^{+0.82}_{-1.04} \pm 0.16$ \\
      OPAL  & \cite{ref:cp_uta:ccs:opal}  & 
      $3.2 \, ^{+1.8}_{-2.0} \pm 0.5$ \\
      CDF   & \cite{ref:cp_uta:ccs:cdf}   & 
      $0.79 \, ^{+0.41}_{-0.44}$ \\
      \hline
      \mc{2}{l}{\bf Average} & 
      $0.675 \pm 0.026$ & $0.012 \pm 0.022 $ \\
      \hline
    \end{tabular*}
    \label{tab:cp_uta:ccs}
  \end{center}
\end{table}

%% file: tables/cp_uta/jpsiKstar.tex
\begin{table}[htb]
	\begin{center}
		\caption{
			Averages from $\Bz \to J/\psi K^{*0}$ transversity analyses.
		}
		\vspace{0.2cm}
		\setlength{\tabcolsep}{0.0pc}
		\begin{tabular*}{\textwidth}{@{\extracolsep{\fill}}lrcccc} \hline
		\mc{2}{l}{Experiment} & $N(B\bar{B})$ & $\sin 2\beta$ & $\cos 2\beta$ & Correlation \\
		\hline
	\babar & \cite{ref:cp_uta:ccs:babar:psi_kstar2} & 88M & $-0.10 \pm 0.57 \pm 0.14$ & $3.32 ^{+0.76}_{-0.96} \pm 0.27$ & $-0.37$ \\

	\belle & \cite{ref:cp_uta:ccs:belle:psi_kstar} & 275M & $0.24 \pm 0.31 \pm 0.05$ & $0.56 \pm 0.79 \pm 0.11$ & $0.22$ \\

        \hline
	\mc{3}{l}{\bf Average} & $0.16 \pm 0.28$ & $1.64 \pm 0.62$ &  -  \\
        \mc{3}{l}{\small Confidence level} & \small $0.61~(0.5\sigma)$ & \small $0.03~(2.2\sigma)$ \\
		\hline
		\end{tabular*}
		\label{tab:cp_uta:ccs:psi_kstar}
	\end{center}
\end{table}

%% file: tables/cp_uta/dstardstarks.tex
\begin{table}[htb]
  \begin{center}
    \caption{
      Results from time-dependent analysis of $\Bz \to \Dstarp \Dstarm \KS$.
    }
    \vspace{0.2cm}
    \setlength{\tabcolsep}{0.0pc}
    \begin{tabular*}{\textwidth}{@{\extracolsep{\fill}}lrcccc} \hline
      \mc{2}{l}{Experiment} & $N(B\bar{B})$ & 
      $\frac{J_c}{J_0}$ & 
      $\frac{2J_{s1}}{J_0} \sin(2\beta)$ &  
      $\frac{2J_{s2}}{J_0} \cos(2\beta)$ \\
      \hline
      \babar & \cite{ref:cp_uta:ccs:ddks:babar} & 230M & 
      $0.76 \pm 0.18 \pm 0.07$ &
      $0.10 \pm 0.24 \pm 0.06$ & 
      $0.38 \pm 0.24 \pm 0.05$ \\
      \hline
    \end{tabular*}
    \label{tab:cp_uta:ccs:dstardstarks}
  \end{center}
\end{table}

%% file: tables/cp_uta/psiphi.tex
\begin{table}[htb]
	\begin{center}
		\caption{
			 Results from time-dependent analysis of $\Bs \to J/\psi \phi$.
		}
		\vspace{0.2cm}
		\setlength{\tabcolsep}{0.0pc}
      \begin{tabular*}{\textwidth}{@{\extracolsep{\fill}}lrccc} 
        \hline
        \mc{2}{l}{Experiment} & $\tau(\Bs)$ & $\Delta\Gamma$ & $\phi_s$ \\
	\dzero & \cite{D01_DGs} & $1.49 \pm 0.08 \, ^{+0.01}_{-0.03}$ & $0.17 \pm 0.09 \pm 0.03$ & $-0.79 \pm 0.56 \, ^{+0.14}_{-0.01}$ \\
        \hline
      \end{tabular*}

      \vspace{2ex}

      \begin{tabular*}{\textwidth}{@{\extracolsep{\fill}}lrcccc} 
        \hline
        \mc{2}{l}{Experiment} & $A_\perp$ & $|A_0|^2 - |A|_\parallel^2$ & $\delta_\parallel$ & $\delta_{0}$ \\
        \hline
	\dzero & \cite{D01_DGs} & $0.46 \pm 0.06 \pm 0.01$ & $0.37 \pm 0.06 \pm 0.01$ & $3.30 \pm 1.10 \pm 0.00$ & $0.70 \pm 1.00 \pm 0.00$ \\
 	\hline
      \end{tabular*}

		\label{tab:cp_uta:ccs:psiphi}
	\end{center}
\end{table}

%% file: tables/cp_uta/dh0.tex
\begin{table}[htb]
  \begin{center}
    \caption{
      Averages from $\Bz \to D^{(*)}h^0$ analyses.
    }
    \vspace{0.2cm}
    \setlength{\tabcolsep}{0.0pc}
    \resizebox{\textwidth}{!}{
      \begin{tabular}{@{\extracolsep{2mm}}lrcccc} \hline 
        \mc{2}{l}{Experiment} & $N(B\bar{B})$ & 
        $\sin(2\beta)$ & $\cos(2\beta)$ & $|\lambda|$ \\
        \hline
        \babar & \cite{ref:cp_uta:cud_beta:babar} & 311M &
        $0.45 \pm 0.36 \pm 0.05 \pm 0.07$ & 
        $0.54 \pm 0.54 \pm 0.08 \pm 0.18$ &
        $0.975 \, ^{+0.093}_{-0.085} \pm 0.012 \pm 0.002$ \\
        \belle & \cite{ref:cp_uta:cud_beta:belle} & 386M &
        $0.78 \pm 0.44 \pm 0.22$ &
        $1.87 \, ^{+0.40}_{-0.53} \, ^{+0.22}_{-0.32}$ \\
        \hline
        \mc{3}{l}{\bf Average} & 
        $0.57 \pm 0.30$ &
        $1.16 \pm 0.42$ \\
        \mc{3}{l}{\small Confidence level} & 
        \small $0.59~(0.5\sigma)$ &
        \small $0.12~(1.6\sigma)$ \\
        \hline
      \end{tabular}
    }
    \label{tab:cp_uta:cud_beta}
  \end{center}
\end{table}

%% file: tables/cp_uta/sPenguinAverages.tex
\begin{table}
	\begin{center}
		\caption{
      Averages of $-\etacp S_{b \to q\bar q s}$ and $C_{b \to q\bar q s}$.
		}
		\vspace{0.2cm}
		\setlength{\tabcolsep}{0.0pc}
		\begin{tabular*}{\textwidth}{@{\extracolsep{\fill}}lrcccc} \hline
		\mc{2}{l}{Experiment} & $N(B\bar{B})$ & $- \etacp S_{b \to q\bar q s}$ & $C_{b \to q\bar q s}$ & Correlation \\
		\hline
      \mc{6}{c}{$\phi \Kz$} \\
	\babar & \cite{ref:cp_uta:qqs:babar:kkk0_tddp} & 347M & $0.12 \pm 0.31 \pm 0.10$ & $0.18 \pm 0.20 \pm 0.10$ &  -  \\

	\belle & \cite{ref:cp_uta:ccs:belle} & 535M & $0.50 \pm 0.21 \pm 0.06$ & $-0.07 \pm 0.15 \pm 0.05$ & $0.05$ \\

	\mc{3}{l}{\bf Average} & $0.39 \pm 0.18$ & $0.01 \pm 0.13$ & $0.03$ \\
        \mc{3}{l}{\small Confidence level} & 
        \mc{3}{c}{\small $\chi^2 = 1.8/2$ dof (CL=0.41 $\to 0.8\sigma$)} \\

		\hline

      \mc{6}{c}{$\etapr \Kz$} \\
	\babar & \cite{ref:cp_uta:qqs:babar:etapks} & 384M & $0.58 \pm 0.10 \pm 0.03$ & $-0.16 \pm 0.07 \pm 0.03$ & $0.03$ \\

	\belle & \cite{ref:cp_uta:ccs:belle} & 535M & $0.64 \pm 0.10 \pm 0.04$ & $0.01 \pm 0.07 \pm 0.05$ & $0.09$ \\

	\mc{3}{l}{\bf Average} & $0.61 \pm 0.07$ & $-0.09 \pm 0.06$ & $0.04$ \\
        \mc{3}{l}{\small Confidence level} & 
        \mc{3}{c}{\small $\chi^2 = 2.3/2$ dof (CL=0.32 $\to 1.0\sigma$)} \\
		\hline

      \mc{6}{c}{$\KS\KS\KS$} \\
	\babar & \cite{ref:cp_uta:qqs:babar:ksksks} & 347M & $0.66 \pm 0.26 \pm 0.08$ & $-0.14 \pm 0.22 \pm 0.05$ & $0.09$ \\

	\belle & \cite{ref:cp_uta:ccs:belle} & 535M & $0.30 \pm 0.32 \pm 0.08$ & $-0.31 \pm 0.20 \pm 0.07$ &  -  \\

	\mc{3}{l}{\bf Average} & $0.51 \pm 0.21$ & $-0.23 \pm 0.15$ & $0.04$ \\
        \mc{3}{l}{\small Confidence level} & 
        \mc{3}{c}{\small $\chi^2 = 1.0/2$ dof (CL=0.61 $\to 0.5\sigma$)} \\

		\hline

      \mc{6}{c}{$\pi^0 \KS$} \\
	\babar & \cite{ref:cp_uta:qqs:babar:pi0ks} & 348M & $0.33 \pm 0.26 \pm 0.04$ & $0.20 \pm 0.16 \pm 0.03$ & $-0.06$ \\

	\belle & \cite{ref:cp_uta:qqs:belle:otherqqs} & 532M & $0.33 \pm 0.35 \pm 0.08$ & $0.05 \pm 0.14 \pm 0.05$ & $-0.08$ \\

	\mc{3}{l}{\bf Average} & $0.33 \pm 0.21$ & $0.12 \pm 0.11$ & $-0.06$ \\
        \mc{3}{l}{\small Confidence level} & 
        \mc{3}{c}{\small $\chi^2 = 0.5/2$ dof (CL=0.79 $\to 0.3\sigma$)} \\

		\hline

		\hline
      \mc{6}{c}{$\rho^0 \KS$} \\
	\babar & \cite{ref:cp_uta:qqs:babar:rhoks} & 227M & $0.20 \pm 0.52 \pm 0.24$ & $0.64 \pm 0.41 \pm 0.20$ &  -  \\

	\mc{3}{l}{\bf Average} & $0.20 \pm 0.57$ & $0.64 \pm 0.46$ &  -  \\

		\hline

      \mc{6}{c}{$\omega \KS$} \\
	\babar & \cite{ref:cp_uta:qqs:babar:omegaks} & 347M & $0.62 ^{+0.25}_{-0.30} \pm 0.02$ & $-0.43 ^{+0.25}_{-0.23} \pm 0.03$ &  -  \\

	\belle & \cite{ref:cp_uta:qqs:belle:otherqqs} & 532M & $0.11 \pm 0.46 \pm 0.07$ & $0.09 \pm 0.29 \pm 0.06$ & $-0.04$ \\

	\mc{3}{l}{\bf Average} & $0.48 \pm 0.24$ & $-0.21 \pm 0.19$ &  -  \\
        \mc{3}{l}{\small Confidence level} & 
        {\small $\chi^2 = 0.9$ (CL=0.35 $\to 0.9\sigma)$} &
        {\small $\chi^2 = 1.8$ (CL=0.18 $\to 1.3\sigma)$} \\

		\hline

      \mc{6}{c}{$f_0 \Kz$} \\
	\babar & \cite{ref:cp_uta:qqs:babar:kkk0_tddp,ref:cp_uta:qqs:babar:f0ks} & -- & $0.62 \pm 0.23$ & $-0.36 \pm 0.23$ &  -  \\

	\belle & \cite{ref:cp_uta:qqs:belle:otherqqs} & 532M & $0.18 \pm 0.23 \pm 0.11$ & $0.15 \pm 0.15 \pm 0.07$ & $-0.01$ \\

	\mc{3}{l}{\bf Average} & $0.42 \pm 0.17$ & $-0.02 \pm 0.13$ & $-0.00$ \\        \mc{3}{l}{\small Confidence level} & 
        \mc{3}{c}{\small $\chi^2 = 4.9/2$ dof (CL=0.09 $\to 1.7\sigma$)} \\

		\hline

      \mc{6}{c}{$\pi^0 \pi^0 \KS$} \\
	\babar & \cite{ref:cp_uta:qqs:babar:pi0pi0ks} & 227M & $-0.84 \pm 0.71 \pm 0.08$ & $0.27 \pm 0.52 \pm 0.13$ &  -  \\

	\mc{3}{l}{\bf Average} & $-0.84 \pm 0.71$ & $0.27 \pm 0.54$ &  -  \\

		\hline

      \mc{6}{c}{$K^+K^- \Kz$} \\
	\babar Q2B & \cite{ref:cp_uta:qqs:babar:kkk0_q2b} & 227M & $0.41 \pm 0.18 \pm 0.07 \pm 0.11$ & $0.23 \pm 0.12 \pm 0.07$ &  -  \\

	\belle & \cite{ref:cp_uta:qqs:belle:otherqqs} & 532M & $0.68 \pm 0.15 \pm 0.03 ^{+0.21}_{-0.13}$ & $0.09 \pm 0.10 \pm 0.05$ & $-0.00$ \\

	\mc{3}{l}{\bf Average} & $0.58 \pm 0.13$ & $0.15 \pm 0.09$ &  -  \\
        \mc{3}{l}{\small Confidence level} & 
        {\small $\chi^2 = 1.6$ (CL=0.21)} &
   	{\small $\chi^2 = 0.6$ (CL=0.43)} \\

		\hline





		\hline
		\end{tabular*}
		\label{tab:cp_uta:qqs}
	\end{center}
\end{table}

%% file: tables/cp_uta/kkk0.tex
 \begin{table}[htb]
  \begin{center}
    \caption{
      Results from time-dependent Dalitz plot analysis of 
      the $\Bz \to K^+K^-\Kz$ decay.
    }
    \vspace{0.2cm}
    \setlength{\tabcolsep}{0.0pc}
    \resizebox{\textwidth}{!}{
      \begin{tabular}{l@{\hspace{2mm}}r@{\hspace{2mm}}c@{\hspace{2mm}}|@{\hspace{2mm}}c@{\hspace{2mm}}c@{\hspace{2mm}}|@{\hspace{2mm}}c@{\hspace{2mm}}c} 
        \hline 
        \mc{2}{l}{Experiment} & $N(B\bar{B})$ &
        \mc{2}{c}{$\phi\Kz$} & \mc{2}{c}{$f_0\Kz$} \\
        & & & $\beta^{\rm eff}$ & $\Acp$ & $\beta^{\rm eff}$ & $\Acp$ \\
        \babar & \cite{ref:cp_uta:qqs:babar:kkk0_tddp} & 347M &
        $0.06 \pm 0.16 \pm 0.05$ & $-0.18 \pm 0.20 \pm 0.10$ &
        $0.18 \pm 0.19 \pm 0.04$ & $0.45 \pm 0.28 \pm 0.10$ \\
        \hline
      \end{tabular}
    }

    \vspace{2ex}

      \begin{tabular}{l@{\hspace{2mm}}r@{\hspace{2mm}}c@{\hspace{2mm}}|@{\hspace{2mm}}c@{\hspace{2mm}}c} 
        \hline 
        \mc{2}{l}{Experiment} & $N(B\bar{B})$ &
        \mc{2}{c}{$K^+K^-\Kz$} \\
        & & & $\beta^{\rm eff}$ & $\Acp$ \\
        \babar & \cite{ref:cp_uta:qqs:babar:kkk0_tddp} & 347M &
        $0.361 \pm 0.079 \pm 0.037$ & $-0.034 \pm 0.079 \pm 0.025$ \\
        \hline
      \end{tabular}

      \label{tab:cp_uta:kkk0_tddp}
  \end{center}
\end{table}

%% file: tables/cp_uta/btoccd.tex
\begin{table}[htb]
	\begin{center}
		\caption{
     Averages for the $b \to c\bar{c}d$ modes,
     $\Bz \to J/\psi \pi^{0}$, $D^+D^-$, $D^{*+}D^{*-}$ and $D^{*\pm}D^\mp$.
		}
		\vspace{0.2cm}
		\setlength{\tabcolsep}{0.0pc}
		\begin{tabular*}{\textwidth}{@{\extracolsep{\fill}}lrcccc} \hline
		\mc{2}{l}{Experiment} & $N(B\bar{B})$ & $S_{CP}$ & $C_{CP}$ & Correlation \\
		\hline
      \mc{6}{c}{$J/\psi \pi^{0}$} \\
	\babar & \cite{ref:cp_uta:ccd:babar:psipi0} & 232M & $-0.68 \pm 0.30 \pm 0.04$ & $-0.21 \pm 0.26 \pm 0.06$ & $0.08$ \\

	\belle & \cite{ref:cp_uta:ccd:belle:psipi0} & 152M & $-0.72 \pm 0.42 \pm 0.09$ & $0.01 \pm 0.29 \pm 0.03$ & $-0.12$ \\

	\mc{3}{l}{\bf Average} & $-0.68 \pm 0.25$ & $-0.11 \pm 0.20$ & $0.00$ \\
        \mc{3}{l}{\small Confidence level} & 
        \mc{3}{c}{\small $\chi^2 = 0.3/2$ dof (CL=0.86 $\to 0.2\sigma$)} \\
		\hline

      \mc{6}{c}{$D^{+} D^{-}$} \\
	\babar & \cite{ref:cp_uta:ccd:babar:dd_dstard} & 232M & $-0.29 \pm 0.63 \pm 0.06$ & $0.11 \pm 0.35 \pm 0.06$ &  -  \\

	\belle & \cite{ref:cp_uta:ccd:belle:dd} & 535M & $-1.12 \pm 0.37 \pm 0.09$ & $-0.92 \pm 0.23 \pm 0.05$ & $0.04$ \\

	\mc{3}{l}{\bf Average} & $-0.89 \pm 0.33$ & $-0.60 \pm 0.20$ & $0.03$ \\
        \mc{3}{l}{\small Confidence level} & 
        \mc{3}{c}{\small $\chi^2 = 7.1/2$ dof (CL=0.03 $\to 2.2\sigma$)} \\

		\hline

      \mc{6}{c}{$D^{*+} D^{*-}$} \\
	\babar & \cite{ref:cp_uta:ccd:babar:dstardstar} & 227M & $-0.75 \pm 0.25 \pm 0.03$ & $0.06 \pm 0.17 \pm 0.03$ & $0.04$ \\

	\belle & \cite{ref:cp_uta:ccd:belle:dstardstar} & 152M & $-0.75 \pm 0.56 \pm 0.10 \pm 0.06$ & $0.26 \pm 0.26 \pm 0.05 \pm 0.01$ &  -  \\

	\mc{3}{l}{\bf Average} & $-0.75 \pm 0.23$ & $0.12 \pm 0.14$ & $0.03$ \\
        \mc{3}{l}{\small Confidence level} & 
        \mc{3}{c}{\small $\chi^2 = 0.4/2$ dof (CL=0.82 $\to 0.2\sigma$)} \\
		\hline
		\end{tabular*}

                \vspace{2ex}

    \resizebox{\textwidth}{!}{
		\begin{tabular}{@{\extracolsep{2mm}}lrcccccc} \hline
		\mc{2}{l}{Experiment} & $N(B\bar{B})$ & $S_{+-}$ & $C_{+-}$ & $S_{-+}$ & $C_{-+}$ & ${\cal A}$ \\
		\hline
      \mc{8}{c}{$D^{*\pm} D^{\mp}$} \\
	\babar & \cite{ref:cp_uta:ccd:babar:dd_dstard} & 232M & $-0.54 \pm 0.35 \pm 0.07$ & $0.09 \pm 0.25 \pm 0.06$ & $-0.29 \pm 0.33 \pm 0.07$ & $0.17 \pm 0.24 \pm 0.04$ \\
        \belle & \cite{ref:cp_uta:ccd:belle:dstard} & 152M & $-0.55 \pm 0.39 \pm 0.12$ & $-0.37 \pm 0.22 \pm 0.06$ & $-0.96 \pm 0.43 \pm 0.12$ & $0.23 \pm 0.25 \pm 0.06$ & $0.07 \pm 0.08 \pm 0.04$ \\
	\mc{3}{l}{\bf Average} & $-0.54 \pm 0.27$ & $-0.17 \pm 0.17$ & $-0.53 \pm 0.27$ & $0.20 \pm 0.18$ & $0.07 \pm 0.09$ \\
        \mc{3}{l}{\small Confidence level} & 
        {\small CL=0.99 $(0.0\sigma)$} &
        {\small CL=0.18 $(1.3\sigma)$} &
        {\small CL=0.23 $(1.2\sigma)$} &
        {\small CL=0.87 $(0.2\sigma)$} \\
        \hline
		\end{tabular}
              }
		\label{tab:cp_uta:ccd}
	\end{center}
\end{table}

%% file: tables/cp_uta/qqdAverages.tex
\begin{table}
	\begin{center}
		\caption{
			Results for $\Bz \to \KS\KS$.
		}
		\vspace{0.2cm}
		\setlength{\tabcolsep}{0.0pc}
		\begin{tabular*}{\textwidth}{@{\extracolsep{\fill}}lrcccc} \hline
		\mc{2}{l}{Experiment} & $N(B\bar{B})$ & $S_{CP}$ & $C_{CP}$ & Correlation \\
		\hline
	\babar & \cite{ref:cp_uta:qqd:babar:ksks} & 350M & $-1.28 \, ^{+0.80}_{-0.73} \, ^{+0.11}_{-0.16}$ & $-0.40 \pm 0.41 \pm 0.06$ & $-0.32$ \\


		\hline
		\end{tabular*}
		\label{tab:cp_uta:qqd}
	\end{center}
\end{table}

%% file: tables/cp_uta/bsgAverages.tex
\begin{table}
	\begin{center}
		\caption{
      Averages for $b \to s \gamma$ modes.
		}
		\vspace{0.2cm}
		\setlength{\tabcolsep}{0.0pc}
		\begin{tabular*}{\textwidth}{@{\extracolsep{\fill}}lrcccc} \hline
		\mc{2}{l}{Experiment} & $N(B\bar{B})$ & $S_{CP} (b \to s \gamma)$ & $C_{CP} (b \to s \gamma)$ & Correlation \\
		\hline
      \mc{6}{c}{$\Kstar(892)\gamma$} \\
	\babar & \cite{ref:cp_uta:bsg:babar:kspi0gamma} & 232M & $-0.21 \pm 0.40 \pm 0.05$ & $-0.40 \pm 0.23 \pm 0.04$ & $0.07$ \\

	\belle & \cite{ref:cp_uta:bsg:belle:kspi0gamma} & 532M & $-0.32 ^{+0.36}_{-0.33} \pm 0.05$ & $0.20 \pm 0.24 \pm 0.05$ & $0.08$ \\

	\mc{3}{l}{\bf Average} & $-0.28 \pm 0.26$ & $-0.11 \pm 0.17$ & $0.07$ \\
        \mc{3}{l}{\small Confidence level} & 
        \mc{3}{c}{\small $\chi^2 = 3.2/2$ dof (CL=0.20 $\to 1.3\sigma$)} \\

		\hline

      \mc{6}{c}{$\KS \pi^0 \gamma$ (including $\Kstar(892)\gamma$)} \\
	\babar & \cite{ref:cp_uta:bsg:babar:kspi0gamma} & 232M & $-0.06 \pm 0.37$ & $-0.48 \pm 0.22$ & $0.05$ \\

	\belle & \cite{ref:cp_uta:bsg:belle:kspi0gamma} & 532M & $-0.10 \pm 0.31 \pm 0.07$ & $0.20 \pm 0.20 \pm 0.06$ & $0.08$ \\

	\mc{3}{l}{\bf Average} & $-0.09 \pm 0.24$ & $-0.12 \pm 0.15$ & $0.06$ \\
        \mc{3}{l}{\small Confidence level} & 
        \mc{3}{c}{\small $\chi^2 = 5.1/2$ dof (CL=0.08 $\to 1.8\sigma$)} \\

		\hline
		\end{tabular*}
		\label{tab:cp_uta:bsg}
	\end{center}
\end{table}

%% file: tables/cp_uta/uudAverages.tex
\begin{table}
	\begin{center}
		\caption{
      Averages for $b \to u \bar u d$ modes.
		}
		\vspace{0.2cm}
		\setlength{\tabcolsep}{0.0pc}
		\begin{tabular*}{\textwidth}{@{\extracolsep{\fill}}lrcccc} \hline
		\mc{2}{l}{Experiment} & $N(B\bar{B})$ & $S_{CP}$ & $C_{CP}$ & Correlation \\
		\hline
      \mc{6}{c}{$\pi^{+} \pi^{-}$} \\
	\babar & ~\cite{ref:cp_uta:uud:babar:pipi} & 350M & $-0.53 \pm 0.14 \pm 0.02$ & $-0.16 \pm 0.11 \pm 0.03$ & $-0.08$ \\

	\belle & ~\cite{ref:cp_uta:uud:belle:pipi} & 532M & $-0.61 \pm 0.10 \pm 0.04$ & $-0.55 \pm 0.08 \pm 0.05$ & $-0.15$ \\

	\mc{3}{l}{\bf Average} & $-0.59 \pm 0.09$ & $-0.39 \pm 0.07$ & $-0.10$ \\
        \mc{3}{l}{\small Confidence level} & 
        \mc{3}{c}{\small $\chi^2 = 7.4/2$ dof (CL=0.02 $\to 2.3\sigma$)} \\

		\hline

      \mc{6}{c}{$\rho^{+} \rho^{-}$} \\
	\babar & ~\cite{ref:cp_uta:uud:babar:rhorho} & 350M & $-0.19 \pm 0.21 ^{+0.05}_{-0.07}$ & $-0.07 \pm 0.15 \pm 0.06$ & $-0.06$ \\

	\belle & ~\cite{ref:cp_uta:uud:belle:rhorho} & 535M & $0.19 \pm 0.30 \pm 0.07$ & $-0.16 \pm 0.21 \pm 0.07$ & $0.10$ \\

	\mc{3}{l}{\bf Average} & $-0.06 \pm 0.18$ & $-0.11 \pm 0.13$ & $-0.00$ \\
        \mc{3}{l}{\small Confidence level} & 
        \mc{3}{c}{\small $\chi^2 = 1.2/2$ dof (CL=0.56 $\to 0.6\sigma$)} \\

		\hline
		\end{tabular*}

                \vspace{2ex}

    \resizebox{\textwidth}{!}{
 		\begin{tabular}{@{\extracolsep{2mm}}lrcccccc} \hline
 		\mc{2}{l}{Experiment} & $N(B\bar{B})$ & $A_{CP}$ & $C$ & $S$ & $\Delta C$ & $\Delta S$ \\
 		\hline
      \mc{8}{c}{$a_1^{\pm} \pi^{\mp}$} \\
	\babar & ~\cite{ref:cp_uta:uud:babar:a1pi} & 384M & $-0.07 \pm 0.07 \pm 0.02$ & $-0.10 \pm 0.15 \pm 0.09$ & $0.37 \pm 0.21 \pm 0.07$ & $0.26 \pm 0.15 \pm 0.07$ & $-0.14 \pm 0.21 \pm 0.06$ \\
 	\hline
		\end{tabular}
              }
		\label{tab:cp_uta:uud}
	\end{center}
\end{table}

%% file: tables/cp_uta/pipipi0_q2b.tex
\begin{table}
	\begin{center}
		\caption{
                  Averages of quasi-two-body parameters extracted
                  from time-dependent Dalitz plot analysis of 
                  $\Bz \to \pi^+\pi^-\pi^0$.
		}
		\vspace{0.2cm}
		\setlength{\tabcolsep}{0.0pc}
    \resizebox{\textwidth}{!}{
		\begin{tabular}{@{\extracolsep{2mm}}lrcccccc} \hline
		\mc{2}{l}{Experiment} & $N(B\bar{B})$ & ${\cal A}_{CP}^{\rho\pi}$ & $C_{\rho\pi}$ & $S_{\rho\pi}$ & $\Delta C_{\rho\pi}$ & $\Delta S_{\rho\pi}$ \\
		\hline
	\babar & \cite{ref:cp_uta:uud:babar:pipipi0} & 347M & $-0.14 \pm 0.04 \pm 0.01$ & $0.15 \pm 0.09 \pm 0.04$ & $0.01 \pm 0.12 \pm 0.03$ & $0.38 \pm 0.09 \pm 0.02$ & $0.06 \pm 0.13 \pm 0.03$ \\
 	\belle & \cite{ref:cp_uta:uud:belle:pipipi0} & 447M & $-0.12 \pm 0.05 \pm 0.03$ & $-0.13 \pm 0.09 \pm 0.06$ & $0.06 \pm 0.13 \pm 0.07$ & $0.35 \pm 0.10 \pm 0.06$ & $-0.12 \pm 0.14 \pm 0.07$ \\
	\mc{3}{l}{\bf Average} & $-0.13 \pm 0.03$ & $0.03 \pm 0.07$ & $0.03 \pm 0.09$ & $0.36 \pm 0.07$ & $-0.02 \pm 0.10$ \\
        \mc{3}{l}{\small Confidence level} & 
        \mc{5}{c}{\small $\chi^2 = 4.7/5$ dof (CL=0.45 $\to 0.8\sigma$)} \\

        \hline
		\end{tabular}
              }

                \vspace{2ex}

		\begin{tabular*}{\textwidth}{@{\extracolsep{\fill}}lrcccc} \hline
		\mc{2}{l}{Experiment} & $N(B\bar{B})$ & ${\cal A}^{-+}_{\rho\pi}$ & ${\cal A}^{+-}_{\rho\pi}$ & Correlation \\
		\hline
	\babar & \cite{ref:cp_uta:uud:babar:pipipi0} & 347M & $-0.38 ^{+0.15}_{-0.16} \pm 0.07$ & $0.03 \pm 0.07 \pm 0.03$ & $0.62$ \\
	\belle & \cite{ref:cp_uta:uud:belle:pipipi0} & 447M & $0.08 \pm 0.17 \pm 0.12$ & $0.22 \pm 0.08 \pm 0.05$ & $0.53$ \\
	\mc{3}{l}{\bf Average} & $-0.19 \pm 0.13$ & $0.11 \pm 0.06$ & $0.46$ \\
        \mc{3}{l}{\small Confidence level} & 
        \mc{3}{c}{\small $\chi^2 = 3.8/2$ dof (CL=0.15 $\to 1.4\sigma$)} \\

		\hline
		\end{tabular*}

                \vspace{2ex}

		\begin{tabular*}{\textwidth}{@{\extracolsep{\fill}}lrcccc} \hline
		\mc{2}{l}{Experiment} & $N(B\bar{B})$ & $C_{\rho^0\pi^0}$ & $S_{\rho^0\pi^0}$ & Correlation \\
		\hline
	\belle & \cite{ref:cp_uta:uud:belle:pipipi0} & 447M & $0.45 \pm 0.35 \pm 0.32$ & $0.15 \pm 0.57 \pm 0.43$ & $0.07$ \\


		\hline
		\end{tabular*}
		\label{tab:cp_uta:uud:rhopi_q2b}
	\end{center}
\end{table}

%% file: tables/cp_uta/cud.tex
\begin{table}
  \begin{center}
    \caption{
      Averages for $b \to c\bar{u}d / u\bar{c}d$ modes.
      Note that the ``\belle (combined)'' result for $D^{*\pm}\pi^{\mp}$
      is a combination of the 
      ``\belle (full rec.)'' and ``\belle (partial rec.)'' results.
    }
    \vspace{0.2cm}
    \setlength{\tabcolsep}{0.0pc}
    \begin{tabular*}{\textwidth}{@{\extracolsep{\fill}}lrccc} \hline 
      \mc{2}{l}{Experiment} & $N(B\bar{B})$ & $a$ & $c$ \\
      \hline
      \mc{5}{c}{$D^{*\pm}\pi^{\mp}$} \\
      \babar (full rec.) & \cite{ref:cp_uta:cud:babar:full} & 232M &
      $ -0.040 \pm 0.023 \pm 0.010$ & $\ph{-}0.049 \pm 0.042 \pm 0.015 $ \\
      \babar (partial rec.) & \cite{ref:cp_uta:cud:babar:partial} & 232M &
      $ -0.034 \pm 0.014 \pm 0.009$ & $-0.019 \pm 0.022 \pm 0.013$ \\
      \belle (full rec.) & \cite{ref:cp_uta:cud:belle} & 386M &
      $-0.039 \pm 0.020 \pm 0.013$ & $-0.011 \pm 0.020 \pm 0.013$ \\
      \belle (partial rec.) & \cite{ref:cp_uta:cud:belle} & 386M &
      $-0.041 \pm 0.019 \pm 0.017$ & $-0.007 \pm 0.019 \pm 0.017$ \\
      {\small \belle (combined)} & {\small \cite{ref:cp_uta:cud:belle}} & {\small 386M} &
      {\small $-0.040 \pm 0.014 \pm 0.011$} & {\small $-0.009 \pm 0.014 \pm 0.011$} \\
      \mc{3}{l}{\bf Average} & 
      $-0.037 \pm 0.011$ & $-0.006 \pm 0.014$ \\
      \mc{3}{l}{\small Confidence level} & 
      \small $0.96~(0.0\sigma)$ & \small $0.41~(0.8\sigma)$ \\
      \hline
      \mc{5}{c}{$D^{\pm}\pi^{\mp}$} \\
      \babar (full rec.) & \cite{ref:cp_uta:cud:babar:full} & 232M &
      $ -0.010 \pm 0.023 \pm 0.007$ & $ -0.033 \pm 0.042 \pm 0.012$ \\
      \belle (full rec.) & \cite{ref:cp_uta:cud:belle} & 386M &
      $ -0.050 \pm 0.021 \pm 0.012$ & $ -0.019 \pm 0.021 \pm 0.012$ \\
      \mc{3}{l}{\bf Average} & $ -0.030 \pm 0.017$ & $ -0.022 \pm 0.021 $ \\
      \mc{3}{l}{\small Confidence level} & 
      \small $0.24~(1.2\sigma)$ & \small $0.78~(0.3\sigma)$ \\
      \hline
      \mc{5}{c}{$D^{\pm}\rho^{\mp}$} \\
      \babar (full rec.) & \cite{ref:cp_uta:cud:babar:full} & 232M &
      $ -0.024 \pm 0.031 \pm 0.009$ & $ -0.098 \pm 0.055 \pm 0.018$ \\
      \mc{3}{l}{\bf Average} & $ -0.024 \pm 0.033 $ & $ -0.098 \pm 0.058$ \\
      \hline 
    \end{tabular*}
    \label{tab:cp_uta:cud}
  \end{center}
\end{table}

%% file: tables/cp_uta/cus_glw.tex
\begin{table}
  \begin{center}
    \caption{
      Averages from GLW analyses of $b \to c\bar{u}s / u\bar{c}s$ modes.
    }
    \vspace{0.2cm}
    \resizebox{\textwidth}{!}{
      \setlength{\tabcolsep}{0.0pc}
      \begin{tabular}{@{\extracolsep{2mm}}lrccccc} \hline 
        \mc{2}{l}{Experiment} & $N(B\bar{B})$ &
        $A_{\CP+}$ & $A_{\CP-}$ & $R_{\CP+}$ & $R_{\CP-}$ \\
        \hline
        \mc{7}{c}{$D_{\CP} K^-$} \\
        \babar & \cite{ref:cp_uta:cus:babar:dcpk} & 232M &
        $\ph{-}0.35 \pm 0.13 \pm 0.04$ & $-0.06 \pm 0.13 \pm 0.04$ & 
        $ 0.90 \pm 0.12 \pm 0.04$ & $ 0.86 \pm 0.10 \pm 0.05$ \\
        \belle & \cite{ref:cp_uta:cus:belle:dcpk} & 275M &
        $\ph{-}0.06 \pm 0.14 \pm 0.05$ & $-0.12 \pm 0.14 \pm 0.05$ & 
        $ 1.13 \pm 0.16 \pm 0.08$ & $ 1.17 \pm 0.14 \pm 0.14$ \\
        \mc{3}{l}{\bf Average} & 
        $ 0.22 \pm 0.10$ & $-0.09 \pm 0.10$ & $ 0.98 \pm 0.10$ & $ 0.94 \pm 0.10$ \\
        \hline
        \mc{7}{c}{$\Dstar_{\CP} K^-$} \\
        \babar & \cite{ref:cp_uta:cus:babar:dstarcpk} & 123M &
        $-0.10 \pm 0.23 \, ^{+0.03}_{-0.04}$ & & 
        $ 1.06 \pm 0.26 \, ^{+0.10}_{-0.09}$ &  \\
        \belle & \cite{ref:cp_uta:cus:belle:dcpk} & 275M &
        $-0.20 \pm 0.22 \pm 0.04$ & $ 0.13 \pm 0.30 \pm 0.08$ & 
        $ 1.41 \pm 0.25 \pm 0.06$ & $ 1.15 \pm 0.31 \pm 0.12$ \\
        \mc{3}{l}{\bf Average} & 
        $-0.15 \pm 0.16$ & $ 0.13 \pm 0.31$ & $ 1.25 \pm 0.19$ & $ 1.15 \pm 0.33$ \\
        \hline
        \mc{7}{c}{$D_{\CP} K^{*-}$} \\
        \babar & \cite{ref:cp_uta:cus:babar:dcpkstar} & 232M &
        $-0.08 \pm 0.19 \pm 0.08$ & $-0.26 \pm 0.40 \pm 0.12$ &
        $ 1.96 \pm 0.40 \pm 0.11$ & $ 0.65 \pm 0.26 \pm 0.08$ \\
        \hline
      \end{tabular}
    }
    \label{tab:cp_uta:cus:glw}
  \end{center}
\end{table}

%% file: tables/cp_uta/cus_ads.tex
\begin{table}
  \begin{center} 
    \caption{
      Averages from ADS analyses of $b \to c\bar{u}s / u\bar{c}s$ and 
      $b \to c\bar{u}d / u\bar{c}d$ modes.
    }
    \vspace{0.2cm}
    \setlength{\tabcolsep}{0.0pc}
    \begin{tabular*}{\textwidth}{@{\extracolsep{\fill}}lrccc} \hline 
      \mc{2}{l}{Experiment} & $N(B\bar{B})$ &
      $A_{\rm ADS}$ & $R_{\rm ADS}$ \\
      \hline
      \mc{5}{c}{$D K^-$, $D \to K^+\pi^-$} \\
      \babar & \cite{ref:cp_uta:cus:babar:dk_ads} & 232M &
      & $ 0.013 \, ^{+0.011}_{-0.009}$ \\ 
      \belle & \cite{ref:cp_uta:cus:belle:dk_ads} & 386M &
      & $ 0.000 \pm 0.008 \pm 0.001$ \\
      \mc{3}{l}{\bf Average} & 
      & $ 0.006 \pm 0.006$ \\
      \hline
      \mc{5}{c}{$\Dstar K^-$, $\Dstar \to D\pi^0$, $D \to K^+\pi^-$} \\
      \babar & \cite{ref:cp_uta:cus:babar:dk_ads} & 232M &
      & $-0.002 \, ^{+0.010}_{-0.006}$ \\
      \hline
      \mc{5}{c}{$\Dstar K^-$, $\Dstar \to D\gamma$, $D \to K^+\pi^-$} \\
      \babar & \cite{ref:cp_uta:cus:babar:dk_ads} & 232M &
      & $ 0.011 \, ^{+0.018}_{-0.013}$ \\
      \hline
      \mc{5}{c}{$D K^{*-}$, $D \to K^+\pi^-$, $K^{*-} \to \KS \pi^-$} \\
      \babar & \cite{ref:cp_uta:cus:babar:dkstar_ads} & 232M &
      $ -0.22 \pm 0.61 \pm 0.17$ & $ 0.046 \pm 0.031 \pm 0.008$ \\
      \hline
      \mc{5}{c}{$D K^{-}$, $D \to K^+\pi^-\pi^0$} \\
      \babar & \cite{ref:cp_uta:cus:babar:dkstar_kpipi0} & 226M &
      & $0.012 \pm 0.012 \pm 0.009$ \\
      \hline \hline
      \mc{5}{c}{$D \pi^-$, $D \to K^+\pi^-$} \\
      \belle & \cite{ref:cp_uta:cus:belle:dk_ads} & 386M &
      $ 0.10 \pm 0.22 \pm 0.06$ & $ 0.0035 \, ^{+0.0008}_{-0.0007} \pm 0.0003$ \\
      \hline 
    \end{tabular*}
    \label{tab:cp_uta:cus:ads}
  \end{center}
\end{table}

%% file: tables/cp_uta/cus_dalitz.tex
\begin{table}[htb]
	\begin{center}
		\caption{
      Averages from Dalitz plot analyses of $b \to c\bar{u}s / u\bar{c}s$ modes.
      Note that the uncertainities assigned to the averages do not include model errors.	
		}
		\vspace{0.2cm}
		\setlength{\tabcolsep}{0.0pc}
    \resizebox{\textwidth}{!}{
		\begin{tabular}{@{\extracolsep{2mm}}lrccccc} \hline
		\mc{2}{l}{Experiment} & $N(B\bar{B})$ & $x_+$ & $y_+$ & $x_-$ & $y_-$ \\
		\hline
                \mc{7}{c}{$D K^-$, $D \to \KS \pi^+\pi^-$} \\
	\babar & \cite{ref:cp_uta:cus:babar:dk_dalitz} & 347M & $-0.072 \pm 0.056 \pm 0.014 \pm 0.029$ & $-0.033 \pm 0.066 \pm 0.007 \pm 0.018$ & $0.041 \pm 0.059 \pm 0.018 \pm 0.011$ & $0.056 \pm 0.071 \pm 0.007 \pm 0.023$ \\
 	\belle & \cite{ref:cp_uta:cus:belle:dk_dalitz} & 386M & $-0.135 ^{+0.069}_{-0.070} \pm 0.017 \pm 0.051$ & $-0.085 ^{+0.090}_{-0.086} \pm 0.009 \pm 0.066$ & $0.025 ^{+0.072}_{-0.080} \pm 0.013 \pm 0.068$ & $0.170 ^{+0.093}_{-0.117} \pm 0.016 \pm 0.049$ \\
	\mc{3}{l}{\bf Average} & $-0.097 \pm 0.045$ & $-0.051 \pm 0.053$ & $0.045 \pm 0.047$ & $0.093 \pm 0.058$ \\
        \mc{3}{l}{\small Confidence level} & 
        \mc{4}{c}{\small $\chi^2 = 1.5/4$ dof (CL=0.83 $\to 0.2\sigma$)} \\
 		\hline

                \mc{7}{c}{$\Dstar K^-$, $\Dstar \to D\pi^0$ or $D\gamma$, $D \to \KS \pi^+\pi^-$} \\
	\babar & \cite{ref:cp_uta:cus:babar:dk_dalitz} & 347M & $0.084 \pm 0.088 \pm 0.015 \pm 0.018$ & $0.096 \pm 0.111 \pm 0.032 \pm 0.017$ & $-0.106 \pm 0.091 \pm 0.020 \pm 0.009$ & $-0.019 \pm 0.096 \pm 0.022 \pm 0.016$ \\
 	\belle & \cite{ref:cp_uta:cus:belle:dk_dalitz} & 386M & $0.032 ^{+0.120}_{-0.116} \pm 0.004 \pm 0.049$ & $0.008 ^{+0.137}_{-0.136} \pm 0.011 \pm 0.074$ & $-0.128 ^{+0.167}_{-0.146} \pm 0.023 \pm 0.071$ & $-0.339 ^{+0.172}_{-0.158} \pm 0.027 \pm 0.053$ \\
	\mc{3}{l}{\bf Average} & $0.067 \pm 0.071$ & $0.061 \pm 0.088$ & $-0.110 \pm 0.080$ & $-0.101 \pm 0.085$ \\
        \mc{3}{l}{\small Confidence level} & 
        \mc{4}{c}{\small $\chi^2 = 3.2/4$ dof (CL=0.52 $\to 0.6\sigma$)} \\
 		\hline

                \mc{7}{c}{$D K^{*-}$, $D \to \KS \pi^+\pi^-$} \\
	\babar & \cite{ref:cp_uta:cus:babar:dkstar_dalitz} & 227M & $-0.070 \pm 0.230 \pm 0.130 \pm 0.030$ & $-0.010 \pm 0.320 \pm 0.180 \pm 0.050$ & $-0.200 \pm 0.200 \pm 0.110 \pm 0.030$ & $0.260 \pm 0.300 \pm 0.160 \pm 0.030$ \\
 	\belle & \cite{ref:cp_uta:cus:belle:dk_dalitz} & 386M & $-0.105 ^{+0.177}_{-0.167} \pm 0.006 \pm 0.088$ & $-0.004 ^{+0.164}_{-0.156} \pm 0.013 \pm 0.095$ & $-0.784 ^{+0.249}_{-0.295} \pm 0.029 \pm 0.097$ & $-0.281 ^{+0.440}_{-0.335} \pm 0.046 \pm 0.086$ \\
	\mc{3}{l}{\bf Average} & $-0.094 \pm 0.144$ & $-0.007 \pm 0.146$ & $-0.480 \pm 0.173$ & $-0.056 \pm 0.253$ \\
        \mc{3}{l}{\small Confidence level} & 
        \mc{4}{c}{\small $\chi^2 = 4.6/4$ dof (CL=0.33 $\to 1.0\sigma$)} \\
 		\hline
		\end{tabular}
              }
		\label{tab:cp_uta:cus:dalitz}
	\end{center}
\end{table}

%% file: slbdecays.tex


\section{Semileptonic $B$ decays}
\label{slbdecays}

Measurements of semileptonic $B$-meson decays are essential to
determining the magnitude of the CKM matrix elements $|V_{cb}|$ and
$|V_{ub}|$.
The ratio $|V_{ub}|/|V_{cb}|$ is an important ingredient in tests of the   
flavor sector of the Standard Model since it limits the precision with
which one side-length of the Unitarity Triangle is known. That side-length
and the precisely known angle $\phi_1$ or $\beta$, opposite to it, provide 
a stringest test of the Kobayashi-Maskawa mechanism for CP violation.

In the following, we provide averages of exclusive and inclusive
 branching fractions, the product of $|V_{cb}|$ and the form factor
 normalisation $F(1)$ and $G(1)$ for  $\BzbDstarlnu$ and
$\BzbDplnu$ decays respectively, and $|V_{ub}|$ as determined from
inclusive and exclusive measurements of $\B\to X_u \ell \nul$ decays.
Brief descriptions of all parameters
and analyses (published or preliminary) relevant for the
determination of the combined results are given.  The descriptions are
based on the information available on the web page at\\
 \centerline{\tt http://www.slac.stanford.edu/xorg/hfag/semi/summer06/summer06.shtml}
 A description of the technique employed for calculating averages can be found in
 section~\ref{sec:method}. 



\subsection{Common set of input parameters}
\label{sec:inputparams}
In the  combination of the  published results, the central  values and
errors are rescaled to a common set of input parameters, summarized in
Table~\ref{tab:common.param}   and   provided   in   the   file   {\tt
common.param} (accessible from the  web-page).  All measurements with a
dependence  on any  of these  parameters are  rescaled to  the central
values  given  in  Table~\ref{tab:common.param},  and their  error  is
recalculated based on the  error provided in the column ``Excursion''.
The detailed dependence for  each measurement  is contained  in files
(provided by the experiments) accessible from the web-page.

\begin{table}[!htb]
\caption{Common input  parameters for the  combination of semileptonic
$B$    decays.   Most    of    the   parameters    are   taken    from
Ref.~\cite{Yao2006PDG}.  This  table  is  encoded in  the  file  {\tt
common.param}. The units are picoseconds for lifetimes and percentage
for branching fractions.}
\begin{center}
\begin{tabular}{|l|c|c|l|}\hline
Parameter    &Assumed Value  &Excursion             &Description\\
\hline\hline 
rb           &$21.629$       &$\pm0.066$            &$R_b$\\
bdst         &$1.27$         &$\pm0.021$            &$\cbf(\Bb\to \Dstar\tau\nub)$\\
bdsd         &$1.62$         &$\pm0.040$            &$\cbf(\Bb\to \Dstar D)$\\
bdst2        &$0.65$         &$\pm0.013$            &$\cbf(b\to \Dstar \tau)$ (OPAL incl) \\
bdsd2        &$4.2$          &$\pm1.5$              &$\cbf(b\to \Dstar D)$ (OPAL incl)\\
bdsd3        &$0.87$         &${}^{+0.23}_{-0.19}$  &$\cbf(b\to \Dstar D)$ (DELPHI incl)\\
xe           &$0.702$        &$\pm0.008$            &$B$ fragmentation:
$\langle E_B\rangle/E_{\rm beam}$\\
bdsi         &$17.3$         &$\pm2.0$              &$\cbf(b\to \Dstarp\ \mathrm{incl})$\\
cdsi         &$22.6$         &$\pm1.4$              &$\cbf(c\to \Dstarp\ \mathrm{incl})$\\
\hline
tb0          &$1.527$        &$\pm0.008$            &$\tau(\Bz)$\\
tbplus       &$1.643$        &$\pm0.010$            &$\tau(\Bp)$\\
tbps         &$1.440$        &$\pm0.036$            &$\tau(\Bs)$\\
\hline
fbd          &$40.2$         &$\pm0.9$              &$\Bz$ fraction at $\sqrt{s} = m_{Z^0}$\\
fbs          &$10.4$         &$\pm0.9$              &$\Bs$ fraction at $\sqrt{s} = m_{Z^0}$\\
fbar         &$9.1$          &$\pm1.5$              &Baryon fraction at $\sqrt{s} = m_{Z^0}$\\
\hline
dst          &$67.7$         &$\pm0.5$              &$\cbf(\Dstarp\to \Dz\pi^+)$\\
dkpp         &$9.51$         &$\pm0.34$              &$\cbf(\Dp \to K^-\pi^+\pi^+)$\\
dkp          &$3.80$         &$\pm0.07$             &$\cbf(\Dz \to K^-\pi^+)$\\
dkpzp        &$14.1$         &$\pm0.5$              &$\cbf(\Dz \to K^-\pi^+\piz)$\\
dkppp        &$7.72$         &$\pm0.28$             &$\cbf(\Dz \to K^-\pi^+\pi^+\pi^-)$\\
dkzpp        &$2.90$         &$\pm0.19$             &$\cbf(\Dz \to K^0\pi^+\pi^-)$\\
dkln         &$6.7$          &$\pm0.19$              &$\cbf(\Dz \to K^-\ell^+\nu)$\\
dkk          &$3.84$         &$\pm0.10$              &$\cbf(\Dz \to K^-K^+)$\\
dkx          &$1.100$        &$\pm0.025$            &$K^-\pi^+X$ rates\\
dkox         &$0.42$         &$\pm0.05$             &$\cbf(\Dz \to K^0X)$\\
dnlx         &$6.71$         &$\pm0.29$             &$\cbf(\Dz \to X\ell\nub)$\\
dkpcl        &$61.9$         &$\pm2.9$              &$\cbf(\Dstarz \to \Dz\piz)$\\
dssR         &$0.64$         &$\pm0.11$             &$\cbf(b\to D^{**}\ell\nub)\times\cbf(D^{**}\to D^{*+}X)$ \\
\hline
fb0          &$49.3$         &$\pm0.7$              &$f^{00} = \cbf(\FourS\to \Bz\Bzb)$\\
chid         &$0.188$        &$\pm0.002$            &$\chi_d$, time-integrated probability for \Bz\ mixing\\
chi          &$0.0925$        &$\pm0.0018$            &$\chi = \chi_d\times (f^{00}/100)$\\
\hline
\end{tabular}
\end{center}
\label{tab:common.param}
\end{table}

\subsection{Exclusive CKM-favored decays}
\label{slbdecays_b2cexcl}
Averages are provided for the branching fractions
$\cbf(\BzbDstarlnu)$ and $\cbf(\BzbDplnu)$.  In addition,
averages are provided for
$\vcb F(1)$ vs $\rho^2$, where $F(1)$ and $\rho^2$
are the normalization and slope of the form factor at zero
recoil in $\BzbDstarlnu$ decays, and for the corresponding quantities
$\vcb G(1)$ vs $\rho^2$ in $\BzbDplnu$ decays.

\mysubsubsection{\BzbDstarlnu}
\label{slbdecays_dstarlnu}
The measurements included in the average, shown in
Table~\ref{tab:dstarlnu}, are scaled to a consistent set of input
parameters and their errors, see Section~\ref{sec:inputparams}.  
Therefore some of the (older)
measurements are subject to considerable adjustments.
Advances have also been made in the determination of $V_{cb}$
from exclusive $B\to D^* l \nu$ decays with substantially improved
measurements of the form factor ratios $R_1$ and $R_2$. The
measurements included in the average have been adjusted to take into
account the new values.
 
In order to reduce the dependence on theoretical error
 estimates, the central values and errors for the form factors $R_1$
 and $R_2$ are taken from the measurement by
 BABAR~\cite{Aubert:2006mb}. 
 The original measurements by CLEO~\cite{Duboscq:1995mv},
 $R_1=1.18\pm0.30\pm0.12$ and $R_2=0.71\pm0.22\pm0.07$
 have been improved
 by BABAR, with results that are consistent with the earlier ones, but
 considerably more precise,
 $R_1=1.417\pm0.061\pm0.044$ and $R_2=0.836\pm0.037\pm0.022$. All
 earlier measurements have been adjusted to these values of $R_1$
 and $R_2$.

\begin{figure}[!ht]
 \begin{center}
  \unitlength1.0cm 
  \begin{picture}(14.,8.0)  
   \put( -0.5,  0.0){\includegraphics[width=7.5cm]{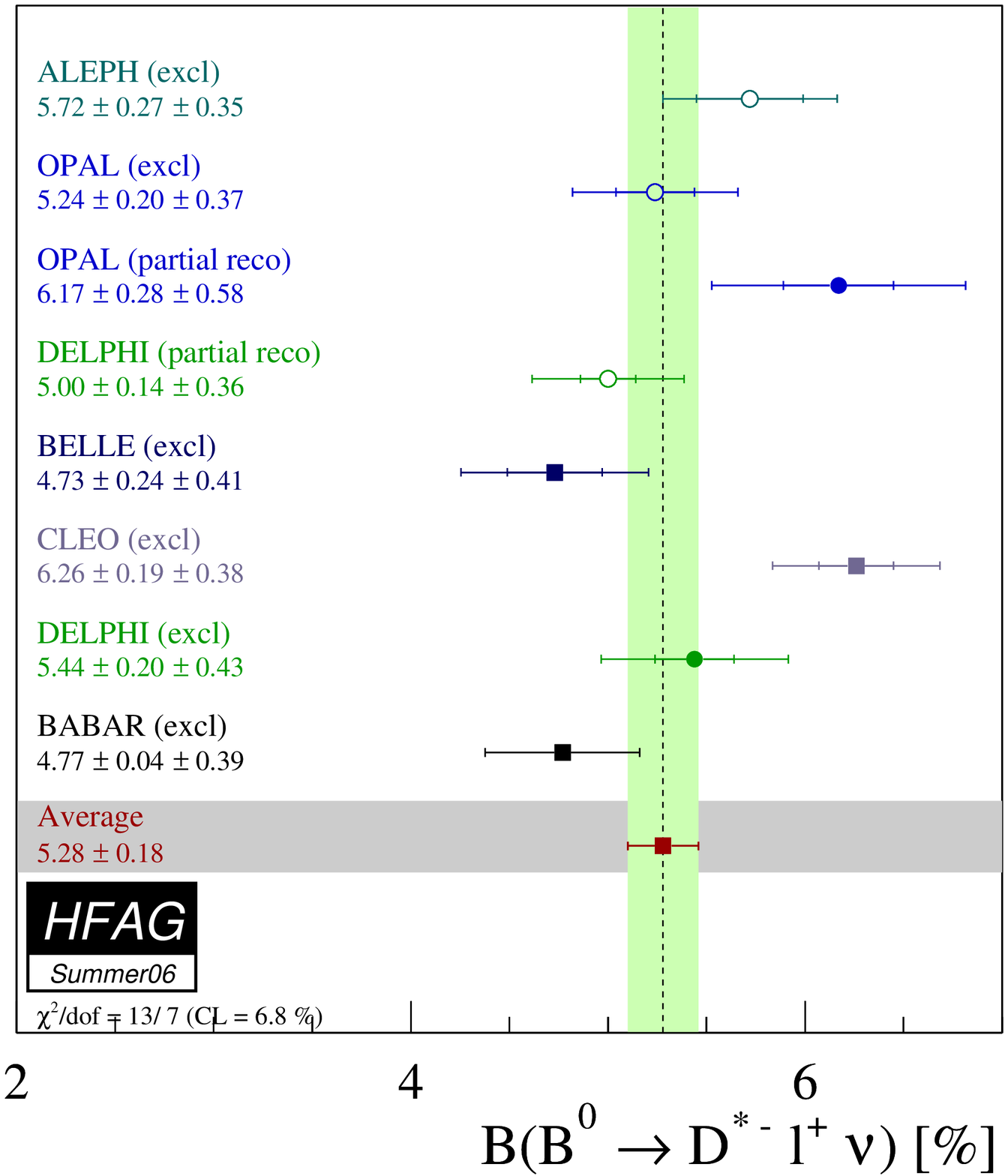}}

   \put(  8.0,  0.0){\includegraphics[width=7.5cm]{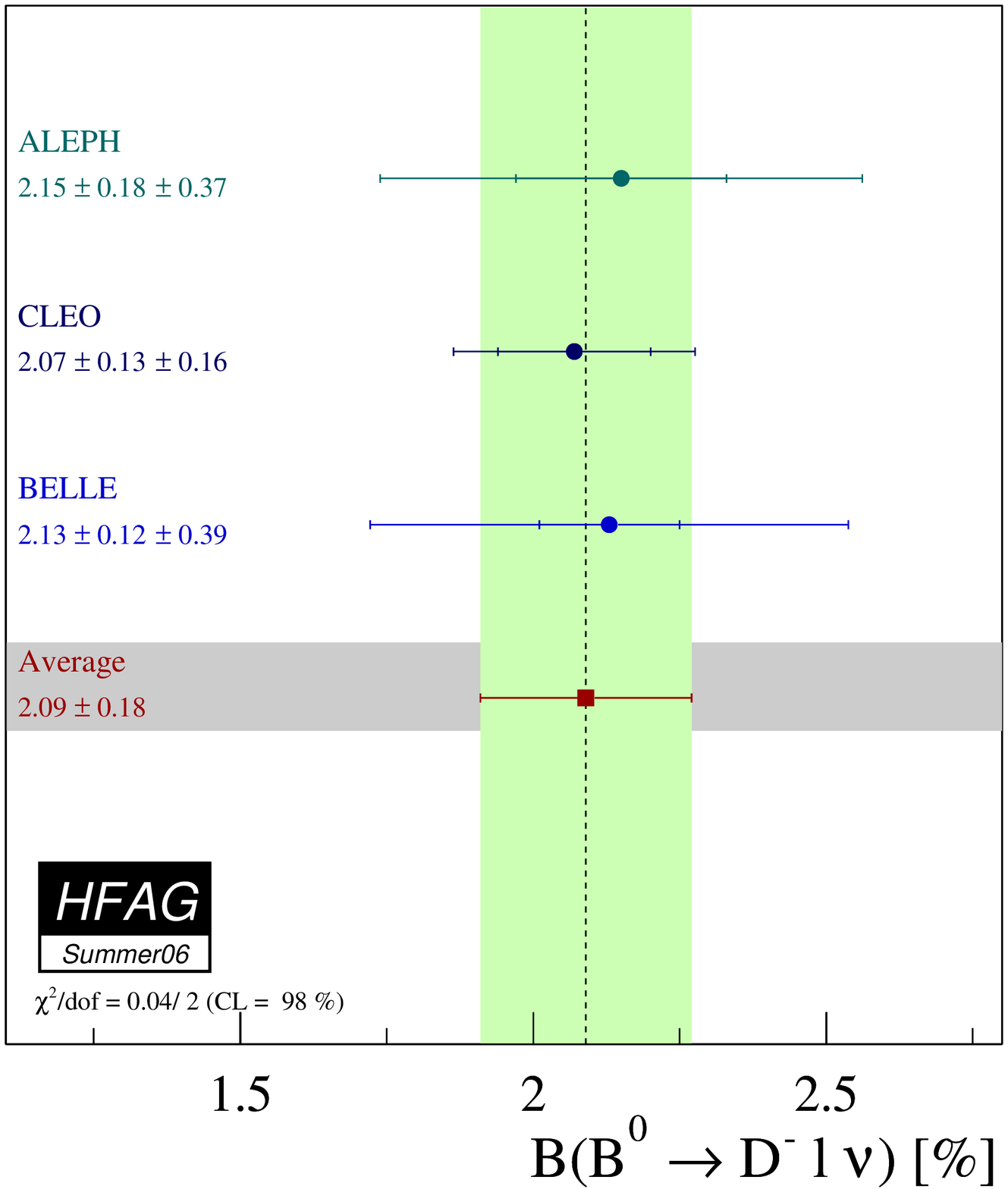}}
    
   \put(  5.5,  6.8){{\large\bf a)}}
   \put( 14.0,  6.8){{\large\bf b)}}
  \end{picture}
  \caption{Average branching fraction  of exclusive semileptonic $B$ decays
(a) \BzbDstarlnu\ and (b) \BzbDplnu\ and individual
  results, where ``excl'' and ``partial reco'' refer to full and
  partial reconstruction of the \BzbDstarlnu\ decay, respectively.}
  \label{fig:brdsl}
 \end{center}
\end{figure}

The average branching fraction $\cbf(\BzbDstarlnu)$ is determined in a
one-dimensional fit from the measurements provided in
Table~\ref{tab:dstarlnu}.  At LEP, the measurements of \BzbDstarlnu\ decays have been done both
with inclusive and exclusive analyses based on a partial and full reconstruction of the
\BzbDstarlnu\ decay, respectively.
Statistical correlations between measurements from the same experiment
have been taken into account. Figure~\ref{fig:brdsl}(a) illustrates the
measurements and the resulting average.  

\begin{table}[!htb]
\caption{Average branching fraction $\cbf(\BzbDstarlnu)$ and individual
  results, where ``excl'' and ``partial reco'' refer to full and
  partial reconstruction of the \BzbDstarlnu\ decay, respectively.}
\begin{center} 
\begin{tabular}{|l|c|c|}\hline
Experiment    &$\cbf(\BzbDstarlnu) [\%]$    (rescaled)     &$\cbf(\BzbDstarlnu) [\%]$  (published) \\
\hline\hline 
ALEPH (excl)~\hfill\cite{Buskulic:1996yq}
&$5.72\pm 0.27_{\rm stat} \pm 0.35_{\rm syst}$
&$5.53\pm 0.26_{\rm stat} \pm 0.52_{\rm syst}$ \\
OPAL (excl)~\hfill\cite{Abbiendi:2000hk}
&$5.24\pm 0.20_{\rm stat} \pm 0.37_{\rm syst}$
&$5.11\pm 0.20_{\rm stat} \pm 0.49_{\rm syst}$ \\
OPAL (partial reco)~\hfill\cite{Abbiendi:2000hk}
&$6.17\pm 0.28_{\rm stat} \pm 0.58_{\rm syst}$
&$5.92\pm 0.28_{\rm stat} \pm 0.68_{\rm syst}$ \\
DELPHI (partial reco)~\hfill\cite{Abreu:2001ic}
&$5.00\pm 0.14_{\rm stat} \pm 0.36_{\rm syst}$
&$4.70\pm 0.14_{\rm stat} \ {}^{+0.36}_{-0.31}\ {}_{\rm syst}$ \\
\belle  (excl)~\hfill\cite{Abe:2001cs}
&$4.73\pm 0.24_{\rm stat} \pm 0.41_{\rm syst}$
&$4.60\pm 0.24_{\rm stat} \pm 0.42_{\rm syst}$ \\
CLEO  (excl)~\hfill\cite{Adam:2002uw}
&$6.26\pm 0.19_{\rm stat} \pm 0.38_{\rm syst}$
&$6.09\pm 0.19_{\rm stat} \pm 0.40_{\rm syst}$ \\
DELPHI (excl)~\hfill\cite{Abdallah:2004rz}
&$5.44\pm 0.20_{\rm stat} \pm 0.43_{\rm syst}$
&$5.90\pm 0.20_{\rm stat} \pm 0.50_{\rm syst}$ \\
\babar\ (excl)~\hfill\cite{Aubert:2006mb}
&$4.77\pm 0.04_{\rm stat}\pm 0.39_{\rm syst}$
&$4.77\pm 0.04_{\rm stat} \pm 0.39_{\rm syst}$ \\
\hline 
{\bf Average}                              &\mathversion{bold}$5.28\pm0.18$           &\mathversion{bold}$\chi^2/\dof = 13/7$ (CL=$6.8\%$)  \\
\hline 
\end{tabular}
\end{center}
\label{tab:dstarlnu}
\end{table}

\begin{figure}[!ht]
 \begin{center}
  \unitlength1.0cm 
  \begin{picture}(14.,8.0)  
   \put(  8.0, -0.2){\includegraphics[width=8.0cm]{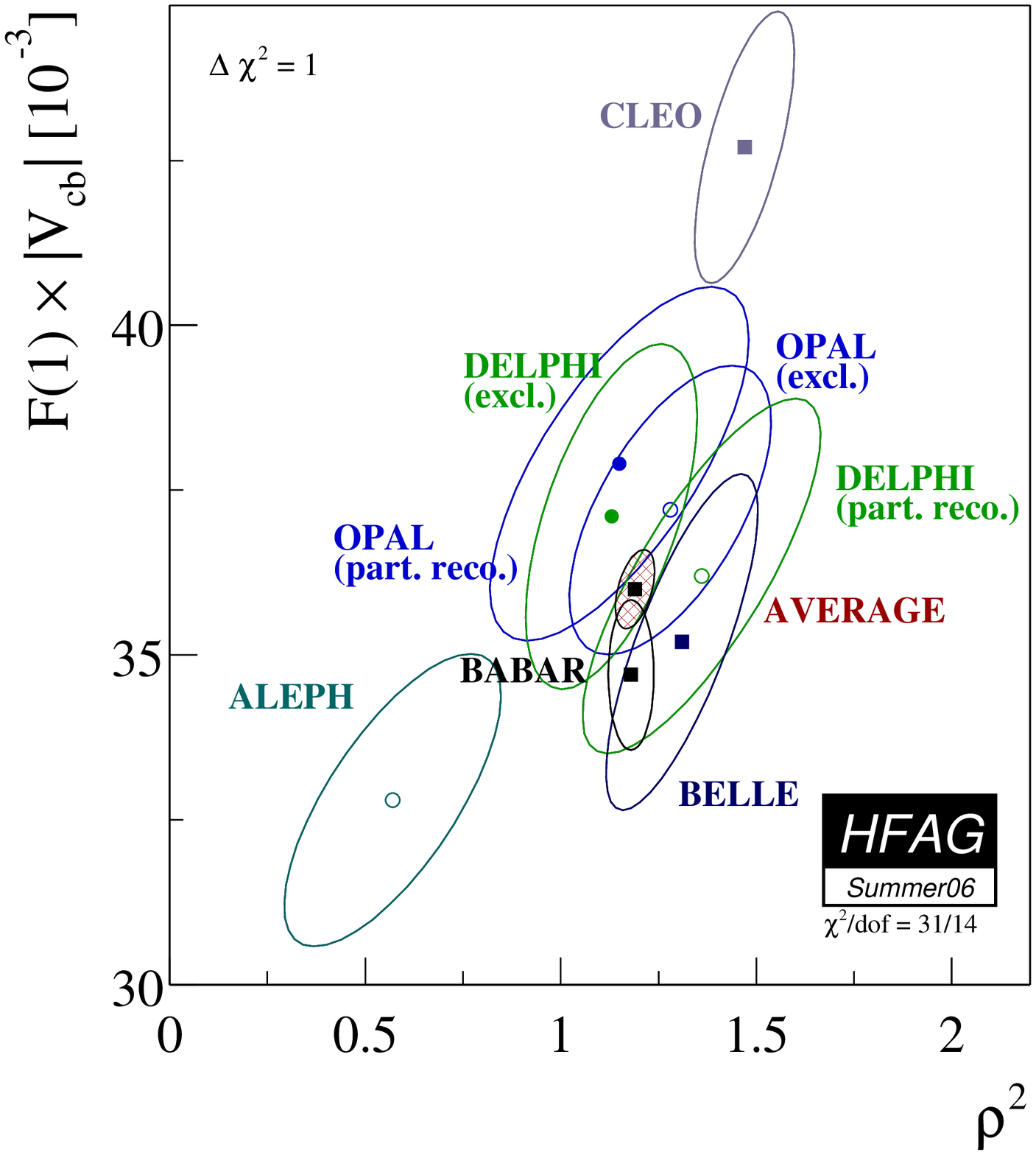}}
   \put( -0.5,  0.0){\includegraphics[width=7.5cm]{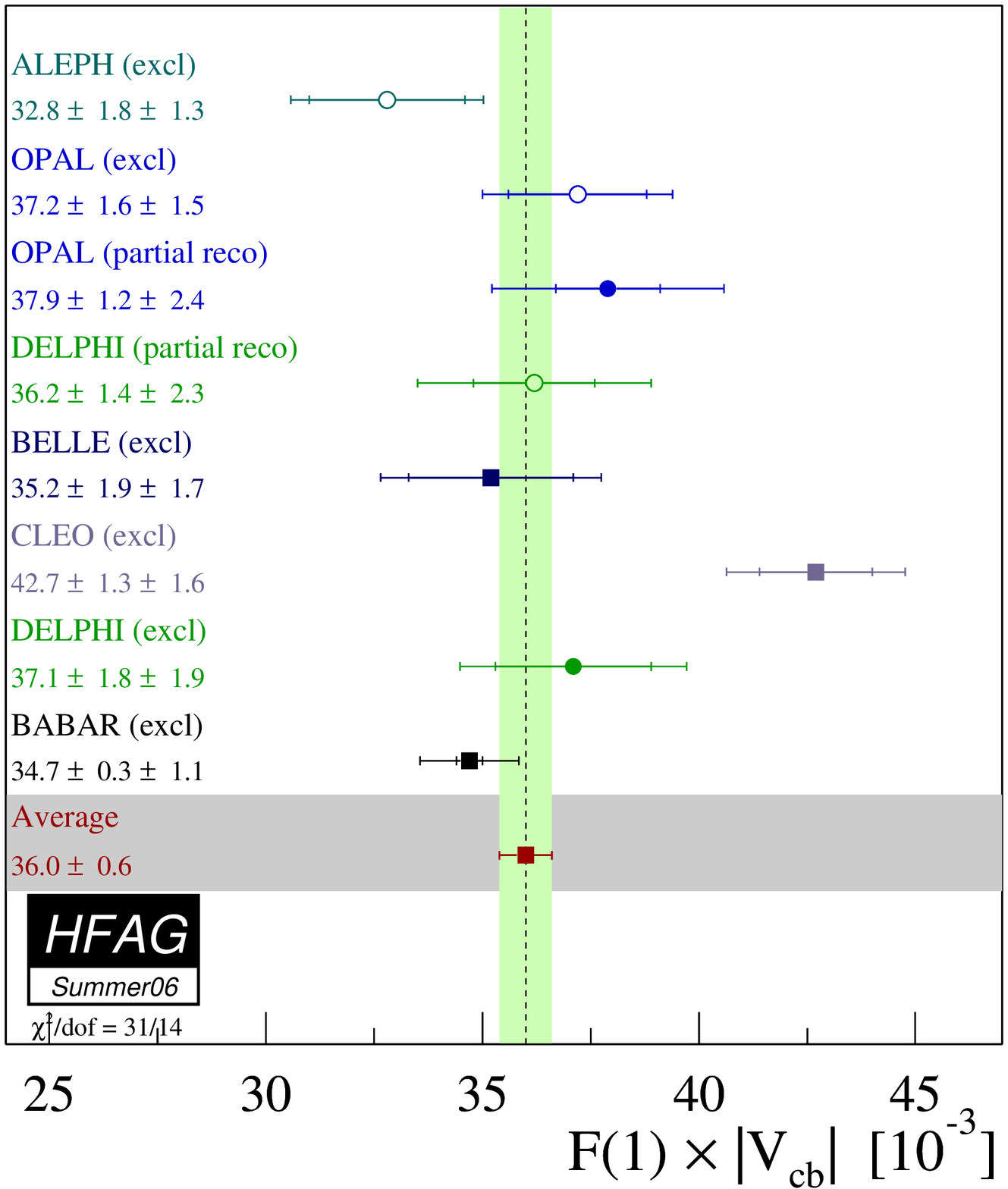}}
   \put(  5.5,  6.8){{\large\bf a)}}  
   \put( 14.4,  6.8){{\large\bf b)}}
   \end{picture} \caption{(a)   Illustration  of
   $F(1)\vcb$   vs.  $\rho^2$.  The   error  ellipses   correspond  to
   $\Delta\chi^2 = 1$ (CL=39\%).
   (b)  Illustration of the  average $F(1)\vcb$
   and   rescaled  measurements   of   exclusive  \BzbDstarlnu\   decays
   determined   in  a  two-dimensional   fit, where ``excl'' 
   and ``partial reco'' refer to full and partial reconstruction.
  }  \label{fig:vcbf1} \end{center}
\end{figure}

The average for $F(1)\vcb$ is determined by the two-dimensional
combination of the results provided in Table~\ref{tab:vcbf1}.  This
allows the correlation between $F(1)\vcb$ and $\rho^2$ to be
maintained. Figure~\ref{fig:vcbf1}(a) illustrates the average
$F(1)\vcb$ and the measurements included in the
average. Figure~\ref{fig:vcbf1}(b) provides a one-dimensional
projection for illustrative purposes. The largest 
systematic errors correlated between measurements are owing to
uncertainties on: $R_b$,
the ratio of production cross-sections
$\sigma_{b\bar{b}}/\sigma_{\rm{had}}$, the $B^0$ fraction at
$\sqrt{s}=m_{Z^0}$, the branching fractions $\cbf(D^0\to K^-\pi^+)$ and
$\cbf(D^0\to K^-\pi^+\pi^0)$, the correlated background from $D^{**}$,
and the $D^*$ form factor ratios $R_1$ and $R_2$.
Together these uncertainties account for about two thirds of the
systematic error. In all the measurements the total systematic errors are
reduced with respect to the published values because the values and
uncertainties assumed for parameters on which these measurements
depend, for example $R_1$ and $R_2$,  have since been better determined.


\begin{table}[!htb]
\caption{Average of $F(1)\vcb$ determined in the decay \BzbDstarlnu\ and
individual  results, where ``excl'' and ``partial reco'' refer to full and
  partial reconstruction of the \BzbDstarlnu\ decay, respectively.
The  fit  for  the average  has  $\chi^2/\dof  =
31/14$ (CL=$0.5\%$).  The total  correlation between  the average  $F(1)\vcb$ and
$\rho^2$ is 0.46.}
\begin{center}
\begin{tabular}{|l|c|c|}\hline
Experiment                                 &$F(1)\vcb [10^{-3}]$ (rescaled) &$\rho^2$ (rescaled) \\
                                           &$F(1)\vcb [10^{-3}]$ (published) &$\rho^2$ (published) \\
\hline\hline 
ALEPH (excl)~\hfill\cite{Buskulic:1996yq}
&$32.8\pm 1.8_{\rm stat} \pm 1.3_{\rm syst}$
&$0.57\pm 0.25_{\rm stat} \pm 0.12_{\rm syst}$ \\
&$31.9\pm 1.8_{\rm stat} \pm 1.9_{\rm syst}$
&$0.37\pm 0.26_{\rm stat} \pm 0.14_{\rm syst}$ \\
\hline
OPAL (excl)~\hfill\cite{Abbiendi:2000hk}
&$37.2\pm 1.6_{\rm stat} \pm 1.5_{\rm syst}$
&$1.28\pm 0.21_{\rm stat} \pm 0.15_{\rm syst}$ \\
&$36.8\pm 1.6_{\rm stat} \pm 2.0_{\rm syst}$
&$1.31\pm 0.21_{\rm stat} \pm 0.16_{\rm syst}$ \\
\hline
OPAL (partial reco)~\hfill\cite{Abbiendi:2000hk}
&$37.9\pm 1.2_{\rm stat} \pm 2.4_{\rm syst}$
&$1.15\pm 0.14_{\rm stat} \pm 0.30_{\rm syst}$ \\
&$37.5\pm 1.2_{\rm stat} \pm 2.5_{\rm syst}$
&$1.12\pm 0.14_{\rm stat} \pm 0.29_{\rm syst}$ \\
\hline
DELPHI (partial reco)~\hfill\cite{Abreu:2001ic}
&$36.2\pm 1.4_{\rm stat} \pm 2.3_{\rm syst}$
&$1.36\pm 0.14_{\rm stat} \pm 0.27_{\rm syst}$ \\
&$35.5\pm1.4_{\rm stat}\ {}^{+2.3}_{-2.4}{}_{\rm syst}$
&$1.34\pm0.14_{\rm stat}\ {}^{+0.24}_{-0.22}{}_{\rm syst}$\\
\hline
\belle  (excl)~\hfill\cite{Abe:2001cs}
&$35.2\pm 1.9_{\rm stat} \pm 1.7_{\rm syst}$
&$1.31\pm 0.16_{\rm stat} \pm 0.11_{\rm syst}$ \\
&$35.8\pm 1.9_{\rm stat} \pm 1.9_{\rm syst}$
&$1.45\pm 0.16_{\rm stat} \pm 0.20_{\rm syst}$ \\
\hline
CLEO  (excl)~\hfill\cite{Adam:2002uw}
&$42.7\pm 1.3_{\rm stat} \pm 1.6_{\rm syst}$
&$1.47\pm 0.09_{\rm stat} \pm 0.09_{\rm syst}$ \\
&$43.1\pm 1.3_{\rm stat} \pm 1.8_{\rm syst}$
&$1.61\pm 0.09_{\rm stat} \pm 0.21_{\rm syst}$ \\
\hline
DELPHI (excl)~\hfill\cite{Abdallah:2004rz}
&$37.2\pm 1.8_{\rm stat} \pm 1.9_{\rm syst}$
&$1.13\pm 0.15_{\rm stat} \pm 0.16_{\rm syst}$ \\
&$39.2\pm 1.8_{\rm stat} \pm 2.3_{\rm syst}$
&$1.32\pm 0.15_{\rm stat} \pm 0.33_{\rm syst}$ \\
\hline
\babar\ (excl)~\hfill\cite{Aubert:2006mb}
&$34.7\pm 0.3_{\rm stat} \pm 1.1_{\rm syst}$
&$1.18\pm 0.05_{\rm stat} \pm 0.03_{\rm syst}$ \\
&$34.7\pm 0.3_{\rm stat} \pm 1.1_{\rm syst}$
&$1.18\pm 0.05_{\rm stat} \pm 0.03_{\rm syst}$ \\
\hline 
    {\bf Average }
    &\mathversion{bold}$36.0\pm0.6$        &\mathversion{bold}$1.19\pm0.05$ \\
\hline 
\end{tabular}
\end{center}
\label{tab:vcbf1}
\end{table}

For a determination of \vcb, the form factor at zero recoil $F(1)$
needs to be computed.  A possible choice is
$F(1) = 0.919^{+0.030}_{-0.035}$~\cite{Hashimoto:2001nb},
which takes into account the QED correction($+0.7\%$), resulting in

\begin{displaymath}
\vcb = (39.2 \pm 0.7_{\rm exp} \pm 1.4_{\rm theo}) \times 10^{-3},
\end{displaymath}
where the errors are from experiment and theory, respectively.

\noindent

\mysubsubsection{\BzbDplnu}
\label{slbdecays_dlnu}

The average branching fraction $\cbf(\BzbDplnu)$ is determined by the
combination of the results provided in Table~\ref{tab:dlnu}.  The
error sources here are the same as discussed for $\cbf(\BzbDstarlnu)$,
but generally higher due to
larger background levels, less stringent kinematic constraints, and
larger kinematic suppression at the endpoint. 
Figure~\ref{fig:brdsl}(b) illustrates the measurements and the
resulting average.

\begin{table}[!htb]
\caption{Average of the branching fraction $\cbf(\BzbDplnu)$ and individual
results. }
\begin{center}
\begin{tabular}{|l|c|c|}\hline
Experiment                                 &$\cbf(\BzbDplnu) [\%]$ (rescaled) &$\cbf(\BzbDplnu) [\%]$ (published) \\
\hline\hline 
ALEPH ~\hfill\cite{Buskulic:1996yq}        &$2.15 \pm0.18_{\rm stat} \pm0.37_{\rm syst}$  &$2.35 \pm0.18_{\rm stat} \pm0.44_{\rm syst}$ \\
CLEO  ~\hfill\cite{Bartelt:1998dq}         &$2.07 \pm0.13_{\rm stat} \pm0.16_{\rm syst}$  &$2.20 \pm0.13_{\rm stat} \pm0.18_{\rm syst}$ \\
\belle  ~\hfill\cite{Abe:2001yf}           &$2.13 \pm0.12_{\rm stat} \pm0.39_{\rm syst}$  &$2.13 \pm0.12_{\rm stat} \pm0.41_{\rm syst}$ \\
\hline 
{\bf Average}                              &\mathversion{bold}$2.09 \pm0.18$      &\mathversion{bold}$\chi^2/\dof = 0.04/2$ (CL=$98\%$)  \\
\hline 
\end{tabular}
\end{center}
\label{tab:dlnu}
\end{table}

\begin{figure}[!ht]
 \begin{center}
  \unitlength1.0cm 
  \begin{picture}(14.,8.) 
   \put(  8.0, -0.2){\includegraphics[width=8.0cm]{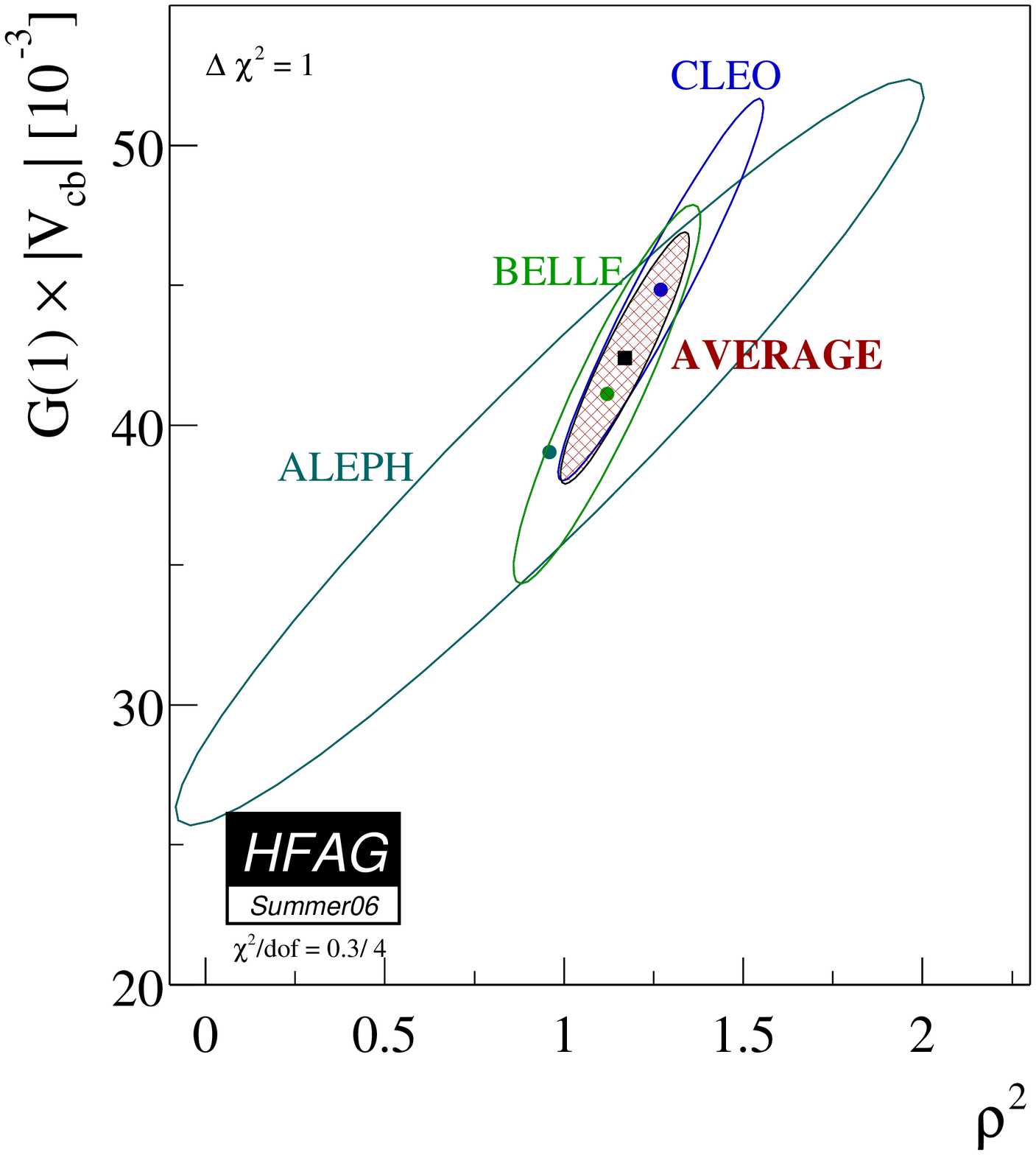}}
   \put( -0.5,  0.0){\includegraphics[width=7.5cm]{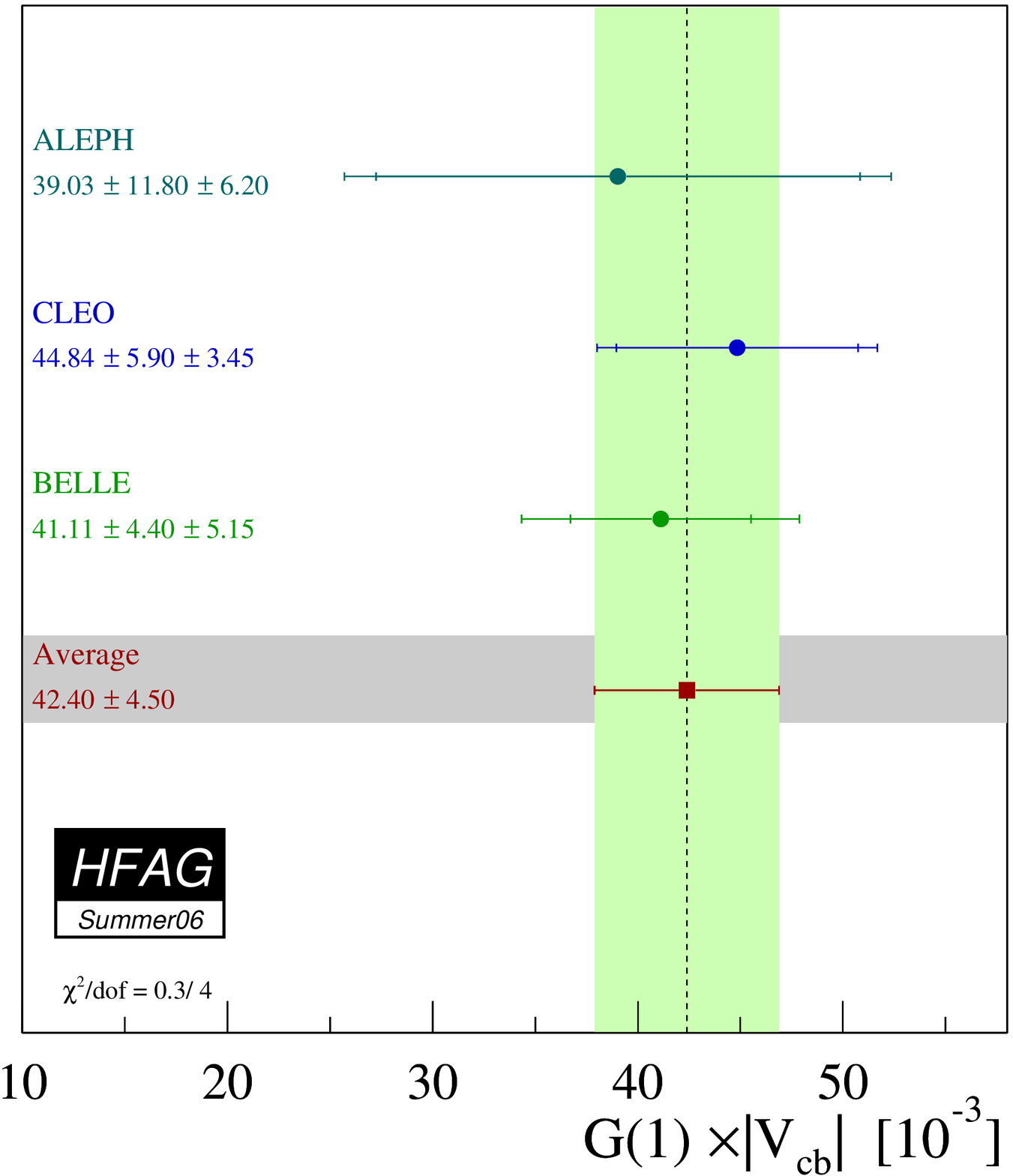}}
   \put(  5.5,  6.8){{\large\bf a)}}
   \put( 14.4,  6.8){{\large\bf b)}}
  \end{picture}
  \caption{(a) Illustration  of
   $G(1)\vcb$   vs.  $\rho^2$.  The   error  ellipses   correspond  to
   $\Delta\chi^2 = 1$. 
  (b) Illustration of the  average $G(1)\vcb$
   and   rescaled  measurements   of   exclusive  \BzbDplnu\   decays
   determined   in  a  two-dimensional   fit.   }
  \label{fig:vcbg1}
 \end{center}
\end{figure}

The average for $G(1)\vcb$ is determined by the two-dimensional
combination of the results provided in Table~\ref{tab:vcbg1}.
Figure~\ref{fig:vcbg1} (a)
provides a one-dimensional projection for illustrative purposes,
(b) illustrates the average $G(1)\vcb$ and the
measurements included in the average. 

\begin{table}[!htb]
\caption{Average  of $G(1)\vcb$  determined  in the  decay \BzbDplnu\  and
individual  results. The  fit  for the  average  has $\chi^2/\dof  =
0.3/4$.   The total  correlation between  the average  $G(1)\vcb$ and
$\rho^2$ is $0.93$.}
\begin{center}
\begin{tabular}{|l|c|c|}\hline
Experiment &$G(1)\vcb [10^{-3}]$ (rescaled)  &$\rho^2$ (rescaled) \\ 
           &$G(1)\vcb [10^{-3}]$ (published) &$\rho^2$ (published) \\
\hline\hline 
ALEPH~\hfill\cite{Buskulic:1996yq}  &$39.0 \pm11.8_{\rm stat} \pm6.2_{\rm syst}$   &$0.96 \pm0.98_{\rm stat} \pm0.36_{\rm syst}$ \\
                                    &$31.1 \pm9.9_{\rm stat}  \pm8.6_{\rm syst}$   &$0.20 \pm0.98_{\rm stat} \pm0.50_{\rm syst}$ \\
\hline
CLEO ~\hfill\cite{Bartelt:1998dq}   &$44.8 \pm5.9_{\rm stat} \pm3.5_{\rm syst}$    &$1.27 \pm0.25_{\rm stat} \pm0.14_{\rm syst}$ \\
                                    &$44.8 \pm6.1_{\rm stat} \pm3.7_{\rm syst}$  &$1.30 \pm0.27_{\rm stat} \pm0.14_{\rm syst}$ \\
\hline
\belle~\hfill\cite{Abe:2001yf}       &$41.1 \pm4.4_{\rm stat} \pm5.2_{\rm syst}$    &$1.12 \pm0.22_{\rm stat} \pm0.14_{\rm syst}$ \\
                                    &$41.1 \pm4.4_{\rm stat} \pm5.1_{\rm syst}$    &$1.12 \pm0.22_{\rm stat} \pm0.14_{\rm syst}$ \\
\hline 
{\bf Average }                      &\mathversion{bold}$42.4 \pm 4.5$       &\mathversion{bold}$1.17 \pm0.18$      \\
\hline 
\end{tabular}
\end{center}
\label{tab:vcbg1}
\end{table}

For a determination of \vcb, the form factor at zero recoil $G(1)$
needs to be computed.  A possible choice is
$G(1) = 1.04\pm0.01_{\rm{power}}\pm0.01_{\rm{pert}}$~\cite{Uraltsev:2003ye},
which takes into account the QED correction($+0.7\%$), resulting in

\begin{displaymath}
\vcb = (40.8 \pm 4.3_{\rm exp} \pm 0.6_{\rm theo}) \times 10^{-3},
\end{displaymath}
where the errors are from experiment and theory, respectively.

\subsection{Inclusive CKM-favored decays}
\label{slbdecays_b2cincl}

\subsubsection{Inclusive Semileptonic Branching Fraction for $\Bz/\Bu\to X\ell\nul$}

The average of the inclusive semileptonic branching fraction has 
undergone revision. We have gone from averaging measurements of full 
branching fractions, which relied on various models, to averaging
partial branching fractions, which are relatively model independent. 
The subsequent average of the partial branching fractions is extrapolated to the full phase
space using a model-independent approach.   

\begin{figure}[!ht]
 \begin{center}
  \unitlength1.0cm 
  \begin{picture}(14.,8.0)  
   \put(  8.0, -0.2){\includegraphics[width=8.0cm]{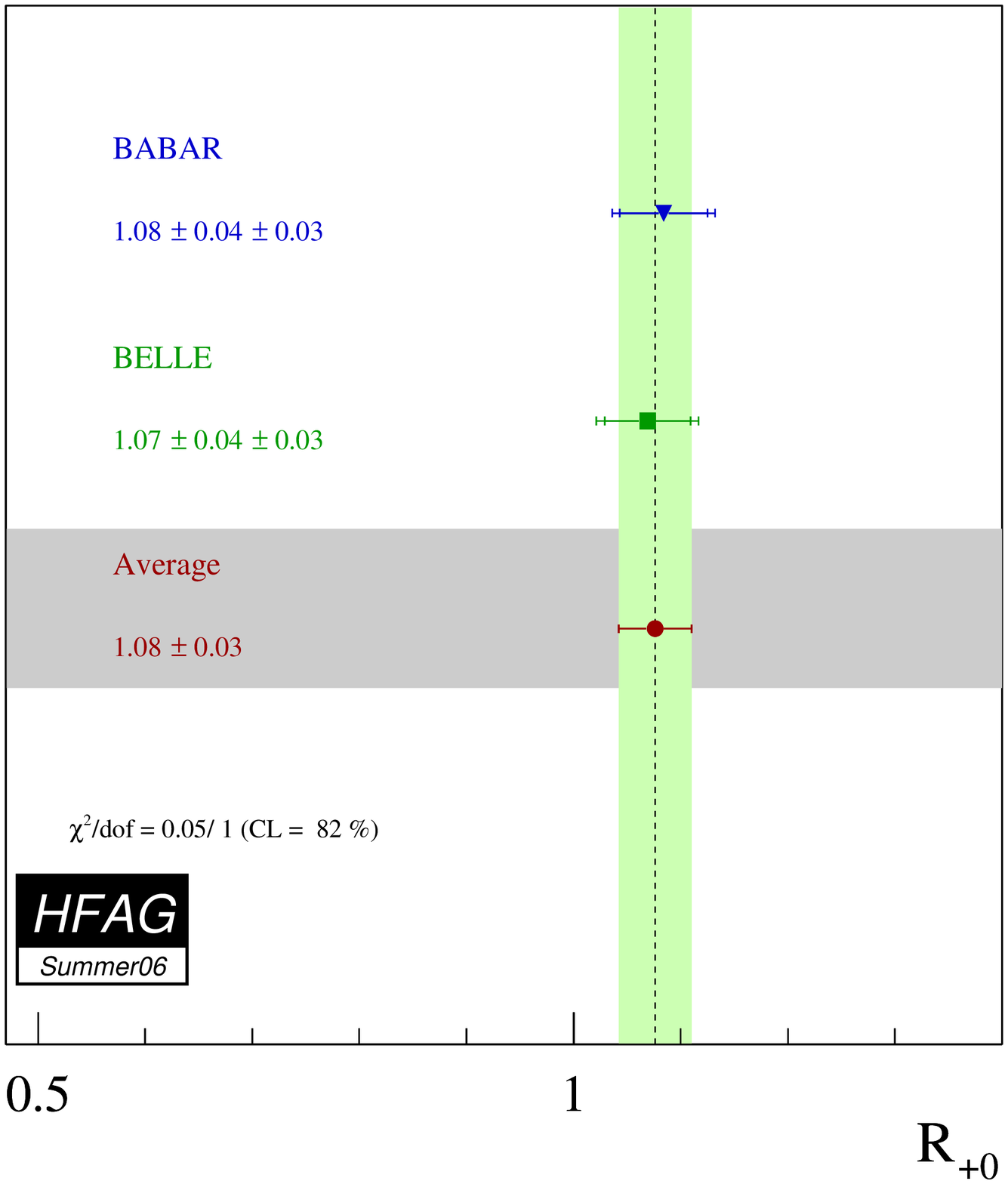}}
   \put( -0.5,  0.0){\includegraphics[width=8.0cm]{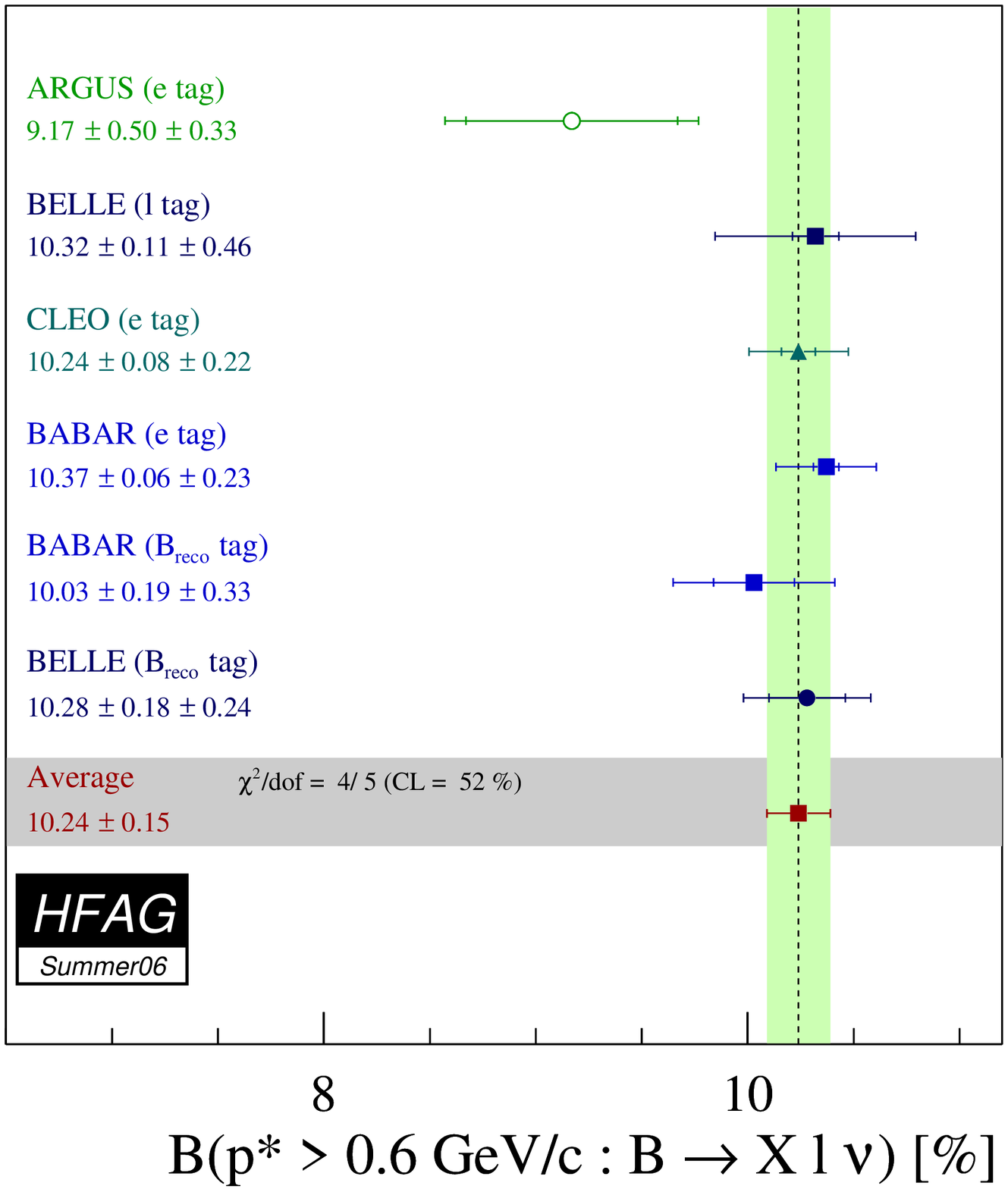}}
   \put(  5.5,  6.8){{\large\bf a)}}  
   \put( 14.4,  6.8){{\large\bf b)}}
   \end{picture} \caption{
(a) Measurements of $\cbf(\Bz/\Bu\to X\ell\nul)$ and their average. 
(b) Measurements of the ratio of the branching fractions $R_{+0}$ and their average.
In parenthesis we identify the type of analysis performed: $l$-tag and
$e$-tag refer to inclusive analyses where both electrons and muons
or just the former are reconstructed, respectively, and
$B_{\rm{reco}}$-tag refer to analyses at the $\FourS$ where one of the
$B$-mesons is fully reconstructed in a hadronic mode with the
lepton reconstructed from the other $B$ decay.}
\label{fig:b2xlnubf}
\end{center}
\end{figure}

We use measurements that require the momentum of
the prompt charged lepton $(p_{\mathrm{cms}})$ to be greater than $0.6$ GeV/$c$,
as measured in the rest frame of
either the $B$-meson or $\Upsilon(4S)$~\footnote{The difference in reference frames has a negligible impact.}. 
The measurements and average are given in Table~\ref{tab:brisltot} and
plotted in Figure~\ref{fig:b2xlnubf}. We extrapolate from the partial to full
branching fraction using a factor  derived from a global fit~\cite{Buchmuller:2005zv}
used to extract HQET parameters in the Kinetic
Scheme~\cite{Gambino:2004qm}. That fit, which doesn't include any 
of the measurements used in our average, yielded the
extrapolation factor $1.0495 \pm
0.0005 \pm 0.0010$. The first error includes experiment and theory
uncertainties as got from the fit.
The second error is assigned as an additional theoretical uncertainty
for the ratio $\Gamma_{\mathrm{sl}}(0.6)/\Gamma_{\mathrm{sl}}(0.0)$ -
for the total width ($\Gamma_{\mathrm{sl}}(0.0)$) this uncertainty is
typically $1.5$\%,
on the recommendation of the authors of the fit, for the ratio it is estimated to be $0.1$\%. 
Finally the full branching fraction is extracted as
$\cbf(\Bz/\Bu\to X\ell\nul) = (10.75 \pm 0.16)\%$ . 

\begin{table}[!htb]
\caption{Average of the partial semileptonic branching fractions
  $\cbf(\Bz/\Bu\to X\ell\nul)(p_{\mathrm{cms}}>0.6\,\rm{GeV}/c)$ and the
  full branching fraction extrapolated from the average. In
  parentheses we identify the type of analysis performed: $\ell$-tag and
  $e$-tag refer to inclusive analyses where both electrons and muons
  or just electrons are reconstructed, respectively, and
  $B_{\rm{reco}}$-tag refers to analyses at the $\FourS$ where one of the
  $B$-mesons is fully reconstructed in a hadronic mode with the
  lepton reconstructed from the other $B$ decay.}
\begin{center}
\begin{tabular}{|l|c|c|}\hline
Experiment                                   &$\cbf(\Bz/\Bu\to X\ell\nul)[\%]$ (rescaled)  \\
       & $(p_{\mathrm{cms}}>0.6\,\rm{GeV}/c)$   \\
\hline\hline 
ARGUS ($e$-tag)~\hfill\cite{Albrecht:1993pu}       & $9.17  \pm0.50 \pm0.33$ \\
\belle ($\ell$-tag)~\hfill\cite{Abe:2002du}        & $10.32 \pm0.11 \pm0.46$ \\
CLEO  ($e$-tag)~\hfill\cite{Mahmood:2004kq}        & $10.24 \pm0.08 \pm0.22$ \\
\babar ($e$-tag)~\hfill\cite{Aubert:2004td}        & $10.37 \pm0.06 \pm0.23$ \\
\babar (\breco-tag)~\hfill\cite{Aubert:2006au}    & $10.03 \pm0.19 \pm0.33$ \\
\belle (\breco-tag)~\hfill\cite{Urquijo:2006wd}    & $10.28 \pm0.18 \pm0.24$ \\
\hline 
    {\bf Average at $(p_{\mathrm{cms}}>0.6\,\rm{GeV}/c)$}
    &\mathversion{bold}$10.24\pm0.15$ \\
    & \mathversion{bold}$\chi^2/\dof = 4.2/5$ (CL=$52\%$)  \\
    \hline
    $\cbf_{tot}(\Bz/\Bu\to X\ell\nul)$ (\%)&
       $\mathbf{10.75 \pm 0.16}$ \\ 
    \hline
\end{tabular}
\end{center}
\label{tab:brisltot}
\end{table}

\subsubsection{Ratio of $\cbf(\Bu\to X\ellp\nul)$ to $\cbf(\Bz\to X^-\ellp\nul)$}
The total width of semileptonic $B$-meson decays is
expected to be the same for both neutral and charged channels.
Therefore the ratio of the branching fractions, $R_{+0}$, 
should be equivalent to the ratio of the B-meson lifetimes
$\tau_{B+}/\tau_{B^0}$, where
\[
R_{+0} = \frac{\cbf(\Bu\to X\ellp\nul)}{\cbf(\Bz\to X^-\ellp\nul)}.
\]
Recently both Babar and Belle reported precise
measurements of $R_{+0}$, using a ``$B_{\rm{reco}}$''-tagged sample, 
for which we provide an average. The
measurements and average are listed in Table~\ref{tab:Rp0} and plotted in
Figure~\ref{fig:b2xlnubf}. The average, $1.076 \pm 0.034$ is in agreement with the
ratio of lifetimes, $1.076 \pm 0.008$~\cite{Barberio:2006bi}.

\begin{table}[!htb]
\caption{Individual measurements and average of the ratio of the branching fractions $R_{+0}$.}
\begin{center}
\begin{tabular}{|l|c|}\hline
Experiment                                   &$R_{+0}$    \\
\hline\hline 
\babar~\hfill\cite{Aubert:2006au}   &$1.084\pm0.041\pm0.025$    \\
\belle~\hfill\cite{Urquijo:2006wd}   &$1.069\pm0.040\pm0.026$    \\
\hline 
    {\bf Average}
    & \mathversion{bold}$1.076 \pm 0.034$ \\
       & \mathversion{bold}$\chi^2/\dof = 0.05/1$  \\
      & {\bf CL=}\mathversion{bold}$82\%$ \\
    \hline
\end{tabular}
\end{center}
\label{tab:Rp0}
\end{table}

\subsubsection{Branching Fractions for $\Bu\to X\ellp\nul$ and $\Bz\to X^-\ellp\nul$}
\begin{figure}[!ht]
 \begin{center}
  \unitlength1.0cm 
  \begin{picture}(14.,8.0)  
   \put( -0.5,  0.0){\includegraphics[width=8.0cm]{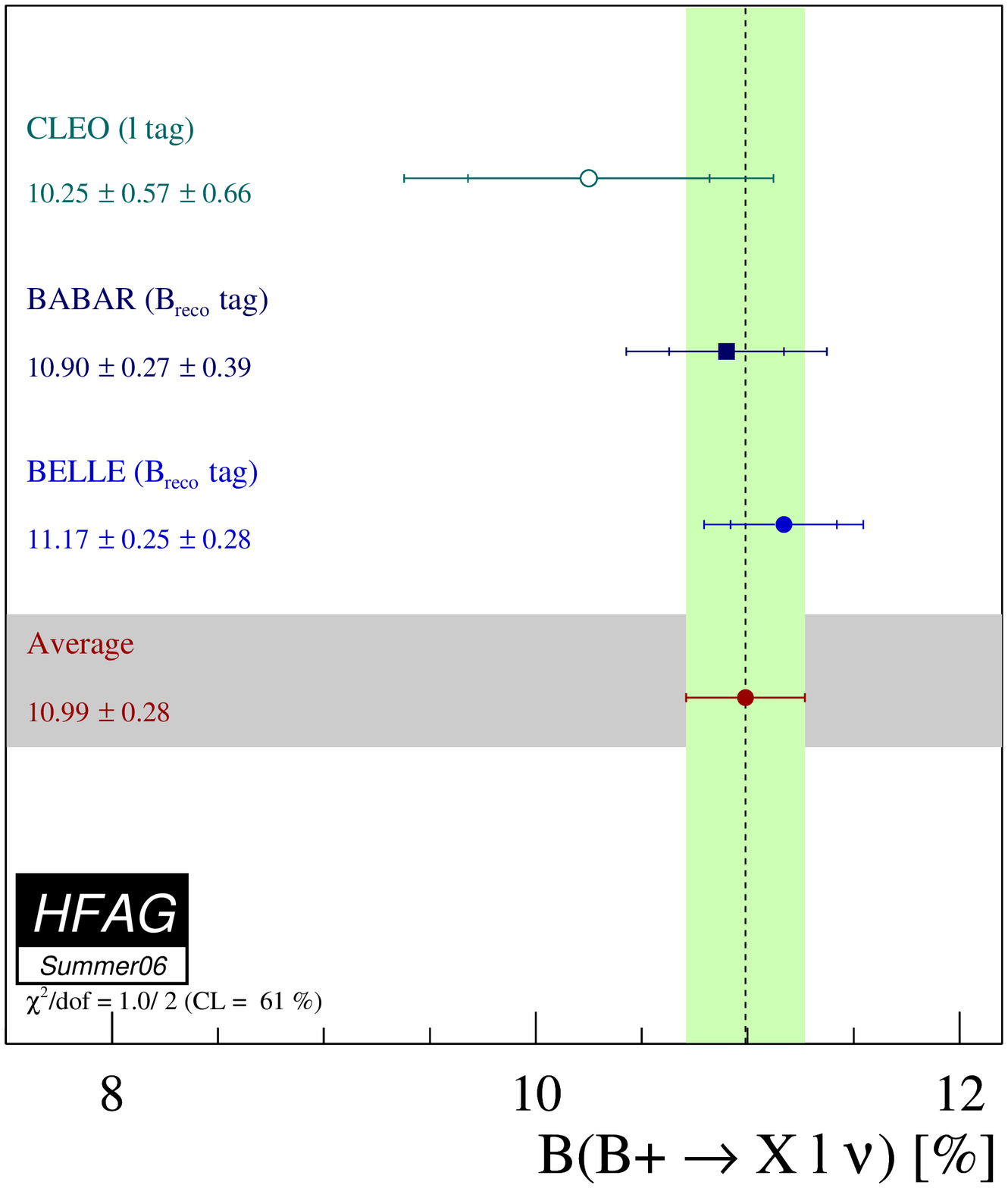}}
   \put(  8.0, -0.2){\includegraphics[width=8.0cm]{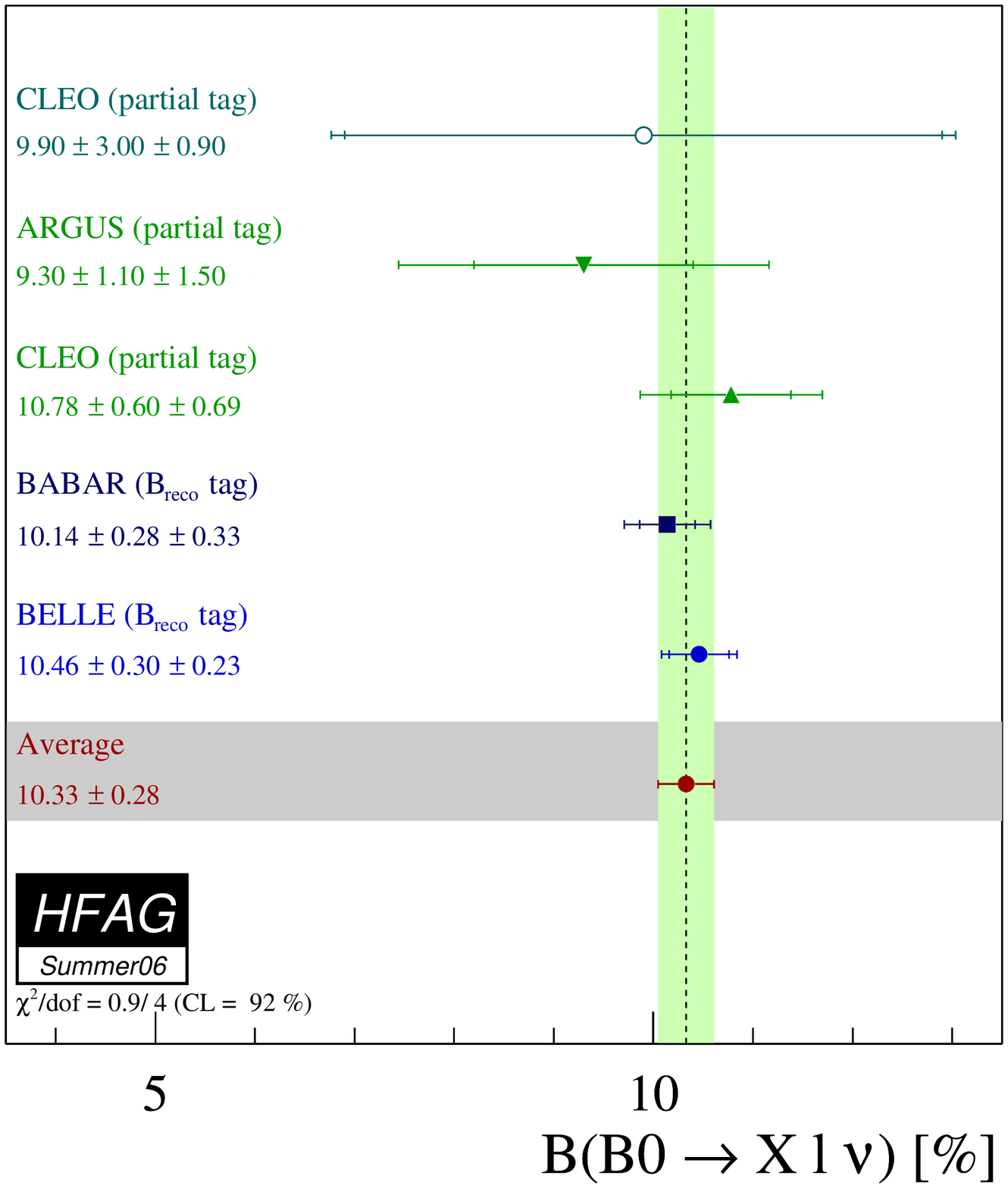}}
   \put(  5.5,  6.8){{\large\bf a)}}  
   \put( 14.4,  6.8){{\large\bf b)}}
   \end{picture} \caption{(a) Measurements of the total semileptonic 
branching fraction $\cbf(\Bu\to X\ellp\nul)$ and their average. 
(b) Individual measurements and average of the total semileptonic branching fraction 
$\cbf(\Bz\to X^-\ellp\nu)$. ``$\ell$-tag'', ``partial-tag'' and ``\breco-tag'' indicate analysis 
experimental technique.}
\label{fig:brisltotbp0}
\end{center}
\end{figure}

For the first time we provide averages of the branching fractions of $\Bu\to X\ellp\nul$ and
$\Bz\to X^-\ellp\nul$ separately,  
using the available measurements at the \FourS. We include the
measurements listed in the Review of Particle Physics~\cite{Yao2006PDG} and
as well as the latest measurements made by Belle and Babar that utilise
the full reconstruction $B$-meson tag~\cite{Aubert:2006au,Urquijo:2006wd}. In contrast to the $B$
admixture average, averages are made of the full branching
fraction since the CLEO and ARGUS papers do not stipulate partial
branching fractions. The measurements and averages are given in
Tables~\ref{tab:brisltotbp} and~\ref{tab:brisltotb0} and plotted in
Figure~\ref{fig:brisltotbp0} for
$\Bu\to X\ellp\nul$ and $\Bz \to X^- l^+ \nul$, respectively.  

\begin{table}[!htb]
  \caption{Individual measurements and average of the total
    semileptonic branching fraction
    $\cbf(\Bu\to X\ellp\nul)$. ``$\ell$-tag'' and ``\breco-tag'' indicate analysis experimental technique.}
\begin{center}
\begin{tabular}{|l|c|}\hline
Experiment                                   &$\cbf_{tot}(\Bu\to X\ellp\nul)[\%]$ (rescaled)    \\
\hline\hline 
CLEO  ($\ell$-tag)~\hfill\cite{Artuso:1997we}          &$10.25 \pm0.57 \pm0.66$    \\
\babar (\breco-tag)~\hfill\cite{Aubert:2006au}        &$10.90 \pm0.27 \pm0.39$    \\
\belle (\breco-tag)~\hfill\cite{Urquijo:2006wd}        &$11.17 \pm0.25 \pm0.28$     \\
\hline 
    {\bf Average}
    & \mathversion{bold}$10.99 \pm 0.28$ \\
       & \mathversion{bold}$\chi^2/\dof = 1.0/2$  \\
       &   {\bf CL=}\mathversion{bold}$61\%$ \\
    \hline
\end{tabular}
\end{center}
\label{tab:brisltotbp}
\end{table}

\begin{table}[!htb]
\caption{Individual measurements and average of the total semileptonic branching fraction $\cbf(\Bz\to X^-\ellp\nu)$. ``partial-tag'' and ``\breco-tag'' indicate analysis experimental technique. }
\begin{center}
\begin{tabular}{|l|c|}\hline
Experiment & $\cbf_{tot}(\Bz\to X^-\ellp\nu)[\%]$ (rescaled)   \\ 
\hline\hline 
CLEO  (partial-tag)~\hfill\cite{Henderson:1991ks}      &$9.9\pm3.0\pm0.9$ \\
ARGUS (partial-tag)~\hfill~\cite{Albrecht:1993gr}      &$9.3\pm1.1\pm 1.5$ \\
CLEO  (partial-tag)~\hfill\cite{Artuso:1997we}         &$10.78\pm0.60\pm0.69$ \\
\babar (\breco-tag)~\hfill\cite{Aubert:2006au}        &$10.14\pm0.28\pm0.33$ \\
\belle (\breco-tag)~\hfill\cite{Urquijo:2006wd}        &$10.46\pm0.30\pm0.23$ \\
\hline 
    {\bf Average} & \mathversion{bold}$10.33\pm0.28$ \\
   &  \mathversion{bold}$\chi^2/\dof =0.9/4$ \\
   &   {\bf CL=} \mathversion{bold}$92\%$ 
       \\
    \hline
\end{tabular}
\end{center}
\label{tab:brisltotb0}
\end{table}

\subsubsection{\vcb\, Determined from $B\to X \ell \nub$}
The magnitude of the CKM matrix $|V_{cb}|$ can be determined from the branching fraction
of inclusive charmed semileptonic $B$-meson decays $\cbf(\Bxclnu)$ and
with parameters that describe the motion of the $b$-quark in the
$B$-meson. These parameters, within the framework of the Heavy Quark
Expansion (HQE), include the $b$-quark mass, $m_b$. Phenomenology  
of these decays is reviewed in many papers~\cite{b2xintros}. In practice $|V_{cb}|$,
$m_b$, and other parameters are determined simultaneously
from a {\it global} fit to data of inclusive semileptonic
and radiative $B$-meson decays. These data include rates
and moments of energy and hadronic mass spectra measured as a function of
cut-offs in those variables. HFAG has yet to quote an average value of $|V_{cb}|$ obtained from
such fits. To date three types of global fits have been performed; these 
differ in the choice of scheme, either {\it pole}, ``1S'' or
{\it kinetic}, used to define the $b$-quark mass $m_b$. T
able~\ref{tab:gfits} lists the results of global fits performed to date. 
We are working to determine $|V_{cb}|$ from global fits in both the
``1S'' and {\it kinetic} schemes. The fits will utilise a common
set of measurements. Fits in the {\t pole} scheme are disfavoured due
to the poor convergence of series expansions in that scheme.

\begin{table}[!htb]
\caption{{\it Global fit} results for the {\it pole}, ``1S'' and {\it
  kinetic} schemes.}
\begin{center}
\begin{tabular}{|l|l|}\hline
  Source (Scheme)& Measurements \\
\hline\hline 
Battaglia {\it et al.} (Kinetic)~\cite{Battaglia:2002tm}
& $|V_{cb}|=(41.9\pm 0.7_{meas}\pm0.6_{fit}\pm0.4_{pert})\times 10^{-3}$ \\
& $m_b^{\rm{kin}}=4.59 \pm 0.08_{fit}\pm 0.01_{syst.}\,\rm{GeV}/c^2$ \\
\hline
Battaglia {\it et al.} (Pole)~\cite{Battaglia:2002tm}
& $|V_{cb}|=(41.3\pm 0.7_{meas}\pm0.7_{fit}\pm0.2_{nl}\pm0.9_{pert})\times 10^{-3}$ \\
& $\mathrm{\bar{\Lambda}}=0.40 \pm 0.10_{fit}\pm
0.02_{syst.}\,\rm{GeV}/c^2$ \\
\hline
CLEO (Pole)~\cite{Mahmood:2002tt}
& $|V_{cb}|=(40.8\pm 0.5_{\Gamma_{\rm{SL}}}\pm0.4_{\lambda_1,\mathrm{\bar{\Lambda}}}\pm0.9_{theory})\times 10^{-3}$ \\
& $\mathrm{\bar{\Lambda}}=0.39 \pm 0.03_{stat}\pm0.06_{syst.}\pm0.12_{theory}\,\rm{GeV}/c^2$ \\
(1S) & $m_b^{\rm{1S}}=4.82 \pm
0.07_{exp}\pm0.11_{theory}\,\rm{GeV}/c^2$ \\
\hline
BABAR (Kinetic)~\cite{Aubert:2004aw}
& $|V_{cb}|=(41.4 \pm 0.4_{exp}\pm0.4_{HQE}\pm0.6_{theory})\times 10^{-3}$ \\
& $m_b^{\rm{kin}}=4.61 \pm 0.05_{exp}\pm0.04_{HQE}\pm0.02_{theory}\,\rm{GeV}/c^2$ \\\hline
Bauer {\it et al.} (1S)~\cite{Bauer:2004ve}
&  $|V_{cb}|=(41.4 \pm 0.6 \pm0.1_{\tau_B})\times 10^{-3}$ \\
&  $m_b^{\rm{1S}}=4.68 \pm 0.03\,\rm{GeV}/c^2$ \\\hline
Buchm\"uller \& Fl\"acher (Kinetic)~\cite{Buchmuller:2005zv}
& $|V_{cb}|=(41.96 \pm 0.23_{exp}\pm0.35_{HQE}\pm0.59_{\Gamma_{\rm{SL}}})\times 10^{-3}$ \\
& $m_b^{\rm{kin}}=4.59 \pm 0.025_{exp}\pm0.030_{HQE}\,\rm{GeV}/c^2$ \\\hline
Belle (Kinetic)~\cite{Urquijo:ICHEP06}
& $|V_{cb}|=(41.93 \pm 0.65_{fit}\pm0.48_{\alpha_s}\pm0.68_{theory})\times 10^{-3}$ \\
& $m_b^{\rm{kin}}=4.564 \pm 0.076\,\rm{GeV}/c^2$ \\\hline
Belle ($1S$)~\cite{Urquijo:ICHEP06}
& $|V_{cb}|=(41.5 \pm 0.5_{fit}\pm0.2_{\tau_B})\times 10^{-3}$ \\
& $m_b^{\rm{1S}}=4.73 \pm 0.05\,\rm{GeV}/c^2$ \\
\hline 
\end{tabular}
\end{center}
\label{tab:gfits}
\end{table}

Updated values for the parameters $m_b$ and $\mu_{\pi}^2$ are taken
from the Buchm\"uller and Fl\"acher global fit and 
used below in the determination of
\vub\ from inclusive decays. Their fit uses many different measurements: 
BABAR~\cite{ref:BABARmoments,ref:BABARmoments1,ref:BABARbclnumomentsfits,ref:BABARbsgexcl,ref:BABARbsgincl},
Belle~\cite{ref:BELLEmoments,ref:BELLEmoments1,ref:BELLEbsg}, CLEO~\cite{ref:CLEOmoments,ref:CLEOmoments1,ref:CLEObsg}
CDF~\cite{ref:CDFmoments}, and DELPHI~\cite{ref:DELPHImoments}.
We use their result since it uses most of the currently available measurements.

\subsection{Exclusive CKM-suppressed decays}
\label{slbdecays_b2uexcl}
Several new measurements of the exclusive decay
$\Bb\to\pi\ell\nub$ were presented
at the 2006 Summer conferences. Their precision is at a level
that calls for improved calculations of the form factors and in particular their
normalization. 

\begin{figure}
\begin{center}
\includegraphics[width=6.0in]{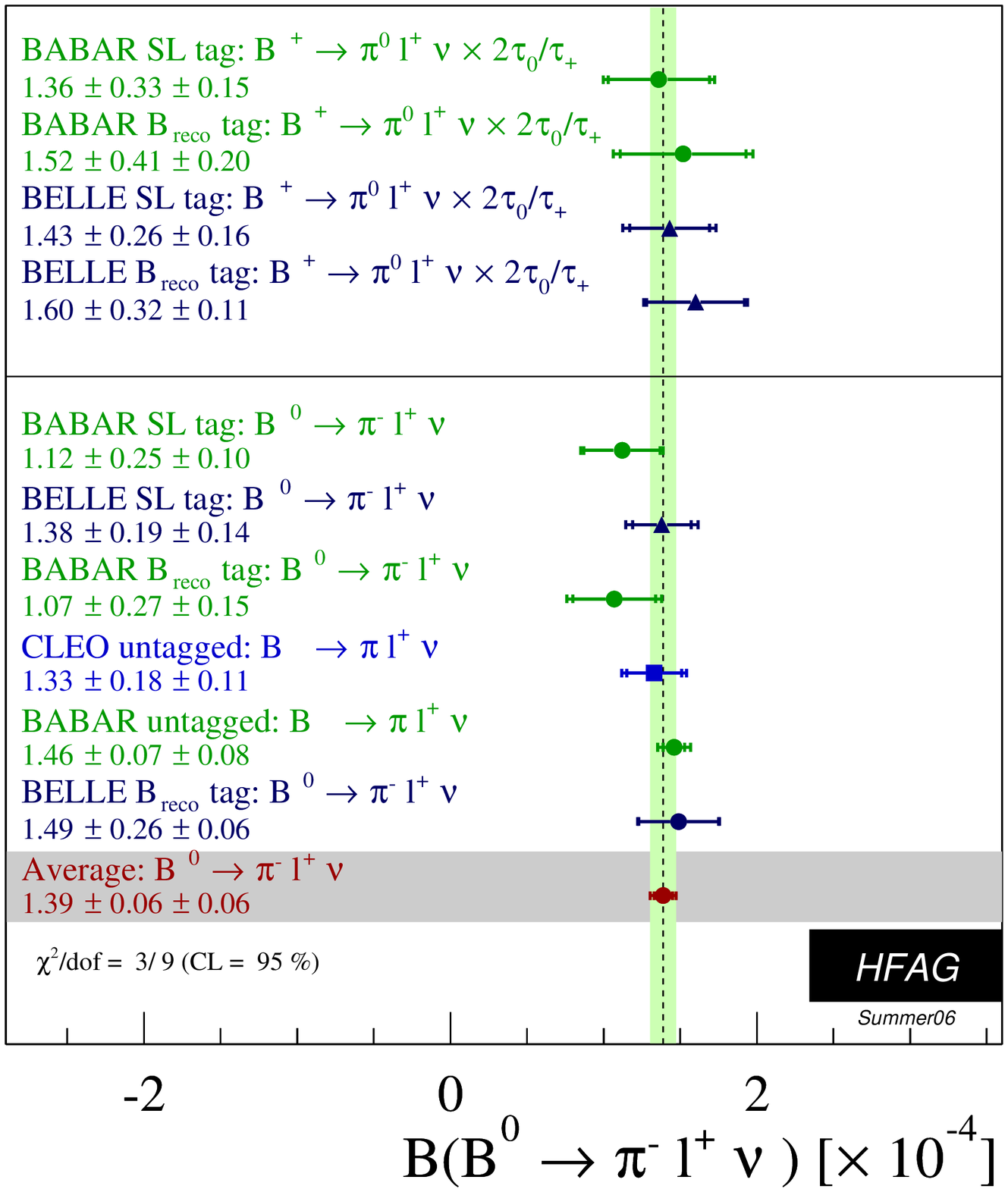}
\end{center}
\caption{Summary of exclusive determinations of $\cbf(\Bb\to\pi
\ell\nub)$ and their average. 
Measured branching fractions for $B\rightarrow \pi^0 l \nu$ have been
multiplied by $2\times \tau_{B^0}/\tau_{B^+}$ in accordance with
isospin symmetry. The labels ``$B_{reco}$'' and ``SL''
refer to type of $B$ decay tag used in a measurement.
``untagged'' refers to an untagged measurement.}
\label{fig:pilnubf}
\end{figure}

Here we list results on exclusive semileptonic branching fractions
and determinations of $\vub$ based on $\Bb\to\pi\ell\nub$ decays.
The measurements are based on two different event selections: tagged
events, in which case the second $B$ meson in the event is fully
reconstructed in either a hadronic decay (``$B_{reco}$'') or in a 
CKM-favored semileptonic
decay (``SL''); and untagged events, in which case the selection infers the momentum
of the undetected neutrino based on measurements of the total 
momentum sum of detected particles and knowledge of the initial state.
The results for the full and partial branching fraction are given
in Table~\ref{tab:pilnubf} and shown in Figure~\ref{fig:pilnubf}.   

When averaging these results, systematic uncertainties due to external
inputs, e.g., form factor shapes and background estimates from the
modeling of $\Bb\to X_c\ell\nub$ and $\Bb\to X_u\ell\nub$ decays, are
treated as fully correlated (in the sense of Eq.~\ref{eq:correlrho}).
Uncertainties due to experimental reconstruction effects are treated
as fully correlated among measurements from a given experiment.  Varying
the assumed dependence of the quoted errors on the measured value (see
Eq.~\ref{eq:chi2that}) for error sources where the
dependence was not obvious had no significant impact.


\begin{table}[!htb]
\caption{Summary of exclusive determinations of $\cbf(\Bb\to\pi
\ell\nub)$. The errors quoted
correspond to statistical and systematic uncertainties, respectively.
Measured branching fractions for $B\rightarrow \pi^0 l \nu$ have been
multiplied by $2\times \tau_{B^0}/\tau_{B^+}$ in accordance with
isospin symmetry. The labels ``$B_{reco}$'' and ``SL'' tags refer to
the type of $B$
decay tag used in a measurement, and ``untagged'' refers to an untagged measurement.}
\begin{center}
\begin{small}
\begin{tabular}{|lccc|}
\hline
& $\cbf [10^{-4}]$
& $\cbf(q^2>16\,\gev^2/c^2) [10^{-4}]$
& $\cbf(q^2<16\,\gev^2/c^2) [10^{-4}]$
\\
\hline\hline
CLEO $\pi^+,\pi^0$~\cite{ref:CLEOpilnu}
& $1.33\pm 0.18\pm 0.11\ $ 
& $0.25\pm 0.09\pm 0.05$
& $1.08\pm 0.16\pm 0.10$
\\ 
BABAR $\pi^+$~\cite{ref:BABARpilnu}
& $1.46\pm 0.07\pm 0.08\ $
& $0.38\pm 0.04\pm 0.03$
& $1.09\pm 0.06\pm 0.07$
\\  \hline
{\bf Average of untagged}
& \mathversion{bold}$1.43 \pm 0.07\pm 0.09\ $
& \mathversion{bold}$0.35\pm 0.04\pm 0.04$
& \mathversion{bold}$1.09\pm 0.06\pm 0.07$
\\ \hline\hline
BELLE SL $\pi^+$~\cite{ref:BELLEpilnuSL}
& $1.38\pm 0.19\pm 0.14\ $
& $0.36\pm 0.10\pm 0.04$
& $1.02\pm 0.16\pm 0.11$
\\ 
BELLE SL $\pi^0$~\cite{ref:BELLEpilnuSL}
& $1.43\pm 0.26\pm 0.16\ $
& $0.37\pm 0.15\pm 0.04$
& $1.06\pm 0.22\pm 0.11$
\\ 
BABAR SL $\pi^+$~\cite{ref:BABARtagged}
& $1.12\pm 0.25\pm 0.12\ $
& $0.29\pm 0.15\pm 0.04$
& $0.83\pm 0.22\pm 0.08$
\\ 
BABAR SL $\pi^0$~\cite{ref:BABARtagged}
& $1.36\pm 0.33\pm 0.19\ $
& $0.19\pm 0.22\pm 0.07$
& $1.17\pm 0.30\pm 0.11$
\\ 
BABAR $B_{reco}$ $\pi^+$~\cite{ref:BABARtagged}
& $1.07\pm 0.27\pm 0.17\ $
& $0.65\pm 0.20\pm 0.13$
& $0.42\pm 0.18\pm 0.05$
\\ 
BABAR $B_{reco}$ $\pi^0$~\cite{ref:BABARtagged}
& $1.52\pm 0.41 \pm0.20\ $
& $0.48\pm 0.22\pm 0.11$
& $1.04\pm 0.35\pm 0.15$
\\ 
BELLE $B_{reco}$ $\pi^+$~\cite{ref:BELLEpilnuBreco}
& $1.49\pm 0.20\pm 0.16\ $
& n/a
& n/a
\\ 
BELLE $B_{reco}$ $\pi^0$~\cite{ref:BELLEpilnuBreco}
& $1.60\pm 0.32\pm0.11\ $
& n/a
& n/a
\\  \hline
{\bf Average of tagged}
& \mathversion{bold}$1.35 \pm 0.10\pm 0.07\ $
& \mathversion{bold}$0.36\pm 0.06\pm 0.03$
& \mathversion{bold}$0.83\pm 0.09\pm 0.06$
\\  \hline\hline
{\bf Average}
& \mathversion{bold}$1.39 \pm 0.06\pm 0.06\ $
& \mathversion{bold}$0.35\pm 0.03\pm 0.03$
& \mathversion{bold}$0.97\pm 0.05\pm 0.05$
\\ 
\hline\hline
\end{tabular}\\
\end{small}
\end{center}
\label{tab:pilnubf}
\end{table}

The determination of \vub\ from the $\Bb\to\pi\ell\nub$ decays is
shown in Table~\ref{tab:pilnuvub} and uses our average for the branching
fraction given in Table~\ref{tab:pilnubf}.  Two theoretical approaches are
used: Lattice QCD (quenched and unquenched) and 
QCD sum rules.
Lattice calculations of the Form Factors (FF) are limited to small hadron momenta, i.e.
large $q^2$, while calculations based on light cone sum rules are restricted
to small $q^2$. More precise calculations of the FF, in particular their
normalization, are needed to reduce the overall uncertainties.


\begin{table}[hbtf]
\caption{\label{tab:pilnuvub}
Determinations of \vub\ based on the average total and partial
$\Bb\to\pi\ell\nub$ decay branching fraction stated in
Table~\ref{tab:pilnubf}. The first
uncertainty is experimental, and the second is from theory.  The
full or partial \cbf\ are used as indicated.
}
\begin{center}
\begin{tabular}{|lc|}
\hline
Method & $\Vub [10^{-3}]$ \\\hline\hline
LCSR, full $q^2$~\cite{BallZwicky} & $3.43\pm 0.10 {}^{+0.67}_{-0.42}$ \\ 
LCSR, $q^2<16\,\gev^2/c^2$~\cite{BallZwicky}   & $3.41\pm 0.13 {}^{+0.56}_{-0.38}$ \\ \hline
HPQCD, full $q^2$~\cite{HPQCDpilnu}& $3.89\pm 0.12 {}^{+0.84}_{-0.51}$ \\ 
HPQCD, $q^2>16\,\gev^2/c^2$~\cite{HPQCDpilnu}  & $3.97\pm 0.25 {}^{+0.59}_{-0.41}$ \\  \hline
FNAL, full $q^2$~\cite{FNALpilnu}  & $3.82\pm 0.12 {}^{+0.88}_{-0.52}$ \\ 
FNAL, $q^2>16\,\gev^2/c^2$~\cite{FNALpilnu}    & $3.55\pm 0.22 {}^{+0.61}_{-0.40}$ \\  \hline
APE, full $q^2$~\cite{Abada:2000ty}  & $3.61\pm 0.11 {}^{+1.11}_{-0.57}$ \\ 
APE, $q^2>16\,\gev^2/c^2$~\cite{Abada:2000ty}    & $3.58\pm 0.22 {}^{+1.37}_{-0.63}$ \\ 
\hline
\end{tabular}
\end{center}
\end{table}

Branching fractions for other $\Bb\to X_u\ell\nub$ decays are given in
Table~\ref{tab:xslother}.  At this time the determination of $\Vub$
from these other channels looks less promising than for
$\Bb\to\pi\ell\nub$.

\begin{table}[!htb]
\caption{Summary of branching fractions to $\cbf(\Bb\to X\ell\nub)$ decays
other than $\Bb\to\pi\ell\nub$. The errors quoted
correspond to statistical and systematic
uncertainties, respectively.  Where a third uncertainty is quoted, it
corresponds to uncertainties from form factor shapes.}
\begin{center}
\begin{small}
\begin{tabular}{|llc|}
\hline
Experiment & Mode & $\cbf [10^{-4}]$  \\\hline\hline
CLEO~\cite{ref:CLEOrholnu} & $\Bz\to\rho^-\ell\nub$ & $2.69\pm 0.41\ {}^{+\ 0.35}_{-\ 0.40}\ \pm 0.50$ \\ 
BABAR~\cite{ref:BABARrholnuBreco} & $\Bz\to\rho^-\ell\nub$ & $2.57\pm 0.52\pm 0.59\ \quad\ \,\,\quad  $ \\ 
BABAR~\cite{ref:BABARrholnuBreco2} & $\Bz\to\rho^-\ell\nub$ & $3.29\pm 0.42\pm 0.47\pm 0.60 $ \\ 
BABAR~\cite{ref:BABARpilnu} & $\Bz\to\rho^-\ell\nub$ & $2.14\pm 0.21\pm 0.51\pm 0.28 $ \\ 
BELLE~\cite{ref:BELLEpilnuSL} & $\Bz\to\rho^-\ell\nub$ & $2.17\pm 0.54\pm 0.31\pm 0.08$ \\ 
BELLE~\cite{ref:BELLEpilnuSL} & $\Bp\to\rho^0\ell\nub$ & $1.33\pm 0.23\pm 0.17\pm 0.03$ \\ 
BELLE~\cite{ref:BELLEomegalnu} & $\Bp\to\omega\ell\nub$ & $1.3\ \,\pm 0.4\ \,\pm 0.2\ \,\pm 0.3\ \,$ \\ 
CLEO~\cite{ref:CLEOpilnu} & $\Bp\to\eta\ell\nub$ & $0.84\pm 0.31\pm 0.16 \pm 0.09$ \\ 
BABAR~\cite{ref:BABARetalnuBreco} & $\Bp\to\eta\ell\nub$ & $0.84\pm 0.27\pm 0.21$ \\ 
BABAR~\cite{ref:BABARetalnuBreco} & $\Bp\to\eta^\prime\ell\nub$ & $0.33\pm 0.60\pm 0.30$ \\ 
\hline
\end{tabular}\\
\end{small}
\end{center}
\label{tab:xslother}
\end{table}



\subsection{Inclusive CKM-suppressed decays}
\label{slbdecays_b2uincl}
Our last update~\cite{Barberio:2006bi}
provided a best value of $|V_{ub}|$ that  
was the first to be based on common inputs in a consistent framework. In
this update we use two theory prescriptions
in addition to BLNP~\cite{ref:BLNP}, namely DGE~\cite{ref:DGE} and
BLL~\cite{ref:BLL}.
We determine an average for \vub\ for each of BLNP, DGE, and BLL, and
do not advocate the use of one method over another.

The large background from $\Bb\to X_c\ell\nub$ decays is the chief
experimental limitation in determinations of $\vub$.  Cuts designed to
reject this background limit the acceptance for $\Bb\to X_u\ell\nub$
decays. The calculation of partial rates for these restricted
acceptances is more complicated and requires substantial theoretical machinery.
BLNP and DGE authors have provided us codes that allows us to perform
the calculation for all the limited regions of phase space covered by
measurements. BLL provides calculations for measurements that cut on the dilepton mass
squared and hadronic mass.

In the averages performed the systematic errors associated with the
modeling of $\Bb\to X_c\ell\nub$ and $\Bb\to X_u\ell\nub$ decays and the theoretical
uncertainties are taken as fully correlated among all measurements
in the sense of Eq.~\ref{eq:correlrho}. Reconstruction-related
uncertainties are taken as fully correlated within a given experiment.
From the three results quoted in Ref.~\cite{ref:belle-mx}, only one is
used in the average, as they are all based on the same dataset and are
highly correlated. Specifically we use the $M_X$ analysis result for
BLNP and DGE averages, and 
the $M_X,q^2$ analysis result for the BLL average.
The other experimental results have negligible
statistical correlations. The assumed dependence of the quoted error
on the measured value was input for each source of error, as
discussed in section~\ref{sec:method}.
Measurements of partial decay rates for $\Bb\to X_u\ell\nub$
transitions from $\Upsilon(4S)$ decays are given in
Table~\ref{tab:bulnu}.  

Recently BABAR measured \vub\ using the prescription of Leibovich,
Low, and Rothstein (LLR)~\cite{Leibovich:1999xf}. LLR reduces model
dependence by combining data of the hadronic mass spectrum from
$\Bxulnu$ decays with that of the photon energy spectrum from
$\Bxsg$ decays~\cite{Aubert:2006qi} thus eliminating the need for a model
of the shape function, which is necessary, for example, in BLNP. 
However shape function model uncertainties are
not altogether eliminated as they still enter via
the signal models used for the determination of efficiency. In
principle other measurements of spectra of $\Bxulnu$ could utilise the
available BABAR measurement of the photon energy spectrum in the
rest frame of the $B$-meson to extract \vub\ using LLR, we've
considered this undertaking, however it is not yet practical. Later, for
completeness, we provide a comparison of the LLR-based \vub\ with our
BLNP, DGE and BLL averages. 

\begin{table}[!htb]
\caption{Summary of inclusive determinations of partial branching
  fractions for $B\rightarrow X_u l \nu$ decays.
  The errors quoted on $\Delta\cbf$ correspond to
  statistical and systematic uncertainties, respectively.  The
  $s_\mathrm{h}^{\mathrm{max}}$ variable is described in Ref.~\cite{ref:shmax} }
\begin{center}
\begin{small}
\begin{tabular}{|llc|}
\hline\hline
Measurement & Accepted region &  $\Delta\cbf [10^{-4}]$ \\
\hline
CLEO~\cite{ref:cleo-endpoint}
& $E_e>2.1\,\gev$ 
& $3.3\pm 0.2\pm 0.7$  \\ 
BABAR~\cite{ref:babar-elq2}
& $E_e>2.0\,\gev$, $s_\mathrm{h}^{\mathrm{max}}<3.5\,\mathrm{GeV^2}$ & $4.4\pm 0.4\pm 0.4$ \\
BABAR~\cite{ref:babar-endpoint}
& $E_e>2.0\,\gev$  & $5.7\pm 0.4\pm 0.5$\\
BELLE~\cite{ref:belle-endpoint}
& $E_e>1.9\,\gev$  & $8.5\pm 0.4\pm 1.5$ \\
BABAR~\cite{ref:babar-q2mx}
& $M_X<1.7\,\gev/c^2, q^2>8\,\gev^2/c^2$ & $8.7\pm 0.9\pm 0.9$\\
BELLE~\cite{ref:belle-mxq2Anneal}
& $M_X<1.7\,\gev/c^2, q^2>8\,\gev^2/c^2$ & $7.4\pm 0.9\pm 1.3$\\
BELLE~\cite{ref:belle-mx}
& $M_X<1.7\,\gev/c^2, q^2>8\,\gev^2/c^2$ & $8.4\pm 0.8\pm 1.0$\\
BELLE~\cite{ref:belle-mx}
& $P_+<0.66\,\gev$  & $11.0\pm 1.0\pm 1.6$ \\
BELLE~\cite{ref:belle-mx}
& $M_X<1.7\,\gev/c^2$ & $12.4\pm 1.1\pm 1.2$ \\ \hline
\end{tabular}\\
\end{small}
\end{center}
\label{tab:bulnu}
\end{table}

\subsubsection{BLNP}
\begin{figure}
\begin{center}
\includegraphics[width=6.0in]{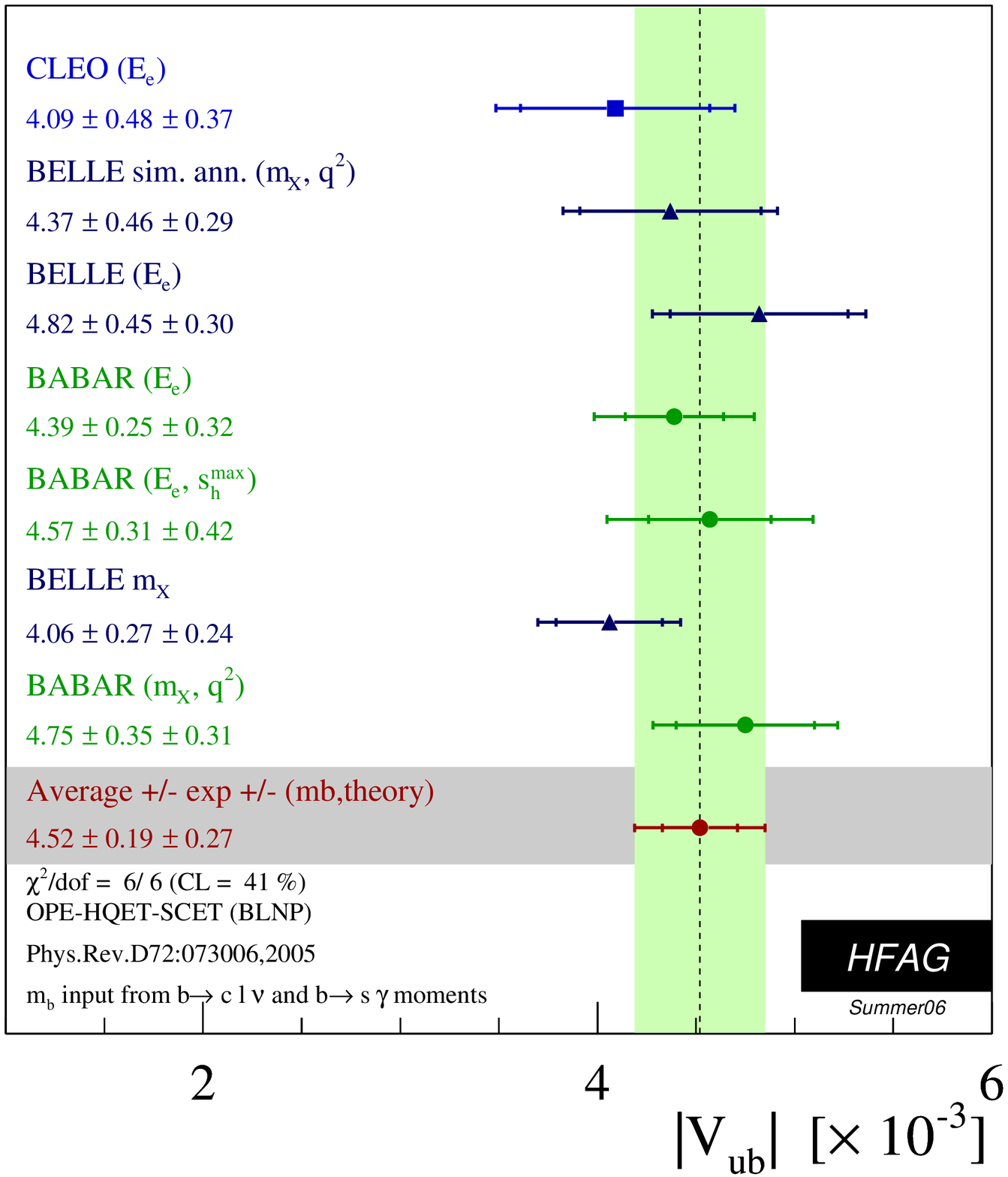}
\end{center}
\caption{Measurements of $\vub$ from inclusive semileptonic decays 
and their average in the BLNP description.
``$E_e$'', ``$M_X$'', ``$(M_X,q^2)$'' and ``($E_e,s^{max}_h$)'' indicate the 
analysis type.}
\label{fig:BLNP}
\end{figure}

BLNP, which stands for Bosch, Lange, Neubert and Paz~\cite{ref:BLNP,
  ref:Neubert-new-1,ref:Neubert-new-2,ref:Neubert-new-3,ref:Neubert-new-4},
provides theoretical expressions for the triple
differential $B\to X_u l^- \nu$ decay rate incorporating all known
contributions whilst smoothly interpolating between the ``shape-function region'' of large hadronic
energy and small invariant mass, and the ``OPE region'' in which all
hadronic kinematical variables scale with $M_B$. BLNP assign
uncertainties for the 
$b$-quark mass which enters through the leading shape function, sub-leading shape function forms, possible weak annihilation
contribution, and for matching scales. The extracted values
of \vub\, for each measurement along with their average are given in
Table~\ref{tab:bulnuBLNP} and illustrated in Figure~\ref{fig:BLNP}.
The breakdown of the uncertainty on the
average is given in Table~\ref{tab:bulnuBLNPerror}. The error on
the $b$-quark mass is the source of the largest uncertainty while the
uncertainty assigned for the matching scales is a close second.

\begin{table}[!htb]
\caption{Summary of inclusive determinations with the investigated acceptance 
region in the analysis and their determination of $\vub$ in the BLNP prescription 
together with their average. The errors quoted on \vub\ correspond to
experimental and theoretical uncertainties, respectively.  The
$s_\mathrm{h}^{\mathrm{max}}$ variable is described in Ref.~\cite{ref:shmax}. }
\begin{center}
\begin{small}
\begin{tabular}{|llcc|}
\hline
& accepted region & $f_u$
&  $\Vub [10^{-3}]$\\
\hline\hline
CLEO~\cite{ref:cleo-endpoint}
& $E_e>2.1\,\gev$ & 0.13
& $4.09\pm 0.48\pm 0.37$ \\ 
BELLE~\cite{ref:belle-endpoint}
& $E_e>1.9\,\gev$  & 0.24
& $4.82\pm 0.45\pm 0.30$ \\ 
BABAR~\cite{ref:babar-endpoint}
& $E_e>2.0\,\gev$  & 0.19
& $4.39\pm 0.25\pm 0.32$ \\ 
BABAR~\cite{ref:babar-elq2}
& $E_e>2.0\,\gev$, $s_\mathrm{h}^{\mathrm{max}}<3.5\,\mathrm{GeV^2}$ & 0.13
& $4.57\pm 0.31\pm 0.42$ \\
BELLE~\cite{ref:belle-mx}
& $M_X<1.7\,\gev/c^2$   & 0.47
& $4.06 \pm 0.27 \pm 0.24$ \\
BELLE~\cite{ref:belle-mxq2Anneal}
& $M_X<1.7\,\gev/c^2, q^2>8\,\gev^2/c^2$ & 0.24
& $4.37 \pm 0.46 \pm 0.29$ \\ 
BABAR~\cite{ref:babar-q2mx}
& $M_X<1.7\,\gev/c^2, q^2>8\,\gev^2/c^2$ & 0.24
& $4.75 \pm 0.35 \pm 0.31$ \\ 
\hline\hline
{\bf Average}
& \mathversion{bold}$\chi^2=6/6$, CL$=0.41$ &
& \mathversion{bold}$4.52 \pm 0.19 \pm 0.27$ \\
\hline
\end{tabular}\\
\end{small}
\end{center}
\label{tab:bulnuBLNP}
\end{table}

\begin{table}[!htb]
\caption{Summary of uncertainties on the $\vub$ average in the BLNP prescription.}
\begin{center}
\begin{small}
\begin{tabular}{|lr|} \hline
  Source & Uncertainty(\%) \\ \hline \hline
  Statistics & $\pm2.1$ \\
  Detector & $\pm2.7$ \\
  $B\to X_c l \nu$ model & $\pm2.1$ \\
  $B\to X_u l \nu$ model & $\pm1.4$ \\
  Heavy quark parameters $m_b$ & $\pm4.1$ \\
  Sub-leading shape functions & $\pm0.9$ \\
  BLNP theory : Matching scales $\mu,\mu_i,\mu_h$  & $\pm3.8$ \\
  Weak annihilation  & $\pm2.1$ \\\hline\hline
  Total &  $\pm7.4$ \\ \hline
\end{tabular}  
\end{small}
\end{center}
\label{tab:bulnuBLNPerror}
\end{table}

\subsubsection{DGE}
\begin{figure}
\begin{center}
\includegraphics[width=6.0in]{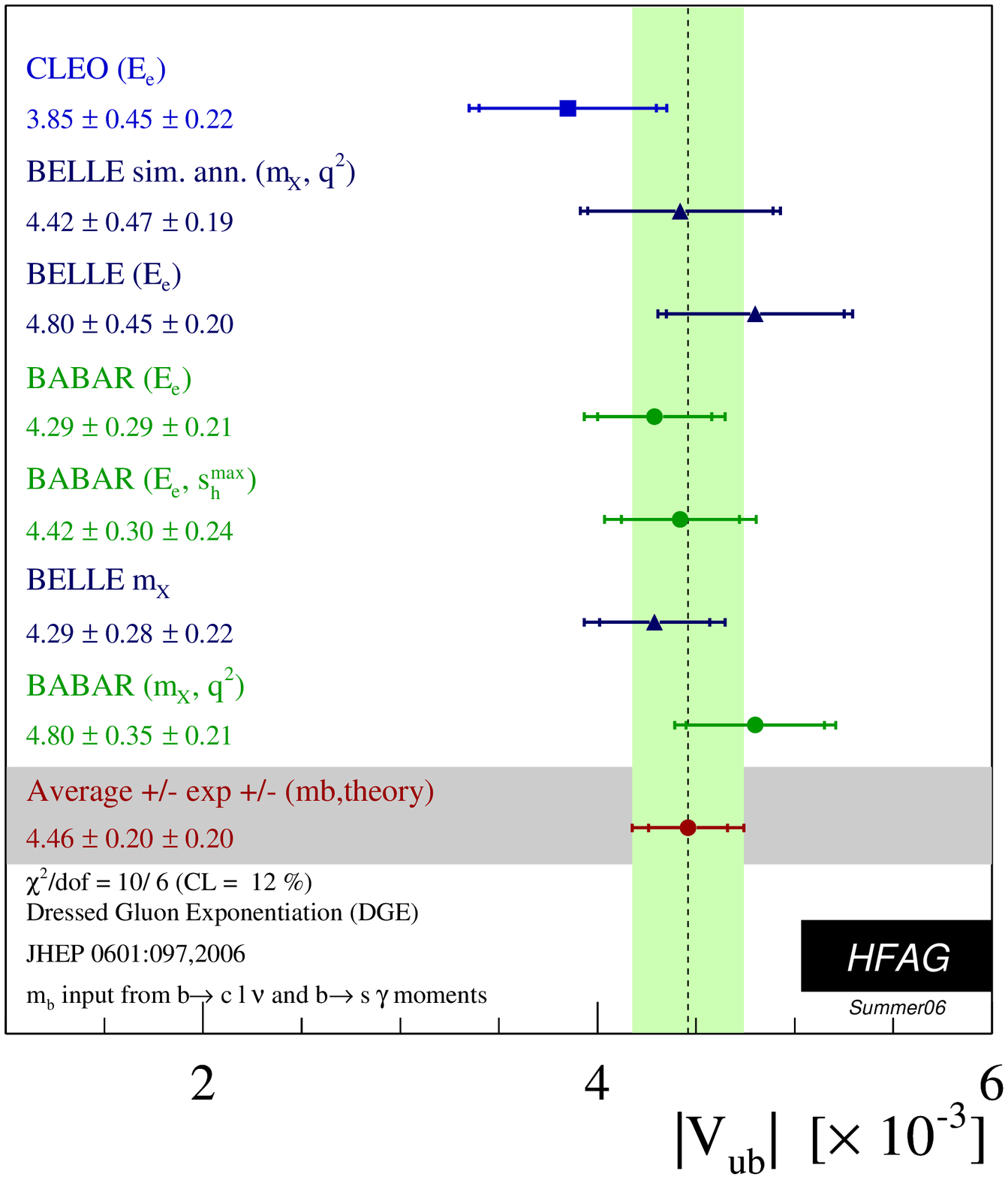}
\end{center}
\caption{Measurements of $\vub$ from inclusive semileptonic decays 
and their average in the DGE description. 
``$E_e$'', ``$M_X$'', ``$(M_X,q^2)$'' and ``($E_e,s^{max}_h$)'' indicate the 
analysis type.}
\label{fig:DGE}
\end{figure}

DGE, which stands for Dressed Gluon Exponentiation~\cite{ref:DGE}, is
a framework where the
on-shell $b$-quark calculation, converted into hadronic variables, is
directly used as an approximation to the meson decay spectrum without
need of a leading-power non-perturbative function (or, in other words,
a shape function). The on-shell mass of the $b$-quark within the $B$-meson ($m_b$) is
required as an input. Theoretical uncertainties are assessed by
varying the inputs of: the $b$-quark mass ($m_b$), the strong coupling constant
($\alpha_s$), the number of light fermion flavours ($N_F$), and the method
and scale of the matching scheme intrinsic to the approach. 
The extracted values
of \vub\, for each measurement along with their average are given in
Table~\ref{tab:bulnuDGE} and illustrated in Figure~\ref{fig:DGE}.
The breakdown of the uncertainty on the
average is given in Table~\ref{tab:bulnuDGEerror}. The error in $m_b$
contributes to the evalution of the total semileptonic width as well
as to the spectral fraction, and is a leading source of uncertainty.
The next largest uncertainty is from the matching scale and scheme.  

\begin{table}[!htb]
\caption{Summary of inclusive determinations with the investigate acceptance 
region in the analysis and their determination of $\vub$ in the DGE prescription 
together with their average. The errors quoted on \vub\ correspond to
experimental and theoretical uncertainties, respectively.  The
$s_\mathrm{h}^{\mathrm{max}}$ variable is described in Ref.~\cite{ref:shmax}. }
\begin{center}
\begin{small}
\begin{tabular}{|llcc|}
\hline
& accepted region & $f_u$
&  $\Vub [10^{-3}]$\\
\hline\hline
CLEO~\cite{ref:cleo-endpoint}
& $E_e>2.1\,\gev$ & 0.22
& $3.85\pm 0.45\pm 0.22$ \\ 
BELLE~\cite{ref:belle-endpoint}
& $E_e>1.9\,\gev$  & 0.37
& $4.80\pm 0.45\pm 0.20$ \\ 
BABAR~\cite{ref:babar-endpoint}
& $E_e>2.0\,\gev$  & 0.30
& $4.29\pm 0.29\pm 0.21$ \\ 
BABAR~\cite{ref:babar-elq2}
& $E_e>2.0\,\gev$, $s_\mathrm{h}^{\mathrm{max}}<3.5\,\mathrm{GeV^2}$ & 0.21
& $4.42\pm 0.30\pm 0.24$ \\
BELLE~\cite{ref:belle-mx}
& $M_X<1.7\,\gev/c^2$   & 0.64
& $4.29 \pm 0.28 \pm 0.22$ \\
BELLE~\cite{ref:belle-mxq2Anneal}
& $M_X<1.7\,\gev/c^2, q^2>8\,\gev^2/c^2$ & 0.36
& $4.42 \pm 0.47 \pm 0.19$ \\ 
BABAR~\cite{ref:babar-q2mx}
& $M_X<1.7\,\gev/c^2, q^2>8\,\gev^2/c^2$ & 0.36
& $4.80 \pm 0.35 \pm 0.21$ \\ 
\hline\hline
{\bf Average}
& \mathversion{bold}$\chi^2=10/6$, \mathversion{bold}CL$=0.12$ &
& \mathversion{bold}$4.46 \pm 0.20 \pm 0.20$ \\
\hline
\end{tabular}\\
\end{small}
\end{center}
\label{tab:bulnuDGE}
\end{table}

\begin{table}[!htb]
\caption{Summary of uncertainties on the $\vub$ average in the DGE prescription.}
\begin{center}
\begin{small}
\begin{tabular}{|lr|} \hline
  Source & Uncertainty(\%) \\\hline \hline
  Statistics & $\pm1.8$ \\
  Detector & $\pm2.5$ \\
  $B\to X_c l \nu$ model & $\pm2.3$ \\
  $B\to X_u l \nu$ model & $\pm2.3$ \\
  Spectral fraction ($m_b$) & $\pm1.2$ \\
  Strong coupling $\alpha_s$  & $\pm1.0$ \\
  Total semileptonic width ($m_b$)  & $\pm3.0$ \\
  DGE theory : matching scales   & $\pm2.9$ \\\hline\hline
  Total &  $\pm6.3$ \\ \hline
\end{tabular}  
\end{small}
\end{center}
\label{tab:bulnuDGEerror}
\end{table}

\subsubsection{BLL}
\begin{figure}
\begin{center}
\includegraphics[width=6.0in]{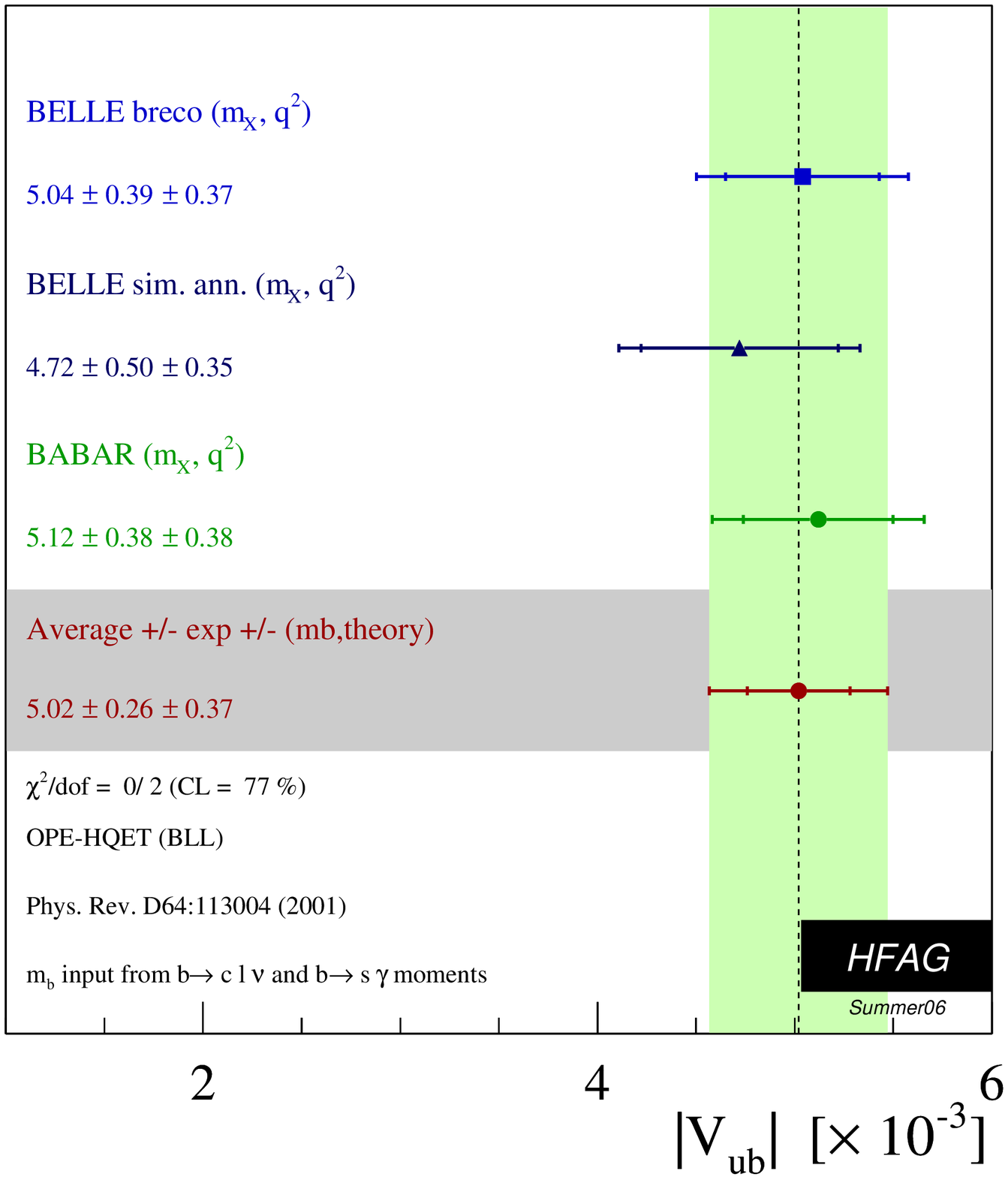}
\end{center}
\caption{Measurements of $\vub$ from inclusive semileptonic decays 
and their average in the BLL prescription.
``$(M_X, q^2)$'' indicates the analysis type.}
\label{fig:BLL}
\end{figure}

BLL, which  stands for Bauer, Ligeti, and Luke~\cite{ref:BLL}, is a
HQET-based prescription that advocates combined cuts on the dilepton invariant mass, $q^2$,
and hadronic mass, $m_X$, to minimise the overall uncertainty on \vub.
In their reckoning a cut on $m_X$ only, although most efficient at
preserving phase space (~80\%), makes the calculation of the partial
rate untenable owing to uncalculable corrections
to the $b$-quark distribution function or shape function. These corrections are
suppressed if a cut on $q^2$ is introduced. The cut combination used
in measurements is $M_x<1.7$ GeV/$c^2$ and $q^2 > 8$ GeV$^2$/$c^2$.  
The extracted values
of \vub\, for each measurement along with their average are given in
Table~\ref{tab:bulnuBLL} and illustrated in Figure~\ref{fig:BLL}.
The breakdown of the uncertainty on the
average is given in Table~\ref{tab:bulnuBLLerror}. The leading
uncertainties, both from theory, are due to residual shape function
effects and third order terms in the OPE expansion. The leading
experimental uncertainty is due to statistics. 

\begin{table}[!htb]
\caption{Summary of inclusive determinations of $\vub$ in the BLL prescription.
  The errors quoted on \vub\ correspond to
experimental and theoretical uncertainties, respectively. }
\begin{center}
\begin{small}
\begin{tabular}{|llcc|}
\hline
& accepted region & $f_u$
&  $\Vub [10^{-3}]$\\
\hline\hline
BELLE~\cite{ref:belle-mxq2Anneal}
& $M_X<1.7\,\gev/c^2, q^2>8\,\gev^2/c^2$ & 0.26
& $4.72 \pm 0.50 \pm 0.35$ \\ 
BABAR~\cite{ref:babar-q2mx}
& $M_X<1.7\,\gev/c^2, q^2>8\,\gev^2/c^2$ & 0.26
& $5.12 \pm 0.38 \pm 0.38$ \\ 
BELLE~\cite{ref:belle-mx}
& $M_X<1.7\,\gev/c^2, q^2>8\,\gev^2/c^2$ & 0.26
& $5.04 \pm 0.39 \pm 0.37$ \\ 
\hline\hline
{\bf Average}
& \mathversion{bold}$\chi^2=0.5/2$, \mathversion{bold} CL$=0.77$ &
& \mathversion{bold}$5.02 \pm 0.26 \pm 0.37$ \\
\hline
\end{tabular}\\
\end{small}
\end{center}
\label{tab:bulnuBLL}
\end{table}

\begin{table}[!htb]
\caption{Summary of uncertainties on $\vub$ in the BLL prescription.}
\begin{center}
\begin{small}
\begin{tabular}{|lr|} \hline
  Source & Uncertainty(\%) \\ \hline\hline
  Statistics & $\pm3.2$ \\
  Detector & $\pm2.9$ \\
  $B\to X_c l \nu$ model & $\pm1.8$ \\
  $B\to X_u l \nu$ model & $\pm2.4$ \\
  Spectral fraction ($m_b$) & $\pm3.0$ \\
  Perturbative : Strong coupling $\alpha_s$  & $\pm3.0$ \\
  Residual shape function & $\pm4.5$ \\
  Third order terms in the OPE  & $\pm4.0$ \\\hline\hline
  Total &  $\pm9.1$ \\ \hline\hline
\end{tabular}  
\end{small}
\end{center}
\label{tab:bulnuBLLerror}
\end{table}

\subsubsection{Summary}
We have performed averages of inclusive \vub\ measurements using the three
frameworks of BLNP, DGE and BLL. Table~\ref{tab:vubcomparison} lists
these averages, as well as additional extractions for 
$m_X/q^2$ measurements only and the LLR-based BABAR measurement. The
values display a remarkable agreement. The uncertainties that
dominate both BLNP and DGE averages arise from $m_b$ and 
matching scales, while for BLL they are due to residual shape function
effects and higher order terms in the OPE. The overall uncertainties
on \vub\ for BLNP, DGE and BLL averages are 7.4\%, 6.3\% and 9.15l4\%.
respectively. A value judgement based on a direct comparison should be
avoided, experimental and theoretical uncertainties play out
differently between the schemes. What is clear is that 
a better determination of $m_b$ and further investigation of matching scale effects and
shape functions will add to our understanding of \vub\, and hopefully
improve the precision to which it is known. The overall uncertainty on
the one measurement employing LLR is
12.0\%, although it is larger than the others it is a first look at
the measurements that we can expect to see more of from $B$-factories. 

\begin{table}[!htb]
\caption{Summary of inclusive determinations of $\vub$.
  The errors quoted on \vub\ correspond to
  experimental and theoretical uncertainties, respectively. }
\begin{center}
\begin{small}
\begin{tabular}{|lc|}
\hline
Framework
&  $\Vub [10^{-3}]$\\
\hline\hline
BLNP
& $4.52 \pm 0.19 \pm 0.27$ \\ 
DGE
& $4.46 \pm 0.20 \pm 0.20$ \\
LLR (BABAR)
& $4.43 \pm 0.45 \pm 0.29$ \\
\hline\hline
BLL ($m_X/q^2$ only)
& $5.02 \pm 0.26 \pm 0.37$ \\ 
BLNP ($m_X/q^2$ only)
& $4.65 \pm 0.25 \pm 0.31$ \\
DGE  ($m_X/q^2$ only)
& $4.70 \pm 0.26 \pm 0.22$ \\
\hline
\end{tabular}\\
\end{small}
\end{center}
\label{tab:vubcomparison}
\end{table}

%% file: rare.tex
\mysection{Charmless \B-decay branching fractions and
               their asymmetries }
\label{sec:rare}

The aim of this section is to provide the branching fractions and
the partial rate asymmetries ($A_{CP}$) of charmless 
$B$ decays. The asymmetry is defined as 
$A_{CP} = \frac{N_{\Bbar} -N_B}{N_{\Bbar} +N_B}$, where $N_{\Bbar}$ 
and $N_B$ are respectively number of $\Bzb/\Bm$ and $\Bz/\Bp$ decaying
into a specific final state. 
Four different $B$ decay categories are considered: 
charmless mesonic, baryonic, radiative and leptonic. Measurements supported 
with  written documents are accepted in  
the averages; written documents could be journal papers, 
conference contributed papers, preprints or conference proceedings.  
Results from  $A_{CP}$ measurements  obtained from time dependent analyses 
are listed and described in Sec.~\ref{sec:cp_uta}. Measurements of  charmful 
baryonic $B$ decays, which were included in our previous averages
\cite{hfag_hepex_endof2005}, are now shown    
in Section 7, which deals with $B$ decays to charm.  

So far all branching fractions assume equal production of charged and
neutral $B$ pairs.  The best measurements to date show that this is
still a good approximation (see Sec.~\ref{sec:life_mix}).
For branching fractions, we provide either averages or the most stringent
90\% confidence level upper limits.  If one or more experiments have
measurements with $>$4$\sigma$ for a decay channel, all available central values
for that channel are used in the averaging.  We also give central values
and errors for cases where the significance of the average value is at
least $3 \sigma$, even if no single measurement is above $4 \sigma$. 
Since a few decay modes are sensitive to the contribution of
 new physics and the current experimental upper limits are not far from the 
Standard Model expectation, we provide the combined upper limits or
averages in these cases.
Their upper limits can be estimated assuming that the errors are 
Gaussian.  For $A_{CP}$ we provide averages in all cases.  

Our averaging is performed by maximizing the likelihood,
   $\displaystyle {\mathcal L} = \prod_i {\mathcal P}_i(x),$  
where ${\mathcal P_i}$ is the probability density function (PDF) of the
$i$th  measurement, and $x$ is the branching fraction or $A_{CP}$.
The PDF is modeled by an asymmetric Gaussian function with the measured
central value as its mean and the quadratic sum of the statistical
and systematic errors as the standard deviations. The experimental
uncertainties are considered to be uncorrelated with each other when the 
averaging is performed. No error scaling is applied when the fit $\chi^2$ is 
greater than 1 since we believe that tends to overestimate the errors
except in cases of extreme disagreement (we have no such cases).
One exception to consider the correlated systematic errors is the inclusive
$b\to s\gamma$ mode, which is sensitive to physics beyond the Standard Model.
We tried to include as many measurements as possible and take the common
systematic errors into account when performing the average. The detail is  
described in Sec. ~\ref{sec:btosg}.

At present, we have measurements of more than 250  decay modes, reported in
more than 150 papers. Because the number of references is so large, we do
not include them with the tables shown here but the full set of
references is available quickly from active gifs at the 
``Winter 2007'' link on 
the rare web page: {\tt http://www.slac.stanford.edu/xorg/hfag/rare/index.html}.
Finally the inclusive measurements of $B\to KX$ branching fractions 
and full angular analysis on  $B^0 \to \phi K_2^*(1430)^0$ are listed
for the first time. 

\mysubsection{Mesonic charmless decays}

\input{charmless}

\mysubsection{Radiative and leptonic decays}

\input radll

\mysubsection{$B\to s\gamma$}
\label{sec:btosg}
\input btosg

\mysubsection{Baryonic decays}

\input{bary}

\clearpage
\mysubsection{$B_s$ decays}

\input{bs}

\mysubsection{Charge asymmetries}

\input{acp}

 \clearpage
\mysubsection{Polarization measurements}
 \input{polar}

%% file: charmless.tex
\begin{table}[!htbp]
\vspace{-1.5cm}
\begin{center}
\caption{
Branching fractions (BF) of charmless mesonic $B^+$ decays
(in units of $10^{-6}$). Upper limits are at 90\% CL.
Values in {\red red} ({\blue blue}) are new {\red published}
({\blue preliminary}) result since PDG2006  [as of March 15, 2007].
}
\scriptsize
\begin{tabular}{|l@{}ccccccc|} 
\sgline
RPP\#   & Mode & PDG2006 Avg. & BABAR & Belle & CLEO & CDF & New Avg. \\
\sglinespb
182                                               & 
$K^0\pi^+$                                        & 
$24.1\pm 1.7$                                     & 
\red{$\err{23.9}{1.1}{1.0}$}                      & 
\blue{$\aerr{22.8}{0.8}{0.7}{1.3}$}               & 
$\aerrsy{18.8}{3.7}{3.3}{2.1}{1.8}$               & 
\nodata                                           & 
$23.1 \pm 1.0$                                    \\

183                                               & 
$K^+\pi^0$                                        & 
$12.1\pm 0.8$                                     & 
\blue{$\err{13.3}{0.6}{0.6}$}                     & 
\blue{$\err{12.4}{0.5}{0.6}$}                     & 
$\aerrsy{12.9}{2.4}{2.2}{1.2}{1.1}$               & 
\nodata                                           & 
$12.8 \pm 0.6$                                    \\

184                                               & 
$\etapr K^+$                                      & 
$70.5\pm3.5$                                      & 
{$\err{68.9}{2.0}{3.2}$}                          & 
\red{$\err{69.2}{2.2}{3.7}$}                      & 
$\aerr{80}{10}{9}{7}$                             & 
\nodata                                           & 
$\cerr{69.7}{2.8}{2.7}$                           \\
\end{tabular}
\end{center}

\vspace{-0.4cm} 
\hspace*{-0.8cm} 
$\dag$Product BF - daughter BF taken to be 100\%; 
~\S $M_{\phi\phi}<2.85$ GeV/$c^2$
\end{table}

\begin{table}[!htbp]
\begin{center}
\caption{
Branching fractions (BF) of charmless mesonic $B^+$ decays (part 2)
(in units of $10^{-6}$). Upper limits are at 90\% CL.
Values in {\red red} ({\blue blue}) are new {\red published}
({\blue preliminary}) result since PDG2006  [as of March 15, 2007].
}
\footnotesize
\begin{tabular}{|l@{}ccccccc|} 
\sgline
RPP\#   & Mode & PDG2006 Avg. & BABAR & Belle & CLEO & CDF & New Avg. \\
\sglinespb

254                                               & 
$\pi^+\pi^0$                                      & 
$5.5 \pm 0.6$                                     & 
\blue{$\err{5.1}{0.5}{0.3}$}                      & 
\blue{$\berr{6.5}{0.4}{0.4}{0.5}$}                & 
$\aerrsy{4.6}{1.8}{1.6}{0.6}{0.7}$                & 
\nodata                                           & 
$5.7 \pm 0.4$                                     \\

255                                               & 
$\pi^+\pi^-\pi^+$                                 & 
$\err{16.2}{1.2}{0.9}$                            & 
{$\err{16.2}{1.2}{0.9}$}                          & 
\nodata                                           & 
\nodata                                           & 
\nodata                                           & 
$16.2 \pm 1.5$                                    \\

256                                               & 
$\rho^0\pi^+$                                     & 
$8.7\pm 1.1$                                      & 
{$\berr{8.8}{1.0}{0.6}{0.9}$}                     & 
$\aerr{8.0}{2.3}{2.0}{0.7}$                       & 
$\aerr{10.4}{3.3}{3.4}{2.1}$                      & 
\nodata                                           & 
$\cerr{8.7}{1.0}{1.1}$                            \\

257                                               & 
$f_0(980)\pi^+$ $\dag$                            & 
$<3.0$                                            & 
{$<3.0$}                                          & 
\nodata                                           & 
\nodata                                           & 
\nodata                                           & 
{$<3.0$}                                          \\

258                                               & 
$f_2(1270)\pi^+$                                  & 
$\err{8.2}{2.1}{1.4}$                             & 
{$\err{8.2}{2.1}{1.4}$}                           & 
\nodata                                           & 
\nodata                                           & 
\nodata                                           & 
$8.2 \pm 2.5$                                     \\

259                                               & 
$\rho^0(1450)\pi^+$                               & 
$<2.3$                                            & 
{$<2.3$}                                          & 
\nodata                                           & 
\nodata                                           & 
\nodata                                           & 
{$<2.3$}                                          \\

260                                               & 
$f_0(1370)\pi^+$                                  & 
$<3.0$                                            & 
{$<3.0$}                                          & 
\nodata                                           & 
\nodata                                           & 
\nodata                                           & 
{$<3.0$}                                          \\

261                                               & 
$f_0(600)\pi^+$                                   & 
$<4.1$                                            & 
{$<4.1$}                                          & 
\nodata                                           & 
\nodata                                           & 
\nodata                                           & 
{$<4.1$}                                          \\

262                                               & 
$\pi^+\pi^-\pi^+ (NR)$                            & 
$<4.6$                                            & 
{$<4.6$}                                          & 
\nodata                                           & 
\nodata                                           & 
\nodata                                           & 
{$<4.6$}                                          \\

264                                               & 
$\rho^+\pi^0$                                     & 
$12.0 \pm 1.9$                                    & 
\blue{$\err{10.2}{1.4}{0.9}$}                     & 
{$\berr{13.2}{2.3}{1.4}{1.9}$}                    & 
$<43$                                             & 
\nodata                                           & 
$\cerr{10.9}{1.4}{1.5}$                           \\

266                                               & 
$\rho^+\rho^0$                                    & 
$26\pm 6$                                         & 
\red{$\err{16.8}{2.2}{2.3}$}                      & 
$\berr{31.7}{7.1}{3.8}{6.7}$                      & 
\nodata                                           & 
\nodata                                           & 
$18.2 \pm 3.0$                                    \\

$~-$                                              & 
$\rho^+f_0(980)$ $\dag$                           & 
New                                               & 
\red{$<1.9$}                                      & 
\nodata                                           & 
\nodata                                           & 
\nodata                                           & 
{$<1.9$}                                          \\

269                                               & 
$\omega\pi^+$                                     & 
$5.9 \pm 1.0$                                     & 
\red{$\err{6.1}{0.7}{0.4}$}                       & 
\red{$\err{6.9}{0.6}{0.5}$}                       & 
$\aerr{11.3}{3.3}{2.9}{1.4}$                      & 
\nodata                                           & 
$6.7 \pm 0.6$                                     \\

270                                               & 
$\omega\rho^+$                                    & 
$\berr{12.6}{3.7}{3.3}{1.6}$                      & 
\red{$\berr{10.6}{2.1}{1.6}{1.0}$}                & 
\nodata                                           & 
$<61$                                             & 
\nodata                                           & 
$\cerr{10.6}{2.6}{2.3}$                           \\

271                                               & 
$\eta\pi^+$                                       & 
$4.9 \pm 0.5$                                     & 
{$\err{5.1}{0.6}{0.3}$}                           & 
\blue{$\err{4.2}{0.4}{0.2}$}                      & 
\cerr{1.2}{2.8}{1.2}                              & 
\nodata                                           & 
$4.4 \pm 0.4$                                     \\

272                                               & 
$\etapr\pi^+$                                     & 
$4.0\pm0.9$                                       & 
{$\err{4.0}{0.8}{0.4}$}                           & 
\red$\aerr{1.8}{0.7}{0.6}{0.1}$                   & 
$\cerr{1.0}{5.8}{1.0}$                            & 
\nodata                                           & 
$\cerr{2.6}{0.6}{0.5}$                            \\

273                                               & 
$\etapr\rho^+$                                    & 
$<22$                                             & 
\red{$\aerrsy{8.7}{3.1}{2.8}{2.3}{1.3}$}          & 
\blue{$<4.7$}                                     & 
$\cerr{11.2}{11.9}{7.0}$                          & 
\nodata                                           & 
$\cerr{9.1}{3.7}{2.8}$                            \\

274                                               & 
$\eta\rho^+$                                      & 
$\err{8.4}{1.9}{1.1}$                             & 
{$\err{8.4}{1.9}{1.1}$}                           & 
\blue{$\aerr{4.1}{1.4}{1.3}{0.4}$}                & 
$\cerr{4.8}{5.2}{3.8}$                            & 
\nodata                                           & 
$5.4 \pm 1.2$                                     \\

275                                               & 
$\phi\pi^+$                                       & 
$<0.41$                                           & 
\red{$<0.24$}                                     & 
\nodata                                           & 
$<5$                                              & 
\nodata                                           & 
{$<0.24$}                                         \\

276                                               & 
$\phi\rho^+$                                      & 
$<16$                                             & 
\nodata                                           & 
\nodata                                           & 
$<16$                                             & 
\nodata                                           & 
$<16$                                             \\

277                                               & 
$a_0^0(980)\pi^+$ $\dag$                          & 
$<5.8$                                            & 
{$<5.8$}                                          & 
\nodata                                           & 
\nodata                                           & 
\nodata                                           & 
{$<5.8$}                                          \\

$~-$                                              & 
$a_0^+(980)\pi^0$ $\dag$                          & 
New                                               & 
\blue{$<1.3$}                                     & 
\nodata                                           & 
\nodata                                           & 
\nodata                                           & 
{$<1.3$}                                          \\

\sglinespt
\end{tabular}
\end{center}

\vspace{-0.4cm} 
\hspace*{-0.8cm} 
$\dag$Product BF - daughter BF taken to be 100\%; 
\end{table}
\clearpage

\begin{table}
\begin{center}
\caption{
Branching fractions of charmless mesonic $B^0$ decays
(in units of $10^{-6}$).
Upper limits are at 90\% CL.
Values in {\red red} ({\blue blue}) are new {\red published}
({\blue preliminary}) result since PDG2006  [as of March 15, 2007].
}
\scriptsize
\begin{tabular}{|lccccccc|} 
\sgline
RPP\#   & Mode & PDG2006 Avg. & BABAR & Belle & CLEO & CDF & New Avg. \\
\hline
168                                               & 
$K^+\pi^-$                                        & 
$18.2\pm 0.8$                                     & 
\red{$\err{19.1}{0.6}{0.6}$}                      & 
\blue{$\err{19.9}{0.4}{0.8}$}                     & 
$\aerrsy{18.0}{2.3}{2.1}{1.2}{0.9}$               & 
\nodata                                           & 
$19.4 \pm 0.6$                                    \\

169                                               & 
$K^0\pi^0$                                        & 
$11.5 \pm 1.0$                                    & 
\blue{$\err{10.5}{0.7}{0.5}$}                     & 
\blue{$\berr{9.2}{0.7}{0.6}{0.7}$}                & 
$\aerrsy{12.8}{4.0}{3.3}{1.7}{1.4}$               & 
\nodata                                           & 
$10.0 \pm 0.6$                                    \\

170                                               & 
$\etapr K^0$                                      & 
$68\pm 4$                                         & 
{$\err{67.4}{3.3}{3.2}$}                          & 
\red{$\aerr{58.9}{3.6}{3.5}{4.3}$}                & 
$\aerr{89}{18}{16}{9}$                            & 
\nodata                                           & 
$64.9 \pm 3.5$                                    \\
\end{tabular}
\end{center}
\vspace{-0.4cm}
$\dag$Product BF - daughter BF taken to be 100\%, 
 $\ddag$Relative BF converted to absolute BF
~\S $M_{\phi\phi}<2.85$ GeV/$c^2$
 $^1$Excludes $M(K_SK_S)$ regions [3.400,3.429] and [3.540,3.585] and $M(K_SK_L)<1.049$ GeV/$c^2$
\end{table}
\clearpage

\begin{table}
\begin{center}
\caption{
Branching fractions of charmless mesonic $B^0$ decays (part 2)
(in units of $10^{-6}$).
Upper limits are at 90\% CL.
Values in {\red red} ({\blue blue}) are new {\red published}
({\blue preliminary}) result since PDG2006  [as of March 15, 2007].
}
\scriptsize
\begin{tabular}{|lccccccc|} 
\sgline
RPP\#   & Mode & PDG2006 Avg. & BABAR & Belle & CLEO & CDF & New Avg. \\
\hline
229                                               & 
$\pi^+\pi^-$                                      & 
$4.6\pm0.4$                                       & 
\red{$\err{5.5}{0.4}{0.3}$}                       & 
\blue{$\err{5.1}{0.2}{0.2}$}                      & 
$\aerrsy{4.5}{1.4}{1.2}{0.5}{0.4}$                & 
\blue{$\err{5.10}{0.33}{0.36}$} $\ddag$           & 
$5.16 \pm 0.22$                                   \\

230                                               & 
$\pi^0\pi^0$                                      & 
$1.5\pm 0.5$                                      & 
\blue{$\err{1.48}{0.26}{0.12}$}                   & 
\blue{$\err{1.1}{0.3}{0.1}$}                      & 
$<4.4$                                            & 
\nodata                                           & 
$1.31 \pm 0.21$                                   \\

231                                               & 
$\eta\pi^0$                                       & 
$<2.5$                                            & 
\red{$<1.3$}                                      & 
{$<2.5$}                                          & 
$<2.9$                                            & 
\nodata                                           & 
{$<1.3$}                                          \\
\end{tabular}
\end{center}
\vspace{-0.4cm}
$\dag$Product BF - daughter BF taken to be 100\%, 
 $\ddag$Relative BF converted to absolute BF
\end{table}

\begin{table}
\begin{center}
\caption{
 Relative branching fractions of $B^0\to K^+K^-, K^+\pi^-, \pi^+\pi^-$.
Values in {\red red} ({\blue blue}) are new {\red published}
({\blue preliminary}) result since PDG2006  [as of March 15, 2007].
}
\small
\begin{tabular}{|lccccc|} \hline
RPP\# &Mode & PDG2006 Avg. &  CDF & D\O\ & New Avg.  \\ \sglinespb
$179$                                             & 
$\mathcal{B}(B^0\rightarrow K^+K^-)/\mathcal{B}(B^0\rightarrow K^+\pi^-)$& 
\nodata                                           & 
\blue{$0.020\pm0.008\pm0.006$}                    & 
\nodata                                           & 
$0.020 \pm 0.010$                                 \\

$229$                                             & 
$\mathcal{B}(B^0\rightarrow\pi^+\pi^-)/\mathcal{B}(B^0\rightarrow K^+\pi^-)$& 
\nodata                                           & 
\blue{$0.259\pm0.017\pm0.016$}                    & 
\nodata                                           & 
$0.259 \pm 0.023$                                 \\

\hline
\end{tabular}
\end{center}
\end{table}
\clearpage

%% file: radll.tex
\begin{table}[!hb]
\begin{center}
\caption{
Branching fractions of semileptonic and radiative $B^+$ decays
(in units of $10^{-6}$). Upper limits are at 90\% CL.
Values in {\red red} ({\blue blue}) are new {\red published}
({\blue preliminary}) result since PDG2006  [as of March 15, 2007].
}
\end{center}
\vspace{-0.3cm}
\scriptsize
\begin{tabular}{|lccccccc|} \hline
RPP\# &Mode & PDG2006 Avg. & BABAR  & Belle & CLEO & CDF
 & New Avg.  \\
 \sglinespb

240                                               & 
$K^*(892)^+\gamma$                                & 
$40.3\pm 2.6$                                     & 
{$38.7\pm 2.8 \pm 2.6$}                           & 
{$42.5\pm 3.1\pm 2.4$}                            & 
$\aerr{37.6}{8.9}{8.3}{2.8} $                     & 
\nodata                                           & 
$40.3 \pm 2.6$                                    \\

241                                               & 
$K_1(1270)^+\gamma$                               & 
$\err{43}{9}{9}$                                  & 
\nodata                                           & 
{$\err{43}{9}{9}$}                                & 
\nodata                                           & 
\nodata                                           & 
$43 \pm 12$                                       \\

242                                               & 
$K^+\eta\gamma$                                   & 
$\aerr{8.4}{1.5}{1.2}{0.9}$                       & 
{\red $\err{10.0}{1.3}{0.5}$}                     & 
{$\aerr{8.4}{1.5}{1.2}{0.9}$}                     & 
\nodata                                           & 
\nodata                                           & 
$9.4 \pm 1.1$                                     \\
\hline
\end{tabular}
~\S~$M_{K\pi\pi} < 2.4$ GeV/$c^2$
\end{table}

\begin{table}[!htbp]
\begin{center}
\caption{
Branching fractions of semileptonic and radiative $B^0$ decays
(in units of $10^{-6}$).  Upper limits are at 90\% CL.
Values in {\red red} ({\blue blue}) are new {\red published}
({\blue preliminary}) result since PDG2006  [as of March 15, 2007].
}
\vspace{0.3cm}
\scriptsize
\begin{tabular}{|lccccccc|} \hline
RPP\# &Mode & PDG2006 Avg. & BABAR  & Belle & CLEO & CDF
 & New Avg.  \\ \sglinespb
213                                               & 
$K^*(892)^0\gamma$                                & 
$40.1\pm 2.0$                                     & 
{$39.2\pm 2.0\pm 2.4$}                            & 
{$40.1\pm 2.1\pm 1.7$}                            & 
$\aerr{45.5}{7.2}{6.8}{3.4}$                      & 
\nodata                                           & 
$40.1 \pm 2.0$                                    \\

214                                               & 
$K^0\eta\gamma$                                   & 
$\aerrsy{8.7}{3.1}{2.7}{1.9}{1.6}$                & 
{\red $\aerr{11.3}{2.8}{2.6}{0.6}$}               & 
{$\aerrsy{8.7}{3.1}{2.7}{1.9}{1.6}$}              & 
\nodata                                           & 
\nodata                                           & 
$\cerr{10.3}{2.3}{2.1}$                           \\

$~-$                                              & 
$K^0\etapr\gamma$                                 & 
New                                               & 
{\red $<6.6$}                                     & 
\nodata                                           & 
\nodata                                           & 
\nodata                                           & 
{ $<6.6$}                                         \\
\hline
\end{tabular}
\end{center}
~\dag~$1.25$ GeV/$c^2 < M_{K\pi} < 1.6$ GeV/$c^2$
\end{table}

\begin{table}
\begin{center}
\caption{
Branching fractions of  semileptonic and radiative $B$ decays
(in units of $10^{-6}$).  Upper limits are at 90\% CL.
Values in {\red red} ({\blue blue}) are new {\red published}
({\blue preliminary}) result since PDG2006  [as of March 15, 2007].
}
\end{center}

\hspace*{-0.0cm}
\scriptsize
\begin{tabular}{|lcccccc|} \hline
RPP\# &Mode & PDG2006 Avg. & BABAR  & Belle & CLEO
 & New Avg.  \\ \sglinespb
~61                                               & 
$K_2^*(1430)\gamma $                              & 
$\err{1.7}{0.6}{0.1}$                             & 
\nodata                                           & 
\nodata                                           & 
$\err{1.7}{0.6}{0.1}$                             & 
$1.7 \pm 0.6$                                     \\

~63                                               & 
$K_3^*(1780)\gamma $                              & 
$<37$                                             & 
\nodata                                           & 
{$<2.8$}                                          & 
\nodata                                           & 
{$<2.8$}                                          \\

~70                                               & 
$s\gamma$                                         & 
$343\pm 29$                                       & 
{$\berr{327}{18}{55}{41}$}                        & 
$\berrsyt{355}{32}{30}{31}{11}{7}$                & 
$\berr{321}{43}{32}{29}$                          & 
$\errede{355}{24}{9}{10}{3}$                      \\
\hline
\end{tabular}
\vspace{-0.7cm}
 ~\dag $E_\gamma > 2.0$ GeV; ~\ddag $M(\ell^+\ell^-)>0.2$ GeV/$c^2$
\end{table}

\begin{table}
\begin{center}
\caption{
Branching fractions of  inclusive $B$ decays
(in units of $10^{-6}$). Values in {\red red} ({\blue blue}) are new {\red published}
({\blue preliminary}) result since PDG2006  [as of March 15, 2007].
}
\vspace{0.3cm}
\begin{tabular}{|lcccccc|} \hline
RPP\# &Mode & PDG2006 Avg. & BABAR  & Belle & CLEO & New Avg.  \\ 
\sglinespb

$~-$                                              & 
$K^+ X$                                           & 
New                                               & 
{\blue\aerrsy{196}{37}{34}{31}{30}}               & 
\nodata                                           & 
\nodata                                           & 
$\cerr{196}{48}{45}$                              \\

$~-$                                              & 
$K^0 X$                                           & 
New                                               & 
{\blue\aerrsy{154}{55}{48}{55}{41}}               & 
\nodata                                           & 
\nodata                                           & 
$\cerr{154}{77}{63}$                              \\

\hline
\end{tabular}
\end{center}
\vspace{-0.3cm}
~~~~~~~~~~~~~~~~~~\dag~$p^* > 2.34$ GeV
\end{table}

\begin{table}[!htbp]
\begin{center}
\caption{
 Branching fractions of leptonic  $B$ decays
(in units of $10^{-6}$). Upper limits are at 90\% CL.
Values in {\red red} ({\blue blue}) are new {\red published}
({\blue preliminary}) result since PDG2006  [as of March 15, 2007].
}
\vspace{0.3cm}
\begin{tabular}{|lcccccccc|} \hline
RPP\# &Mode & PDG2006 Avg. & BABAR  & Belle & CLEO &CDF & D\O\
 & New Avg.  \\ \sglinespb
~15                                               & 
$e^+\nu$                                          & 
$<15$                                             & 
{\blue $<7.9$}                                    & 
{\blue $<1.0$}                                    & 
$<15$                                             & 
\nodata                                           & 
\nodata                                           & 
{ $<1.0$}                                         \\

~16                                               & 
$\mu^+\nu$                                        & 
$<6.6$                                            & 
{\blue $<6.2$}                                    & 
{\blue $<1.7$}                                    & 
$<21$                                             & 
\nodata                                           & 
\nodata                                           & 
{ $<1.7$}                                         \\

~17                                               & 
$\tau^+\nu$                                       & 
$<260$                                            & 
{\blue $\err{88}{68}{11}$}                        & 
{\red $\aerrsy{179}{56}{49}{46}{51}$}             & 
$<840$                                            & 
\nodata                                           & 
\nodata                                           & 
$132 \pm 49$                                      \\
\hline
\end{tabular}
\end{center}
\end{table}

\clearpage

%% file: btosg.tex
The decay $b \to s\gamma$ proceeds through a process of 
flavor changing neutral current. Since the charged Higgs or SUSY particles may
contribute in the penguin loop, the branching fraction is sensitive to physics
beyond the Standard Model. Experimentally, the branching fraction is measured
using either a semi-inclusive or an inclusive approach. A minimum 
photon energy requirement is applied in the analysis and the branching fraction
is corrected based on the theoretical model for the photon energy spectrum 
(shape function). Although there are several experimental results   
available, only one measurement each for \babar, Belle and CLEO   
is used in the HFAG average to avoid dealing with correlated 
 errors for results reported from the same experiment. Furthermore, the 
 model uncertainties from the shape function should be highly 
correlated but no proper action was made in our previous averages. 
To perform the average with better precision and good accuracy, it is 
important to use as many experimental 
results as possible and to handle the shape function issue in a proper
way. In this note, we report the updated average of $b\to s\gamma$ branching
fraction by implementing a common  shape function.  

Several shape function schemes are commonly used.
Usually one is chosen to obtain the extrapolation factor, 
defined as the ratio of the $b\to s\gamma$ branching fractions 
with minimum photon energies above and at 1.6 GeV,
and the difference between various schemes are treated as the 
model uncertainty. Recently O. Buchm\"uller and H. Fl\"acher have calculated the
extrapolation factors \cite{Buchmuller:2005zv}.
Table \ref{tab:factor} lists the extrapolation factors with various photon
energy cuts for three different schemes and the average. The appropriate
approach to average the experimental results is to first convert them 
according to the average extrapolation factors and then perform the average,
assuming that the errors of the extrapolation factors are 100\% correlated. 

\begin{table}[h]
\caption{Extrapolation factor in various scheme with various minimum
   photon energy requirement (in GeV).}
\begin{tabular}{lccccc} \hline\hline
Scheme & $E_\gamma < 1.7$ & $E_\gamma < 1.8$ & $E_\gamma < 1.9$ & 
$E_\gamma < 2.0$ & $E_\gamma < 2.242$ \\ \hline
Kinetic & $0.986\pm 0.001$ & $0.968\pm 0.002$ & $0.939\pm 0.005$ & $0.903\pm0.009$ & $0.656\pm 0.031$ \\
Neubert SF & $0.982\pm 0.002$ & $0.962\pm 0.004$ & $0.930\pm 0.008$ & $0.888\pm 0.014$ & $0.665\pm 0.035$ \\
Kagan-Neubert & $0.988\pm 0.002$ & $0.970\pm 0.005$ & $0.940\pm 0.009$ & $0.892\pm 0.014$ & $0.643\pm 0.033$\\ \hline
Average & $0.985\pm 0.004$ & $0.967\pm 0.006$ & $0.936\pm 0.010$ & $0.894\pm 0.016$ &  $0.655\pm 0.037$ \\ \hline
\end{tabular}
\label{tab:factor}
\end{table}        
                  
After surveying all available experimental results, the five shown in 
Table \ref{tab:measurement} are selected for the average.  They
have provided in their papers either the $b\to s\gamma$ branching fraction at 
a certain photon energy cut or the extrapolation factor used.
Therefore we are able to convert them to the values at $E_{\rm min}= 1.6$ 
GeV using the information in Table \ref{tab:factor}.  The errors are, 
in order, statistical, systematic and shape-function systematic,
except for the \babar\ inclusive where there is a second systematic
error (third quoted error) due to theoretical uncertainties.
Moreover, in the three inclusive analyses a possible $b\to d\gamma$ 
contamination has been considered according to the theoretical expectation
of $(4.0\pm 1.6)$\%. The uncertainty from the $b\to d\gamma$ 
fraction in the three inclusive measurements should not be considered 
independently.  For those three measurements, a fourth uncertainty for
the $b\to d\gamma$ fraction is included.
We perform the average assuming that the systematic errors of the shape 
function and the $d\gamma$ fraction are correlated, and the other systematic
errors and the statistical errors are Gaussian and uncorrelated.     
The obtained average is 
${\cal B}(b\to s\gamma) = (355\pm 24^{+9}_{-10}\pm3)\times 10^{-6}$ with
a $\chi^2$/DOF$= 0.74/4$, where the 
errors are combined statistical and systematic, systematic due to the shape 
function, and the $d\gamma$ fraction. The last two errors are estimated to 
be the difference of the average after simultaneously varying the central
value of each experimental result by $\pm 1\sigma$. Although  a small fraction
of events was used in both the semi-inclusive and inclusive analyses in the
same experiment, we neglect their statistical correlations. Some
other correlated systematic errors, such as photon detection and the background
suppression, are not considered in our new average. 
In the future it would be better if each collaboration would provide a single 
combined result so that the average can be performed more accurately and easily.

\begin{table}[h]
\caption{Reported branching fraction, minimum photon energy, branching fraction
at minimum photon energy  and converted branching fraction for the 
decay $b\to s\gamma$. All the branching fractions are in units of $10^{-6}$.
See text for an explanation of the errors.
} 
\begin{tabular}{lccccc} \hline\hline
Mode & Reported $\cal{B}$ & $E_{\rm min}$ & $\cal{B}$ at $E_{\rm min}$ & Modified ${\cal{B}}~(E_{\rm min}=1.6)$ \\ \hline
CLEO Inc. \cite{cleo}& $321 \pm 43\pm 27^{+18}_{-10}$ & 2.0 & $306\pm 41\pm 26$ & $329\pm 44\pm 28 \pm 6\pm 6$ \\  
Belle Semi.\cite{belle1} & $336\pm 53 \pm 42^{+50}_{-54}$ & 2.24 & $-$ & 
$369\pm 58 \pm 46^{+56}_{-60}$\\
Belle Inc.\cite{belle2} & $355\pm 32^{+30+11}_{-31-7}$ & 1.8 &$351\pm32\pm29$&
$350\pm 32^{+30}_{-31} \pm 2\pm 2$ \\
\babar\ Semi.\cite{babar1} & $327\pm 18^{+55+4}_{-40-9}$ & 1.9 & 
$327\pm 18^{+55+4}_{-40-9}$ & $349\pm 20^{+59+4}_{-46-3}$ \\
\babar\ Inc.\cite{babar2} & $-$ & 1.9 & $367\pm 29\pm 34\pm 29$ & 
$392\pm 31\pm 36\pm 30 \pm 4\pm 6$ \\\hline   
\end{tabular}
\label{tab:measurement}
\end{table}

%% file: bary.tex
\begin{table}[!htbp]
\begin{center}
\caption{
 Branching fractions of  baryonic $B^+$ decays (in units of $10^{-6}$).
Upper limits are at 90\% CL.
values in {\red red} ({\blue blue}) are new {\red published}
({\blue preliminary}) result since PDG2006  [as of March 15, 2007].
}
\end{center}
\begin{center}
\footnotesize
\hspace*{-1.5cm} \vspace*{-0.5cm}
\begin{tabular}{|lcccccc|} 
\sgline
RPP\#   & Mode & PDG2006 Avg. & BABAR & Belle & CLEO & New Avg. \\
\sgline
286                                               & 
$p\overline p \pi^+$                              & 
$\cerr{3.1}{0.8}{0.7}$~\S                         & 
\nodata                                           & 
{\blue $\aerr{1.89}{0.28}{0.24}{0.14}~\ddag$ }    & 
$<160$                                            & 
$\cerr{3.06}{0.82}{0.72}$                         \\

289                                               & 
$p\overline p K^+$                                & 
$5.6\pm1.0$~\S                                    & 
{$\err{6.7}{0.5}{0.4}~\dag$}                      & 
{\blue $\aerr{5.98}{0.29}{0.27}{0.46}~\ddag$}     & 
\nodata                                           & 
$6.10 \pm 0.48$                                   \\

290                                               & 
$\Theta^{++} \overline p$ $^*$                    & 
$<0.091$                                          & 
{$<0.09$}                                         & 
{$<0.091$}                                        & 
\nodata                                           & 
{$<0.09$}                                         \\

291                                               & 
$f_J(2221) K^+$ $^*$                              & 
$<0.41$                                           & 
\nodata                                           & 
{$<0.41$}                                         & 
\nodata                                           & 
{$<0.41$}                                         \\

292                                               & 
$p \overline\Lambda(1520)$                        & 
$< 1.5$                                           & 
{$< 1.5$}                                         & 
\nodata                                           & 
\nodata                                           & 
{$< 1.5$}                                         \\

294                                               & 
$p\overline p K^{*+}$                             & 
$\aerrsy{10.3}{3.6}{2.8}{1.3}{1.7}~\ddag$         & 
\nodata                                           & 
{$\aerrsy{10.3}{3.6}{2.8}{1.3}{1.7}~\ddag$}       & 
\nodata                                           & 
$\cerr{10.3}{3.8}{3.3}$                           \\

295                                               & 
$p \overline\Lambda$                              & 
$<0.49$                                           & 
\nodata                                           & 
{\blue $< 0.32$}                                  & 
$< 1.5$                                           & 
{ $< 0.29$}                                       \\

New                                               & 
$p \overline\Lambda \pi^0$                        & 
\nodata                                           & 
\nodata                                           & 
{\blue $\aerr{3.00}{0.61}{0.53}{0.33}$}           & 
\nodata                                           & 
{$\cerr{3.00}{0.69}{0.62}$}                                       \\

299                                               & 
$\Lambda \overline{\Lambda} \pi^+$                & 
$<2.8~\ddag$                                      & 
\nodata                                           & 
{$<2.8~\ddag$ }                                   & 
\nodata                                           & 
{$<2.8~\ddag$ }                                   \\

300                                               & 
$\Lambda \overline{\Lambda} K^+$                  & 
$\aerr{2.9}{0.9}{0.7}{0.4}~\ddag$                 & 
\nodata                                           & 
{$\aerr{2.9}{0.9}{0.7}{0.4}~\ddag$}               & 
\nodata                                           & 
$\cerr{2.9}{1.0}{0.8}$                            \\

301                                               & 
$\overline{\Delta}^0 p$                           & 
$<380$                                            & 
\nodata                                           & 
{\blue $<1.42$}                                   & 
$<380$                                            & 
$<1.42$                                          \\

302                                               & 
$\Delta^{++} \overline p$                         & 
$<150$                                            & 
\nodata                                           & 
{\blue $<0.14$}                                   & 
$<150$                                            & 
$<0.14$                                        \\


\hline
\end{tabular}
\vspace*{0.6cm}
\end{center}
\hspace{1.cm}$\dag$ Charmonium decays to $p\bar p$ have been
statistically subtracted.
                                                                                
\hspace{1.cm}$\ddag$ The charmonium mass region has been vetoed.
                                                                                
\hspace{1.cm}$^*$ Product BF - daughter BF taken to be 100\%:
                                                                                
\hspace{1.cm}$~~~~ \Theta(1540)^{++}\to K^+p$ (pentaquark candidate);
                                                                                
\hspace{1.cm}$~~~~ {\mathcal G}(2220)\to p\overline p$ (glueball candidate).\\
\end{table}

\begin{table}[!htbp]
\begin{center}
\caption{
Branching fractions of  baryonic $B^0$ decays
(in units of $10^{-6}$). Upper limits are at 90\% CL.
values in {\red red} ({\blue blue}) are new {\red published}
({\blue preliminary}) result since PDG2006  [as of March 15, 2007].
}
\vskip 0.25cm
\end{center}
\begin{center}
\footnotesize
\begin{tabular}{|lcccccc|} 
\sgline
RPP\#   & Mode & PDG2006 Avg. & BABAR & Belle & CLEO & New Avg. \\
\sgline
266                                               & 
$p \overline{p}$                                  & 
$< 0.27$                                          & 
{$<0.27$}                                         & 
{\blue $<0.11$}                                   & 
$<1.4$                                            & 
{ $<0.11$}                                        \\

268                                               & 
$p \overline{p} K^0$                              & 
$\cerr{2.1}{0.6}{0.4}$~\S                         & 
\nodata                                           & 
{$\aerr{2.40}{0.64}{0.44}{0.28}~\ddag$}           & 
\nodata                                           & 
$\cerr{2.40}{0.70}{0.52}$                         \\

269                                               & 
$\Theta^+ K^0$~$\dag$                             & 
$<0.23$                                           & 
\nodata                                           & 
{$<0.23$}                                         & 
\nodata                                           & 
{$<0.23$}                                         \\

270                                               & 
$p \overline{p} K^{*0}$                           & 
$<7.6~\ddag$                                      & 
\nodata                                           & 
{$<7.6~\ddag$}                                    & 
\nodata                                           & 
{$<7.6~\ddag$}                                    \\

271                                               & 
$p \overline\Lambda \pi^-$                        & 
$2.6\pm0.5$~\S                                    & 
\blue {$\err{3.30}{0.53}{0.31}$}                  & 
{\blue $\aerr{3.23}{0.33}{0.29}{0.33}$}           & 
$<13$                                             & 
$\cerr{3.29}{0.47}{0.44}$                         \\

272                                               & 
$p \overline\Lambda K^-$                          & 
$<0.82$                                           & 
\nodata                                           & 
$< 0.82$                                          & 
\nodata                                           & 
$< 0.82$                                          \\

273                                               & 
$p \overline\Sigma^0 \pi^-$                       & 
$<3.8$                                            & 
\nodata                                           & 
$< 3.8$                                           & 
\nodata                                           & 
$< 3.8$                                           \\

274                                               & 
$\Lambda \overline\Lambda$                        & 
$<0.69$                                           & 
\nodata                                           & 
{\blue $<0.32$}                                   & 
$<1.2$                                            & 
{ $<0.32$}                                        \\

            
\hline
\end{tabular}
\end{center}
\vspace*{0.6cm}
\hspace{1.cm}$\dag$ Product BF - daughter BF taken to be 100\%:
$\ddag$ The charmonium mass region has been
 
\hspace{1.2cm} vetoed. $\Theta(1540)^+\to p K_S^0$ (pentaquark candidate).
\end{table}

%% file: bs.tex
\begin{table}[!htbp]
\caption{
 $B_s$   
branching fractions (in units of $10^{-6}$). Upper limits are at 90\% CL.
Values in {\red red} ({\blue blue}) are new {\red published}
({\blue preliminary}) result since PDG2006  [as of March 15, 2007].
}
\vskip 0.25cm

\begin{center}
\begin{tabular}{|lcccccc|} \hline
RPP\# &Mode & PDG2006 Avg. &  Belle & CDF & D\O\
 & New Avg.  \\ \sglinespb
$~9$                                              & 
$\pi^+\pi^-$                                      & 
$<170 $                                           & 
\nodata                                           & 
\blue$0.53\pm0.31\pm0.40$                         & 
\nodata                                           & 
$0.53 \pm 0.51$                                   \\

$15$                                              & 
$\phi\phi$                                        & 
$14\pm8$                                          & 
\nodata                                           & 
$\aerr{14}{6}{5}{6}$ $\dag$                       & 
\nodata                                           & 
$\cerr{14}{8}{7}$                                 \\

$16$                                              & 
$\pi^+K^-$                                        & 
$<210 $                                           & 
\nodata                                           & 
\blue $5.0\pm0.75\pm1.0$                          & 
\nodata                                           & 
$5.00 \pm 1.25$                                   \\

$17$                                              & 
$K^+K^-$                                          & 
$<59 $                                            & 
\blue $<340$                                      & 
\blue $\err{24.4}{1.4}{4.6}$                      & 
\nodata                                           & 
$24.4 \pm 4.8$                                    \\
\hline
\end{tabular}
\end{center}
\hspace{1.cm} $\dag$Relative BF converted to absolute BF
\end{table}

\begin{table}[h]
\begin{center}
\caption{
 $B_s$ rare relative
branching fractions.
Values in {\red red} ({\blue blue}) are new {\red published}
({\blue preliminary}) result since PDG2006  [as of March 15, 2007].
}
\vskip 0.25cm

{\scriptsize

\begin{tabular}{|lccccc|} \hline
RPP\# &Mode & PDG2006 Avg. &  CDF & D\O\
 & New Avg.  \\ \sglinespb
$~9$                                              & 
$\it{f}_s\mathcal{B}(B^0_s\rightarrow\pi^+\pi^-)/\it{f}_d\mathcal{B}(B^0\rightarrow K^+\pi-)$& 
\nodata                                           & 
\blue$0.007\pm0.004\pm0.005 $                     & 
\nodata                                           & 
$0.007 \pm 0.006$                                 \\

$15$                                              & 
$\mathcal{B}(B^0_s\rightarrow\phi\phi)/\mathcal{B}(B^0_s\rightarrow J/\psi\phi)$& 
\nodata                                           & 
$(10^{+5}_{-4}\pm1)\times 10^{-3} $               & 
\nodata                                           & 
$\cerr{10}{7}{6}$                                 \\

$16$                                              & 
$\it{f}_s\mathcal{B}(B^0_s\rightarrow K^+\pi^-)/\it{f}_d\mathcal{B}(B^0_d\rightarrow K^+\pi^-)$& 
\nodata                                           & 
\blue$0.066\pm0.010\pm0.010 $                     & 
\nodata                                           & 
$0.066 \pm 0.014$                                 \\

$17$                                              & 
$\it{f}_s\mathcal{B}(B^0_s\rightarrow K^+K^-)/\it{f}_d\mathcal{B}(B^0_d\rightarrow K^+\pi^-)$& 
\nodata                                           & 
\blue$0.324\pm0.019\pm0.041 $                     & 
\nodata                                           & 
$0.324 \pm 0.045$                                 \\
\hline
\end{tabular}
}
\end{center}
\end{table}

%% file: acp.tex
\begin{sidewaystable}[!htbp]
\parbox{8.2in}{\caption{
 $CP$ asymmetries for charmless hadronic charged $B$ decays.
Values in {\red red} ({\blue blue}) are 
new {\red published}
({\blue preliminary}) result since PDG2006 [as of March 15, 2007].
}}
                                                                                
\vspace*{ 0.5cm}
\scriptsize
\begin{tabular}{|lccccccc|} 
\hline
RPP\#   & Mode & PDG2006 Avg. & BABAR & Belle & CLEO & CDF & New Avg. \\
\sglinespb

182                                               & 
$K^0\pi^+$                                        & 
$-0.02\pm 0.07$                                   & 
{\red$\err{-0.029}{0.039}{0.010}$}                & 
{\blue$\err{0.03}{0.03}{0.01}$}                   & 
$\err{0.18}{0.24}{0.02}$                          & 
\nodata                                           & 
$0.009 \pm 0.025$                                 \\

183                                               & 
$K^+\pi^0$                                        & 
$0.04\pm 0.04$                                    & 
{\blue $\err{0.016}{0.041}{0.012}$}               & 
{\blue $\err{0.07}{0.03}{0.01}$}                  & 
$\err{-0.29}{0.23}{0.02}$                         & 
\nodata                                           & 
$0.047 \pm 0.026$                                 \\
\end{tabular}
\end{sidewaystable}

\begin{sidewaystable}[!htbp]
\begin{center}
\parbox{7.5in}{\caption{
Charmless hadronic $CP$ asymmetries for $B^\pm/B^0$ admixtures.
Values in {\red red} ({\blue blue}) are new {\red published}
({\blue preliminary}) result since PDG2006  [as of March 15, 2007].
}}

\vspace*{ 0.5cm}
\scriptsize
\hspace*{-1.4cm}
\begin{tabular}{|lccccccc|}
\hline
RPP\#    & Mode & PDG2006 Avg. & BABAR & Belle & CLEO & CDF & New Avg. \\
\sglinespb
~58                                               & 
$K^*\gamma$                                       & 
$-0.01\pm 0.07$                                   & 
{$-0.013\pm 0.036\pm 0.010$}                      & 
{$-0.015\pm 0.044\pm 0.012$}                      & 
$0.08\pm 0.13\pm 0.03$                            & 
\nodata                                           & 
$-0.010 \pm 0.028$                                \\

~70                                               & 
$s\gamma$                                         & 
$0.00\pm0.04$                                     & 
$\err{0.025}{0.050}{0.015}$                       & 
$\err{0.002}{0.050}{0.030}$                       & 
$\err{-0.079}{0.108}{0.022}$                      & 
\nodata                                           & 
$0.004 \pm 0.037$                                 \\

$~-$                                              & 
$(s+d)\gamma$                                     & 
New                                               & 
\red$\err{-0.11}{0.12}{0.02}$                     & 
\nodata                                           & 
\nodata                                           & 
\nodata                                           & 
$-0.11 \pm 0.12$                                  \\
\hline
\end{tabular}
\end{center}

\vskip -1cm
\begin{center}
\parbox{8.5in}{\caption{
 $CP$ asymmetries for charmless hadronic neutral $B$ decays.
Values in {\red red} ({\blue blue}) are new {\red published}
({\blue preliminary}) result since PDG2006 [as of March 15, 2007].
}}

\vspace*{ 0.5cm}
\scriptsize

\begin{tabular}{|lccccccc|} 
\sgline
RPP\#   & Mode & PDG2006 Avg. & BABAR & Belle & CLEO & CDF & New Avg. \\
\sglinespb

168                                               & 
$K^+\pi^-$                                        & 
$-0.113\pm 0.020$                                 & 
{\blue $\berr{-0.107}{0.018}{0.007}{0.004}$}      & 
{\blue $\err{-0.093}{0.018}{0.008}$}              & 
$\err{-0.04}{0.16}{0.02}$                         & 
{\blue $\err{-0.086}{0.023}{0.009}$}              & 
$-0.095 \pm 0.013$                                \\

171                                               & 
$\etapr K^{*0}$                                   & 
New                                               & 
\red{$\err{-0.08}{0.25}{0.02}$}                   & 
\nodata                                           & 
\nodata                                           & 
\nodata                                           & 
$-0.08 \pm 0.25$                                  \\
\end{tabular}
\vskip 0.5cm\end{center}

\large
\hspace{-1.9cm}
\dag~Measurements of time-dependent $CP$ asymmetries are listed on 
the Unitarity Triangle home page. (http://www.slac.stanford.edu/xorg/hfag/triangle/index.html) 

\end{sidewaystable}

%% file: polar.tex
%

%

\begin{table}[!htbp]
\caption{
 Longitudinal polarization fraction $f_L$ for $B^+$ decays.
Values in {\red red} ({\blue blue}) are new {\red published}
({\blue preliminary}) result since PDG2006 [as of March 15, 2007].
\vspace{0.3cm}
}

\begin{center}
\begin{tabular}{|lccccc|} 
\sgline
RPP\#   & Mode & PDG2006 Avg. & BABAR & Belle & New Avg. \\
\sglinespb

%

213                                               & 
$K^{*+}\rho^0$                                    & 
$\aerr{0.96}{0.04}{0.15}{0.04}$                   & 
$\aerr{0.96}{0.04}{0.15}{0.05}$                   & 
\nodata                                           & 
$\cerr{0.96}{0.06}{0.16}$                         \\

214                                               & 
$K^{*0}\rho^+$                                    & 
$\berr{0.43}{0.11}{0.05}{0.02}$                   & 
\red{$\err{0.52}{0.10}{0.04}$}                    & 
{$\berr{0.43}{0.11}{0.05}{0.02}$}                 & 
$0.48 \pm 0.08$                                   \\

236                                               & 
$\phi K^{*+}$                                     & 
$0.50\pm0.07$                                     & 
$\err{0.46}{0.12}{0.03}$                          & 
{$\err{0.52}{0.08}{0.03}$}                        & 
$0.50 \pm 0.07$                                   \\

266                                               & 
$\rho^+\rho^0$                                    & 
$0.96\pm0.06$                                     & 
\red{$\berr{0.905}{0.042}{0.023}{0.027}$}         & 
$\err{0.95}{0.11}{0.02}$                          & 
$\cerr{0.912}{0.044}{0.045}$                      \\

270                                               & 
$\omega\rho^+$                                    & 
$\aerr{0.88}{0.12}{0.15}{0.03}$                   & 
\red{$\err{0.82}{0.11}{0.02}$}                    & 
\nodata                                           & 
$0.82 \pm 0.11$                                   \\

\sglinespt
\end{tabular}
\end{center}
\end{table}

\begin{table}
\caption{
Full angular analysis of $B^+\to\phi K^{*+}$.
Values in {\red red} ({\blue blue}) are new {\red published}
({\blue preliminary}) result since PDG2006 [as of March 15, 2007].
\vspace{0.3cm}
}

\begin{center}
\begin{tabular}{|lccccc|} 
\sgline
\RPP & Parameter & PDG2006 Avg. & BABAR & Belle & New Avg. \\
\sglinespb
\nodata                                           & 
$f_\perp$                                         & 
$\err{0.19}{0.08}{0.02}$                          & 
\nodata                                           & 
{$\err{0.19}{0.08}{0.02}$}                        & 
$0.19 \pm 0.08$                                   \\

\nodata                                           & 
$\phi_\parallel$                                  & 
$\err{2.10}{0.28}{0.04}$                          & 
\nodata                                           & 
{$\err{2.10}{0.28}{0.04}$}                        & 
$2.10 \pm 0.28$                                   \\

\nodata                                           & 
$\phi_\perp$                                      & 
$\err{2.31}{0.30}{0.07}$                          & 
\nodata                                           & 
{$\err{2.31}{0.30}{0.07}$}                        & 
$2.31 \pm 0.31$                                   \\

\sglinespt
\end{tabular}
\end{center}
\hspace*{2.0cm}BR, $f_L$ and $A_{CP}$ are tabulated separately.
\end{table}

\large

\begin{table}[!htbp]
\caption{
Longitudinal polarization fraction $f_L$ for $B^0$ decays.
Values in {\red red} ({\blue blue}) are new {\red published}
({\blue preliminary}) result since PDG2006  [as of March 15, 2007].
\vspace{0.3cm}
}
\scriptsize

\begin{center}
\begin{tabular}{|lcccccc|} 
\sgline
RPP\#   & Mode & PDG2006 Avg. & BABAR & Belle & CDF & New Avg. \\
\hline

199                                               & 
$K^{*0}\rho^0$                                    & 
New                                               & 
\red{$\err{0.57}{0.09}{0.08}$}                    & 
\nodata                                           & 
\nodata                                           & 
$0.57 \pm 0.12$                                   \\

204                                               & 
$\phi K^{*0}$                                     & 
$0.48\pm0.04$                                     & 
{\red$\err{0.506}{0.040}{0.015}$}                 & 
{$\err{0.45}{0.05}{0.02}$}                        & 
\blue{$\err{0.57}{0.10}{0.05}$}                   & 
$0.491 \pm 0.032$                                 \\

$~-$                                              & 
$\phi K_2^*(1430)^0$                              & 
New                                               & 
{\red$\err{0.85}{0.07}{0.04}$}                    & 
\nodata                                           & 
\nodata                                           & 
$0.85 \pm 0.08$                                   \\

253                                               & 
$\rho^0\rho^0$                                    & 
New                                               & 
\blue{$\aerr{0.86}{0.11}{0.13}{0.05}$}            & 
\nodata                                           & 
\nodata                                           & 
$\cerr{0.86}{0.12}{0.14}$                         \\

257                                               & 
$\rho^+\rho^-$                                    & 
$\cerr{0.967}{0.022}{0.027}$                      & 
\blue{$\berr{0.977}{0.024}{0.015}{0.013}$}        & 
{$\aerr{0.941}{0.034}{0.040}{0.030}$}             & 
\nodata                                           & 
$0.968 \pm 0.023$                                 \\

\sglinespt
\end{tabular}
\end{center}
\end{table}

\begin{table}[!htbp]
\caption{
Full angular analysis of $B^0 \to \phi K^{*0}$.
Values in {\red red} ({\blue blue}) are new {\red published}
({\blue preliminary}) result since PDG2006  [as of March 15, 2007].
\vspace{0.3cm}
}

\begin{center}
\scriptsize
\begin{tabular}{|lcccccc|} 
\sgline
\RPP & Parameter & PDG2006 Avg. & BABAR & Belle & CDF & New Avg. \\
\sglinespb
\nodata                                           & 
$f_\perp = \Lambda_{\perp\perp}$                  & 
$0.26 \pm0.04$                                    & 
{\red$\err{0.227}{0.038}{0.013}$}                 & 
{$\aerr{0.31}{0.06}{0.05}{0.02}$}                 & 
\blue{$\err{0.20}{0.10}{0.05}$}                   & 
$0.252 \pm 0.031$                                 \\

\nodata                                           & 
$\phi_\parallel$                                  & 
$\cerr{2.36}{0.18}{0.16}$                         & 
{\red$\err{2.31}{0.14}{0.08}$}                    & 
{$\aerr{2.40}{0.28}{0.24}{0.07}$}                 & 
\blue{$\err{2.97}{0.52}{0.26}$}                   & 
$\cerr{2.37}{0.14}{0.13}$                         \\

\nodata                                           & 
$\phi_\perp$                                      & 
$2.49\pm0.18$                                     & 
{\red$\err{2.24}{0.15}{0.09}$}                    & 
{$\err{2.51}{0.25}{0.06}$}                        & 
\blue{$\err{2.77}{0.37}{0.37}$}                   & 
$2.36 \pm 0.14$                                   \\

\nodata                                           & 
$A_{CP}^0$                                        & 
$0.01\pm0.09$                                     & 
{\red$\err{-0.03}{0.08}{0.02}$}                   & 
{$\err{0.13}{0.12}{0.04}$}                        & 
\nodata                                           & 
$0.02 \pm 0.07$                                   \\

\nodata                                           & 
$A_{CP}^\perp$                                    & 
$-0.16\pm0.15$                                    & 
{\red$\err{-0.03}{0.16}{0.05}$}                   & 
{$\err{-0.20}{0.18}{0.04}$}                       & 
\nodata                                           & 
$-0.11 \pm 0.12$                                  \\

\nodata                                           & 
$\Delta\phi_\parallel$                            & 
$0.02\pm0.28$                                     & 
{\red$\err{0.24}{0.14}{0.08}$}                    & 
{$\err{-0.32}{0.27}{0.07}$}                       & 
\nodata                                           & 
$0.10 \pm 0.14$                                   \\

\nodata                                           & 
$\Delta\phi_\perp$                                & 
$0.03\pm0.33$                                     & 
{\red$\err{0.19}{0.15}{0.08}$}                    & 
{$\err{-0.30}{0.25}{0.06}$}                       & 
\nodata                                           & 
$0.04 \pm 0.14$                                   \\

\sglinespt
\end{tabular}
\end{center}
BR, $f_L$ and $A_{CP}$ are tabulated separately.
\end{table}

\begin{table}[!htbp]
\caption{
Full angular analysis of $B^0 \to \phi K_2^*(1430)^0$.
Values in {\red red} ({\blue blue}) are new {\red published}
({\blue preliminary}) result since PDG2006  [as of March 15, 2007].
\vspace{0.3cm}
}

\begin{center}
\scriptsize
\begin{tabular}{|lcccccc|} 
\sgline
\RPP & Parameter & PDG2006 Avg. & BABAR & Belle & CDF & New Avg. \\
\sglinespb
\nodata                                           & 
$f_\perp = \Lambda_{\perp\perp}$                  & 
New                                               & 
{\red$\aerr{0.045}{0.049}{0.040}{0.013}$}         & 
\nodata                                           & 
\nodata                                           & 
$\cerr{0.045}{0.051}{0.042}$                      \\

\nodata                                           & 
$\phi_\parallel$                                  & 
New                                               & 
{\red$\err{2.90}{0.39}{0.06}$}                    & 
\nodata                                           & 
\nodata                                           & 
$2.90 \pm 0.40$                                   \\

\nodata                                           & 
$\phi_\perp$                                      & 
New                                               & 
{\red$\aerr{5.7}{0.6}{0.9}{0.1}$}                 & 
\nodata                                           & 
\nodata                                           & 
$\cerr{5.7}{0.6}{0.9}$                            \\

\sglinespt
\end{tabular}
\end{center}
BR, $f_L$ and $A_{CP}$ are tabulated separately.
\end{table}

%% file: b2charm.tex
%
			 
\newpage\normalsize
\mysection{$B$ Decays to open charm and charmonium final states }\label{sec:BtoCharm}
\input{b2charm_summer2006_v009/b2charm_summer2006_v009_blurb}

\clearpage
\input{b2charm_summer2006_v009/b2charm_summer2006_v009_tables}
\clearpage

%% file: b2charm_summer2006_v009/b2charm_summer2006_v009_blurb.tex

This section reports the updated contribution to the HFAG report from the ``$B \to$ charm " group\footnote{The HFAG/BtoCharm group was formed in the spring of 2005; it performs its work using an XML database backed web application.}.
The mandate of the group is to compile measurements and perform averages of all available quantities related to 
$B$ decays to charmed particles, excluding CP related quantities. To date the 
group has analyzed a total of 431 measurements reported in 131 papers, principally branching fractions.
The group aims to organize and present the copious information on $B$ decays to charmed
particles obtained from a combined sample of more than one billion $B$ mesons
from the BABAR, Belle and CDF Collaborations. 

Branching fractions for rare $B$-meson decays or decay chains of a few $10^{-7}$ are being measured with 
statistical uncertainties typically below $30\%$. 
Results for more common decay chains, with branching fractions around $10^{-4}$, are becoming precision measurements, 
with uncertainties typically at the $3\%$ level. 
Some decays have been observed for the first time, for example $B^- \to \Lambda_c^- \Lambda_c^+ K^-$ 
or $B^-\to \chi_{c1} \pi^-$, with a branching fraction of $(6.5\pm3.7)\times 10^{-4}$ and $(2.2\pm0.5)\times 10^{-5}$, respectively.

Among the many results, we highlight the great improvements that have been attained towards a deeper understanding 
of recently discovered new states with either hidden or open charm content. 
The $J/\psi\gamma$ decay mode of the $X(3870)$ has been observed by both BABAR and Belle and the average branching fraction is $\hfcBR(B^-\to X(3870) K^-)\times \hfcBR(X(3870\to J/\psi\gamma)=(2.4\pm0.5)\times 10^{-6}$: this final state allowed to unambiguously establish the positive $C$ parity of the $X(3870)$.
Branching fractions for several $B$ decays to $D_{sJ}^{*-}(2317)$ and $D_{sJ}^-(2460)$ have been measured, and also absolute branching fraction measurements have been reported: $\hfcBR(\bar{B}^0\to D_{sJ}^-(2460) D^+) = (0.26\pm0.17)\times 10^{-2}$, $\hfcBR(\bar{B}^0\to D_{sJ}^-(2460) D^{*+}) = (0.88\pm0.24)\times 10^{-2}$, $\hfcBR(B^-\to D_{sJ}^-(2460) D^0) = (0.43\pm0.21)\times 10^{-2}$ and $\hfcBR(B^-\to D_{sJ}^-(2460) D^{*0}) = (1.12\pm0.33)\times 10^{-2}$. The abundance of measurements with many different final states for these particles is of the greatest importance for quantum number assignments, and already some of the proposed theoretical interpretations have been ruled out.

The measurements are classified according to the decaying particle: Charged B, Neutral B or Miscellaneous; the decay products 
and the type of quantity: branching fraction, product of branching fractions, ratio of branching fractions or other quantities. 
For the decay product classification the below precedence order is used to ensure that each measurement appears in only one category. 
\begin{itemize}\addtolength{\itemsep}{-0.4\baselineskip}
\item new particles
\item strange $D$ mesons
\item baryons
\item $J/\psi$
\item charmonium other than $J/\psi$
\item multiple $D$, $D^{*}$ or $D^{**}$ mesons
\item a single $D^{*}$ or $D^{**}$ meson
\item a single $D$ meson
\item other particles
\end{itemize}
  
Within each table the measurements are color coded according to the
publication status and age. Table~\ref{tab:hfc99999} provides a key to the
color scheme and categories used. When viewing the tables with most pdf
viewers every number, label and average provides hyperlinks to the corresponding 
reference and individual quantity web pages on the HFAG/Charm group website
\hfhref{}{http://hfag.phys.ntu.edu.tw}.
The links provided in the captions of the table lead to the corresponding compilation
pages.  Both the individual and compilation webpages provide a graphical view
of the results, in a variety of formats.

Tables \ref{tab:hfc01101} to \ref{tab:hfc03300} provide either limits at 90\%
confidence level or measurements with statistical and systematic uncertainties 
and in some cases a third error corresponding to correlated systematics. 
For details on the meanings of the uncertainties and access to the references 
click on the numbers to visit the corresponding web pages.  Where there are
multiple determinations of the same quantity by one experiment the table
footnotes act to distinguish the methods or datasets used; such cases are
visually highlighted in the table by presenting the measurements on the lines
beneath the quantity label.
Where both limits and measured values of a quantity are available the limits 
are presented in the tables but are not used in the determination of the
average. Where only limits are available the most stringent is presented in
the Average column of the tables.
Where available the PDG 2006 result is also presented.

%% file: b2charm_summer2006_v009/b2charm_summer2006_v009_tables.tex

\begin{\hftabletype}[\hftableposn]

\begin{center}
\caption{Key to the colors used to classify the results presented in tables \ref{tab:hfc01101} to \ref{tab:hfc03300}. When viewing these tables in a pdf viewer each number, label and average provides a hyperlink to the corresponding online version provided by the charm subgroup website \hfurl{}. Where an experiment has multiple determinations of a single quantity they are distinguished by the table footnotes.}
 \label{tab:hfc99999} 
 
\hfmetadata{  }

\begin{tabular}{ll}  
Class
&  Definition
\\\hline 

{ \hfwaitingtext{waiting} } \hftabletlcell 
 & { \hfdeftext{Results without a preprint available } } 
\\%

{ \hfpubhottext{pubhot} }  
 & { \hfdeftext{Results published during or after 2006} } 
\\%

{ \hfprehottext{prehot} }  
 & { \hfdeftext{Preprint released during or after  2006} } 
\\%

{ \hfpubtext{pub} }  
 & { \hfdeftext{Results published after or during the last PDG year} } 
\\%

{ \hfpretext{pre} }  
 & { \hfdeftext{Preprint released after or during the last PDG year} } 
\\%

{ \hfpuboldtext{pubold} }  
 & { \hfdeftext{Results published before the last PDG year} } 
\\%

{ \hfpreoldtext{preold} }  
 & { \hfdeftext{Preprint released before the last PDG year} } 
\\%

{ \hferrortext{error} }  
 & { \hfdeftext{Incomplete information to classify } } 
\\%

{ \hfsuperceededtext{superceeded} }  
 & { \hfdeftext{Results superceeded by more recent measurements from the same experiment } } 
\\%

{ \hfinactivetext{inactive} }  
 & { \hfdeftext{Results in the process of being entered into the database} } 
\\%

{ \hfnoquotext{noquo} }  
 & { \hfdeftext{Results without quotes} } 
\\\hline 

\end{tabular}

\end{center}

\end{\hftabletype}
\clearpage  
\begin{\hftabletype}[\hftableposn]

\begin{center}
\hfnewcaption{Branching fractions}{charged B}{\hfnewp}{in units of $10^{-3}$}{upper limits are at 90\% CL}{00101.html}{tab:hfc01101}
\hfmetadata{ [ .tml created 2007-04-06T22:30:37.653+08:00] }

\begin{tabular}{lccccc}  
\hflmode
&  \hflpdg2006
&  \hflbelle
&  \hflbabar
&  \hflcdf
&  \hflavg
\\\hline 

 \hfhref{BR_-521_-321+81.html}{ \hflabel{X(3872) K^-} } \hftabletlcell 
 &  \hfhref{BR_-521_-321+81.html}{ \hfpdg{<0.32} }  
 &  { \,}  
 &  \hfhref{0505003.html}{ \hfpubhot{<0.32} }  
 &  { \,}  
 &  \hfhref{BR_-521_-321+81.html}{ \hfavg{<0.32} } 
\\%

 \hfhref{BR_-521_-85+421.html}{ \hflabel{D_{sJ}^{-}(2460) D^0} }  
 &  { \,}  
 &  { \,}  
 &  \hfhref{0604004.html}{ \hfpubhot{4.3\pm1.6\pm1.3} }  
 &  { \,}  
 &  \hfhref{BR_-521_-85+421.html}{ \hfavg{4.3\pm2.1} } 
\\%

 \hfhref{BR_-521_-85+423.html}{ \hflabel{D_{sJ}^{-}(2460) D^{*0}(2007)} } \hftableblcell 
 &  { \,}  
 &  { \,}  
 &  \hfhref{0604004.html}{ \hfpubhot{11.2\pm2.6\pm2.0} }  
 &  { \,}  
 &  \hfhref{BR_-521_-85+423.html}{ \hfavg{11.2\pm3.3} } 
\\\hline 

\end{tabular}

\end{center}


\begin{center}
\hfnewcaption{Product branching fractions}{charged B}{\hfnewp}{in units of $10^{-4}$}{upper limits are at 90\% CL}{00101.html}{tab:hfc02101}
\hfmetadata{ [ .tml created 2007-04-06T22:30:50.619+08:00] }

\begin{tabular}{lccccc}  
\hflmode
&  \hflpdg2006
&  \hflbelle
&  \hflbabar
&  \hflcdf
&  \hflavg
\\\hline 

 \hfhref{BR_-521_-321+81xBR_81_22+443.html}{ \hflabel{K^- X(3872) [ \gamma J/\psi(1S) ]} } \hftabletlcell 
 &  { \,}  
 &  \hfhref{0506060.html}{ \hfpre{0.0180\pm0.0060\pm0.0010} }  
 &  \hfhref{0607015.html}{ \hfpubhot{0.033\pm0.010\pm0.003} }  
 &  { \,}  
 &  \hfhref{BR_-521_-321+81xBR_81_22+443.html}{ \hfnewavg{0.022\pm0.005} } 
\\%

 \hfhref{BR_-521_-321+81xBR_81_221+443.html}{ \hflabel{K^- X(3872) [ J/\psi(1S) \eta ]} }  
 &  \hfhref{BR_-521_-321+81xBR_81_221+443.html}{ \hfpdg{<0.077} }  
 &  { \,}  
 &  \hfhref{0505002.html}{ \hfpub{<0.077} }  
 &  { \,}  
 &  \hfhref{BR_-521_-321+81xBR_81_221+443.html}{ \hfavg{<0.077} } 
\\%

 \hfhref{BR_-521_-321+81xBR_81_-211+211+443.html}{ \hflabel{K^- X(3872) [ \pi^+ \pi^- J/\psi(1S) ]} }  
 &  \hfhref{BR_-521_-321+81xBR_81_-211+211+443.html}{ \hfpdg{0.11\pm0.02} }  
 &  \hfhref{0506061.html}{ \hfpre{0.13\pm0.02\pm0.01} }  
 &  \hfhref{0507002.html}{ \hfpubhot{0.101\pm0.025\pm0.010} }  
 &  { \,}  
 &  \hfhref{BR_-521_-321+81xBR_81_-211+211+443.html}{ \hfavg{0.12\pm0.02} } 
\\%

 \hfhref{BR_-521_-321+83xBR_83_22+443.html}{ \hflabel{K^- Y(3940) [ J/\psi(1S) \gamma ]} }  
 &  { \,}  
 &  { \,}  
 &  \hfhref{0607015.html}{ \hfpubhot{<0.140} }  
 &  { \,}  
 &  \hfhref{BR_-521_-321+83xBR_83_22+443.html}{ \hfnewavg{<0.140} } 
\\%

 \hfhref{BR_-521_-321+86xBR_86_-211+211+443.html}{ \hflabel{K^- Y(4260) [ J/\psi(1S) \pi^+ \pi^- ]} }  
 &  \hfhref{BR_-521_-321+86xBR_86_-211+211+443.html}{ \hfpdg{<0.29} }  
 &  { \,}  
 &  \hfhref{0507002.html}{ \hfpubhot{0.20\pm0.07\pm0.02} }  
 &  { \,}  
 &  \hfhref{BR_-521_-321+86xBR_86_-211+211+443.html}{ \hfavg{0.20\pm0.07} } 
\\
\hline 

\end{tabular}

\end{center}

\end{\hftabletype}
\clearpage  
\begin{\hftabletype}[\hftableposn]

\begin{center}
\hfnewcaption{Branching fractions}{charged B}{\hfdstr}{in units of $10^{-4}$}{upper limits are at 90\% CL}{00102.html}{tab:hfc01102}
\hfmetadata{ [ .tml created 2007-04-06T22:47:18.396+08:00] }

\begin{tabular}{lccccc}  
\hflmode
&  \hflpdg2006
&  \hflbelle
&  \hflbabar
&  \hflcdf
&  \hflavg
\\\hline 

 \hfhref{BR_-521_-431+333.html}{ \hflabel{D_s^- \phi(1020)} } \hftabletlcell 
 &  \hfhref{BR_-521_-431+333.html}{ \hfpdg{<0.019} }  
 &  { \,}  
 &  \hfhref{0506020.html}{ \hfpubhot{<0.019} }  
 &  { \,}  
 &  \hfhref{BR_-521_-431+333.html}{ \hfavg{<0.019} } 
\\%

 \hfhref{BR_-521_-433+333.html}{ \hflabel{D_s^{*-} \phi(1020)} }  
 &  \hfhref{BR_-521_-433+333.html}{ \hfpdg{<0.120} }  
 &  { \,}  
 &  \hfhref{0506020.html}{ \hfpubhot{<0.120} }  
 &  { \,}  
 &  \hfhref{BR_-521_-433+333.html}{ \hfavg{<0.120} } 
\\%

 \hfhref{BR_-521_-431+111.html}{ \hflabel{D_s^- \pi^0} }  
 &  \hfhref{BR_-521_-431+111.html}{ \hfpdg{<1.70} }  
 &  { \,}  
 &  \hfhref{0506015.html}{ \hfwaiting{<0.28} }  
 &  { \,}  
 &  \hfhref{BR_-521_-431+111.html}{ \hfavg{<0.28} } 
\\%

 \hfhref{BR_-521_-321+-211+433.html}{ \hflabel{D_s^{*+} K^- \pi^-} }  
 &  \hfhref{BR_-521_-321+-211+433.html}{ \hfpdg{<9.8} }  
 &  { \,}  
 &  \hfhref{0607001.html}{ \hfprehot{1.84\pm0.19\pm0.40\pm0.06} }  
 &  { \,}  
 &  \hfhref{BR_-521_-321+-211+433.html}{ \hfnewavg{1.84\pm0.45} } 
\\%

 \hfhref{BR_-521_-321+-211+431.html}{ \hflabel{D_s^+ K^- \pi^-} }  
 &  \hfhref{BR_-521_-321+-211+431.html}{ \hfpdg{<7.0} }  
 &  { \,}  
 &  \hfhref{0607001.html}{ \hfprehot{1.88\pm0.13\pm0.41\pm0.06} }  
 &  { \,}  
 &  \hfhref{BR_-521_-321+-211+431.html}{ \hfnewavg{1.88\pm0.43} } 
\\%

 \hfhref{BR_-521_-433+421.html}{ \hflabel{D_s^{*-} D^0} }  
 &  \hfhref{BR_-521_-433+421.html}{ \hfpdg{72\pm26} }  
 &  { \,}  
 &  \hfhref{0604004.html}{ \hfpubhot{93\pm18\pm19} }  
 &  { \,}  
 &  \hfhref{BR_-521_-433+421.html}{ \hfavg{93\pm26} } 
\\%

 \hfhref{BR_-521_-431+423.html}{ \hflabel{D_s^- D^{*0}(2007)} }  
 &  \hfhref{BR_-521_-431+423.html}{ \hfpdg{100\pm40} }  
 &  { \,}  
 &  \hfhref{0604004.html}{ \hfpubhot{121\pm23\pm20} }  
 &  { \,}  
 &  \hfhref{BR_-521_-431+423.html}{ \hfavg{121\pm30} } 
\\
\hline 

\end{tabular}

\end{center}


\begin{center}
\hfnewcaption{Product branching fractions}{charged B}{\hfdstr}{in units of $10^{-4}$}{upper limits are at 90\% CL}{00102.html}{tab:hfc02102}
\hfmetadata{ [ .tml created 2007-04-06T22:47:30.087+08:00] }

\begin{tabular}{lccccc}  
\hflmode
&  \hflpdg2006
&  \hflbelle
&  \hflbabar
&  \hflcdf
&  \hflavg
\\\hline 

 \hfhref{BR_-521_-431+423xBR_-431_-211+333.html}{ \hflabel{D^{*0}(2007) D_s^- [ \phi(1020) \pi^- ]} } \hftabletlcell 
 &  \hfhref{BR_-521_-431+423xBR_-431_-211+333.html}{ \hfpdg{4.4\pm1.7} }  
 &  { \,}  
 &  \hfhref{0604004.html}{ \hfpubhot{2.95\pm0.65\pm0.36} }  
 &  { \,}  
 &  \hfhref{BR_-521_-431+423xBR_-431_-211+333.html}{ \hfavg{2.95\pm0.74} } 
\\%

 \hfhref{BR_-521_-433+421xBR_-431_-211+333.html}{ \hflabel{D_s^{*-} D^0 [ D_s^- \to \phi(1020) \pi^- ]} }  
 &  \hfhref{BR_-521_-433+421xBR_-431_-211+333.html}{ \hfpdg{3.2\pm1.1} }  
 &  { \,}  
 &  \hfhref{0604004.html}{ \hfpubhot{3.13\pm1.19\pm0.58} }  
 &  { \,}  
 &  \hfhref{BR_-521_-433+421xBR_-431_-211+333.html}{ \hfavg{3.1\pm1.3} } 
\\%

 \hfhref{BR_-521_-431+421xBR_-431_-211+333.html}{ \hflabel{D^0 D_s^- [ \phi(1020) \pi^- ]} }  
 &  \hfhref{BR_-521_-431+421xBR_-431_-211+333.html}{ \hfpdg{4.80\pm1.00} }  
 &  { \,}  
 &  \hfhref{0604004.html}{ \hfpubhot{4.00\pm0.61\pm0.61} }  
 &  { \,}  
 &  \hfhref{BR_-521_-431+421xBR_-431_-211+333.html}{ \hfavg{4.00\pm0.86} } 
\\%

 \hfhref{BR_-521_-433+423xBR_-431_-211+333.html}{ \hflabel{D_s^{*-} D^{*0}(2007) [ D_s^- \to \phi(1020) \pi^- ]} } \hftableblcell 
 &  \hfhref{BR_-521_-433+423xBR_-431_-211+333.html}{ \hfpdg{9.7\pm2.8} }  
 &  { \,}  
 &  \hfhref{0604004.html}{ \hfpubhot{8.6\pm1.5\pm1.1} }  
 &  { \,}  
 &  \hfhref{BR_-521_-433+423xBR_-431_-211+333.html}{ \hfavg{8.6\pm1.9} } 
\\\hline 

\end{tabular}

\end{center}

\end{\hftabletype}
\clearpage  
\begin{\hftabletype}[\hftableposn]

\begin{center}
\hfnewcaption{Branching fractions}{charged B}{\hfbary}{in units of $10^{-5}$}{upper limits are at 90\% CL}{00103.html}{tab:hfc01103}
\hfmetadata{ [ .tml created 2007-04-06T23:00:31.754+08:00] }

\begin{tabular}{lccccc}  
\hflmode
&  \hflpdg2006
&  \hflbelle
&  \hflbabar
&  \hflcdf
&  \hflavg
\\\hline 

 \hfhref{BR_-521_-2212+443+3212.html}{ \hflabel{J/\psi(1S) \Sigma^0 \bar{p}} } \hftabletlcell 
 &  \hfhref{BR_-521_-2212+443+3212.html}{ \hfpdg{<1.10} }  
 &  \hfhref{0607005.html}{ \hfpub{<1.10} }  
 &  { \,}  
 &  { \,}  
 &  \hfhref{BR_-521_-2212+443+3212.html}{ \hfnewavg{<1.10} } 
\\%

 \hfhref{BR_-521_-2212+443+3122.html}{ \hflabel{J/\psi(1S) \Lambda \bar{p}} }  
 &  \hfhref{BR_-521_-2212+443+3122.html}{ \hfpdg{1.18\pm0.31} }  
 &  \hfhref{0607005.html}{ \hfpub{1.16\pm0.28\pm^{0.18}_{0.23}} }  
 &  \hfhref{0506010.html}{ \hfpubold{1.16\pm^{0.74}_{0.53}\pm^{0.42}_{0.18}} }  
 &  { \,}  
 &  \hfhref{BR_-521_-2212+443+3122.html}{ \hfnewavg{1.16\pm0.31} } 
\\%

 \hfhref{BR_-521_-2212+413+2212.html}{ \hflabel{D^{*+}(2010) p \bar{p}} }  
 &  { \,}  
 &  \hfhref{0506083.html}{ \hfpubold{<1.50} }  
 &  { \,}  
 &  { \,}  
 &  \hfhref{BR_-521_-2212+413+2212.html}{ \hfavg{<1.50} } 
\\%

 \hfhref{BR_-521_-2212+411+2212.html}{ \hflabel{D^+ p \bar{p}} }  
 &  { \,}  
 &  \hfhref{0506083.html}{ \hfpubold{<1.50} }  
 &  { \,}  
 &  { \,}  
 &  \hfhref{BR_-521_-2212+411+2212.html}{ \hfavg{<1.50} } 
\\%

 \hfhref{BR_-521_-2212+4114.html}{ \hflabel{\Sigma_c^{*0} \bar{p}} }  
 &  \hfhref{BR_-521_-2212+4114.html}{ \hfpdg{<4.6} }  
 &  \hfhref{0506086.html}{ \hfpubold{<4.6} }  
 &  { \,}  
 &  { \,}  
 &  \hfhref{BR_-521_-2212+4114.html}{ \hfavg{<4.6} } 
\\%

 \hfhref{BR_-521_-2212+4112.html}{ \hflabel{\Sigma_c^0 \bar{p}} }  
 &  \hfhref{BR_-521_-2212+4112.html}{ \hfpdg{<8.0} }  
 &  \hfhref{0506086.html}{ \hfpubold{<9.3} }  
 &  { \,}  
 &  { \,}  
 &  \hfhref{BR_-521_-2212+4112.html}{ \hfavg{<9.3} } 
\\
\hline 

\end{tabular}

\end{center}


\begin{center}
\hfnewcaption{Product branching fractions}{charged B}{\hfbary}{in units of $10^{-5}$}{upper limits are at 90\% CL}{00103.html}{tab:hfc02103}
\hfmetadata{ [ .tml created 2007-04-06T23:00:47.489+08:00] }

\begin{tabular}{lccccc}  
\hflmode
&  \hflpdg2006
&  \hflbelle
&  \hflbabar
&  \hflcdf
&  \hflavg
\\\hline 

 \hfhref{BR_-521_-321+441xBR_441_-3122+3122.html}{ \hflabel{K^- \eta_c(1S) [ \Lambda \bar{\Lambda} ]} } \hftabletlcell 
 &  { \,}  
 &  \hfhref{0607009.html}{ \hfpubhot{0.095\pm^{0.025}_{0.022}\pm^{0.008}_{0.011}} }  
 &  { \,}  
 &  { \,}  
 &  \hfhref{BR_-521_-321+441xBR_441_-3122+3122.html}{ \hfnewavg{0.10\pm0.03} } 
\\%

 \hfhref{BR_-521_-321+441xBR_441_-2212+2212.html}{ \hflabel{K^- \eta_c(1S) [ p \bar{p} ]} }  
 &  \hfhref{BR_-521_-321+441xBR_441_-2212+2212.html}{ \hfpdg{0.12\pm0.04} }  
 &  \hfhref{0607009.html}{ \hfpubhot{0.14\pm0.01\pm^{0.02}_{0.02}} }  
 &  \hfhref{0507001.html}{ \hfpub{0.18\pm^{0.03}_{0.02}\pm0.02} }  
 &  { \,}  
 &  \hfhref{BR_-521_-321+441xBR_441_-2212+2212.html}{ \hfnewavg{0.15\pm0.02} } 
\\%

 \hfhref{BR_-521_-321+443xBR_443_-3122+3122.html}{ \hflabel{K^- J/\psi(1S) [ \Lambda \bar{\Lambda} ]} }  
 &  { \,}  
 &  \hfhref{0607009.html}{ \hfpubhot{0.20\pm^{0.03}_{0.03}\pm0.03} }  
 &  { \,}  
 &  { \,}  
 &  \hfhref{BR_-521_-321+443xBR_443_-3122+3122.html}{ \hfnewavg{0.20\pm0.05} } 
\\%

 \hfhref{BR_-521_-321+443xBR_443_-2212+2212.html}{ \hflabel{K^- J/\psi(1S) [ p \bar{p} ]} }  
 &  \hfhref{BR_-521_-321+443xBR_443_-2212+2212.html}{ \hfpdg{0.22\pm0.01} }  
 &  \hfhref{0607009.html}{ \hfpubhot{0.22\pm0.01\pm0.01} }  
 &  \hfhref{0507001.html}{ \hfpub{0.22\pm0.02\pm0.01} }  
 &  { \,}  
 &  \hfhref{BR_-521_-321+443xBR_443_-2212+2212.html}{ \hfnewavg{0.22\pm0.01} } 
\\%

 \hfhref{BR_-521_-4122+4132xBR_4132_211+3312.html}{ \hflabel{\Lambda_c^- \Xi_c^0 [ \Xi^- \pi^+ ]} } \hftableblcell 
 &  { \,}  
 &  \hfhref{0607006.html}{ \hfpre{4.80\pm^{1.00}_{0.90}\pm1.10\pm1.20} }  
 &  { \,}  
 &  { \,}  
 &  \hfhref{BR_-521_-4122+4132xBR_4132_211+3312.html}{ \hfnewavg{4.8\pm1.9} } 
\\\hline 

\end{tabular}

\end{center}


\begin{center}
\hfnewcaption{Ratios of branching fractions}{charged B}{\hfbary}{in units of $10^{0}$}{upper limits are at 90\% CL}{00103.html}{tab:hfc03103}
\hfmetadata{ [ .tml created 2007-04-06T23:00:57.754+08:00] }

\begin{tabular}{lccccc}  
\hflmode
&  \hflpdg2006
&  \hflbelle
&  \hflbabar
&  \hflcdf
&  \hflavg
\\\hline 

 \hfhref{BR_-521_-2212+-211+4122xoBR_-511_-2212+4122.html}{ \hflabel{\frac{{\cal{B}} ( B^- \to \Lambda_c^+ \bar{p} \pi^- )}{{\cal{B}} ( \bar{B}^0 \to \Lambda_c^+ \bar{p} )}} } \hftabletlcell\hftableblcell 
 &  { \,}  
 &  { \,}  
 &  \hfhref{0604003.html}{ \hfprehot{16.4\pm2.9\pm1.3} }  
 &  { \,}  
 &  \hfhref{BR_-521_-2212+-211+4122xoBR_-511_-2212+4122.html}{ \hfavg{16.4\pm3.2} } 
\\\hline 

\end{tabular}

\end{center}

\end{\hftabletype}
\clearpage  
\begin{\hftabletype}[\hftableposn]

\begin{center}
\hfnewcaption{Branching fractions}{charged B}{\hfjpsi}{in units of $10^{-4}$}{upper limits are at 90\% CL}{00104.html}{tab:hfc01104}
\hfmetadata{ [ .tml created 2007-04-06T23:15:15.107+08:00] }

\begin{tabular}{lccccc}  
\hflmode
&  \hflpdg2006
&  \hflbelle
&  \hflbabar
&  \hflcdf
&  \hflavg
\\\hline 

 \hfhref{BR_-521_-211+421+443.html}{ \hflabel{J/\psi(1S) D^0 \pi^-} } \hftabletlcell 
 &  \hfhref{BR_-521_-211+421+443.html}{ \hfpdg{<0.25} }  
 &  \hfhref{0506038.html}{ \hfpub{<0.25} }  
 &  \hfhref{0506005.html}{ \hfpub{<0.52} }  
 &  { \,}  
 &  \hfhref{BR_-521_-211+421+443.html}{ \hfavg{<0.25} } 
\\%

 \hfhref{BR_-521_-321+333+443.html}{ \hflabel{J/\psi(1S) \phi(1020) K^-} }  
 &  \hfhref{BR_-521_-321+333+443.html}{ \hfpdg{0.52\pm0.17} }  
 &  { \,}  
 &  \hfhref{0506012.html}{ \hfpubold{0.44\pm0.14\pm0.05\pm0.01} }  
 &  { \,}  
 &  \hfhref{BR_-521_-321+333+443.html}{ \hfavg{0.44\pm0.15} } 
\\%

 \hfhref{BR_-521_-211+443.html}{ \hflabel{J/\psi(1S) \pi^-} }  
 &  \hfhref{BR_-521_-211+443.html}{ \hfpdg{0.49\pm0.06} }  
 &  \hfhref{0506076.html}{ \hfpubold{0.38\pm0.06\pm0.03} }  
 &  \hfhref{0506004.html}{ \hfpub{0.54\pm0.04\pm0.02} }  
 &  { \,}  
 &  \hfhref{BR_-521_-211+443.html}{ \hfavg{0.48\pm0.04} } 
\\%

 \hfhref{BR_-521_-321+221+443.html}{ \hflabel{J/\psi(1S) \eta K^-} }  
 &  \hfhref{BR_-521_-321+221+443.html}{ \hfpdg{1.08\pm0.33} }  
 &  { \,}  
 &  \hfhref{0505002.html}{ \hfpub{1.08\pm0.23\pm0.24\pm0.03} }  
 &  { \,}  
 &  \hfhref{BR_-521_-321+221+443.html}{ \hfavg{1.08\pm0.33} } 
\\%

 \hfhref{BR_-521_-411+443.html}{ \hflabel{J/\psi(1S) D^-} }  
 &  \hfhref{BR_-521_-411+443.html}{ \hfpdg{<1.20} }  
 &  { \,}  
 &  \hfhref{0506003.html}{ \hfpub{<1.20} }  
 &  { \,}  
 &  \hfhref{BR_-521_-411+443.html}{ \hfavg{<1.20} } 
\\%

 \hfhref{BR_-521_-321+443.html}{ \hflabel{J/\psi(1S) K^-} }  
 &  \hfhref{BR_-521_-321+443.html}{ \hfpdg{10.08\pm0.35} }  
 &  { \,}  
 &  { \,}  
 &  { \,}  
 &  \hfhref{BR_-521_-321+443.html}{ \hfavg{10.26\pm0.37} } 
\\%

{ \hflabel{\,} }  
 &  { \,}  
 &  \hfhref{0506076.html}{ \hfpubold{10.10\pm0.20\pm0.70\pm0.20} }  
 &  \hfhref{0506109.html}{ \hfpub{10.61\pm0.15\pm0.44\pm0.18} } 
\hffootnotemark{1}
 
 &  { \,}  
 &  { \,} 
\\
\hline 

\end{tabular}

\begin{description}\addtolength{\itemsep}{\hffootitemsep\baselineskip}

	  \item[ \hffootnotemark{1} ] \hffootnotetext{ MEASUREMENT OF BRANCHING FRACTIONS AND CHARGE ASYMMETRIES FOR EXCLUSIVE B DECAYS TO CHARMONIUM (\hfbb{124}) } 
     ;  \hffootnotetext{ $B^- \rightarrow J/\psi K^-$ with $J/\psi$ to leptons } 
	  \item[ \hffootnotemark{2} ] \hffootnotetext{ MEASUREMENT OF THE $B^+ \rightarrow p \overline{p} K^+$ BRANCHING FRACTION AND STUDY OF THE DECAY DYNAMICS (\hfbb{232}) } 
     ;  \hffootnotetext{ $B^- \rightarrow J/\psi K^-$ with $J/\psi \rightarrow p\overline{p}$ } 
	  \item[ \hffootnotemark{3} ] \hffootnotetext{ Measurements of the absolute branching fractions of $B^\pm \rightarrow K^\pm X_{c\overline{c}}$ (\hfbb{231.8}) } 
     ;  \hffootnotetext{ $B^- \rightarrow J/\psi K^-$ (inclusive) } 
\end{description}

\end{center}


\begin{center}
\hfnewcaption{Product branching fractions}{charged B}{\hfjpsi}{in units of $10^{-4}$}{upper limits are at 90\% CL}{00104.html}{tab:hfc02104}
\hfmetadata{ [ .tml created 2007-04-06T23:15:31.21+08:00] }

\begin{tabular}{lccccc}  
\hflmode
&  \hflpdg2006
&  \hflbelle
&  \hflbabar
&  \hflcdf
&  \hflavg
\\\hline 

 \hfhref{BR_-521_-321+10443xBR_10443_-211+211+443.html}{ \hflabel{K^- h_c(1P) [ J/\psi(1S) \pi^+ \pi^- ]} } \hftabletlcell\hftableblcell 
 &  \hfhref{BR_-521_-321+10443xBR_10443_-211+211+443.html}{ \hfpdg{<0.034} }  
 &  { \,}  
 &  \hfhref{0506005.html}{ \hfpub{<0.034} }  
 &  { \,}  
 &  \hfhref{BR_-521_-321+10443xBR_10443_-211+211+443.html}{ \hfavg{<0.034} } 
\\\hline 

\end{tabular}

\end{center}

\end{\hftabletype}
\clearpage  
\begin{\hftabletype}[\hftableposn]

\begin{center}
\hfnewcaption{Ratios of branching fractions}{charged B}{\hfjpsi}{in units of $10^{0}$}{upper limits are at 90\% CL}{00104.html}{tab:hfc03104}
\hfmetadata{ [ .tml created 2007-04-06T23:15:51.023+08:00] }

\begin{tabular}{lccccc}  
\hflmode
&  \hflpdg2006
&  \hflbelle
&  \hflbabar
&  \hflcdf
&  \hflavg
\\\hline 

 \hfhref{BR_-521_-211+443xoBR_-521_-321+443.html}{ \hflabel{\frac{{\cal{B}} ( B^- \to J/\psi(1S) \pi^- )}{{\cal{B}} ( B^- \to J/\psi(1S) K^- )}} } \hftabletlcell 
 &  \hfhref{BR_-521_-211+443xoBR_-521_-321+443.html}{ \hfpdg{0.049\pm0.006} }  
 &  { \,}  
 &  \hfhref{0506004.html}{ \hfpub{0.054\pm0.004\pm0.001} }  
 &  \hfhref{0506095.html}{ \hfpubold{0.0500\pm^{0.0190}_{0.0170}\pm0.0010} }  
 &  \hfhref{BR_-521_-211+443xoBR_-521_-321+443.html}{ \hfavg{0.053\pm0.004} } 
\\%

 \hfhref{BR_-521_-20323+443xoBR_-521_-10323+443.html}{ \hflabel{\frac{{\cal{B}} ( B^- \to J/\psi(1S) K_1^-(1400) )}{{\cal{B}} ( B^- \to J/\psi(1S) K_1^-(1270) )}} }  
 &  \hfhref{BR_-521_-20323+443xoBR_-521_-10323+443.html}{ \hfpdg{<0.30} }  
 &  \hfhref{0506072.html}{ \hfpubold{<0.30} }  
 &  { \,}  
 &  { \,}  
 &  \hfhref{BR_-521_-20323+443xoBR_-521_-10323+443.html}{ \hfavg{<0.30} } 
\\%

 \hfhref{BR_-521_-321+10441xoBR_-521_-321+443.html}{ \hflabel{\frac{{\cal{B}} ( B^- \to \chi_{c0}(1P) K^- )}{{\cal{B}} ( B^- \to J/\psi(1S) K^- )}} }  
 &  \hfhref{BR_-521_-321+10441xoBR_-521_-321+443.html}{ \hfpdg{0.60\pm0.20} }  
 &  \hfhref{0506080.html}{ \hfpubold{0.60\pm^{0.21}_{0.18}\pm0.05\pm0.08} }  
 &  { \,}  
 &  { \,}  
 &  \hfhref{BR_-521_-321+10441xoBR_-521_-321+443.html}{ \hfavg{0.60\pm^{0.23}_{0.20}} } 
\\%

 \hfhref{BR_-521_-321+441xoBR_-521_-321+443.html}{ \hflabel{\frac{{\cal{B}} ( B^- \to \eta_c(1S) K^- )}{{\cal{B}} ( B^- \to J/\psi(1S) K^- )}} }  
 &  \hfhref{BR_-521_-321+441xoBR_-521_-321+443.html}{ \hfpdg{1.33\pm0.44} }  
 &  { \,}  
 &  { \,}  
 &  { \,}  
 &  \hfhref{BR_-521_-321+441xoBR_-521_-321+443.html}{ \hfavg{1.12\pm0.20} } 
\\%

{ \hflabel{\,} }  
 &  { \,}  
 &  { \,}  
 &  \hfhref{0506001.html}{ \hfpub{1.28\pm0.10\pm0.38} } 
\hffootnotemark{1}
 
 &  { \,}  
 &  { \,} 
\\%

{ \hflabel{\,} }  
 &  { \,}  
 &  { \,}  
 &  \hfhref{0505003.html}{ \hfpubhot{1.06\pm0.23\pm0.04} } 
\hffootnotemark{2}
 
 &  { \,}  
 &  { \,} 
\\%

 \hfhref{BR_-521_-323+443xoBR_-521_-321+443.html}{ \hflabel{\frac{{\cal{B}} ( B^- \to J/\psi(1S) K^{*-}(892) )}{{\cal{B}} ( B^- \to J/\psi(1S) K^- )}} }  
 &  \hfhref{BR_-521_-323+443xoBR_-521_-321+443.html}{ \hfpdg{1.39\pm0.09} }  
 &  { \,}  
 &  \hfhref{0506109.html}{ \hfpub{1.37\pm0.05\pm0.08} }  
 &  \hfhref{0506094.html}{ \hfpubold{1.92\pm0.60\pm0.17} }  
 &  \hfhref{BR_-521_-323+443xoBR_-521_-321+443.html}{ \hfavg{1.38\pm0.09} } 
\\%

 \hfhref{BR_-521_-10323+443xoBR_-521_-321+443.html}{ \hflabel{\frac{{\cal{B}} ( B^- \to J/\psi(1S) K_1^-(1270) )}{{\cal{B}} ( B^- \to J/\psi(1S) K^- )}} } \hftableblcell 
 &  { \,}  
 &  \hfhref{0506072.html}{ \hfpubold{1.80\pm0.34\pm0.34} }  
 &  { \,}  
 &  { \,}  
 &  \hfhref{BR_-521_-10323+443xoBR_-521_-321+443.html}{ \hfavg{1.80\pm0.48} } 
\\\hline 

\end{tabular}

\begin{description}\addtolength{\itemsep}{\hffootitemsep\baselineskip}

	  \item[ \hffootnotemark{1} ] \hffootnotetext{ Branching Fraction Measurements of $B \rightarrow \eta_c K$ Decays (\hfbb{86.1}) } 
     ;  \hffootnotetext{ Ratio $B^- \rightarrow \eta_{c} K^-$ to $B^- \rightarrow J/\psi K^-$ with $\eta_c \rightarrow K\overline{K}\pi$ } 
	  \item[ \hffootnotemark{2} ] \hffootnotetext{ Measurements of the absolute branching fractions of $B^\pm \rightarrow K^\pm X_{c\overline{c}}$ (\hfbb{231.8}) } 
     ;  \hffootnotetext{ Ratio $B^- \rightarrow \eta_c K^-$ to $B^- \rightarrow J/\psi K^-$  (inclusive analysis) } 
\end{description}

\end{center}

\end{\hftabletype}
\clearpage  
\begin{\hftabletype}[\hftableposn]

\begin{center}
\hfnewcaption{Branching fractions}{charged B}{\hfochm}{in units of $10^{-4}$}{upper limits are at 90\% CL}{00105.html}{tab:hfc01105}
\hfmetadata{ [ .tml created 2007-04-06T23:27:47.326+08:00] }

\begin{tabular}{lccccc}  
\hflmode
&  \hflpdg2006
&  \hflbelle
&  \hflbabar
&  \hflcdf
&  \hflavg
\\\hline 

 \hfhref{BR_-521_-321+10443.html}{ \hflabel{h_c(1P) K^-} } \hftabletlcell 
 &  { \,}  
 &  \hfhref{0607008.html}{ \hfpubhot{<0.038} }  
 &  { \,}  
 &  { \,}  
 &  \hfhref{BR_-521_-321+10443.html}{ \hfnewavg{<0.038} } 
\\%

 \hfhref{BR_-521_-323+445.html}{ \hflabel{\chi_{c2}(1P) K^{*-}(892)} }  
 &  \hfhref{BR_-521_-323+445.html}{ \hfpdg{<0.120} }  
 &  { \,}  
 &  \hfhref{0506006.html}{ \hfpub{<0.120} }  
 &  { \,}  
 &  \hfhref{BR_-521_-323+445.html}{ \hfavg{<0.120} } 
\\%

 \hfhref{BR_-521_-211+20443.html}{ \hflabel{\chi_{c1}(1P) \pi^-} }  
 &  { \,}  
 &  \hfhref{0607011.html}{ \hfpubhot{0.22\pm0.04\pm0.03} }  
 &  { \,}  
 &  { \,}  
 &  \hfhref{BR_-521_-211+20443.html}{ \hfnewavg{0.22\pm0.05} } 
\\%

 \hfhref{BR_-521_-321+445.html}{ \hflabel{\chi_{c2}(1P) K^-} }  
 &  \hfhref{BR_-521_-321+445.html}{ \hfpdg{<0.29} }  
 &  { \,}  
 &  \hfhref{0506006.html}{ \hfpub{<0.30} }  
 &  { \,}  
 &  \hfhref{BR_-521_-321+445.html}{ \hfavg{<0.30} } 
\\%

 \hfhref{BR_-521_-211+10441.html}{ \hflabel{\chi_{c0}(1P) \pi^-} }  
 &  { \,}  
 &  { \,}  
 &  \hfhref{0506014.html}{ \hfpub{<0.61} }  
 &  { \,}  
 &  \hfhref{BR_-521_-211+10441.html}{ \hfavg{<0.61} } 
\\%

 \hfhref{BR_-521_-321+10441.html}{ \hflabel{\chi_{c0}(1P) K^-} }  
 &  \hfhref{BR_-521_-321+10441.html}{ \hfpdg{1.60\pm0.50} }  
 &  { \,}  
 &  { \,}  
 &  { \,}  
 &  \hfhref{BR_-521_-321+10441.html}{ \hfnewavg{1.88\pm0.30} } 
\\%

{ \hflabel{\,} }  
 &  { \,}  
 &  \hfhref{0506080.html}{ \hfpubold{6.00\pm^{2.10}_{1.80}\pm0.70\pm0.90} }  
 &  \hfhref{0506008.html}{ \hfpub{2.70\pm0.70} } 
\hffootnotemark{2}
 
 &  { \,}  
 &  { \,} 
\\%

{ \hflabel{\,} }  
 &  { \,}  
 &  { \,}  
 &  \hfhref{0603006.html}{ \hfpubhot{1.84\pm0.32\pm0.14\pm0.28} } 
\hffootnotemark{1}
 
 &  { \,}  
 &  { \,} 
\\%

{ \hflabel{\,} }  
 &  { \,}  
 &  { \,}  
 &  \hfhref{0506013.html}{ \hfpub{1.34\pm0.45\pm0.15\pm0.14} } 
\hffootnotemark{3}
 
 &  { \,}  
 &  { \,} 
\\%

{ \hflabel{\,} }  
 &  { \,}  
 &  { \,}  
 &  \hfhref{0505003.html}{ \hfpubhot{<1.80} } 
\hffootnotemark{6b}
 
 &  { \,}  
 &  { \,} 
\\%

 \hfhref{BR_-521_-321+100441.html}{ \hflabel{\eta_c(2S) K^-} }  
 &  \hfhref{BR_-521_-321+100441.html}{ \hfpdg{3.4\pm1.8} }  
 &  { \,}  
 &  \hfhref{0505003.html}{ \hfpubhot{3.40\pm1.80\pm0.30} }  
 &  { \,}  
 &  \hfhref{BR_-521_-321+100441.html}{ \hfavg{3.4\pm1.8} } 
\\%

 \hfhref{BR_-521_-323+20443.html}{ \hflabel{\chi_{c1}(1P) K^{*-}(892)} }  
 &  \hfhref{BR_-521_-323+20443.html}{ \hfpdg{3.60\pm0.90} }  
 &  \hfhref{0607004.html}{ \hfpubhot{4.10\pm0.60\pm0.90} }  
 &  \hfhref{0506109.html}{ \hfpub{2.94\pm0.95\pm0.93\pm0.31} }  
 &  { \,}  
 &  \hfhref{BR_-521_-323+20443.html}{ \hfnewavg{3.65\pm0.85} } 
\\%

 \hfhref{BR_-521_-321+30443.html}{ \hflabel{\psi(3770) K^-} }  
 &  \hfhref{BR_-521_-321+30443.html}{ \hfpdg{4.9\pm1.3} }  
 &  \hfhref{0506075.html}{ \hfpub{4.80\pm1.10\pm0.70} }  
 &  \hfhref{0505003.html}{ \hfpubhot{3.50\pm2.50\pm0.30} }  
 &  { \,}  
 &  \hfhref{BR_-521_-321+30443.html}{ \hfavg{4.5\pm1.2} } 
\\%

 \hfhref{BR_-521_-321+20443.html}{ \hflabel{\chi_{c1}(1P) K^-} }  
 &  \hfhref{BR_-521_-321+20443.html}{ \hfpdg{5.30\pm0.70} }  
 &  { \,}  
 &  { \,}  
 &  { \,}  
 &  \hfhref{BR_-521_-321+20443.html}{ \hfnewavg{5.01\pm0.37} } 
\\%

{ \hflabel{\,} }  
 &  { \,}  
 &  \hfhref{0607004.html}{ \hfpubhot{4.50\pm0.20\pm0.70} }  
 &  \hfhref{0505003.html}{ \hfpubhot{8.00\pm1.40\pm0.70} } 
\hffootnotemark{6c}
 
 &  \hfhref{0506091.html}{ \hfpubold{15.5\pm5.4\pm2.0} }  
 &  { \,} 
\\%

{ \hflabel{\,} }  
 &  { \,}  
 &  { \,}  
 &  \hfhref{0607015.html}{ \hfpubhot{4.90\pm0.20\pm0.40} } 
\hffootnotemark{4}
 
 &  { \,}  
 &  { \,} 
\\%

 \hfhref{BR_-521_-321+100443.html}{ \hflabel{\psi(2S) K^-} }  
 &  \hfhref{BR_-521_-321+100443.html}{ \hfpdg{6.48\pm0.35} }  
 &  { \,}  
 &  { \,}  
 &  { \,}  
 &  \hfhref{BR_-521_-321+100443.html}{ \hfavg{6.32\pm0.37} } 
\\%

{ \hflabel{\,} }  
 &  { \,}  
 &  \hfhref{0506076.html}{ \hfpubold{6.90\pm0.60} }  
 &  \hfhref{0506109.html}{ \hfpub{6.17\pm0.32\pm0.38\pm0.23} } 
\hffootnotemark{5}
 
 &  \hfhref{0506092.html}{ \hfpubold{5.50\pm1.00\pm0.60} }  
 &  { \,} 
\\%

{ \hflabel{\,} }  
 &  { \,}  
 &  { \,}  
 &  \hfhref{0505003.html}{ \hfpubhot{4.90\pm1.60\pm0.40} } 
\hffootnotemark{6a}
 
 &  { \,}  
 &  { \,} 
\\%

 \hfhref{BR_-521_-323+100443.html}{ \hflabel{\psi(2S) K^{*-}(892)} }  
 &  \hfhref{BR_-521_-323+100443.html}{ \hfpdg{6.7\pm1.4} }  
 &  \hfhref{0506077.html}{ \hfpreold{8.13\pm0.77\pm0.89} }  
 &  \hfhref{0506109.html}{ \hfpub{5.92\pm0.85\pm0.86\pm0.22} }  
 &  { \,}  
 &  \hfhref{BR_-521_-323+100443.html}{ \hfavg{7.07\pm0.85} } 
\\%

 \hfhref{BR_-521_-321+441.html}{ \hflabel{\eta_c(1S) K^-} }  
 &  \hfhref{BR_-521_-321+441.html}{ \hfpdg{9.1\pm1.3} }  
 &  { \,}  
 &  { \,}  
 &  { \,}  
 &  \hfhref{BR_-521_-321+441.html}{ \hfavg{9.8\pm1.3} } 
\\%

{ \hflabel{\,} }  
 &  { \,}  
 &  \hfhref{0506074.html}{ \hfpubold{12.50\pm1.40\pm^{1.00}_{1.20}\pm3.80} }  
 &  \hfhref{0506001.html}{ \hfpub{12.90\pm0.90\pm1.30\pm3.60} } 
\hffootnotemark{7}
 
 &  { \,}  
 &  { \,} 
\\%

{ \hflabel{\,} }  
 &  { \,}  
 &  { \,}  
 &  \hfhref{0507001.html}{ \hfpub{13.8\pm^{2.3}_{1.5}\pm1.5\pm4.2} } 
\hffootnotemark{8}
 
 &  { \,}  
 &  { \,} 
\\%

{ \hflabel{\,} }  
 &  { \,}  
 &  { \,}  
 &  \hfhref{0505003.html}{ \hfpubhot{8.7\pm1.5} } 
\hffootnotemark{6d}
 
 &  { \,}  
 &  { \,} 
\\%

 \hfhref{BR_-521_-323+10441.html}{ \hflabel{\chi_{c0}(1P) K^{*-}(892)} } \hftableblcell 
 &  \hfhref{BR_-521_-323+10441.html}{ \hfpdg{<29} }  
 &  { \,}  
 &  \hfhref{0506006.html}{ \hfpub{<29} }  
 &  { \,}  
 &  \hfhref{BR_-521_-323+10441.html}{ \hfavg{<29} } 
\\\hline 

\end{tabular}

\begin{description}\addtolength{\itemsep}{\hffootitemsep\baselineskip}

	  \item[ \hffootnotemark{1} ] \hffootnotetext{ Dalitz plot analysis of the decay $B^\pm\rightarrow K^\pm K^\pm K^\mp$ (\hfbb{226}) } 
     ;  \hffootnotetext{ $B^\pm\rightarrow K^\pm \chi_{c0}$, with $chi_c0\rightarrow K^+K^-$ (Dalitz analysis) } 
	  \item[ \hffootnotemark{2} ] \hffootnotetext{ MEASUREMENT OF THE BRANCHING FRACTION FOR $B^\pm \rightarrow \chi_{c0} K^\pm$. (\hfbb{88.9}) } 
     ;  \hffootnotetext{ $B^- \rightarrow \chi_{c0} K^-$ with $\chi_{c0}\rightarrow K^+ K^-, \pi^+ \pi^-$ } 
	  \item[ \hffootnotemark{3} ] \hffootnotetext{ Dalitz-plot analysis of the decays $B^\pm \rightarrow K^\pm \pi^\mp \pi^\pm$ (\hfbb{226}) } 
     ;  \hffootnotetext{ $B^- \rightarrow \chi_{c0} K^-$ with $\chi_{c0} \rightarrow \pi^+ \pi^-$ (Dalitz analysis) } 
	  \item[ \hffootnotemark{4} ] \hffootnotetext{ Search for $B\rightarrow X(3872) K, X(3872)\rightarrow J/\psi\gamma$ (\hfbb{287}) } 
     ;  \hffootnotetext{ $B^- \rightarrow \chi_{c1} K^-$ with $\chi_{c1}$ to $J/\psi \gamma$ } 
	  \item[ \hffootnotemark{5} ] \hffootnotetext{ MEASUREMENT OF BRANCHING FRACTIONS AND CHARGE ASYMMETRIES FOR EXCLUSIVE B DECAYS TO CHARMONIUM (\hfbb{124}) } 
     ;  \hffootnotetext{ $B^- \rightarrow \psi(2S) K^-$ with $\psi(2S)$ to leptons } 
	  \item[ \hffootnotemark{6} ] \hffootnotetext{ Measurements of the absolute branching fractions of $B^\pm \rightarrow K^\pm X_{c\overline{c}}$ (\hfbb{231.8}) } 
     ;  \hffootnotemark{6a}  \hffootnotetext{ $B^- \rightarrow \psi(2S) K^-$ (inclusive) }  ;  \hffootnotemark{6b}  \hffootnotetext{ $B^- \rightarrow \chi_{c0} K^-$ (inclusive) }  ;  \hffootnotemark{6c}  \hffootnotetext{ $B^- \rightarrow \chi_{c1} K^-$ (inclusive) }  ;  \hffootnotemark{6d}  \hffootnotetext{ $B^- \rightarrow \eta_c K^-$ (inclusive) } 
	  \item[ \hffootnotemark{7} ] \hffootnotetext{ Branching Fraction Measurements of $B \rightarrow \eta_c K$ Decays (\hfbb{86.1}) } 
     ;  \hffootnotetext{ $B^- \rightarrow \eta_{c} K^-$ with $\eta_c \rightarrow K\overline{K}\pi$ } 
	  \item[ \hffootnotemark{8} ] \hffootnotetext{ MEASUREMENT OF THE $B^+ \rightarrow p \overline{p} K^+$ BRANCHING FRACTION AND STUDY OF THE DECAY DYNAMICS (\hfbb{232}) } 
     ;  \hffootnotetext{ $B^- \rightarrow \eta_c K^-$ with $\eta_c \rightarrow p\overline{p}$ } 
\end{description}

\end{center}

\end{\hftabletype}
\clearpage  
\begin{\hftabletype}[\hftableposn]

\begin{center}
\hfnewcaption{Ratios of branching fractions}{charged B}{\hfochm}{in units of $10^{-1}$}{upper limits are at 90\% CL}{00105.html}{tab:hfc03105}
\hfmetadata{ [ .tml created 2007-04-06T23:27:58.376+08:00] }

\begin{tabular}{lccccc}  
\hflmode
&  \hflpdg2006
&  \hflbelle
&  \hflbabar
&  \hflcdf
&  \hflavg
\\\hline 

 \hfhref{BR_-521_-211+20443xoBR_-521_-321+20443.html}{ \hflabel{\frac{{\cal{B}} ( B^- \to \chi_{c1}(1P) \pi^- )}{{\cal{B}} ( B^- \to \chi_{c1}(1P) K^- )}} } \hftabletlcell 
 &  { \,}  
 &  \hfhref{0607011.html}{ \hfpubhot{0.43\pm0.08\pm0.03} }  
 &  { \,}  
 &  { \,}  
 &  \hfhref{BR_-521_-211+20443xoBR_-521_-321+20443.html}{ \hfnewavg{0.43\pm0.09} } 
\\%

 \hfhref{BR_-521_-323+20443xoBR_-521_-321+20443.html}{ \hflabel{\frac{{\cal{B}} ( B^- \to \chi_{c1}(1P) K^{*-}(892) )}{{\cal{B}} ( B^- \to \chi_{c1}(1P) K^- )}} }  
 &  \hfhref{BR_-521_-323+20443xoBR_-521_-321+20443.html}{ \hfpdg{5.1\pm2.3} }  
 &  { \,}  
 &  \hfhref{0506109.html}{ \hfpub{5.1\pm1.7\pm1.6} }  
 &  { \,}  
 &  \hfhref{BR_-521_-323+20443xoBR_-521_-321+20443.html}{ \hfavg{5.1\pm2.3} } 
\\%

 \hfhref{BR_-521_-323+100443xoBR_-521_-321+100443.html}{ \hflabel{\frac{{\cal{B}} ( B^- \to \psi(2S) K^{*-}(892) )}{{\cal{B}} ( B^- \to \psi(2S) K^- )}} } \hftableblcell 
 &  \hfhref{BR_-521_-323+100443xoBR_-521_-321+100443.html}{ \hfpdg{9.6\pm1.7} }  
 &  { \,}  
 &  \hfhref{0506109.html}{ \hfpub{9.60\pm1.50\pm0.90} }  
 &  { \,}  
 &  \hfhref{BR_-521_-323+100443xoBR_-521_-321+100443.html}{ \hfavg{9.6\pm1.7} } 
\\\hline 

\end{tabular}

\end{center}

\end{\hftabletype}
\clearpage  
\begin{\hftabletype}[\hftableposn]


\begin{center}
\hfnewcaption{Product branching fractions}{charged B}{\hfmuld}{in units of $10^{-4}$}{upper limits are at 90\% CL}{00106.html}{tab:hfc02106}
\hfmetadata{ [ .tml created 2007-04-06T23:39:37.965+08:00] }

\begin{tabular}{lccccc}  
\hflmode
&  \hflpdg2006
&  \hflbelle
&  \hflbabar
&  \hflcdf
&  \hflavg
\\\hline 

 \hfhref{BR_-521_-211+10423xBR_10423_-211+211+423.html}{ \hflabel{\pi^- D_1^0(2420) [ D^{*0}(2007) \pi^- \pi^+ ]} } \hftabletlcell 
 &  \hfhref{BR_-521_-211+10423xBR_10423_-211+211+423.html}{ \hfpdg{<0.060} }  
 &  \hfhref{0506030.html}{ \hfpub{<0.060} }  
 &  { \,}  
 &  { \,}  
 &  \hfhref{BR_-521_-211+10423xBR_10423_-211+211+423.html}{ \hfavg{<0.060} } 
\\%

 \hfhref{BR_-521_-211+425xBR_425_-211+211+423.html}{ \hflabel{\pi^- D_2^{*0}(2460) [ D^{*0}(2007) \pi^- \pi^+ ]} }  
 &  \hfhref{BR_-521_-211+425xBR_425_-211+211+423.html}{ \hfpdg{<0.22} }  
 &  \hfhref{0506030.html}{ \hfpub{<0.22} }  
 &  { \,}  
 &  { \,}  
 &  \hfhref{BR_-521_-211+425xBR_425_-211+211+423.html}{ \hfavg{<0.22} } 
\\%

 \hfhref{BR_-521_-211+425xBR_425_-211+413.html}{ \hflabel{\pi^- D_2^{*0}(2460) [ D^{*+}(2010) \pi^- ]} }  
 &  \hfhref{BR_-521_-211+425xBR_425_-211+413.html}{ \hfpdg{1.80\pm0.50} }  
 &  \hfhref{0506071.html}{ \hfpub{1.80\pm0.30\pm0.30\pm0.20} }  
 &  \hfhref{0506113.html}{ \hfpreold{1.80\pm0.30\pm0.50} }  
 &  { \,}  
 &  \hfhref{BR_-521_-211+425xBR_425_-211+413.html}{ \hfavg{1.80\pm0.36} } 
\\%

 \hfhref{BR_-521_-211+10423xBR_10423_-211+211+421.html}{ \hflabel{\pi^- D_1^0(2420) [ D^0 \pi^- \pi^+ ]} }  
 &  \hfhref{BR_-521_-211+10423xBR_10423_-211+211+421.html}{ \hfpdg{1.90\pm0.60} }  
 &  \hfhref{0506030.html}{ \hfpub{1.85\pm0.29\pm0.35\pm^{0.00}_{0.46}} }  
 &  { \,}  
 &  { \,}  
 &  \hfhref{BR_-521_-211+10423xBR_10423_-211+211+421.html}{ \hfavg{1.85\pm^{0.45}_{0.65}} } 
\\%

 \hfhref{BR_-521_-211+425xBR_425_-211+411.html}{ \hflabel{\pi^- D_2^{*0}(2460) [ D^+ \pi^- ]} }  
 &  \hfhref{BR_-521_-211+425xBR_425_-211+411.html}{ \hfpdg{3.40\pm0.80} }  
 &  \hfhref{0506071.html}{ \hfpub{3.40\pm0.30\pm0.60\pm0.40} }  
 &  \hfhref{0506113.html}{ \hfpreold{2.90\pm0.20\pm0.50} }  
 &  { \,}  
 &  \hfhref{BR_-521_-211+425xBR_425_-211+411.html}{ \hfavg{3.06\pm0.44} } 
\\%

 \hfhref{BR_-521_-211+20423xBR_20423_-211+413.html}{ \hflabel{\pi^- D_1^0(H) [ D^{*+}(2010) \pi^- ]} }  
 &  { \,}  
 &  \hfhref{0506071.html}{ \hfpub{5.00\pm0.40\pm1.00\pm0.40} }  
 &  { \,}  
 &  { \,}  
 &  \hfhref{BR_-521_-211+20423xBR_20423_-211+413.html}{ \hfavg{5.0\pm1.1} } 
\\%

 \hfhref{BR_-521_-211+10421xBR_10421_-211+411.html}{ \hflabel{\pi^- D_0^{*0} [ D^+ \pi^- ]} }  
 &  { \,}  
 &  \hfhref{0506071.html}{ \hfpub{6.10\pm0.60\pm0.90\pm1.60} }  
 &  { \,}  
 &  { \,}  
 &  \hfhref{BR_-521_-211+10421xBR_10421_-211+411.html}{ \hfavg{6.1\pm1.9} } 
\\%

 \hfhref{BR_-521_-211+10423xBR_10423_-211+413.html}{ \hflabel{\pi^- D_1^0(2420) [ D^{*+}(2010) \pi^- ]} } \hftableblcell 
 &  \hfhref{BR_-521_-211+10423xBR_10423_-211+413.html}{ \hfpdg{6.8\pm1.5} }  
 &  \hfhref{0506071.html}{ \hfpub{6.80\pm0.70\pm1.30\pm0.30} }  
 &  \hfhref{0506113.html}{ \hfpreold{5.90\pm0.30\pm1.10} }  
 &  { \,}  
 &  \hfhref{BR_-521_-211+10423xBR_10423_-211+413.html}{ \hfavg{6.23\pm0.91} } 
\\\hline 

\end{tabular}

\end{center}

\end{\hftabletype}
\clearpage  
\begin{\hftabletype}[\hftableposn]

\begin{center}
\hfnewcaption{Ratios of branching fractions}{charged B}{\hfmuld}{in units of $10^{0}$}{upper limits are at 90\% CL}{00106.html}{tab:hfc03106}
\hfmetadata{ [ .tml created 2007-04-06T23:39:46.912+08:00] }

\begin{tabular}{lccccc}  
\hflmode
&  \hflpdg2006
&  \hflbelle
&  \hflbabar
&  \hflcdf
&  \hflavg
\\\hline 

 \hfhref{BR_-521_-321+421xoBR_-521_-211+421.html}{ \hflabel{\frac{{\cal{B}} ( B^- \to D^0 K^- )}{{\cal{B}} ( B^- \to D^0 \pi^- )}} } \hftabletlcell 
 &  \hfhref{BR_-521_-321+421xoBR_-521_-211+421.html}{ \hfpdg{0.083\pm0.010} }  
 &  \hfhref{0506082.html}{ \hfpubold{0.077\pm0.005\pm0.006} }  
 &  \hfhref{0601002.html}{ \hfpub{0.083\pm0.003\pm0.002} }  
 &  \hfhref{0607021.html}{ \hfwaiting{0.065\pm0.007\pm0.004} }  
 &  \hfhref{BR_-521_-321+421xoBR_-521_-211+421.html}{ \hfnewavg{0.079\pm0.003} } 
\\%

 \hfhref{BR_-521_-321+423xoBR_-521_-211+423.html}{ \hflabel{\frac{{\cal{B}} ( B^- \to D^{*0}(2007) K^- )}{{\cal{B}} ( B^- \to D^{*0}(2007) \pi^- )}} }  
 &  \hfhref{BR_-521_-321+423xoBR_-521_-211+423.html}{ \hfpdg{0.08\pm0.02} }  
 &  \hfhref{0506081.html}{ \hfpubold{0.078\pm0.019\pm0.009} }  
 &  \hfhref{0601003.html}{ \hfpub{0.081\pm0.004\pm^{0.004}_{0.003}} }  
 &  { \,}  
 &  \hfhref{BR_-521_-321+423xoBR_-521_-211+423.html}{ \hfavg{0.081\pm0.005} } 
\\%

 \hfhref{BR_-521_-211+425xoBR_-521_-211+10423.html}{ \hflabel{\frac{{\cal{B}} ( B^- \to D_2^{*0}(2460) \pi^- )}{{\cal{B}} ( B^- \to D_1^0(2420) \pi^- )}} }  
 &  { \,}  
 &  { \,}  
 &  \hfhref{0506113.html}{ \hfpreold{0.80\pm0.07\pm0.16} }  
 &  { \,}  
 &  \hfhref{BR_-521_-211+425xoBR_-521_-211+10423.html}{ \hfavg{0.80\pm0.17} } 
\\%

 \hfhref{BR_-521_-211+423xoBR_-521_-211+421.html}{ \hflabel{\frac{{\cal{B}} ( B^- \to D^{*0}(2007) \pi^- )}{{\cal{B}} ( B^- \to D^0 \pi^- )}} }  
 &  { \,}  
 &  { \,}  
 &  \hfhref{0607016.html}{ \hfpubhot{1.14\pm0.07\pm0.04} }  
 &  { \,}  
 &  \hfhref{BR_-521_-211+423xoBR_-521_-211+421.html}{ \hfnewavg{1.14\pm0.08} } 
\\%

 \hfhref{BR_-521_-211+93xoBR_-521_-211+421.html}{ \hflabel{\frac{{\cal{B}} ( B^- \to D^{**+} \pi^- )}{{\cal{B}} ( B^- \to D^0 \pi^- )}} }  
 &  { \,}  
 &  { \,}  
 &  \hfhref{0607016.html}{ \hfpubhot{1.22\pm0.13\pm0.23} }  
 &  { \,}  
 &  \hfhref{BR_-521_-211+93xoBR_-521_-211+421.html}{ \hfnewavg{1.22\pm0.26} } 
\\%

 \hfhref{BR_-521_-211+421xoBR_-511_-211+411.html}{ \hflabel{\frac{{\cal{B}} ( B^- \to D^0 \pi^- )}{{\cal{B}} ( \bar{B}^0 \to D^+ \pi^- )}} } \hftableblcell 
 &  { \,}  
 &  { \,}  
 &  { \,}  
 &  \hfhref{0506099.html}{ \hfpubhot{1.97\pm0.10\pm0.21} }  
 &  \hfhref{BR_-521_-211+421xoBR_-511_-211+411.html}{ \hfavg{1.97\pm0.23} } 
\\\hline 

\end{tabular}

\end{center}

\end{\hftabletype}
\clearpage  
\begin{\hftabletype}[\hftableposn]

\begin{center}
\hfnewcaption{Branching fractions}{charged B}{\hfsgdx}{in units of $10^{-4}$}{upper limits are at 90\% CL}{00107.html}{tab:hfc01107}
\hfmetadata{ [ .tml created 2007-04-06T23:57:49.6+08:00] }

\begin{tabular}{lccccc}  
\hflmode
&  \hflpdg2006
&  \hflbelle
&  \hflbabar
&  \hflcdf
&  \hflavg
\\\hline 

 \hfhref{BR_-521_-413+-311.html}{ \hflabel{D^{*-}(2010) \bar{K}^0} } \hftabletlcell 
 &  \hfhref{BR_-521_-413+-311.html}{ \hfpdg{<0.090} }  
 &  { \,}  
 &  \hfhref{0506016.html}{ \hfpub{<0.090} }  
 &  { \,}  
 &  \hfhref{BR_-521_-413+-311.html}{ \hfavg{<0.090} } 
\\%

 \hfhref{BR_-521_-321+423.html}{ \hflabel{D^{*0}(2007) K^-} }  
 &  \hfhref{BR_-521_-321+423.html}{ \hfpdg{3.70\pm0.40} }  
 &  \hfhref{0506081.html}{ \hfpubold{3.59\pm0.87\pm0.41\pm0.31} }  
 &  { \,}  
 &  { \,}  
 &  \hfhref{BR_-521_-321+423.html}{ \hfavg{3.6\pm1.0} } 
\\%

 \hfhref{BR_-521_-323+423.html}{ \hflabel{D^{*0}(2007) K^{*-}(892)} }  
 &  \hfhref{BR_-521_-323+423.html}{ \hfpdg{8.1\pm1.4} }  
 &  { \,}  
 &  \hfhref{0506105.html}{ \hfpub{8.30\pm1.10\pm0.96\pm0.27} }  
 &  { \,}  
 &  \hfhref{BR_-521_-323+423.html}{ \hfavg{8.3\pm1.5} } 
\\%

 \hfhref{BR_-521_-321+311+423.html}{ \hflabel{D^{*0}(2007) K^- K^0} }  
 &  \hfhref{BR_-521_-321+311+423.html}{ \hfpdg{<10.6} }  
 &  \hfhref{0506090.html}{ \hfpubold{<10.6} }  
 &  { \,}  
 &  { \,}  
 &  \hfhref{BR_-521_-321+311+423.html}{ \hfavg{<10.6} } 
\\%

 \hfhref{BR_-521_-211+-211+413.html}{ \hflabel{D^{*+}(2010) \pi^- \pi^-} }  
 &  \hfhref{BR_-521_-211+-211+413.html}{ \hfpdg{13.5\pm2.2} }  
 &  \hfhref{0506071.html}{ \hfpub{12.50\pm0.80\pm2.20} }  
 &  \hfhref{0506113.html}{ \hfpreold{12.20\pm0.50\pm1.80} }  
 &  { \,}  
 &  \hfhref{BR_-521_-211+-211+413.html}{ \hfavg{12.3\pm1.5} } 
\\%

 \hfhref{BR_-521_-321+313+423.html}{ \hflabel{D^{*0}(2007) K^- K^{*0}(892)} }  
 &  \hfhref{BR_-521_-321+313+423.html}{ \hfpdg{15.0\pm4.0} }  
 &  \hfhref{0506090.html}{ \hfpubold{15.3\pm3.1\pm2.9} }  
 &  { \,}  
 &  { \,}  
 &  \hfhref{BR_-521_-321+313+423.html}{ \hfavg{15.3\pm4.2} } 
\\%

 \hfhref{BR_-521_-211+-211+-211+211+413.html}{ \hflabel{D^{*+}(2010) \pi^- \pi^+ \pi^- \pi^-} }  
 &  \hfhref{BR_-521_-211+-211+-211+211+413.html}{ \hfpdg{26.0\pm4.0} }  
 &  \hfhref{0506028.html}{ \hfpub{25.6\pm2.6\pm3.3} }  
 &  { \,}  
 &  { \,}  
 &  \hfhref{BR_-521_-211+-211+-211+211+413.html}{ \hfavg{25.6\pm4.2} } 
\\%

 \hfhref{BR_-521_-211+423.html}{ \hflabel{D^{*0}(2007) \pi^-} }  
 &  \hfhref{BR_-521_-211+423.html}{ \hfpdg{46.0\pm4.0} }  
 &  { \,}  
 &  { \,}  
 &  { \,}  
 &  \hfhref{BR_-521_-211+423.html}{ \hfnewavg{52.8\pm2.8} } 
\\%

{ \hflabel{\,} }  
 &  { \,}  
 &  { \,}  
 &  \hfhref{0607012.html}{ \hfpubhot{55.20\pm1.70\pm4.20\pm0.20} } 
\hffootnotemark{2}
 
 &  { \,}  
 &  { \,} 
\\%

{ \hflabel{\,} }  
 &  { \,}  
 &  { \,}  
 &  \hfhref{0607016.html}{ \hfpubhot{51.3\pm2.2\pm2.8} } 
\hffootnotemark{1}
 
 &  { \,}  
 &  { \,} 
\\%

 \hfhref{BR_-521_-211+93.html}{ \hflabel{D^{**+} \pi^-} }  
 &  { \,}  
 &  { \,}  
 &  \hfhref{0607016.html}{ \hfpubhot{55.0\pm5.2\pm10.4} }  
 &  { \,}  
 &  \hfhref{BR_-521_-211+93.html}{ \hfnewavg{55\pm12} } 
\\%

 \hfhref{BR_-521_-211+-211+-211+211+211+423.html}{ \hflabel{D^{*0}(2007) \pi^- \pi^+ \pi^- \pi^+ \pi^-} }  
 &  { \,}  
 &  \hfhref{0506028.html}{ \hfpub{56.7\pm9.1\pm8.5} }  
 &  { \,}  
 &  { \,}  
 &  \hfhref{BR_-521_-211+-211+-211+211+211+423.html}{ \hfavg{57\pm12} } 
\\%

 \hfhref{BR_-521_-211+-211+211+423.html}{ \hflabel{D^{*0}(2007) \pi^- \pi^+ \pi^-} } \hftableblcell 
 &  \hfhref{BR_-521_-211+-211+211+423.html}{ \hfpdg{103\pm12} }  
 &  \hfhref{0506028.html}{ \hfpub{105.5\pm4.7\pm12.9} }  
 &  { \,}  
 &  { \,}  
 &  \hfhref{BR_-521_-211+-211+211+423.html}{ \hfavg{106\pm14} } 
\\\hline 

\end{tabular}

\begin{description}\addtolength{\itemsep}{\hffootitemsep\baselineskip}

	  \item[ \hffootnotemark{1} ] \hffootnotetext{ Measurement of the Absolute Branching Fractions $B \rightarrow D^{(*,**)}\pi$ with a Missing Mass method (\hfbb{231}) } 
     ;  \hffootnotetext{ $B^- \rightarrow D^{*0} \pi^-$ } 
	  \item[ \hffootnotemark{2} ] \hffootnotetext{ Branching fraction measurements and isospin analyses for $\bar{B} \rightarrow D^{(*)}\pi^-$
decays (\hfbb{65}) } 
     ;  \hffootnotetext{ $B^- \rightarrow D^{*0}\pi^-$ } 
\end{description}

\end{center}

\end{\hftabletype}
\clearpage  
\begin{\hftabletype}[\hftableposn]

\begin{center}
\hfnewcaption{Branching fractions}{charged B}{\hfsgld}{in units of $10^{-4}$}{upper limits are at 90\% CL}{00108.html}{tab:hfc01108}
\hfmetadata{ [ .tml created 2007-04-07T00:05:06.795+08:00] }

\begin{tabular}{lccccc}  
\hflmode
&  \hflpdg2006
&  \hflbelle
&  \hflbabar
&  \hflcdf
&  \hflavg
\\\hline 

 \hfhref{BR_-521_-411+-311.html}{ \hflabel{D^- \bar{K}^0} } \hftabletlcell 
 &  \hfhref{BR_-521_-411+-311.html}{ \hfpdg{<0.050} }  
 &  { \,}  
 &  \hfhref{0506016.html}{ \hfpub{<0.050} }  
 &  { \,}  
 &  \hfhref{BR_-521_-411+-311.html}{ \hfavg{<0.050} } 
\\%

 \hfhref{BR_-521_-321+421.html}{ \hflabel{D^0 K^-} }  
 &  \hfhref{BR_-521_-321+421.html}{ \hfpdg{4.08\pm0.24} }  
 &  \hfhref{0506082.html}{ \hfpubold{3.83\pm0.25\pm0.30\pm0.22} }  
 &  { \,}  
 &  { \,}  
 &  \hfhref{BR_-521_-321+421.html}{ \hfavg{3.83\pm0.45} } 
\\%

 \hfhref{BR_-521_-323+421.html}{ \hflabel{D^0 K^{*-}(892)} }  
 &  \hfhref{BR_-521_-323+421.html}{ \hfpdg{6.30\pm0.80} }  
 &  { \,}  
 &  { \,}  
 &  { \,}  
 &  \hfhref{BR_-521_-323+421.html}{ \hfavg{5.29\pm0.45} } 
\\%

{ \hflabel{\,} }  
 &  { \,}  
 &  { \,}  
 &  \hfhref{0603001.html}{ \hfpubhot{5.29\pm0.30\pm0.34} } 
\hffootnotemark{1}
 
 &  { \,}  
 &  { \,} 
\\%

{ \hflabel{\,} }  
 &  { \,}  
 &  { \,}  
 &  \hfhref{0506104.html}{ \hfsuperceeded{6.30\pm0.70\pm0.50} } 
\hffootnotemark{2}
 
 &  { \,}  
 &  { \,} 
\\%

 \hfhref{BR_-521_-321+311+421.html}{ \hflabel{D^0 K^- K^0} }  
 &  \hfhref{BR_-521_-321+311+421.html}{ \hfpdg{5.5\pm1.6} }  
 &  \hfhref{0506090.html}{ \hfpubold{5.50\pm1.40\pm0.80} }  
 &  { \,}  
 &  { \,}  
 &  \hfhref{BR_-521_-321+311+421.html}{ \hfavg{5.5\pm1.6} } 
\\%

 \hfhref{BR_-521_-321+313+421.html}{ \hflabel{D^0 K^- K^{*0}(892)} }  
 &  \hfhref{BR_-521_-321+313+421.html}{ \hfpdg{7.5\pm1.7} }  
 &  \hfhref{0506090.html}{ \hfpubold{7.5\pm1.3\pm1.1} }  
 &  { \,}  
 &  { \,}  
 &  \hfhref{BR_-521_-321+313+421.html}{ \hfavg{7.5\pm1.7} } 
\\%

 \hfhref{BR_-521_-211+-211+411.html}{ \hflabel{D^+ \pi^- \pi^-} }  
 &  \hfhref{BR_-521_-211+-211+411.html}{ \hfpdg{10.2\pm1.6} }  
 &  \hfhref{0506071.html}{ \hfpub{10.20\pm0.40\pm1.50} }  
 &  \hfhref{0506113.html}{ \hfpreold{8.70\pm0.40\pm1.30} }  
 &  { \,}  
 &  \hfhref{BR_-521_-211+-211+411.html}{ \hfavg{9.4\pm1.0} } 
\\%

 \hfhref{BR_-521_-211+421.html}{ \hflabel{D^0 \pi^-} }  
 &  \hfhref{BR_-521_-211+421.html}{ \hfpdg{49.2\pm2.0} }  
 &  { \,}  
 &  { \,}  
 &  { \,}  
 &  \hfhref{BR_-521_-211+421.html}{ \hfnewavg{47.5\pm1.9} } 
\\%

{ \hflabel{\,} }  
 &  { \,}  
 &  { \,}  
 &  \hfhref{0607012.html}{ \hfpubhot{49.00\pm0.70\pm2.20\pm0.06} } 
\hffootnotemark{3}
 
 &  { \,}  
 &  { \,} 
\\%

{ \hflabel{\,} } \hftableblcell 
 &  { \,}  
 &  { \,}  
 &  \hfhref{0607016.html}{ \hfpubhot{44.9\pm2.1\pm2.3} } 
\hffootnotemark{4}
 
 &  { \,}  
 &  { \,} 
\\\hline 

\end{tabular}

\begin{description}\addtolength{\itemsep}{\hffootitemsep\baselineskip}

	  \item[ \hffootnotemark{1} ] \hffootnotetext{ Measurement of the $B^- \rightarrow D^0 K^{*-}$ branching fraction (\hfbb{232}) } 
     ;  \hffootnotetext{ Measurement of of the $B^- \rightarrow D^0 K^{*-}$ branching fraction } 
	  \item[ \hffootnotemark{2} ] \hffootnotetext{ Measurement of the Branching Fraction for $B^- \rightarrow D^0K^{*-}$ (\hfbb{86}) } 
     ;  \hffootnotetext{ $B^- \rightarrow D^0K^{*-}$ } 
	  \item[ \hffootnotemark{3} ] \hffootnotetext{ Branching fraction measurements and isospin analyses for $\bar{B} \rightarrow D^{(*)}\pi^-$
decays (\hfbb{65}) } 
     ;  \hffootnotetext{ $B^- \rightarrow D^0\pi^-$ } 
	  \item[ \hffootnotemark{4} ] \hffootnotetext{ Measurement of the Absolute Branching Fractions $B \rightarrow D^{(*,**)}\pi$ with a Missing Mass method (\hfbb{231}) } 
     ;  \hffootnotetext{ $B^- \rightarrow D^0\pi^-$ } 
\end{description}

\end{center}

\end{\hftabletype}
\clearpage  
\begin{\hftabletype}[\hftableposn]

\begin{center}
\hfnewcaption{Branching fractions}{neutral B}{\hfnewp}{in units of $10^{-3}$}{upper limits are at 90\% CL}{00201.html}{tab:hfc01201}
\hfmetadata{ [ .tml created 2007-04-07T00:10:05.542+08:00] }

\begin{tabular}{lccccc}  
\hflmode
&  \hflpdg2006
&  \hflbelle
&  \hflbabar
&  \hflcdf
&  \hflavg
\\\hline 

 \hfhref{BR_-511_-321+82.html}{ \hflabel{X^{+}(3872) K^-} } \hftabletlcell 
 &  \hfhref{BR_-511_-321+82.html}{ \hfpdg{<0.50} }  
 &  { \,}  
 &  \hfhref{0505003.html}{ \hfpubhot{<0.50} }  
 &  { \,}  
 &  \hfhref{BR_-511_-321+82.html}{ \hfavg{<0.50} } 
\\%

 \hfhref{BR_-511_-85+411.html}{ \hflabel{D_{sJ}^{-}(2460) D^+} }  
 &  { \,}  
 &  { \,}  
 &  \hfhref{0604004.html}{ \hfpubhot{2.60\pm1.50\pm0.70} }  
 &  { \,}  
 &  \hfhref{BR_-511_-85+411.html}{ \hfavg{2.6\pm1.7} } 
\\%

 \hfhref{BR_-511_-85+413.html}{ \hflabel{D_{sJ}^{-}(2460) D^{*+}(2010)} } \hftableblcell 
 &  { \,}  
 &  { \,}  
 &  \hfhref{0604004.html}{ \hfpubhot{8.8\pm2.0\pm1.4} }  
 &  { \,}  
 &  \hfhref{BR_-511_-85+413.html}{ \hfavg{8.8\pm2.4} } 
\\\hline 

\end{tabular}

\end{center}


\begin{center}
\hfnewcaption{Product branching fractions}{neutral B}{\hfnewp}{in units of $10^{-4}$}{upper limits are at 90\% CL}{00201.html}{tab:hfc02201}
\hfmetadata{ [ .tml created 2007-04-07T00:10:15.455+08:00] }

\begin{tabular}{lccccc}  
\hflmode
&  \hflpdg2006
&  \hflbelle
&  \hflbabar
&  \hflcdf
&  \hflavg
\\\hline 

 \hfhref{BR_-511_-85+211xBR_-85_-431+22.html}{ \hflabel{\pi^+ D_{sJ}^{-}(2460) [ D_s^- \gamma ]} } \hftabletlcell 
 &  \hfhref{BR_-511_-85+211xBR_-85_-431+22.html}{ \hfpdg{<0.040} }  
 &  \hfhref{0506032.html}{ \hfpub{<0.040} }  
 &  { \,}  
 &  { \,}  
 &  \hfhref{BR_-511_-85+211xBR_-85_-431+22.html}{ \hfavg{<0.040} } 
\\%

 \hfhref{BR_-511_-311+81xBR_81_-211+211+443.html}{ \hflabel{\bar{K}^0 X(3872) [ J/\psi(1S) \pi^+ \pi^- ]} }  
 &  \hfhref{BR_-511_-311+81xBR_81_-211+211+443.html}{ \hfpdg{<0.103} }  
 &  { \,}  
 &  \hfhref{0507002.html}{ \hfpubhot{0.051\pm0.028\pm0.007} }  
 &  { \,}  
 &  \hfhref{BR_-511_-311+81xBR_81_-211+211+443.html}{ \hfavg{0.05\pm0.03} } 
\\%

 \hfhref{BR_-511_-321+82xBR_82_111+211+443.html}{ \hflabel{K^- X^{+}(3872) [ J/\psi(1S) \pi^+ \pi^0 ]} }  
 &  \hfhref{BR_-511_-321+82xBR_82_111+211+443.html}{ \hfpdg{<0.054} }  
 &  { \,}  
 &  \hfhref{0506009.html}{ \hfpub{<0.054} }  
 &  { \,}  
 &  \hfhref{BR_-511_-321+82xBR_82_111+211+443.html}{ \hfavg{<0.054} } 
\\%

 \hfhref{BR_-511_-321+85xBR_85_22+431.html}{ \hflabel{K^- D_{sJ}^{+}(2460) [ D_s^+ \gamma ]} }  
 &  \hfhref{BR_-511_-321+85xBR_85_22+431.html}{ \hfpdg{<0.094} }  
 &  \hfhref{0506051.html}{ \hfpre{<0.086} }  
 &  { \,}  
 &  { \,}  
 &  \hfhref{BR_-511_-321+85xBR_85_22+431.html}{ \hfavg{<0.086} } 
\\%

 \hfhref{BR_-511_-84+211xBR_-84_-431+111.html}{ \hflabel{\pi^+ D_{sJ}^{*}(2317)^{-} [ D_s^- \pi^0 ]} }  
 &  \hfhref{BR_-511_-84+211xBR_-84_-431+111.html}{ \hfpdg{<0.25} }  
 &  \hfhref{0506032.html}{ \hfpub{<0.25} }  
 &  { \,}  
 &  { \,}  
 &  \hfhref{BR_-511_-84+211xBR_-84_-431+111.html}{ \hfavg{<0.25} } 
\\%

 \hfhref{BR_-511_-321+84xBR_84_111+431.html}{ \hflabel{K^- D_{sJ}^{*}(2317)^{+} [ D_s^+ \pi^0 ]} }  
 &  \hfhref{BR_-511_-321+84xBR_84_111+431.html}{ \hfpdg{0.43\pm0.15} }  
 &  \hfhref{0506051.html}{ \hfpre{0.44\pm0.08\pm0.06\pm0.11} }  
 &  { \,}  
 &  { \,}  
 &  \hfhref{BR_-511_-321+84xBR_84_111+431.html}{ \hfavg{0.44\pm0.15} } 
\\%

 \hfhref{BR_-511_-85+411xBR_-85_-431+-211+211.html}{ \hflabel{D^+ D_{sJ}^{-}(2460) [ D_s^- \pi^+ \pi^- ]} }  
 &  \hfhref{BR_-511_-85+411xBR_-85_-431+-211+211.html}{ \hfpdg{<2.00} }  
 &  \hfhref{0506040.html}{ \hfpubold{<2.00} }  
 &  { \,}  
 &  { \,}  
 &  \hfhref{BR_-511_-85+411xBR_-85_-431+-211+211.html}{ \hfavg{<2.00} } 
\\%

 \hfhref{BR_-511_-85+411xBR_-85_-431+111.html}{ \hflabel{D^+ D_{sJ}^{-}(2460) [ D_s^- \pi^0 ]} }  
 &  \hfhref{BR_-511_-85+411xBR_-85_-431+111.html}{ \hfpdg{<3.6} }  
 &  \hfhref{0506040.html}{ \hfpubold{<3.6} }  
 &  { \,}  
 &  { \,}  
 &  \hfhref{BR_-511_-85+411xBR_-85_-431+111.html}{ \hfavg{<3.6} } 
\\%

 \hfhref{BR_-511_-85+411xBR_-85_-433+22.html}{ \hflabel{D^+ D_{sJ}^{-}(2460) [ D_s^{*-} \gamma ]} }  
 &  \hfhref{BR_-511_-85+411xBR_-85_-433+22.html}{ \hfpdg{<6.0} }  
 &  \hfhref{0506040.html}{ \hfpubold{<6.0} }  
 &  { \,}  
 &  { \,}  
 &  \hfhref{BR_-511_-85+411xBR_-85_-433+22.html}{ \hfavg{<6.0} } 
\\%

 \hfhref{BR_-511_-85+411xBR_-85_-431+22.html}{ \hflabel{D^+ D_{sJ}^{-}(2460) [ D_s^- \gamma ]} }  
 &  \hfhref{BR_-511_-85+411xBR_-85_-431+22.html}{ \hfpdg{6.6\pm1.7} }  
 &  \hfhref{0506040.html}{ \hfpubold{8.2\pm^{2.2}_{1.9}\pm2.5} }  
 &  \hfhref{0506025.html}{ \hfpub{8.00\pm2.00\pm1.00\pm^{3.00}_{2.00}} }  
 &  { \,}  
 &  \hfhref{BR_-511_-85+411xBR_-85_-431+22.html}{ \hfavg{8.1\pm^{2.2}_{2.5}} } 
\\%

 \hfhref{BR_-511_-84+411xBR_-84_-433+22.html}{ \hflabel{D^+ D_{sJ}^{*}(2317)^{-} [ D_s^{*-} \gamma ]} }  
 &  \hfhref{BR_-511_-84+411xBR_-84_-433+22.html}{ \hfpdg{<9.5} }  
 &  \hfhref{0506040.html}{ \hfpubold{<9.5} }  
 &  { \,}  
 &  { \,}  
 &  \hfhref{BR_-511_-84+411xBR_-84_-433+22.html}{ \hfavg{<9.5} } 
\\%

 \hfhref{BR_-511_-84+411xBR_-84_-431+111.html}{ \hflabel{D^+ D_{sJ}^{*}(2317)^{-} [ D_s^- \pi^0 ]} }  
 &  \hfhref{BR_-511_-84+411xBR_-84_-431+111.html}{ \hfpdg{9.7\pm3.7} }  
 &  \hfhref{0506040.html}{ \hfpubold{8.6\pm^{3.3}_{2.6}\pm2.6} }  
 &  \hfhref{0506025.html}{ \hfpub{18.0\pm4.0\pm3.0\pm^{6.0}_{4.0}} }  
 &  { \,}  
 &  \hfhref{BR_-511_-84+411xBR_-84_-431+111.html}{ \hfavg{10.4\pm^{3.2}_{3.5}} } 
\\%

 \hfhref{BR_-511_-84+413xBR_-84_-431+111.html}{ \hflabel{D^{*+}(2010) D_{sJ}^{*}(2317)^{-} [ D_s^- \pi^0 ]} }  
 &  \hfhref{BR_-511_-84+413xBR_-84_-431+111.html}{ \hfpdg{15.0\pm6.0} }  
 &  { \,}  
 &  \hfhref{0506025.html}{ \hfpub{15.0\pm4.0\pm2.0\pm^{5.0}_{3.0}} }  
 &  { \,}  
 &  \hfhref{BR_-511_-84+413xBR_-84_-431+111.html}{ \hfavg{15.0\pm^{6.7}_{5.4}} } 
\\%

 \hfhref{BR_-511_-85+413xBR_-85_-431+22.html}{ \hflabel{D^{*+}(2010) D_{sJ}^{-}(2460) [ D_s^- \gamma ]} }  
 &  \hfhref{BR_-511_-85+413xBR_-85_-431+22.html}{ \hfpdg{23.0\pm8.0} }  
 &  { \,}  
 &  \hfhref{0506025.html}{ \hfpub{23.0\pm3.0\pm3.0\pm^{8.0}_{5.0}} }  
 &  { \,}  
 &  \hfhref{BR_-511_-85+413xBR_-85_-431+22.html}{ \hfavg{23.0\pm^{9.1}_{6.6}} } 
\\%

 \hfhref{BR_-511_-85+411xBR_-85_-433+111.html}{ \hflabel{D^+ D_{sJ}^{-}(2460) [ D_s^{*-} \pi^0 ]} }  
 &  \hfhref{BR_-511_-85+411xBR_-85_-433+111.html}{ \hfpdg{20.0\pm5.0} }  
 &  \hfhref{0506040.html}{ \hfpubold{22.7\pm^{7.3}_{6.2}\pm6.8} }  
 &  \hfhref{0506025.html}{ \hfpub{28.0\pm8.0\pm5.0\pm^{10.0}_{6.0}} }  
 &  { \,}  
 &  \hfhref{BR_-511_-85+411xBR_-85_-433+111.html}{ \hfavg{24.6\pm^{7.2}_{8.2}} } 
\\%

 \hfhref{BR_-511_-85+413xBR_-85_-433+111.html}{ \hflabel{D^{*+}(2010) D_{sJ}^{-}(2460) [ D_s^{*-} \pi^0 ]} } \hftableblcell 
 &  \hfhref{BR_-511_-85+413xBR_-85_-433+111.html}{ \hfpdg{55\pm23} }  
 &  { \,}  
 &  \hfhref{0506025.html}{ \hfpub{55.0\pm12.0\pm10.0\pm^{19.0}_{12.0}} }  
 &  { \,}  
 &  \hfhref{BR_-511_-85+413xBR_-85_-433+111.html}{ \hfavg{55\pm^{25}_{20}} } 
\\\hline 

\end{tabular}

\end{center}


\begin{center}
\hfnewcaption{Ratios of branching fractions}{neutral B}{\hfnewp}{in units of $10^{0}$}{upper limits are at 90\% CL}{00201.html}{tab:hfc03201}
\hfmetadata{ [ .tml created 2007-04-07T00:10:27.853+08:00] }

\begin{tabular}{lccccc}  
\hflmode
&  \hflpdg2006
&  \hflbelle
&  \hflbabar
&  \hflcdf
&  \hflavg
\\\hline 

 \hfhref{BR_-511_-311+81xoBR_-521_-321+81.html}{ \hflabel{\frac{{\cal{B}} ( \bar{B}^0 \to X(3872) \bar{K}^0 )}{{\cal{B}} ( B^- \to X(3872) K^- )}} } \hftabletlcell\hftableblcell 
 &  { \,}  
 &  { \,}  
 &  \hfhref{0507002.html}{ \hfpubhot{0.50\pm0.30\pm0.05} }  
 &  { \,}  
 &  \hfhref{BR_-511_-311+81xoBR_-521_-321+81.html}{ \hfavg{0.50\pm0.30} } 
\\\hline 

\end{tabular}

\end{center}

\end{\hftabletype}
\clearpage  
\begin{\hftabletype}[\hftableposn]

\begin{center}
\hfnewcaption{Branching fractions}{neutral B}{\hfdstr}{in units of $10^{-3}$}{upper limits are at 90\% CL}{00202.html}{tab:hfc01202}
\hfmetadata{ [ .tml created 2007-04-07T00:24:43.433+08:00] }

\begin{tabular}{lccccc}  
\hflmode
&  \hflpdg2006
&  \hflbelle
&  \hflbabar
&  \hflcdf
&  \hflavg
\\\hline 

 \hfhref{BR_-511_-431+211.html}{ \hflabel{D_s^- \pi^+} } \hftabletlcell 
 &  \hfhref{BR_-511_-431+211.html}{ \hfpdg{0.022\pm0.007} }  
 &  { \,}  
 &  { \,}  
 &  { \,}  
 &  \hfhref{BR_-511_-431+211.html}{ \hfavg{0.014\pm0.003} } 
\\%

{ \hflabel{\,} }  
 &  { \,}  
 &  \hfhref{0506089.html}{ \hfpubold{0.024\pm^{0.010}_{0.008}\pm0.004\pm0.006} }  
 &  \hfhref{0505001.html}{ \hfsuperceeded{0.032\pm0.009\pm0.007\pm0.008} } 
\hffootnotemark{1c}
 
 &  { \,}  
 &  { \,} 
\\%

{ \hflabel{\,} }  
 &  { \,}  
 &  { \,}  
 &  \hfhref{0604001.html}{ \hfpubhot{0.013\pm0.003\pm0.001\pm0.002} } 
\hffootnotemark{2c}
 
 &  { \,}  
 &  { \,} 
\\%

 \hfhref{BR_-511_-431+213.html}{ \hflabel{D_s^- \rho^+(770)} }  
 &  \hfhref{BR_-511_-431+213.html}{ \hfpdg{<0.60} }  
 &  { \,}  
 &  \hfhref{0506022.html}{ \hfpre{<0.019} }  
 &  { \,}  
 &  \hfhref{BR_-511_-431+213.html}{ \hfavg{<0.019} } 
\\%

 \hfhref{BR_-511_-431+9000211.html}{ \hflabel{D_s^- a_0^+(980)} }  
 &  { \,}  
 &  { \,}  
 &  \hfhref{0506017.html}{ \hfpubhot{<0.019} }  
 &  { \,}  
 &  \hfhref{BR_-511_-431+9000211.html}{ \hfavg{<0.019} } 
\\%

 \hfhref{BR_-511_-321+433.html}{ \hflabel{D_s^{*+} K^-} }  
 &  \hfhref{BR_-511_-321+433.html}{ \hfpdg{<0.025} }  
 &  { \,}  
 &  { \,}  
 &  { \,}  
 &  \hfhref{BR_-511_-321+433.html}{ \hfavg{0.020\pm0.006} } 
\\%

{ \hflabel{\,} }  
 &  { \,}  
 &  { \,}  
 &  \hfhref{0604001.html}{ \hfpubhot{0.020\pm0.005\pm0.003\pm0.003} } 
\hffootnotemark{2b}
 
 &  { \,}  
 &  { \,} 
\\%

{ \hflabel{\,} }  
 &  { \,}  
 &  { \,}  
 &  \hfhref{0505001.html}{ \hfsuperceeded{<0.025} } 
\hffootnotemark{1b}
 
 &  { \,}  
 &  { \,} 
\\%

 \hfhref{BR_-511_-321+431.html}{ \hflabel{D_s^+ K^-} }  
 &  \hfhref{BR_-511_-321+431.html}{ \hfpdg{0.031\pm0.008} }  
 &  { \,}  
 &  { \,}  
 &  { \,}  
 &  \hfhref{BR_-511_-321+431.html}{ \hfavg{0.027\pm0.005} } 
\\%

{ \hflabel{\,} }  
 &  { \,}  
 &  \hfhref{0506089.html}{ \hfpubold{0.046\pm^{0.012}_{0.011}\pm0.006\pm0.012} }  
 &  \hfhref{0505001.html}{ \hfsuperceeded{0.032\pm0.010\pm0.007\pm0.008} } 
\hffootnotemark{1a}
 
 &  { \,}  
 &  { \,} 
\\%

{ \hflabel{\,} }  
 &  { \,}  
 &  { \,}  
 &  \hfhref{0604001.html}{ \hfpubhot{0.025\pm0.004\pm0.002\pm0.003} } 
\hffootnotemark{2a}
 
 &  { \,}  
 &  { \,} 
\\%

 \hfhref{BR_-511_-433+211.html}{ \hflabel{D_s^{*-} \pi^+} }  
 &  \hfhref{BR_-511_-433+211.html}{ \hfpdg{<0.041} }  
 &  { \,}  
 &  { \,}  
 &  { \,}  
 &  \hfhref{BR_-511_-433+211.html}{ \hfavg{0.028\pm0.008} } 
\\%

{ \hflabel{\,} }  
 &  { \,}  
 &  { \,}  
 &  \hfhref{0604001.html}{ \hfpubhot{0.028\pm0.006\pm0.004\pm0.003} } 
\hffootnotemark{2d}
 
 &  { \,}  
 &  { \,} 
\\%

{ \hflabel{\,} }  
 &  { \,}  
 &  { \,}  
 &  \hfhref{0505001.html}{ \hfsuperceeded{<0.041} } 
\hffootnotemark{1d}
 
 &  { \,}  
 &  { \,} 
\\%

 \hfhref{BR_-511_-433+9000211.html}{ \hflabel{D_s^{*-} a_0^+(980)} }  
 &  { \,}  
 &  { \,}  
 &  \hfhref{0506017.html}{ \hfpubhot{<0.036} }  
 &  { \,}  
 &  \hfhref{BR_-511_-433+9000211.html}{ \hfavg{<0.036} } 
\\%

 \hfhref{BR_-511_-2212+431+3122.html}{ \hflabel{D_s^+ \Lambda \bar{p}} }  
 &  { \,}  
 &  \hfhref{0607019.html}{ \hfwaiting{0.036\pm0.009\pm0.006\pm0.009} }  
 &  { \,}  
 &  { \,}  
 &  \hfhref{BR_-511_-2212+431+3122.html}{ \hfnewavg{0.04\pm0.01} } 
\\%

 \hfhref{BR_-511_-431+433.html}{ \hflabel{D_s^- D_s^{*+}} }  
 &  \hfhref{BR_-511_-431+433.html}{ \hfpdg{<0.130} }  
 &  { \,}  
 &  \hfhref{0506021.html}{ \hfpub{<0.130} }  
 &  { \,}  
 &  \hfhref{BR_-511_-431+433.html}{ \hfavg{<0.130} } 
\\%

 \hfhref{BR_-511_-431+431.html}{ \hflabel{D_s^- D_s^+} }  
 &  \hfhref{BR_-511_-431+431.html}{ \hfpdg{<0.100} }  
 &  \hfhref{0506066.html}{ \hfpre{<0.200} }  
 &  \hfhref{0506021.html}{ \hfpub{<0.100} }  
 &  { \,}  
 &  \hfhref{BR_-511_-431+431.html}{ \hfavg{<0.100} } 
\\%

 \hfhref{BR_-511_-431+215.html}{ \hflabel{D_s^- a_2^+(1320)} }  
 &  { \,}  
 &  { \,}  
 &  \hfhref{0506017.html}{ \hfpubhot{<0.190} }  
 &  { \,}  
 &  \hfhref{BR_-511_-431+215.html}{ \hfavg{<0.190} } 
\\%

 \hfhref{BR_-511_-433+215.html}{ \hflabel{D_s^{*-} a_2^+(1320)} }  
 &  { \,}  
 &  { \,}  
 &  \hfhref{0506017.html}{ \hfpubhot{<0.200} }  
 &  { \,}  
 &  \hfhref{BR_-511_-433+215.html}{ \hfavg{<0.200} } 
\\%

 \hfhref{BR_-511_-433+433.html}{ \hflabel{D_s^{*+} D_s^{*-}} }  
 &  \hfhref{BR_-511_-433+433.html}{ \hfpdg{<0.24} }  
 &  { \,}  
 &  \hfhref{0506021.html}{ \hfpub{<0.24} }  
 &  { \,}  
 &  \hfhref{BR_-511_-433+433.html}{ \hfavg{<0.24} } 
\\%

 \hfhref{BR_-511_-433+411.html}{ \hflabel{D_s^{*-} D^+} }  
 &  \hfhref{BR_-511_-433+411.html}{ \hfpdg{8.6\pm3.4} }  
 &  { \,}  
 &  \hfhref{0604004.html}{ \hfpubhot{6.7\pm2.0\pm1.1} }  
 &  { \,}  
 &  \hfhref{BR_-511_-433+411.html}{ \hfavg{6.7\pm2.3} } 
\\%

 \hfhref{BR_-511_-431+411.html}{ \hflabel{D_s^- D^+} }  
 &  \hfhref{BR_-511_-431+411.html}{ \hfpdg{6.5\pm2.1} }  
 &  \hfhref{0506066.html}{ \hfpre{7.42\pm0.23\pm1.36} }  
 &  \hfhref{0604004.html}{ \hfpubhot{9.0\pm1.8\pm1.4} }  
 &  { \,}  
 &  \hfhref{BR_-511_-431+411.html}{ \hfavg{7.8\pm1.2} } 
\\%

 \hfhref{BR_-511_-431+413.html}{ \hflabel{D_s^- D^{*+}(2010)} }  
 &  \hfhref{BR_-511_-431+413.html}{ \hfpdg{8.8\pm1.6} }  
 &  { \,}  
 &  { \,}  
 &  { \,}  
 &  \hfhref{BR_-511_-431+413.html}{ \hfavg{6.8\pm1.6} } 
\\%

{ \hflabel{\,} }  
 &  { \,}  
 &  { \,}  
 &  \hfhref{0604004.html}{ \hfpubhot{5.70\pm1.60\pm0.90} } 
\hffootnotemark{3a}
 
 &  { \,}  
 &  { \,} 
\\%

{ \hflabel{\,} }  
 &  { \,}  
 &  { \,}  
 &  \hfhref{0506110.html}{ \hfpubold{10.3\pm1.4\pm1.3\pm2.6} } 
\hffootnotemark{4a}
 
 &  { \,}  
 &  { \,} 
\\%

 \hfhref{BR_-511_-433+413.html}{ \hflabel{D_s^{*-} D^{*+}(2010)} }  
 &  \hfhref{BR_-511_-433+413.html}{ \hfpdg{17.9\pm1.6} }  
 &  { \,}  
 &  { \,}  
 &  { \,}  
 &  \hfhref{BR_-511_-433+413.html}{ \hfavg{18.2\pm1.6} } 
\\%

{ \hflabel{\,} }  
 &  { \,}  
 &  { \,}  
 &  \hfhref{0506019.html}{ \hfpub{18.80\pm0.90\pm1.60\pm0.60} } 
\hffootnotemark{5}
 
 &  { \,}  
 &  { \,} 
\\%

{ \hflabel{\,} }  
 &  { \,}  
 &  { \,}  
 &  \hfhref{0604004.html}{ \hfpubhot{16.5\pm2.3\pm1.9} } 
\hffootnotemark{3b}
 
 &  { \,}  
 &  { \,} 
\\%

{ \hflabel{\,} }  
 &  { \,}  
 &  { \,}  
 &  \hfhref{0506110.html}{ \hfpubold{19.7\pm1.5\pm3.0\pm4.9} } 
\hffootnotemark{4b}
 
 &  { \,}  
 &  { \,} 
\\%

 \hfhref{BR_-511_-10433+413.html}{ \hflabel{D_{s1}^-(2536) D^{*+}(2010)} } \hftableblcell 
 &  { \,}  
 &  { \,}  
 &  \hfhref{0609001.html}{ \hfpubhot{92.00\pm24.00\pm1.00} }  
 &  { \,}  
 &  \hfhref{BR_-511_-10433+413.html}{ \hfnewavg{92\pm24} } 
\\\hline 

\end{tabular}

\begin{description}\addtolength{\itemsep}{\hffootitemsep\baselineskip}

	  \item[ \hffootnotemark{1} ] \hffootnotetext{ A study of the rare decays $\bar{B}^0 \rightarrow D_s^{-(*)}\pi^+$ and 
$\bar{B}^0 \rightarrow D_s^{+(*)} K^-$ (\hfbb{84.3}) } 
     ;  \hffootnotemark{1a}  \hffootnotetext{ $\bar{B}^0 \rightarrow D_s^+ K^-$ }  ;  \hffootnotemark{1b}  \hffootnotetext{ $\bar{B}^0 \rightarrow D_s^{*+} K^-$ }  ;  \hffootnotemark{1c}  \hffootnotetext{ $\bar{B}^0 \rightarrow D_s^- \pi^+$ }  ;  \hffootnotemark{1d}  \hffootnotetext{ $\bar{B}^0 \rightarrow D_s^{*-} \pi^+$ } 
	  \item[ \hffootnotemark{2} ] \hffootnotetext{ Observation of Decays $\bar{B}^0 \rightarrow D_s^{-(*)}\pi^+$ and $\bar{B}^0 \rightarrow D_s^{+(*)}K^-$ (\hfbb{230}) } 
     ;  \hffootnotemark{2a}  \hffootnotetext{ $\bar{B}^0 \rightarrow D_s^+ K^-$ }  ;  \hffootnotemark{2b}  \hffootnotetext{ $\bar{B}^0 \rightarrow D_s^{*+} K^-$ }  ;  \hffootnotemark{2c}  \hffootnotetext{ $\bar{B}^0 \rightarrow D_s^- \pi^+$ }  ;  \hffootnotemark{2d}  \hffootnotetext{ $\bar{B}^0 \rightarrow D_s^{*-} \pi^+$ } 
	  \item[ \hffootnotemark{3} ] \hffootnotetext{ Study of $\bar{B} \rightarrow D^{(*)+,-}X^{-}$ and $\bar{B} \rightarrow D_s^{(*)-}X^{+,0}$ decays and measurement of $D_s^-$ and $D_{sJ}^-(2460)$ absolute branching fractions (\hfbb{230}) } 
     ;  \hffootnotemark{3a}  \hffootnotetext{ $\bar{B}^0 \rightarrow D_s^-D^{*+}$ }  ;  \hffootnotemark{3b}  \hffootnotetext{ $\bar{B}^0 \rightarrow D_s^{*-}D^{*+})$ } 
	  \item[ \hffootnotemark{4} ] \hffootnotetext{ Measurement of $\bar{B}^0 \rightarrow D_s^{(*)}D^*$ Branching Fractions and $D_s^*D^*$
 Polarization with a Partial Reconstruction technique (\hfbb{22.7}) } 
     ;  \hffootnotemark{4a}  \hffootnotetext{ $\bar{B}^0 \rightarrow D_s^- D^{*+}$ }  ;  \hffootnotemark{4b}  \hffootnotetext{ $\bar{B}^0 \rightarrow D_s^{*-} D^{*+}$ } 
	  \item[ \hffootnotemark{5} ] \hffootnotetext{ Measurement of the $\bar{B}^0 \rightarrow D_s^{*-} D^+$ and $D_s^+ \rightarrow \phi \pi^+$ branching
fractions (\hfbb{123}) } 
     ;  \hffootnotetext{ $\bar{B}^0 \rightarrow D_s^{*-} D^{*+}$ } 
\end{description}

\end{center}

\end{\hftabletype}
\clearpage  
\begin{\hftabletype}[\hftableposn]

\begin{center}
\hfnewcaption{Product branching fractions}{neutral B}{\hfdstr}{in units of $10^{-4}$}{upper limits are at 90\% CL}{00202.html}{tab:hfc02202}
\hfmetadata{ [ .tml created 2007-04-07T00:24:55.429+08:00] }

\begin{tabular}{lccccc}  
\hflmode
&  \hflpdg2006
&  \hflbelle
&  \hflbabar
&  \hflcdf
&  \hflavg
\\\hline 

 \hfhref{BR_-511_-431+411xBR_-431_-211+333xBR_333_-321+321.html}{ \hflabel{D^+ D_s^- [ \pi^- \phi(1020) [ K^+ K^- ] ]} } \hftabletlcell 
 &  \hfhref{BR_-511_-431+411xBR_-431_-211+333xBR_333_-321+321.html}{ \hfpdg{1.41\pm0.41} }  
 &  \hfhref{0506066.html}{ \hfpre{1.47\pm0.05\pm0.21} }  
 &  { \,}  
 &  { \,}  
 &  \hfhref{BR_-511_-431+411xBR_-431_-211+333xBR_333_-321+321.html}{ \hfavg{1.47\pm0.22} } 
\\%

 \hfhref{BR_-511_-431+411xBR_-431_-211+333.html}{ \hflabel{D^+ D_s^- [ \phi(1020) \pi^- ]} }  
 &  \hfhref{BR_-511_-431+411xBR_-431_-211+333.html}{ \hfpdg{2.90\pm0.80} }  
 &  { \,}  
 &  \hfhref{0604004.html}{ \hfpubhot{2.67\pm0.61\pm0.47} }  
 &  { \,}  
 &  \hfhref{BR_-511_-431+411xBR_-431_-211+333.html}{ \hfavg{2.67\pm0.77} } 
\\%

 \hfhref{BR_-511_-433+411xBR_-431_-211+333.html}{ \hflabel{D_s^{*-} D^+ [ D_s^- \to \phi(1020) \pi^- ]} }  
 &  \hfhref{BR_-511_-433+411xBR_-431_-211+333.html}{ \hfpdg{3.8\pm1.4} }  
 &  { \,}  
 &  \hfhref{0604004.html}{ \hfpubhot{4.14\pm1.19\pm0.94} }  
 &  { \,}  
 &  \hfhref{BR_-511_-433+411xBR_-431_-211+333.html}{ \hfavg{4.1\pm1.5} } 
\\%

 \hfhref{BR_-511_-431+413xBR_-431_-211+333.html}{ \hflabel{D^{*+}(2010) D_s^- [ \phi(1020) \pi^- ]} }  
 &  \hfhref{BR_-511_-431+413xBR_-431_-211+333.html}{ \hfpdg{3.90\pm0.50} }  
 &  { \,}  
 &  \hfhref{0604004.html}{ \hfpubhot{5.11\pm0.94\pm0.72} }  
 &  { \,}  
 &  \hfhref{BR_-511_-431+413xBR_-431_-211+333.html}{ \hfavg{5.1\pm1.2} } 
\\%

 \hfhref{BR_-511_-433+413xBR_-431_-211+333.html}{ \hflabel{D_s^{*-} D^{*+}(2010) [ D_s^- \to \phi(1020) \pi^- ]} } \hftableblcell 
 &  \hfhref{BR_-511_-433+413xBR_-431_-211+333.html}{ \hfpdg{7.9\pm1.3} }  
 &  { \,}  
 &  \hfhref{0604004.html}{ \hfpubhot{12.2\pm2.2\pm2.2} }  
 &  { \,}  
 &  \hfhref{BR_-511_-433+413xBR_-431_-211+333.html}{ \hfavg{12.2\pm3.1} } 
\\\hline 

\end{tabular}

\end{center}


\begin{center}
\hfnewcaption{Ratios of branching fractions}{neutral B}{\hfdstr}{in units of $10^{0}$}{upper limits are at 90\% CL}{00202.html}{tab:hfc03202}
\hfmetadata{ [ .tml created 2007-04-07T00:25:07.012+08:00] }

\begin{tabular}{lccccc}  
\hflmode
&  \hflpdg2006
&  \hflbelle
&  \hflbabar
&  \hflcdf
&  \hflavg
\\\hline 

 \hfhref{BR_-511_-433+411xoBR_-511_-431+411.html}{ \hflabel{\frac{{\cal{B}} ( \bar{B}^0 \to D_s^{*-} D^+ )}{{\cal{B}} ( \bar{B}^0 \to D_s^- D^+ )}} } \hftabletlcell 
 &  { \,}  
 &  { \,}  
 &  { \,}  
 &  \hfhref{0601004.html}{ \hfwaiting{0.90\pm0.20\pm0.10} }  
 &  \hfhref{BR_-511_-433+411xoBR_-511_-431+411.html}{ \hfavg{0.90\pm0.22} } 
\\%

 \hfhref{BR_-511_-431+413xoBR_-511_-431+411.html}{ \hflabel{\frac{{\cal{B}} ( \bar{B}^0 \to D_s^- D^{*+}(2010) )}{{\cal{B}} ( \bar{B}^0 \to D_s^- D^+ )}} }  
 &  { \,}  
 &  { \,}  
 &  { \,}  
 &  \hfhref{0601004.html}{ \hfwaiting{1.50\pm0.50\pm0.10} }  
 &  \hfhref{BR_-511_-431+413xoBR_-511_-431+411.html}{ \hfavg{1.50\pm0.51} } 
\\%

 \hfhref{BR_-511_-431+411xoBR_-511_-211+-211+211+411.html}{ \hflabel{\frac{{\cal{B}} ( \bar{B}^0 \to D_s^- D^+ )}{{\cal{B}} ( \bar{B}^0 \to D^+ \pi^+ \pi^- \pi^- )}} }  
 &  { \,}  
 &  { \,}  
 &  { \,}  
 &  \hfhref{0601004.html}{ \hfwaiting{1.99\pm0.13\pm0.11\pm0.45} }  
 &  \hfhref{BR_-511_-431+411xoBR_-511_-211+-211+211+411.html}{ \hfavg{1.99\pm0.48} } 
\\%

 \hfhref{BR_-511_-433+413xoBR_-511_-431+411.html}{ \hflabel{\frac{{\cal{B}} ( \bar{B}^0 \to D_s^{*-} D^{*+}(2010) )}{{\cal{B}} ( \bar{B}^0 \to D_s^- D^+ )}} } \hftableblcell 
 &  { \,}  
 &  { \,}  
 &  { \,}  
 &  \hfhref{0601004.html}{ \hfwaiting{2.60\pm0.50\pm0.20} }  
 &  \hfhref{BR_-511_-433+413xoBR_-511_-431+411.html}{ \hfavg{2.60\pm0.54} } 
\\\hline 

\end{tabular}

\end{center}

\end{\hftabletype}
\clearpage  
\begin{\hftabletype}[\hftableposn]

\begin{center}
\hfnewcaption{Branching fractions}{neutral B}{\hfbary}{in units of $10^{-5}$}{upper limits are at 90\% CL}{00203.html}{tab:hfc01203}
\hfmetadata{ [ .tml created 2007-04-07T00:43:29.953+08:00] }

\begin{tabular}{lccccc}  
\hflmode
&  \hflpdg2006
&  \hflbelle
&  \hflbabar
&  \hflcdf
&  \hflavg
\\\hline 

 \hfhref{BR_-511_-2212+443+2212.html}{ \hflabel{J/\psi(1S) \bar{p} p} } \hftabletlcell 
 &  \hfhref{BR_-511_-2212+443+2212.html}{ \hfpdg{<0.083} }  
 &  \hfhref{0607005.html}{ \hfpub{<0.083} }  
 &  \hfhref{0506010.html}{ \hfpubold{<0.190} }  
 &  { \,}  
 &  \hfhref{BR_-511_-2212+443+2212.html}{ \hfnewavg{<0.083} } 
\\%

 \hfhref{BR_-511_-2212+4122.html}{ \hflabel{\Lambda_c^+ \bar{p}} }  
 &  \hfhref{BR_-511_-2212+4122.html}{ \hfpdg{2.20\pm0.80} }  
 &  \hfhref{0506087.html}{ \hfpubold{2.19\pm^{0.56}_{0.49}\pm0.32\pm0.57} }  
 &  \hfhref{0604003.html}{ \hfprehot{2.15\pm0.36\pm0.13\pm0.56} }  
 &  { \,}  
 &  \hfhref{BR_-511_-2212+4122.html}{ \hfavg{2.17\pm0.53} } 
\\%

 \hfhref{BR_-511_-2212+211+4114.html}{ \hflabel{\Sigma_c^{*0} \bar{p} \pi^+} }  
 &  \hfhref{BR_-511_-2212+211+4114.html}{ \hfpdg{<12.1} }  
 &  { \,}  
 &  { \,}  
 &  { \,}  
 &  \hfhref{BR_-511_-2212+211+4114.html}{ \hfnewavg{<3.3} } 
\\%

{ \hflabel{\,} }  
 &  { \,}  
 &  \hfhref{0506086.html}{ \hfpubold{<12.1} } 
\hffootnotemark{1}
 
 &  { \,}  
 &  { \,}  
 &  { \,} 
\\%

{ \hflabel{\,} }  
 &  { \,}  
 &  \hfhref{0607018.html}{ \hfpubhot{<3.3} } 
\hffootnotemark{2}
 
 &  { \,}  
 &  { \,}  
 &  { \,} 
\\%

 \hfhref{BR_-511_-2212+423+2212.html}{ \hflabel{D^{*0}(2007) p \bar{p}} }  
 &  { \,}  
 &  { \,}  
 &  { \,}  
 &  { \,}  
 &  \hfhref{BR_-511_-2212+423+2212.html}{ \hfavg{11.1\pm1.3} } 
\\%

{ \hflabel{\,} }  
 &  { \,}  
 &  \hfhref{0506083.html}{ \hfpubold{12.0\pm^{3.3}_{2.9}\pm2.1} }  
 &  \hfhref{0506024.html}{ \hfsuperceeded{6.70\pm2.10\pm0.82\pm0.36} } 
\hffootnotemark{3c}
 
 &  { \,}  
 &  { \,} 
\\%

{ \hflabel{\,} }  
 &  { \,}  
 &  { \,}  
 &  \hfhref{0603003.html}{ \hfpubhot{11.00\pm1.00\pm0.90} } 
\hffootnotemark{4c}
 
 &  { \,}  
 &  { \,} 
\\%

 \hfhref{BR_-511_-2212+421+2212.html}{ \hflabel{D^0 p \bar{p}} }  
 &  { \,}  
 &  { \,}  
 &  { \,}  
 &  { \,}  
 &  \hfhref{BR_-511_-2212+421+2212.html}{ \hfavg{11.39\pm0.91} } 
\\%

{ \hflabel{\,} }  
 &  { \,}  
 &  \hfhref{0506083.html}{ \hfpubold{11.8\pm1.5\pm1.6} }  
 &  \hfhref{0506024.html}{ \hfsuperceeded{12.40\pm1.40\pm1.16\pm0.30} } 
\hffootnotemark{3b}
 
 &  { \,}  
 &  { \,} 
\\%

{ \hflabel{\,} }  
 &  { \,}  
 &  { \,}  
 &  \hfhref{0603003.html}{ \hfpubhot{11.30\pm0.60\pm0.80} } 
\hffootnotemark{4b}
 
 &  { \,}  
 &  { \,} 
\\%

 \hfhref{BR_-511_-2212+-211+4224.html}{ \hflabel{\Sigma_c^{*++} \bar{p} \pi^-} }  
 &  \hfhref{BR_-511_-2212+-211+4224.html}{ \hfpdg{16.0\pm7.0} }  
 &  { \,}  
 &  { \,}  
 &  { \,}  
 &  \hfhref{BR_-511_-2212+-211+4224.html}{ \hfnewavg{12.9\pm^{3.3}_{3.4}} } 
\\%

{ \hflabel{\,} }  
 &  { \,}  
 &  \hfhref{0506086.html}{ \hfpubold{16.3\pm^{5.7}_{5.1}\pm2.8\pm4.2} } 
\hffootnotemark{1}
 
 &  { \,}  
 &  { \,}  
 &  { \,} 
\\%

{ \hflabel{\,} }  
 &  { \,}  
 &  \hfhref{0607018.html}{ \hfpubhot{12.0\pm1.0\pm2.0\pm3.0} } 
\hffootnotemark{2}
 
 &  { \,}  
 &  { \,}  
 &  { \,} 
\\%

 \hfhref{BR_-511_-2212+211+4112.html}{ \hflabel{\Sigma_c^0 \bar{p} \pi^+} }  
 &  \hfhref{BR_-511_-2212+211+4112.html}{ \hfpdg{10.0\pm8.0} }  
 &  { \,}  
 &  { \,}  
 &  { \,}  
 &  \hfhref{BR_-511_-2212+211+4112.html}{ \hfnewavg{14.0\pm4.9} } 
\\%

{ \hflabel{\,} }  
 &  { \,}  
 &  \hfhref{0607018.html}{ \hfpubhot{14.0\pm2.0\pm2.0\pm4.0} } 
\hffootnotemark{2}
 
 &  { \,}  
 &  { \,}  
 &  { \,} 
\\%

{ \hflabel{\,} }  
 &  { \,}  
 &  \hfhref{0506086.html}{ \hfpubold{<15.9} } 
\hffootnotemark{1}
 
 &  { \,}  
 &  { \,}  
 &  { \,} 
\\%

 \hfhref{BR_-511_-2212+-211+4222.html}{ \hflabel{\Sigma_c^{++} \bar{p} \pi^-} }  
 &  \hfhref{BR_-511_-2212+-211+4222.html}{ \hfpdg{28.0\pm9.0} }  
 &  { \,}  
 &  { \,}  
 &  { \,}  
 &  \hfhref{BR_-511_-2212+-211+4222.html}{ \hfnewavg{21.8\pm^{5.1}_{5.2}} } 
\\%

{ \hflabel{\,} }  
 &  { \,}  
 &  \hfhref{0506086.html}{ \hfpubold{23.8\pm^{6.3}_{5.5}\pm4.1\pm6.2} } 
\hffootnotemark{1}
 
 &  { \,}  
 &  { \,}  
 &  { \,} 
\\%

{ \hflabel{\,} }  
 &  { \,}  
 &  \hfhref{0607018.html}{ \hfpubhot{21.0\pm2.0\pm3.0\pm5.0} } 
\hffootnotemark{2c}
 
 &  { \,}  
 &  { \,}  
 &  { \,} 
\\%

 \hfhref{BR_-511_-2212+-211+411+2212.html}{ \hflabel{D^+ p \bar{p} \pi^-} }  
 &  { \,}  
 &  { \,}  
 &  { \,}  
 &  { \,}  
 &  \hfhref{BR_-511_-2212+-211+411+2212.html}{ \hfavg{33.8\pm3.2} } 
\\%

{ \hflabel{\,} }  
 &  { \,}  
 &  { \,}  
 &  \hfhref{0506024.html}{ \hfsuperceeded{38.00\pm3.50\pm4.50\pm0.95} } 
\hffootnotemark{3a}
 
 &  { \,}  
 &  { \,} 
\\%

{ \hflabel{\,} }  
 &  { \,}  
 &  { \,}  
 &  \hfhref{0603003.html}{ \hfpubhot{33.8\pm1.4\pm2.9} } 
\hffootnotemark{4a}
 
 &  { \,}  
 &  { \,} 
\\%

 \hfhref{BR_-511_-2212+-211+413+2212.html}{ \hflabel{D^{*+}(2010) p \bar{p} \pi^-} }  
 &  \hfhref{BR_-511_-2212+-211+413+2212.html}{ \hfpdg{65\pm16} }  
 &  { \,}  
 &  { \,}  
 &  { \,}  
 &  \hfhref{BR_-511_-2212+-211+413+2212.html}{ \hfavg{48.1\pm4.9} } 
\\%

{ \hflabel{\,} }  
 &  { \,}  
 &  { \,}  
 &  \hfhref{0506024.html}{ \hfsuperceeded{56.1\pm5.9\pm6.4\pm3.6} } 
\hffootnotemark{3d}
 
 &  { \,}  
 &  { \,} 
\\%

{ \hflabel{\,} }  
 &  { \,}  
 &  { \,}  
 &  \hfhref{0603003.html}{ \hfpubhot{48.1\pm2.2\pm4.4} } 
\hffootnotemark{4d}
 
 &  { \,}  
 &  { \,} 
\\%

 \hfhref{BR_-511_-4122+-311+4122.html}{ \hflabel{\Lambda_c^+ \Lambda_c^- \bar{K}^0} }  
 &  { \,}  
 &  \hfhref{0607007.html}{ \hfpre{79\pm^{29}_{23}\pm12\pm41} }  
 &  { \,}  
 &  { \,}  
 &  \hfhref{BR_-511_-4122+-311+4122.html}{ \hfnewavg{79\pm^{52}_{49}} } 
\\%

 \hfhref{BR_-511_-2212+-211+211+4122.html}{ \hflabel{\Lambda_c^+ \bar{p} \pi^+ \pi^-} } \hftableblcell 
 &  \hfhref{BR_-511_-2212+-211+211+4122.html}{ \hfpdg{130\pm40} }  
 &  \hfhref{0506086.html}{ \hfpubold{110\pm^{12}_{12}\pm19\pm29} }  
 &  { \,}  
 &  { \,}  
 &  \hfhref{BR_-511_-2212+-211+211+4122.html}{ \hfavg{110\pm37} } 
\\\hline 

\end{tabular}

\begin{description}\addtolength{\itemsep}{\hffootitemsep\baselineskip}

	  \item[ \hffootnotemark{1} ] \hffootnotetext{ STUDY OF EXCLUSIVE B DECAYS TO CHARMED BARYONS AT BELLE. (\hfbb{31.7}) } 
    
	  \item[ \hffootnotemark{2} ] \hffootnotetext{ Study of the charmed baryonic decays $\bar{B}^0\to\Sigma_c^{++}\bar{p}\pi^-$ and $\bar{B}^0\to\Sigma_c^{0}\bar{p}\pi^+$ (\hfbb{386}) } 
     ;  \hffootnotemark{2c}  \hffootnotetext{ B0bar to Sigmac(2455)++ pbar pi } 
	  \item[ \hffootnotemark{3} ] \hffootnotetext{ Measurement of the Branching Fraction for the decays
$\bar{B}^0 \rightarrow D^{*+}p\bar{p}\pi^-$, $\bar{B}^0 \rightarrow D^+ p \bar{p} \pi^-$,
$\bar{B}^0 \rightarrow \bar{D}^{*0} p \bar{p}$, $\bar{B}^0 \rightarrow \bar{D}^0 p \bar{p}$ (\hfbb{124}) } 
     ;  \hffootnotemark{3a}  \hffootnotetext{ $\bar{B}^0 \rightarrow D^{+}p\bar{p}\pi^-$ }  ;  \hffootnotemark{3b}  \hffootnotetext{ $\bar{B}^0 \rightarrow \bar{D}^{0}p\bar{p}$ }  ;  \hffootnotemark{3c}  \hffootnotetext{ $\bar{B}^0 \rightarrow \bar{D}^{*0}p\bar{p}$ }  ;  \hffootnotemark{3d}  \hffootnotetext{ $\bar{B}^0 \rightarrow D^{*+}p\bar{p}\pi^-$ } 
	  \item[ \hffootnotemark{4} ] \hffootnotetext{ Measurements of the Decays $B^0 \rightarrow \bar{D}^0 p \bar{p}$,  $B^0 \rightarrow \bar{D}^{*0} p \bar{p}$, 
$B^0 \rightarrow D^-p\bar{p}\pi+-$, and $B^0 \rightarrow D^- p \bar{p} \pi^+$ (\hfbb{232}) } 
     ;  \hffootnotemark{4a}  \hffootnotetext{ $\bar{B}^0 \rightarrow D^{+}p\bar{p}\pi^-$ }  ;  \hffootnotemark{4b}  \hffootnotetext{ $\bar{B}^0 \rightarrow \bar{D}^{0}p\bar{p}$ }  ;  \hffootnotemark{4c}  \hffootnotetext{ $\bar{B}^0 \rightarrow \bar{D}^{*0}p\bar{p}$ }  ;  \hffootnotemark{4d}  \hffootnotetext{ $\bar{B}^0 \rightarrow D^{*+}p\bar{p}\pi^-$ } 
\end{description}

\end{center}

\end{\hftabletype}
\clearpage  
\begin{\hftabletype}[\hftableposn]

\begin{center}
\hfnewcaption{Product branching fractions}{neutral B}{\hfbary}{in units of $10^{-5}$}{upper limits are at 90\% CL}{00203.html}{tab:hfc02203}
\hfmetadata{ [ .tml created 2007-04-07T00:43:51.712+08:00] }

\begin{tabular}{lccccc}  
\hflmode
&  \hflpdg2006
&  \hflbelle
&  \hflbabar
&  \hflcdf
&  \hflavg
\\\hline 

 \hfhref{BR_-511_-4122+4232xBR_4232_211+211+3312.html}{ \hflabel{\Lambda_c^- \Xi_c^+ [ \Xi^- \pi^+ \pi^+ ]} } \hftabletlcell\hftableblcell 
 &  { \,}  
 &  \hfhref{0607006.html}{ \hfpre{9.3\pm^{3.7}_{2.8}\pm1.9\pm2.4} }  
 &  { \,}  
 &  { \,}  
 &  \hfhref{BR_-511_-4122+4232xBR_4232_211+211+3312.html}{ \hfnewavg{9.3\pm^{4.8}_{4.1}} } 
\\\hline 

\end{tabular}

\end{center}

\end{\hftabletype}
\clearpage  
\begin{\hftabletype}[\hftableposn]

\begin{center}
\hfnewcaption{Branching fractions}{neutral B}{\hfjpsi}{in units of $10^{-4}$}{upper limits are at 90\% CL}{00204.html}{tab:hfc01204}
\hfmetadata{ [ .tml created 2007-04-07T00:53:09.223+08:00] }

\begin{tabular}{lccccc}  
\hflmode
&  \hflpdg2006
&  \hflbelle
&  \hflbabar
&  \hflcdf
&  \hflavg
\\\hline 

 \hfhref{BR_-511_22+443.html}{ \hflabel{J/\psi(1S) \gamma} } \hftabletlcell 
 &  \hfhref{BR_-511_22+443.html}{ \hfpdg{<0.016} }  
 &  { \,}  
 &  \hfhref{0506002.html}{ \hfpub{<0.016} }  
 &  { \,}  
 &  \hfhref{BR_-511_22+443.html}{ \hfavg{<0.016} } 
\\%

 \hfhref{BR_-511_333+443.html}{ \hflabel{J/\psi(1S) \phi(1020)} }  
 &  \hfhref{BR_-511_333+443.html}{ \hfpdg{<0.092} }  
 &  { \,}  
 &  \hfhref{0506012.html}{ \hfpubold{<0.090} }  
 &  { \,}  
 &  \hfhref{BR_-511_333+443.html}{ \hfavg{<0.090} } 
\\%

 \hfhref{BR_-511_221+443.html}{ \hflabel{J/\psi(1S) \eta} }  
 &  \hfhref{BR_-511_221+443.html}{ \hfpdg{<0.27} }  
 &  \hfhref{0607020.html}{ \hfpubhot{0.096\pm0.017\pm0.007} }  
 &  \hfhref{0506012.html}{ \hfpubold{<0.27} }  
 &  { \,}  
 &  \hfhref{BR_-511_221+443.html}{ \hfnewavg{0.10\pm0.02} } 
\\%

 \hfhref{BR_-511_225+443.html}{ \hflabel{J/\psi(1S) f_2(1270)} }  
 &  { \,}  
 &  \hfhref{0506078.html}{ \hfpre{0.10\pm0.04\pm0.02} }  
 &  { \,}  
 &  { \,}  
 &  \hfhref{BR_-511_225+443.html}{ \hfavg{0.10\pm0.04} } 
\\%

 \hfhref{BR_-511_421+443.html}{ \hflabel{J/\psi(1S) D^0} }  
 &  \hfhref{BR_-511_421+443.html}{ \hfpdg{<0.130} }  
 &  \hfhref{0506038.html}{ \hfpub{<0.200} }  
 &  \hfhref{0506003.html}{ \hfpub{<0.130} }  
 &  { \,}  
 &  \hfhref{BR_-511_421+443.html}{ \hfavg{<0.130} } 
\\%

 \hfhref{BR_-511_111+443.html}{ \hflabel{J/\psi(1S) \pi^0} }  
 &  \hfhref{BR_-511_111+443.html}{ \hfpdg{0.22\pm0.04} }  
 &  \hfhref{0506076.html}{ \hfpubold{0.23\pm0.05\pm0.02} }  
 &  \hfhref{0607014.html}{ \hfpubhot{0.19\pm0.02\pm0.02} }  
 &  { \,}  
 &  \hfhref{BR_-511_111+443.html}{ \hfnewavg{0.20\pm0.02} } 
\\
\hline 

\end{tabular}

\end{center}

\end{\hftabletype}
\clearpage  
\begin{\hftabletype}[\hftableposn]

\begin{center}
\hfnewcaption{Ratios of branching fractions}{neutral B}{\hfjpsi}{in units of $10^{0}$}{upper limits are at 90\% CL}{00204.html}{tab:hfc03204}
\hfmetadata{ [ .tml created 2007-04-07T00:53:29.612+08:00] }

\begin{tabular}{lccccc}  
\hflmode
&  \hflpdg2006
&  \hflbelle
&  \hflbabar
&  \hflcdf
&  \hflavg
\\\hline 

 \hfhref{BR_-511_-10313+443xoBR_-521_-321+443.html}{ \hflabel{\frac{{\cal{B}} ( \bar{B}^0 \to J/\psi(1S) \bar{K}_1^0(1270) )}{{\cal{B}} ( B^- \to J/\psi(1S) K^- )}} } \hftabletlcell 
 &  { \,}  
 &  \hfhref{0506072.html}{ \hfpubold{1.30\pm0.34\pm0.28} }  
 &  { \,}  
 &  { \,}  
 &  \hfhref{BR_-511_-10313+443xoBR_-521_-321+443.html}{ \hfavg{1.30\pm0.44} } 
\\%

 \hfhref{BR_-511_-311+441xoBR_-511_-311+443.html}{ \hflabel{\frac{{\cal{B}} ( \bar{B}^0 \to \eta_c(1S) \bar{K}^0 )}{{\cal{B}} ( \bar{B}^0 \to J/\psi(1S) \bar{K}^0 )}} }  
 &  \hfhref{BR_-511_-311+441xoBR_-511_-311+443.html}{ \hfpdg{1.39\pm0.49} }  
 &  { \,}  
 &  \hfhref{0506001.html}{ \hfpub{1.34\pm0.19\pm0.13\pm0.38} }  
 &  { \,}  
 &  \hfhref{BR_-511_-311+441xoBR_-511_-311+443.html}{ \hfavg{1.34\pm0.44} } 
\\%

 \hfhref{BR_-511_-313+443xoBR_-511_-311+443.html}{ \hflabel{\frac{{\cal{B}} ( \bar{B}^0 \to J/\psi(1S) \bar{K}^{*0}(892) )}{{\cal{B}} ( \bar{B}^0 \to J/\psi(1S) \bar{K}^0 )}} } \hftableblcell 
 &  \hfhref{BR_-511_-313+443xoBR_-511_-311+443.html}{ \hfpdg{1.50\pm0.09} }  
 &  { \,}  
 &  \hfhref{0506109.html}{ \hfpub{1.51\pm0.05\pm0.08} }  
 &  \hfhref{0506094.html}{ \hfpubold{1.39\pm0.36\pm0.10} }  
 &  \hfhref{BR_-511_-313+443xoBR_-511_-311+443.html}{ \hfavg{1.50\pm0.09} } 
\\\hline 

\end{tabular}

\end{center}


\begin{center}
\hfnewcaption{Miscellaneous quantities}{neutral B}{\hfjpsi}{in units of $10^{0}$}{upper limits are at 90\% CL}{00204.html}{tab:hfc04204}
\hfmetadata{ [ .tml created 2007-04-07T00:53:37.302+08:00] }

\begin{tabular}{lccccc}  
\hflmode
&  \hflpdg2006
&  \hflbelle
&  \hflbabar
&  \hflcdf
&  \hflavg
\\\hline 

 \hfhref{Azero_-511_313+443xoAzero_511_313+443.html}{ \hflabel{\frac{\vert{\cal{A}}_{0}\vert^{2} ( \bar{B}^0 \to J/\psi(1S) K^{*0}(892) )}{\vert{\cal{A}}_{0}\vert^{2} ( B^0 \to J/\psi(1S) K^{*0}(892) )}} } \hftabletlcell 
 &  { \,}  
 &  { \,}  
 &  \hfhref{0506007.html}{ \hfpub{<0.26} }  
 &  { \,}  
 &  \hfhref{Azero_-511_313+443xoAzero_511_313+443.html}{ \hfavg{<0.26} } 
\\%

 \hfhref{Azero_511_-313+443xoAzero_-511_-313+443.html}{ \hflabel{\frac{\vert{\cal{A}}_{0}\vert^{2} ( B^0 \to J/\psi(1S) \bar{K}^{*0}(892) )}{\vert{\cal{A}}_{0}\vert^{2} ( \bar{B}^0 \to J/\psi(1S) \bar{K}^{*0}(892) )}} } \hftableblcell 
 &  { \,}  
 &  { \,}  
 &  \hfhref{0506007.html}{ \hfpub{<0.32} }  
 &  { \,}  
 &  \hfhref{Azero_511_-313+443xoAzero_-511_-313+443.html}{ \hfavg{<0.32} } 
\\\hline 

\end{tabular}

\end{center}

\end{\hftabletype}
\clearpage  
\begin{\hftabletype}[\hftableposn]

\begin{center}
\hfnewcaption{Branching fractions}{neutral B}{\hfochm}{in units of $10^{-3}$}{upper limits are at 90\% CL}{00205.html}{tab:hfc01205}
\hfmetadata{ [ .tml created 2007-04-07T01:15:07.927+08:00] }

\begin{tabular}{lccccc}  
\hflmode
&  \hflpdg2006
&  \hflbelle
&  \hflbabar
&  \hflcdf
&  \hflavg
\\\hline 

 \hfhref{BR_-511_-313+445.html}{ \hflabel{\chi_{c2}(1P) \bar{K}^{*0}(892)} } \hftabletlcell 
 &  \hfhref{BR_-511_-313+445.html}{ \hfpdg{<0.036} }  
 &  { \,}  
 &  \hfhref{0506006.html}{ \hfpub{<0.036} }  
 &  { \,}  
 &  \hfhref{BR_-511_-313+445.html}{ \hfavg{<0.036} } 
\\%

 \hfhref{BR_-511_-311+445.html}{ \hflabel{\chi_{c2}(1P) \bar{K}^0} }  
 &  \hfhref{BR_-511_-311+445.html}{ \hfpdg{<0.026} }  
 &  { \,}  
 &  \hfhref{0506006.html}{ \hfpub{<0.041} }  
 &  { \,}  
 &  \hfhref{BR_-511_-311+445.html}{ \hfavg{<0.041} } 
\\%

 \hfhref{BR_-511_-313+20443.html}{ \hflabel{\chi_{c1}(1P) \bar{K}^{*0}(892)} }  
 &  \hfhref{BR_-511_-313+20443.html}{ \hfpdg{0.32\pm0.06} }  
 &  \hfhref{0607004.html}{ \hfpubhot{0.31\pm0.03\pm0.07} }  
 &  \hfhref{0506109.html}{ \hfpub{0.33\pm0.04\pm0.05\pm0.03} }  
 &  { \,}  
 &  \hfhref{BR_-511_-313+20443.html}{ \hfnewavg{0.32\pm0.05} } 
\\%

 \hfhref{BR_-511_-311+20443.html}{ \hflabel{\chi_{c1}(1P) \bar{K}^0} }  
 &  \hfhref{BR_-511_-311+20443.html}{ \hfpdg{0.39\pm0.04} }  
 &  \hfhref{0607004.html}{ \hfpubhot{0.35\pm0.03\pm0.05} }  
 &  \hfhref{0506109.html}{ \hfpub{0.45\pm0.04\pm0.02\pm0.05} }  
 &  { \,}  
 &  \hfhref{BR_-511_-311+20443.html}{ \hfnewavg{0.40\pm0.04} } 
\\%

 \hfhref{BR_-511_-311+100443.html}{ \hflabel{\psi(2S) \bar{K}^0} }  
 &  \hfhref{BR_-511_-311+100443.html}{ \hfpdg{0.62\pm0.06} }  
 &  \hfhref{0506076.html}{ \hfpubold{0.67\pm0.11} }  
 &  \hfhref{0506109.html}{ \hfpub{0.65\pm0.06\pm0.04\pm0.02} }  
 &  { \,}  
 &  \hfhref{BR_-511_-311+100443.html}{ \hfavg{0.65\pm0.07} } 
\\
\hline 

\end{tabular}

\end{center}


\begin{center}
\hfnewcaption{Ratios of branching fractions}{neutral B}{\hfochm}{in units of $10^{0}$}{upper limits are at 90\% CL}{00205.html}{tab:hfc03205}
\hfmetadata{ [ .tml created 2007-04-07T01:15:20.335+08:00] }

\begin{tabular}{lccccc}  
\hflmode
&  \hflpdg2006
&  \hflbelle
&  \hflbabar
&  \hflcdf
&  \hflavg
\\\hline 

 \hfhref{BR_-511_-313+20443xoBR_-511_-311+20443.html}{ \hflabel{\frac{{\cal{B}} ( \bar{B}^0 \to \chi_{c1}(1P) \bar{K}^{*0}(892) )}{{\cal{B}} ( \bar{B}^0 \to \chi_{c1}(1P) \bar{K}^0 )}} } \hftabletlcell 
 &  \hfhref{BR_-511_-313+20443xoBR_-511_-311+20443.html}{ \hfpdg{0.72\pm0.16} }  
 &  { \,}  
 &  \hfhref{0506109.html}{ \hfpub{0.72\pm0.11\pm0.12} }  
 &  { \,}  
 &  \hfhref{BR_-511_-313+20443xoBR_-511_-311+20443.html}{ \hfavg{0.72\pm0.16} } 
\\%

 \hfhref{BR_-511_-311+441xoBR_-521_-321+441.html}{ \hflabel{\frac{{\cal{B}} ( \bar{B}^0 \to \eta_c(1S) \bar{K}^0 )}{{\cal{B}} ( B^- \to \eta_c(1S) K^- )}} }  
 &  { \,}  
 &  { \,}  
 &  \hfhref{0506001.html}{ \hfpub{0.87\pm0.13\pm0.07} }  
 &  { \,}  
 &  \hfhref{BR_-511_-311+441xoBR_-521_-321+441.html}{ \hfavg{0.87\pm0.15} } 
\\%

 \hfhref{BR_-511_-313+100443xoBR_-511_-311+100443.html}{ \hflabel{\frac{{\cal{B}} ( \bar{B}^0 \to \psi(2S) \bar{K}^{*0}(892) )}{{\cal{B}} ( \bar{B}^0 \to \psi(2S) \bar{K}^0 )}} }  
 &  \hfhref{BR_-511_-313+100443xoBR_-511_-311+100443.html}{ \hfpdg{1.00\pm0.17} }  
 &  { \,}  
 &  \hfhref{0506109.html}{ \hfpub{1.00\pm0.14\pm0.09} }  
 &  { \,}  
 &  \hfhref{BR_-511_-313+100443xoBR_-511_-311+100443.html}{ \hfavg{1.00\pm0.17} } 
\\%

 \hfhref{BR_-511_-313+441xoBR_-511_-311+441.html}{ \hflabel{\frac{{\cal{B}} ( \bar{B}^0 \to \eta_c(1S) \bar{K}^{*0}(892) )}{{\cal{B}} ( \bar{B}^0 \to \eta_c(1S) \bar{K}^0 )}} } \hftableblcell 
 &  \hfhref{BR_-511_-313+441xoBR_-511_-311+441.html}{ \hfpdg{1.30\pm0.40} }  
 &  \hfhref{0506074.html}{ \hfpubold{1.33\pm0.36\pm^{0.24}_{0.33}} }  
 &  { \,}  
 &  { \,}  
 &  \hfhref{BR_-511_-313+441xoBR_-511_-311+441.html}{ \hfavg{1.33\pm^{0.43}_{0.49}} } 
\\\hline 

\end{tabular}

\end{center}

\end{\hftabletype}
\clearpage  
\begin{\hftabletype}[\hftableposn]

\begin{center}
\hfnewcaption{Branching fractions}{neutral B}{\hfmuld}{in units of $10^{-3}$}{upper limits are at 90\% CL}{00206.html}{tab:hfc01206}
\hfmetadata{ [ .tml created 2007-04-07T01:29:58.176+08:00] }

\begin{tabular}{lccccc}  
\hflmode
&  \hflpdg2006
&  \hflbelle
&  \hflbabar
&  \hflcdf
&  \hflavg
\\\hline 

 \hfhref{BR_-511_-421+421.html}{ \hflabel{D^0 \bar{D}^0} } \hftabletlcell 
 &  { \,}  
 &  { \,}  
 &  \hfhref{0604002.html}{ \hfpubhot{<0.060} }  
 &  { \,}  
 &  \hfhref{BR_-511_-421+421.html}{ \hfavg{<0.060} } 
\\%

 \hfhref{BR_-511_-423+423.html}{ \hflabel{D^{*0}(2007) \bar{D}^{*0}(2007)} }  
 &  \hfhref{BR_-511_-423+423.html}{ \hfpdg{<27} }  
 &  { \,}  
 &  \hfhref{0604002.html}{ \hfpubhot{<0.090} }  
 &  { \,}  
 &  \hfhref{BR_-511_-423+423.html}{ \hfavg{<0.090} } 
\\%

 \hfhref{BR_-511_-421+-311+111+421.html}{ \hflabel{D^0 \bar{D}^0 \pi^0 \bar{K}^0} }  
 &  { \,}  
 &  \hfhref{0607010.html}{ \hfpubhot{0.17\pm0.07\pm^{0.03}_{0.05}} }  
 &  { \,}  
 &  { \,}  
 &  \hfhref{BR_-511_-421+-311+111+421.html}{ \hfnewavg{0.17\pm0.08} } 
\\%

 \hfhref{BR_-511_-423+421.html}{ \hflabel{D^0 \bar{D}^{*0}(2007)} }  
 &  { \,}  
 &  { \,}  
 &  \hfhref{0604002.html}{ \hfpubhot{<0.29} }  
 &  { \,}  
 &  \hfhref{BR_-511_-423+421.html}{ \hfavg{<0.29} } 
\\%

 \hfhref{BR_-511_-411+411.html}{ \hflabel{D^- D^+} }  
 &  \hfhref{BR_-511_-411+411.html}{ \hfpdg{0.19\pm0.06} }  
 &  \hfhref{0506037.html}{ \hfpub{0.32\pm0.06\pm0.05} }  
 &  \hfhref{0604002.html}{ \hfpubhot{0.28\pm0.04\pm0.03\pm0.04} }  
 &  { \,}  
 &  \hfhref{BR_-511_-411+411.html}{ \hfavg{0.30\pm0.05} } 
\\%

 \hfhref{BR_-511_-413+411.html}{ \hflabel{D^{*-}(2010) D^+} }  
 &  \hfhref{BR_-511_-413+411.html}{ \hfpdg{<0.63} }  
 &  \hfhref{0506084.html}{ \hfpubold{1.17\pm0.26\pm^{0.20}_{0.24}\pm0.08} }  
 &  \hfhref{0604002.html}{ \hfpubhot{0.57\pm0.07\pm0.06\pm0.04} }  
 &  { \,}  
 &  \hfhref{BR_-511_-413+411.html}{ \hfavg{0.62\pm0.09} } 
\\%

 \hfhref{BR_-511_-413+413.html}{ \hflabel{D^{*+}(2010) D^{*-}(2010)} }  
 &  \hfhref{BR_-511_-413+413.html}{ \hfpdg{0.83\pm0.11} }  
 &  { \,}  
 &  { \,}  
 &  { \,}  
 &  \hfhref{BR_-511_-413+413.html}{ \hfavg{0.81\pm0.08} } 
\\
\hline 

\end{tabular}

\begin{description}\addtolength{\itemsep}{\hffootitemsep\baselineskip}

	  \item[ \hffootnotemark{1} ] \hffootnotetext{ Measurement of Branching Fraction and CP-violating charge
asymmetried for B meson decays to $D^{(*)}D^{(*)}$ and 
implications for the CKM angle $\gamma$ (\hfbb{232}) } 
     ;  \hffootnotetext{ $\bar{B}^0 \rightarrow D^{*+}D^{*-}$ } 
	  \item[ \hffootnotemark{2} ] \hffootnotetext{ Measurement of the branching fraction and CP content for the 
decay $B^0$ to $D^*D^*$ (\hfbb{23}) } 
     ;  \hffootnotetext{ $\bar{B}^0 \rightarrow D^{*-}D^{*+}$ } 
\end{description}

\end{center}

\end{\hftabletype}
\clearpage  
\begin{\hftabletype}[\hftableposn]

\begin{center}
\hfnewcaption{Product branching fractions}{neutral B}{\hfmuld}{in units of $10^{-4}$}{upper limits are at 90\% CL}{00206.html}{tab:hfc02206}
\hfmetadata{ [ .tml created 2007-04-07T01:30:28.496+08:00] }

\begin{tabular}{lccccc}  
\hflmode
&  \hflpdg2006
&  \hflbelle
&  \hflbabar
&  \hflcdf
&  \hflavg
\\\hline 

 \hfhref{BR_-511_-321+415xBR_415_211+421.html}{ \hflabel{K^- D_2^{*+}(2460) [ D^0 \pi^+ ]} } \hftabletlcell 
 &  \hfhref{BR_-511_-321+415xBR_415_211+421.html}{ \hfpdg{0.18\pm0.05} }  
 &  { \,}  
 &  \hfhref{0506018.html}{ \hfpubhot{0.18\pm0.04\pm0.03} }  
 &  { \,}  
 &  \hfhref{BR_-511_-321+415xBR_415_211+421.html}{ \hfavg{0.18\pm0.05} } 
\\%

 \hfhref{BR_-511_-211+415xBR_415_-211+211+413.html}{ \hflabel{\pi^- D_2^{*+}(2460) [ D^{*+}(2010) \pi^- \pi^+ ]} }  
 &  \hfhref{BR_-511_-211+415xBR_415_-211+211+413.html}{ \hfpdg{<0.24} }  
 &  \hfhref{0506030.html}{ \hfpub{<0.24} }  
 &  { \,}  
 &  { \,}  
 &  \hfhref{BR_-511_-211+415xBR_415_-211+211+413.html}{ \hfavg{<0.24} } 
\\%

 \hfhref{BR_-511_-211+10413xBR_10413_-211+211+413.html}{ \hflabel{\pi^- D_1^+(2420) [ D^{*+}(2010) \pi^- \pi^+ ]} }  
 &  \hfhref{BR_-511_-211+10413xBR_10413_-211+211+413.html}{ \hfpdg{<0.33} }  
 &  \hfhref{0506030.html}{ \hfpub{<0.33} }  
 &  { \,}  
 &  { \,}  
 &  \hfhref{BR_-511_-211+10413xBR_10413_-211+211+413.html}{ \hfavg{<0.33} } 
\\%

 \hfhref{BR_-511_-211+20413xBR_20413_211+423.html}{ \hflabel{\pi^- D_1^+(H) [ D^{*0}(2007) \pi^+ ]} }  
 &  { \,}  
 &  \hfhref{0506070.html}{ \hfpre{<0.70} }  
 &  { \,}  
 &  { \,}  
 &  \hfhref{BR_-511_-211+20413xBR_20413_211+423.html}{ \hfavg{<0.70} } 
\\%

 \hfhref{BR_-511_-211+10413xBR_10413_-211+211+411.html}{ \hflabel{\pi^- D_1^+(2420) [ D^+ \pi^- \pi^+ ]} }  
 &  \hfhref{BR_-511_-211+10413xBR_10413_-211+211+411.html}{ \hfpdg{0.89\pm0.29} }  
 &  \hfhref{0506030.html}{ \hfpub{0.89\pm0.15\pm0.17\pm^{0.00}_{0.26}} }  
 &  { \,}  
 &  { \,}  
 &  \hfhref{BR_-511_-211+10413xBR_10413_-211+211+411.html}{ \hfavg{0.89\pm^{0.23}_{0.34}} } 
\\%

 \hfhref{BR_-511_-211+10411xBR_10411_211+421.html}{ \hflabel{\pi^- D_0^{*+} [ D^0 \pi^+ ]} }  
 &  { \,}  
 &  \hfhref{0506070.html}{ \hfpre{<1.20} }  
 &  { \,}  
 &  { \,}  
 &  \hfhref{BR_-511_-211+10411xBR_10411_211+421.html}{ \hfavg{<1.20} } 
\\%

 \hfhref{BR_-511_-211+415xBR_415_211+423.html}{ \hflabel{\pi^- D_2^{*+}(2460) [ D^{*0}(2007) \pi^+ ]} }  
 &  { \,}  
 &  \hfhref{0506070.html}{ \hfpre{2.45\pm0.42\pm^{0.35}_{0.45}\pm^{0.39}_{0.17}} }  
 &  { \,}  
 &  { \,}  
 &  \hfhref{BR_-511_-211+415xBR_415_211+423.html}{ \hfavg{2.45\pm^{0.67}_{0.64}} } 
\\%

 \hfhref{BR_-511_-211+415xBR_415_211+421.html}{ \hflabel{\pi^- D_2^{*+}(2460) [ D^0 \pi^+ ]} }  
 &  { \,}  
 &  \hfhref{0506070.html}{ \hfpre{3.08\pm0.33\pm0.09\pm^{0.15}_{0.02}} }  
 &  { \,}  
 &  { \,}  
 &  \hfhref{BR_-511_-211+415xBR_415_211+421.html}{ \hfavg{3.08\pm^{0.37}_{0.34}} } 
\\%

 \hfhref{BR_-511_-211+10413xBR_10413_211+423.html}{ \hflabel{\pi^- D_1^+(2420) [ D^{*0}(2007) \pi^+ ]} }  
 &  { \,}  
 &  \hfhref{0506070.html}{ \hfpre{3.68\pm0.60\pm^{0.71}_{0.40}\pm^{0.65}_{0.30}} }  
 &  { \,}  
 &  { \,}  
 &  \hfhref{BR_-511_-211+10413xBR_10413_211+423.html}{ \hfavg{3.68\pm^{1.13}_{0.78}} } 
\\%

 \hfhref{BR_-511_223+20423xBR_20423_-211+413.html}{ \hflabel{\omega(782) D_1^0(H) [ D^{*+}(2010) \pi^- ]} } \hftableblcell 
 &  { \,}  
 &  { \,}  
 &  \hfhref{0603002.html}{ \hfpubhot{4.10\pm1.20\pm1.00\pm0.40} }  
 &  { \,}  
 &  \hfhref{BR_-511_223+20423xBR_20423_-211+413.html}{ \hfavg{4.1\pm1.6} } 
\\\hline 

\end{tabular}

\end{center}


\begin{center}
\hfnewcaption{Ratios of branching fractions}{neutral B}{\hfmuld}{in units of $10^{0}$}{upper limits are at 90\% CL}{00206.html}{tab:hfc03206}
\hfmetadata{ [ .tml created 2007-04-07T01:30:44.936+08:00] }

\begin{tabular}{lccccc}  
\hflmode
&  \hflpdg2006
&  \hflbelle
&  \hflbabar
&  \hflcdf
&  \hflavg
\\\hline 

 \hfhref{BR_-511_-321+411xoBR_-511_-211+411.html}{ \hflabel{\frac{{\cal{B}} ( \bar{B}^0 \to D^+ K^- )}{{\cal{B}} ( \bar{B}^0 \to D^+ \pi^- )}} } \hftabletlcell 
 &  \hfhref{BR_-511_-321+411xoBR_-511_-211+411.html}{ \hfpdg{0.07\pm0.02} }  
 &  \hfhref{0506081.html}{ \hfpubold{0.068\pm0.015\pm0.007} }  
 &  { \,}  
 &  { \,}  
 &  \hfhref{BR_-511_-321+411xoBR_-511_-211+411.html}{ \hfavg{0.07\pm0.02} } 
\\%

 \hfhref{BR_-511_-321+413xoBR_-511_-211+413.html}{ \hflabel{\frac{{\cal{B}} ( \bar{B}^0 \to D^{*+}(2010) K^- )}{{\cal{B}} ( \bar{B}^0 \to D^{*+}(2010) \pi^- )}} }  
 &  \hfhref{BR_-511_-321+413xoBR_-511_-211+413.html}{ \hfpdg{0.07\pm0.02} }  
 &  \hfhref{0506081.html}{ \hfpubold{0.074\pm0.015\pm0.006} }  
 &  \hfhref{0506018.html}{ \hfpubhot{0.078\pm0.003\pm0.003} }  
 &  { \,}  
 &  \hfhref{BR_-511_-321+413xoBR_-511_-211+413.html}{ \hfavg{0.077\pm0.004} } 
\\%

 \hfhref{BR_-511_-211+92xoBR_-511_-211+411.html}{ \hflabel{\frac{{\cal{B}} ( \bar{B}^0 \to D^{**0} \pi^- )}{{\cal{B}} ( \bar{B}^0 \to D^+ \pi^- )}} }  
 &  { \,}  
 &  { \,}  
 &  \hfhref{0607016.html}{ \hfpubhot{0.77\pm0.22\pm0.29} }  
 &  { \,}  
 &  \hfhref{BR_-511_-211+92xoBR_-511_-211+411.html}{ \hfnewavg{0.77\pm0.36} } 
\\%

 \hfhref{BR_-511_-211+413xoBR_-511_-211+411.html}{ \hflabel{\frac{{\cal{B}} ( \bar{B}^0 \to D^{*+}(2010) \pi^- )}{{\cal{B}} ( \bar{B}^0 \to D^+ \pi^- )}} }  
 &  { \,}  
 &  { \,}  
 &  \hfhref{0607016.html}{ \hfpubhot{0.99\pm0.11\pm0.08} }  
 &  { \,}  
 &  \hfhref{BR_-511_-211+413xoBR_-511_-211+411.html}{ \hfnewavg{0.99\pm0.14} } 
\\%

 \hfhref{BR_-511_113+421xoBR_-511_223+421.html}{ \hflabel{\frac{{\cal{B}} ( \bar{B}^0 \to D^0 \rho^0(770) )}{{\cal{B}} ( \bar{B}^0 \to D^0 \omega(782) )}} }  
 &  { \,}  
 &  \hfhref{0506088.html}{ \hfpubold{1.60\pm0.80} }  
 &  { \,}  
 &  { \,}  
 &  \hfhref{BR_-511_113+421xoBR_-511_223+421.html}{ \hfavg{1.60\pm0.80} } 
\\%

 \hfhref{BR_-511_-14+13+411xoBR_-511_-211+411.html}{ \hflabel{\frac{{\cal{B}} ( \bar{B}^0 \to D^+ \mu^- \bar{\nu}_\mu )}{{\cal{B}} ( \bar{B}^0 \to D^+ \pi^- )}} }  
 &  { \,}  
 &  { \,}  
 &  { \,}  
 &  \hfhref{0506102.html}{ \hfwaiting{9.80\pm1.00\pm0.60\pm1.20} }  
 &  \hfhref{BR_-511_-14+13+411xoBR_-511_-211+411.html}{ \hfavg{9.8\pm1.7} } 
\\%

 \hfhref{BR_-511_-14+13+413xoBR_-511_-211+413.html}{ \hflabel{\frac{{\cal{B}} ( \bar{B}^0 \to D^{*+}(2010) \mu^- \bar{\nu}_\mu )}{{\cal{B}} ( \bar{B}^0 \to D^{*+}(2010) \pi^- )}} } \hftableblcell 
 &  { \,}  
 &  { \,}  
 &  { \,}  
 &  \hfhref{0506102.html}{ \hfwaiting{17.70\pm2.30\pm0.60\pm1.20} }  
 &  \hfhref{BR_-511_-14+13+413xoBR_-511_-211+413.html}{ \hfavg{17.7\pm2.7} } 
\\\hline 

\end{tabular}

\end{center}

\end{\hftabletype}
\clearpage  
\begin{\hftabletype}[\hftableposn]

\begin{center}
\hfnewcaption{Branching fractions}{neutral B}{\hfsgdx}{in units of $10^{-4}$}{upper limits are at 90\% CL}{00207.html}{tab:hfc01207}
\hfmetadata{ [ .tml created 2007-04-07T01:55:21.097+08:00] }

\begin{tabular}{lccccc}  
\hflmode
&  \hflpdg2006
&  \hflbelle
&  \hflbabar
&  \hflcdf
&  \hflavg
\\\hline 

 \hfhref{BR_-511_-311+423.html}{ \hflabel{D^{*0}(2007) \bar{K}^0} } \hftabletlcell 
 &  \hfhref{BR_-511_-311+423.html}{ \hfpdg{<0.66} }  
 &  { \,}  
 &  { \,}  
 &  { \,}  
 &  \hfhref{BR_-511_-311+423.html}{ \hfavg{0.36\pm0.12} } 
\\%

{ \hflabel{\,} }  
 &  { \,}  
 &  \hfhref{0506085.html}{ \hfpubold{<0.66} }  
 &  \hfhref{0506023.html}{ \hfsuperceeded{0.45\pm0.19\pm0.05} } 
\hffootnotemark{2}
 
 &  { \,}  
 &  { \,} 
\\%

{ \hflabel{\,} }  
 &  { \,}  
 &  { \,}  
 &  \hfhref{0603004.html}{ \hfpubhot{0.36\pm0.12\pm0.03} } 
\hffootnotemark{1}
 
 &  { \,}  
 &  { \,} 
\\%

 \hfhref{BR_-511_-423+-313.html}{ \hflabel{\bar{D}^{*0}(2007) \bar{K}^{*0}(892)} }  
 &  \hfhref{BR_-511_-423+-313.html}{ \hfpdg{<0.40} }  
 &  \hfhref{0506085.html}{ \hfpubold{<0.40} }  
 &  { \,}  
 &  { \,}  
 &  \hfhref{BR_-511_-423+-313.html}{ \hfavg{<0.40} } 
\\%

 \hfhref{BR_-511_-313+423.html}{ \hflabel{D^{*0}(2007) \bar{K}^{*0}(892)} }  
 &  \hfhref{BR_-511_-313+423.html}{ \hfpdg{<0.69} }  
 &  \hfhref{0506085.html}{ \hfpubold{<0.69} }  
 &  { \,}  
 &  { \,}  
 &  \hfhref{BR_-511_-313+423.html}{ \hfavg{<0.69} } 
\\%

 \hfhref{BR_-511_331+423.html}{ \hflabel{D^{*0}(2007) \eta^\prime(958)} }  
 &  \hfhref{BR_-511_331+423.html}{ \hfpdg{1.23\pm0.35} }  
 &  \hfhref{0506035.html}{ \hfpub{1.21\pm0.34\pm0.22} }  
 &  \hfhref{0506106.html}{ \hfpub{<2.6} }  
 &  { \,}  
 &  \hfhref{BR_-511_331+423.html}{ \hfavg{1.21\pm0.40} } 
\\%

 \hfhref{BR_-511_111+423.html}{ \hflabel{D^{*0}(2007) \pi^0} }  
 &  \hfhref{BR_-511_111+423.html}{ \hfpdg{2.70\pm0.50} }  
 &  \hfhref{0607017.html}{ \hfpubhot{1.39\pm0.18\pm0.26} }  
 &  \hfhref{0506106.html}{ \hfpub{2.90\pm0.40\pm0.46\pm0.19} }  
 &  { \,}  
 &  \hfhref{BR_-511_111+423.html}{ \hfnewavg{1.69\pm0.28} } 
\\%

 \hfhref{BR_-511_221+423.html}{ \hflabel{D^{*0}(2007) \eta} }  
 &  \hfhref{BR_-511_221+423.html}{ \hfpdg{2.60\pm0.60} }  
 &  \hfhref{0607017.html}{ \hfpubhot{1.40\pm0.28\pm0.26} }  
 &  \hfhref{0506106.html}{ \hfpub{2.60\pm0.40\pm0.37\pm0.16} }  
 &  { \,}  
 &  \hfhref{BR_-511_221+423.html}{ \hfnewavg{1.77\pm0.32} } 
\\%

 \hfhref{BR_-511_225+423.html}{ \hflabel{f_2(1270) D^{*0}(2007)} }  
 &  { \,}  
 &  \hfhref{0506070.html}{ \hfpre{1.86\pm0.65\pm0.60\pm^{0.80}_{0.52}} }  
 &  { \,}  
 &  { \,}  
 &  \hfhref{BR_-511_225+423.html}{ \hfavg{1.9\pm^{1.2}_{1.0}} } 
\\%

 \hfhref{BR_-511_-321+413.html}{ \hflabel{D^{*+}(2010) K^-} }  
 &  \hfhref{BR_-511_-321+413.html}{ \hfpdg{2.14\pm0.20} }  
 &  \hfhref{0506081.html}{ \hfpubold{2.04\pm0.41\pm0.17\pm0.16} }  
 &  { \,}  
 &  { \,}  
 &  \hfhref{BR_-511_-321+413.html}{ \hfavg{2.04\pm0.47} } 
\\%

 \hfhref{BR_-511_223+423.html}{ \hflabel{D^{*0}(2007) \omega(782)} }  
 &  \hfhref{BR_-511_223+423.html}{ \hfpdg{4.2\pm1.1} }  
 &  \hfhref{0607017.html}{ \hfpubhot{2.29\pm0.39\pm0.40} }  
 &  \hfhref{0506106.html}{ \hfpub{4.20\pm0.70\pm0.86\pm0.27} }  
 &  { \,}  
 &  \hfhref{BR_-511_223+423.html}{ \hfnewavg{2.66\pm0.50} } 
\\%

 \hfhref{BR_-511_-211+311+413.html}{ \hflabel{D^{*+}(2010) K^0 \pi^-} }  
 &  \hfhref{BR_-511_-211+311+413.html}{ \hfpdg{3.00\pm0.80} }  
 &  { \,}  
 &  \hfhref{0506026.html}{ \hfpub{3.00\pm0.70\pm0.22\pm0.20} }  
 &  { \,}  
 &  \hfhref{BR_-511_-211+311+413.html}{ \hfavg{3.00\pm0.76} } 
\\%

 \hfhref{BR_-511_-323+413.html}{ \hflabel{D^{*+}(2010) K^{*-}(892)} }  
 &  \hfhref{BR_-511_-323+413.html}{ \hfpdg{3.30\pm0.60} }  
 &  { \,}  
 &  \hfhref{0506026.html}{ \hfpub{3.20\pm0.60\pm0.27\pm0.12} }  
 &  { \,}  
 &  \hfhref{BR_-511_-323+413.html}{ \hfavg{3.20\pm0.67} } 
\\%

 \hfhref{BR_-511_113+423.html}{ \hflabel{D^{*0}(2007) \rho^0(770)} }  
 &  \hfhref{BR_-511_113+423.html}{ \hfpdg{<5.1} }  
 &  { \,}  
 &  { \,}  
 &  { \,}  
 &  \hfhref{BR_-511_113+423.html}{ \hfavg{3.73\pm0.99} } 
\\%

{ \hflabel{\,} }  
 &  { \,}  
 &  \hfhref{0506070.html}{ \hfpre{3.73\pm0.87\pm0.46\pm^{0.18}_{0.08}} } 
\hffootnotemark{4}
 
 &  { \,}  
 &  { \,}  
 &  { \,} 
\\%

{ \hflabel{\,} }  
 &  { \,}  
 &  \hfhref{0506088.html}{ \hfpubold{<5.1} } 
\hffootnotemark{3}
 
 &  { \,}  
 &  { \,}  
 &  { \,} 
\\%

 \hfhref{BR_-511_-321+311+413.html}{ \hflabel{D^{*+}(2010) K^- K^0} }  
 &  \hfhref{BR_-511_-321+311+413.html}{ \hfpdg{<4.7} }  
 &  \hfhref{0506090.html}{ \hfpubold{<4.7} }  
 &  { \,}  
 &  { \,}  
 &  \hfhref{BR_-511_-321+311+413.html}{ \hfavg{<4.7} } 
\\%

 \hfhref{BR_-511_-211+211+423.html}{ \hflabel{D^{*0}(2007) \pi^+ \pi^-} }  
 &  \hfhref{BR_-511_-211+211+423.html}{ \hfpdg{6.2\pm2.2} }  
 &  { \,}  
 &  { \,}  
 &  { \,}  
 &  \hfhref{BR_-511_-211+211+423.html}{ \hfavg{9.0\pm1.4} } 
\\%

{ \hflabel{\,} }  
 &  { \,}  
 &  \hfhref{0506088.html}{ \hfpubold{6.2\pm1.2\pm1.8} } 
\hffootnotemark{3}
 
 &  { \,}  
 &  { \,}  
 &  { \,} 
\\%

{ \hflabel{\,} }  
 &  { \,}  
 &  \hfhref{0506070.html}{ \hfpre{10.90\pm0.80\pm1.60} } 
\hffootnotemark{4}
 
 &  { \,}  
 &  { \,}  
 &  { \,} 
\\%

 \hfhref{BR_-511_-321+313+413.html}{ \hflabel{D^{*+}(2010) K^- K^{*0}(892)} }  
 &  \hfhref{BR_-511_-321+313+413.html}{ \hfpdg{12.9\pm3.3} }  
 &  \hfhref{0506090.html}{ \hfpubold{12.9\pm2.2\pm2.5} }  
 &  { \,}  
 &  { \,}  
 &  \hfhref{BR_-511_-321+313+413.html}{ \hfavg{12.9\pm3.3} } 
\\%

 \hfhref{BR_-511_-211+92.html}{ \hflabel{D^{**0} \pi^-} }  
 &  { \,}  
 &  { \,}  
 &  \hfhref{0607016.html}{ \hfpubhot{23.4\pm6.5\pm8.8} }  
 &  { \,}  
 &  \hfhref{BR_-511_-211+92.html}{ \hfnewavg{23\pm11} } 
\\%

 \hfhref{BR_-511_-211+-211+211+211+423.html}{ \hflabel{D^{*0}(2007) \pi^- \pi^+ \pi^- \pi^+} }  
 &  \hfhref{BR_-511_-211+-211+211+211+423.html}{ \hfpdg{27.0\pm5.0} }  
 &  \hfhref{0506028.html}{ \hfpub{26.0\pm4.7\pm3.7} }  
 &  { \,}  
 &  { \,}  
 &  \hfhref{BR_-511_-211+-211+211+211+423.html}{ \hfavg{26.0\pm6.0} } 
\\%

 \hfhref{BR_-511_-211+413.html}{ \hflabel{D^{*+}(2010) \pi^-} }  
 &  \hfhref{BR_-511_-211+413.html}{ \hfpdg{27.6\pm2.1} }  
 &  { \,}  
 &  { \,}  
 &  { \,}  
 &  \hfhref{BR_-511_-211+413.html}{ \hfnewavg{26.2\pm1.3} } 
\\%

{ \hflabel{\,} }  
 &  { \,}  
 &  \hfhref{0506070.html}{ \hfpre{23.00\pm0.60\pm1.90} }  
 &  \hfhref{0607012.html}{ \hfpubhot{27.90\pm0.80\pm1.70\pm0.05} } 
\hffootnotemark{5}
 
 &  { \,}  
 &  { \,} 
\\%

{ \hflabel{\,} }  
 &  { \,}  
 &  { \,}  
 &  \hfhref{0607016.html}{ \hfpubhot{29.9\pm2.3\pm2.4} } 
\hffootnotemark{6}
 
 &  { \,}  
 &  { \,} 
\\%

 \hfhref{BR_-511_-211+223+413.html}{ \hflabel{D^{*+}(2010) \omega(782) \pi^-} }  
 &  { \,}  
 &  { \,}  
 &  \hfhref{0603002.html}{ \hfpubhot{28.8\pm2.1\pm2.8\pm1.4} }  
 &  { \,}  
 &  \hfhref{BR_-511_-211+223+413.html}{ \hfavg{28.8\pm3.8} } 
\\%

 \hfhref{BR_-511_-211+-211+-211+211+211+413.html}{ \hflabel{D^{*+}(2010) \pi^- \pi^+ \pi^- \pi^+ \pi^-} }  
 &  { \,}  
 &  \hfhref{0506028.html}{ \hfpub{47.2\pm5.9\pm7.1} }  
 &  { \,}  
 &  { \,}  
 &  \hfhref{BR_-511_-211+-211+-211+211+211+413.html}{ \hfavg{47.2\pm9.2} } 
\\%

 \hfhref{BR_-511_-211+-211+211+413.html}{ \hflabel{D^{*+}(2010) \pi^- \pi^+ \pi^-} } \hftableblcell 
 &  \hfhref{BR_-511_-211+-211+211+413.html}{ \hfpdg{76\pm18} }  
 &  \hfhref{0506028.html}{ \hfpub{68.1\pm2.3\pm7.2} }  
 &  { \,}  
 &  { \,}  
 &  \hfhref{BR_-511_-211+-211+211+413.html}{ \hfavg{68.1\pm7.6} } 
\\\hline 

\end{tabular}

\begin{description}\addtolength{\itemsep}{\hffootitemsep\baselineskip}

	  \item[ \hffootnotemark{1} ] \hffootnotetext{ A study of the  $\bar{B}^0 \rightarrow D^{(*)0} K^{(*)0}$ decays (\hfbb{226}) } 
     ;  \hffootnotetext{ $\bar{B}^0 \rightarrow D^{*0} \bar{K}^{0}$ } 
	  \item[ \hffootnotemark{2} ] \hffootnotetext{ A study of the  $\bar{B}^0 \rightarrow D^{(*)0} K^{(*)0}$ decays (\hfbb{124}) } 
     ;  \hffootnotetext{ $B \rightarrow D^{*0} \bar{K}^{0}$ } 
	  \item[ \hffootnotemark{3} ] \hffootnotetext{ Study of $\bar{B^{0}} \to D^{(*) 0} \pi^+ \pi^-$ Decays (\hfbb{31.3}) } 
    
	  \item[ \hffootnotemark{4} ] \hffootnotetext{ Study of $\bar{B}^{0} \to  D^{(*)0} \pi^{+} \pi^{-}$ decays ; Dalitz fit analysis (\hfbb{152}) } 
    
	  \item[ \hffootnotemark{5} ] \hffootnotetext{ Branching fraction measurements and isospin analyses for $\bar{B} \rightarrow D^{(*)}\pi^-$
decays (\hfbb{65}) } 
     ;  \hffootnotetext{ $\bar{B}^0 \rightarrow D^{*+}\pi^-$ } 
	  \item[ \hffootnotemark{6} ] \hffootnotetext{ Measurement of the Absolute Branching Fractions $B \rightarrow D^{(*,**)}\pi$ with a Missing Mass method (\hfbb{231}) } 
     ;  \hffootnotetext{ $\bar{B}^0 \rightarrow D^{*+}\pi^-$ } 
\end{description}

\end{center}

\end{\hftabletype}
\clearpage  
\begin{\hftabletype}[\hftableposn]

\begin{center}
\hfnewcaption{Branching fractions}{neutral B}{\hfsgld}{in units of $10^{-4}$}{upper limits are at 90\% CL}{00208.html}{tab:hfc01208}
\hfmetadata{ [ .tml created 2007-04-07T02:09:33.683+08:00] }

\begin{tabular}{lccccc}  
\hflmode
&  \hflpdg2006
&  \hflbelle
&  \hflbabar
&  \hflcdf
&  \hflavg
\\\hline 

 \hfhref{BR_-511_-421+-313.html}{ \hflabel{\bar{D}^0 \bar{K}^{*0}(892)} } \hftabletlcell 
 &  \hfhref{BR_-511_-421+-313.html}{ \hfpdg{<0.180} }  
 &  { \,}  
 &  { \,}  
 &  { \,}  
 &  \hfhref{BR_-511_-421+-313.html}{ \hfavg{<0.110} } 
\\%

{ \hflabel{\,} }  
 &  { \,}  
 &  \hfhref{0506085.html}{ \hfpubold{<0.180} }  
 &  \hfhref{0506023.html}{ \hfsuperceeded{<0.41} } 
\hffootnotemark{2c}
 
 &  { \,}  
 &  { \,} 
\\%

{ \hflabel{\,} }  
 &  { \,}  
 &  { \,}  
 &  \hfhref{0603004.html}{ \hfpubhot{<0.110} } 
\hffootnotemark{1c}
 
 &  { \,}  
 &  { \,} 
\\%

 \hfhref{BR_-511_-421+-321+211.html}{ \hflabel{\bar{D}^0 K^- \pi^+} }  
 &  \hfhref{BR_-511_-421+-321+211.html}{ \hfpdg{<0.190} }  
 &  { \,}  
 &  \hfhref{0506018.html}{ \hfpubhot{<0.190} }  
 &  { \,}  
 &  \hfhref{BR_-511_-421+-321+211.html}{ \hfavg{<0.190} } 
\\%

 \hfhref{BR_-511_-313+421.html}{ \hflabel{D^0 \bar{K}^{*0}(892)} }  
 &  \hfhref{BR_-511_-313+421.html}{ \hfpdg{0.53\pm0.08} }  
 &  { \,}  
 &  { \,}  
 &  { \,}  
 &  \hfhref{BR_-511_-313+421.html}{ \hfavg{0.42\pm0.06} } 
\\%

{ \hflabel{\,} }  
 &  { \,}  
 &  \hfhref{0506085.html}{ \hfpubold{0.48\pm^{0.11}_{0.10}\pm0.05} }  
 &  \hfhref{0506023.html}{ \hfsuperceeded{0.62\pm0.14\pm0.06} } 
\hffootnotemark{2b}
 
 &  { \,}  
 &  { \,} 
\\%

{ \hflabel{\,} }  
 &  { \,}  
 &  { \,}  
 &  \hfhref{0603004.html}{ \hfpubhot{0.40\pm0.07\pm0.03} } 
\hffootnotemark{1b}
 
 &  { \,}  
 &  { \,} 
\\%

 \hfhref{BR_-511_-311+421.html}{ \hflabel{D^0 \bar{K}^0} }  
 &  \hfhref{BR_-511_-311+421.html}{ \hfpdg{0.50\pm0.14} }  
 &  { \,}  
 &  { \,}  
 &  { \,}  
 &  \hfhref{BR_-511_-311+421.html}{ \hfavg{0.52\pm0.07} } 
\\%

{ \hflabel{\,} }  
 &  { \,}  
 &  \hfhref{0506085.html}{ \hfpubold{0.50\pm^{0.13}_{0.12}\pm0.06} }  
 &  \hfhref{0506023.html}{ \hfsuperceeded{0.62\pm0.12\pm0.04} } 
\hffootnotemark{2a}
 
 &  { \,}  
 &  { \,} 
\\%

{ \hflabel{\,} }  
 &  { \,}  
 &  { \,}  
 &  \hfhref{0603004.html}{ \hfpubhot{0.53\pm0.07\pm0.03} } 
\hffootnotemark{1a}
 
 &  { \,}  
 &  { \,} 
\\%

 \hfhref{BR_-511_-321+211+421.html}{ \hflabel{D^0 K^- \pi^+} }  
 &  \hfhref{BR_-511_-321+211+421.html}{ \hfpdg{0.88\pm0.17} }  
 &  { \,}  
 &  \hfhref{0506018.html}{ \hfpubhot{0.88\pm0.15\pm0.09} }  
 &  { \,}  
 &  \hfhref{BR_-511_-321+211+421.html}{ \hfavg{0.88\pm0.17} } 
\\%

 \hfhref{BR_-511_331+421.html}{ \hflabel{D^0 \eta^\prime(958)} }  
 &  \hfhref{BR_-511_331+421.html}{ \hfpdg{1.25\pm0.23} }  
 &  \hfhref{0506035.html}{ \hfpub{1.14\pm0.20\pm^{0.10}_{0.13}} }  
 &  \hfhref{0506106.html}{ \hfpub{1.70\pm0.40\pm0.18\pm0.10} }  
 &  { \,}  
 &  \hfhref{BR_-511_331+421.html}{ \hfavg{1.26\pm0.21} } 
\\%

 \hfhref{BR_-511_225+421.html}{ \hflabel{f_2(1270) D^0} }  
 &  { \,}  
 &  \hfhref{0506070.html}{ \hfpre{1.95\pm0.34\pm0.38\pm^{0.32}_{0.02}} }  
 &  { \,}  
 &  { \,}  
 &  \hfhref{BR_-511_225+421.html}{ \hfavg{1.95\pm^{0.60}_{0.51}} } 
\\%

 \hfhref{BR_-511_221+421.html}{ \hflabel{D^0 \eta} }  
 &  \hfhref{BR_-511_221+421.html}{ \hfpdg{2.20\pm0.50} }  
 &  \hfhref{0607017.html}{ \hfpubhot{1.77\pm0.16\pm0.21} }  
 &  \hfhref{0506106.html}{ \hfpub{2.50\pm0.20\pm0.29\pm0.11} }  
 &  { \,}  
 &  \hfhref{BR_-511_221+421.html}{ \hfnewavg{2.02\pm0.21} } 
\\%

 \hfhref{BR_-511_-321+411.html}{ \hflabel{D^+ K^-} }  
 &  \hfhref{BR_-511_-321+411.html}{ \hfpdg{2.00\pm0.60} }  
 &  \hfhref{0506081.html}{ \hfpubold{2.04\pm0.45\pm0.21\pm0.27} }  
 &  { \,}  
 &  { \,}  
 &  \hfhref{BR_-511_-321+411.html}{ \hfavg{2.04\pm0.57} } 
\\%

 \hfhref{BR_-511_223+421.html}{ \hflabel{D^0 \omega(782)} }  
 &  \hfhref{BR_-511_223+421.html}{ \hfpdg{2.50\pm0.60} }  
 &  \hfhref{0607017.html}{ \hfpubhot{2.37\pm0.23\pm0.28} }  
 &  \hfhref{0506106.html}{ \hfpub{3.00\pm0.30\pm0.38\pm0.13} }  
 &  { \,}  
 &  \hfhref{BR_-511_223+421.html}{ \hfnewavg{2.59\pm0.29} } 
\\%

 \hfhref{BR_-511_111+421.html}{ \hflabel{D^0 \pi^0} }  
 &  \hfhref{BR_-511_111+421.html}{ \hfpdg{2.91\pm0.28} }  
 &  \hfhref{0607017.html}{ \hfpubhot{2.25\pm0.14\pm0.35} }  
 &  \hfhref{0506106.html}{ \hfpub{2.90\pm0.20\pm0.27\pm0.13} }  
 &  { \,}  
 &  \hfhref{BR_-511_111+421.html}{ \hfnewavg{2.59\pm0.26} } 
\\%

 \hfhref{BR_-511_113+421.html}{ \hflabel{D^0 \rho^0(770)} }  
 &  \hfhref{BR_-511_113+421.html}{ \hfpdg{2.9\pm1.1} }  
 &  { \,}  
 &  { \,}  
 &  { \,}  
 &  \hfhref{BR_-511_113+421.html}{ \hfavg{2.91\pm^{0.58}_{0.40}} } 
\\%

{ \hflabel{\,} }  
 &  { \,}  
 &  \hfhref{0506088.html}{ \hfpubold{2.90\pm1.00\pm0.40} } 
\hffootnotemark{3}
 
 &  { \,}  
 &  { \,}  
 &  { \,} 
\\%

{ \hflabel{\,} }  
 &  { \,}  
 &  \hfhref{0506070.html}{ \hfpre{2.91\pm0.28\pm0.33\pm^{0.08}_{0.54}} } 
\hffootnotemark{4}
 
 &  { \,}  
 &  { \,}  
 &  { \,} 
\\%

 \hfhref{BR_-511_-321+311+411.html}{ \hflabel{D^+ K^- K^0} }  
 &  \hfhref{BR_-511_-321+311+411.html}{ \hfpdg{<3.1} }  
 &  \hfhref{0506090.html}{ \hfpubold{<3.1} }  
 &  { \,}  
 &  { \,}  
 &  \hfhref{BR_-511_-321+311+411.html}{ \hfavg{<3.1} } 
\\%

 \hfhref{BR_-511_-323+411.html}{ \hflabel{D^+ K^{*-}(892)} }  
 &  \hfhref{BR_-511_-323+411.html}{ \hfpdg{4.50\pm0.70} }  
 &  { \,}  
 &  \hfhref{0506026.html}{ \hfpub{4.60\pm0.60\pm0.47\pm0.16} }  
 &  { \,}  
 &  \hfhref{BR_-511_-323+411.html}{ \hfavg{4.60\pm0.78} } 
\\%

 \hfhref{BR_-511_-211+311+411.html}{ \hflabel{D^+ K^0 \pi^-} }  
 &  \hfhref{BR_-511_-211+311+411.html}{ \hfpdg{4.90\pm0.90} }  
 &  { \,}  
 &  \hfhref{0506026.html}{ \hfpub{4.90\pm0.70\pm0.38\pm0.32} }  
 &  { \,}  
 &  \hfhref{BR_-511_-211+311+411.html}{ \hfavg{4.90\pm0.86} } 
\\%

 \hfhref{BR_-511_-321+313+411.html}{ \hflabel{D^+ K^- K^{*0}(892)} }  
 &  \hfhref{BR_-511_-321+313+411.html}{ \hfpdg{8.8\pm1.9} }  
 &  \hfhref{0506090.html}{ \hfpubold{8.8\pm1.1\pm1.5} }  
 &  { \,}  
 &  { \,}  
 &  \hfhref{BR_-511_-321+313+411.html}{ \hfavg{8.8\pm1.9} } 
\\%

 \hfhref{BR_-511_-211+211+421.html}{ \hflabel{D^0 \pi^+ \pi^-} }  
 &  \hfhref{BR_-511_-211+211+421.html}{ \hfpdg{8.0\pm1.6} }  
 &  { \,}  
 &  { \,}  
 &  { \,}  
 &  \hfhref{BR_-511_-211+211+421.html}{ \hfavg{9.78\pm0.95} } 
\\%

{ \hflabel{\,} }  
 &  { \,}  
 &  \hfhref{0506088.html}{ \hfpubold{8.00\pm0.60\pm1.50} } 
\hffootnotemark{3}
 
 &  { \,}  
 &  { \,}  
 &  { \,} 
\\%

{ \hflabel{\,} }  
 &  { \,}  
 &  \hfhref{0506070.html}{ \hfpre{10.70\pm0.60\pm1.00} } 
\hffootnotemark{4}
 
 &  { \,}  
 &  { \,}  
 &  { \,} 
\\%

 \hfhref{BR_-511_-211+411.html}{ \hflabel{D^+ \pi^-} }  
 &  \hfhref{BR_-511_-211+411.html}{ \hfpdg{34.0\pm9.0} }  
 &  { \,}  
 &  { \,}  
 &  { \,}  
 &  \hfhref{BR_-511_-211+411.html}{ \hfnewavg{26.5\pm1.5} } 
\\%

{ \hflabel{\,} }  
 &  { \,}  
 &  { \,}  
 &  \hfhref{0607012.html}{ \hfpubhot{25.50\pm0.50\pm1.60\pm0.10} } 
\hffootnotemark{6}
 
 &  { \,}  
 &  { \,} 
\\%

{ \hflabel{\,} } \hftableblcell 
 &  { \,}  
 &  { \,}  
 &  \hfhref{0607016.html}{ \hfpubhot{30.3\pm2.3\pm2.3} } 
\hffootnotemark{5}
 
 &  { \,}  
 &  { \,} 
\\\hline 

\end{tabular}

\begin{description}\addtolength{\itemsep}{\hffootitemsep\baselineskip}

	  \item[ \hffootnotemark{1} ] \hffootnotetext{ A study of the  $\bar{B}^0 \rightarrow D^{(*)0} K^{(*)0}$ decays (\hfbb{226}) } 
     ;  \hffootnotemark{1a}  \hffootnotetext{ $\bar{B}^0 \rightarrow D^{0} \bar{K}^{0}$ }  ;  \hffootnotemark{1b}  \hffootnotetext{ $\bar{B}^0 \rightarrow D^{0} \bar{K}^{*0}$ }  ;  \hffootnotemark{1c}  \hffootnotetext{ $\bar{B}^0 \rightarrow \bar{D}^{0} \bar{K}^{*0}$ } 
	  \item[ \hffootnotemark{2} ] \hffootnotetext{ A study of the  $\bar{B}^0 \rightarrow D^{(*)0} K^{(*)0}$ decays (\hfbb{124}) } 
     ;  \hffootnotemark{2a}  \hffootnotetext{ $B \rightarrow D^{0} \bar{K}^{0}$ }  ;  \hffootnotemark{2b}  \hffootnotetext{ $\bar{B}^0 \rightarrow D^{0} K^{*0}$ }  ;  \hffootnotemark{2c}  \hffootnotetext{ $\bar{B}^0 \rightarrow \bar{D}^{0} \bar{K}^{*0}$ } 
	  \item[ \hffootnotemark{3} ] \hffootnotetext{ Study of $\bar{B^{0}} \to D^{(*) 0} \pi^+ \pi^-$ Decays (\hfbb{31.3}) } 
    
	  \item[ \hffootnotemark{4} ] \hffootnotetext{ Study of $\bar{B}^{0} \to  D^{(*)0} \pi^{+} \pi^{-}$ decays ; Dalitz fit analysis (\hfbb{152}) } 
    
	  \item[ \hffootnotemark{5} ] \hffootnotetext{ Measurement of the Absolute Branching Fractions $B \rightarrow D^{(*,**)}\pi$ with a Missing Mass method (\hfbb{231}) } 
     ;  \hffootnotetext{ $\bar{B}^0 \rightarrow D^+\pi^-$ } 
	  \item[ \hffootnotemark{6} ] \hffootnotetext{ Branching fraction measurements and isospin analyses for $\bar{B} \rightarrow D^{(*)}\pi^-$
decays (\hfbb{65}) } 
     ;  \hffootnotetext{ $\bar{B}^0 \rightarrow D^+\pi^-$ } 
\end{description}

\end{center}

\end{\hftabletype}
\clearpage  
\begin{\hftabletype}[\hftableposn]

\begin{center}
\hfnewcaption{Product branching fractions}{neutral B}{\hfsgld}{in units of $10^{-5}$}{upper limits are at 90\% CL}{00208.html}{tab:hfc02208}
\hfmetadata{ [ .tml created 2007-04-07T02:09:47.858+08:00] }

\begin{tabular}{lccccc}  
\hflmode
&  \hflpdg2006
&  \hflbelle
&  \hflbabar
&  \hflcdf
&  \hflavg
\\\hline 

 \hfhref{BR_-511_-313+421xBR_-313_-321+211.html}{ \hflabel{D^0 \bar{K}^{*0}(892) [ K^- \pi^+ ]} } \hftabletlcell\hftableblcell 
 &  { \,}  
 &  { \,}  
 &  \hfhref{0506018.html}{ \hfpubhot{3.80\pm0.60\pm0.40} }  
 &  { \,}  
 &  \hfhref{BR_-511_-313+421xBR_-313_-321+211.html}{ \hfavg{3.80\pm0.72} } 
\\\hline 

\end{tabular}

\end{center}

\end{\hftabletype}
\clearpage  
\begin{\hftabletype}[\hftableposn]

\begin{center}
\hfnewcaption{Branching fractions}{miscellaneous}{charmed particles }{in units of $10^{-3}$}{upper limits are at 90\% CL}{00300.html}{tab:hfc01300}
\hfmetadata{ [ .tml created 2007-04-07T02:24:42.198+08:00] }

\begin{tabular}{lccccc}  
\hflmode
&  \hflpdg2006
&  \hflbelle
&  \hflbabar
&  \hflcdf
&  \hflavg
\\\hline 

 \hfhref{BR_97_-421+96+111+421.html}{ \hflabel{{\cal{B}} ( B \to D^0 \bar{D}^0 \pi^0 K )} } \hftabletlcell 
 &  { \,}  
 &  \hfhref{0607010.html}{ \hfpubhot{0.13\pm0.03\pm^{0.02}_{0.04}} }  
 &  { \,}  
 &  { \,}  
 &  \hfhref{BR_97_-421+96+111+421.html}{ \hfnewavg{0.13\pm0.04} } 
\\%

 \hfhref{BR_-5122_-3122+443.html}{ \hflabel{{\cal{B}} ( \bar{\Lambda}_b^0 \to J/\psi(1S) \bar{\Lambda} )} }  
 &  \hfhref{BR_-5122_-3122+443.html}{ \hfpdg{0.47\pm0.28} }  
 &  { \,}  
 &  { \,}  
 &  \hfhref{0506097.html}{ \hfpubold{0.47\pm0.21\pm0.19} }  
 &  \hfhref{BR_-5122_-3122+443.html}{ \hfavg{0.47\pm0.28} } 
\\%

 \hfhref{BR_-421_-411+423.html}{ \hflabel{{\cal{B}} ( \bar{D}^0 \to D^{*0}(2007) D^- )} }  
 &  { \,}  
 &  { \,}  
 &  \hfhref{0604002.html}{ \hfpubhot{0.63\pm0.14\pm0.08\pm0.06} }  
 &  { \,}  
 &  \hfhref{BR_-421_-411+423.html}{ \hfavg{0.63\pm0.17} } 
\\%

 \hfhref{BR_-531_333+443.html}{ \hflabel{{\cal{B}} ( \bar{B}_s^0 \to J/\psi(1S) \phi(1020) )} } \hftableblcell 
 &  \hfhref{BR_-531_333+443.html}{ \hfpdg{0.93\pm0.33} }  
 &  { \,}  
 &  { \,}  
 &  \hfhref{0506094.html}{ \hfpubold{0.93\pm0.28\pm0.17} }  
 &  \hfhref{BR_-531_333+443.html}{ \hfavg{0.93\pm0.33} } 
\\\hline 

\end{tabular}

\end{center}


\begin{center}
\hfnewcaption{Product branching fractions}{miscellaneous}{charmed particles }{in units of $10^{-5}$}{upper limits are at 90\% CL}{00300.html}{tab:hfc02300}
\hfmetadata{ [ .tml created 2007-04-07T02:24:49.681+08:00] }

\begin{tabular}{lccccc}  
\hflmode
&  \hflpdg2006
&  \hflbelle
&  \hflbabar
&  \hflcdf
&  \hflavg
\\\hline 

 \hfhref{BR_97_83+96xBR_83_223+443.html}{ \hflabel{{\cal{B}} ( B \to K Y(3940) [ \omega(782) J/\psi(1S) ] )} } \hftabletlcell\hftableblcell 
 &  \hfhref{BR_97_83+96xBR_83_223+443.html}{ \hfpdg{7.1\pm3.4} }  
 &  \hfhref{0506033.html}{ \hfpub{7.1\pm1.3\pm3.1} }  
 &  { \,}  
 &  { \,}  
 &  \hfhref{BR_97_83+96xBR_83_223+443.html}{ \hfavg{7.1\pm3.4} } 
\\\hline 

\end{tabular}

\end{center}


\begin{center}
\hfnewcaption{Ratios of branching fractions}{miscellaneous}{charmed particles }{in units of $10^{0}$}{upper limits are at 90\% CL}{00300.html}{tab:hfc03300}
\hfmetadata{ [ .tml created 2007-04-07T02:25:01.657+08:00] }

\begin{tabular}{lccccc}  
\hflmode
&  \hflpdg2006
&  \hflbelle
&  \hflbabar
&  \hflcdf
&  \hflavg
\\\hline 

 \hfhref{BR_-531_333+100443xoBR_-531_333+443.html}{ \hflabel{\frac{{\cal{B}} ( \bar{B}_s^0 \to \psi(2S) \phi(1020) )}{{\cal{B}} ( \bar{B}_s^0 \to J/\psi(1S) \phi(1020) )}} } \hftabletlcell 
 &  { \,}  
 &  { \,}  
 &  { \,}  
 &  \hfhref{0506101.html}{ \hfpubhot{0.52\pm0.13\pm0.07} }  
 &  \hfhref{BR_-531_333+100443xoBR_-531_333+443.html}{ \hfavg{0.52\pm0.15} } 
\\%

 \hfhref{BR_-531_-211+-211+211+431xoBR_-511_-211+-211+211+411.html}{ \hflabel{\frac{{\cal{B}} ( \bar{B}_s^0 \to D_s^+ \pi^+ \pi^- \pi^- )}{{\cal{B}} ( \bar{B}^0 \to D^+ \pi^+ \pi^- \pi^- )}} }  
 &  { \,}  
 &  { \,}  
 &  { \,}  
 &  \hfhref{0506098.html}{ \hfpubhot{1.05\pm0.10\pm0.23} }  
 &  \hfhref{BR_-531_-211+-211+211+431xoBR_-511_-211+-211+211+411.html}{ \hfavg{1.05\pm0.25} } 
\\%

 \hfhref{BR_-531_-211+431xoBR_-511_-211+411.html}{ \hflabel{\frac{{\cal{B}} ( \bar{B}_s^0 \to D_s^+ \pi^- )}{{\cal{B}} ( \bar{B}^0 \to D^+ \pi^- )}} }  
 &  { \,}  
 &  { \,}  
 &  { \,}  
 &  \hfhref{0601001.html}{ \hfpubhot{1.13\pm0.08\pm0.05\pm0.15} }  
 &  \hfhref{BR_-531_-211+431xoBR_-511_-211+411.html}{ \hfavg{1.13\pm0.18} } 
\\%

 \hfhref{BR_-531_-431+431xoBR_-511_-431+411.html}{ \hflabel{\frac{{\cal{B}} ( \bar{B}_s^0 \to D_s^- D_s^+ )}{{\cal{B}} ( \bar{B}^0 \to D_s^- D^+ )}} }  
 &  { \,}  
 &  { \,}  
 &  { \,}  
 &  \hfhref{0601004.html}{ \hfwaiting{1.67\pm0.41\pm0.12\pm0.46} }  
 &  \hfhref{BR_-531_-431+431xoBR_-511_-431+411.html}{ \hfavg{1.67\pm0.63} } 
\\%

 \hfhref{BR_-5122_-4122+211xoBR_-511_-211+411.html}{ \hflabel{\frac{{\cal{B}} ( \bar{\Lambda}_b^0 \to \Lambda_c^- \pi^+ )}{{\cal{B}} ( \bar{B}^0 \to D^+ \pi^- )}} }  
 &  { \,}  
 &  { \,}  
 &  { \,}  
 &  \hfhref{0506103.html}{ \hfpubhot{3.30\pm0.30\pm0.40\pm1.10} }  
 &  \hfhref{BR_-5122_-4122+211xoBR_-511_-211+411.html}{ \hfavg{3.3\pm1.2} } 
\\%

 \hfhref{BR_-5122_-4122+-13+14xoBR_-5122_-4122+211.html}{ \hflabel{\frac{{\cal{B}} ( \bar{\Lambda}_b^0 \to \Lambda_c^- \mu^+ \nu_\mu )}{{\cal{B}} ( \bar{\Lambda}_b^0 \to \Lambda_c^- \pi^+ )}} } \hftableblcell 
 &  { \,}  
 &  { \,}  
 &  { \,}  
 &  \hfhref{0506102.html}{ \hfwaiting{20.00\pm3.00\pm1.20\pm^{0.90}_{2.20}} }  
 &  \hfhref{BR_-5122_-4122+-13+14xoBR_-5122_-4122+211.html}{ \hfavg{20.0\pm^{3.4}_{3.9}} } 
\\\hline 

\end{tabular}

\end{center}

\end{\hftabletype}
\clearpage  
\begin{\hftabletype}[\hftableposn]

\begin{center}
\hfnewcaption{Miscellaneous quantities}{miscellaneous}{charmed particles }{in units of $10^{0}$}{upper limits are at 90\% CL}{00300.html}{tab:hfc04300}
\hfmetadata{ [ .tml created 2007-04-07T02:25:12.515+08:00] }

\begin{tabular}{lccccc}  
\hflmode
&  \hflpdg2006
&  \hflbelle
&  \hflbabar
&  \hflcdf
&  \hflavg
\\\hline 

 \hfhref{Deltapara_97_94+443.html}{ \hflabel{\delta_{\parallel} ( B \to J/\psi(1S) K^{*} )} } \hftabletlcell 
 &  { \,}  
 &  \hfhref{0611001.html}{ \hfpub{-2.887\pm0.090\pm0.008} }  
 &  \hfhref{0607013.html}{ \hfprehot{-2.93\pm0.08\pm0.04} }  
 &  { \,}  
 &  \hfhref{Deltapara_97_94+443.html}{ \hfnewavg{-2.91\pm0.06} } 
\\%

 \hfhref{Deltapara_97_94+100443.html}{ \hflabel{\delta_{\parallel} ( B \to \psi(2S) K^{*} )} }  
 &  { \,}  
 &  { \,}  
 &  \hfhref{0607013.html}{ \hfprehot{-2.80\pm0.40\pm0.10} }  
 &  { \,}  
 &  \hfhref{Deltapara_97_94+100443.html}{ \hfnewavg{-2.80\pm0.41} } 
\\%

 \hfhref{Deltapara_97_94+20443.html}{ \hflabel{\delta_{\parallel} ( B \to \chi_{c1}(1P) K^{*} )} }  
 &  { \,}  
 &  { \,}  
 &  \hfhref{0607013.html}{ \hfprehot{0.00\pm0.30\pm0.10} }  
 &  { \,}  
 &  \hfhref{Deltapara_97_94+20443.html}{ \hfnewavg{0.00\pm0.32} } 
\\%

 \hfhref{Aperp_97_94+20443.html}{ \hflabel{\vert{\cal{A}}_{\perp}\vert^{2} ( B \to \chi_{c1}(1P) K^{*} )} }  
 &  { \,}  
 &  { \,}  
 &  \hfhref{0607013.html}{ \hfprehot{0.03\pm0.04\pm0.02} }  
 &  { \,}  
 &  \hfhref{Aperp_97_94+20443.html}{ \hfnewavg{0.03\pm0.04} } 
\\%

 \hfhref{Apara_97_94+20443.html}{ \hflabel{\vert{\cal{A}}_{\parallel}\vert^{2} ( B \to \chi_{c1}(1P) K^{*} )} }  
 &  { \,}  
 &  { \,}  
 &  \hfhref{0607013.html}{ \hfprehot{0.20\pm0.07\pm0.04} }  
 &  { \,}  
 &  \hfhref{Apara_97_94+20443.html}{ \hfnewavg{0.20\pm0.08} } 
\\%

 \hfhref{Aperp_97_94+443.html}{ \hflabel{\vert{\cal{A}}_{\perp}\vert^{2} ( B \to J/\psi(1S) K^{*} )} }  
 &  { \,}  
 &  \hfhref{0611001.html}{ \hfpub{0.195\pm0.012\pm0.008} }  
 &  \hfhref{0607013.html}{ \hfprehot{0.233\pm0.010\pm0.005} }  
 &  { \,}  
 &  \hfhref{Aperp_97_94+443.html}{ \hfnewavg{0.219\pm0.009} } 
\\%

 \hfhref{Apara_97_94+443.html}{ \hflabel{\vert{\cal{A}}_{\parallel}\vert^{2} ( B \to J/\psi(1S) K^{*} )} }  
 &  { \,}  
 &  \hfhref{0611001.html}{ \hfpub{0.231\pm0.012\pm0.008} }  
 &  \hfhref{0607013.html}{ \hfprehot{0.211\pm0.010\pm0.006} }  
 &  { \,}  
 &  \hfhref{Apara_97_94+443.html}{ \hfnewavg{0.219\pm0.009} } 
\\%

 \hfhref{Apara_97_94+100443.html}{ \hflabel{\vert{\cal{A}}_{\parallel}\vert^{2} ( B \to \psi(2S) K^{*} )} }  
 &  { \,}  
 &  { \,}  
 &  \hfhref{0607013.html}{ \hfprehot{0.22\pm0.06\pm0.02} }  
 &  { \,}  
 &  \hfhref{Apara_97_94+100443.html}{ \hfnewavg{0.22\pm0.06} } 
\\%

 \hfhref{Aperp_97_94+100443.html}{ \hflabel{\vert{\cal{A}}_{\perp}\vert^{2} ( B \to \psi(2S) K^{*} )} }  
 &  { \,}  
 &  { \,}  
 &  \hfhref{0607013.html}{ \hfprehot{0.30\pm0.06\pm0.02} }  
 &  { \,}  
 &  \hfhref{Aperp_97_94+100443.html}{ \hfnewavg{0.30\pm0.06} } 
\\%

 \hfhref{Azero_97_94+100443.html}{ \hflabel{\vert{\cal{A}}_{0}\vert^{2} ( B \to \psi(2S) K^{*} )} }  
 &  { \,}  
 &  { \,}  
 &  \hfhref{0607013.html}{ \hfprehot{0.48\pm0.05\pm0.02} }  
 &  { \,}  
 &  \hfhref{Azero_97_94+100443.html}{ \hfnewavg{0.48\pm0.05} } 
\\%

 \hfhref{Azero_97_94+443.html}{ \hflabel{\vert{\cal{A}}_{0}\vert^{2} ( B \to J/\psi(1S) K^{*} )} }  
 &  { \,}  
 &  \hfhref{0611001.html}{ \hfpub{0.574\pm0.012\pm0.009} }  
 &  \hfhref{0607013.html}{ \hfprehot{0.556\pm0.009\pm0.010} }  
 &  { \,}  
 &  \hfhref{Azero_97_94+443.html}{ \hfnewavg{0.56\pm0.01} } 
\\%

 \hfhref{Azero_97_94+20443.html}{ \hflabel{\vert{\cal{A}}_{0}\vert^{2} ( B \to \chi_{c1}(1P) K^{*} )} }  
 &  { \,}  
 &  { \,}  
 &  \hfhref{0607013.html}{ \hfprehot{0.77\pm0.07\pm0.04} }  
 &  { \,}  
 &  \hfhref{Azero_97_94+20443.html}{ \hfnewavg{0.77\pm0.08} } 
\\%

 \hfhref{Deltaperp_97_94+100443.html}{ \hflabel{\delta_{\perp} ( B \to \psi(2S) K^{*} )} }  
 &  { \,}  
 &  { \,}  
 &  \hfhref{0607013.html}{ \hfprehot{2.80\pm0.30\pm0.10} }  
 &  { \,}  
 &  \hfhref{Deltaperp_97_94+100443.html}{ \hfnewavg{2.80\pm0.32} } 
\\%

 \hfhref{Deltaperp_97_94+443.html}{ \hflabel{\delta_{\perp} ( B \to J/\psi(1S) K^{*} )} } \hftableblcell 
 &  { \,}  
 &  \hfhref{0611001.html}{ \hfpub{2.938\pm0.064\pm0.010} }  
 &  \hfhref{0607013.html}{ \hfprehot{2.91\pm0.05\pm0.03} }  
 &  { \,}  
 &  \hfhref{Deltaperp_97_94+443.html}{ \hfnewavg{2.92\pm0.04} } 
\\\hline 

\end{tabular}

\end{center}

\end{\hftabletype}
\clearpage  

%% file: hfag_summary.tex
\section{Summary}
\label{sec:summary}

This article provides the updated world averages for 
$b$-hadron properties as of at the end of 2006.
A small selection of highlights of the results described in Sections
\ref{sec:life_mix}-\ref{sec:BtoCharm} is given in 
Table~\ref{tab_summary}.

\begin{table}
\caption{ Brief summary of the world averages as of at the end of 2006.}
\label{tab_summary}
\begin{center}
\begin{tabular}{|l|c|}
\hline
 {\bf\boldmath \b-hadron lifetimes} &   \\
 ~~$\tau(\Bd)$  & \hfagTAUBD \\
 ~~$\tau(\Bu)$  & \hfagTAUBU \\
 ~~$\tau(\Bs\to~\mbox{flavour specific})$  & \hfagTAUBSSL \\
 ~~$\bar{\tau}(\Bs) = 1/\Gs$  & \hfagTAUBSMEANCON \\
 ~~$\tau(\Bc)$  & \hfagTAUBC \\
 ~~$\tau(\Lb)$  & \hfagTAULB \\
\hline
 {\bf\boldmath \b-hadron fractions} &   \\
 ~~$f^{+-}/f^{00}$ in \Ups decays  & \hfagFF \\ 
 ~~\fBs in \Upsfive decays & \hfagFSFIVE \\
 ~~$\fBd=\fBu$ at high energy & \hfagWFBD \\
 ~~\fBs at high energy  & \hfagWFBS \\
 ~~\fbb at high energy  & \hfagWFBB \\
\hline
 {\bf\boldmath \Bd\ and \Bs\ mixing parameters} &   \\
 ~~\dmd &  \hfagDMDWU \\
 ~~$|q/p|_{\particle{d}}$ & \hfagQPDB  \\
 ~~\dms  &  \hfagDMS \\
 ~~$\DGGs = (\Gamma_{\rm L} - \Gamma_{\rm H})/\Gs$ & \hfagDGSGSCON \\
 ~~$|q/p|_{\particle{s}}$ & \hfagQPS   \\
\hline
 {\bf Measurements related to Unitarity Triangle angles} & \\
 ~~ $\stwob \equiv \sin\! 2\phi_1$ & $0.675 \pm 0.026$ \\
 ~~ $\beta \equiv \phi_1$          & $\left( 21.2 \pm 1.0 \right)^\circ$ \\
 ~~ $-\etacp S_{\etapr \Kz}$       & $0.61 \pm 0.07$ \\
 ~~ $S_{\pi^+\pi^-}$               & $-0.59 \pm 0.09$ \\  
\hline
 {\bf\boldmath Semileptonic \B decay parameters} &   \\
 ~~${\cal B}(\BzbDstarlnu)$  & $( 5.28 \pm 0.18)\%$ \\
 ~~${\cal B}(\BzbDplnu)$      & $( 2.09 \pm 0.18)\%$ \\
 ~~${\cal B}(\Bb\to  X\ell\nub)$   & $(10.95 \pm 0.15)\%$ \\
 ~~$|V_{\particle{cb}}|F(1)\ (\BzbDstarlnu)$   
       & $ (36.0 \pm 0.6) \times 10^{-3}$ \\
 ~~$|V_{\particle{cb}}|G(1) \ (\BzbDplnu)$   
       & $(42.4 \pm 4.5) \times 10^{-3}$ \\
 ~~${\cal B}(\Bb\to  \pi\ell\nub)$   & $(1.39 \pm 0.08) \times 10^{-4}$ \\
\hline
 {\bf\boldmath Rare \B decays} &   \\
~~${\cal B}(\Bp\to\tau^+\nu)$ & $ (132 \pm 49)\times 10^{-6}$ \\
\hline
\end{tabular}
\end{center}
\end{table}

Concerning the lifetime and mixing averages, the most significant changes
since the end of 2005~\cite{hfag_hepex_endof2005}
are due to new measurements from the Tevatron experiments, 
mainly in the areas of heavy \b-baryon lifetimes and \Bs mixing
parameters. After more than a decade of effort at LEP, SLC, and Tevatron, 
\Bs oscillations have been observed by CDF with a frequency 
in agreement with the Standard Model prediction. First 
results on \CP violation in \Bs mixing have also been obtained 
from Tevatron data, 
consistent with no \CP violation. In the mean time, \Bs production
at the \Upsfive has been established. 

Measurements by \babar and \belle
of the time-dependent \CP violation parameter
$S_{b \to c\bar c s}$ in \B decays to charmonium and a neutral kaon
have established \CP violation in \B decays,
and allow a precise extraction of
the Unitarity Triangle parameter $\stwob \equiv \sin\! 2\phi_1$.
Recent studies of $\B \to \jpsi \Kstar$ (Sec.~\ref{sec:cp_uta:ccs:vv}),
$\B \to D^{(*)}h^0$, where $h^0 = \pi^0$ \etc. (Sec.~\ref{sec:cp_uta:cud_beta})
and $\B \to D^{*+}D^{*-}\KS$ (Sec.~\ref{sec:cp_uta:ccs:DstarDstarKs})
allow to resolve the ambiguity in the solutions for $\beta \equiv \phi_1$ from
the measurement of $\stwob \equiv \sin\! 2\phi_1$.
Measurements of
time-dependent \CP asymmetries in hadronic $b \to s$ penguin decays
continue to provide insight into possible new physics.
In this update, results from both \babar and \belle have been updated.
A particularly notable change is that the \CP violation effect in 
$\B \to \eta^\prime K^0$ is now established 
with more than $5\sigma$ significance in both experiments.
Also noteworthy is that \babar has performed the first time-dependent
Dalitz plot analysis of the $B \to K^+K^-K^0$ decay,
and quasi-two-body \CP violation parameters for the decays 
$B \to \phi K^0$ and $B \to f_0 K^0$ are obtained from this analysis.
Compared to the previous round of averages,
the consistency with the Standard Model expectation remains at about 
the same level, in terms of significance.
Results from time-dependent analyses
with the decays $\Bz \to \pi^+\pi^-, \rho^\pm\pi^\mp$ and 
$\rho^+\rho^-$ provide constraints on the 
Unitarity Triangle angle $\alpha \equiv \phi_2$ (Sec.~\ref{sec:cp_uta:uud}).
Both \babar\ and \belle\ now observed \CP violation in 
$\Bz \to \pi^+\pi^-$ with more than $5\sigma$ significance,
and both experiments have now performed time-dependent Dalitz plot analyses
of $\Bz \to (\rho\pi)^0 \to \pi^+\pi^-\pi^0$.
Progress continues to constrain the third Unitarity Triangle angle 
$\gamma \equiv \phi_3$.
Both \babar and \belle are 
using $\Bm \to \DorDstar \Km$ decays (Sec.~\ref{sec:cp_uta:cus}),
with multiple $\DorDstar$ decays to final states accessible to both
$D^{(*)0}$ and $\bar{D}^{(*)0}$.
At present, the most constraining results
arise from the Dalitz plot analysis of the $D \to \KS\pi^+\pi^-$ decay.


Progress in the determination of properties of semileptonic
$B$-meson decays has been steady over the last year.
\babar made new measurements of the $D^*$ form factor ratios
$R_1$ and $R_2$, which were a factor of $5$ more precise than the
older CLEO measurements. They caused the value of $|V_{cb}|F(1)$ to shift by
$-1.7\sigma$ but had no impact on it's precision, which is dominated
by the branching fraction. For the first time we computed averages for
the inclusive semileptonic branching fraction of neutral and charged
$B$-mesons separately and their ratio. The latter was found to be, as
one would expect, consistent with the corresponding lifetime ratio. We
extracted \vub\ from inclusive measurements using in addition to BLNP,
DGE and BLL. The average found using the DGE method was found to be in
very good agreement with that of BLNP, while BLL, which makes use of a
subsample of the available measurements, is also in agreement with DGE
and BLNP. New and
updated measurements of the $B\to\pi l\nu$ branching fraction have
reduced the error on the average by 25\%. As was the case last year,
the precision with which \vub\ is known is limited by the uncertainty on the form
factor, which is at the level of 15--20\%. Significant strides have been made in
reducing uncertainties from the shape of the $q^2$ spectrum, with a
new measurement made in $q^2$ bins of size $2\,$GeV$^2$.      

For rare \B\ decays, branching fractions and charge asymmetries of many
new decay modes have been measured recently, mostly by \babar\ and Belle.
There are several hundred measurements in the tables in \Sec{rare}.
Particularly noteworthy is the measurement of $B\to\tau\nu$; Belle sees
evidence for this decay while \babar\ reports a somewhat smaller
branching fraction.  There are new results from \babar\ on the
vector-vector decays $B\to\phi K^*$ and $B\to\rho K^*$.  They confirm
earlier indications that the longitudinal polarization is about 0.5 for
the penguin-dominated modes, while it is $\sim$1.0 for tree-dominated
decays.  This pattern still has no theoretical explanation.
The branching fractions for inclusive charmless $B$ decays with a kaon 
and the full angular analysis of $B$ decays into a vector-tensor final state
are added for this update. 


In the sector of $B$ decays to charmed particles, reduction in the
uncertainties and new measurements have steadily continued to be
obtained. Branching fractions for rare $B$-meson decays or decay chains
of a few $10^{-7}$ are being measured with statistical uncertainties
typically below $30\%$. Results for more common decay chains, with
branching fractions around $10^{-4}$, are becoming precision
measurements, with uncertainties typically at the $3\%$ level. Some
decays have been observed for the first time, for example $B^- \to
\Lambda_c^- \Lambda_c^+ K^-$ or $B^-\to \chi_{c1} \pi^-$, with a
branching fraction of $(6.5\pm3.7)\times 10^{-4}$ and $(2.2\pm0.5)\times
10^{-5}$, respectively. Great improvements have been attained towards a
deeper understanding of recently discovered new states with either
hidden or open charm content. The $J/\psi\gamma$ decay mode of the
$X(3870)$ has been observed by both \babar and \belle and the average
branching fraction is $\hfcBR(B^-\to X(3870) K^-)\times \hfcBR(X(3870\to
J/\psi\gamma)=(2.4\pm0.5)\times 10^{-6}$: this final state allowed to
unambiguously establish the positive $C$ parity of the
$X(3870)$. Branching fractions for several $B$ decays to
$D_{sJ}^{*-}(2317)$ and $D_{sJ}^-(2460)$ have been measured, and also
absolute branching fraction measurements have been reported:
$\hfcBR(\bar{B}^0\to D_{sJ}^-(2460) D^+) = (0.26\pm0.17)\times 10^{-2}$,
$\hfcBR(\bar{B}^0\to D_{sJ}^-(2460) D^{*+}) = (0.88\pm0.24)\times
10^{-2}$, $\hfcBR(B^-\to D_{sJ}^-(2460) D^0) = (0.43\pm0.21)\times
10^{-2}$ and $\hfcBR(B^-\to D_{sJ}^-(2460) D^{*0}) = (1.12\pm0.33)\times
10^{-2}$. The abundance of measurements with many different final states
for these particles is of the greatest importance for quantum number
assignments, and already some of the proposed theoretical
interpretations have been ruled out. 

\section*{ Acknowledgments }

We would like to thank collaborators of \babar, \belle, CDF, CLEO, \dzero,
LEP, and SLD experiments who provided fruitful results on \b-hadron
properties and cooperated with the HFAG for averaging.
These results are thanks to the excellent operations of the
accelerators and collaborations with experimental groups by the
accelerator groups of PEP-II, KEKB, CESR, Tevatron, LEP, and SLC.